\documentclass[11pt]{book}
\usepackage{latexsym}
\usepackage{amsfonts}
\usepackage{a4wide} 
\usepackage{hyperref}
\usepackage{multicol}
\usepackage{epsfig}
\usepackage{color}
\usepackage{fancybox}
\usepackage{color}
\usepackage{alltt}

\definecolor{c1114}{rgb}{1,0.2156862745098,0.75294117647059}
\definecolor{c1112}{rgb}{1,0,0}
\definecolor{c1121}{rgb}{0,0,0.54509803921569}
\definecolor{c1123}{rgb}{0,0.5,0}

\definecolor{c1082}{rgb}{0.,0.,0.}
\definecolor{c1083}{rgb}{1,0,0}

\newcommand{\Proofitemm}[1]{\medskip \noindent $#1\;$}
\newcommand{\Proofitemf}[1]{\noindent $#1\;$}
\newcommand{\Defitem}[1]{\smallskip \noindent $#1\;$}

\newcommand{\mod}{ \ \w{mod} \ }

\newcommand{\hbra}{\noindent\hbox to \textwidth{\leaders\hrule height1.8mm depth-1.5mm\hfill}}
\newcommand{\hket}{\noindent\hbox to \textwidth{\leaders\hrule height0.3mm\hfill}}
\newcommand{\ratio}{.3}

\newtheorem{definition}{Definition}

\newtheorem{corollary}{Corollary}
\newtheorem{proposition}{Proposition}
\newtheorem{example}{Example}
\newtheorem{exercise}{Exercise}

\newtheorem{remark}{Remark}

\counterwithin{theorem}{chapter}
\counterwithin{fact}{chapter}
\counterwithin{definition}{chapter}
\counterwithin{lemma}{chapter}
\counterwithin{corollary}{chapter}
\counterwithin{proposition}{chapter}
\counterwithin{example}{chapter}
\counterwithin{exercise}{chapter}
\counterwithin{remark}{chapter}
\counterwithin{proviso}{chapter}
\counterwithin{solve}{chapter}

\newcommand{\Proof}{\noindent {\sc Proof}. }

\newcommand{\qed}{\hfill${\Box}$}

\newcommand{\Figbar}{{\center \rule{\hsize}{0.3mm}}}    %horizontal thiner line for figures

\newcommand{\cmp}[1]{{\cal C}(#1)}
\newcommand{\rb}[1]{{\cal R}(#1)}             % readback translation
\newcommand{\co}[1]{{\cal C}_{{\it o}}(#1)} % optimistic atomic transformation
\newcommand{\cp}[1]{{\cal C}_{{\it p}}(#1)} % pessimistic  atomic transformation

\newcommand{\aGives}{\vdash^{{\it vn}}} % type judgement in the administrative calculus

\newcommand{\cl}[1]{{\cal #1}}          % \cl{R} to make R calligraphic
\newcommand{\la}{\langle}               % the brackets for pairing (see also \pair)
\newcommand{\ra}{\rangle}

\newcommand{\ul}[1]{\underline{#1}}     % to underline
\newcommand{\ol}[1]{\overline{#1}}      % to overline
\newcommand{\Gives}{\vdash}             % in a type judgment
\newcommand{\sGives}{\vdash^{{\it syn}}}        % syntax directed
\newcommand{\lGives}{\vdash^{{\it let}}}        % let system directed
\newcommand{\piGives}{\vdash^{\pi}}             % in a type judgment
\newcommand{\sub}[2]{[#2 / #1]}         % substitution \sub{x}{U} gives [U/x]   
\newcommand{\lsub}[2]{#2 / #1}          % substitution \lsub{x}{U} gives U/x, to be used in a list.

\newcommand{\arrow}{\rightarrow}        % right thin arrow
\newcommand{\trarrow}{\stackrel{*}{\rightarrow}}        % trans closure
\newcommand{\rtrarrow}{\stackrel{*}{\rightarrow}}
\newcommand{\ltrarrow}{\stackrel{*}{\leftarrow}}
\newcommand{\lrtrarrow}{\stackrel{*}{\leftrightarrow}}
\newcommand{\Nat}{\mathbf{N}}                 % natural numbers
\newcommand{\Bool}{\mathbf{B}} % booleans
\newcommand{\Alt}{ \mid\!\!\mid  }

\newcommand{\infer}[2]{\begin{array}{c} #1 \\ \hline #2 \end{array}}
\newcommand{\hoare}[3]{\{#1\} \ #2 \ \{#3 \}}
\newcommand{\jones}[5]{#1:(#2,#3,#4,#5)}

\newcommand{\Or}{\vee}                  % disjunction
\newcommand{\OR}{\bigvee}               % big disjunction
\newcommand{\AND}{\wedge}               % conjunction
\newcommand{\Arrow}{\Rightarrow}        % right double arrow
\newcommand{\IFF}{\mbox{~~iff~~}}       % iff in roman and with spaces
\newcommand{\dl}{[\![}                  % semantic [[
\newcommand{\dr}{]\!]}                  % semantic ]]
\newcommand{\prt}[1]{{\cal P}(#1)}      % Parts of a set
\newcommand{\finprt}[1]{{\cal P}_{fin}(#1)}% Finite parts
\newcommand{\union}{\cup}               % union
\newcommand{\inter}{\cap}               % intersection
\newcommand{\Union}{\bigcup}            % big union
\newcommand{\Inter}{\bigcap}            % big intersection
\newcommand{\cpl}[1]{#1^{c}}            % complement
\newcommand{\card}{\sharp}              % cardinality
\newcommand{\minus}{\backslash}         % set difference
\newcommand{\sequence}[2]{\{#1\}_{#2}}  % ex. \sequence{d_n}{n\in \omega}
\newcommand{\comp}{\circ}               % functional composition
\newcommand{\set}[1]{\{#1\}}            % set enumeration
\newcommand{\pset}[1]{\{\! | #1 |\!\}}  % pseudo-set notation {| |}
\newcommand{\join}{\vee}                % join
\newcommand{\JOIN}{\bigvee}             % big join 
\newcommand{\MEET}{\bigwedge}           % big meet
\newcommand{\dcl}{\downarrow}           % down closure
\newcommand{\pres}[2]{(#1\triangleright #2)} %TCCS present
\newcommand{\tick}{{\sf tick} \ }          %tick action
\newcommand{\tr}[1]{{\it tr}(#1)}           %trace

\newcommand{\parrow}{\rightharpoonup}   % partial function space
\newcommand{\rel}[1]{\;{\cal #1}\;}     % infix relation (calligraphic)
\newcommand{\stable}[2]{{\cl S}(#1,#2)}  %stability of a predicate under a relation

\newcommand{\pair}[2]{\langle #1 , #2 \rangle} % pairing \pair{x}{y}, do not use < >.

\newcommand{\bv}[1]{{\sf bv}(#1)}                       % bound names
\newcommand{\ite}[3]{{\sf if~} #1 {\sf ~then~} #2 {\sf ~else~} #3}      %if then else
\newcommand{\while}[2]{{\sf while ~}#1{\sf ~do~}#2} %while
\newcommand{\parimp}{{\sf Imp}_{\parallel}}   %par imp
\newcommand{\cJ}{{\sf cJ}}
\newcommand{\sJ}{{\sf J}}
\newcommand{\var}[1]{{\sf var}(#1)} % variables
\newcommand{\fv}[1]{{\sf fv}(#1)} %free variables
\newcommand{\fvt}[1]{{\sf ftv}(#1)} %free type variables
\newcommand{\ftv}[1]{{\sf ftv}(#1)} %idem
\newcommand{\pos}[2]{\langle #1 \rangle #2} %diamond
\newcommand{\whn}[2]{[#1 ] #2}              %box
\newcommand{\await}[2]{{\sf await ~}#1{\sf ~do~}#2}
\newcommand{\awaitt}[2]{{\sf await' ~}#1{\sf ~do~}#2}
\newcommand{\atomic}[1]{{\sf atomic}(#1)}
\newcommand{\sk}{{\sf skip}}
\newcommand{\lets}[3]{{\sf let} \ #1 = #2 \ {\sf in} \ #3}    % simple let
\newcommand{\letonce}[3]{{\sf letonce} \ #1 = #2 \ {\sf in} \ #3}    % use once let
\newcommand{\letjoin}[3]{{\sf letjoin} \ (#1,#2) \ {\sf in} \ #3}    % let join
\newcommand{\cps}[1]{{\cal C}_{{\it cps}}(#1)}

\newcommand{\vn}[1]{{\cal C}_{{\it vn}}(#1)}     % value named
\newcommand{\evn}[2]{{\cal E}_{{\it vn}}(#1,#2)}   % vn for eval cxt

\newcommand{\mand}{\mbox{ and }}
\newcommand{\w}[1]{{\it #1}}    %To write in math style

\newcommand{\vcb}[1]{{#1^{*}}}

\newcommand{\qqs}[2]{\forall\, #1\;\: #2}
\newcommand{\xst}[2]{\exists\, #1\;\: #2}

\newcommand{\s}[1]{{\sf #1}}    % sans-serif 
\newcommand{\vc}[1]{#1^{*}}
\newcommand{\lnorm}{\lbrack\!\lbrack}
\newcommand{\rnorm}{\rbrack\!\rbrack}
\newcommand{\sem}[1]{\lnorm #1\rnorm}
\newcommand{\sm}[2]{\sem{#1}^{#2}}

\newcommand{\act}[1]{\stackrel{#1}{\rightarrow}} %labelled action low high
\newcommand{\eval}{\Downarrow}

\newcommand{\Z}{{\bf Z}}
\newcommand{\Real}{\mathbb{R}^{+}}

\newcommand{\isum}{\oplus}

\newcommand{\bem}[1]{{\em #1}}

\newcommand{\wact}[1]{\stackrel{#1}{\Rightarrow}} %weak labelled action low %high
\newcommand{\lact}[1]{\stackrel{#1}{\leftarrow}}
\newcommand{\lwact}[1]{\stackrel{#1}{\Leftarrow}} 

\newcommand{\wbis}{\approx}  %weak bisimulation
\newcommand{\bis}{\sim} %strong bisimulation
\newcommand{\prob}[1]{\mathbb{P}(#1)}
\newcommand{\prb}{\mathbb{P}}

\newcommand{\bbis}{\sim_{BB}}
\newcommand{\bebis}{\sim_{BE}}
\newcommand{\wbbis}{\approx_{BB}}
\newcommand{\wbebis}{\approx_{BE}}

\newcommand{\cbis}{\sim_{C}}
\newcommand{\lbis}{\sim_{L}}

\newcommand{\cwbis}{\approx_{C}}
\newcommand{\lwbis}{\approx_{L}}

\newcommand{\new}[2]{\nu #1 \ #2}
\newcommand{\local}[3]{\s{var} \ #1=#2 \ #3}

\newcommand{\chtype}[1]{{\it Ch}(#1)}

\newcommand{\impl}{\supset}
\newcommand{\gexp}{\succeq}
\newcommand{\lexp}{\preceq}

\newcommand{\inp}[1]{#1_{\downarrow}}
\newcommand{\out}[1]{#1_{\uparrow}}

\newcommand{\ml}{{\it ML}}
\newcommand{\java}{{\it Java}}

\newcommand{\emb}{\stackrel{\triangleright}{\_}}
\newcommand{\semb}{\triangleright}
\newcommand{\mset}[1]{\{\!| #1 |\!\}}
\newcommand{\cngr}{\equiv}

\newcommand{\depth}[1]{{\it depth}(#1)}
\newcommand{\pt}[2]{{\it PT}(#1,#2)} %principal type
\newcommand{\refe}{{\sf ref} \ }
\newcommand{\Refe}{{\sf Ref} \ }

\newcommand{\Let}[3]{{\sf let } \ #1 = #2 \ {\sf in } \ #3}

\newcommand{\er}[1]{{\it er}(#1)}
\newcommand{\C}{{\it C}}
\newcommand{\ccs}{{\it CCS}}
\newcommand{\tccs}{{\it TCCS}}

\newcommand{\SL}{{\it SL}}
\newcommand{\esterel}{{\it Esterel}}
\newcommand{\sco}{\mid}     % symbol to replace the semi-colon in cps 
\newcommand{\h}[1]{{\cal C}_{{\it h}}(#1)}      % hoisting conversion
\newcommand{\cc}[1]{{\cal C}_{{\it cc}}(#1)}      % closure conversion
\newcommand{\prj}[2]{\pi_{#1}(#2)}     % projection 
\newcommand{\pack}[1]{{\sf pack}(#1)}
\newcommand{\unpack}[1]{{\sf unpack}(#1)}
\newcommand{\imp}{{\sf Imp}}   %par imp
\newcommand{\vm}{{\sf Vm}}              %virtual machine language
\newcommand{\access}[1]{\stackrel{#1}{\leadsto}}
\newcommand{\text}[1]{\mbox{#1}}
\newcommand{\letarrow}{\rightarrow_{{\sf let}}} 

\newcommand{\intind}[2]{\int #1 \mathrm{d}#2}

\newcommand{\may}{{\sf may}}

\newcommand{\must}{{\sf must}}

\newcommand{\mustnot}{{\sf mustnot}}
\newcommand{\tmayleq}{\leq_{{\sf may}}^{t}}
\newcommand{\tmustleq}{\leq_{{\sf must}}^{t}}

\newcommand{\mayleq}{\leq_{{\sf may}}}
\newcommand{\mustleq}{\leq_{{\sf must}}}
\newcommand{\acset}[1]{A(#1)}
\newcommand{\acsset}[1]{A_{S}(#1)}

\newcommand{\trace}[1]{T(#1)}
\newcommand{\inftrace}[1]{T^\omega(#1)}

\newcommand{\otleq}{\leq_{T}^{o}}   %trace object leq  
\makeindex

\begin{document}

\bibliographystyle{alpha}

\title{{\Huge \bf Operational methods in semantics}}

\author{{\huge Roberto M. Amadio}\\~\\
{\Large Universit\'e Paris Cité, CNRS, IRIF, F-75013, France}}

\maketitle

{\footnotesize \tableofcontents}

\chapter*{Preface}
\addcontentsline{toc}{chapter}{Preface}
\markboth{{\it Preface}}{{\it Preface}}

The focus of these lecture notes is on abstract models and basic ideas
and results that relate to the {\em operational semantics} of
programming languages largely conceived.  The approach is to start
with an abstract description of the computation steps of programs and
then to build on top semantic equivalences, specification languages,
and static analyses.  While other approaches to the semantics of
programming languages are possible, it appears that the operational
one is particularly effective in that it requires a moderate level of
mathematical sophistication and scales reasonably well to a large
variety of programming features.  In practice, operational semantics
is a suitable framework to build portable language implementations and
to specify and test program properties (see, {\em e.g.},
\cite{DBLP:books/daglib/0069232}).  It is also used routinely to
tackle more ambitious tasks such as proving the correctness of a
compiler or a static analyzer (see, {\em e.g.},
\cite{DBLP:conf/popl/Leroy06}).

These lecture notes contain a selection of the material taught by the
author over several years in courses on the semantics of programming
languages, foundations of programming, compilation, and concurrency
theory.
They are oriented towards master students interested in
fundamental research in computer science. The reader is supposed to be
familiar with the main programming paradigms (imperative, functional,
object-oriented,$\ldots$) and to have been exposed to the notions
of concurrency and synchronization as usually discussed in a course
on operating systems.
The reader is also expected to have attended introductory courses on
automata, formal languages, mathematical logic, and compilation of
programming languages. 

Our goal 
is to provide a compact reference for
grasping the basic ideas of a rapidly evolving field. This means that
we concentrate on the simple cases and we give a self-contained
presentation of the proof techniques. Following this approach, we manage
to cover a rather large spectrum of topics within a coherent terminology
and to shed some light, we hope, on the connections among apparently 
different formalisms.

Chapter \ref{intro-sec} introduces, in the setting of a very simple
{\em imperative programming language}, some of the main ideas and
applications of operational semantics.  A {\em sequential} programming
language is a formalism to define a system of computable functions;
the closer the formalism to the notion of function, the simpler the
semantics.  The first formalism we consider is the one of {\em term
  rewriting systems} (chapters
\ref{rewrite-sec}--\ref{firstfun-sec}). On one hand, (term) rewriting
is ubiquitous in operational semantics and so it seems to be a good
idea to set on solid foundations the notions of {\em termination} and 
{\em confluence}. 
On the other hand, under suitable
conditions, term rewriting is a way of defining {\em first-order functions}
on inductively defined data structures.

The second formalism we introduce 
(chapters \ref{lambda-sec}--\ref{equiv-sec}) 
is the {\em $\lambda$-calculus}, which is a notation
to represent higher-order functions.  In this setting, a function is
itself a datum that can be passed as an argument or returned as a
result.  We spend some time to explain the mechanisms needed
for correctly implementing the $\lambda$-calculus via the notion of
{\em closure}. 
We then address the issue of program equivalence (or refinement)
and claim that the notion of {\em contextual equivalence} provides 
a natural answer to this issue. We also show that 
the co-inductively defined notion of 
{\em simulation} provides an effective method 
to reason about contextual equivalence.

Chapters \ref{simple-type-sec}--\ref{systemF-sec} introduce increasingly
expressive {\em type systems} for the $\lambda$-calculus. 
The general idea is that types express properties of 
program expressions which are invariant under execution.
As such, types are a way of documenting the way a program expression
can be used and by combining program expressions according
to their types we can avoid many run-time errors.
In their purest form, types can be connected with logical propositions
and this leads to a fruitful interaction with a branch
of mathematical logic known as {\em proof theory}.
Sometimes types lead to verbose programs. 
To address this issue we introduce 
{\em type inference} techniques which are automatic methods
discharging the programmer from the task of explicitly 
writing the types of the program expressions.
Also sometimes types are a bit of a straight jacket in that 
they limit the way programs can be combined or reused.
We shall see that {\em polymorphic} types (and later {\em subtyping})
address, to some extent, these issues.

Chapters \ref{transform-sec}--\ref{typing-sec}, introduce various
standard {\em program transformations} that chained together allow
to compile a higher-order (functional) language into a basic assembly language.
We also show that the type systems presented in the previous chapters
shed light on the program transformations.

Chapters \ref{record-sec}--\ref{obj-sec} consider the problem of
formalizing the operational semantics of {\em imperative} and
{\em object-oriented} programming languages.  We show that the notion of
higher-order computable function is still useful to understand the
behavior of programs written in these languages.  We start by
enriching the functional languages considered with record and variant
data types. This is an opportunity to discuss the notion of {\em subtyping}
which is another way of making a type system more flexible.
Concerning functions with {\em side-effects}, we show that they can be
compiled to ordinary functions by expliciting the fact that each
computation takes a memory as argument and returns a new memory as
result.  Concerning objects, we show that they can be understood as
a kind of recursively defined records.

Starting from chapter \ref{introduction-sec}, 
we move from {\em sequential} to {\em
concurrent} programming models where several threads/processes {\em
compete} for the same resources (e.g. write a variable or a
channel).  Most of the time, this results into {\em non-deterministic}
behavior which means that with the same input the system can move to
several (incomparable) states.
Chapters \ref{trace-shared-sec}--\ref{rely-guarantee-sec} focus on a
concurrent extension of the simple model of imperative programming
introduced in chapter \ref{intro-sec}. In particular, we introduce 
a compositional {\em trace semantics}, {\em rely-guarantee} assertions,
and mechanisms to implement {\em atomic} execution.

Chapters \ref{lts-sec}--\ref{det-sec} take a more abstract look at
concurrency in the framework of {\em labelled transition systems}.
We develop the notion of {\em bisimulation} and we consider its
logical characterization through a suitable {\em modal logic}.
Labelled transition systems extended with a {\em rendez-vous} synchronization
mechanism lead to a simple calculus of concurrent systems known as
$\ccs$. We consider a method to compare processes through
may and must {\em testing pre-orders} and finally
rely on the $\ccs$ calculus to explore the connections between
{\em determinacy} and {\em confluence}.

Chapters \ref{time-sec}--\ref{proba-sec} describe two relevant
extensions of non-deterministic computation. In the first one, we consider
the notion of {\em timed (or synchronous)} computation where processes 
proceed in lockstep (at the same speed)
and the computation is regulated by a notion of
instant. In the second one, we consider systems  which 
exhibit both non-deterministic  and {\em probabilistic} behaviors.

Chapter \ref{pi-sec} introduces an extension
of $\ccs$, known as {\em $\pi$-calculus}, where processes can communicate 
channel names. We show that the theory of equivalence developed
for $\ccs$ can be lifted to the $\pi$-calculus and that the $\pi$-calculus
can be regarded as a concurrent extension of the $\lambda$-calculus.

Finally, chapter \ref{conc-obj-sec} builds on chapter \ref{obj-sec} 
to formalize a fragment of the concurrency available in the $\java$
programming language and to discuss the notion of 
{\em linearization} of concurrent data structures.

While the choice of the topics is no doubt biased by the 
interests of the author, it still provides a fair
representation of the possibilities offered by {\em operational semantics}.
Links between operational semantics and more `mathematical' semantics
based, {\em e.g.}, on domain and/or category theory are not
developed at all in these lecture notes;
we refer the interested reader to, {\em e.g.}, \cite{AC98,Gunter92,Winskel93}.

Most topics discussed in these lecture notes can form the basis of
interesting {\em programming experiences} such as the construction of a
compiler, a static type analyzer, or a verification condition
generator.  The {\em proofs} sketched in these lecture notes can also
become the object of a programming experience in the sense that they
can be formalized and checked in suitable {\em proof assistants} 
(experiments in this direction can be found, {\em
  e.g.}, in the books
\cite{DBLP:books/daglib/0035083,DBLP:books/sp/NipkowK14,Pierce:SF}).

Each chapter ends with a summary of the main concepts and results
introduced and a few bibliographic references.  These references are
suggestions for further study possibly leading to research problems.
Quite often we prefer to quote the `classic' papers that introduced a
concept than the most recent ones which elaborated on it.  Reading the
`classics' is a very valuable exercise which helps in building some
historical perspective especially in a discipline like computer
science where history is so short.  Sections marked with a $(*)$ contain
more advanced and/or technical material. 

These lecture notes contain enough material for a two semesters course; 
however, there are many possible shortcuts to fit just one semester
course. The chapters \ref{intro-sec}--\ref{obj-sec} cover {\em sequential
  languages}.  Chapters \ref{intro-sec}--\ref{rewrite-sec}, and some
of \ref{termination-trs-sec} are a recommended introduction and the
chapters \ref{lambda-sec}--\ref{simple-type-sec} constitute the
backbone on the $\lambda$-calculus. The remaining chapters can be
selected according to the taste of the instructor and the interests of
the students. Topics covered include: term rewriting systems (chapters
\ref{unification-sec}, \ref{confluence-trs-sec}, \ref{firstfun-sec}), type systems (chapters \ref{type-inf-sec}, \ref{systemF-sec}, \ref{record-sec}), type
inference (chapters \ref{unification-sec}, \ref{hindley-inf-sec},
\ref{type-inf-sec}), program transformations (chapters
\ref{transform-sec}, \ref{typing-sec}), and imperative and object-oriented
languages (chapters \ref{lamref-sec}, \ref{obj-sec}).
The chapters \ref{introduction-sec}--\ref{conc-obj-sec} focus on 
{\em concurrent languages} and assume some familiarity with the 
basic material mentioned above.
Chapter \ref{introduction-sec} is a recommended introduction to 
concurrency.
Chapters \ref{trace-shared-sec}--\ref{rely-guarantee-sec} cover 
a simple model of shared memory concurrency while
chapters \ref{lts-sec}--\ref{testing-sec} lead to the calculus $\ccs$, a basic
model of message passing concurrency.
The following chapters explore the notions of deterministic (chapter \ref{det-sec}),
timed (chapter \ref{time-sec}), and probabilistic (chapter \ref{proba-sec}) computation.
The final chapters move towards models of concurrency that integrate the
complexity of a sequential language. In particular, we discuss
the $\pi$-calculus (chapter \ref{pi-sec}) which extends a compiled form
of the  $\lambda$-calculus (chapter \ref{transform-sec}) and
a concurrent object  oriented language (chapter \ref{conc-obj-sec})
which extends the object-oriented language presented in chapter \ref{obj-sec}.

{\footnotesize
\chapter*{Notation}\label{Notation}
\addcontentsline{toc}{chapter}{Notation}
\markboth{{\it Notation}}{{\it Notation}}

\begin{multicols}{2}
We collect the notations most frequently used in these
lecture notes.

\subsection*{Set theoretical}
\[
\begin{array}{ll}

\emptyset 	&\mbox{empty set} \\

\Nat 		&\mbox{natural numbers}\\

\Z              &\mbox{integers} \\

\union, \inter  &\mbox{union, intersection of two sets} \\

\Union, \Inter  &\mbox{union, intersection of a family of sets} \\

\cpl{X}		&\mbox{complement of }X \\

\prt{X}\mbox{ or }2^X 	&\mbox{subsets of }X \\

\finprt{X}      &\mbox{finite subsets of }X \\

\card X		&\mbox{cardinality of }X \\

R^{*}		&\mbox{reflexive and transitive closure of } R %\\

\end{array}
\]

\subsection*{Order theoretical}
\[
\begin{array}{ll}

(P,<)           &\mbox{strict partial order} \\

(P,\leq)	&\mbox{partial order} \\

\JOIN X		&\mbox{least upper bound (lub)} \\

\MEET X		&\mbox{greatest lower bound (glb)} \\

\end{array}
\]

\subsection*{Bound variables and substitution}
We introduce a number of operators that {\em bound}
variables. The rules for renaming bound variables and 
for substituting a term in a term with bound variables are
the same that apply in, say, first-order logic.
If $T,S,\ldots$ are terms and $x$ is a variable, 
we denote with $\fv{T}$ the set of variables occurring free in $T$ 
and with $[T/x]S$ the substitution of $T$ for $x$ in $S$.

\subsection*{Function update}
\[
\begin{array}{ll}

f[e/d](x)       &= \left\{
			\begin{array}{ll}
			e	&\mbox{if } x=d \\
			f(x)	&\mbox{otherwise}
		   	\end{array} \right.
\end{array}
\]

\subsection*{Simple imperative languages}
\[
\begin{array}{ll}

S  &\mbox{statement} \\ 
K  &\mbox{continuation} \\
C  &\mbox{context} \\
s:\w{Id} \arrow \Z   &\mbox{state} \\
\Downarrow           &\mbox{evaluation (big-step semantics)} \\
A,B,\ldots           &\mbox{logical assertions} \\
\hoare{A}{S}{B}        &\mbox{partial correctness assertion}\\
\jones{P}{A}{R}{G}{B}  &\mbox{rely guarantee assertion}
\end{array}
\]

\subsection*{Rewriting}
\[
\begin{array}{ll}
\arrow &\mbox{rewriting relation} \\
\Sigma &\mbox{signature} \\
V=\set{x,y,\ldots} &\mbox{set of variables} \\
T_\Sigma(V)         &\mbox{terms built over $\Sigma$ and $V$} \\
t,s,\ldots         &\mbox{first-order terms} \\
\var{t}            &\mbox{set of variables occurring in term $t$} \\
C                  &\mbox{context} \\
S:V\arrow T_\Sigma(V) &\mbox{substitution} \\
R\leq S            &\mbox{substitution pre-order $\exists T \ (T\comp R=S)$} \\
>_p                &\mbox{product order}\\
>_{\w{lex}}          &\mbox{lexicographic order}\\
>_m                &\mbox{multi-set order} \\
>_r                &\mbox{recursive path order}
\end{array}
\]

\subsection*{$\lambda$-calculus}
\[
\begin{array}{ll}
M,N,\ldots     &\mbox{$\lambda$-terms} \\
C              &\mbox{context} \\
V              &\mbox{value} \\
E              &\mbox{evaluation context} \\
M[\eta]        &\mbox{closure} \\
A,B,\ldots     &\mbox{types} \\
\sigma         &\mbox{type schema} \\
A\leq B        &\mbox{subtyping}

\end{array}
\]

\subsection*{Objects}
\[
\begin{array}{ll}
C,D,\ldots &\mbox{classes} \\
e  &\mbox{expressions} \\
v  &\mbox{values} \\
C\leq D &\mbox{class inheritance} \\
E &\mbox{evaluation context} \\
\Delta &\mbox{redex}\\
h &\mbox{heap} \\
\end{array}
\]

\subsection*{Processes}
\[
\begin{array}{ll}
P,Q,\ldots   &\mbox{processes} \\
\alpha,\beta,\ldots       &\mbox{actions} \\
\tau         &\mbox{internal (silent) action} \\
a,b,\ldots   &\mbox{observable actions} \\
s            &\mbox{(finite) word of observable actions} \\
\act{\alpha} &\mbox{strong labelled transition} \\
\wact{\alpha}&\mbox{weak labelled transition} \\
\bis         &\mbox{strong bisimulation} \\
\wbis        &\mbox{weak bisimulation}\\
A,B,\ldots   &\mbox{modal formulae} \\
\pos{\alpha}{A} &\mbox{diamond modality} \\
\whn{\alpha}{A} &\mbox{box modality} \\
P\downarrow s &\mbox{hereditary termination} \\
\mayleq         &\mbox{may pre-order} \\
\mustleq        &\mbox{must pre-order} \\
\Delta          &\mbox{distribution} \\
\prb            &\mbox{probability}
\end{array}
\]
\end{multicols}
}

\chapter{Introduction to operational semantics}\label{intro-sec}
\markboth{{\it Introduction}}{{\it Introduction}}

The goal of this introductory chapter is to present 
at an elementary level some ideas of the operational
approach to the semantics of programming languages
and to illustrate some of their applications.

To this end, we shall focus on a standard toy imperative language
called $\imp$.  As a first step we describe formally and at an
abstract level the computations of $\imp$ programs.  In doing this, we
identify two styles  known as {\em big-step} and {\em small-step}.
 Then, based on this specification, we introduce a suitable
notion of pre-order on statements and check that this pre-order is
preserved by the operators of the $\imp$ language.

As a second step, we introduce a specification formalism for $\imp$ programs
that relies on so called {\em partial correctness assertions} (pca's). 
We present sound rules for reasoning about such assertions and 
a structured methodology to reduce reasoning about pca's
to ordinary reasoning in a suitable theory of (first-order) logic.
We also show that the pre-order previously defined on statements
coincides with the one induced by pca's.

As a third and final step, we specify a toy compiler from the $\imp$ language
to an hypothetical {\em virtual machine} whose semantics is also defined
using operational techniques. We then 
apply the developed framework to prove the correctness of the compiler.

\section{A simple imperative language}\label{imp-sec}\index{$\imp$ language}
We assume the reader is familiar with the idea that the {\em syntax}
of a programming language can be specified via a context-free
grammar. 
The syntax of the $\imp$ language is described 
in Table \ref{synt-imp-tab} where we distinguish
the syntactic categories of identifiers (or variables),
integers, values, numerical expressions, boolean conditions,
statements, and programs.
We shall not dwell on questions of 
grammar ambiguity and priority of operators. Whenever we look 
at a syntactic expression
we assume it contains enough parentheses so that no ambiguity
arises on the order of application of the operators. 

We also assume the reader is familiar with the notion of {\em formal system}.
A formal system is composed of {\em formulae} specified by a certain syntax and
{\em inference rules} to derive formulae from other formulae. Depending on the context,
the formulae may be called assertions or judgments.
We often rely on the following suggestive notation to describe inference rules:
\[
\infer{A_1,\ldots, A_n}{B}~,
\]
which means that if we can infer formulae $A_1,\ldots,A_n$ (the hypotheses) then we can also infer formula $B$
(the conclusion). To bootstrap the inference process we need some rule with no hypothesis, {\em i.e.}, where $n=0$. 
Such rules are called {\em axioms}. A rule with $m$ conclusions is regarded as an abbreviation for 
$m$ rules which share the same hypotheses:
\[
\infer{A_1,\ldots,A_n}{B_1,\ldots,B_m} \quad \mbox{is equivalent to} \quad \infer{A_1,\ldots,A_n}{B_1}, \cdots ,  \infer{A_1,\ldots,A_n}{B_m}~.
\]

\begin{table}[b]
{\footnotesize
\[
\begin{array}{lll}

\w{id}&::= x\Alt y\Alt \ldots       &\mbox{(identifiers)} \\
\w{n} &::= 0 \Alt -1 \Alt +1 \Alt \ldots &\mbox{(integers)} \\
v &::= n \Alt \s{true} \Alt \s{false}  &\mbox{(values)} \\
e &::= \w{id} \Alt n \Alt e+e    &\mbox{(numerical expressions)} \\
b &::= e<e  &\mbox{(boolean conditions)} \\
S &::= \s{skip} \Alt \w{id}:=e \Alt S;S \Alt 
  \s{if} \ b \ \s{then} \ S \ \s{else} \ S 
   \Alt \s{while} \ b \ \s{do} \ S  &\mbox{(statements)} \\
P &::= \s{prog} \ S      &\mbox{(programs)}

\end{array}
\]}
\caption{Syntax of the $\imp$ language}\label{synt-imp-tab}\index{$\imp$, language}
\end{table}

The $\imp$ language is a rather standard {\em imperative language} with 
\s{while} loops and \s{if-then-else}. 
We call the language {\em imperative} because
the execution of a program is understood as the execution of
a sequence of statements whose effect is to modify a global 
entity known as the {\em state}. 
We can regard the {\em state} as an abstract model of the 
computer's memory.
As in every modeling activity, the name of the game
is to have a simple model but not too simple. In other words, the model
should not contain too many details and still be able to make useful
predictions on programs' behaviors. For the $\imp$ language, we 
shall assume the state is a total function from identifiers 
to integers. Notice that by representing the state as a {\em total} function 
we avoid some technicalities, namely we make sure that the evaluation
of a variable in a state is always defined.

If $s$ is a state, $x$  an identifier, and $n$ an
integer, then we denote with $s[n/x]$  an elementary 
{\em state update} defined as follows:
\begin{equation}\label{state-update-store}
s[n/x](y) =\left\{\begin{array}{ll}
                    n &\mbox{if }x=y\\
                    s(y) &\mbox{otherwise.}
\end{array}\right.
\end{equation}
A first approach at specifying the execution of 
 $\imp$ programs relies on the following judgments (or assertions):
\[
(e,s)\eval v,  \quad (b,s) \eval v, \qquad (S,s)\eval s',\quad (P,s)\eval s',
\]
and it is described in Table \ref{semantics-imp}.
The defined predicates $\eval$ are often called {\em evaluations}.
They specify the final result of the execution (if any)
while neglecting the intermediate steps.
Thus, for a given state, a (boolean) expression evaluates to a value 
while a statement (or a program) evaluates to a state.
This specification style is called {\em big-step}.
\begin{table}
{\footnotesize
\[
\begin{array}{c}

\infer{}
{(v,s)\eval v}

\qquad

\infer{}
{(x,s) \eval s(x)}

\qquad

\infer{(e,s)\eval v\quad (e',s)\eval v'}
{(e+e',s)\eval (v+_{\Z} v')} 

\qquad
\infer{(e,s)\eval v\quad (e',s)\eval v'}
{(e<e',s) \eval (v<_{\Z} v')} \\ \\ 

\infer{}
{(\s{skip},s)\eval s}

\qquad 

\infer{(e,s) \eval v}
{(x:=e,s) \eval s[v/x]} 

\qquad

\infer{(S_1,s)\eval s'\quad (S_2,s')\eval s''}
{(S_1;S_2,s)\eval s''} \\\\

\infer{(b,s)\eval \s{true} \quad (S,s)\eval s'}
{ (\s{if} \ b \ \s{then} \ S \ \s{else} \ S',s) \eval  s'}
\qquad
\infer{(b,s)\eval \s{false} \quad (S',s)\eval s'}
{ (\s{if} \ b \ \s{then} \ S \ \s{else} \ S',s) \eval  s'}
\\ \\ 

\infer{(b,s) \eval \s{false}}
{(\s{while} \ b \ \s{do} \  S ,s)\eval s}
\qquad

\infer{(b,s) \eval \s{true} \quad (S;\s{while} \ b \ \s{do} \  S,s) \eval s'}
{(\s{while} \ b \ \s{do} \  S ,s)\eval s'} \\ \\ 

\infer{(S,s)\eval s'}
{(\s{prog} \ S,s)\eval s'}

\end{array}
\]}
\caption{Big-step reduction rules of $\imp$}\label{semantics-imp}
\index{$\imp$, big-step reduction rules}
\end{table}

By opposition, the {\em small-step} approach is based
on the definition of `elementary' reduction rules.
The final result, if any, is obtained by iteration
of the reduction rules. In order to describe the intermediate
steps of the computation, we introduce an additional
syntactic category of continuations. 
A {\em continuation} $K$ is a 
list of statements which terminates with a special
symbol \s{halt}: 
\[
K::= \s{halt} \Alt S \cdot K \qquad\mbox{(continuation).}
\]
A continuation keeps track of the statements that still need
to be executed.
Table \ref{small-step-imp} defines small-step reduction rules for
$\imp$ statements  whose basic judgment has the shape:
\[
(S,K,s) \arrow  (S',K',s')~.
\]
Note that we still rely on the big-step reduction of
(boolean) expressions; the definition of a small step reduction
for (boolean) expressions is left to the reader.
We define the reduction of a program $\s{prog} \ S$ as the reduction
of the statement $S$ with continuation $\s{halt}$.
We can derive a big-step reduction from the small-step one as follows:
\[
\begin{array}{ll}
(S,s) \eval s'                   &\mbox{if }(S,\s{halt},s) \trarrow (\s{skip},\s{halt},s') ~,
\end{array}
\]
where $\trarrow$ denotes the reflexive and transitive closure of the relation $\arrow$.
\begin{table}
{\footnotesize
\[
\begin{array}{lll}
(x:=e,K,s) &\arrow &(\s{skip},K,s[v/x]) \qquad\mbox{if }(e,s)\eval v \\ \\

(S;S',K,s) &\arrow &(S,S'\cdot K,s) \\ \\

(\s{if} \ b \ \s{then} \ S \ \s{else} \ S',K,s) &\arrow 
&\left\{
\begin{array}{ll}
(S,K,s) &\mbox{if }(b,s)\eval \s{true} \\
(S',K,s) &\mbox{if }(b,s)\eval \s{false}
\end{array}
\right. \\ \\

(\s{while} \ b \ \s{do} \ S ,K,s) &\arrow 
&\left\{
\begin{array}{ll}
(S,(\s{while} \ b \ \s{do} \ S)\cdot K,s) &\mbox{if }(b,s)\eval \s{true} \\
(\s{skip},K,s) &\mbox{if }(b,s)\eval \s{false}
\end{array}
\right. \\ \\

(\s{skip},S\cdot K,s) &\arrow
&(S,K,s)

\end{array}
\]}
\caption{Small-step reduction rules of $\imp$ statements}\label{small-step-imp}\index{$\imp$, small-step reduction rules}
\end{table}

Let us pause to consider some properties
of both the big-step and the small-step reductions.
In both cases, reduction is driven by the syntax of the 
object (program, statement,$\ldots$) under consideration.
Moreover it is easy to check that for each state and program 
(or statement, or expression, or boolean expression)
at most one rule applies. This entails that the computation
is {\em deterministic}.
In some situations the computation is stuck, {\em i.e.}, no rule applies.
This happens if we try to add or compare two expressions 
whose values are {\em not} integers.
Also in some situations the computation diverges and
this happens because of the unfolding of the \s{while} loop.

To summarize, given a program and a state, $3$ mutually exclusive situations
may arise: (1) the computation terminates producing a new state, (2)
the computation is stuck in a situation where no rule applies,
and (3) the computation diverges.
Because, our computation rules
are {\em deterministic}, in situation (1) there is exactly
one state which is the outcome of the computation. 
Situation (2) corresponds to an erroneous configuration. Rather 
than leaving the computation stuck it is always possible to add rules
and values so that the computation actually terminates 
returning some significant {\em error message}.
As for situation (3), 
in the big-step approach, 
it arises as an infinite regression in the proof tree.
For instance, assuming $S =  \s{while} \ \s{true} \ \s{do} \ \sk$,
we have:
\[
\begin{array}{c}
(\sk,s)\eval s\quad (S,s)\eval \ ? \\\hline
(\sk;S,s)\eval \ ? \\\hline
(S,s)\eval \ ?
\end{array}
\]
In the small-step approach, a diverging computation is an infinite
reduction as in:
\[
(S,\s{halt},s) \arrow 
(\sk, S\cdot \s{halt},s) \arrow (S,\s{halt},s) \arrow \cdots
\]
Specifying the way programs compute is actually only the first
step in the definition of an operational semantics.
The second step consists in defining a notion of 
{\em program equivalence}.
To answer this question,
we need to decide what exactly is {\em observable}
in the computation of a program.  In general, {\em sequential programs}
are regarded as functions that transform input data into output data.
In particular, sequential {\em imperative} programs such as those of the
$\imp$ language  can be interpreted
as partial functions from states to states. 
Following this idea, we also interpret statements as
partial functions from states to states and (boolean) expressions
as total functions from states to numerical (boolean) values.

\begin{definition}[IO interpretation]\label{io-interp} \index{IO interpretation}
The IO interpretation of $\imp$ programs, statements, and (boolean) expressions
is defined as follows:
\[
\begin{array}{llll}
\sm{P}{IO}   &=\set{(s,s') \mid (P,s) \eval s'}, \qquad
&\sm{S}{IO}  & =\set{(s,s') \mid (S,s) \eval s'}, \\
\sm{b}{IO}   &=\set{(s,v) \mid  (b,s)\eval v}, 
&\sm{e}{IO}  &=\set{(s,v) \mid (e,s) \eval v}~.
\end{array}
\]
\end{definition}

A third step consists in checking the {\em compositionality properties} 
of the proposed interpretation.
For instance,
suppose we have shown that the IO-interpretations of two statements
$S$ and $S'$ coincide. Does this guarantee
that we can always replace any occurrence of the statement $S$ 
in a program with the statement $S'$ without affecting the overall
behavior of the program? 
To make this idea precise we introduce the notion of {\em statement context}.

\begin{definition}[context]\index{$\imp$, context}
A {\em statement context} (or {\em context} for short) $C$ 
is defined by the following grammar:
\[
C ::= [~] \Alt C;S \Alt S;C \Alt
  \s{if} \ b \ \s{then} \ C \ \s{else} \ S  \Alt
   \s{if} \ b \ \s{then} \ S \ \s{else} \ C 
   \Alt \s{while} \ b \ \s{do} \ C  
\]
where $[~]$ is a fresh symbol which stands for a placeholder (or a hole).
\end{definition}

If $C$ is a context and $S$ a statement then $C[S]$
is the statement resulting from replacing the special symbol $[~]$ 
with $S$ in $C$. For instance, if $C= S';[~]$ then $C[S]=S';S$.

\begin{proposition}[compositionality]\label{comp-io-imp-prop}
For all statements $S$ and $S'$ and  context $C$,
if $\sm{S}{IO}\subseteq \sm{S'}{IO}$ then 
$\sm{C[S]}{IO} \subseteq \sm{C[S']}{IO}$.
\end{proposition}
\Proof 
We proceed by induction on the height of the
proof of the judgment $(C[S],s)\eval s'$ and
case analysis on the shape of the context $C$.
For instance, suppose $C=\s{while} \ b \ \s{do} \ C'$.
We distinguish two cases.

\begin{itemize}

\item If $(b,s)\eval \s{false}$ then $s'=s$ and $(C[S'],s)\eval s$.

\item If $(b,s)\eval \s{true}$, $(C'[S],s)\eval s''$, and
$(C[S],s'')\eval s'$.
Then, by inductive hypothesis, we have that:
$(C'[S'],s)\eval s''$ and $(C[S'],s'')\eval s'$.
Hence $(C[S'],s)\eval s'$. \qed
\end{itemize}

\begin{exercise}\label{program-big-small-step}
Implement in your favorite programming language the big-step and small-step
reduction rules (Tables \ref{semantics-imp} and \ref{small-step-imp}) of the $\imp$ language.
\end{exercise}

\begin{exercise}\label{break-continue-ex}
Suppose we extend the $\imp$ language with the commands  
\s{break} and \s{continue}. Their informal semantics is as follows:
\begin{description}

\item[\s{break}]  causes execution of the smallest enclosing \s{while} 
statement to be terminated. 
Program control is immediately transferred to the point
just beyond the terminated statement. It is an error for 
a \s{break} statement to appear where there is no enclosing \s{while}
statement.

\item[\s{continue}]  causes  execution of the smallest enclosing \s{while}
statement to be terminated. Program control is immediately transferred to the
end of the body, and the execution of the affected \s{while}
statement continues from that point with a reevaluation of the loop test. 
It is an error for \s{continue} to appear where there is no enclosing 
\s{while}  statement.

\end{description}
Define the big-step and small-step reduction rules for the extended language.
{\em Hint:} for the big-step, consider extended judgments of the shape 
$(S,s)\eval (o,s')$ where $o$ is an additional information indicating 
the {\em mode} of the result, for the small-step consider 
a new continuation $\s{endloop}(K)$, where $K$ is an arbitrary continuation.
\end{exercise}

\section{Partial correctness assertions}\label{pca-sec}
Most programming languages support the insertion of logical assertions
in the control flow. At run time, whenever a logical assertion is crossed
its validity is checked and an exception is raised if the check fails.
Inserting assertions in programs is an excellent way of documenting
the expectations on the input (pre-conditions) 
and the guarantees on the output (post-conditions).
Moreover, assertions are quite helpful in nailing down bugs.
In the following, we consider systematic methods to 
compose pre and post conditions and possibly prove for
a given statement and pre-condition that a certain post-condition 
will always hold.
We denote with $A,B,\ldots$ {\em assertions}. When we regard
them as {\em syntax} they are formulae with variables ranging over the 
set of program variables. For instance:
\begin{equation}\label{asrt-ex}
\xst{y}{(x=3 \AND z>y>x)}~.
\end{equation}
We write $s\models A$ if the assertion $A$ holds in the interpretation (state) $s$.
Thus a syntactic assertion such as (\ref{asrt-ex}) is {\em semantically} the set of states that satisfy it, namely:
\[
\set{s \mid s(x)=3, s(z)\geq 5}~.
\]

\begin{definition}[pca]\label{pca-def} \index{partial correctness assertion}
A {\em partial correctness assertion} (pca) is a triple
$\hoare{A}{S}{B}$. We say that it is {\em valid}  and write
$\models \hoare{A}{S}{B}$ if:
\[
\qqs{s}{(s\models A \mbox{ and } 
(P,s) \eval s' \mbox{ implies }s'\models B)}~.
\]
\end{definition}
The assertion is {\em partial} because it puts no constraint on the
behavior of a non-terminating, {\em i.e.}, partial, statement.
Table~\ref{hoare-seq} \index{Floyd-Hoare rules} 
describes the so called Floyd-Hoare rules (logic).
The rules are formulated assuming
that $A,B,\ldots$ are {\em sets of states}. 
It is possible to go one step further and replace the sets by predicates
in, say, first-order logic, however this is not essential to understand
the essence of the rules.

\begin{table}[b]
{\footnotesize
\[
\begin{array}{cc}

\infer{A\subseteq A' \quad \hoare{A'}{S}{B'} \quad B'\subseteq B}
{\hoare{A}{S}{B}} 

&\infer{\hoare{A}{S_1}{C}\quad \hoare{C}{S_2}{B}}
{\hoare{A}{S_1;S_2}{B}} \\\\

\infer{\hoare{A\inter b}{S_1}{B}\quad \hoare{A\inter \neg b}{S_2}{B}}
{\hoare{A}{\ite{b}{S_1}{S_2}}{B}} \qquad

&\infer{A\subseteq B}
{\hoare{A}{\sk}{B}} \\ \\ 

\infer{A;\sm{x:=e}{IO}\subseteq B}
{\hoare{A}{x:=e}{B}}  

&\infer{(A\inter \neg b) \subseteq B \quad \hoare{A \inter b}{S}{A}}
{\hoare{A}{\while{b}{S}}{B}}

\end{array}
\]
}
\caption{Floyd-Hoare rules for $\imp$}\label{hoare-seq}
\end{table}

We recall that if $S$ is a statement then $\sm{S}{IO}$ 
is its input-output interpretation (definition \ref{io-interp}).
This is a binary relation on states which for $\imp$ statements
happens to be the graph of a {\em partial} function on states.
In particular, notice that for  
an assignment $x:=e$ we have:
\[
\sm{x:=e}{IO} = \set{(s,s[v/x]) \mid (e,s)\eval v}~,
\]
which is the graph of a {\em total} function.
In the assertions, we identify a boolean predicate $b$ 
with the set of states that satisfy it, thus $b$ stands for
$\set{s \mid s\models b}$.
We denote  with $A,B,\ldots$ unary relations (predicates) 
on the set of states and
with $R,S,\ldots$
binary relations on the set of states.
We combine unary and binary relations as follows:
\[
\begin{array}{lll}

A;R  &= \set{s' \mid \xst{s}{s\in A \mand (s,s')\in R}} &\mbox{(image)}\\
R;A  &= \set{s \mid \xst{s'}{s'\in A \mand (s,s')\in R}} &\mbox{(pre-image).}\\
\end{array}
\]

The first rule in Table~\ref{hoare-seq} allows to weaken the pre-condition
and strengthen the post-condition while the following rules 
are associated with the operators of the language.
The rules are {\em sound} in the sense that if the hypotheses are valid
then the conclusion is valid too. 

\begin{proposition}[soundness pca rules]\index{Floyd-Hoare rules, soundness}
The assertions derived in the system described in Table~\ref{hoare-seq} 
are valid.
\end{proposition}
\Proof 
We just look at the case for the \s{while} rule.
Suppose $s\in A$ and $(\while{b}{S},s)\eval s'$.
We show by induction on the height of the derivation that $s'\in B$.
For the basic case, we have $s\in \neg b$ and we know 
$A\inter  \neg b \subseteq B$.
On the other hand, suppose $s\in b$ and
\[
(S;\while{b}{S},s) \eval s'~.
\]
This means $(S,s)\eval s''$ and 
$(\while{b}{S},s'') \eval s'$.
By hypothesis, we know $s''\in A$ and by inductive hypothesis
$s'\in B$. \qed

\begin{exercise}\label{hoare-rule-assign-ex}
Suppose $A$ is a first-order formula. Show the
validity of the pca \ 
$\hoare{[e/x]A}{x:=e}{A}$.
On the other hand, show that the pca $\ \hoare{A}{x:=e}{[e/x]A} \ $ is
{\em not} valid.
\end{exercise}

Interestingly, one can read the rules {\em bottom up} and show that if the
conclusion is valid then the hypotheses are valid up to an application
of the first `logical' rule.  This allows to reduce the proof of a pca
$\hoare{A}{S}{B}$ to the proof of a purely set-theoretic/logical
statement. The task of traversing the program $S$ and producing
logical assertions can be completely automated once the loops are
annotated with suitable invariants. This is the job of so-called {\em
  verification condition generators}.

\begin{proposition}[inversion pca rules]\label{rule-inv-prop}\index{Floyd-Hoare rules, inversion}
The following properties hold:

\begin{enumerate}

\item If $\hoare{A}{S_1;S_2}{B}$ is valid then 
$\hoare{A}{S_1}{C}$ and $\hoare{C}{S_2}{B}$ are valid where
$C= (A;\sm{S_1}{IO}) \inter (\sm{S_2}{IO};B)$.

\item If $\hoare{A}{\ite{b}{S_1}{S_2}}{B}$ is valid then
$\hoare{A\inter b}{S_1}{B}$ and $\hoare{A\inter \neg b}{S_2}{B}$ are valid.

\item If $\hoare{A}{\sk}{B}$ is valid then $A\subseteq B$ holds.

\item If $\hoare{A}{x:=e}{B}$ is valid then $A;\sm{x:=e}{IO} \subseteq B$ holds.

\item If $\hoare{A}{\while{b}{S}}{B}$ is valid then 
there is $A'\supseteq A$ such that (i) $A'\inter \neg b \subseteq B$ and
(ii) $\hoare{A'\inter b}{S}{A'}$ is valid.

\end{enumerate}
\end{proposition}
\Proof The case for \s{while} is the interesting one. We define:
\[
\begin{array}{lll}

A_0 = A~,
&A_{n+1} = (A_n \inter b);\sm{S}{IO}~,
&A'     = \Union_{n\geq 0} A_n~.

\end{array}
\]
We must have:
$\qqs{n\geq 0}{A_n \inter \neg b \subseteq B}$.
Then for the first condition, we notice that:
\[
\begin{array}{lll}
A' \inter \neg b &= (\Union_{n\geq 0} A_n)\inter \neg b 
                 &= \Union_{n\geq 0} (A_n \inter \neg b) \\
                 &\subseteq \Union_{n\geq 0} B 
                 &= B~.
\end{array}
\]
For the second, we have:
\[
\begin{array}{lll}

(A'\inter b);\sm{S}{IO} &= (\Union_{n\geq 0} A_n \inter b);\sm{S}{IO}
                &= (\Union_{n\geq 0}(A_n \inter b));\sm{S}{IO} \\
                &= \Union_{n \geq 0} (A_n \inter b);\sm{S}{IO} 
                &= \Union_{n\geq 1} A_n \\
                &\subseteq  \Union_{n\geq 0} A_n
                &= A'~.
\end{array}
\]
An assertion such as $A'$ 
is called an {\em invariant} of the loop.
In the proof, $A'$ is defined as the 
limit of an iterative process where
at each step we run the body of the loop. While in theory 
$A'$ does the job, in practice it may be hard to reason 
on its properties; finding a {\em usable} invariant may require 
some creativity. \qed \\

Given a specification language on, say, statements, we can consider
two statements {\em logically} equivalent if they satisfy
exactly the same specifications. We can apply this idea to pca's.

\begin{definition}[pca interpretation]\index{partial correctness assertion, interpretation}
The pca interpretation of a process $P$ is:
\[
\sm{P}{pca}=
\set{(A,B) \mid \quad \models \hoare{A}{P}{B}}~.
\]
\end{definition}

So now we have two possible notions of equivalence for statements: 
one based on the input-output behavior and another based on partial
correctness assertions. 
However, it is not too difficult to show that they coincide.

\begin{proposition}[IO vs. pca]\label{IO-pca}
Let $S_1,S_2$ be statements. Then: 
\[
\sm{S_1}{IO} = \sm{S_2}{IO}  \ 
\mbox{ iff }\ 
\sm{S_1}{pca}=\sm{S_2}{pca} ~.
\]
\end{proposition}
\Proof $(\Rightarrow)$ \ 
Suppose $(A,B)\in \sm{S_1}{pca}$, $s\models A$, 
$(S_2,s) \eval s'$.
Then $(s,s')\in \sm{S_2}{IO}=\sm{S_1}{IO}$.
Hence $s'\models B$ and $(A,B)\in \sm{S_2}{pca}$. 

\Proofitemm{(\Leftarrow)} First a remark. Let us write
$s=_X s'$ if $\qqs{x\in X}{s(x)=s'(x)}$.
Further suppose $X\supseteq \fv{S}$ and $(S,s)\eval s'$. Then:

\begin{enumerate}

\item The variables outside $X$ are untouched: $s=_{\cpl{X}}s'$.

\item If $s=_X s_1$ then $(S,s_1) \eval s'_1$ and $s'=_X s'_1$.

\end{enumerate}

We now move to the proof.  Given a state and a finite set of variables $X$,  define:
\[
IS(s,X) = \MEET_{x\in X} (x=s(x))~.
\]
Notice that: $s'\models IS(s,X)$ iff $s'=_X s$.
We proceed by contradiction, assuming $(s,s')\in \sm{S_1}{IO}$ and 
$(s,s')\notin \sm{S_2}{IO}$. 
Let $X$ be the collection of variables occurring 
in the commands $S_1$ or $S_2$.  Then check that: 
\[
(IS(s,X),\neg IS(s',X))\in \sm{S_{2}}{pca}~.
\]
On the other hand: $(IS(s,X),\neg IS(s',X))\notin \sm{S_{1}}{pca}$.\qed 

\begin{exercise}\label{vcg-generator-ex}
Let $S$ be a statement and $B$ an assertion.
The {\em weakest precondition} of $S$ with respect to $B$
is a predicate that we denote with $\w{wp}(S,B)$ such that:
(i) $\hoare{\w{wp}(S,B)}{S}{B}$ is valid and (ii)
if $\hoare{A}{S}{B}$ is valid then $A\subseteq \w{wp}(S,B)$.
Let us assume the statement $S$ does {\em not} contain \s{while} loops.
Propose a strategy to compute $\w{wp}(S,B)$
and derive a method to reduce the validity of a pca
$\hoare{A}{S}{B}$ to the validity of a logical assertion.
\end{exercise}

\section{A toy compiler (*)}\label{toy-compiler-sec}
This section applies some ideas of operational
semantics to the formal analysis of a toy compiler. The reader should try
to grasp the structure of the formalization and can
certainly skip over the technical details.

We introduce a simple virtual machine $\vm$
to execute $\imp$ programs.
The machine includes the following elements:
(1) a fixed code $C$ (a possibly empty sequence of instructions),
(2) a program counter \w{pc},
(3) a state $s$ (identical to the one of $\imp$ programs),
(4) a stack of integers $\sigma$ which, intuitively, is used
to evaluate boolean and numerical expressions.
The machine includes the following instructions with the associated
informal semantics where `push' and `pop' act on the stack:

{\footnotesize\[
\begin{array}{ll}
\s{cnst}(n)        &\mbox{push \s{n}} \\
\s{var}(x)         &\mbox{push value }x   \\
\s{setvar}(x)      &\mbox{pop value and assign it to }x \\
\s{add}            &\mbox{pop 2 values and push their sum} \\
\s{branch}(k)      &\mbox{jump with offset }k        \\
\s{bge}(k)         &\mbox{pop 2 values and jump if greater or equal with offset }k \\
{\tt halt}           &\mbox{stop computation}
\end{array}
\]}

In the branching instructions, $k$ is an integer that has to be added
to the current program counter in order to determine the following
instruction to be executed. 
Given a sequence $C$, we denote with $|C|$ its length and 
with  $C[i]$ its $i^{\w{th}}$ element (the leftmost element
being the $0^{\w{th}}$ element).
The (small-step) reduction rules
of the instructions are formalized by rules of the shape:
\[
C \Gives (i,\sigma,s) \arrow (j,\sigma',s')~,
\]
and are fully described in Table \ref{semantics-vm}.
As already mentioned, the $\imp$ and $\vm$ reduction rules
share the same notion of state.
We write, {\em e.g.}, $n\cdot \sigma$ to stress that the top element 
of the stack exists and is $n$. We denote with $\epsilon$ an empty stack
or an empty sequence of $\vm$ instructions.
We write $(C,s)\eval s'$ if 
$C\Gives (0,\epsilon,s) \trarrow (i,\epsilon,s')$ and $C[i]=\s{halt}$.
\begin{table}
{\footnotesize
\[
\begin{array}{l|l}

\mbox{Rule}                                                 & C[i]= \\\hline

C \Gives (i, \sigma, s) \arrow (i+1, n\cdot \sigma,s)  & \s{cnst}(n) \\ 
C \Gives (i, \sigma, s) \arrow (i+1,s(x)\cdot \sigma,s)  & \s{var}(x) \\ 
C \Gives (i, n \cdot \sigma, s) \arrow (i+1,\sigma,s[n/x])  & \s{setvar}(x) \\
C \Gives (i, n \cdot n' \cdot \sigma, s) \arrow (i+1, (n+_{\Z} n')\cdot \sigma,s)  & {\tt add} \\
C \Gives (i, \sigma, s) \arrow (i+k+1, \sigma,s)  & \s{branch}(k) \\
C \Gives (i, n \cdot n' \cdot \sigma, s) \arrow (i+1, \sigma,s)  & \s{bge}(k) \mbox{ and }n<_{\Z} n'\\
C \Gives (i, n \cdot n' \cdot \sigma, s) \arrow (i+k+1, \sigma,s)  & \s{bge}(k) \mbox{ and }n\geq_{\Z} n'\\

\end{array}
\]}
\caption{Small-step reduction rules of $\vm$ programs}\label{semantics-vm}\index{virtual machine, reduction rules}
\end{table}

In Table \ref{compil-imp-vm}, 
we define compilation functions $\cl{C}$ from $\imp$ to $\vm$ which
operate on expressions, boolean conditions, statements, and programs.
We write $\w{sz}(e)$, $\w{sz}(b)$, $\w{sz}(S)$ for the number of instructions
the compilation function associates with the expression $e$, the boolean
condition $b$, and the statement $S$, respectively.
For instance, the statement $\s{while}\ (0<1) \ \s{do} \ \sk$ is compiled
as: 
\[
(\s{cnst}(0))(\s{cnst}(1))(\s{bge}(1))(\s{branch}(-4))~.
\]

\begin{table}
{\footnotesize
\[
\begin{array}{c}

\cl{C}(x) = {\tt var}(x)   \qquad
\cl{C}(n)  ={\tt cnst}(n)   \qquad
\cl{C}(e+e') =\cl{C}(e) \cdot \cl{C}(e') \cdot {\tt add} \\ \\

\cl{C}(e<e',k) = \cl{C}(e) \cdot \cl{C}(e') \cdot \s{bge}(k) \\ \\ 

\cl{C}(\sk) = \epsilon 

\qquad  \cl{C}(x:=e) = \cl{C}(e) \cdot {\tt setvar(x)} 

\qquad \cl{C}(S;S') = \cl{C}(S) \cdot \cl{C}(S')\\ \\

\cl{C}(\s{if} \ b \ \s{then} \ S\  \s{else} \ S') = 
\cl{C}(b,k) \cdot \cl{C}(S) \cdot (\s{branch}(k')) \cdot \cl{C}(S')  \\
\mbox{where: } k= \w{sz}(S)+1, \quad k'=\w{sz}(S')

\\ \\

\cl{C}(\s{while} \ b \ \s{do} \  S )
= \cl{C}(b,k)  \cdot \cl{C}(S)  \cdot \s{branch}(k') \\
\mbox{where: } k=\w{sz}(S)+1, \quad k'=-(\w{sz}(b)+\w{sz}(S)+1)\\\\

\cl{C}(\s{prog} \ S) = 
\cl{C}(S) \cdot \s{halt}

\end{array}
\]}

\caption{Compilation from $\imp$ to $\vm$}\label{compil-imp-vm}\index{$\imp$, compilation}
\end{table}

We now consider the question of proving the `correctness' of the
compilation function. The following proposition relates
the big-step reduction of $\imp$ programs to the execution of
the compiled code.

\begin{proposition}[soundness, big-step]\label{soundness-big-step-prop}
The following properties hold:

\Defitem{(1)}
If $(e,s)\eval v$ then 
$C \cdot \cl{C}(e) \cdot C' 
\Gives (i,\sigma,s) \trarrow (j,v\cdot \sigma,s)$
where $i=|C|$ and $j=|C \cdot \cl{C}(e)|$.

\Defitem{(2)}
If $(b,s)\eval \s{true}$ then 
$C \cdot \cl{C}(b,k) \cdot C' 
\Gives (i,\sigma,s) \trarrow (j+k,\sigma,s)$
where  $i=|C|$ and $j=|C \cdot \cl{C}(b,k)|$.

\Defitem{(3)}
If $(b,s)\eval \s{false}$
then $C \cdot \cl{C}(b,k) \cdot C' 
\Gives (i,\sigma,s) \trarrow (j,\sigma,s)$
where  $i=|C|$ and $j=|C \cdot \cl{C}(b,k)|$.

\Defitem{(4)}
If $(S,s)\eval s'$ then
$C \cdot \cl{C}(S) \cdot C' 
\Gives (i,\sigma,s) \trarrow (j,\sigma,s')$
where  $i=|C|$ and $j=|C \cdot \cl{C}(S)|$.
\end{proposition}

\begin{exercise}\label{proof-soundness-big-step-prop}
  Prove proposition \ref{soundness-big-step-prop}.
\end{exercise}

We can obtain similar results working with the 
small-step reduction  of the $\imp$ language.
To this end, given a $\vm$ code $C$, we define an `accessibility relation' 
$\access{C}$ as the least binary relation on $\set{0,\ldots,|C|-1}$ such that:
\[
\begin{array}{ll}

\infer{}
{i\access{C} i}
\quad
&\infer{C[i]=\s{branch}(k)\quad (i+k+1)\access{C} j}
{i\access{C} j}~.

\end{array}
\]
Thus $i\access{C}j$ if in the code $C$ we can go from $i$ to $j$ following
a sequence of unconditional jumps.
We also introduce a ternary relation $R(C,i,K)$ which relates 
a $\vm$ code $C$, a number $i\in\set{0,\ldots,|C|-1}$, and a continuation
$K$. The intuition is that relative to the code $C$, 
the instruction $i$ can be regarded as having continuation $K$. 

\begin{definition}
The ternary relation $R$ is 
the least one that satisfies the following conditions:
\[
\begin{array}{ll}

\infer{
\begin{array}{c}
\\
i\access{C} j\qquad C[j]=\s{halt}
\end{array}}
{R(C,i,\s{halt})}

\qquad
&\infer{
\begin{array}{c}
i\access{C} i'\quad C=C_1 \cdot \cl{C}(S) \cdot C_2\\ 
i'=|C_1|\quad j=|C_1\cdot \cl{C}(S)| \quad R(C,j,K)
\end{array}}
{R(C,i,S\cdot K)}~.

\end{array}
\]
\end{definition}

We can then state the correctness of the compilation function as follows.

\begin{proposition}[soundness, small-step]\label{soundness-small-step}
If $(S,K,s) \arrow (S',K',s')$ and $R(C,i,S\cdot K)$ then 
$C\Gives (i,\sigma,s) \trarrow (j,\sigma,s')$ and
$R(C,j,S'\cdot K')$.
\end{proposition}

\begin{exercise}\label{proof-soundness-small-step}
Prove proposition \ref{soundness-small-step}.
\end{exercise}

\begin{remark}
We have already noticed that an $\imp$ program  
has $3$ possible behaviors: (1) it returns
a (unique) result, (2) it is stuck in an erroneous situation, (3) it diverges.
Proposition \ref{soundness-big-step-prop} guarantees that the compiler
preserves behaviors of type (1). Using the small-step reduction rules
(proposition \ref{soundness-small-step}), we can also 
conclude that if the source program diverges then the compiled code
diverges too. On the other hand, when the source program is stuck
in an erroneous situation the compiled code is allowed to 
have an arbitrary behavior. The following example justifies
this choice. Suppose at source level
we have an error due to the addition of an integer and a boolean. 
Then this error does not need to be reflected at the implementation
level where the same data type may well be used to represent
both integers and booleans.
\end{remark}

\begin{exercise}[stack height]\label{stack-height-ex}
The $\vm$ code coming from the compilation of 
$\imp$ programs has very specific properties.
In particular, for every instruction of the compiled
code it is possible to predict statically, {\em i.e.}, at compile time,
the {\em height of the stack}
whenever the instruction is executed. 
We say that a sequence of instructions $C$ is well formed if there
is a function $h:\set{0,\ldots,|C|} \arrow \Nat$ which satisfies the
conditions listed in Table \ref{well-formed-instr} for $0\leq i \leq |C|-1$.
In this case we write $C:h$.
The conditions defining the predicate $C:h$ are strong enough 
to entail that $h$ correctly predicts the stack height 
and to guarantee the uniqueness of $h$ up to the initial condition.
Show that:
(1)  If $C:h$, $C\Gives (i,\sigma,s) \trarrow (j,\sigma',s')$, and 
$h(i)=|\sigma|$ then $h(j)=|\sigma'|$.
(2) 
If $C:h$, $C:h'$ and $h(0)=h'(0)$ then $h=h'$.
Next prove that the result of the compilation
is a well-formed code. Namely, 
for any expression $e$, statement $S$, and program $P$ the
following assertions hold.
(3) For any $n\in\Nat$ there is a unique $h$ such that 
$\cl{C}(e):h$, $h(0)=n$, and $h(|\cl{C}(e)|)=h(0)+1$.
(4) 
For any $S$, 
there is a unique $h$ such that $\cl{C}(S):h$, $h(0)=0$, and 
$h(|\cl{C}(S)|)=0$.
(5) There is a unique $h$ such that $\cl{C}(P):h$.
\end{exercise}

\begin{table}
{\footnotesize
\[
\begin{array}{l|l}

C[i]=         & \mbox{Conditions for }C:h \\\hline

\s{cnst}(n)
\mbox{ or } \s{var}(x)
&h(i+1) = h(i)+1 \\

\s{add}            
& h(i)\geq 2, \quad h(i+1)=h(i)-1 \\

\s{setvar}(x)      
& h(i)=1, \quad h(i+1)=0 \\

\s{branch}(k)
& 0\leq i+k+1 \leq |C|, \quad h(i)=h(i+1)=h(i+k+1)=0 \\

\s{bge}(k)
& 0\leq i+k+1 \leq |C|, \quad h(i)=2, \quad h(i+1)=h(i+k+1)=0 \\

\s{halt}    
&i=|C|-1, \quad h(i)=h(i+1)=0

\end{array}
\]}
\caption{Conditions for well-formed code}\label{well-formed-instr}
\end{table}

\section{Summary and references}
The first step in defining the operational semantics
of a programming language amounts to specify 
the way a program computes. The following steps are
the specification of the observables (of a computation) and 
the definition of a compositional pre-order (or equivalence)
on programs.  

An alternative and related approach amounts to introduce (partial)
correctness assertions on programs and deem two programs equivalent if
they satisfy the same assertions.  Also the 
validity of a program's assertion can be reduced to the validity of
an ordinary logical statement in a suitable theory of first order logic.

The formal analysis of compilers is a natural application target for
operational semantics.  Each language in the compilation chain is
given a formal semantics and the behavior of the source code is
related to the behavior of its representation in intermediate
languages, and down to object code.

The lecture notes \cite{DBLP:journals/jlp/Plotkin04a} 
are an early (first version appeared in 1981) 
systematic presentation of an operational
approach to the semantics of programming languages.
Rules for reasoning on partial correctness assertions 
of simple imperative programs are presented in 
\cite{FLO67} and \cite{DBLP:journals/cacm/Hoare69} while
\cite{McCarthyPainter67} is an early example 
of mechanized verification of a simple
compiler. The presented case study builds on that example and 
is partially based on  \cite{Leroy2009}.

\chapter{Rewriting systems}\label{rewrite-sec}
\markboth{{\it Rewriting systems}}{{\it Rewriting systems}}

In computer science, a set equipped with a binary reduction relation is 
an ubiquitous structure arising, {\em e.g.}, when formalizing the computation 
rules of an automaton, the generation step of a grammar, or 
the reduction rules of a programming language (such as the rules
for the $\imp$ language in Table \ref{small-step-imp}).

\begin{definition}[rewriting system]\label{reduction-system-def}\index{rewriting system}
A {\em rewriting system} is a pair $(A,\arrow)$ where $A$ is a set
and $\arrow \subseteq A\times A$ is a reduction relation. 
We write $a\arrow b$ for $(a,b)\in \arrow$.
\end{definition}

If we regard the reduction relation as an edge relation, we can also
say that a rewriting system is a (possibly infinite)
directed graph.

Next we introduce some notation.
If $R$ is a binary relation we denote with $R^{-1}$ its inverse
and with $R^*$ its reflexive and transitive closure.
In particular, if $\arrow$ is a reduction relation we also
write $\leftarrow$ for $\arrow^{-1}$,
$\trarrow$ for $(\arrow)^*$ and  $\ltrarrow$ for $(\leftarrow)^*$. 
Finally, $\lrtrarrow$ is defined as 
$(\arrow \union \arrow^{-1})^{*}$. This is the equivalence relation
induced by the rewriting system.

\section{Basic properties}
Termination and confluence are two relevant properties of rewriting
systems. Let us start with {\em termination}, namely 
the fact that all reduction sequences terminate.

\begin{definition}[termination]\label{termination-def}\index{rewriting system, terminating}
A rewriting system $(A,\arrow)$ is {\em terminating} if all sequences
of the shape $a_0 \arrow a_1 \arrow a_2 \arrow \cdots$ are finite. 
\end{definition}

In this definition, we require the sequence (not the set) to be
finite.  In particular, a rewriting system composed of a singleton set
$A$ where $\arrow =A\times A$ is {\em not} terminating.

When the rewriting system corresponds to the reduction rules of a programming
language the termination property is connected to the termination of programs.
This is a fundamental property in program verification. As a matter of fact,
the verification of a program is often decomposed into the proof of a partial
correctness assertion (cf. section \ref{pca-sec}) and a proof of termination.

\begin{example}
Let $A$ be the set of words composed of a (possibly empty) sequence of 
`function symbols' $f$ and an integer $n$. 
Write $f^k$ for $f\cdots f$, $k\geq 0$ times, and define a rewriting relation
$\arrow$ on $A$ as follows:
\[
\begin{array}{ll}
f^{k+1} (n) \arrow \left\{ 
\begin{array}{ll}
f^k (n-10) &\mbox{if }n>100 \\
f^{k+2} (n+11) &\mbox{otherwise.}
\end{array}\right.

\end{array}
\]
This is known as McCarthy's function $91$.  For instance:
\[
f(100) \arrow f(f(111)) \arrow f(101) \arrow 91 \not\arrow
\]
Proving its termination is not trivial, 
but the name of the function gives a hint. 
For another example, consider
the following rewriting relation on {\em positive} natural numbers:
\[
n\arrow \left\{
\begin{array}{ll}
n/2 &\mbox{if }n \mbox{ even} \\
3n+1 &\mbox{if }n\mbox{ odd and }n>1.
\end{array}\right.
\]
This is known as Collatz's function and its termination is a long standing
open problem.
\end{example}

\begin{exercise}\label{l-u-ex}
Consider the following $\imp$ command (extended with integer addition
and division) where $b$ is an arbitrary boolean condition: 
\[
\begin{array}{l}
\s{while} \ (u>l+1) \ \s{do} \
(r:= (u+l)/2 \ ;\  \s{if}\ b \ \s{then} \ u:=r \ \s{else}\  l:=r)~.
\end{array}
\]
Show that the evaluation of the command starting from a state
satisfying $u,l\in \Nat$ terminates.
\end{exercise}

\begin{definition}[normalizing]\label{normal-def}\index{rewriting system, normalizing}
We say that $a\in A$ is a {\em normal form} if there is no $b\in A$
such that $a\arrow b$.
We also say that the rewriting system is {\em normalizing} if 
for all $a\in A$ there is a finite reduction sequence leading
to a normal form.
\end{definition}

A terminating rewriting system is normalizing, 
but the converse fails. For instance, consider:
$A=\set{a,b}$ with $a \arrow a$ and $a\arrow b$.
In some contexts ({\em e.g.}, proof theory), a terminating 
rewriting system is also called {\em strongly normalizing}.
A second property of interest is {\em confluence}.

\begin{definition}[confluence]\label{confluence-def}\index{confluence}
A rewriting system $(A,\arrow)$ is {\em confluent} if for all $a\in A$:
\[
\infer{\qqs{b,c}{(b \ltrarrow a \rtrarrow c)}}{\xst{d}{(b \rtrarrow d \ltrarrow c)}}~.
\]
We also write $b\dcl c$ if $\xst{d}{(b \rtrarrow d \ltrarrow c)}$.
\end{definition}

A property related to confluence is the property called {\em Church-Rosser},
after the logicians who introduced the terminology in the
framework of the $\lambda$-calculus (cf. chapter \ref{lambda-sec}).

\begin{definition}[Church-Rosser]\index{Church-Rosser property}
A rewriting system $(A,\arrow)$ is {\em Church-Rosser} if 
for all $a,b\in A$, $a\lrtrarrow b$ implies $a\dcl b$.
\end{definition}

\begin{proposition}
A rewriting system is {\em Church-Rosser}  iff it is {\em confluent}.
\end{proposition}
\Proof
\Proofitemf{(\Rightarrow)}
If $a\trarrow b$ and $a\trarrow c$ then $b\lrtrarrow c$. 
Hence $\xst{d}{(b\trarrow d, c\trarrow d)}$.

\Proofitemm{(\Leftarrow)}
If $a\lrtrarrow b$ then $a$ and $b$ are connected by a finite sequence 
of `picks and valleys'. For instance:
\[ a \rtrarrow c_1 \ltrarrow c_2 \rtrarrow c_3 \cdots \ltrarrow c_n \rtrarrow b ~.\]
Using confluence, we can then find a common reduct. To show this,
proceed by induction on the number of picks and valleys. \qed \\

Let us look at possible interactions of the introduced properties.

\begin{proposition}
Let $(A,\arrow)$ be a rewriting system. 
\begin{enumerate}

\item If the rewriting system is confluent then every element
has {\em at most one} normal form.

\item If moreover the rewriting system is normalizing 
then every element has a {\em unique normal form}.
\end{enumerate}
\end{proposition}
\Proof
\Proofitemf{(1)} If an element reduces to two distinct normal 
forms then we contradict confluence.

\Proofitemm{(2)} By normalization, there exists a normal form and by (1)
there cannot be two different ones. \qed

\begin{exercise}\label{confluent-commute}
Let $(A,\arrow_1)$ and $(A,\arrow_2)$ be two rewriting systems.
We say that they {\em commute} if 
$a\trarrow_{1} b \mand a\trarrow_{2} c$ implies
$\xst{d}{(b\trarrow_2 {d} \mand c\trarrow_{1}d)}$.
Show that if $\arrow_1$ and $\arrow_2$ are {\em confluent} and {\em commute} 
then $\arrow_1\union \arrow_2$ is {\em confluent} too.
\end{exercise}

\section{Termination and well-founded orders}
Terminating rewriting systems and {\em well-founded orders} are
two sides of the same coin.

\begin{definition}[well-founded order]\index{partial order}\index{well-founded order}
A {\em partial order} $(P,>)$ is a set 
$P$ with a {\em transitive relation} $>$.
A partial order $(P,>)$ is {\em well-founded} if it 
is not possible to define a sequence $\set{x_i}_{i\in \Nat}\subseteq P$ such that:
\[
x_0 >x_1 > x_2 > \cdots
\]
\end{definition}

Notice that in a well-founded order we cannot have an element $x$
such that $x>x$ for otherwise we can define a sequence $x>x>x>\cdots$
(a similar remark concerned the definition of terminating rewriting system).

\begin{exercise}\label{non-wf-ex}
Let $\Nat$ be the set of natural numbers, $\Nat^{k}$ the cartesian product
$\Nat\times \cdots \times \Nat$, $k$-times, and
$A =  \Union\set{\Nat^{k} \mid  k\geq1 }$.
Let $>$ be a binary relation on $A$ such that : 
\[
\begin{array}{ll}
(x_1,\ldots,x_m) > (y_1,\ldots ,y_n) \quad \mbox{ iff }\quad
\xst{k}{(k\leq min(n,m), x_1=y_1,\ldots , x_{k-1}=y_{k-1}, x_k> y_k)}~.
\end{array}
\]
Prove or disprove the assertion that $>$ is a well-founded order.
\end{exercise}

Clearly, every well-founded partial order is a terminating rewriting
system if we regard the order $>$ as the reduction relation.
Conversely, every terminating rewriting system, say $(P,\arrow)$,
induces the well-founded partial order $(P,\stackrel{+}{\arrow})$ where 
$\stackrel{+}{\arrow}$ is the transitive (but {\em not} reflexive) 
closure of $\arrow$.

\begin{proposition}[induction principle]\label{ind-prop}
\index{induction principle}
Let $(P,>)$ be a well-founded partial order and for $x\in P$ let
$\downarrow(x)=\set{y \mid x>y}$.
Then the following {\em induction principle} holds:
\begin{equation}\label{induction-principle}
\infer{\qqs{x}{((\downarrow(x)\subseteq B) \mbox{ implies } x\in B)}}
{B=P}
\end{equation}
\end{proposition}
\Proof
If $x$ is {\em minimal} then the principle requires $x\in B$.
Otherwise, suppose $x_0$ is {\em not minimal} and $x_0\notin B$.
Then there must be $x_1<x_0$ such that $x_1\notin B$.
Again $x_1$ is {\em not minimal} and we can go on to build:
$x_0 > x_1 > x_2 > \cdots$
which {\em contradicts} the hypothesis that $P$ is well-founded. \qed

\begin{exercise}\label{ind-refl-ex}
  Explain why the induction principle (\ref{induction-principle})
  fails if $P$ is a singleton set and $>$ is
reflexive. 
\end{exercise}

\begin{remark}
On the {\em natural numbers} the induction principle can be
stated as:
\[
\begin{array}{c}
\infer{\qqs{n}{(\qqs{n'<n}{n'\in B}\mbox{ implies } n\in B)}}{\qqs{n}{n\in B}}~,

\end{array}
\]
which is equivalent to the 
usual reasoning principle:
\[
 \infer{0\in B \mbox{ and } (\qqs{n}{(n\in B\mbox{ implies } (n+1)\in B)})}{\qqs{n}{n\in B}}~.
\]
\end{remark}

We have shown that on a well-founded order the induction principle
holds. The converse holds too in the following sense.

\begin{proposition}
Let $(P,>)$ be a partial order for which the 
{\em induction principle} (\ref{induction-principle}) holds.
Then $(P,>)$ is {\em well-founded}.
\end{proposition}
\Proof
Define:
$Z=\set{x\in P \mid \mbox{there is no infinite descending chain from }x}$
and $\downarrow(x)=\set{y\mid y<x}$.
The set $Z$ satisfies the condition:
$\qqs{x}{(\downarrow(x) \subseteq Z \mbox{ implies }x\in Z)}$.
Hence by the induction principle $Z=P$.
Thus $P$ is well-founded. \qed

\section{Lifting well-foundation}\label{lifting-well-found-sec}
We examine three ways to lift an order to tuples so as to preserve 
{\em well-foundation}, namely the product order, the lexicographic
order, and the multi-set order.

\begin{definition}[product order]\index{order, product}\index{order, lexicographic}
Let $(P,>)$ be a partial order and let $P^n=P\times \cdots \times P$, 
$n$ times, be the cartesian product ($n\geq 2$).
The {\em product order} on $P^n$ is defined
by $(x_1,\ldots,x_n) >_p (y_1,\ldots,y_n)$ if:
\[
x_i \geq y_i, i=1,\ldots,n \ \mand \ \xst{j\in \set{1,\ldots,n}}{x_j>y_j}~.
\]
The {\em lexicographic order} (from left to right) on $P^n$ is defined by
$(x_1,\ldots,x_n) >_{\w{lex}} (y_1,\ldots,y_n)$ if:
\[
\xst{j\in\set{1,\ldots,n}}{x_1=y_1,\ldots, x_{j-1}=y_{j-1}, x_j>y_j} ~.
\]
\end{definition}

Notice that $(x_1,\ldots,x_n) >_p (y_1,\ldots,y_n)$ implies 
$(x_1,\ldots,x_n) >_{\w{lex}} (y_1,\ldots,y_n)$ but that the converse fails.

\begin{proposition}
If $(P,>)$ is well-founded and $n\geq 2$ then 
$(P^n,>_p)$ and $(P^n,>_{\w{lex}})$ are well-founded.
\end{proposition}
\Proof For the product order, 
suppose there is an infinite descending chain in the product order. 
Then one component must be strictly decreasing infinitely often which
contradicts the hypothesis that $(P,>)$ is well-founded.
As for the lexicographic order,  we proceed by induction on $n$. 
For the induction step, notice that the first component must eventually stabilize  and then apply induction on the remaining components. \qed \\

A third way to compare a finite collection of elements is to consider
them as {\em multi-sets} which we introduce next.

\begin{definition}[multi-sets]\index{multi-set}
A {\em multi-set} $M$ over a set $A$ is a function
$M:A \arrow \Nat$. 
If $M(a)=k$ then $a$ occurs $k$ times in the multi-set.
\end{definition}

\begin{definition}[finite multi-sets]\index{multi-set, finite}
A {\em finite} multi-set is a multi-set $M$ such that
$\set{a \mid M(a)\neq 0}$ is finite.
Let $\cl{M}_{\w{fin}}(X)$ denote the {\em finite multi-sets} over
a set $X$.
\end{definition}

\begin{definition}[multi-set replacement]
Assume $(X,>)$ is a partial order and $M,N\in \cl{M}_{\w{fin}}(X)$.
We write $M >_{1,m} N$ if $N$ is obtained from $M$ by {\em replacing} an element
by a finite multi-set of elements which are strictly smaller.
\end{definition}

\begin{example}
If $X=\Nat$ then 
$\mset{1,3} >_{1,m} \mset{1,2,2,1} >_{1,m} \mset{0,2,2,1}>_{1,m} \mset{0,1,1,2,1}$.
\end{example}

\begin{exercise}\label{cnt-ex-trans-1mset-ex}
Find a counter-example to the transitivity of the relation $>_{1,m}$.
\end{exercise}

\begin{definition}[multi-set order]\index{order, multi-set}
Let  $(X,>)$ be a partial order.
We define the {\em multi-set order} $>_m$ on  $\cl{M}_{\w{fin}}(X)$.
as the {\em transitive closure} of $>_{1,m}$.
\end{definition}

We want to show that if $(X,>)$ is well-founded then $>_m$ is well-founded.
First we recall a classical result known in the literature as {\em K\"onig's lemma}.

\begin{proposition}[König]\label{koenig-prop}\index{K\"onig lemma}
A finitely branching tree with an infinite number of nodes admits an
infinite path.
\end{proposition}
\Proof
First let us make our statement precise.
A {\em tree} can be seen as a subset $D$ of $\Nat^*$ 
(finite words of natural numbers) satisfying
the following properties.
\begin{enumerate}

\item  If $w\in D$ and $w'$ is a prefix of  $w$  then $w'\in D$.

\item If $wi\in D$ and $j<i$ then $wj\in D$.

\end{enumerate}

Notice that this representation is quite general in that it includes trees with a countable number of nodes
and even trees with nodes having a countable number of children
({\em e.g.}, $\Nat^*$ is a tree).
We say that a tree is {\em finitely branching} if every node has a finite
number of children (this is strictly weaker than being able to bound the number of children of every node!).

Now suppose $D$ is a finitely branching tree with infinitely many nodes. 
If $\pi\in \Nat^*$ let $\uparrow(\pi)$ be the set of paths that start with $\pi$.
We show that it is always possible to extend a path $\pi$ such that
$\uparrow(\pi)\inter D$ is infinite to a longer path $\pi\cdot i$ 
with the same property, {\em i.e.}, $\uparrow(\pi\cdot i)\inter D$ is infinite.
Indeed, the hypothesis that $D$ is finitely branching entails that there are finitely many 
$i_1,\ldots,i_k$ such that $\pi i_j\in D$. Since $\uparrow(\pi)\inter D$ is infinite one of these
branches, say $i$, must be used infinitely often. So we have that $\uparrow(\pi\cdot i)\inter D$ is infinite. \qed

\begin{proposition}
If $(P,>)$ is {\em well-founded}
then $(\cl{M}_{\w{fin}}(P),>_m)$ is well-founded.
\end{proposition}
\Proof 
By contradiction suppose we have an infinitely descending chain:
\[
X_0 >_m X_1 >_m \cdots
\]
Because $>_m$ is the transitive closure of $>_{1,m}$ this gives  
an infinitely descending chain:
\[
Y_0 >_{1,m} Y_1 >_{1,m} \cdots
\]
where $X_0=Y_0$.
By definition of $>_{1,m}$, the step from $Y_i$ to $Y_{i+1}$ consists
in taking an element of $Y_i$, say $y$, and replacing it by a {\em finite} multi-set
of elements $\mset{y_1,\ldots,y_k}$ which are strictly smaller. 
Suppose we have drawn a tree whose leaves correspond to the elements of $Y_i$
(if needed we may add a special root node).
Then to move to $Y_{i+1}$ we have to take a leaf of $Y_i$, 
which corresponds to the element $y$, 
and add $k$ branches labelled with the elements $y_1,\ldots,y_k$ (if $k=0$ we may just
add one branch leading to a special `sink node' from which no further expansion is possible).
The tree we build in this way is finitely branching and is infinite.
Then by K\"onig's lemma 
(proposition \ref{koenig-prop}) 
there must be an infinite path in it which corresponds to 
an infinitely descending chain in $(P,>)$. This is a contradiction 
since $(P,>)$ is
supposed to be well-founded. \qed 

\begin{exercise}\label{while-term-ex}
Does the evaluation of the following $\imp$ commands terminate
assuming initially a state where $m,n$  are positive natural numbers?
\[
\begin{array}{l}

\s{while} \ (m\neq n) \ \s{do}  \
(\s{if} \ (m>n)\ \s{then} \ m:=m-n; \ \s{else} \ n:=n-m;)~, \\  

\s{while} \ (m\neq n) \ \s{do} \ 
(\s{if} \ (m>n)\  \s{then } \ m:=m-n; \ \s{else} \ (h:=m;m:=n;n:=h;))~.

\end{array}
\]
\end{exercise}

\begin{exercise}\label{mon-emb-nat-ex}
Let $(A,\arrow)$ be a rewriting system and let $\Nat$ be the set
of natural numbers. A {\em monotonic embedding} is a function $\mu:A\arrow \Nat$
such that if $a\arrow b$ then $\mu(a)>_\Nat \mu(b)$.
Define the set of {\em immediate successors} of $a\in A$ as:
$\w{suc}(a)=\set{b\mid a\arrow b}$,
and say that $A$ is finitely branching if for all elements $a\in A$,
$\w{suc}(a)$ is a finite set.
Prove that:
(1) If a rewriting system has a monotonic embedding then it terminates.
(2) If a rewriting system is finitely branching and terminating then
it has a monotonic embedding.
(3) The following rewriting system $(\Nat \times \Nat,\arrow)$ where:
$(i+1,j) \arrow (i,k)$ and 
$(i,j+1) \arrow (i,j)$, for 
$i,j,k\in \Nat$, 
is terminating, {\em not} finitely branching, and {\em does not} have
a monotonic embedding.
\end{exercise}

\section{Termination and local confluence}
In general, it is hard to prove confluence because we have to consider 
arbitrary long reductions. It is much simpler to reason {\em locally}.

\begin{definition}[local confluence]\index{local confluence}
A rewriting system $(A,\arrow)$ is {\em locally confluent} if for all $a\in A$:
\[
\infer{\qqs{b,c\in A}{(b \leftarrow a \rightarrow c)}}{\xst{d\in A}{(b \rtrarrow d \ltrarrow c)}}~.
\]
\end{definition}

\begin{proposition}[Newman]\label{newman-proposition}\index{Newman}
If a rewriting system $(A,\arrow)$
 is {\em locally confluent} and {\em terminating} then it is {\em confluent}.
\end{proposition}
\Proof
We apply the principle of {\em well-founded induction} to 
$(A,\stackrel{+}{\arrow})$~!
Suppose: 
\[
c_1 \ltrarrow b_1 \leftarrow a \arrow b_2 \rtrarrow c_2~.
\]
By {\em local confluence}: $\xst{d}{(b_1 \rtrarrow d \ltrarrow b_2)}$.
Also, by {\em induction hypothesis} on $b_1$ and $b_2$ we have:
\[
\xst{d'}{(c_1 \rtrarrow d' \ltrarrow d)}~,
\qquad \xst{d''}{(d'\rtrarrow d'' \ltrarrow c_2)}~.
\]
But then $c_1 \dcl c_2$. 
Thus by the {\em principle of well-founded induction}, the rewriting
system is {\em confluent}. \qed

\begin{example}
Let $A=\Nat\union\set{a,b}$ and $\arrow$ such that for $i\in \Nat$:
$i\arrow i+1$, $2\cdot i \arrow a$, and $2\cdot i+1\arrow b$. 
This rewriting system is locally confluent and normalizing, but
{\em not} terminating and {\em not} confluent.
\end{example}

\begin{exercise}\label{rewriting-words-ex}
Let $\Sigma^*$ denote the set of finite words over the
alphabet $\Sigma=\set{f, g_1, g_2}$ with generic elements $w,w',\ldots$
As usual, $\epsilon$ denotes the empty word.
Let $\arrow$ denote the smallest binary relation on $\Sigma^*$
such that for all $w\in \Sigma^*$:
\[
\begin{array}{llllll}
(1)        &f g_1 w \ \arrow  \ g_1 g_1 f f w~,
&(2)       &f g_2 w \ \arrow \ g_2 f w~,  
&(3)       &f \epsilon    \ \arrow \ \epsilon~,
\end{array}
\]
and such that if $w \arrow w'$ and $a\in \Sigma$ then $aw \arrow aw'$. 
This is an example of {\em word} rewriting; a more general notion of 
{\em term} rewriting will be considered in the following section \ref{trs-sec}.
Prove or give a counter-example to the following assertions:
\begin{enumerate}
\item
If $w \trarrow w_1$ and $w \trarrow w_2$ then there exists $w'$ such that $w_1 
\trarrow w'$ and $w_2 \trarrow w'$.

\item
The rewriting system $(\Sigma^*,\arrow)$ is terminating.
\item
Replacing rule $(1)$ with the rule $\ f g_1 w \arrow g_1 g_1 f w$,
the answers to the previous questions are unchanged.
\end{enumerate}
\end{exercise}

\section{Term rewriting systems}\label{trs-sec}
When rewriting systems are defined on sets with structure,
we can exploit this structure, {\em e.g.}, to represent in a more succinct
way the reduction relation and to reason on its properties.
A situation of this type arises when dealing with
sets of {\em first-order terms} (in the sense of first-order logic).
Let us fix some notation. 
A  {\em signature} $\Sigma$ 
is a finite set of function symbols  $\set{f_1,\ldots,f_n}$ 
where each function symbol has an {\em arity}, $\w{ar}(f_i)$, which is
a natural number indicating the number of arguments
of the function.
Let $V$ denote  a countable set of {\em variables} with generic
elements $x,y,z,\ldots$  If $V'\subseteq V$ 
then $T_\Sigma(V')$ is the set of {\em first order terms} over the 
variables $V'$ with generic elements $t,s,\ldots$ (respecting the arity).
So  $T_\Sigma(V')$ is the least set which contains the variables $V'$ and
such that if $f\in \Sigma$, $n=\w{ar}(f)$, and 
$t_1,\ldots,t_n\in T_\Sigma(V')$ then $f(t_1,\ldots,t_n)\in T_\Sigma(V')$.
If $t$ is a term we denote with $\var{t}$ the set of {\em variables}
occurring in the term.

A natural operation we may perform on terms is to 
{\em substitute} terms for variables. \index{term substitution}
Formally, a {\em substitution} is a function $S:V\arrow T_\Sigma(V)$ 
which is the identity almost everywhere. We represent
with the notation $\ [t_1/x_1,\ldots,t_n/x_n] \ $
the substitution $S$ such that $S(x_i)=t_i$ for $i=1,\ldots,n$ and which
is the identity elsewhere. Notice that we always assume $x_i\neq x_j$ if
$i\neq j$. We use \w{id} to denote a substitution which is the identity
everywhere.
We extend $S$ to $T_\Sigma(V)$ by defining, for $f\in \Sigma$:
\[
S(f(t_1,\ldots,t_n)) = f(S(t_1),\ldots,S(t_n))\qquad\mbox{(extension of substitution to terms).}
\]
Thanks to this extension, it is possible to {\em compose} substitutions:
$(T\comp S)$ is the substitution defined by the equation:
\[
(T\comp S)(x) = T(S(x))\qquad\mbox{(composition of substitutions).}
\]
As expected, composition is associative and the identity substitution
behaves as a left and right identity: $\w{id}\comp S = S\comp \w{id}= S$.

\begin{example}
If $t=f(x,y)$, $S=[g(y)/x]$, and $T=[h/y]$ then:
\[
\begin{array}{c}
(T\comp S)(t) = T(S(t)) = T(f(g(y),y)) = f(g(h),h)~.
\end{array}
\]
\end{example}

Next we aim to define the reduction relation schematically 
exploiting the structure of first-order terms.
A {\em context} $C$ is a term with exactly one occurrence of a 
special symbol $[~]$ called {\em hole} and of arity $0$. 
We denote with $C[t]$ the term resulting from the replacement of the hole
$[~]$ by $t$ in $C$.
A {\em term-rewriting rule} (or rule for short)  is a pair of terms 
$(l,r)$ that we write $l\arrow r$ such that $\var{r}\subseteq \var{l}$;
the variables on the right hand side of the rule must occur on the
left hand-side too.

\begin{definition}[term rewriting system]\label{trs-definition}\index{term rewriting system}
A set of term rewriting rules $R=\set{l_1\arrow r_1,\ldots,l_n\arrow r_n}$,
where $l_i,r_i$, $i=1,\ldots,n$ are terms over some signature $\Sigma$,
induces a rewriting system $(T_\Sigma(V),\arrow_R)$ where $\arrow_R$
is the least binary relation such that is $l\arrow r\in R$ is a rule, 
$C$ is a context,
and $S$ is a substitution then:
\[
C[Sl] \arrow_R C[Sr]~.
\]
\end{definition}

\begin{example}
Assume the set of rules $R$ is as follows:
\[
\begin{array}{ccc}
f(x) 
 \ \arrow  \ g(f(s(x)))~,\qquad
&i(0,y,z) \ \arrow  \ y~,\qquad
&i(1,y,z) \ \arrow \ z~.
\end{array}
\]
Then, for instance:
\[
\begin{array}{lllllll}

f(s(y))     &\arrow_R &g(f(s(s(y)))) &\arrow_R &g(g(f(s(s(s(y)))))) &\arrow_R &\cdots \\ 

i(0,1,f(y)) &\arrow_R &1 \\

i(0,1,f(0)) &\arrow_R &i(0,1,g(f(s(0)))) &\arrow_R &\cdots
\end{array}
\]
\end{example}

There is a natural interplay between equational and term rewriting
systems. We illustrate this situation with a few examples.

\begin{example}\label{addition-sum-ex}
Suppose we have a set of equations dealing with natural numbers:
\[
\begin{array}{ll}
+(x,Z) = +(Z,x) =x,
&+(S(x),y) = +(x,S(y)) = S(+(x,y)), \\
+(+(x,y),z)  = +(x,+(y,z))~.
\end{array}
\]
Here the numbers are written in unary notation with a zero $Z$ and 
a successor $S$ function symbols,
and the equations are supposed to capture the behavior of a binary
addition symbol $+$. 
Now it is tempting to {\em orient the equations}
so as to simplify the expression. E.g.
$\ +(x,Z) \arrow x \ $,
but this is {\em not} always obvious! For instance, what is the orientation
of: 
\[
+(S(x),y) = S(+(x,y))\  \mbox{ or }\  +(+(x,y),z)  = +(x,+(y,z)) \ ?
\]
One proposal could be:
\[
\begin{array}{lll}
+(x,Z) \arrow x, \quad
&+(Z,x) \arrow x, \quad
&+(S(x),y) \arrow S(+(x,y)), \\
+(x,S(y)) \arrow S(+(x,y)), 
&+(+(x,y),z)  \arrow +(x,+(y,z))~. 
\end{array}
\]
Thus we have defined a {\em term rewriting system} 
and some interesting and natural questions arise.
Is there a reduction  strategy always leading
to a normal form?
Does {\em any} reduction strategy reach a normal form?
Suppose we apply different reduction strategies,
is it always possible to reach a common reduct?

In our case we are lucky. Termination (and therefore normalization)
is guaranteed. Moreover the system is confluent and therefore each
term has a unique normal form. These properties can be verified
automatically by state of the art tools dealing with term
rewriting systems. Once these properties are verified, we 
have a strategy to decide the equality of two terms: we
reduce the terms to their normal forms and check whether they
are identical.
\end{example}

\begin{example}\label{group-ex}
In this example we look at the equations of {\em group theory}:
\[
\begin{array}{ccccc}
 *(e,x) = x,  
& *(x,e) = x, 
& *(i(x),x) = e, 
& *(x,i(x)) = e, 
&*(*(x,y),z) = *(x,*(y,z))~.
\end{array}
\]
Here $e$ is the identity, $i$ is the inverse function, and 
$*$ is the binary operation of a group.
If we orient the equations from left to right we obtain
a term rewriting system and again automatic tools can 
check that the system is terminating.
However the system as it stands is {\em not} confluent.
In this case, a procedure known as {\em completion} tries to 
add rewriting rules to the system which are compatible with
the equations and preserve termination. 
A possible outcome of this analysis is to add the following rules:
\[
\begin{array}{lll}
i(e) \arrow e,\quad
&*(i(x),*(x,y)) \arrow y,\quad
&i(i(x)) \arrow x, \\

 *(x,*(i(x),y)) \arrow y~,       \quad  
&i(*(x,y)) \arrow *(i(y),i(x))~.

\end{array}
\]
\end{example}

The previous examples may give the impression that checking
termination and confluence is a task that can be automatized.
While this is true in many practical cases, the reader should
keep in mind that in general these properties
are undecidable. Term rewriting systems constitute a powerful
computational model and it is easy to reduce, {\em e.g.},
the halting problem for Turing machines to a termination 
problem for term rewriting systems.

\section{Summary and references}
We have shown that the following concepts are `equivalent':
(1) terminating rewriting system, 
(2) well-founded set, and (3)
partial order with well-founded induction principle.
Also, whenever working in a terminating rewriting system
we have shown that to prove confluence it suffices 
to prove local confluence.
We have also introduced the notion of term rewriting system
which is a way of presenting schematically a rewriting system
using first-order terms. Term rewriting systems are tightly
connected to equational theories and can provide 
procedures to decide when two expressions are equated.
The book \cite{BaaderNipkow99} is a standard and quite readable 
introduction to term rewriting.
Proposition \ref{koenig-prop} is a special case of a theorem due to K\"onig \cite{Koenig26} while proposition \ref{newman-proposition} is due to 
Newman \cite{Newman42}.

\chapter{Syntactic unification}\label{unification-sec}
\markboth{{\it Syntactic unification}}{{\it Syntactic unification}}

Syntactic unification is about solving equations on terms, or equivalently
on finite labelled trees.
We introduce some notation and terminology.
We write $t= s$ if the terms $t$ and $s$ are {\em syntactically equal}.
We define a {\em pre-order on substitutions} as follows:
\[
R\leq S		\IFF 	\xst{T}{T\comp R= S}~.
\]
Thus $R\leq S$ if $S$ is an {\em instance} of $R$ or, equivalently,
if $R$ is {\em more general} than $S$ (note that 
$\w{id}\leq S$, for any $S$).

\begin{exercise}\label{sub-preorder-ex}
Give an example of two substitutions $S,T$ such that:
$S\neq T$, $S\leq T$, and $T \leq S$.
\end{exercise}

A {\em system of equations} $E$ is a finite set of pairs
$\set{t_1=s_1,\ldots,t_n=s_n}$.
A substitution $S$ {\em unifies} a system of equations $E$, written
$S\models E$,
if $St=Ss$ (here $=$ means
identity on $T_\Sigma(V)$) for all $t=s\in E$.
Notice that we are {\em abusing notation} 
by using $=$ both for the identity on terms (semantic level) 
and for a constraint relation (syntactic level).

\begin{exercise}\label{comp-sub-ex}
Show that if $S$ is a substitution unifying the system
$\set{s_1=s_2, x=t}$ then $S$ unifies $\set{[t/x]s_1=[t/x]s_2}$ too.
\end{exercise}

\section{A basic unification algorithm}
A {\em basic algorithm} for unification 
is presented in table \ref{unification-algo} as 
a rewriting system over pairs $(E,S)$ and a special symbol $\bot$ 
(the symmetric rules for $(\s{vt}_i)$, $i=1,2$, are omitted).
This `abstract' presentation of the algorithm is instrumental to the 
proof of its properties. The idea is that we {\em transform} 
the system leaving the set of its solutions unchanged till either
the solution is explicit or it appears that no solution exists.
This is a standard methodology for solving systems of constraints,
{\em e.g.}, consider 
Gaussian elimination for solving systems of linear equations.

\begin{table}[b]
{\footnotesize
\[
\begin{array}{llll}

({\sf v})
&(E\union \set{x=x},S)	&\arrow (E,S) \\

({\sf vt}_1)
&(E\union \set{x=t},S)	&\arrow ([t/x]E,[t/x]\comp S) &\mbox{if }x\notin \var{t}  \\ 

({\sf vt}_2)
&(E\union \set{x=t},S)	&\arrow \bot		      &\mbox{if }x\neq t, x\in \var{t} \\

({\sf f}_1)
&(E\union \set{f(t_1,\ldots,t_n)=f(s_1,\ldots,s_n)},S) 
			&\arrow (E\union \set{t_1=s_1,\ldots,t_n=s_n},S) \\

({\sf f}_2)
&(E\union \set{f(t_1,\ldots,t_n)=g(s_1,\ldots,s_m)},S)
			&\arrow \bot		      &\mbox{if }f\neq g 
\end{array}
\]}
\caption{Unification algorithm} \label{unification-algo}\index{unification algorithm}
\end{table}

\begin{example}
Applying the unification algorithm to the system:
\[
\set{f(x)=f(f(z)),g(a,y)=g(a,x)}~,
\]
leads to the substitution:
$S= [f(z)/y]\comp [f(z)/x] = [f(z)/x]\comp [x/y] = [f(z)/x,f(z)/y]$.
\end{example}

\begin{exercise}\label{apply-unif-algo-ex}
Apply the unification algorithm to the systems of equations:
$\set{f(x,f(x,y)) = f(g(y),f(g(a),z))}$, $a$ constant, and 
$\set{f(x,f(y)) = f(y,f(f(x)))}$.
\end{exercise}

\section{Properties of the algorithm}
We analyse formally the unification algorithm.

\begin{proposition}
The following properties of the algorithm specified 
in table \ref{unification-algo} hold:
\begin{enumerate}

\item  The reduction relation $\arrow$ terminates.

\item If $(E,\w{id})\arrow^*(\emptyset,S)$ then $S$ unifies $E$.

\item If $T$ unifies $E$ then 
all reductions starting from $(E,\w{id})$ terminate with
some $(\emptyset,S)$ such that $S\leq T$.
\end{enumerate}
\end{proposition}
\Proof
\Proofitemf{(1)}
We define a measure on a set of equations as
$\mu(E)=(m,n)$ where pairs are lexicographically
ordered from left to right (cf. section \ref{lifting-well-found-sec}), 
$m$ is the number of variables in $E$,
and $n$ is the number of symbols in the terms in $E$.
The measure is extended to pairs $(E,S)$ and $\bot$ 
by defining $\mu(E,S)=\mu(E)$ and $\mu(\bot)=(0,0)$.
Then we check that $(E,S)\arrow U$ implies 
$\mu(E,S)>\mu(U)$.

\Proofitemm{(2)} We start with a preliminary remark.
In $(E,S)$, the second component
$S$ is just used to {\em accumulate} the substitutions.
Therefore:
\[
(E,S)\arrow^m (\emptyset, S_n\comp \ldots\comp S_1 \comp S) 
\IFF
(E,\w{id})\arrow^m (\emptyset, S_n\comp \ldots\comp S_1)~,
\]
where $m\geq 1$, $n\geq 0$ and the $S_i$ are the elementary
substitutions of the shape $[t/x]$ introduced by rule $({\sf vt}_1)$.
Next we prove the assertion by induction on the length of
the derivation. For instance, suppose:
\[
	(E\union\set{x=t},\w{id}) \arrow
	([t/x]E,[t/x]) \arrow^* (\emptyset, S\comp [t/x])
\]
Then, by the preliminary remark,
the inductive hypothesis applies to
$([t/x]E,\w{id})$. 
Thus $S\models [t/x]E$. Which entails
$S\comp [t/x]\models E$. Moreover, since $x\notin \var{t}$,
$S\comp [t/x](x) = S(t) = S\comp [t/x](t)$.

\Proofitemm{(3)}
By (1), all reduction sequences terminate.
We proceed by induction on the length of the reduction
sequence. We observe that if $E$ is not empty then
at least one rule applies. 
Since $T\models E$ it is easily
checked that rules $({\sf vt}_2)$ and $({\sf f}_2)$ do
{\em not} apply. 
Now suppose, for instance, that:
\[
(E\union\set{x=t},\w{id}) \arrow ([t/x]E,[t/x])
\]
applying rule $({\sf vt}_1)$. 
We recall (exercise \ref{comp-sub-ex}) that if 
$T\models E\union\set{x=t}$ 
then $T\models [t/x]E$ and $T=T \comp [t/x]$.
Then, from $T\models [t/x]E$ 
and the inductive hypothesis,  we conclude that
$([t/x]E,\w{id})\arrow^* (\emptyset,S)$ and $S\leq T$.
Hence: 
$S\comp [t/x] \leq T\comp [t/x] = T$. \qed

\begin{exercise}\label{size-solution}
Let the {\em size of a term} be the number of nodes in its tree
representation. Consider the following unification problem:
\begin{equation}\label{unif-prob}
\{x_1 = f(x_0,x_0), x_2 = f(x_1,x_1),\ldots, x_n = f(x_{n-1},x_{n-1})\}~.
\end{equation}
Compute the most general unifier $S$. 
Show that the size of $S(x_n)$ is exponential in $n$.
\end{exercise}

In view of exercise \ref{size-solution}, we could expect unification
algorithms to be hopelessly inefficient. However a closer look
at the solution of the unification problem (\ref{unif-prob}) 
reveals that the solution can be represented compactly as soon as 
we move from a tree representation to a directed acyclic graph ({\em dag}) 
representation.
This change of perspective allows to share terms and keep the size
of $S(x_n)$ linear in $n$.
Indeed, unification algorithms based on a dag representation
can be implemented to run in 
{\em quasi-linear} time.

\begin{exercise}\label{one-eq-one-binary-symbol-unif-ex}
Propose a method to transform a unification problem of the shape:
\[
E=\set{t_1=s_1,\ldots,t_n=s_n}
\]
over a signature $\Sigma=\set{g_1,\ldots,g_m}$ with $n,m\geq 1$ 
into a unification problem $E'$ with the following properties:

\begin{enumerate}

\item The problem $E'$ contains exactly one equation.

\item The terms in $E'$ are built over a signature  $\Sigma'$
containing exactly one binary symbol $f$.

\item The problem $E$ has a solution if and only if the problem
$E'$ has a solution.

\item  Apply the method to the system:
$E=\set{x=h(y), \quad g(c,x,z)= g(y,z,z)}$, where $x,y,z$ are  variables.
\end{enumerate}
\end{exercise}

\begin{exercise}\label{filter-ex}
Let  $t,s,\ldots$ be terms over a signature $\Sigma$. 
We say that $t$ is a {\em filter}  (or pattern) for $s$  if there is a 
substitution $S$ such that $St = s$. In this case we write: $t \leq s$. 
Show or give a counter-example to the following assertions:

\begin{enumerate}

\item If $t \leq s$ then $t$ and $s$ are unifiable.

\item If $t$ and $s$ are unifiable then $t\leq s$ and $s\leq t$.

\item If $t\leq s$ and $s\leq t$ then $s$ and $t$ are unifiable.

\item For all $t,s$ one can find $r$ such that  $r \leq t$ and $r\leq s$.

\item For all $t,s$ one can find $r$ such that $r \geq t$ and $r\geq s$.
\end{enumerate}
\end{exercise}

\section{Summary and references}
We have shown that there is a simple algorithm to solve the
unification problem on first-order terms. The algorithm either
shows that no solution exists or computes a {\em most general one}.
Moreover the algorithm is efficient as soon as terms are represented
as directed acyclic graphs.
The unification algorithm was brought to the limelight by Robinson's \index{Robinson} work on the resolution principle and its application to 
theorem proving \cite{DBLP:journals/jacm/Robinson65}.

\chapter{Termination of term rewriting systems}\label{termination-trs-sec}
\markboth{{\it Termination of TRS}}{{\it Termination of TRS}}

We introduce two methods to prove {\em termination of TRS}.
The {\em interpretation method}, where we regard the function 
symbols as certain strictly monotonic functions, and
the  {\em recursive path order} (RPO) method which 
is based on a syntactic criterion to compare terms.
We give two proofs that RPO's guarantee termination.
The first relies on {\em reducibility candidates}, 
a technique imported from proof theory,
and the second on the notion of {\em well-partial order} 
and a combinatorial result on the embedding of trees known as 
{\em Kruskal's theorem}. 
The interpretation and the RPO methods 
are examples of {\em reduction orders} which are defined as follows.

\begin{definition}[reduction order]\index{reduction order}
A {\em reduction order} $>$ is a {\em well-founded order} on
$T_\Sigma(V)$ that is {\em closed under context and substitution}:
\[
\infer{t>s}
{C[t]>C[s] \ , \ S t > S s}~,
\]
where $C$ is any one hole context and $S$ is any substitution.
\end{definition}

The notion of reduction order is quite general.

\begin{proposition}\label{red-order-prop}
A TRS $R$ {\em terminates} iff there is a {\em reduction order} $>$ such
that $l\arrow r\in R$ implies $l>r$. 
\end{proposition}
\Proof
\Proofitemf{(\Arrow)} 
If the system terminates then the transitive closure of the reduction relation provides a reduction order.

\Proofitemm{(\Leftarrow)} If we have a reduction order then well-foundedness enforces termination. \qed

\section{Interpretation method}\index{termination, interpretation method}
Suppose the TRS is given over a {\em signature} $\Sigma$.
Fix a {\em well-founded set} $(A,>)$ and assume that 
for each function symbol $f\in\Sigma$, with arity $n$,
we select a function $f^A: A^n\arrow A$ which is 
{\em strictly monotonic}. That is, for all $a_1,\ldots,a_n,a'_i$ if
$a'_i>a_i$ then
\[
f^A(a_1,\ldots,a_{i-1},a'_i,a_{i+1},\ldots,a_{n}) > f^A(a_1,\ldots,a_{i-1},a_i,a_{i+1},\ldots,a_{n})~.
\]
Now if we fix  an {\em assignment} $\theta: V\arrow A$,
for every $t\in T_\Sigma(V)$ there is a unique interpretation
in $A$  which is defined as follows:
\[
\begin{array}{ll}
 \sem{x}\theta = \theta(x)~, \qquad
&\sem{f(t_1,\ldots,t_n)}\theta =f^A(\sem{t_1}\theta,\ldots,\sem{t_n}\theta)~.
\end{array}
\]
Incidentally, this is the usual interpretation of terms in {\em first-order logic}: a term $t$ with variables $x_1,\ldots,x_n$ induces a function
$g_t:A^n\arrow A$ such that:
\[
g_t(a_1,\ldots,a_n)=\sem{t}[a_1/x_1,\ldots,a_n/x_n]~.
\]
In particular, a variable $x$ is interpreted as the identity function: 
$g_x(a) = \sem{x}[a/x]=a$.

\begin{proposition}
Under the hypotheses  described above, the interpretation
induces a {\em reduction order} $>_A$ on $T_\Sigma(V)$ defined by:
$t>_A s$  if $\ \qqs{\theta}{\sem{t}\theta >_A \sem{s}\theta}$.
\end{proposition}
\Proof
First, let us show $>_A$ is well founded. 
Suppose by contradiction:
\[
 t_0>_A t_1 >_A \cdots
\]
Then by taking an arbitrary assignment $\theta$ we have:
$\dl t_0 \dr\theta >_A \dl t_1 \dr\theta >_A \cdots$
But this contradicts the hypothesis that $(A,>)$ is well-founded.

Second, let us check that $>_A$ is preserved  by substitution.
Suppose $t>_A t'$. 
For any $s,x$ we show $[s/x]t>_A [s/x]t'$ 
(the  generalization to a substitution $[s_1/x_1,\ldots,s_n/x_n]$ is left
to the reader).
In other terms, we have to show that for any assignment $\theta$:
\[
     \dl [s/x]t \dr\theta >_A \dl [s/x]t' \dr \theta~.
\]
We note that: $\dl [s/x]t \dr \theta = \dl t \dr \theta[\dl s \dr \theta/x]$.
Thus taking $\theta'=\theta[\dl s \dr \theta/x]$ we have:
\[
     \dl [s/x]t \dr \theta = 
     \dl t \dr \theta ' >_A
     \dl t' \dr \theta' 
     = \dl [s/x]t' \dr \theta~.
\]
Third, we check that $>_A$ is preserved by contexts. 
To do this, we proceed by induction on the context.
The case for the empty context is immediate.
For the inductive step, suppose $C=f(\cdots, C',\cdots)$.
By inductive hypothesis, $C'[t]>_A C'[s]$ if $t>_A s$.
Then we conclude by using the fact that $f^A$ is strictly monotonic
in every argument. \qed

\begin{corollary}
Let $R$  be a TRS and $A$ be an interpretation as specified above. 
Then the  TRS terminates if for all $l\arrow r \in R$ we have:
$l>_A r$.
\end{corollary}
\Proof
We have shown that $>_A$ is a reduction order and we have previously
observed (proposition \ref{red-order-prop}) 
that a system is terminating if all its rules are compatible with
a reduction order. \qed

\begin{example}
Polynomial interpretations are an important and 
popular class of interpretations.
Take $A=\set{n\in \Nat \mid n\geq a\geq 1}$.
With $f^n\in\Sigma$ associate a {\em multivariate polynomial}
  $p_f(x_1,\ldots,x_n)$ such that:

\begin{enumerate}

\item Coefficients range over the natural numbers.
Thus there are {\em no negative coefficients} 
and the polynomials are {\em monotonic}. 

\item $p_f(a,\ldots,a)\in A$. Thus $p_f$ defines a function over the
      domain $A$.

\item  Every variable appears in a monomial with a non-zero multiplicative
      coefficient Thus we have  {\em strictly monotonic functions}.
\end{enumerate}

By extension, we associate with a term $t$ with variables $x_1,\ldots,x_n$ 
a multivariate polynomial $p_t$ with variables $x_1,\ldots,x_n$.
Notice that by taking $a\geq 1$, we make sure multiplication is 
a strictly monotonic function.
\end{example}

\begin{example}
Consider the following rules for {\em addition} and {\em multiplication}
over {\em natural numbers in unary notation}:
\[
\begin{array}{c}

a(\s{z},y) \arrow y~,
\qquad a(x,\s{z}) \arrow x~,
\qquad a(\s{s}(x),\s{s}(y))\arrow \s{s}(\s{s}(a(x,y))~,
 \\ 

m(\s{z},x) \arrow \s{z}~,
\qquad m(\s{s}(x),y) \arrow a(y,m(x,y))~.
\end{array}
\]
A polynomial interpretation showing the termination of this TRS is:
\[
\begin{array}{llll}
p_{\s{z}} = 1~,
&p_{\s{s}} = x+2~,
&p_a = 2x+y+1~,
&p_m = (x+1)(y+1)~.
\end{array}
\]
\end{example}

\begin{exercise}\label{poly-int-ex}
Find a polynomial interpretation showing the termination 
of the TRS:
\[
\begin{array}{ll}

f(f(x,y),z) \arrow f(x,f(y,z))~,
\qquad

&f(x,f(y,z)) \arrow f(y,y)~.
\end{array}
\]
\end{exercise}

\begin{exercise}\label{poly-double-exp}
(1) Find a polynomial interpretation for the TRS:
\[
\begin{array}{lll}
x+\s{0} \arrow x~,   &x+\s{s}(y) \arrow \s{s}(x+y)~,  &\mbox{(addition)}\\
d(0)\arrow 0~,   &d(\s{s}(x))\arrow \s{s}(\s{s}(d(x)))~,  &\mbox{(double)}\\
q(\s{0})\arrow \s{0}~,
\qquad & q(\s{s}(x))\arrow q(x)+\s{s}(d(x)) &\mbox{(square).}
\end{array}
\]
\Defitem{(2)} Consider the term $t\equiv q^{n+1}(\s{s}^2 0)$ whose size is linear in $n$.
Show that there is a reduction:
\[
t \trarrow q(\s{s}^{2^{2^{n}}}(0)),
\]
and derive from this fact the existence of a reduction from $t$
whose length is doubly exponential in $n$.
\end{exercise}

While polynomial interpretations are a conceptually
simple method to prove termination the reader should keep
in mind that they suffer of a couple of limitations.
First, polynomial interpretations are hard to find.
Indeed in general even checking 
whether a polynomial interpretation
is valid is {\em undecidable}.
This follows from the undecidability of the so-called 
Hilbert's $10^{\w{th}}$ problem\index{Hilbert}. This is the problem 
of recognizing the multivariate polynomials with integers
coefficients which have a zero.
The problem was stated in 1900, and finally in 1970
Matiyasevich\index{Matiyasevich} proved that the problem is undecidable.
Second, polynomial interpretations cannot  handle fast growing functions. 
Indeed it can be shown that the length of reductions 
of TRS proven terminating by a polynomial interpretation 
can be at most {\em double exponential}. 
Exercise \ref{poly-double-exp}(2) provides a lower bound, and 
the upper bound is not too hard to obtain. 
In theory one could then consider interpretations based on faster
growing functions such as exponentials, towers of exponentials,$\ldots$
however in practice most automatic systems just look for {\em low degree}
polynomial interpretations.

\section{Recursive path order}
Recursive path orders are a family of reduction orders which
are defined by induction on the structure of the terms.
The way to compare terms is rather simple. First we assume
a strict partial order $>_\Sigma$ on the function symbols in $\Sigma$
(since $\Sigma$ is supposed finite, $>_\Sigma$ is well-founded).
If $f>_\Sigma g$, proving that:
\[
t=f(t_1,\ldots,t_n)>_{r} g(s_1,\ldots,s_m)=s~,
\]
reduces to proving:
$t >_{r} s_i \mbox{ for }i=1,\ldots,m$.
On the other hand, proving that:
\[
t=f(t_1,\ldots,t_n)>_{r} f(s_1,\ldots,s_m)=s~,
\]
reduces to proving that:
$(t_1,\ldots,t_n)>_{r} (s_1,\ldots,s_m)$,
according to one of the orders that preserve well-foundation we
have considered in chapter \ref{rewrite-sec}, namely product order,
lexicographic order, or multi-set order.
What we have described is {\em almost} the official definition
of recursive path order which is given in Table \ref{rpo-table}.

\begin{table}
{\small
\[
\begin{array}{lc}

(R_1) &\infer{s\geq_r t}
{f(\ldots s\ldots) >_r t} \\ \\

(R_2) &\infer{f>_{\Sigma}g~~~f(s_1,\ldots,s_m)>_r t_i ~~i=1,\ldots,n}
{f(s_1,\ldots,s_m) >_r g(t_1,\ldots,t_n)}   \\ \\

(R_3) &\infer{
\begin{array}{c}
(s_1,\ldots,s_m) >_{r}^{\tau(f)} (t_1,\ldots,t_m)\\
f(s_1,\ldots,s_m)>_r t_i ~~i=1,\ldots,m
\end{array}}
{f(s_1,\ldots,s_m) >_r f(t_1,\ldots,t_m)} 

\end{array}
\]}
\caption{Recursive path-order}\label{rpo-table}\index{recursive path-order}
\end{table}

In this definition, we assume that every function symbol $f$ is
assigned a status $\tau(f)$ which determines how $f$'s arguments
are to be compared (product, lexicographic, multi-set,$\ldots$)
Indeed, this is necessary to guarantee termination. For instance,
consider the non-terminating TRS:
\[
\begin{array}{ll}
f(a,b) \arrow f(b,a)~,
&f(b,a) \arrow f(a,b)~,
\end{array}
\]
with $\Sigma=\set{f,a,b}$. Assume $a>_\Sigma b$. If $f$'s arguments 
could be compared with a lexicographic order from left to right {\em or}
from right to left then we could prove both $f(a,b)>_r f(b,a)$ {\em and}
$f(b,a)>_r f(a,b)$.

\begin{exercise}\label{rpo-term-ex}
Consider the TRS:
\[
 \begin{array}{llllll}
(x+y)+z~         &\arrow &x+(y+z)~,  \qquad
&x * \s{s}(y)  & \arrow &x+(y*x)~.
\end{array}
\]
Find a status for the function symbols that allows to prove:
\[
 \begin{array}{llllll}
(x+y)+z         &>_r &x+(y+z)~,  \qquad
&x * \s{s}(y)  &>_r &x+(y*x)~.
\end{array}
\]
\end{exercise}

Another point that deserves to be stressed is that in the rule $(R_3)$
we also require that the term on the left is larger than all the arguments
of the term on the right. To see the necessity of this condition,
consider the non-terminating TRS:
\[
f(a,y) \arrow f(b,f(a,y))~,
\]
where $\Sigma=\set{f,a,b}$, $a>_\Sigma b$, and the status of $f$
is lexicographic from left to right. 

Finally, we notice that there is an additional rule (rule $R_1$)
that entails that a term is larger than all its proper subterms.
This `subterm property' is characteristic of an important class
of orders known as {\em simplification order} that we define next.

\begin{definition}[simplification order]\index{simplification order}
A {\em strict} order $>$ on $T_\Sigma(V)$ is a {\em  simplification order} 
if it is {\em closed under context and substitution} and 
moreover for all functions $f\in \Sigma$ it satisfies:
\[
	f(x_1,\ldots,x_n) > x_i \mbox{ for }i=1,\ldots,n~.
\]
\end{definition}

\begin{exercise}\label{simp-order-cxt-ex}
Show that if  $>$ is a simplification order and $C$ is a one hole context with $C\neq [~]$ then $C[t] > t$.
\end{exercise}

We prove next that the recursive path order is a simplification order.
Further it will be proven in section \ref{simp-wf-sec} that every simplification
order is well-founded. This proof relies on a classical combinatorial
argument known as {\em Kruskal's theorem}.
This is enough to guarantee that the recursive
path order is a {\em reduction order} and therefore 
can be used to prove the termination of TRS. 
We will also give in section \ref{rpo-wf-sec} a direct proof of the
fact that the recursive path order is well-founded 
that avoids the detour through
Kruskal's theorem by using a so called {\em reducibility argument}
(a standard method to prove termination of typed $\lambda$-calculi introduced in chapter \ref{simple-type-sec}).

\begin{proposition}
The {\em recursive path order} is a {\em simplification order}
on $T_\Sigma(V)$. 
\end{proposition}

\Proof To fix the ideas, we consider a particular case where we 
always compare tuples via the 
{\em product order}. We prove the following properties:
(1)  $>$ is strict,
(2) $s>t$ implies $\var{s}\supseteq \var{t}$,
(3) transitivity,
(4) subterm property,
(5) closure under substitution, and
(6) closure under context.

Before proceeding, we formulate  in Table \ref{rpo-product-tab}
a simplified definition of recursive path order for
functions having {\em product status}.
Notice that in $(R_3)$ we drop the condition
$f(s_1,\ldots,s_m)>_r t_i$ for $i=1,\ldots,m$.
It turns out that in this case the condition can be derived
from the transitivity property and the fact that for 
$i=1,\ldots,m$: $f(s_1,\ldots,s_m) >_r s_i \geq_r t_i$.

\begin{table}
{\footnotesize
\[
\begin{array}{lc}

(R_1)
&\infer{s\geq_r t}
{f(\ldots s\ldots) >_r t} \\ \\

(R_2)
&\infer{f>_{\Sigma} g~~~f(s_1,\ldots,s_m)>_r t_i ~~i=1,\ldots,n}
{f(s_1,\ldots,s_m) >_r g(t_1,\ldots,t_n)}   \\ \\

(R_3)
&\infer{%f\approx g~~~
s_i \geq t_i\mbox{ for }i\in \set{1,\ldots,m}, \mbox{ and }
\xst{j\in \set{1,\ldots,m}}{s_j>t_j}}
{f(s_1,\ldots,s_m) >_r f(t_1,\ldots,t_m)} ~.

\end{array}
\]}
\caption{RPO, for functions with product status}\label{rpo-product-tab}
\end{table}

\begin{description}

\item[$>$ is strict]
By induction on $s$ show that $s>s$ is impossible.
Note in particular that $x>t$ and $f>f$ are impossible.

\item[$s>t$ implies $\var{s}\supseteq \var{t}$.]
By induction on the proof of $s>t$.

\item[Transitivity]
Suppose $s_1>s_2$ and $s_2>s_3$. Show
$s_1>s_3$ by induction on $|s_1|+|s_2|+|s_3|$
analyzing the last rules applied in the proof
of $s_1>s_2$ and $s_2>s_3$ ($9$ cases).

\item[Subterm property]
Check that $f(x_1,\ldots,x_n)>x_i$ for $i=1,\ldots,n$.

\item[Closure under substitution]
Show that $t>r$ implies $[s/x]t>[s/x]r$ by induction on
$|t|+|r|$.

\item[Closure under context]
Show by induction on the structure of a one hole context that 
$t>s$ implies $C[t]>C[s]$. \qed
\end{description}

\begin{exercise}\label{ackermann-ex}
Consider the following TRS:
\[
 \begin{array}{llllll}

\w{ack}(\s{z},n) &\arrow &\s{s}(\s{z})~,

&\w{ack}(\s{s}(\s{z}),\s{z})    &\arrow &\s{s}^2(\s{z})~, \\

\w{ack}(\s{s}^2(m),\s{z})  &\arrow &\s{s}^2(m)~,

&\w{ack}(\s{s}(m),\s{s}(n)) &\arrow &\w{ack}(\w{ack}(m,\s{s}(n)),n)~.
 
\end{array}
\]
This TRS corresponds to a very fast growing function known as
Ackermann's function\index{Ackermann}. For instance, this function
grows faster than any tower of exponentials\footnote{Technically,
Ackermann showed that this function cannot be defined by {\em primitive recursion}.} and no polynomial interpretation can prove its termination.
In practice, running ${\it ack(4,4)}$ will produce an {\em out-of-memory} 
exception on most computers. 
Prove the termination by RPO. 
\end{exercise}

\begin{exercise}\label{poly-beats-rpo-ex}
The previous exercise \ref{ackermann-ex} marks a point for
RPO. However, sometimes 
the (polynomial) interpretation method beats the RPO method.
Consider the TRS:
\[
\begin{array}{llllll}
  b(x) &\arrow &r(\s{s}(x))~,\qquad     &r(\s{s}(\s{s}(x))) &\arrow &b(x)~.
\end{array}
\]
(1) Show that the TRS terminates by polynomial interpretation. 
(2) Show that there is no RPO on $\Sigma$ that can prove its termination.
(3) RPO is a particular type of simplification order. 
Is there a simplification order that shows termination of the TRS above?
\end{exercise}

\begin{exercise}\label{no-simp-ord-ex}
The previous exercise \ref{poly-beats-rpo-ex} shows that 
the termination of certain TRS cannot be proven by RPO. 
It turns out that using an arbitrary simplification order
does not change this state of affairs.
Consider the TRS:
\[
f(f(x))\arrow f(g(f(x)))~.
\]
(1) Show that the TRS is terminating.
(2) Show that there is no simplification order $>$ that contains $\arrow$.
\end{exercise}

We terminate with a few remarks concerning the complexity of working
with RPO. Once the order on the signature and the status of the function
is fixed, deciding whether $t>_r s$ can be done in 
{\em time polynomial} in the size of the terms.
However, it is possible to come out with rather artificial examples
where the choice of the order on the signature is not obvious.
In fact it can be shown that deciding whether $t>_r s$ with respect
to {\em some order} on the signature is an {\em {\sc np}-complete} problem.

\section{Recursive path order is well-founded (*)}\label{rpo-wf-sec}
We know that RPO is a {\em simplification order}, {\em i.e.}, a strict
order, closed under context and substitution.
We want to show that it is {\em well-founded} (and therefore a
{\em reduction order}).
To this end, we apply the {\em reducibility candidates method}: 
a proof technique 
developed first to prove termination of typed $\lambda$-calculi.
To simplify the argument, we shall assume that function
arguments are always compared with the lexicographic order from left to right.
The corresponding 
specialized definition of RPO is given in Table \ref{RPO-lex-left-right}.

\begin{table}
{\footnotesize
\[
\begin{array}{c}

\infer{s\geq_r t}
{f(\ldots s\ldots) >_r t} \\ \\

\infer{f>_{\Sigma}g~~~f(s_1,\ldots,s_m)>_r t_i ~~i=1,\ldots,n}
{f(s_1,\ldots,s_m) >_r g(t_1,\ldots,t_n)}   \\ \\

\infer{
\begin{array}{c}
(s_1,\ldots,s_m) >_{r}^{\w{lex}} (t_1,\ldots,t_m)\\
f(s_1,\ldots,s_m)>_r t_i ~~i=1,\ldots,m
\end{array}}
{f(s_1,\ldots,s_m) >_r f(t_1,\ldots,t_m)} 

\end{array}
\]}
\caption{RPO for functions with lexicographic, left-to right status} \label{RPO-lex-left-right}
\end{table}

\begin{definition}
We work on the set of terms $T_\Sigma(V)$ and define:
\[
\begin{array}{lll}
\w{WF}  &= &\set{t\in T_\Sigma(V) \mid \mbox{ there is no infinite sequence } t=t_0 >_r t_1 >_r \cdots}~, \\
\w{Red}(t) &=&\set{s \mid t>_r s}~.
\end{array}
\]
\end{definition}

\begin{exercise}\label{basic-prop-wf}
Show that:
\begin{enumerate}

\item $(\w{WF},>_r)$ is a well-founded set.

\item If $\w{Red}(t)\subseteq \w{WF}$ then $t\in \w{WF}$.

\item If $s\in \w{WF}$ and $s>_r t$ then $t\in \w{WF}$.
\end{enumerate}
\end{exercise}

Let $>_{r}^{lex}$ be the lexicographic ordered induced by $>_r$ on vectors of 
$n$ terms in $\w{WF}$.  The {\em key property} follows.

\begin{proposition}\label{key}
If $s_1,\ldots,s_n\in \w{WF}$ and $f(s_1,\ldots,s_n)>_r t$ then $t\in \w{WF}$. 
\end{proposition}
\Proof By {\em induction on the triple}:
\[
(f,(s_1,\ldots,s_n),|t|)~,
\]
with the lexicographic order from left to right where:

\begin{itemize}

\item The first component is a function symbol ordered by  $>_\Sigma$.

\item The third is the size of the term with the usual order on natural numbers.

\item For the second, consider the set $\Union_{f\in \Sigma} \w{WF}^{\w{ar}(f)}$
ordered by:
\[
(s_1,\ldots,s_n) > (t_1,\ldots,t_m) 
\mbox{ iff }
n=m \mbox{ and } (s_1,\ldots,s_n) >_{r}^{lex} (t_1,\ldots,t_m)~.
\]
Notice that two vectors of different lengths are {\em incomparable}.
Also, $(WF,>_r)$ well-founded implies 
$(WF^n,>_{r}^{lex})$ is well-founded too.
\end{itemize}

\begin{description}

\item[Case $f(s_1,\ldots,s_n)>_r t$ as $s_i=t$ or $s_i>_r t$.]~

\begin{itemize}

\item If $s_i=t$ the conclusion is immediate as $s_i\in \w{WF}$ by hypothesis.

\item If  $s_i>_r t$ then $t\in \w{WF}$ as $s_i \in \w{WF}$.

\end{itemize}

\item[Case $t=g(t_1,\ldots,t_m)$, $f>_\Sigma g$, $f(s_1,\ldots,s_n)>_r t_i$ for $i=1,\ldots,m$.]~

\begin{itemize}

\item We notice that $(f, (s_1,\ldots,s_n), |t|) > (f, (s_1,\ldots,s_n), |t_i|)$ for 
$i=1,\ldots, m$. Hence, by inductive hypothesis, $t_i\in \w{WF}$. 

\item Suppose $g(t_1,\ldots,t_m) >_r u$. We remark $(f,(s_1\ldots,s_n),|t|)> (g,(t_1,\ldots,t_m),|u|)$.
Hence, by inductive hypothesis, $u\in \w{WF}$, 
and by exercise \ref{basic-prop-wf}, $g(t_1,\ldots,t_m)\in \w{WF}$.

\end{itemize}

\item[Case $t=f(t_1,\ldots,t_n)$, $f(s_1,\ldots,s_n)>_r t_i$ for $i=1,\ldots,n$,
$(s_1,\ldots,s_n)>_{r}^{lex} (t_1,\ldots,t_n)$.]~
This case is similar to the previous one.

\begin{itemize}

\item  We remark that
$(f, (s_1,\ldots,s_n), |t|) > (f, (s_1,\ldots,s_n), |t_i|)$ for
$i=1,\ldots, n$. Hence, by inductive hypothesis, $t_i\in \w{WF}$. 

\item Suppose $f(t_1,\ldots,t_n) >_r u$. We notice $(f,(s_1\ldots,s_n),|t|)> (f,(t_1,\ldots,t_n),|u|)$
(second component decreases!).
By inductive hypothesis, $u\in \w{WF}$, and by exercise \ref{basic-prop-wf}, $f(t_1,\ldots,t_n)\in \w{WF}$. \qed

\end{itemize}
\end{description}

\begin{corollary}
All terms are in $\w{WF}$.
\end{corollary}
\Proof By induction on the structure of the terms. \qed

\section{Simplification orders are well-founded (*)}\label{simp-wf-sec}
We prove that all simplification orders (in particular RPO) 
are well-founded.
As already mentioned, the proof goes through a classical combinatorial
result known as {\em Kruskal's theorem}. This result concerns a 
natural binary relation on terms (or labelled trees),
known as {\em homeomorphic embedding}, that we denote $\semb$;
we also denote with $\emb$ the reflexive closure of $\semb$.
The appearance of the embedding relation is justified by the simple observation
that every simplification order contains it.

Kruskal's theorem states that when considering the embedding relation on 
the collection of terms built out of a finite signature and a finite
set of variables there is no infinite descending chain 
$t_0 \semb t_1 \semb \cdots$ (the order is well-founded) and moreover
it is not possible to find an infinite set of terms which are all
incomparable (an infinite anti-chain). Technically, one says 
that the collection of terms with the embedding relation 
is a {\em well partial order} (wpo).

\begin{definition}[homeomorphic embedding]\index{homeomorphic embedding}
Let $\arrow$ be the TRS induced by the rules:
\[
	f(x_1,\ldots,x_n)\arrow x_i \mbox{ for }i=1,\ldots,n~.
\]
We write $t \emb s$, read $t$ embeds $s$, if
$t\trarrow s$, {\em i.e.}, if we can rewrite $t$ in $s$ in a
finite number of steps (possibly $0$).
\end{definition}

\begin{example}
Here is an example of homeomorphic embedding:
\[
 f(f(h(a),h(x)),f(h(x),a)) \emb f(f(a,x),x)~.
\]
\end{example}

\begin{exercise}\label{alternative-def-hom-emb-ex}
Here is another definition of homeomorphic embedding:
\[
\begin{array}{ccc}
\infer{~}{x \emb x}~, \qquad

&\infer{s_i\emb t_i, i=1,\ldots,n}{f(s_1,\ldots,s_n) \emb
 f(t_1,\ldots,t_n)} ~,
\qquad

&\infer{s_i \emb t \mbox{ for some }i}{f(s_1,\ldots,s_n)\emb t}~.

\end{array}
\]
Check that this definition is equivalent to the previous one.
\end{exercise}

\begin{exercise}\label{hom-emb-simp-order-ex}
Show that if $>$ is a simplification order and $\geq$ is its reflexive closure then $t\emb s$  implies $t\geq s$ (in other terms, if $t\emb s$ and $t\neq s$ then
$t>s$).
\end{exercise}

\begin{exercise}[Dickson]\label{Dickson-ex}\index{Dickson}
We consider a relatively simple situation, known as Dickson's lemma,
where we have a well-founded order and moreover all sets of 
incomparable elements are finite.
Consider the {\em product order} $\geq$ on $\Nat^k$ (vectors of natural numbers):
\[
(n_1,\ldots,n_k)\geq (m_1,\ldots,m_k) \mbox{ if }n_i\geq m_i, i=1,\ldots,k~.
\]
\begin{enumerate}

\item Show that $>$ (the strict part of $\geq$) is well-founded.

\item Show by induction on $k$, that from every sequence  
$\sequence{v_n}{n\in \Nat}$ in $\Nat^k$ we can extract a 
  {\em growing subsequence}, namely numbers $i_0<i_1<i_2<\cdots$ such that
  for all $n$, $v_{i_{n}} \geq v_{i_{n+1}}$.

\item Show that every set of incomparable elements in $\Nat^k$ (an {\em anti-chain}) is finite.

\end{enumerate}
\end{exercise}

\begin{definition}[well partial order]\index{well partial order}
A {\em well partial order} 
$(A,>)$ is a strict ($\qqs{a}{a \not> a}$) partial order such that  for any sequence $\set{a_i\mid i\in \Nat}$ in $A$,
\[
\xst{i,j\in \Nat} {j>i \mand a_j \geq a_i}~.
\] 
Such a sequence is called {\em good}.
Otherwise, we call the sequence {\em bad}. This means:
\[
\qqs{i,j\in \Nat}{(j>i \mbox{ implies } a_j\not\geq a_i)}~.
\]
Note that if a sequence is bad all its subsequences are.
\end{definition}

\begin{remark}
In this chapter a partial order by default is {\em strict}.
The reflexive closure of a well partial order is called a
{\em well quasi-ordering} (wqo).
\end{remark}

\begin{proposition}
Well partial orders are the {\em well-founded orders} that have
{\em no infinite anti-chain}.
\end{proposition}
\Proof
\Proofitemf{(\Arrow)}
A wpo must be well-founded for a strictly descending chain gives a bad sequence.
For the same reason, a wpo cannot contain an infinite anti-chain.

\Proofitemm{(\Leftarrow)}
Vice versa, take a well-founded set without infinite anti-chain.
Given an infinite sequence, the set of minimal elements of the sequence
must be finite. Therefore there is a minimal element such that the sequence 
is infinitely often above it. \qed

\begin{proposition}\label{extract-ascending-wpo}
Given a sequence in a wpo, it is always possible to extract an ascending
subsequence.
\end{proposition}
\Proof
Consider a sequence $\sequence{a_i}{i\in \Nat}$.
We want to show that there is  an {\em ascending subsequence}:
\[
i_1<i_2<i_3<\ldots   \mand  a_{i_{1}} \leq a_{i_{2}} \leq a_{i_{3}} \leq \cdots
\]
We notice that in a good sequence there are {\em finitely many} $a_i$ such that
$\qqs{j}{j>i \mbox{ implies } a_j \not\geq a_i}$. Otherwise, the sequence composed  of all such elements is {\em bad}.
Thus starting from a certain point $i_1$, if $i\geq i_1$ then 
$\xst{j>i}{a_j\geq a_i}$. Now starting from $i_1$ we can 
inductively build a sequence 
$i_1<i_2<\ldots$ such that $a_{i_{1}} \leq a_{i_{2}} \leq \cdots$ \qed

\begin{proposition}\label{prod-wpo-prop}
The {\em product} $A\times B$ of wpo's $A,B$, ordered component-wise (product order) is a wpo.
\end{proposition}
\Proof
Consider $\set{(a_i,b_i)\mid i\in \Nat}$ and suppose
$\set{a_i\mid i\in \Nat}$ and $\set{b_i\mid i\in \Nat}$ are
both infinite (otherwise it is easy).
Then consider the subsequence $i_0<i_1<i_2<\cdots$ such that
$a_{i_{0}} \leq a_{i_{1}} \leq a_{i_{2}}\leq \cdots$ (cf. previous
       proposition \ref{extract-ascending-wpo}).
Then find $k>l$  such that $b_{i_{k}}\geq b_{i_{l}}$. \qed \\

Recall that $\emb$ is the {\em homeomorphic embedding} and that the
{\em strict part} of $\emb$, say $\semb$, is contained in every
simplification order.

\begin{proposition}[Kruskal]\label{kruskal-prop}\index{Kruskal}
Suppose $\Sigma$ and $V$ {\em finite}. Then
the {\em strict homeomorphic embedding} $\semb$ on $T_\Sigma(V)$ 
is a {\em well partial order}. 
\end{proposition}
\Proof
We pause to notice that if $\Sigma$ or $V$ are 
{\em infinite} then $(T_\Sigma(V),\semb)$ contains
an {\em infinite anti-chain} and the proposition does {\em not} hold.
The proof proceeds by {\em contradiction}. 
Suppose there is a {\em bad sequence}
in $T_\Sigma(V)$.
Extract from the bad sequence a {\em minimal}
one with respect to the size of the terms, 
say $t_1,t_2,\cdots$ This means that 
having built the  sequence $t_1,\ldots,t_i$, we pick a term
$t_{i+1}$ of minimal size among those that follow $t_i$.
Define:
\[
\begin{array}{l|l}
S_i = \left\{ \begin{array}{ll}
      \emptyset               &\mbox{if }t_i \mbox{ variable} \\
      \set{s_1,\ldots,s_n}    &\mbox{if }t_i =f(s_1,\ldots,s_n) 
       \end{array}\right.    

&S=\Union_{i\geq 0} S_i~.
\end{array}
\]
For the time being, assume  $(S, \semb)$  is a {\em wpo};
this is a tricky point whose proof is postponed.
Since $\Sigma$ and $X$ are {\em finite}, there must
be a symbol that occurs infinitely often as the root of the
minimal bad sequence $t_1,t_2,\ldots$
If it is a variable or a constant we derive a contradiction.
Otherwise, we have $i_0<i_1<\ldots$ with:
\[
t_{i_{k}} = f(s_{1}^{i_{k}}, \ldots, s_{n}^{i_{k}})~.
\]
Now $(S,\semb)$ is a wpo and the product of wpo's is a wpo (proposition 
\ref{prod-wpo-prop}).
Therefore, the sequence:
\[
\sequence{(s_{1}^{i_{k}}, \ldots, s_{n}^{i_{k}})}{k\geq 0}
\]
is {\em good}.
So $\xst{p,q}{q>p \mand s_{l}^{i_{q}} \emb s_{l}^{i_{p}}, l=1,\ldots,n}$.
And this entails $t_{i_{q}} \emb t_{i_{p}}$. Contradiction!

We now come back to the tricky point.
Suppose $(S,\semb)$ is {\em not} a wpo, 
and let $s_1,s_2,s_3\ldots$ be a {\em bad} sequence.
The $s_i$ must be all distinct.
Suppose $s_1\in S_k$. This entails $t_k\semb s_1$.
Let $S_{<k} = S_1\union \cdots \union S_{k-1}$.
There is an index $l$ such that $s_i\notin S_{<k}$ for $i\geq l$.
Consider:
\[
 t_1,\ldots,t_{k-1},s_1, s_l,s_{l+1},\ldots
\]
Since $s_1$ is smaller than $t_k$, {\em by minimality} (!)
this sequence must be {\em good}.
Since $t_1,t_2,\ldots$ and $s_1,s_2,\ldots$ are
{\em bad}, this entails
$s_j \emb t_i$ for some $i\in \set{1,\ldots,k-1}$ and 
$j\in\set{1,l,l+1,\ldots}$.
We distinguish two cases, both leading to a contradiction.
\begin{description}

\item[$j=1$] $t_k \semb s_j \emb t_i$. Contradiction!

\item[$j\geq l$] Suppose $s_j\in S_m\minus S_{<k}$.
Thus $m\geq k > i$ and $t_m\semb s_j \emb t_i$. Contradiction! \qed

\end{description}

\begin{remark}
The presented result is an interesting case study for logicians.
First, the proof we have presented is {\em non-constructive} 
(two nested arguments by contradiction). 
The literature contains proposals for constructive versions of
the proof.
Second, the theorem is a simple example of a combinatorial statement that
{\em cannot be proved} in Peano's Arithmetic (a standard formalization of
arithmetic in first-order logic).
\end{remark}

\begin{exercise}[Higman]\label{Higman-ex}
The following is a special case of Kruskal's theorem on {\em words}
known as {\em Higman's lemma}\index{Higman}.
Let $\Sigma$ be a finite set (alphabet).
Given two words $w,w'\in \Sigma^*$ we 
say that $w'$ is a {\em subsequence} of $w$, and write $w > w'$, if
the word $w'$ can be obtained from the word $w$ by erasing some (at least one)
of its characters. 
Apply Kruskal's theorem to conclude that $>$ is a well partial order.
\end{exercise}

Incidentally, there is also a famous generalization of 
Kruskal's theorem to {\em graphs} known as the {\em graph minor theorem}.
An {\em edge contraction} of a graph consists in removing an
edge while merging the two vertices. 
A graph $G$ is a {\em minor} of the graph $H$ if it can
be obtained from $H$ by a sequence of edge contractions.
It turns out that the {\em minor relation} is a {\em well partial order}.

Next, we present two relevant applications of Kruskal's theorem
to the termination problem of TRS. The first one is another proof
that RPO is well-founded.

\begin{proposition}
Every {\em simplification order} on $T_\Sigma(V)$ is {\em well founded}.
Hence, every {\em simplification order} (in particular RPO) 
is a {\em reduction order}.
\end{proposition}
\Proof
By {\em contradiction}, suppose $t_1 > t_2 > \ldots$
First, we prove by contradiction that $\var{t_1} \supseteq \var{t_2} \supseteq \ldots$ Suppose $x\in \var{t_{i+1}}\minus \var{t_{i}}$ and consider
the {\em substitution} $S=[t_i/x]$.
Then $S t_i = t_i > S t_{i+1}$.
By the {\em subterm property}, we have $St_{i+1} \geq t_i$.
Thus $t_i > t_i$ which {\em contradicts} the hypothesis that $>$ is strict.

Thus we can take $X=\var{t_1}$ which is {\em finite}. 
Then we can {\em apply Kruskal's theorem} to the sequence and conclude:
\[
\xst{i,j}{j>i \mand  t_j \emb t_i}~.
\]
Thus we have both $t_i > t_j$ and $t_j \geq t_i$.
Hence, $t_i > t_i$, which contradicts the hypothesis that 
a simplification order is strict. 
Notice
that we used the fact that $(T_\Sigma(X),\semb)$ is a {\em wpo}, not just a well-founded set. \qed \\

The second application concerns the introduction of 
the interpretation method over the {\em reals}.
Take as domain $A=\set{r \in \Real \mid r\geq a\geq 1}$.
This may appear as a wrong start as the set $A$ is {\em not} well founded!
However, suppose that we associate  with every $f^n\in\Sigma$ 
a multivariate polynomial $p_f(x_1,\ldots,x_n)$ such that:

\begin{enumerate}

\item Coefficients range over the {\em non-negative reals}.

\item $p_f(a,\ldots,a)\in A$: thus $p_f$ defines a function over the
      domain $A$.

\item $p_f(a_1,\ldots,a_n) > a_i$ for $i=1,\ldots,n$ (the new condition!).

\end{enumerate}
Write $s>_At$ if over the domain $A$ 
the polynomial associated with the term $s$ is strictly larger 
than the one associated with the term $t$.
It is easily checked that this is a {\em simplification order}, 
hence a {\em reduction order}.
The fact that we move from integers to real has an interesting
consequence as the first-order theory of reals 
is {\em decidable} ({\em e.g.}, it is decidable whether a first-order assertion
in analytic geometry is valid).

A corollary of this result is that we can decide whether there is a
polynomial interpretation over the reals where the polynomials have a
{\em bounded degree}.
However, we stress that this is a rather {\em theoretical advantage} 
because of the {\em high complexity} of the decision procedures.

\section{Summary and references}
Proving the termination of a TRS amounts to find a reduction order
that is compatible with the rules of the TRS.
One method consists in interpreting the function symbols as functions
over the positive integers with certain strictness properties.
Another method consists in applying the rules of the recursive path orders.
It turns out that the recursive path orders are an instance of
the simplification orders and that the latter are well-founded.
Many other methods for proving termination of TRS have been proposed and
their implementation is available in several tools.
Recursive path orders were introduced in \cite{DBLP:journals/tcs/Dershowitz82} and the presented termination proof is 
based on \cite{DBLP:conf/rta/Raamsdonk01}.
Kruskal's tree theorem is in \cite{Kruskal60} 
with a shorter proof in  \cite{nash-williams63} which is the one 
we present. Its special case for
words is presented in \cite{Higman52} and its generalization
to graphs is presented in a long series of papers starting with  
\cite{RobertsonSeymour}.
Hilbert's $10^{\w{th}}$ problem was among $23$ open problems
put forward at a 1900 international conference on mathematics.
The problem was eventually shown to be undecidable 
by  Matiyasevich in 1970. The decidability of the first-order
theory of real numbers was shown by Tarski\index{Tarski} around 1950.

\chapter{Confluence and completion of term rewriting systems}\label{confluence-trs-sec}
\markboth{{\it Confluence and completion of TRS}}{{\it Confluence and completion of TRS}}

In general the confluence of a term rewriting system (TRS for short) 
is an {\em undecidable} property.
However, if the TRS is {\em terminating and finite}
then the property is {\em decidable}.
By proposition \ref{newman-proposition}, we know that checking {\em local confluence} is 
enough and it turns out that to do that it is enough to consider a {\em finite}
number of cases known as {\em critical pairs}.

\section{Confluence of terminating term rewriting systems}
The definition of critical pair captures the most general
way in which two term rewriting rules can superpose and thus
possibly compromise the local confluence of the TRS.

\begin{definition}[critical pair]\index{critical pair}
Let $l_i\arrow r_i$ for $i=1,2$ be two rules of the TRS
(possibly equal) and assume that the variables in each
rule are {\em renamed} so that $\var{l_1}\inter \var{l_2}=\emptyset$.
Further suppose $l_1=C[l'_{1}]$ where $l'_{1}$ is {\em not} a variable and
let $S$ be the {\em most general unifier} of $l'_{1}$ and $l_2$ (if it exists).
Then $(S(r_1), S(C[r_2]))$ is a {\em critical pair}.
\end{definition}

The pair $(S(r_1), S(C[r_2]))$ is {\em critical} because if we take
the terms $l_1$ and $C[l_2]$ then:
\[
S(r_1) \leftarrow S(l_1) = S(C)[S(l'_1)] = S(C)[S(l_2)] \arrow S(C)[S(r_2)] = S(C[r_2])~.
\]
Thus the two terms in the critical pair must be joinable (have a common reduct).
The main insight is that this is enough to guarantee (local) confluence.
Let us start with a preliminary remark. Let the {\em domain} of a 
substitution $S$ be the set:
\[
\w{dom}(S)=\set{x\in V \mid S(x)\neq x}~.
\]
Given two substitutions $S_1,S_2$ let us define their {\em union} as follows:
\[
(S_1\union S_2)(x)=\left\{ \begin{array}{ll}
                           S_1(x) &\mbox{if }x\in \w{dom}(S_1)\minus \w{dom}(S_2) \\
                           S_2(x) &\mbox{if }x\in \w{dom}(S_2)\minus \w{dom}(S_1) \\
                           x      &\mbox{otherwise.}
\end{array}\right.
\]

\begin{exercise}  \label{disjoint-union-sub-ex}
Suppose $\var{t}\inter \var{s}=\emptyset$, 
               $\w{dom}(S_1)\subseteq \var{t}$, and 
               $\w{dom}(S_2)\subseteq \var{s}$.
Then show that $S_1(t)=S_2(s)$ entails that $S_1\union S_2 \models t=s$.
Notice that if $\var{t}\inter \var{s}\neq \emptyset$ then the 
assertion is false. 
E.g., take: $t=x$, $s=f(x)$, $S_1=[f(x)/x]$, $S_2=\w{id}$, $S_1\union S_2=S_1$ 
but $S_1(t)\neq S_1(s)$.
\end{exercise}

\begin{proposition}
Suppose given a {\em finite} and {\em terminating} TRS.
Then the TRS is {\em confluent} iff all the {\em critical pairs}
induced by its rules are {\em joinable} (and the 
latter is a {\em decidable} condition).
\end{proposition}
\Proof The test is {\em necessary} as explained above.
Because the TRS is terminating, it is enough to
show that the test guarantees {\em local confluence}.
To check local confluence a {\em finite case analysis}
suffices.
If $s\arrow t_1$ and $s \arrow t_2$, then we can find
rules $l_1\arrow r_1, l_2\arrow r_2$, contexts
$C_1,C_2$ and substitutions $S_1,S_2$ such that
\[
\begin{array}{lll}
           s=C_1[S_1 l_1] = C_2[S_2 l_2],
           &t_1=C_1[S_1 r_1],
           &t_2=C_2[S_2 r_2]~.
\end{array}         
\]
We sketch and provide concrete examples for the main
cases to consider.
\begin{description}

\item[Case 1]
The paths corresponding to the contexts $C_1$ and
$C_2$ are incomparable (neither is a prefix of the other). 
In this case one can close the diagram 
in one step. For instance, assume the rules:
\[
g_i(x)\arrow k_i(x),\quad  i=1,2~,
\]
and consider $h(g_1(x),g_2(x))$.

\item[Case 2]
 There is a variable $x$ in $l_1$ 
such that $S_2 l_2$ is actually a subterm of $S_1(x)$.
In this case one can always close the diagram, though it may take
{\em several} steps. For instance, assume the rules:
\[
f(x,x,x)\arrow h(x,x), \quad g(x)\arrow k(x)~.
\]
and consider $f(g(x),g(x),g(x))$.

\item[Case 3]
We can decompose $l_1$ in $C[l'_{1}]$ so that:
\[
l'_{1}\mbox{ is not a variable and }
S_1 l'_{1} = S_2 l_2~.
\]
One can show that this situation is always an instance of a
{\em critical pair}. For instance, 
assume the rules:
\[
f(f(x,y),z) \arrow f(x,f(y,z)), \quad f(i(x),x) \arrow e~,
\]
and consider $f(f(i(x),x),z)$. \qed
\end{description}

\begin{exercise}\label{confluence-termination-ex}
Consider the TRS with rules:
\[
\begin{array}{llllll}
f(x,g(y,z)) &\arrow &g(f(x,y),f(x,z))~,\qquad
&g(g(x,y),z) &\arrow &g(x,g(y,z)) ~.
\end{array}
\]
Is the resulting reduction system terminating and/or confluent?
\end{exercise}

\section{Completion of term rewriting systems (*)}
The test for local confluence is the basis for an
iterative symbolic computation method
known as {\em Knuth-Bendix completion}.
Given an equational theory, the {\em goal}
is to obtain a confluent and terminating
term rewriting system for it. The main steps
in Knuth-Bendix completion are as follows:

\begin{enumerate}

\item Orient the equations thus obtaining a TRS.

\item Check {\em termination} of the TRS.

\item Then check local {\em confluence}.

\item If a critical pair {\em cannot be joined}, then we add the corresponding 
equation and we repeat the process.
\end{enumerate}

Notice that there is no guarantee that the process terminates! 
At various places, one may require a human intervention:
orientation of the rules, well-founded order to check termination, 
selection of the rules to add,$\ldots$

\begin{example}
The following law describes so called `central grupoids':
\[
(x*y)*(y*z) = y~.
\]
\item Any simplification ordering $>$ satisfies:
$(x*y)*(y*z) > y$.
So we orient the equation from left to right.
A {\em critical pair} is:
\[
y'*((y'*z')*z) \leftarrow ((x'*y')*(y'*z'))*((y'*z')*z) \arrow (y'*z')~.
\]
Any {\em simplification ordering} satisfies: $y'*((y'*z')*z)>(y'*z')$.
Another critical pair is:
\[
(x*(x'*y'))*y'\leftarrow (x*(x'*y'))*((x'*y')*(y'*z')) \arrow (x'*y')~.
\]
Again any {\em simplification ordering} satisfies: 
$(x*(x'*y'))*y'>(x'*y')$.
Thus we get a terminating TRS with three rules.
In the next iteration all critical pairs turn out to be joinable and thus the
completion {\em terminates successfully}.
\end{example}

\begin{example}
The equations for left/right distributivity of $*$ over $+$ are:
\[
\begin{array}{llllll}
x*(y+z) &= &(x*y)+(x*z)~,\qquad  &(u+v)*w &= &(u*w)+(v*w)~.
\end{array}
\]
Ordering from left to right, a {\em critical pair} is:
\[
(u*(y+z))+(v*(y+z)) \leftarrow (u+v)*(y+z) \arrow ((u+v)*y)+((u+v)*z)~.
\]
If we normalize the two terms we get:
\[
((u*y)+(v*y))+((u*z)+(v*z)), \quad
((u*y)+(u*z))+((v*y)+(v*z))~,
\]
and it is problematic to order them.
\end{example}

\begin{example}
Consider the equations:
\[
x+\s{z}=x, \quad \s{s}(x+y)=x+\s{s}(y), \quad x+\s{s}(\s{z})=\s{s}(x)~.
\]
It can be easily checked that by orienting them from left to right we obtain
a terminating TRS. However, there is a critical pair between 
the second and third rule:
\[
\s{s}(\s{s}(x)) \leftarrow \s{s}(x+\s{s}(\s{z})) \arrow x+\s{s}(\s{s}(\s{z}))~.
\]
In turn this forces the rule:
$ x+\s{s}(\s{s}(\s{z})) \arrow \s{s}(\s{s}(x))$.
In this case, a simple completion method {\em may diverge} 
as one has to add all the rules of the shape:
\[
x+\s{s}^n(\s{z}) \arrow \s{s}^n(x)~.
\]
However, an alternative completion strategy succeeds by orienting 
the second rule in the opposite direction:
$x+\s{s}(y)\arrow \s{s}(x+y)$.
\end{example}

\begin{exercise}\label{ter-conf-examples-ex}
Prove termination and confluence of the TRS considered in examples
\ref{addition-sum-ex} and \ref{group-ex}.
\end{exercise}

\begin{exercise}\label{ex-summary}
Consider the TRS with rules:
\[
\begin{array}{llllll}
f(f(x,y),z) &\arrow &f(x,f(y,z)),\qquad
&f(i(x),x) &\arrow &e~.
\end{array}
\]
(1) Can you show termination by RPO?
(2) Can you show termination by polynomial interpretation?
(3) Is the system confluent?
Next consider the TRS:
\[
f(f(x)) \arrow g(x)~.
\]
(4) Is it confluent?
(5) Add the rule $f(g(x)) \arrow g(f(x))$. Is this terminating by RPO ?
(6) And by polynomial interpretation? 
(7) Is the system confluent?
(8) Same questions if we add the rule $g(f(x)) \arrow f(g(x))$.
\end{exercise}

\begin{exercise}\label{non-modular-termination-ex}
Let $R_1$ be a TRS with rule ($x$ is a variable):
\[
f(a,b,x) \arrow f(x,x,x)~,
\]
and let $R_2$ be another TRS with rules ($x,y$ variables):
\[
\begin{array}{cccccc}
g(x,y) &\arrow &x, \quad &g(x,y) &\arrow &y~.
\end{array}
\]
(1) Show that the systems $R_1$ and $R_2$ terminate.
(2) Prove or give a counter-example to the confluence of the TRS 
$R_1$ and $R_2$.
(3) Show that the TRS $R_1\union R_2$ does {\em not} terminate.
(4) However, show that the TRS $R_1\union R_2$ is normalizing
(every term has a normal form).
\end{exercise}

\section{Summary and references}
The {\em critical pair test} is a practical test to check
the {\em (local) confluence} of TRS and the basis of an iterative
method known as {\em Knuth-Bendix completion} \cite{KB70}.
The method starts with a TRS  which is typically derived 
from a set of equations. It then checks the TRS for termination and 
local confluence. If local confluence fails, then we try to orient
the critical pairs and start the verification again.
Many sophisticated {\em refinements} of the completion procedure
have been proposed and implemented in a variety of tools.
Also similar ideas have been developed in parallel and
independently in the area of computer algebra 
where a technique known as {\em Gr\"obner bases} 
is used to solve decision problems in rings of polynomials.
Examples in this chapter are based on \cite{BaaderNipkow99}
and the reader is invited to check them with one of the tools
available online.

\chapter{Term rewriting systems as functional programs}\label{firstfun-sec}
\markboth{{\it TRS as functional programs}}{{\it TRS as functional programs}}

We focus on term rewriting systems whose symbols 
can be partitioned in {\em constructors}
and {\em functions} and whose term rewriting rules guarantee 
a deterministic evaluation. Such systems can be regarded as 
rudimentary {\em first-order} functional programs. 
We then consider a recursive definition mechanism known
as {\em primitive recursion} which guarantees termination.
Going beyond termination,  we present conditions
that guarantee {\em termination in polynomial time}. 
More precisely, we define 
a restricted form of primitive recursion on binary words,
also known as {\em bounded recursion on notation},
in which one can program exactly the {\em functions} 
computable in polynomial time.

\section{A class of term rewriting systems}
We consider term rewriting systems whose signature $\Sigma$ is partitioned into 
{\em constructor symbols} denoted with $\s{c},\s{d},\ldots$ and
{\em function symbols} denoted with $f,g,\ldots$
A {\em value} is a term composed of constructor symbols and
denoted with $v,v',\ldots$ while a {\em pattern} is a term
composed of constructor symbols and variables and denoted
with $p,p',\ldots$ To make sure the collection of values is not
empty, we assume that there is at least a {\em constant} (a symbol with
arity $0$) among the constructors.
We assume all term rewriting rules have the shape:
\[
f(p_1,\ldots,p_n) \arrow e~.
\]
Moreover given two distinct rules:
\[
f(p_1,\ldots,p_n) \arrow e~, \qquad f(p'_1,\ldots,p'_n) \arrow e~,
\]
which refer to the same function symbol $f$, we assume
that they cannot superpose, {\em i.e.}, it is {\em not} possible to
find values $v_1,\ldots,v_n$ and substitutions $S$ and $S'$ such that
$v_i=S(p_i) =S'(p'_i)$ for $i=1,\ldots,n$.
Under these hypotheses, closed terms are evaluated according to the following rules:
\[
\begin{array}{cc}

\infer{e_j \eval v_j~~j=1,\ldots,n}
{\s{c}(e_1,\ldots,e_n) \eval \s{c}(v_1\ldots,v_n)}

\qquad
&\infer{\begin{array}{c}
       e_j \eval v_j,~f(p_1,\ldots,p_n) \arrow e,\\
       S p_{j}= v_j, j=1,\ldots,n,\quad S(e)\eval v
       \end{array}}
      {f(e_1,\ldots,e_n)\eval v}~.

\end{array}
\]
Notice that the first rule guarantees that for all values $v$, we
have: $v\eval v$.
If the term is not a value, then we look for the innermost-leftmost
term of the shape $f(v_1,\ldots,v_n)$ and look for a rule
$f(p_1,\ldots,p_n)\arrow e$ which applies to it (by hypothesis, there is
at most one). If no rule applies then the evaluation is stuck.
If $v$ is a value its size $|v|$ is a natural number defined by:
\[
|c(v_1,\ldots,v_n)|=1+\Sigma_{i=1,\ldots,n} |v_i|~.
\]
By extension, if $t\eval v$ then $|t|=|v|$. Thus, in this chapter,
the notation $|t|$ denotes the size of the unique value to which the
closed term $t$ evaluates (if any).

\begin{example}
We introduce some {\em constructors} along with their arity.

{\footnotesize\[
\begin{array}{ll}
\s{t}^0, \s{f}^0 &\mbox{(boolean values)} \\
\s{z}^0, \s{s}^{1}&\mbox{(tally (unary) natural numbers)} \\
\s{nil}^0, \s{c}^{1} &\mbox{(lists)} \\
\s{\epsilon}^{0},\s{0}^{1},\s{1}^{1} &\mbox{(binary words)}
\end{array}
\]}

and some {\em functions}:

{\footnotesize\[
\begin{array}{llll}

\w{ite}(\s{t},x,y) &\arrow &x            &\mbox{(if-then-else)} \\
\w{ite}(\s{f},x,y) &\arrow &y                                 \\ \\

\w{leq}(\s{z},y)            &\arrow & \s{t} &\mbox{(less-equal)} \\
\w{leq}(\s{s}(x), \s{z})    &\arrow & \s{f} \\
\w{leq}(\s{s}(x), \s{s}(y)) &\arrow &\w{leq}(x,y)  \\ \\

\w{ins}(x,\s{nil})        &\arrow &\s{c}(x,\s{nil})  &\mbox{(list insertion)}\\
\w{ins}(x,\s{c}(y,l))     &\arrow &\w{ite}(\w{leq}(x,y),\s{c}(x,\s{c}(y,l)),  
 \s{c}(y,\w{ins}(x,l)))
\end{array}
\]}
\end{example}

\begin{exercise}\label{sorting-firstorder-ex}
Continue the previous example by defining
functions to sort lists of tally natural number 
according to various standard algorithms such as
insertion sort, quick sort,$\ldots$
\end{exercise}

Notice that it is straightforward to program the functions above in a language
with pattern-matching such as $\ml$.

\section{Primitive recursion}\label{prim-rec-sec}\index{primitive recursive function}
We restrict further the class of term rewriting systems
so that termination is guaranteed. Assume the following constructor symbols
for tally natural numbers:
$\s{z}^{0}$ and $\s{s}^{1}$.
Also assume the following {\em basic function symbols} $z,s,p^{n}_{i}$ 
for $i=1,\ldots,n$ with the following rules:
\[
\begin{array}{llll}
z(x) &\arrow &\s{z}	         &\mbox{(zero)} \\
s(x) &\arrow &\s{s}(x)          &\mbox{(successor)} \\
p^{n}_{i}(x_1,\ldots,x_n)&\arrow &x_i  &\mbox{(projections).} 
\end{array}
\]
New function symbols can be introduced according to the 
composition and primitive recursion rules which are described below.
We shall use $x^*$ to denote a (possibly empty) sequence of variables
$x_1,\ldots,x_n$.

\begin{description}

\item[Composition] Given $g$ of arity $k$ and  
$h_i$ of arity $n$ for $i=1,\ldots,k$ introduce a new function 
$f$ of arity $n$ with the rule:
\[
f(\vc{x}) \arrow g(h_1(\vc{x}),\ldots,h_k(\vc{x}))~.
\] 

\item[Primitive Recursion]
Given $g$ of arity $n$ and $h$ of arity $n+2$ introduce a 
new function $f$ of arity $n+1$ with the rules:
\[
\begin{array}{llllll}
  f(\s{z},\vc{y}) &\arrow &g(\vc{y})~, \qquad
  
&f(\s{s}(x),\vc{y}) &\arrow &h(f(x,y),x,\vc{y})~.
\end{array}
\]
\end{description}

\begin{example}
We practice primitive recursion by defining a few arithmetic function. 

{\footnotesize\[
\begin{array}{llll}
\w{add}(\s{z},y)  &\arrow &y  &\mbox{(addition)}\\
\w{add}(\s{s}(x),y) &\arrow &\s{s}(\w{add}(x,y))   \\ 

\w{mul}(\s{z},y)  &\arrow &\s{z} &\mbox{(multiplication)}\\
\w{mul}(\s{s}(x),y)     &\arrow &\w{add}(\w{mul}(x,y), y) \\

\w{exp}(\s{z},x)  &\arrow &\s{s}(\s{z})  &\mbox{(exponentiation)}\\
\w{exp}(\s{s}(x),y)&\arrow &\w{mul}(\w{exp}(x,y),y)~.
\end{array}
\]}

We can go on to describe towers of exponentials,$\ldots$ 
The complexity of the programmable functions is still very high!
\end{example}

\begin{exercise}\label{prim-rec-ex}
Define primitive recursive functions to: 
(1) {\em decrement} by  one (with $0-1 = 0$),
(2) {\em subtract} (with $x-y = 0$ if $y>x$),
(3) compute an {\em if-then-else},
(4) compute the {\em minimum} of two numbers.
\end{exercise}

We notice that there is  a {\em trade-off} between 
primitive recursion and full recursion. 
Namely, in the former {\em termination} is for free but some 
{\em functions} cannot be represented and some {\em algorithms}
are more difficult or impossible to represent.
For instance, it can be shown by a diagonalization argument that
the universal function (the interpreter) for primitive recursive
functions is a total function but not a primitive recursive one.
It can also be shown that the {\em natural}
algorithm that computes the minimum of two tally natural number
cannot be expressed by primitive recursion 
(details in the following example).

\begin{example}
Primitive recursion is a bit of a straight-jacket to guarantee
termination. For instance, the following rules could be used
to define the minimum of two tally natural numbers.

{\footnotesize\[
\begin{array}{lll}
\w{min}(\s{z},y)           &\arrow &\s{z} \\
\w{min}(\s{s}(x),\s{z})     &\arrow &\s{z} \\
\w{min}(\s{s}(x),\s{s}(y))  &\arrow &\s{s}(\w{min}(x,y))~.

\end{array}
\]}

The rules scan the two numbers in parallel and stop as soon as
they reach the end of the smallest one.
However this definition of \w{min} is not primitive recursive.
Worse, it can be shown that no primitive recursive definition 
of $\w{min}$ produces an algorithm computing $\w{min}(v,u)$ in time
$\w{min}(|v|,|u|)$. 
\end{example}

We are soon going to address complexity issues and it is well known
that in this case unary notation is rather odd.
Indeed unary notation requires {\em too much space} and because
representation of the input is so  {\em  large} 
complexities of the operations can be unexpectedly {\em  low}.
For instance, in unary notation we can compute the addition in constant
time: it is enough to regard numbers as lists and concatenate them.
So we revise the notion of primitive recursion by working with
the following constructors which correspond to binary words:
$\s{\epsilon}^0, \s{0}^{1} ,\s{1}^{1}$.
The {\em basic functions} are now $e^1$, $s_i^{1}$ for $i=0,1$,
and $p_{i}^{n}$ for $i=1,\ldots,n$ with the following rules:
\[
\begin{array}{llll}
e(x) &\arrow &\s{\epsilon}	         &\mbox{(empty word)} \\
s_i(x) &\arrow &\s{i}x                   &\mbox{(successors)} \\
p_i(x_1,\ldots,x_n)& \arrow &x_i           &\mbox{(projections).} 
\end{array}
\]
As before, we can introduce new functions according to two mechanisms.
\begin{description}

\item[Composition]  Given $g$ of arity $k$ and $h_i$ of arity $n$ for 
$i=1,\ldots,k$ we introduce a new function symbol $f$ with the rule:
\[
f(\vc{x}) \arrow g(h_1(\vc{x}),\ldots,h_k(\vc{x}))~.
\] 
\item[Primitive recursion on binary notation]
Given $g$ of arity $n$ and 
$h_i$ of arity $n+2$ for $i=0,1$ we introduce a new function
symbol $f$ with the rules:
\begin{equation}\label{prim-rec-not-def}
\begin{array}{lll}
  f(\s{\epsilon},\vc{y})       	&\arrow &g(\vc{y}), \\
  f(\s{0}(x),\vc{y})            &\arrow &h_0(f(x,\vc{y}),x,\vc{y}),\\
  f(\s{1}(x),\vc{y})            &\arrow &h_1(f(x,\vc{y}),x,\vc{y})~.
\end{array}
\end{equation}
\end{description}
The class of functions definable in this way are the primitive recursive
functions on binary notation, also known as functions defined by
{\em recursion on notation}.

\begin{exercise}\label{brn-ex}
Assume binary numbers are represented as 
binary words where the 
{\em least significant  digit is on the left}.
We consider the problem of defining some standard arithmetic functions by primitive recursion 
on binary words.

\begin{enumerate}
\item Define a function that takes a binary word and 
removes all $`0'$ that do not occur on the left of a $1$ 
(hence $\epsilon$ can be taken as the {\em canonical representation of zero}).

\item Show that the functions {\em division by 2}, {\em modulo 2}, 
{\em successor}, {\em if-then-else}, {\em predecessor}, {\em number of digits}
can be defined  by primitive recursion.

\item Suppose a function that implements  
{\em addition} is given 
(known definitions of this function are quite technical).
Implement multiplication by primitive recursion on binary notation.

\end{enumerate}
\end{exercise}

\section{Functional programs computing in polynomial time (*)}
How can we compute a function defined by primitive recursion?
Suppose $v_k=\s{i}_k\ldots \s{i}_1\s{\epsilon}$.
Here is a simple loop that computes $f(v_k,\vc{v})$:
\begin{equation}\label{loop-prim-rec}
\begin{array}{l}
r=g(\vc{v}); \\
\s{for}~(j=1;j\leq k;j=j+1)\{ r=h_{i_{j}}(r, v_{j-1},\vc{v}); \}\\
\s{return}~r ~.
\end{array}
\end{equation}
Problem: suppose that $h_i$ and $g$ can be computed in {\em polynomial
time}. Can we conclude that $f$ can be computed in {\em polynomial time}?
Well, here is what can go wrong.
Consider first the function $d$ doubling the size of its input:
\[
\begin{array}{llllll}
d(\epsilon)   &\arrow &\s{1}\s{\epsilon}~, \qquad
&d(\s{i}(x))  &\arrow &\s{i}(\s{i} (d(x)))\quad \s{i}=\s{0},\s{1}~.
\end{array}
\]
Then consider the function $e$:
\[
\begin{array}{llllll}

e(\s{\epsilon}) &\arrow &\s{1}\epsilon~, \qquad
&e(\s{i}(x))    &\arrow &d(e(x))\quad \s{i}=\s{0},\s{1}~.
\end{array}
\]
These functions are definable by primitive recursion on binary 
notation (exercise!)
and $|e(v)|$ is exponential in $|v|$. 
Iterating $|v|$ times polynomial time operations 
can generate data whose size is {\em  not} polynomial in $|v|$.

We now introduce a notion of definition by bounded recursion on notation (BRN). 
This is an ordinary primitive recursion on binary words as in (\ref{prim-rec-not-def})
with the additional requirement that there exists 
a {\em polynomial} 
$S_f$ with non-negative coefficients such that:
\[
|f(v,\vc{v})| \leq S_f(|v|,|\vc{v}|)~\qquad\mbox{(polynomial bound on value size)}.
\]
It turns out that the  {\em functions} computable by an algorithm in BRN
are exactly those computable in PTIME. This result decomposes in
the following two propositions.

\begin{proposition}\label{Cobham-prop}\index{Cobham}
If we can define an {\em algorithm by BRN} then we can compute its result
in {\em PTIME}.
\end{proposition}
\Proof 
First we prove by induction on the definition of a function $f$ in BRN
that there is a polynomial $S_f$ such that for all $v_1,\ldots,v_n$:
\[
|f(v_1,\ldots,v_n)|\leq S_f(|v_1|,\ldots,|v_n|)~.
\]
This is clear for the {\em basic functions} and for {\em BRN}. 
For the {\em composition}, say $h$, of $f$ with $(g_1,\ldots,g_k)$ we have:
\[
h(\vc{v}) \arrow f(g_1(\vc{v}),\ldots,g_k(\vc{v}))~.
\]
Then by inductive hypothesis:
\[
|g_{i}(\vc{v})| \leq S_{g_{i}}(|\vc{v}|)~.
\]
Let $S$ be a polynomial that bounds all $S_{g_{i}}$.
Applying again the inductive hypothesis:
\[
\begin{array}{lll}
|f(g_1(\vc{v}),\ldots,g_k(\vc{v}))| 
          &\leq  &S_f(|g_1(\vc{v})|,\ldots,|g_k(\vc{v})|) \\
          &\leq &S_f(S_{g_{1}}(|\vc{v}|), \ldots,S_{g_{k}}(|\vc{v}|)) \\
          &\leq &S_f(S(|\vc{v}|),\ldots,S(|\vc{v}|))~,
\end{array}
\] 
and the composition of polynomials is a polynomial.
Thus data computed by BRN has {\em size polynomial in the size of the input}.

Next, we prove by induction on the definition of a function $f$ in BRN
that there is a polynomial $T_f$ such that
$f(v_1,\ldots,v_n)$ can be computed in time
$T_f(|v_1|,\ldots,|v_n|)$.  
{\em Recursion} is the interesting case.
Consider again the loop (\ref{loop-prim-rec})
computing primitive recursion.
For all steps $j=1,\ldots,k$, we have that: $|r| \leq S_f(k,|\vc{v}|)$.
Let $T_h$ be a polynomial that bounds both $T_{h_{0}}$ and $T_{h_{1}}$.
Then the computation of the $k$ steps is performed in at most:
\[
k \cdot T_h(S_f(k, |\vc{v}|), k, |\vc{v}|)  = |v_k| \cdot  T_h(S_f(|v_k|, |\vc{v}|), |v_k|, |\vc{v}|)~,
\]
which is a polynomial in $|v_k|,|\vc{v}|$.  \qed

\begin{proposition}\label{Cobham2-prop}
If there is a {\em PTIME algorithm} in some Turing-equivalent formalism 
then we can compile it to an {\em algorithm in BRN that 
computes the same function}.
\end{proposition}
\Proof
Let $M=(\Sigma, Q, q_o, F,\delta)$ be a {\em Turing machine} (TM) with:
(i) $\Sigma$ alphabet, 
(ii) $Q$ states, $q_o\in Q$ initial state,
(iii) $F\subseteq Q$ final states, and
(iv) $\delta:\Sigma \times Q \arrow  \Sigma \times Q \times \set{L,R}$ transition function.\footnote{The reader may be familiar with
slightly different definitions of TM. The details of the definition
are not essential for the following discussion.}

If $x$ is a real number let $\lceil x \rceil$ be the least integer $n$ such that $x\leq n$.
Obviously, elements in $\Sigma$ and $Q$ can be encoded as {\em binary
words} of length $\lceil \w{log}_{2}(\card \Sigma) \rceil$ and  
$\lceil \w{log}_{2}(\card Q ) \rceil$, respectively.
The {\em configuration of a TM} can be described by a tuple
$(q,h,l,r)$ where:
(i) $q$ is the current {\em state},
(ii) $h$ is the {\em character read},
(iii) $l$ are the characters on the {\em left hand side of the head},
and (iv)
$r$ are the characters on the {\em right hand side of the head}.

Next, we have to define a {\em step function}
that simulates one step of a Turing machine while working 
on the encodings of states and characters.
Informally, the function \w{step} is a {\em case analysis} 
corresponding to
the finite table defining the transitions of the TM. {\em E.g.}, the rule:
\[
\begin{array}{ll}
\w{step}(\s{0}\s{1}\s{\epsilon}, \s{1}\s{\epsilon},\s{0}l', r) \arrow
(\s{1}\s{1}\s{\epsilon}, \s{0}\s{\epsilon}, l', \s{0}r)~,
\end{array}
\]
describes the situation where being in state $01$ and reading $1$, we go
in state $11$, write $0$, and move to the left. 
The only technical difficulty here is that {\em tuples} are not a primitive
data structures in our formalization. 
However, using the  arithmetic functions, we can program
{\em pairing of natural numbers} and the {\em related projections}.
Alternatively (and more naturally), one could extend the framework 
with a {\em pairing} constructor.

We {\em ignore} these problems and assume a function $\w{step}$
that takes a tuple $(q,h,l,r)$ and returns the tuple $(q',h',l',r')$
describing the following state.
Now comes a {\em key idea} which we present first using a simplified
notation. We have an {\em initial configuration} $v_0$ and
a function $\w{step}$ such that:
\[
|\w{step}(v)| \leq |v|+1~.
\]
We want to {\em iterate} $\w{step}$ on $v_o$ at least $P(|v_o|)$ times 
where $P$ is a polynomial of {\em degree $k$}.
W.l.o.g., we may assume the TM loops after reaching the final state so that
running it longer does not hurt.

We assume an {\em expansion function} $\w{exp}$ 
such that for all $k$ there is an $m$ such that:
\[
|\w{exp}^m(v)| \geq |v|^k~.
\]
To do this, it is enough to iterate a function that squares the size of its
entry. Remember that $k$ and therefore $m$ are {\em constants}, {\em i.e.},
they do not depend on the size of the input.
Then we define a function $\w{it}$ as:
\begin{equation}\label{it-def}
\begin{array}{llllll}
\w{it}(\epsilon, v) &\arrow &v~,
&\w{it}(\s{i}\cdot u,v) &\arrow &\w{step}(\w{it}(u,v))\quad \s{i}=\s{0},\s{1}~.
\end{array} 
\end{equation}
This is a definition by {\em bounded recursion on notation} since 
assuming $S_{it}(n,m)=n+m$ (a polynomial!) we have:
\[
|\w{it}(u,v)| \leq S_{it}(|u|,|v|)~.
\]
Then to iterate the \w{step} function at least $|v_0|^k$ times on the
initial configuration we run:
$\w{it}(\w{exp}^{m}(v_0),v_0)$.

We can now go back to TM. All we have to do is to rewrite the
\w{it} function above (\ref{it-def}) as follows:
\[
\begin{array}{llllll}
\w{it}(\epsilon,q,h,l,r)      &\arrow &(q,h,l,r)~,\qquad
&\w{it}(\s{i}u,q,h,l,r)        &\arrow &\w{step}(\w{it}(u,q,h,l,r))\quad \s{i}=\s{0},\s{1}~.
\end{array}
\]
To summarize, given a TM running in $P$ time ($P$ fixed polynomial), 
for any input $v_0$ we:
(i) initialize a {\em counter} to a value $u$ such that  $|u|\geq P(|v_0|)$ and
(ii) perform a BRN on the counter thus iterating 
the step function $|u|$ times.
Notice that here the iteration works on the 
{\em length of the counter} and not on its binary
representation. Otherwise,  the definition would not be by BRN and  
termination could take exponential time! \qed

\begin{remark}
\Proofitemf{(1)}
It is quite possible to program a function that takes 
{\em exponential time} and runs in {\em polynomial space}
(never going twice through the same configuration!).
For instance, take a function that {\em counts}
from $\s{0}^n\epsilon$ to $\s{1}^n\epsilon$.
Implicitly, proposition \ref{Cobham-prop} states that as long as
we stick with BRN such function {\em cannot} be programmed.
In the counting function, 
the problem is {\em not} the size of the data (the identity
function gives the bound!) but the fact that the recursion mechanism
is not compatible with primitive recursion on notation.

\Proofitemm{(2)}
The proof of proposition \ref{Cobham2-prop} 
suggests that there is a {\em trivial way}
of building PTIME algorithms.
Take {\em any program} and {\em instrument it} so that it 
keeps a counter that stops after a number of steps which 
is polynomial in the size of the input  (for a fixed polynomial).
Of course, the problem with this `time-out' approach is that 
we have no idea whether the program will produce interesting answers
{\em before} running it. 

\Proofitemm{(3)}
The reader should keep in mind that while it is possible to 
build a programming language (a decidable syntax) 
that  computes exactly the {\em PTIME functions}, 
it is not possible to build one 
that contains exactly the {\em PTIME programs}, {\em e.g.},
the set of Turing's machines computing in \w{PTIME} is undecidable.
\end{remark}

Proposition \ref{Cobham-prop} 
restricts the programmer to primitive recursion
and it provides no clue on {\em how to find a polynomial bound} on the size.
Is it possible to find a {\em syntactic criterion} that guarantees the
existence of a polynomial bound?
The {\em high-complexity} of programs defined by primitive recursion on
binary notation   depends on the fact that we have {\em nested recursions}, 
{\em i.e.}, the result of a primitive recursion
can be used as the main argument of another primitive recursion as in:
\begin{equation}\label{bad-recursion-ex}
\begin{array}{lllllll}
d(\epsilon)   &\arrow &\s{1}\s{\epsilon}~, \qquad
&d(\s{i}(x))   &\arrow &\s{i}(\s{i} (d(x))) &\s{i}=\s{0},\s{1}~, \\

e(\s{\epsilon}) &\arrow &\s{1}\epsilon~, \qquad
&e(\s{i}(x))    &\arrow &d(e(x)) &\s{i}=\s{0},\s{1}      \qquad \Leftarrow \quad\mbox{(`Bad' recursion!)}
\end{array}
\end{equation}
A key insight is that if we forbid this by a 
{\em syntactic mechanism} then {\em data size stays polynomial} and 
moreover it is still possible to define all
functions computable in PTIME. To this end,
functions' arguments  are partitioned into {\em two zones} 
(syntactically separated by a semi-colon):
\[
f(x_1,\ldots,x_n;y_1,\ldots,y_m) \quad n,m\geq 0~.
\]
The ones on the left are called {\em normal} and those on the right {\em safe}.
Let us refer to the functions in this new class as SRN ({\em Safe Recursion on
  Notation}) functions.
The {\em invariant} one maintains on SRN functions 
is that there is a polynomial $P_f$ such that:
\[
|f(v_1,\ldots,v_n;u_1,\ldots,u_m)|\leq P_f(|v_1|,\ldots,|v_n|)+\s{max}(|u_1|,\ldots,|u_m|)~.
\]
In particular, if $f$ has no normal arguments then the size of its result is bound by the size  of its arguments up to an {\em additive} constant.

Unlike in BRN,  the existence of the polynomial is {\em guaranteed}
by the way recursion and composition are restricted.
Assuming $g,h_0,h_1$ are SRN functions, we can define a new SRN function 
$f$ with the rules:
\[
\begin{array}{lll}
f(\epsilon,\vc{x};\vc{y})  &\arrow &g(\vc{x};\vc{y}) \\
f(\s{0}x,\vc{x};\vc{y})    &\arrow &h_0(x,\vc{x};\vc{y},f(x,\vc{x};\vc{y}))\\
f(\s{1}x,\vc{x};\vc{y})    &\arrow &h_1(x,\vc{x};\vc{y},f(x,\vc{x};\vc{y}))~.
\end{array}
\]
The main argument lies in the normal zone (on the left) while the
recursive calls take place in the safe zone (on the right).
The way SRN functions are composed is also restricted so that
expressions plugged in the normal zone do not depend on arguments
in the safe zone.
Specifically, assuming, $f,g_1,\ldots,g_k,h_1,\ldots,h_l$ are SRN functions
we define their {\em safe} composition as:
\[
\begin{array}{c}
f(\ g_1(\vc{x};),\ldots,g_k(\vc{x};)\ ; \ 
    h_1(\vc{x};\vc{y}),\ldots, h_l(\vc{x};\vc{y}) \ )~.
\end{array}
\]
With such restrictions it is possible
to define a function that doubles the size of its argument but
it is {\em not} possible to iterate it as in (\ref{bad-recursion-ex}).

\section{Summary and references}
Functions defined by primitive recursion on unary or binary
notation are guaranteed to terminate; 
the book \cite{Rose} is a compact reference on 
hierarchies of total recursive functions.
If moreover, we restrict the size of the computed values 
to be polynomial in  the size of the input then we can program exactly
the functions computable in {\em polynomial time}.
This is an early result in complexity theory \cite{Cobham64}.
The fact that the size bounds can be obtained through 
a syntactic discipline has been observed more recently
in \cite{DBLP:journals/cc/BellantoniC92}. 
The reader is warned that this syntactic discipline is quite restrictive
and hardly practical.

\chapter{$\lambda$-calculus}\label{lambda-sec}
\markboth{{\it $\lambda$-calculus}}{{\it $\lambda$-calculus}}

The $\lambda$-calculus is a compact notation to
represent (higher-order) functions.  It turns out that this notation
embodies directly many concepts arising in programming languages
such as: (higher-order) functions, recursive definitions, scoping
rules, and evaluation strategies. Moreover, it is sufficiently expressive
to describe a number of programming features such as:
control flow operators,  side-effects, records, 
and objects which will be discussed in the following chapters.
When enriched with types, the terms of the $\lambda$-calculus can 
be regarded as proofs in a (constructive) logic. This connection
sheds light on the design of type systems for programming languages
and explains the role of the $\lambda$-calculus in (higher-order)
proof assistants.

In this chapter, we start the technical development by introducing 
an equational theory on $\lambda$-terms known as $\beta$-conversion 
and we prove the
confluence of the related reduction rule. We also prove similar
results for a stronger theory known as $\beta\eta$-conversion.
Next, we show that the $\lambda$-calculus
is sufficiently expressive to represent partial recursive
functions (the $\lambda$-calculus is Turing equivalent). Finally we
introduce a term rewriting system known as {\em combinatory logic}
which simulates, to some extent, the $\lambda$-calculus.

\section{Syntax} \index{$\lambda$-calculus, type-free}
The (type-free) $\lambda$-calculus is composed
of the $\lambda$-terms defined by the following
grammar:
\[   
	M::= \w{id} \Alt (\lambda \w{id}.M) \Alt (MM)\qquad \mbox{($\lambda$-terms)}
\]
where $\w{id}::= x \Alt y \Alt \ldots$ is the syntactic category of
variables.
This is a minimal language
where the only operations allowed are {\em abstraction}
$\lambda x.M$ and {\em application} $MN$. 
In a language such as $\ml$, one would write $\lambda x.M$ as
{\small {\tt function x -> M}}.

It is important to notice that the abstraction $\lambda x.M$ 
{\em binds} the variable $x$
in the $\lambda$-term $M$ just as the quantified first-order formula
$\forall x.A$ binds $x$ in $A$. Consequently, in the $\lambda$-calculus
a variable can occur free or bound. We denote with $\fv{M}$ the set of
variables occurring {\em free} in the $\lambda$-term $M$.

When writing $\lambda$-terms we shall take some freedom.
First, we may write $\lambda x_1,\ldots,x_n.M$ for 
$\lambda x_1\ldots\lambda x_n.M$.
Second, we assume application associates to the left, and therefore write
$M_1 M_2 \ldots M_n$ for $(\cdots (M_1 M_2) \cdots M_n)$.
Third, we suppose application binds more than $\lambda$-abstraction and
write $\lambda x.MN$ for $\lambda x.(MN)$.

A number of programming operations can be introduced as {\em syntactic
sugar}. For instance, the operation
${\sf let}~x=M~{\sf~in~} N$ that binds the $\lambda$-term $M$ to the variable
$x$ and runs $N$  can be represented as $(\lambda x.N)M$.

$\lambda$-terms, like first-order logic formulae or integrals, 
are always manipulated up to the {\em renaming 
of bound variables}. For instance, we identify the $\lambda$-terms
$\lambda x.x$ and $\lambda y.y$, just as we would identify the
formulae $\qqs{x}{x=x}$ and $\qqs{y}{y=y}$, or the integrals
$\intind{x \ }{x}$ and $\intind{y \ }{y}$.

We remark that renaming involves a substitution of variables for variables. On
the other hand, the operation of substitution is really defined up to
renaming. For instance, to define $[y/x](\lambda y.xy)$ we start by
renaming the abstraction as $\lambda z.[z/y](xy) = \lambda z.xz$, where
$z$ is a fresh variable, and then we apply the substitution $[y/x]$
under the abstraction to obtain $\lambda z.yz$. 
More generally, to define a substitution $[N/x](\lambda y.M)$ we
have first to rename the bound variable $y$ as a (fresh) variable
$z$ which does not occur free either in $N$ or in $\lambda y.M$ and
then we can define the substitution as $\lambda z.[N/x][z/y]M$.
As such, the substitution is {\em not} a function
since countably many (equivalent) choices of the fresh variable $z$
are possible. However, we can make it into a function by
assuming an enumeration of the variables and picking up, for instance,
the first fresh variable that appears in the enumeration.

So we proceed as follows: first we define a substitution function 
on $\lambda$-terms, second we define the relation of $\alpha$-conversion, and
third we assume that $\lambda$-terms are handled up to $\alpha$-conversion.
In particular, in the proofs we shall distribute a substitution under
a $\lambda$-abstraction by silently assuming that an appropriate
renaming has been carried on.

\begin{definition}[size]\label{size-lambda-term-def}\index{size, $\lambda$-term}
If $M$ is a $\lambda$-term then its size $|M|$  is a natural number defined
as follows:
\[
\begin{array}{lll}
|x|=1, &|\lambda x.M| = 1+|M|, &|MN|=1+|M|+|N|~.
\end{array}
\]
\end{definition}

\begin{definition}[substitution]
\label{sub-def}\index{substitution, $\lambda$-calculus}
The substitution of a $\lambda$-term $N$ for a variable $x$ in the 
$\lambda$-term $M$ is denoted by $[N/x]M$ and it is defined as follows:
\[
\begin{array}{lll}

\sub{x}{N}y &=\left\{\begin{array}{ll}
                    N &\mbox{if }x=y \\
                    y &\mbox{otherwise}
                    \end{array}\right. \\ \\

\sub{x}{N}(M_1 M_2) &=\sub{x}{N}M_1 \sub{x}{N}M_2 \\ \\
\sub{x}{N}(\lambda y.M) 

&=\left\{\begin{array}{ll}
\lambda y.M &\mbox{if $x\notin \fv{\lambda y.M}$} \\
\lambda y.\sub{x}{N}M &\mbox{o.w., and $y\notin \fv{N}$} \\
\lambda z.\sub{x}{N}\sub{y}{z}M &\mbox{o.w., and $z$ first variable s.t. $z\notin \fv{MN}$.}
\end{array}\right.

\end{array}
\]
\end{definition}

To show that this definition makes sense consider
first the definition restricted to the case where 
$N$ is a variable and check that the substitution of a variable
for a variable in a $\lambda$-term leaves the size of the $\lambda$-term unchanged.

\begin{definition}[context]\label{context-def}\index{context, $\lambda$-calculus}
A {\em (one-hole) context} $C$ is defined by:
\[
C::= [~] \Alt \lambda \w{id}.C \Alt CM \Alt MC~.
\]
We write $C[N]$ for the $\lambda$-term obtained by
{\em replacing the hole} $[~]$ with the $\lambda$-term $N$
without paying attention to the potential capture
of variables. Formally:
\[
\begin{array}{llll}
[N]  = N,
&(\lambda x.C)[N] =\lambda x.C[N], 
&(CM)[N] =C[N]M,
&(MC)[N] = MC[N]~.
\end{array}
\]
\end{definition}

We are now ready to define the relation of renaming which is called
$\alpha$-conversion in the $\lambda$-calculus.
Henceforth $\lambda$-terms are considered up to $\alpha$-conversion.

\begin{definition}[$\alpha$-conversion]\label{alpha-conversion-def} \index{$\alpha$-conversion}
$\alpha$-conversion is the least equivalence relation $\equiv$ 
on $\lambda$-terms 
such that for any context $C$, $\lambda$-term $M$, and variables $x,y$
such that $y\notin \fv{M}$ we have:
\[
C[\lambda x.M] \equiv C[\lambda y.\sub{x}{y}M]~.
\]
\end{definition}

\begin{remark}
As already mentioned, the replacement operation does not pay attention
to the bound variables. For instance, if $C=\lambda x.[~]$ and
$N=x$ then $C[N]=\lambda x.x$. For this reason, contexts, unlike
$\lambda$-terms, should not be considered up to renaming.
\end{remark}

\begin{definition}[$\beta$-reduction]\label{beta-rule}\index{$\beta$-rule}
The {\em $\beta$-rule}
is the following reduction relation between $\lambda$-terms:
$$\begin{array}{ll}
(\beta) & C[(\lambda x.M)N]\arrow C[\sub{x}{N}M]\; ,
\end{array}$$
where $C$ is a context, $M,N$ are $\lambda$-terms, and $x$ is a variable.
\end{definition}

The subterm $(\lambda x.M)N$ which is transformed by the $\beta$-rule 
is called the {\em redex} (or $\beta$-redex). 
We may also refer to the $\lambda$-term resulting from the application
of the rule as the {\em reduced} $\lambda$-term.
Notice that definition $\ref{beta-rule}$ 
is schematic but does {\em not} quite define a TRS since $\lambda$-terms 
are not quite first-order terms. The equivalence induced by
$\beta$-reduction is called $\beta$-conversion and it is defined
as follows.

\begin{definition}[$\beta$-conversion]\label{beta-conversion}\index{$\beta$-conversion}
We denote with $=_\beta$ the equivalence 
relation $\stackrel{*}{\leftrightarrow}_{\beta}$.
\end{definition}

\begin{example}\label{ex-lambda-term}
Here are some {\em $\lambda$-terms} which are used often enough
to deserve a specific name:
\[
\begin{array}{lllll}
I \equiv \lambda x.x,
&K\equiv \lambda x, y.x,
&S \equiv \lambda x,y, z. xz(yz),
&\Delta \equiv  \lambda x.xx,
&\Delta_f \equiv \lambda x.f(xx)~.
\end{array}
\]
And here are some {\em examples of $\beta$-reduction} (up to $\alpha$-conversion!):
\[
\begin{array}{lllll}
II  \arrow  I,
&KMN \trarrow M,
&SKK \trarrow I, 
&\Delta\Delta \arrow \Delta\Delta,
&\Delta_f \Delta_f \arrow f(\Delta_f \Delta_f)~.
\end{array}
\]
\end{example}

\begin{exercise}[$\beta$-normal forms]\label{ex-normal-form}
Let $\w{NF}$ be the smallest set of $\lambda$-terms such
that:
\[
\infer{M_i\in \w{NF} \quad i=1,\ldots,k\quad k\geq 0}
{\lambda x_1\ldots x_n.xM_1\ldots M_k\in \w{NF}}~.
\]
Show that $\w{NF}$ is exactly the set of $\lambda$-terms
in {\em $\beta$-normal form}.
\end{exercise}

\begin{exercise}[Curry fixed point]\label{ex-curry-fixpoint}\index{fixed point, Curry}
Let $Y\equiv\lambda f.\Delta_f\Delta_f$ where 
$\Delta_f\equiv \lambda x.f(xx)$.
Show that:
\[
YM =_\beta M(YM)~.
\]
This is known as {\em Curry's fixed point combinator}.
\end{exercise}

\begin{exercise}[Turing fixed point]
\label{ex-turing-fixpoint}\index{fixed point, Turing}
{\em Turing's fixed point combinator} is defined by:
\[
Y_T\equiv (\lambda x,y.y(xxy))(\lambda x,y.y(xxy))~.
\]
Show that $Y_T f$ is not only 
convertible to, but {\em reduces to}: $f(Y_Tf)$. 
\end{exercise}

\section{Confluence}\index{confluence, $\lambda$-calculus}\label{confluence-lambda-sec}
Clearly, there are many possible ways of reducing a $\lambda$-term.
Are they confluent? Let us first examine the case for {\em local confluence}.

\begin{proposition}[local confluence]\label{local-conf-beta-prop}
Let $M$ be a $\lambda$-term. Then the following holds:

\begin{enumerate}

\item  If $M\arrow M'$ then $\sub{x}{M}N\trarrow\sub{x}{M'}N$. 

\item  If $N\arrow N'$ then $\sub{x}{M}N \arrow \sub{x}{M}N'$.

\item  $\beta$-reduction is {\em locally confluent}, that is:
\[
\infer{\qqs{M,N,P}{(M\arrow N, \qquad M\arrow P)}}
{\xst{Q}{(N\trarrow Q,  \qquad P\trarrow Q)}}~.
\]
\end{enumerate}
\end{proposition}
\Proof \Proofitemf{(1)} By induction on $N$.

\Proofitemm{(2)} Suppose $N=C[(\lambda y.N_1)N_2]$. We notice:
\[
\sub{x}{M}((\lambda y.N_1)N_2) \equiv (\lambda y.\sub{x}{M}N_1)\sub{x}{M}N_2 
\arrow \sub{y}{\sub{x}{M}N_2}(\sub{x}{M}N_1) \equiv
\sub{x}{M}(\sub{y}{N_2}N_1)~.
\]
\Proofitemf{(3)} The interesting case arises if one redex is contained
in the other.  Suppose $\Delta$ is a $\beta$-redex.
If $M\equiv C[(\lambda x.M')C'[\Delta]]$ 
apply (1), and if 
$M\equiv C[(\lambda x.C'[\Delta])M']$ apply (2). \qed \\

Let us notice that $\beta$ reduction may both {\em erase} a redex as
in $(\lambda x.I)(II) \arrow I$ and {\em duplicate} it as in 
$\Delta(II) \arrow (II)(II)$. 
It turns out that it is possible to define a notion 
of {\em parallel reduction} $\Arrow$ with the following properties.
\begin{itemize}

\item $\arrow  \ \subset \ \Arrow \ \subset \ \trarrow$.

\item A {\em strong confluence property} holds for $\Arrow$:
if $M\Arrow N$ and $M\Arrow N'$ then there is $P$ such that $N\Arrow P$ and 
$N'\Arrow P$.

\item The relation $\Arrow$ is simple enough to be analyzed. 

\end{itemize}

The idea is that in a parallel reduction we are allowed to reduce at once
the redexes that are in the $\lambda$-term but not those which are created
by the reductions. For instance, we have:
$(II)(II) \Arrow II$ but $(II)(II)\not\Arrow I$.

\begin{definition}[parallel $\beta$-reduction]\label{parallel-red-lambda} \index{$\beta$-reduction, parallel}
Parallel $\beta$-reduction is defined as follows:
\[
\begin{array}{cc}

\infer{}{M\Arrow M}

&\infer{M\Arrow M'\quad N\Arrow N'}
{(\lambda x.M)N\Arrow \sub{x}{N'}M'} \\ \\

\infer{M\Arrow M'\quad N\Arrow N'}
{MN\Arrow M'N'} 

&\infer{M\Arrow M'}
{\lambda x.M\Arrow\lambda x.M'}~.

\end{array}
\]
\end{definition}

\begin{exercise}\label{min-par-red-ex}
Let $M \equiv (\lambda x.Ix)(II)$ where
$I\equiv \lambda z.z$.
What is the minimum number of parallel reductions
needed to reduce $M$ to $I$?
\end{exercise}

First we notice the following structural and substitution properties
of parallel reduction.

\begin{proposition}
Parallel reduction enjoys 
the following {\em structural properties}:
\[
\begin{array}{c}

\infer{\lambda x.M\Arrow N}
{N\equiv \lambda x.M',\qquad M\Arrow M'} \\ \\

\infer{MN \Arrow L}
{\begin{array}{c}
(L\equiv M'N',\qquad M\Arrow M', \qquad N\Arrow N') \ \mbox{ or }\\
(M\equiv \lambda x.P,\quad P\Arrow P', \quad N\Arrow N', \quad L\equiv [N'/x]P')
\end{array}}
\end{array}
\]
\end{proposition}
\Proof By case analysis on the definition of parallel reduction. \qed

\begin{proposition}
Parallel reduction enjoys the following
{\em substitution property}:
\[
\infer{M\Arrow M' \qquad N\Arrow N'}
{[N/x]M\Arrow [N'/x]M'}~.
\]
\end{proposition}
\Proof  By induction 
on the definition of $M\Arrow M'$.
For the base case we also need an induction on the structure of $M$. \qed \\

We are then ready to prove {\em strong confluence} of parallel reduction.

\begin{proposition}\label{strong-conf-par-red-prop}
Parallel reduction enjoys the following {\em strong confluence} property:
\[
\infer{\qqs{M,N_1,N_2}{(N_1 \Leftarrow M \Arrow N_2)}}
{\xst{P}{(N_1\Arrow P \Leftarrow N_2 )}}~.
\]
\end{proposition}
\Proof 
One can proceed by induction on $M\Arrow N_1$ and case analysis
on $M\Arrow N_2$ to close the diagram. \qed

\begin{corollary}[confluence, $\beta$]\index{confluence, $\beta$-reduction}
$\beta$-reduction is confluent.
\end{corollary}
\Proof
We have: 
\[
\arrow_\beta \ \subset  \ \Arrow\  \subset \ \trarrow_\beta~.
\] 
If $M\arrow_\beta \cdots \arrow_\beta N_i$, $i=1,2$ then 
$M \Arrow \cdots \Arrow N_i$, $i=1,2$.
Now apply strong confluence to close the diagram and build $P$ such
that $N_i \Arrow \cdots \Arrow P$, $i=1,2$.
This implies $N_i \trarrow_\beta \cdots \trarrow_\beta P$, $i=1,2$ 
and, by transitivity of $\trarrow_\beta$, we conclude that
$N_i\trarrow_\beta P$, $i=1,2$. \qed \\

The $\beta$-rule is the basic rule of the $\lambda$-calculus.
The second most popular rule is the $\eta$-rule.

\begin{definition}[$\eta$-reduction]\label{eta-rule-def}\index{$\eta$-rule}
The $\eta$-rule is defined by:
$$\begin{array}{ll}
(\eta) & C[\lambda x.Mx]\arrow C[M]\qquad\mbox{if }x\not\in \fv{M}~,
\end{array}$$
for $C$ context, $M$ $\lambda$-term, and $x$ variable.
\end{definition}

The $\eta$-rule is a kind of {\em extensionality rule}.
If we read it backwards, it asserts that `every $\lambda$-term is a 
function'. This intuition can actually be made precise in
the model theory of $\lambda$-calculus.

\begin{proposition}[confluence $\beta\eta$]\label{confluence-eta-rule}\index{$\eta$-rule, confluence}
The following properties hold:
\begin{enumerate}

\item $\eta$ reduction is strongly {\em confluent} in the following
sense:
\[
\infer{M \arrow_\eta N_i\quad i=1,2\quad N_1\not\equiv N_2}
{\xst{P}{(N_i \arrow_\eta P,\quad i=1,2)}}~.
\]

\item The $\beta$ and $\eta$ reductions {\em commute} 
in the following sense:
\[
\infer{M \arrow_\beta N_1\quad M\arrow_\eta N_2\quad N_1\not\equiv N_2}
{\xst{P}{(N_1 (\arrow_\eta)^* P,\quad N_2 \arrow_\beta P)}}~.
\]

\item The $(\arrow_\beta)^*$ and $(\arrow_\eta)^*$ reductions commute in the
following sense:
\[
\infer{M (\arrow_\beta)^* N_1\quad M (\arrow_\eta)^* N_2}
{\xst{P}{(N_1 (\arrow_\eta)^* P,\quad N_2 (\arrow_\beta)^* P)}}~.
\]

\item $\beta\eta$ reduction is {\em confluent}.

\end{enumerate}
\end{proposition}
\Proof
\Proofitemf{(1)} 
Two redexes that superpose have the shape: $\lambda x.C[\lambda y.My]x$.
Analyze what can happen.

\Proofitemm{(2)} If the $\beta$ redex contains the $\eta$ redex we can have the following situations:
\begin{itemize}
\item  $(\lambda x.Mx)N$: the reduced are identical.
\item  $(\lambda x.C[\lambda y.My])N$: close the diagram in one step. 
\item  $(\lambda x.M)C[\lambda y.Ny]$: it may take $0$, $1$ or more $\eta$ steps to close the diagram.
\end{itemize}
On the other hand, if the $\eta$ redex contains the $\beta$ redex we can have:
\begin{itemize}

\item  $\lambda x.(\lambda y.M)x$: the reduced are identical.

\item $\lambda x.C[(\lambda y.M_1)M_2]x$: close in one step.
\end{itemize}

\Proofitemm{(3)} 
First show commutation of $(\arrow_\eta)^*$ with respect to $\arrow_\beta\union \w{Id}$. Then proceed by induction on the number  of $\beta$ reductions.

\Proofitemm{(4)} 
Consider the number of alternations of $(\arrow_\beta)^*$ 
and $(\arrow_\eta)^*$. \qed

\begin{example}\label{non-confluence-ex}
Here is an extension of the $\lambda$-calculus that
{\em does not preserve confluence} (we refer to \cite{Barendregt84} 
for a proof).
We add to the language a constant $D$ and the rule:
\[
Dxx \arrow x~.
\]
This rule may seem artificial, but it is actually 
a simplification of a natural rule called
{\em surjective paring} (an extensionality rule for pairs) 
which also leads to a non-confluent system:
\[
D(Fx)(Sx)\arrow x~.
\]
Here $D$, $F$, $S$ are constants where intuitively $D$
is the pairing while $F$ and $S$ are the first and second projection.
We stress that here the property that fails is just confluence (not local confluence). Indeed a surjective pairing rule is introduced in {\em terminating}
typed $\lambda$-calculi. By proposition \ref{newman-proposition}, 
surjective pairing  in these calculi {\em is} confluent.
\end{example}

\section{Programming}\label{program-lambda-sec}
All {\em partial recursive functions} can be represented in the (type free)
$\lambda$-calculus. Thus the $\lambda$-calculus,
regarded as a computational model, is {\em Turing equivalent}.
Proving this result is a matter of programming in the $\lambda$-calculus.
The proof we outline below relies on the following 
{\em definition} of the partial recursive functions.

\begin{definition}[minimalisation]\label{minimalisation-def}\index{minimization}
Given a {\em total function} $f:\Nat^{k+1}\arrow \Nat$ a
{\em partial function} $\mu(f):\Nat^k\parrow \Nat$ is defined 
by {\em minimization} as follows:
\[
\mu(f)(x_1,\ldots,x_k) = \left\{
\begin{array}{ll}
x_0 &\mbox{if }x_0=\w{min}\set{x\in \Nat \mid  f(x,x_1,\ldots,x_k)=0} \\
\uparrow &\mbox{if }\qqs{x}{f(x,x_1,\ldots,x_k)>0}
\end{array}
\right.
\]
where $\uparrow$ means that the function is undefined.
\end{definition}

\begin{definition}[partial recursive functions]\label{pr-fun-def}\index{partial recursive functions}
The set of partial recursive functions is the smallest set of functions on 
(vectors of) natural numbers which contains the {\em basic functions} (zero, successor, projections) and
is closed under {\em function composition}, {\em primitive recursion} 
(see chapter \ref{prim-rec-sec}), and  {\em minimization}.
\end{definition}

We discuss next the representation of partial recursive functions
in the $\lambda$-calculus.

\begin{definition}[Church numerals]\label{church-num-def}\index{Church numerals}
A natural number $n$ is represented by the following $\lambda$-term $\ul{n}$
known as Church numeral:
\begin{equation}
\ul{n} \equiv \lambda f.\lambda x.(f\cdots (fx)\cdots) \qquad \mbox{(Church numerals)}
\end{equation}
where $f$ is applied $n$ times. 
\end{definition}

In a sense this is similar to the {\em tally natural numbers}
considered in chapter \ref{prim-rec-sec}.
We shall see in chapter \ref{systemF-sec} that the inductive definition 
of natural numbers actually {\em suggests} their 
representation in the $\lambda$-calculus as Church numerals.

We also have to fix a class of $\lambda$-terms that represent a 
diverging computation. A natural choice is to consider the
$\lambda$-terms that do {\em not} have a {\em head normal form}.

\begin{definition}[head normal form] \index{head normal form} \label{hnf-def}
A $\lambda$-term is (has) 
a head normal if it has the shape (it reduces to a $\lambda$-term
of the shape):
\[
\lambda x_1,\ldots,x_n.xM_1\cdots M_m\qquad n,m\geq 0.
\]
\end{definition}

\begin{definition}[function representation]\label{fun-rep-def}\index{function representation, $\lambda$-calculus}
A $\lambda$-term $F$ {\em represents} 
a partial function $f:\Nat^k \arrow \Nat$ if
for all $n_1,\ldots, n_k \in \Nat$:
\[
\begin{array}{lll}
f(n_1,\ldots,n_k)=m  &\mbox{ iff } &F\ul{n_{1}} \cdots \ul{n_{k}}=_\beta \ul{m} 
\\ 
f(n_1,\ldots,n_k)\uparrow &\mbox{ iff } &F\ul{n_{1}} \cdots \ul{n_{k}} 
\mbox{ has no head normal form}.
\end{array}
\]
\end{definition}

We can represent the arithmetic functions {\em addition, successor, 
and multiplication} with the following $\lambda$-terms:
\[\begin{array}{lll}
A &\equiv \lambda n.\lambda m.\lambda f.\lambda x.(n\ f)(m\ f\ x) &\mbox{(addition)}\\
S &\equiv \lambda n.A\ n \ \ul{1}                          &\mbox{(successor)}\\
M & \equiv \lambda n.\lambda m.\lambda f .n(m \ f)       &\mbox{(multiplication).}
\end{array}
\]
To represent {\em boolean values} we introduce the following
$\lambda$-terms:
\[
\begin{array}{llll}
T \equiv \lambda x.\lambda y.x &\mbox{(true)}~, \qquad
F \equiv \lambda x.\lambda  y.y &\mbox{(false).}
\end{array}
\]
Then an {\em if-then-else} $\lambda$-term can be defined as follows:
\[
\begin{array}{lll}
\w{C} &\equiv \lambda x.\lambda y.\lambda z.x \ y\ z &\mbox{(if-then-else).} 
\end{array}\]
The reader may check that:
$C  T x y \trarrow_\beta x$ and $C F x y \trarrow_\beta y$.
A {\em test-for-zero} $\lambda$-term on Church numerals can
be defined as follows:
\[
\begin{array}{lll}
\w{Z} &\equiv \lambda n.n(\lambda x.F)T &\mbox{(test-for-zero).}
\end{array}
\]
We can also introduce $\lambda$-terms to {\em build pairs} and
to {\em project pairs} as follows:
\[
\begin{array}{lll}
P &\equiv \lambda x.\lambda y.\lambda z.z\ x \ y\qquad  &\mbox{(pairing)}  \\
P_1 &\equiv \lambda p.p(\lambda x,y.x)              &\mbox{(first projection)} \\
P_2 &\equiv \lambda p.p(\lambda x,y.y)              &\mbox{(second projection).}
\end{array}
\]
Again, the reader may check that $P_i(P M_1 M_n)\trarrow_\beta M_i$ for $i=1,2$.

\begin{exercise}\label{pd-sub-exp-ex}
Check that the $\lambda$-term:
\[
\w{Pd} \ \equiv \ \lambda n,f, x.n(\lambda g,h.h(gf))(\lambda y.x)(\lambda z.z)
\]
represents the predecessor function where it is assumed that the predecessor
of $0$ is $0$ (chapter \ref{systemF-sec} provides 
a rational reconstruction of this complicated $\lambda$-term).
Define $\lambda$-terms to represent the subtraction function,
where $m-n=0$ if $n>m$, and the exponential function $n^m$.
\end{exercise}

Let us now consider the $3$ {\em composition mechanisms}, namely:
function composition, primitive recursion, and minimization.
It should be clear that {\em function composition} can be directly represented
in the $\lambda$-calculus. Primitive recursion can be regarded as a particular
case of recursive function definition.
In turn, a {\em recursive function definition} such as:
\[
\s{letrec}\  g(x) = M \ \s{in} \ N~,
\]
where $g$ may appear in $M$ and $N$ is coded in the $\lambda$-calculus as:
\[
(\lambda g.N)(Y(\lambda g.\lambda x.M))~,
\]
where $Y$ is the fixed point combinator of exercise \ref{ex-curry-fixpoint} or \ref{ex-turing-fixpoint}.
Moreover, recursive definitions provide a direct
mechanism to mimick definitions by minimization.
Given a function $f$, consider the following recursive 
definition of the function $g$:
\[
g(x_0,x_1,\ldots,x_k) = 
\s{if} \ (f(x_0,x_1,\ldots,x_k)=0) \ \s{then} \ x_0 \ \s{else} \ g(x_0+1,x_1,\ldots,x_k)~.
\]
Then $\mu(f)(x_1,\ldots,x_k) = g(0,x_1,\ldots,x_k)$. 
Putting all together, we have the following result.

\begin{proposition}
For all partial recursive functions $f:\Nat^k\arrow \Nat$ there 
is a closed $\lambda$-term $F$ which represents $f$ in the
sense of definition \ref{fun-rep-def}.
\end{proposition}

\section{Combinatory logic (*)}
Combinatory logic is a relative of the $\lambda$-calculus
which can be presented as a term rewriting system. 

\begin{definition}[combinatory logic]\label{comb-logic-def} \index{combinatory logic}
We consider a binary application operation $@$ and 
two constants $K$ and $S$. 
As in the $\lambda$-calculus, we write $MN$ for $@(M,N)$
and let application associate to the left.
The system comes with two term rewriting rules:
\[
\begin{array}{cc}
K\ x\ y \arrow x~, \qquad &S\ x\ y\ z \arrow x\ z(y\ z)~.
\end{array}
\]
\end{definition}

It turns out that in combinatory logic there
is a way to {\em simulate} $\lambda$-abstraction.

\begin{definition}\label{abs-def-comb-logic} \index{abstraction, combinatory logic}
We define a function $\lambda$ that takes a variable and a
term of combinatory logic and produces a term of combinatory
logic. We abbreviate $SKK$ as $I$.
\[
\begin{array}{lll}
\lambda(x,x)  &= I  \\
\lambda(x,M)  &= KM                             &\mbox{if }x\notin \var{M}\\
\lambda(x,MN) &= S(\lambda(x,M))(\lambda(x,N))~.
\end{array}
\]
\end{definition}

The fact that we called the function above $`\lambda'$ is justified
by the following proposition.

\begin{proposition}
If $M,N$ are terms of combinatory logic and $x$ is a variable
then:
\[
\lambda(x,M)N \trarrow [N/x]M~.
\]
\end{proposition}
\Proof
By induction on $M$ following the definition of the translation. \qed

\begin{exercise}\label{cl-confulence-ex}
Using the fact that combinatory logic (CL) is a TRS prove
local confluence of CL. 
Then adapt the
method of parallel reduction presented in section \ref{confluence-lambda-sec}
to prove the confluence of CL.
\end{exercise}

Combinatory logic seems mathematically simpler than the $\lambda$-calculus.
Why is it not used? One reason is that terms written in combinatory logic
tend to be unreadable. Another deeper reason is that the notion
of conversion induced by the rules $S$ and $K$ is weaker than the one induced
by the $\beta$ rule. For instance, the reader may check that 
the translations in CL of the $\lambda$-terms
$\lambda z.(\lambda x.x)z$ and $\lambda z.z$
do not have a common reduct. 
An alternative approach goes through the notion of {\em closure} 
(see following chapter \ref{env-closure-sec}). This is quite
appropriate for discussing implementation techniques,
but as in combinatory logic, the notation tends to become less
manageable.

\section{Summary and references}
The $\lambda$-calculus is a minimal notation to represent
higher-order functions. The $\lambda$-terms are transformed
according to one basic rewriting rule: the $\beta$-rule.
The $\lambda$-calculus with the $\beta$-rule 
is a confluent rewriting system and it is sufficiently
expressive to represent all partial recursive functions.
A second rule, the $\eta$-rule, can be added to the system while
preserving confluence.
The $\lambda$-calculus is {\em not} a term rewriting system but
there are term rewriting systems such as {\em combinatory logic} 
which can mimick to some extent  the behavior of $\lambda$-terms.

The $\lambda$-calculus was introduced by Church \index{Church} as
part of an investigation in the formal foundations of
mathematics and logic \cite{DBLP:journals/jsyml/Church40}.  
At the time, the $\lambda$-calculus provided one of 
the concurrent formalizations
of {\em partial recursive functions}, {\em i.e.}, computable
functions, along with, {\em e.g.}, Turing machines.
The foundational character of the language is even stronger
when it is enriched with {\em types}. We shall start addressing this point 
in chapter \ref{simple-type-sec}. 
The related system of combinatory logic is based on work
by Sch\"onfinkel and Curry.\index{Curry}\index{Sch\"{o}nfinkel}
The book \cite{Barendregt84} is the basic reference for the
type-free $\lambda$-calculus. It is enough to skim the first
introductory chapters to have an idea of the great variety
of results connected to the formalism.

\chapter{Weak reduction strategies, closures, and abstract machines}\label{env-closure-sec}
\markboth{{\it Abstract machines}}{{\it Abstract machines}}

Full $\beta(\eta)$-reduction is the basis for the symbolic manipulation
of $\lambda$-terms, {\em e.g.}, in proof assistants, in program
transformations, and in higher-order unification and pattern-matching.
However, when the $\lambda$-calculus  is regarded as the core
of a programming language it is sensible to consider {\em weaker} reduction
strategies. This chapter focuses on these weaker reduction strategies and their
implementation.

\section{Weak reduction strategies}
A {\em weak} reduction strategy is a strategy to reduce $\lambda$-terms
that does {\em not} reduce under functional abstractions. Thus in 
a weak reduction strategy all $\lambda$-terms of the form $\lambda x.M$
are normal forms.  

\begin{definition}[weak reduction]\index{weak $\beta$-reduction}
We define the weak $\beta$-reduction relation $\arrow_w$ as 
the least binary relation on $\lambda$-terms such that:
\[
\begin{array}{lll}
\infer{}{(\lambda x.M)N \arrow_w [N/x]M}
\quad
&\infer{M\arrow_w M'}{MN\arrow_w M'N}
\quad
&\infer{N\arrow_w N'}{MN \arrow_w MN'}~.
\end{array}
\]
\end{definition}

As such weak reduction is {\em not} confluent. For instance, 
we have:
\[
K(II) \arrow_w KI~, \qquad K(II) \arrow_w \lambda x.II~,
\]
and  $KI$ and $\lambda x.II$ have no common reduct. The problem here is
that the redex $II$ is under a $\lambda$ and cannot be reduced.
When considering the $\lambda$-calculus as the core
of a programming language, the usual approach is to fix
a particular
{\em deterministic} weak reduction strategy.
Two popular ones we discuss next are known as 
{\em call-by-name} and {\em call-by-value}.
The definition of these strategies relies on a notion of {\em value}.

\begin{definition}[value]\index{value, $\lambda$-calculus}
A {\em value} $V$ is a closed $\lambda$-term of the shape
$\lambda x.M$ (a $\lambda$-abstraction).
\end{definition}

In the following, the call-by-name and call-by-value reduction
strategies are defined on {\em closed} $\lambda$-terms. We
actually define the reduction strategies in $3$ different ways
which turn out to be equivalent.

\begin{definition}[call-by-name]\label{cbn-def}\index{call-by-name}
We define the call-by-name reduction relation $\arrow_n$ as 
the least binary relation on closed $\lambda$-terms such that:
\[
\begin{array}{cc}
\infer{}{(\lambda x.M)N \arrow_n [N/x]M}
\qquad
&\infer{M\arrow_n M'}{MN\arrow_n M'N}~.

\end{array}
\]
\end{definition}

\begin{definition}[call-by-value]\label{cbv-def}\index{call-by-value}
We define the call-by-value reduction relation $\arrow_v$ as 
the least binary relation on closed $\lambda$-terms such that:
\[
\begin{array}{lll}
\infer{}{(\lambda x.M)V \arrow_v [V/x]M}
\qquad
&\infer{M\arrow_v M'}{MN\arrow_v M'N}
\qquad
&\infer{N\arrow_v N'}{VN \arrow_v VN'}~.
\end{array}
\]
\end{definition}

\begin{remark}
The basic difference between call-by-name and call-by-value is that
in the latter we insist that the term passed to the function is a
value.
Also notice that in the definitions above, we have taken the convention that
the function is reduced before the argument. Of course, an alternative
definition where the argument is reduced before the function 
is possible. This choice only matters if the language has
side-effects (cf. chapter \ref{lamref-sec}).
\end{remark}

The definitions \ref{cbn-def} and \ref{cbv-def} give a strategy 
to look for a subterm which is a redex of the right shape. 
The one-hole context which sourrounds the redex is called 
{\em evaluation context}. 

\begin{definition}[evaluation contexts]\label{ev-cxt-def}\index{evaluation context}
Call-by-name and call-by-value evaluation contexts are denoted with $E,E',\ldots$ and
are defined as follows:
\[
\begin{array}{ll}
E::= [~] \Alt EM  &\mbox{(call-by-name evaluation context)} \\
E::= [~] \Alt EM \Alt VE &\mbox{(call-by-value evaluation context).}
\end{array}
\]
\end{definition}

\begin{proposition}[decomposition]\label{decomp-ev-cxt}
Let $M$ be a closed $\lambda$-term. Then either $M$ is a value or there
is a unique call-by-name (call-by-value) evaluation context $E$
such that:
\[
M\equiv E[(\lambda x.M_1)M_2] \qquad  (M\equiv E[(\lambda x.M_1)V])~.
\]
\end{proposition}
\Proof
By induction on the structure of $M$. 
$M$ cannot be a variable because it is closed.
If $M$ is a $\lambda$-abstraction then it is a value.
Suppose, $M\equiv M'M''$. 
We consider the case for {\em call-by-name}.
If $M'$ is a value then
it must be a $\lambda$-abstraction and $E\equiv [~]$.
Otherwise, by inductive hypothesis $M'\equiv E'[\Delta]$ where
$\Delta$ is a $\beta$-redex and we take $E\equiv E'M''$. \qed \\

We can rely on evaluation contexts to provide alternative and 
equivalent definitions of call-by-name and call-by-value.

\begin{definition}\label{cbn-cbv-ev-cxt}
Let $\arrow_{en}$ be the least binary reduction relation on closed 
$\lambda$-terms such
that:
\[
M\arrow_{en} N \mbox{if }M\equiv E[(\lambda x.M_1)M_2] \mbox{ and }N\equiv E[[M_2/x]M_1], E \mbox{ call-by-name evaluation context.}
\]
Let $\arrow_{ev}$ be the least binary reduction relation on closed
$\lambda$-terms such that:
\[
M\arrow_{ev} N \mbox{if }M\equiv E[(\lambda x.M_1)V] \mbox{ and }N\equiv E[[V/x]M_1], E \mbox{ call-by-value evaluation context.}
\]
\end{definition}

\begin{proposition}
The call-by-name reduction relation $\arrow_n$ coincides with 
the relation $\arrow_{en}$ and the call-by-value reduction relation
$\arrow_v$ coincides with the relation $\arrow_{ev}$.
\end{proposition}
\Proof 
We consider the proof for call-by-name. 
To show that $\arrow_n\subseteq \arrow_{en}$, 
we proceed by induction of the
proof height of $M\arrow_n N$. 
For the base case take $E=[~]$.
For the inductive case, suppose $MN\arrow_n M'N$ because
$M\arrow_n M'$. Then by inductive hypothesis, there are
$E'$ and $\Delta\equiv (\lambda x.M_1)M_2$ such that
$M\equiv E'[\Delta]$ and $M'\equiv E'[[M_2/x]M_1]$.
Then take $E=E'N$, $MN \equiv E[\Delta]$, $M'N\equiv E[\Delta']$, 
and $\Delta'\equiv [M_2/x]M_1$. 

In the other direction, suppose $E[\Delta]\arrow_{\w{en}} E[\Delta']$
where $\Delta\equiv (\lambda x.M_1)M_2$ and $\Delta'\equiv [M_2/x]M_1$.
We proceed by induction on the
structure of the evaluation context $E$.
If $E=[~]$ then $\Delta \arrow_n \Delta'$.
If $E=E'N$ then by induction hypothesis, $E'[\Delta]\arrow_n E'[\Delta']$
and therefore $E[\Delta]\arrow_n E[\Delta']$. \qed \\

Yet another presentation of call-by-name and call-by-value 
consists in defining a big-step (cf. section \ref{imp-sec}) 
 evaluation relation $\eval$.

\begin{definition}[call-by-name evaluation]\label{cbn-eval}
The call-by-name evaluation relation $\eval_n$ is the least binary relation 
on closed $\lambda$-terms such that:
\[
\begin{array}{ll}
\infer{}{V\eval_n V}\qquad

&\infer{M\eval_n \lambda x.M'\qquad [N/x]M\eval_n V}
{MN \eval_n V}~.
\end{array}
\]
\end{definition}

\begin{definition}[call-by-value evaluation]\label{cbv-eval}
The call-by-value evaluation relation $\eval_v$ is the least binary relation 
on closed $\lambda$-terms such that:
\[
\begin{array}{ll}
\infer{}{V\eval_v V}
\qquad
&\infer{M\eval_v \lambda x.M'\qquad N\eval_v V'\qquad [V'/x]M\eval V}
{MN \eval_v V}~.
\end{array}
\]
\end{definition}

\begin{proposition}
Let $M$ be a closed $\lambda$-term.  Then:
\begin{enumerate}
\item If $M\eval_n V$ then $M(\arrow_n)^* V$.

\item If $M\arrow_n M'$ and $M'\eval_n V$ then $M\eval_n V$.

\item If $M(\arrow_n)^* M'\not\arrow_n$ then $M\eval_n M'$.
\end{enumerate}
The same properties hold if we replace $\eval_n$ with $\eval_v$
and $\arrow_n$ with $\arrow_v$, respectively.
\end{proposition}
\Proof
\Proofitemf{(1)} By induction on the proof height of the judgment
$M\eval_n V$. The base case follows by reflexivity of $(\arrow_n)^*$.
For the inductive step, suppose $MN\eval_n V$ because 
$M\eval_n \lambda x.M_1$ and $[N/x]M_1\eval_n V$. By inductive hypothesis,
$M(\arrow_n)^* \lambda x.M_1$. Then:
\[
MN (\arrow_n)^* (\lambda x.M_1)N \arrow_n [N/x]M_1~,
\]
and by inductive hypothesis $[N/x]M_1 (\arrow_n)^* V$.

\Proofitemm{(2)} If $M\arrow_n M'$ then 
$M\equiv (\lambda x.M_1)M_2 N_1\cdots N_k$ and 
$M'\equiv [M_2/x]M_1N_1\cdots N_k$.
If $M'\eval_n V$ we must have:
\[
[M_2/x]M_1\eval_n \lambda x_1.P_1, \quad
[N_1/x_1]P_1 \eval_n \lambda x_2.P_2, \quad \cdots \quad
[N_k/x_k]P_k \eval_n V~.
\]
Then to  prove $M\eval V$  it suffices to extend
the proof for $M'$ with an additional step 
$\lambda x.M_1 \eval_n \lambda x.M_1$.

\Proofitemm{(3)}
By proposition \ref{decomp-ev-cxt}, if $M'$ does not reduce then it is
a value and we have $M'\eval_n M'$. If $M$ reduces to $M'$ in $k$ steps
then we apply property (2) $k$ times starting from $M'$. \qed

\section{Static vs. dynamic binding}\label{dynamic-static-sec}\index{binding, static or dynamic}
The implementation of the reduction of a $\beta$-redex such as 
$(\lambda x.M)N$ is usually decomposed in two steps.
\begin{itemize}

\item The formal parameter $x$ is bound to the {\em argument} $N$. 
The collection of bindings is called an {\em environment}.

\item When the formal parameter $x$ is {\em used} in the body of the function 
$M$, the argument bound to it is retrieved from the environment.

\end{itemize}

This high-level description leaves many design choices unspecified.
One basic issue is what exactly constitutes
an `argument'.  Indeed, in the programming languages jargon,
one speaks of {\em static} vs. {\em dynamic} binding.
This issue already arises in a very simple language
of expressions whose syntax is as follows, where as usual
$\w{id}::=x\Alt y \Alt \cdots$:
\[
     e::= \bot \Alt n \Alt \w{id} \Alt \s{let}~\w{id}=e~\s{in}~e \Alt 
     \s{quote}(e) \Alt \s{unquote}(e)~\qquad \mbox{(expressions).}
\]
Here $\bot$ represents a computation that \bem{diverges},  
$n$ is an \bem{integer}, \s{quote} allows to \bem{freeze} the
evaluation of an expression and \s{unquote} to \bem{unfreeze} it.
We denote with $\w{Exp}$ the set of expressions in this language.
The collection of \bem{values} is defined by:
\[
v::= n \Alt \s{quote}(e)~\qquad\mbox{(values).}
\]
It is possible to encode this simple language in the $\lambda$-calculus
and reproduce the same phenomena we describe next.

\begin{definition}[dynamic environment]\index{environment, dynamic}
A {\em dynamic} environment is a 
partial function $\eta: \w{Id} \parrow \w{Exp}$ 
with finite domain mapping
identifiers to expressions.
\end{definition}

Table \ref{eval-dynamic-tab} introduces two evaluation relations
for this language of expressions with dynamic binding following
either a {\em by-name} or a {\em  by-value}  strategy.
The basic assertion $e[\eta]\eval v[\eta']$ states that an expression
$e$ in an environment $\eta$ evaluates to a value $v$ in an environment
$\eta'$. 
The first $3$ rules defining the assertion 
are shared while two distinct rules, one for by-name and 
the other for by-value, cover expressions of the shape
$\s{let}~x=e'~\s{in}~e$.

\begin{table}
{\small
\[
\begin{array}{ccc}

\infer{}{v[\eta] \eval v[\eta]} \qquad

&\infer{\eta(x)[\eta] \eval v[\eta']}{x[\eta]\eval v[\eta']} 

&\infer{e[\eta] \eval \s{quote}(e')[\eta'] \quad 
        e'[\eta'] \eval v[\eta'']}
{\s{unquote}(e)[\eta] \eval v[\eta'']} 
\end{array}
\]
\[
\begin{array}{cc}
\begin{array}{c}
\mbox{By-name:} \\ 
\infer{e[\eta[e'/x]]\eval v[\eta'],}
{(\s{let}~x=e'~\s{in}~e)[\eta]) \eval v[\eta']} 
\end{array}

&\begin{array}{c}
\mbox{By-value:}\\ 

\infer{e'[\eta]\eval v'[\eta']\quad e[\eta[v'/x]] \eval v[\eta]}
{(\s{let}~x=e'~\s{in}~e)[\eta] \eval v[\eta]}
\end{array}

\end{array}
\]}
\caption{Dynamic binding with by-name and by-value evaluation}
\label{eval-dynamic-tab}
\end{table}

In {\em dynamic binding}, an environment binds an expression with an identifier.
However, in turn the expression may contain identifiers and their
binding with other expressions may be lost. 
In {\em static binding}, we introduce a more complex object
which is called a {\em closure}. This is an expression along with
an {\em environment} that associates identifiers with {\em closures}.
This looks like a circular definition of closure and environment 
but things can be well-defined in an inductive style as follows.

\begin{definition}[static environment]\index{environment, static}
The set of environments $\w{Env}$ is the smallest set of 
partial functions on  \w{Id} such that 
if 
$e_i\in \w{Exp}$, $\eta_i\in \w{Env}$ and $\fv{e_i}\subseteq \w{dom}(\eta_i)$
for $i=1,\ldots, n$ ($n\geq 0$) then
\[
[e_1[\eta_1]/x_1,\ldots,e_n[\eta_n]/x_n]\in \w{Env}~.
\]
We denote with $\emptyset$ the empty environment.
\end{definition}

\begin{definition}[closure]\index{closure}
A {\em closure} is a pair $e[\eta]$ composed of an expression $e\in \w{Exp}$ and an 
environment $\eta\in \w{Env}$ such that $\fv{e}\subseteq \w{dom}(\eta)$.
\end{definition}

Table \ref{static-eval-tab} defines an evaluation relation  whose
basic assertion is $e[\eta]\eval v[\eta]$.  The first $3$ rules are
formally identical to those for dynamic binding but recall that
now $\eta$ is a {\em static} environment.

\begin{table}
{\small
\[
\begin{array}{ccc}

\infer{}{v[\eta] \eval v[\eta]} \qquad

&\infer{\eta(x) \eval v[\eta']}{x[\eta]\eval v[\eta']} \qquad

&\infer{\begin{array}{c}
e[\eta] \eval \s{quote}(e')[\eta'] \qquad
        e'[\eta'] \eval v[\eta'']\end{array}}
{\s{unquote}(e)[\eta] \eval v[\eta'']} 
\end{array}
\]

\[
\begin{array}{cc}

\begin{array}{c}
\mbox{By-name:} \\ 
\infer{e[\eta[e'[\eta]/x] \eval v[\eta']}
{(\s{let}~x=e'~\s{in}~e)[\eta] \eval v[\eta']} 
\end{array}

&\begin{array}{c}
\mbox{By-value:} \\ 
\infer{\begin{array}{c}
e'[\eta]\eval u[\eta'']\qquad
        e[\eta[u[\eta'']/x]) \eval v[\eta']\end{array}}
{( \s{let}~x=e'~\s{in}~e)[\eta] \eval v[\eta']}
\end{array}

\end{array}
\]}

\caption{Static binding with by-name and by-value evaluation}\label{static-eval-tab}
\end{table}

We have presented four possible semantics of our language of expressions:
dynamic by-name, dynamic by-value, static by-name, and static by-value.
We can deem that two of them are different if we can a find
a closed expression where one produces a value and the other another
value or no value at all. 

\begin{proposition}
The four presented semantics are different.
\end{proposition}
\Proof
The expression $e\equiv \s{let} \ x=\bot \ \s{in} \ 3$ 
distinguishes evaluation \bem{by-name} and \bem{by-value} in
both static and dynamic binding.
Indeed, the evaluation {\em by-name} returns a value and
the one {\em by-value} does not.
Next consider the following expressions and evaluations:
\[
\begin{array}{lll}
e_1 \equiv \s{let} \ x=3\ \s{in} \ e_2, 
&e_2 \equiv \s{let}\ y=x\ \s{in} \ e_3, 
&e_3 \equiv \s{let} \ x=5\ \s{in} \ y ~.
\end{array}
\]
\[
\begin{array}{llll}
e_1[\emptyset]\eval 5[\emptyset] & \mbox{(dynamic, by-name)} 
&e_1[\emptyset] \eval 3[\emptyset] &\mbox{(dynamic, by-value)} \\
e_1[\emptyset]\eval 3[\emptyset] &\mbox{(static, by-name)} 
&e_1[\emptyset] \eval 3[\emptyset] &\mbox{(static, by-value).}
\end{array}
\]
Thus it remains to distinguish dynamic and static binding with
a by-value evaluation. To do this, we rely on the 
\s{quote}, \s{unquote} operations and 
modify the expressions above as follows:
\[
\begin{array}{lll}
e_1 \equiv \s{let} \ x=3\ \s{in} \ e_2, 

&e_2\equiv \s{let}\ y=\s{quote}(x)\ \s{in} \ e_3, 

&e_3\equiv \s{let} \ x=5\ \s{in} \ \s{unquote}(y)~.
\end{array}
\]
Now we have:
$e_1[\emptyset]\eval 5[\emptyset]$ with dynamic binding, by-value and 
$e_1[\emptyset] \eval 3[\emptyset]$ with static binding, by-value. \qed\\

The examples in the previous proof show that the correct
implementation of $\lambda$-calculus relies on static
binding; henceforth dynamic binding will be ignored.

\section{Environments and closures}
We adapt to the call-by-name and call-by-value 
$\lambda$-calculus the notions of environment and closure 
we have discussed in the previous section \ref{dynamic-static-sec}.
To this end, we reuse the notion of environment 
modulo the replacement of the expressions
(denoted $e$) by the $\lambda$-terms (denoted $M$).
A {\em closure}, denoted with $c,c',\ldots$,
is now a pair composed of a $\lambda$-term and an environment that
we shall write as  $M[\eta]$.
A (closure) {\em value}, denoted with $v,v',\ldots$,
is a closure whose term is a $\lambda$-abstraction.
Table \ref{big-step-closure-tab} describes the evaluation rules for
closures according to a call-by-name and a call-by-value strategy.
The first two rules are shared by both strategies.
Notice that the $\beta$-rule is now decomposed in a rule where the
argument is bound as a closure to the formal parameter in the environment  
and a rule where the closure associated with the formal parameter is retrieved
from the environment.

\begin{table}[b]
\[
\begin{array}{cc}

\infer{}{v \eval v} \qquad

&\infer{\eta(x) \eval v}{x[\eta] \eval v} 
 \\ \\

\infer{
\begin{array}{c}
\mbox{By-name:}\\
M[\eta] \eval_n \lambda x.M_{1}[\eta']\\
M_1[\eta'[M'[\eta]/x]] \eval_n v
\end{array}}
{(MM')[\eta] \eval_n v}

\qquad
&\infer{
\begin{array}{c}
\mbox{By-value:}\\
M[\eta] \eval_v \lambda x.M_{1}[\eta']\quad
		M'[\eta] \eval_v v'\\	
M_{1}[\eta'[v'/x]] \eval_v v
\end{array}}
{(MM')[\eta] \eval_v v}

\end{array}
\]
\caption{Evaluation of closures: call-by-name and call-by-value}
\label{big-step-closure-tab}
\end{table}

In the presentation of the evaluation relations, at each reduction step, we have
to traverse the evaluation context in order to reach the redex
to be reduced. A more efficient approach consists in 
storing the traversed evaluation context in a {\em stack} and then
to push and pop elements on the stack as needed (cf. small-step reduction rules
for $\imp$ in chapter \ref{intro-sec}).
The form of the stack depends on the reduction strategy.
In call-by-name, the evaluation context is the composition
of {\em elementary contexts} of the shape $[~]N$, where $N$ is 
a $\lambda$-term.
Then a stack representation of the evaluation context is
just a list of closures (arguments with their environment):
\[
s = c_1 : \ldots : c_n \qquad \mbox{(stack for call-by-name).}
\]
The reduction relation presented in table \ref{machine-cbn-tab} 
now operates on pairs 
$(M[\eta],s)$ composed of a closure and a stack.
Initially $\lambda$-terms are supposed closed and 
the stack is supposed empty.

\begin{table}
\[
\begin{array}{lll}

(x[\eta],s) 		&\arrow& (\eta(x),s) \\

((MM')[\eta],s)		&\arrow& (M[\eta], M'[\eta]:s) \\

((\lambda x.M)[\eta], c:s)  	&\arrow& (M[\eta[c/x]],s)

\end{array}
\]
\caption{Abstract machine for call-by-name}\label{machine-cbn-tab}\index{abstract machine, call-by-name}
\end{table}

A similar approach works for call-by-value.
This time an evaluation context can be regarded
as the composition of {\em elementary contexts} of the 
shape: $[~]N$ or $V[~]$. We code these elementary contexts
as a list as follows:
\[
\begin{array}{cc}
  [~]c      \ \equiv \ r:c \quad \mbox{($r$  for right),} 
& v[~] \ \equiv \ l:v  \quad \mbox{($l$ for left).}
\end{array}
\]
Then the stack $s$ has the shape:
\[
s = m_1 : c_1 : \ldots m_n : c_n \qquad \mbox{where }m \in \set{l,r}
\qquad \mbox{(stack for call-by-value).}\]
The reduction rules are described in table \ref{machine-cbv-tab}.

\begin{table}
\[
\begin{array}{lll}

(x[\eta],s) 			&\arrow& (\eta(x),s) \\

((MM')[\eta], s)		&\arrow& (M[\eta], r:M'[\eta]:s) \\

(v, r:c:s)  			&\arrow& (c,l:v:s) \\

(v, l:(\lambda x.M)[\eta]:s)	&\arrow& (M[\eta[v/x]], s)

\end{array}
\]
\caption{Abstract machine for call-by-value} \label{machine-cbv-tab}
\index{abstract machine, call-by-value}
\end{table}

\begin{exercise}\label{cbv-machine-op-ex}
Suppose we add to the $\lambda$-calculus with call-by-value
a certain number of operators 
$\w{op}_1,\ldots,\w{op}_m$ with 
arity $n_1,\ldots,n_m$, $n_j\geq 0$.
(1) What are the new evaluation contexts?
(2) How is the abstract machine to be modified?
\end{exercise}

The rules in the abstract machines described in tables \ref{machine-cbn-tab} 
and  \ref{machine-cbv-tab} form the basis for an implementation.
As usual in the implementation of term rewriting rules, one
can avoid the costly duplication of terms by using pointers.
Specifically, in the rule for application  one just needs to duplicate
the pointer to the environment rather than the whole environment.
In a machine implementation, variables can be replaced by
de Brujin indexes \index{de Brujin indexes} 
which express the number of $\lambda$'s that 
one has to traverse in the syntax tree to go from the variable
to the binder. For instance, the $\lambda$-term $\lambda x.x(\lambda y.xy)$
is represented by $\lambda.0(\lambda.10)$. We can rely on this notation
for closures too. In this case, we regard the environment as a list and
let a de Brujin index express the number of $\lambda$'s and 
elements in the environment that one has to traverse to go from
the variable to the term bound to the variable. For instance, in
$(\lambda.20)[c;c']$ the variable $2$ refers to the closure $c'$ while
the variable $0$ refers to the $\lambda$.
Using this notation, {\em e.g.},
the last rule of Table \ref{machine-cbn-tab} can be written as:
\[
((\lambda.M)[\eta], c:s)  	\ \arrow \ (M[c:\eta],s)~.
\]
It is interesting to notice that during the (abstract) machine computation 
(de Brujin) indexes are {\em never} modified. This simple remark 
makes manifest that the number of closures
in an environment is bounded by the largest index (plus $1$)
of the initial $\lambda$-term to be reduced. In practice, the
inputs of a functional program have indexes of bounded size and
therefore, in this case, the selection of an element in an environment
can be done in constant time. More generally, assuming that lists have
bounded length and that duplicated environments are shared each
computation step described by the rules in Tables \ref{machine-cbn-tab} and 
\ref{machine-cbv-tab} can be implemented in costant time, thus justfying
the `abstract machine' terminology. Notice however, that this analysis
ignores the hidden cost of garbage collection.

\begin{exercise}\label{debrujin-abs-machine-ex}
Implement the abstract machines in Tables
\ref{machine-cbn-tab} and \ref{machine-cbv-tab}
using De Brujin notation.
\end{exercise}

\begin{exercise}\label{control-abort-ex}
Suppose we add to the call-by-name $\lambda$-calculus two
monadic operators: $\cl{C}$ for {\em control} and $\cl{A}$ for {\em abort}.
If $M$ is a term then $\cl{C}M$ and $\cl{A}M$ are $\lambda$-terms.
An evaluation context $E$ is always defined as:
$E::= [~] \Alt EM$,
and the reduction of the control and abort operators is 
governed by the following rules:
\[
\begin{array}{llllll}
E[\cl{C}M]  &\arrow &M(\lambda x.\cl{A}E[x])~, \qquad
E[\cl{A}M]  &\arrow &M~.
\end{array}
\]
Adapting the rules in Table \ref{machine-cbn-tab}, design an abstract machine
to execute the terms in this extended language (a similar exercise can
be carried on for call-by-value).
{\em Hint:} assume an operator \w{ret} which takes a {\em whole stack} and 
{\em retracts} it into a closure; then, {\em e.g.}, the rule for the control 
operator can be
formulated as:
$\ ((\cl{C}M)[\eta],s) \ \arrow  \ (M[\eta], \w{ret}(s))$.
\end{exercise}

\section{Summary and references}
We have focused on two popular weak reduction strategies:
call-by-name and call-by-value and considered alternative and
equivalent presentations via {\em evaluation contexts} and 
via (big-step) {\em evaluation relations}.
We have highlighted the distinction between static and dynamic
binding and shown that the implementation of the former relies
on the notions of {\em closure} and {\em environment}.
Static binding leads to a correct implementation of the 
call-by-name and call-by-value  $\lambda$-calculus.
Further we have shown that {\em evaluation contexts} can be implemented
as stacks and that this leads to abstract machines for call-by-name
and call-by-value. The notion of 
call-by-name and call-by-value evaluation strategy in the 
$\lambda$-calculus is studied in \cite{DBLP:journals/tcs/Plotkin75}
and an early notion of abstract machine is presented in \cite{Landin64}.
de Brujin notation for $\lambda$-terms is introduced in \cite{DeBrujin72}.
The implementation techniques studied for weak
reduction strategies can be extended to the (non-weak)
$\beta$-reduction presented in chapter \ref{lambda-sec} (see, {\em e.g.}, 
\cite{DBLP:journals/jacm/CurienHL96}).

\chapter{Contextual equivalence and simulation}\label{equiv-sec}
\markboth{{\it Equivalence}}{{\it Equivalence}}

We look for a notion of {\em pre-order} (and a derived {\em equivalence}) 
among program expressions (not necessarily full programs). 
It should be {\em natural} and {\em usable}. 
To this end, we introduce first a notion of {\em contextual pre-order} which is
{\em natural} and then we show that it can be characterized as a
certain {\em simulation} which is easier to reason about.
The notion of simulation is an example of {\em co-inductively defined
relation}. We take the opportunity to put on solid grounds 
some basic notions on fixed points of monotonic functions and (co-)inductive
definitions.

\section{Observation pre-order and equivalence}
We focus on a (deterministic) {\em call-by-name $\lambda$-calculus}
as presented in chapter \ref{env-closure-sec}. However, the approach applies 
to programming languages in general  (including non-deterministic ones).
We work with the evaluation relation for call-by-name 
in definition \ref{cbn-eval} and simply write $\eval$ 
rather than $\eval_n$ since no confusion with call-by-value can arise.
Also, in this chapter, all terms are $\lambda$-terms.

We write $M\eval$ and say that $M$ converges if $\xst{V}{M\eval V}$.
Note that for every closed term $M$ either there is a unique value 
$V$ to which $M$ evaluates or the evaluation diverges (the derivation
tree is infinite). The situation were the evaluation is stuck
cannot arise.

In order to define a pre-order (or an equivalence) among two terms
we have to {\em decide} (cf. chapter \ref{intro-sec})
in which {\em contexts} the terms can be placed and
which {\em observations} can be performed on the terms
once they are placed in the contexts.
Our {\em hypotheses} are as follows.
\begin{itemize}

\item All contexts $C$ such that $C[M]$ and $C[N]$ are {\em closed terms}.
We insist on closing contexts because reduction is defined on closed terms.

\item We observe the {\em termination} of the term placed in a closing context.
Observing natural numbers or booleans would not change the 
state of affairs.
\end{itemize}

\begin{definition}[contextual pre-order and equivalence]\label{context-pre-order-def}\index{contextual pre-order, call-by-name $\lambda$-calculus}
We define the contextual {\em pre-order} on terms as:
\[
M\leq_C N \mbox{ if for all closing }C \ (C[M]\eval \mbox{ implies } C[N]\eval \ )~.
\]
Contextual {\em equivalence} is derived by defining:
\[
M \approx_C N \mbox{ if }M\leq_C N \mbox{ and } N\leq_C M~.
\]
\end{definition}

Thus two terms are deemed `equivalent' if 
from the point of view of the admitted observation they are 
indistinguishable in any closing context.

\begin{exercise}\label{prop-cxt-preorder}
Prove the following properties:
\begin{enumerate}

\item $\leq_C$ is a pre-order (reflexive and transitive).

\item If $M\leq_C N$ then for all contexts $C$ (not necessarily closing)
$C[M]\leq_C C[N]$.

\item $\lambda x.\lambda y.x \not \leq_C \lambda x.\lambda y.y$.

\item If $\ul{n}, \ul{m}$ are Church numerals (cf. definition \ref{church-num-def}) with $n\neq m$ then $\ul{n} \not\leq_C \ul{m}$.

\item Find a pair of terms $M,N$ such that
$M \not=_\beta N$ and you expect $M\approx_C N$.
\end{enumerate}
\end{exercise}

To prove that $M\not \leq_C N$ it suffices to find a context
such  that  $C[M]\eval$ and $C[N]\not\eval$. On the  other hand,
to prove that $M\leq_C N$ we have to consider {\em all closing contexts}.
For instance, proving $(\lambda x.M)N \leq_C [N/x]M$ is not so easy!
This motivates the quest for a more practical proof method
based on the notion of simulation which will be discussed in section
\ref{sim-sec}.

\begin{exercise}\label{io-lambda-ex}
Let $\leq_{\w{IO}}$ be a relation on closed terms defined by:
\[
M\leq_{\w{IO}} N \mbox{ if }\qqs{P \mbox{ closed}}{MP\eval \mbox{ implies } NP\eval}~
\]
Show that $\leq_{\w{IO}}$ is a pre-order and that it is {\em not} preserved by contexts.
\end{exercise}

\section{Fixed points}\label{fixed-point-sec}
In this and the following section, 
we make a pause to state and prove some {\em general facts}
on partial orders, monotonic/continuous functions,
fixed points, and (co-)inductive definitions.
These facts are {\em used all the time} when manipulating
programming languages, formal languages, logics,$\ldots$
The reader would be well-advised to become {\em acquainted} with these concepts.
We start by recalling some standard definitions on partial orders
(notice that, unlike in chapter \ref{rewrite-sec}, 
partial orders are supposed to be reflexive).

\begin{definition}[partial order]\label{partial-order-def}\index{partial order}
A {\em partial order} $(L,\leq)$ is a set $L$ equipped with
a binary relation $\leq$ which is 
reflexive, anti-symmetric, and transitive.
\end{definition}

\begin{definition}[upper/lower bounds]\index{upper bound}\index{lower bound} \index{infimum} \index{supremum}
Suppose $(L,\leq)$ is a partial order and 
let $X\subseteq L$ be a subset (possibly empty). 
An element $y\in L$ is an {\em upper bound} for $X$ 
if $\qqs{x\in X}{x\leq y}$.  
An element $y\in L$ is the {\em supremum} (sup) of $X$ if it is the 
least upper bound.
The notions of {\em lower bound} and {\em infimum} (inf) 
are defined in a dual way.
\end{definition}

\begin{definition}[lattice]\label{lattice-def}\index{lattice} \index{lattice, complete}
A {\em lattice}  is a partial order $(L,\leq)$ such that
every pair of elements of $L$ has a sup and an inf.
 A {\em complete lattice} is a partial order $(L,\leq)$ such that
every subset of $L$ has a sup (the existence of the inf follows).
\end{definition}

\begin{exercise}\label{fixed-special-ex}
Show that:
(1) The {\em subsets of a set} with the inclusion relation
as partial order form a complete lattice.
(2) Every subset of a complete lattice has an {\em inf}.
(3) Every {\em finite lattice} is {\em complete}. 
\end{exercise}

Next we introduce the notion of monotonic, {\em i.e.}, order-preserving
function and consider the structure of its fixed points in a complete lattice.

\begin{definition}[monotonic function]\label{monotonice-fun-def}\index{monotonic function}
A {\em monotonic function} $f:L\arrow L$ on a partial order $L$ is a
function respecting the  order:
\[
\qqs{x,y}{(\ x\leq y \mbox{ implies } f(x)\leq f(y) \ )}~.
\]
We say that $x$ is a {\em fixed point} of $f$ if $f(x)=x$. 
\end{definition}

\begin{proposition}[Tarski]\label{Tarsky-prop}\index{fixed points monotonic functions}\index{Tarski}
Let $f:L\arrow L$ be a monotonic function on a complete lattice.
Then $f$ has  a {\em greatest} and a {\em least fixed point} expressed by:
\[
\w{sup}\set{x \mid x\leq f(x)} \qquad \mbox{ and }\qquad
\w{inf}\set{x \mid f(x)\leq x}~.
\]
\end{proposition}
\Proof
Set $z=\w{sup}\set{x \mid x\leq f(x)}$. If $f(y)=y$ then $y\leq z$.
Hence it remains to show that $z$ is a fixed point.
First, we show:
\[
z\leq \w{sup}\set{f(x) \mid x\leq f(x)} \leq f(z)~.
\]
Then by monotonicity:  $f(z)\leq f(f(z))$.
And by definition of $z$, we derive $f(z)\leq z$. \qed

\begin{exercise}\label{apl-tarsky-ex}
Let $(\Nat \union \set{\infty},\leq)$ be the  set of natural numbers
with an added maximum element $\infty$, $0<1<2<\ldots <\infty$. 
Show that every monotonic function $f$ on this order has 
a fixed point.
\end{exercise}

The following exercises consider two situations which often arise in practice.

\begin{exercise}[fixed points on finite lattices]\label{fix-finite-lat}
Let $(L,\leq)$ be a {\em finite} lattice and $f:L\arrow L$ be a 
monotonic function. Let $\bot$ ($\top$) be 
the {\em least}  ({\em greatest}) element of $L$. 
If  $x\in L$ then let $f^n(x)$ be the {\em $n$-time iteration} of $f$ on $x$,
where $f^0(x)=x$. 

\begin{enumerate}

\item Show that there is an $n\geq 0$ such that  the 
{\em least fixed point} of $f$ equals $f^n(\bot)$.

\item State and prove a dual property for the {\em greatest fixed point}.

\item Show that these properties fail to hold if one removes the hypothesis
that the  lattice is {\em finite}.

\end{enumerate}
\end{exercise}

\begin{exercise}[fixed points of continuous functions]\label{fix-cont-fun}
A subset $X$ of a partial order is {\em directed} if
\[
\qqs{x,y\in X}\xst{z\in}{(x\leq z) \mbox{ and }(y\leq z)}~.
\]
A function on a complete lattice is {\em continuous} if it preserves
the \w{sup} of {\em directed sets}:
\[
f(\w{sup}(X)) = \w{sup}(f(X))\qquad \mbox{(if }X\mbox{ directed).}
\] 
\begin{enumerate}
\item Show that a {\em continuous} function is {\em monotonic}.

\item Give an example of a function on a complete
lattice which is continuous but does {\em not} 
preserve the \w{sup} of a (non-directed) set.

\item Show that the {\em least fixed point} of a continuous function $f$ 
is expressed by:
\[
\w{sup}\set{f^n(\bot) \mid n\geq 0}~.
\]
\end{enumerate}
\end{exercise}

We can summarize the exercises 
\ref{fix-finite-lat} and \ref{fix-cont-fun} as follows.
If the the lattice is finite,  to compute the  least 
(greatest) fixed point it suffices to iterate
a finite number of times the monotonic function starting from the
least (greatest) element.
Otherwise, if the function is continuous (preserves directed sets),
then the {\em least fixed point}  is the \w{sup} 
of the  (countable) iteration of the function starting from the  least
element.
Similar remarks apply to the greatest fixed point modulo suitable
definitions of the notions of co-directed set and co-continuous function.

In general, it is possible to build the least or greatest fixed point
of a monotonic function on a complete lattice as an {\em iterative}
process provided one accepts a {\em transfinite number of iterations}.
To do this, we can rely on the notion of {\em ordinal} in set theory. \index{ordinals} Intuitively, ordinals are obtained by iterating the operations of 
successor and supremum; the former are
 called {\em successor ordinals}
and the latter {\em limit  ordinals}:
\[
0,1,2,\ldots,\omega,\omega+1,\omega+2,\ldots,\omega+\omega,\ldots
\qquad\mbox{(first few ordinals).}
\]
Formally, in {\em set theory}, the set $X$ is an ordinal if:
\begin{itemize}
\item $Z\in Y \in X$ implies $Z\in X$.

\item  All sequences such that $X_0\ni X_1 \ni X_2 \ni \cdots$ are finite.
\end{itemize}
Notice that if we read $X\in Y$ as $X<Y$ the first property corresponds
to {\em transitivity} and the second to {\em well-foundation}.
So an ordinal is a set which is transitive and well-founded with respect to
the $\in$-relation.
In set theory, the role of $0$ is played by the empty set, the successor of an ordinal set
$\kappa$ is the set $\set{\kappa}\union \kappa$, and the limit of a sequence
of ordinals is their union.
Now given a complete lattice $L$ and a monotonic function $f:L\arrow L$,
we can define the (transfinite) sequence:
\[
\begin{array}{llllll}
f_{0}        &=\bot~, 
&f_{\kappa+1}  &=f(f_\kappa)~, 
&f_{\kappa}   &=\JOIN_{\kappa'<\kappa} f_{\kappa'}\qquad\kappa \mbox{ limit ordinal.}
\end{array}
\]
This defines an increasing sequence which must reach the least fixed point 
when the cardinality of the ordinal $\kappa$ is greater than the
cardinality of the complete lattice $L$; for otherwise, we would have
a subset of $L$ whose cardinality is greater than $L$.
A dual argument shows that we can approximate the greatest fixed
point starting from the top element of the lattice.

\section{(Co-)Inductive definitions}\label{co-ind-def-sec}
We discuss  examples of inductive and co-inductive set definitions.
Behind these definitions there is a complete lattice and
a monotonic function, and the (co-)inductive set
which is defined is nothing but the least (the greatest) fixed point
of the monotonic function.

\begin{example}[an inductive definition]
Let $\Z$ be the set of integer numbers and  $\w{suc}$ and $+$ the
standard successor and addition operations, respectively.
We could define:

\begin{quote}

The least subset of $\Z$ which contains $\set{0,2}$ 
and is closed under the addition operation.

\end{quote}

\noindent
It is not so obvious that we are indeed defining a set.
One has to make sure that the `least set' 
does exist.  
To do this, we explicit a function $f:2^{\Z} \arrow 2^{\Z}$,
\[
f(X) =\set{0,2}\union \set{x+y\mid x,y\in X}~,
\]
such that ``$X$ contains the set $\set{0,2}$ and $X$ is closed under
the  addition operation'' iff $f(X)\subseteq X$. 
Then one remarks that $2^{\Z}$ is a complete lattice
and $f$ is monotonic (indeed continuous, see exercise \ref{fix-cont-fun}). 
Hence the least fixed point exists and is expressed by:
\[
\Inter\set{X \mid f(X)\subseteq X} = \Union_{n\geq 0} f^n(\emptyset)~.
\]
\end{example}

\begin{example}[another inductive definition]
Let $R$ be a  {\em binary relation} on a set $D$.
The  {\em reflexive and transitive closure} $R^*$ is the 
least relation that contains the identity relation, the relation
$R$ and such that if $(x,y),(y,z)\in R^*$ then $(x,z)\in R^*$. 
Let us show that we can regard $R^*$ as a least fixed point.
As complete lattice, we take the binary relations on the set
$D$ ordered by inclusion. As monotonic function $f$, we define:
\[
f(S)=\w{Id}_D \union  R \union S\comp S
\]
where $\w{Id}_D$ is the identity relation on $D$ and $S\comp S$ is
the (relational) composition of $S$ with itself.
\end{example}

\begin{example}[buggy inductive definition]
Monotonicity is a key property. 
As usual, let us write $(x\cngr y)\mod 2$ if the integers
$x,y$ have the property that $(x-y)$ is a multiple of $2$.
Suppose we `define' $X$ as the least set of
integers such that:
(1) $0\in X$ and 
(2) if $x\in X$ then $\qqs{y\in X}{( \ (x\cngr (y+1))\mod 2 \ ) }$.
Unfortunately such a set does {\em not} exist. We should have 
a set of integers $X$ such that  $0\in X$ and $(0\cngr 1)\mod 2$.
\end{example}

The notion of {\em co-inductive definition} is obtained by {\em dualization}.
Rather than looking for the {\em least} set such that$\ldots$, we now look 
for the {\em greatest} set such that$\ldots$

\begin{example}[co-inductive definition]
A typical example of co-inductive definition 
arises in the theory of finite automata. 
Let $M=(Q,\Sigma,q_0,F,\delta)$ be a 
{\em finite deterministic automaton} with 
$Q$ set of states, $\Sigma$ input alphabet, $q_0$ initial state,
$F$ set of accepting states, and 
$\delta: \Sigma \times Q\arrow Q$ transition function.
Consider the function $f:2^{(Q\times Q)} \arrow 2^{(Q\times Q)}$ defined by
 $(q,q') \in f(R)$  if
 \begin{enumerate}
 
 \item $q\in F$ iff $q'\in F$.
 
 \item $\qqs{a\in \Sigma}{(\delta(a,q),\delta(a,q'))\in R}$.
 
 \end{enumerate}

It is easy to check that the function $f$ is monotonic on the 
set of binary relations on $Q$ ordered by inclusion.
The least $R$ such that $R=f(R)$ is simply the empty relation
which is not very interesting. However, the greatest $R$ such
that $f(R)=R$ is the relation that corresponds to 
state equivalence in finite automata. Indeed, since $2^{Q\times Q}$ is
a finite lattice, the definition gives a way to compute the equivalence
on states (see exercise \ref{fix-finite-lat}). 
Start with the full relation $Q\times Q$ and iterate the function $f$
till you reach a fixed point.
\end{example}

Co-inductive definitions are quite useful in defining various
notions of diverging computation. We illustrate this point
in the following example.

\begin{example}[another co-inductive definition]\label{ex-coind}
Let $(S,\arrow)$ be a set of {\em states} and 
$\arrow \subseteq S\times S$ be a {\em transition relation}.
Define $D$ as the {\em greatest} subset of $S$
such that if $s\in D$ then:
$\xst{s'}{s\arrow s' \mbox{ and } s'\in D}$.
We take as complete lattice the parts of $S$
ordered by inclusion. The monotonic function $f$ associated with
the definition is for $X\subseteq S$:
\[
f(X) = \set{s \mid \xst{s'}{(s\arrow s' \mbox{ and }s'\in X)}}~.
\]
To see the definition at work, suppose $S=\set{1,2,3,4}$ with transitions:
$1\arrow 2,3,4$, $3\arrow 1$, $4\arrow 4$.
The greatest fixed point of $f$ is $\set{1,3,4}$ on the other hand
the least fixed point is just the empty set.
Intuitively, the greatest fixed point of $f$ is the collection of elements
starting from which there is an infinite reduction sequence.
\end{example}

To summarize a (co-)inductive definition is well-defined
if the associated function is {\em monotonic}. In this case, 
the defined set corresponds to a least
(greatest) {\em fixed point} of the associated function.

\begin{exercise}\label{coind-examples-modified-ex}
Modify example \ref{ex-coind} so as to define the collection of elements
which are {\em not} normalizing, {\em i.e.}, there is {\em no} 
reduction sequence  leading to an element in normal form 
(cf. definition \ref{normal-def}).
\end{exercise}

\section{Simulation}\label{sim-sec}
Simulation is a standard example of co-inductive definition
of a binary relation which can be used to compare programs' behaviors.

\begin{definition}[simulation]\label{sim-def-lambda} \index{simulation, $\lambda$-calculus}
We say that a binary relation on {\em closed terms} 
$S$ is a {\em simulation}
if whenever $(M,N)\in S$ we have: (1) if $M\eval$ then $N\eval$ and 
(2) for all $P$ closed $(MP,NP)\in S$.
We shall also use the infix notation $M\ S \ N$ for $(M,N)\in S$.
We define $\leq_S$ as the  largest simulation.
\end{definition}

\begin{exercise}\label{sim-gfp}
Show that $\leq_S$ is the largest fixed point of the following function
on binary relations:
\[
\begin{array}{lll }
f(S)  &= \{(M,N) \mid  &M\eval \mbox{ implies } N \eval~, \quad  \qqs{P \mbox{ closed}}{(MP,NP)\in S} \}~.
\end{array}
\]
\end{exercise}

\begin{definition}
We extend $\leq_S$ to {\em open terms} by defining:
\[
M\leq_S N \mbox{ if for all closing substitutions } \sigma \ (\sigma M \leq_S \sigma N)~.
\]
We also write $M=_S N$ if $M\leq_S N$ and $N\leq_S M$.
\end{definition}

To prove that $M\leq_S N$ ($M,N$ closed) 
it suffices to find a relation $S$ which is a
simulation and such that $M \ S \ N$. 
The following proof contains several examples of this technique.

\begin{proposition}
The following properties of simulation hold:
\begin{enumerate}

\item $\leq_S$ is a pre-order (on open terms).

\item If $M\leq_S N$ then for any substitution $\sigma$ (not necessarily closed)
$\sigma M \leq_S \sigma N$.

\item If $M\eval V$ and $N\eval V$, $M,N$ closed, then  $M=_S N$.

\item $(\lambda x.M)N =_S [N/x]M$ ($M,N$ can be open).

\item If $M\leq_C N$ then $M\leq_S N$ ($M,N$ can be open).
\end{enumerate}
\end{proposition}

\Proof \Proofitemf{(1)}
On closed terms $\leq_S$ is reflexive and transitive.
If $M$ is an open term then $M\leq_S M$ because
for every closing substitutions $\sigma M\leq_S \sigma M$.
Suppose $M\leq_S N$ and $N\leq_S P$. Given a closing substitution
$\sigma$ for $M,P$ we can always extend it to a closing substitution 
$\sigma'$ for $N$. Then we have:
\[
\sigma M \equiv \sigma' M \leq_S \sigma' N \leq_S \sigma' P  \equiv \sigma P~.
\]
\Proofitemf{(2)} Suppose $M\leq_S N$ and $\sigma$ is a substitution.
To prove $\sigma M\leq_S \sigma N$, we have to check that for all
closing substitution $\sigma'$, $\sigma'(\sigma M) \leq_S \sigma'(\sigma N)$.
And this holds because $(\sigma' \comp \sigma)$ is a closing substitution 
for $M,N$.

\Proofitemm{(3)} We check that the following relation on closed terms
is a simulation:
\[
S= \set{(MP_1\cdots P_n, NP_1\cdots P_n) \mid M\eval V, N\eval V, n\geq 0}~.
\]
If $MP_1\cdots P_n \eval$ then we must have $VP_1\cdots P_n\eval$ and
therefore $NP_1\cdots P_n\eval$.

\noindent Also if $(MP_1\cdots P_n, NP_1\cdots P_n)\in S$ then for all $P$ closed, 
$(MP_1\cdots P_n P,NP_1\cdots P_n P)\in S$. 

\Proofitemf{(4)} Both the following relation and its inverse are simulations:
\[
S= \set{((\lambda x.M)NP_1\cdots P_n,[N/x]MP_1\cdots P_n) \mid n\geq 0}~.
\]
Thus $(\lambda x.M)N=_S [N/x]M$ holds if $(\lambda x.M)N$ is closed.
If $(\lambda x.M)N$ is open and $\sigma$ is a closing substitution then
we observe:
\[
\sigma (\lambda x.M)N \equiv (\lambda x.\sigma M)\sigma N =_S
[\sigma N/x]\sigma M \equiv \sigma([N/x]M)~.
\]
\Proofitemf{(5)} First check that $\leq_C$ on closed terms is
a simulation. Thus $\leq_C\subseteq \leq_S$ on closed terms. 
For open terms, suppose $M\leq_C N$, 
let $\vc{x}$ be the list of variables free in $M,N$, and
let $\sigma$ be any closing substitution.
Then take the closed terms:
$M'\equiv (\lambda \vc{x}.M)\sigma(\vc{x})$ 
and $N'\equiv (\lambda \vc{x}.N)\sigma(\vc{x})$.
We have $M'\leq_C N'$ and therefore $M'\leq_S N'$.
Moreover $\sigma M =_S M'$ and $\sigma N=_S N'$.
Therefore: $\sigma M \leq_S \sigma N$. \qed

\begin{exercise}\label{simulation-exercises}
Prove that:
\begin{enumerate}

\item If $M\not\eval$ and $N\not\eval$ then $M=_S N$.

\item Let $\Omega_n \equiv \lambda x_1.\ldots.\lambda x_n.\Omega$.
Then $\Omega_n <_S \Omega_{n+1}$ (strictly) and, for all $M$,
$\Omega_0 \leq_S M$.

\item Let $K^{\infty} \equiv YK$. Then for all $M$, $M\leq_S K^{\infty}$.

\item $\lambda x,y.xy \not\leq_S\lambda x.x$ (thus $\eta$-conversion is
unsound).
\end{enumerate}
\end{exercise}

So it seems {\em easier} proving $M\leq_S N$ than proving $M\leq_C N$.
However, we still need to check that $\leq_S$ is {\em preserved
by contexts}. If this property holds, then it is easy to 
conclude that the largest simulation coincides with the contextual pre-order.

\begin{proposition}\label{cxt-sim-prop}
Let $x$ be a variable and $M,N,P$ be terms.
If $M\leq_S N$ then: (1) $MP\leq_S NP$ and
(2) $\lambda x.M\leq_S \lambda x.N$.
\end{proposition} \Proofitemf{(1)} Because $\leq_S$ is a simulation.

\Proofitemm{(2)} Suppose $M\leq_S N$ and let $\sigma$ be a closing substitution
for $\lambda x.M, \lambda x.N$. As usual, suppose $\sigma$ commutes
with $\lambda$ up to renaming. Now for all closed $P$, define
$\sigma'$ as the substitution that extends $\sigma$ so that
$\sigma'(x)=P$. Then, by hypothesis:
\[
[P/x]\sigma M = \sigma' M \leq_S \sigma' N = [P/x]\sigma N~.
\]
Then we have:
\[
(\lambda x.\sigma M)P =_S [P/x]\sigma M \leq_S [P/x]\sigma N = (\lambda x.\sigma N)P~.
\]
Clearly $(\lambda x.\sigma M)\eval$ implies $(\lambda x.\sigma N) \eval$.
With reference to the function $f$ defined in exercise \ref{sim-gfp},
we have shown $(\lambda x.\sigma M)\ f(\leq_S) \ (\lambda x.\sigma N)$,
and we know $f(\leq_S)=\leq_S$. \qed 

\begin{exercise}\label{io-lambda-revised-ex}
Let us revise the pre-order considered in exercise \ref{io-lambda-ex} by
defining a relation $\leq_{\w{IO}^*}$ on closed terms as:
\[
M\leq_{\w{IO}^*} N \mbox{ if for all $n\geq 0$, $P_1,\ldots,P_n$ closed, 
$MP_1\cdots P_n\eval$ \mbox{ implies }$NP_1\cdots P_n\eval$.}
\]
Prove that $\leq_{\w{IO}^*}$ coincides with $\leq_S$.
\end{exercise}

\section{Simulation is a congruence (*)}
Unfortunately, it is {\em not} so easy to prove 
that  $M\leq_S N$ implies $PM \leq_S PN$.
The proof plan is  to introduce an {\em auxiliary relation} $\leq_A$
on {\em open} terms, which includes $\leq_S$, is preserved by contexts, 
and, with some work, turns out to coincide with $\leq_S$.

\begin{definition}[auxiliary simulation relation]\label{aux-sim-rel}\index{Howe}
The auxiliary relation $M\leq_A N$ is defined {\em inductively} on $M$
by the following rules:
\[
\begin{array}{ccc}

\infer{x\leq_S N}{x\leq_A N} 

&\infer{M\leq_A M'\quad  \lambda x.M' \leq_S N}{\lambda x.M\leq_A N} 

&\infer{M_1\leq_A M'_1\quad M_2\leq_A M'_2\quad M'_1M'_2\leq_S N}{M_1M_2\leq_A N}~.

\end{array}
\]
\end{definition}

The definition of the auxiliary relation seems rather {\em mysterious}. 
To have a clue, let us look at its {\em properties}.

\begin{proposition}
The auxiliary relation $\leq_A$ enjoys the following properties:

\begin{enumerate}

\item $\leq_A$ is {\em reflexive}.

\item $\leq_A \comp \leq_S \ \subseteq \ \leq_A$.

\item $\leq_S \ \subseteq \ \leq_A$.

\end{enumerate}

\end{proposition}
\Proof
\Proofitemf{(1)} By  induction on the structure of the term.

\Proofitemm{(2)}
Suppose $M\leq_A N \leq_S P$ and proceed by induction
on the proof of $M\leq_A N$.

\Proofitemm{(3)} By the previous property (2), 
using the fact that $\leq_A$ is reflexive. \qed \\

The next proposition introduces the key properties of 
the auxiliary relation.

\begin{proposition}[key properties]\label{key-prop-aux}
Let $M,M',N,N'$ be terms. Then:
\begin{enumerate}
\item If $M\leq_A M'$ and  $N\leq_A N'$ then $[N/x]M\leq_A [N'/x]M'$.

\item If $M\leq_A M'$ and $N\leq_A N'$ then $MN \leq_A M'N'$.

\item If $M\eval V$ and $M\leq_A N$ then $V\leq_A N$.

\end{enumerate}
\end{proposition}
\Proof \Proofitemf{(1)} By induction on the proof of $M\leq_A M'$. For instance,
suppose:
\[
\infer{M\leq_A M''\quad \lambda y.M''\leq_S M'}
{\lambda y.M \leq_A M'}~.
\]
We have to prove:
$\lambda y.[N/x]M \leq_A [N'/x]M'$.
By inductive hypothesis, we know:
$[N/x]M \leq_A [N'/x]M''$.
Also, by substitutivity of $\leq_S$ we have:
\[
[N'/x](\lambda y.M'')\equiv \lambda y.[N'/x]M'' \leq_S [N'/x]M'~.
\]
Hence, by definition of $\leq_A$, we conclude.

\Proofitemm{(2)} Consider the terms $Mx$ and $M'x$, with $x$ fresh.
From $M\leq_A M'$ we can derive $Mx\leq_A M'x$.
Then by property (1) above, we know that:
\[
N\leq_A N' \mbox{ implies } [N/x](Mx)\equiv MN \leq_A [N'/x](M'x)\equiv M'N'~.
\]
\Proofitemf{(3)} We proceed by induction on $M\eval V$. We detail the main case. Suppose:
\[
\begin{array}{cc}

\infer{M_1 \eval \lambda x.M''_1\quad [M_2/x]M''_1 \eval V}
{M_1M_2 \eval V} 

&\infer{M_1\leq_A M'_1\quad M_2\leq_A M'_2\quad M'_1M'_2 \leq_S N}
{M_1M_2 \leq_A N}~.
\end{array}
\]
By induction hypothesis on $M_1\eval \lambda x.M''_1$ we derive:
\begin{equation}\label{s1-aux}
\lambda x.M''_1 \leq_A M'_1~.
\end{equation}
The proof of the property (\ref{s1-aux}) above must have the following shape:
\begin{equation}\label{s2-aux}
\infer{M''_1 \leq_A M'''_1\quad \lambda x.M'''_1 \leq_S M'_1}
{\lambda x.M''_1 \leq_A M'_1}~.
\end{equation}
By the substitutivity property (proposition \ref{key-prop-aux}.1), 
we derive:
\begin{equation}\label{s3-aux}
[M_2/x]M''_1 \leq_A [M'_2/x]M'''_1~.
\end{equation}
Also by proposition \ref{cxt-sim-prop}, we know that:
\begin{equation}\label{s4-aux}
[M'_2/x]M'''_1 =_S (\lambda x.M'''_1)M'_2 \leq_S M'_1 M'_2 (\leq_S N)~.
\end{equation}
So we have:
$[M_2/x]M''_1 \eval V$, $[M'_2/x]M''_1\leq_A N$,
and by inductive hypothesis, we conclude: $V \leq_A N$.~\qed\\

We can now prove the announced result: the largest simulation
coincides with the contextual pre-order.

\begin{proposition}
Let $M,N$ be terms. Then:

\begin{enumerate}
\item $\leq_A$ is a simulation (and therefore $\leq_A \subseteq \leq_S$).

\item $M \leq_A N$ implies $M\leq_C N$.
\end{enumerate}
\end{proposition}
\Proof
\Proofitemf{(1)}  If $M \leq_A N$ and $M\eval V$ then $V\leq_A N$ by 
proposition \ref{key-prop-aux}.3. Since $V$ has the shape
$\lambda x.M'$ we must have:
\[
\infer{M'\leq_A M''\quad \lambda x.M''\leq_S N}
{\lambda x.M'\leq_A N}~,
\]
and, by definition of simulation, $N\eval$.
Also by proposition \ref{key-prop-aux}.2, we know that $M\leq_A N$ implies
$MP\leq_A NP$. 

\Proofitemm{(2)} Suppose $M\leq_A N$ and $C$ one hole, closing context. Then 
$C[M]\leq_A C[N]$, and this implies $C[M]\leq_S C[N]$. So if 
$C[M]\eval$ then $C[N]\eval$ too. \qed

\begin{exercise}\label{sim-cbv-ex}
We define a notion of contextual pre-order $\leq_C$ for the call-by-value
$\lambda$-calculus simply by taking definition \ref{context-pre-order-def} and 
considering that the predicate $\eval$ corresponds to call-by-value evaluation.
We also say that a (call-by-value) simulation is a binary relation
$S$ on closed $\lambda$ terms such that whenever $(M,N)\in S$ we have:
(1) if $M\eval$ then $N\eval$ and (2) for all closed values $V$, $(MV,NV)\in S$.
Denote with $\leq_S$ the largest simulation. If $M,N$ are terms 
(possibly open) say that $M\leq_S N$ if for all closing substitutions $\sigma$ 
mapping variables to values  $\sigma M\leq_S \sigma N$. 
Adapt the theory developed in this chapter to prove that 
the pre-orders $\leq_C$ and $\leq_S$ coincide.
\end{exercise}

\section{Summary and references}
The contextual pre-order is a natural compositional way to 
compare terms and the simulation pre-order is an effective
method to reason on this relation. 
Simulation is a typical example of co-inductive definition
and corresponds to the greatest fixed point of a monotonic
function.   Dually, inductive definitions correspond to the
least fixed point of a monotonic function. Such fixed points
are guaranteed to exist for monotonic functions over
complete lattices and in many practical situations they
can be effectively computed or at least approximated.
The notion of simulation (and bisimulation, see chapter \ref{lts-sec}) 
was introduced in \cite{DBLP:conf/tcs/Park81} in the context of the
semantics of concurrent processes where it is extensively
used.  The proof that simulation is preserved
by contexts is based on \cite{DBLP:journals/iandc/Howe96}
and it can be extended to a number of other calculi,
including, {\em e.g.}, the  call-by-value $\lambda$-calculus 
(exercise \ref{sim-cbv-ex} and \cite{PittsAM:howmho}).

\chapter{Propositional types}\label{simple-type-sec}
\markboth{{\it Propositional types}}{{\it Propositional types}}

The reader is supposed to be familiar with the usage
of types in programming languages. 
Then, if we regard the $\lambda$-calculus as the kernel
of a programming language, it is natural to wonder what
kind of types could be associated with $\lambda$-terms.

\section{Simple types}
For the time being, we shall focus on a collection of
{\em propositional} types which include basic types such as 
integers, booleans,$\ldots$ and functional, product, and sum types.
We may also refer to these types as {\em simple} types as opposed
to more complex types including quantifications we shall discuss
in chapters \ref{type-inf-sec} and \ref{systemF-sec}.

\begin{definition}[types]\index{types, propositional}
We define the collection of (functional) {\em propositional types} as follows:
\[
	A:: = b \Alt \w{tid} \Alt (A \arrow A)~,
\]
where $b$ is a {\em basic type} (there can be more) and 
$\w{tid}::= t\Alt s \Alt \ldots$ are {\em type variables}.
\end{definition}

By default, we assume $\arrow$ associates to the right; so
$A\arrow B \arrow C$ stands for $A\arrow (B\arrow C)$.

\begin{definition}[contexts]\index{type context}
 A {\em type context} $\Gamma$ is a set of pairs
$\set{x_1:A_1,\ldots,x_n:A_n}$ where
all variables $x_1,\ldots,x_n$ are {\em 
distinct.} 
\end{definition}

We use $\Gamma,x:A$ as an
{\em abbreviation} for $\Gamma \union \set{x:A}$
where $x$ does not occur in $\Gamma$. Also we abbreviate type context
to context whenever no confusion may arise with {\em term} contexts.
Table \ref{simple-types-tab} presents a first system to assign
types to $\lambda$-terms. In this formulation, the variable
of a $\lambda$-abstraction is decorated with a type as in 
$\lambda x:A.M$. As in the usual programming practice,
the type $A$ specifies the type of the parameter of the function.
The presented system is composed of
a rule (\w{asmp}) to discharge an {\em assumption} from the context,
a rule $(\arrow_I)$ which {\em introduces} a functional type, 
and a rule $(\arrow_E)$ which {\em eliminates} a functional type.
This presentation style where the rules associated with the type
operators are split into introduction and elimination rules
comes from logic where it  is called {\em natural deduction}. % (Dag Prawitz).
The following exercises are a first illustration of the
connection between type systems and logic.

\begin{table}[b]
{\footnotesize
\[
\begin{array}{c}

(\w{asmp})
\quad \infer{x:A\in \Gamma}
{\Gamma \Gives x:A}  \\ \\

(\arrow_I)
\quad \infer{\Gamma,x:A \Gives M:B}
{\Gamma \Gives \lambda x:A.M:A \arrow B} 
\qquad
(\arrow_E)
\quad \infer{\Gamma \Gives M:A\arrow B\qquad
	\Gamma \Gives N:A}
	{\Gamma \Gives MN :B}

\end{array}
\]}
\caption{Assignment of propositional types to $\lambda$-terms}\label{simple-types-tab}
\index{type assignment, Church-style}
\end{table}

\begin{exercise}\label{taut-type-ex0}
Show that if $x_1:A_1,\ldots,x_n:A_n\Gives M:B$
is derivable then $(A_1\arrow \cdots (A_n \arrow B)\cdots)$ is
a {\em tautology} of propositional logic where we interpret $\arrow$ as
implication and atomic types as propositional variables.
Conclude that there are types $A$ which are {\em not inhabited},
{\em i.e.}, there is no (closed) $\lambda$-term $M$ such that $\emptyset \Gives M:A$.
\end{exercise}

\begin{exercise}\label{taut-type-ex}
Show that there is no $\lambda$-term $M$ such that:
$\emptyset \Gives M:(b\arrow b)\arrow b$.
Write $A\arrow b$ as $\neg A$.
Show that there are $\lambda$-terms $N_1$ and $N_2$ such that:
\[
\begin{array}{ll}

\emptyset \Gives N_1 : A \arrow (\neg \neg A) ~,
&\emptyset \Gives N_2:(\neg\neg\neg A) \arrow (\neg A)~.

\end{array}
\]
On the other hand, there are tautologies which are not inhabited!
For instance, consider:
$A \equiv ((t\arrow s) \arrow t)\arrow t$.
Show that there is no $\lambda$-term $M$ in normal form such that $\emptyset
\Gives M:A$ is derivable. This is enough because later we shall show that all
typable $\lambda$-terms normalize to a $\lambda$-term of the same type.
For another example, show that there is no $\lambda$-term $M$ in normal
  form such that $\emptyset \Gives M: \neg \neg t \arrow t$ is
  derivable (the intuitionistic/constructive negation is not
  involutive!).
\end{exercise}

Next we review a few alternative presentations of the type system.
In Table \ref{simple-types-tab}, $\lambda$-abstractions are decorated with
types. However, we can also consider a presentation where types
are assigned to  {\em pure}, {\em i.e.}, type-less, $\lambda$-terms.
Then one speaks of a presentation in  {\em Curry-style}, as opposed
to the previous one which is in {\em Church-style}.
In our case, the only difference between the two is that 
the rule $(\arrow_I)$ in Curry-style becomes:\index{type-assignment, Curry-style}
\[
\infer{\Gamma,x:A \Gives M:B}{\Gamma \Gives \lambda x.M:A \arrow B}~.
\]
An important consequence of this change is that now 
a $\lambda$-term may have more than one type.
This makes the {\em type inference problem} (see chapters \ref{hindley-inf-sec} and  \ref{type-inf-sec}) more interesting and so this problem
is often studied for systems in Curry-style.

\begin{table}
{\footnotesize
\[
\begin{array}{ll}
\w{id}  &::= x\Alt y \Alt \ldots \\
M       &::= \w{id}^A \Alt \lambda \w{id}^A.M \Alt MM
\end{array}
\]
\[
\infer{}{x^A:A}
\qquad
\infer{M:A  \arrow B \quad  N:A}{MN:B}
\qquad
\infer{M:B}{\lambda x^A.M:A\arrow B}
\]}
\caption{System with type labelled variables}\label{var-lab-tab}
\index{type assignment, with type labelled variables}
\end{table}

Yet another presentation of the type system is possible by 
{\em labeling every variable} with its type and by {\em dropping the context}.
This gives the system presented in Table \ref{var-lab-tab}.
Finally, we may decide  to {\em label every $\lambda$-term} 
(not just the variables) 
with its type and in this case we can {\em drop the type} since the type
of a $\lambda$-term is just the outermost label. The resulting system 
is given in Table \ref{term-lab-tab}.

\begin{table}
{\footnotesize
\[
\begin{array}{ll}
\w{id}  &::= x\Alt y \Alt \ldots \\
M       &::= \w{id}^A \Alt (\lambda \w{id}^A.M)^A \Alt (MM)^A
\end{array}
\]
\[
\infer{}{x^A}
\qquad
\infer{M^{A \arrow B} \quad N^{A}}
{(M^{A \arrow B} \ N^{A})^{B}}
\qquad
\infer{M^{B}}{(\lambda x^A.M^{B})^{(A\arrow B)}}
\]}
\caption{System with type-labelled $\lambda$-terms}\label{term-lab-tab}
\index{type assignment, with type-labelled $\lambda$-terms}
\end{table}

The previous exercises \ref{taut-type-ex0} and \ref{taut-type-ex} 
suggest that the functional type constructor can be regarded as a logical
implication. It turns out that one may push this connection further
by regarding the product (sum) type constructor as a logical
conjunction (disjunction). The resulting type system in natural deduction
style is presented in Table \ref{logical-tab}. Later in chapters
\ref{type-inf-sec} and \ref{systemF-sec}, we shall see that this connection
can be extended even further to types with universal and existential 
quantifications.

\begin{table}
{\footnotesize
\[
\begin{array}{c}

(\times_{I}) 
\quad
\infer{\Gamma \Gives M_1:A_1\quad \Gamma \Gives M_2:A_2}
{\Gamma \Gives \pair{M_1}{M_2}:A_1\times A_2} \\ \\

(\times_{E,1})
\quad \infer{\Gamma \Gives M: A_1\times A_2}
{\Gamma \Gives \pi_1(M):A_1} 
\qquad
(\times_{E,2})
\quad \infer{\Gamma \Gives M:A_1\times A_2}
{\Gamma \Gives \pi_2(M):A_2}  \\ \\  \\

(+_{I,1})
\quad \infer{\Gamma \Gives M:A_1}
{\Gamma \Gives \s{in_{1}}^{A_{1}+A_{2}}(M): A_1 + A_2}
\qquad 
(+_{I,2})
\quad \infer{\Gamma \Gives M:A_2}
{\Gamma \Gives \s{in_{2}}^{A_{1}+A_{2}}(M): A_1 + A_2} \\ \\

(+_E)\quad 
\infer{\Gamma \Gives M:(A_1+A_2)\quad \Gamma \Gives N_i:A_i\arrow B\quad i=1,2}
{\Gamma \Gives \s{case} (M, N_1, N_2):B}

\end{array}
\]}
\caption{Typing rules for product and sum}\label{logical-tab}
\index{type assignment, product and sum}
\end{table}

Not all term constructors found in a programming language have
a logical interpretation. For instance, Table \ref{non-logical-tab}
introduces typing rules for a constant zero \s{Z}, a successor
function \s{S}, and a fixed point combinator \s{Y}.
While the fixed point combinator allows to define general 
recursive functions (see chapter \ref{lambda-sec}) its 
logical interpretation is problematic.
Indeed, with the rule $(\s{Y})$, every  type $A$ is {\em inhabited} 
by the closed $\lambda$-term $Y(\lambda x:A.x)$. Thus the typing rule for $Y$ 
is definitely incompatible with logic as it leads to inconsistency!

\begin{table}
{\footnotesize
\[
\begin{array}{ccc}

(\s{Z})
\quad
\infer{}{\Gamma \Gives \s{Z}:\w{nat}} \quad

&(\s{S})\quad
\infer{\Gamma \Gives M:\w{nat}}{\Gamma \Gives \s{S}M:\w{nat}} \quad

&(\s{Y}) 
\quad
\infer{\Gamma \Gives M:(A \arrow A)}
{\Gamma \Gives \s{Y}M: A} %\\ \\

\end{array}
\]}
\caption{Non-logical extension of the type system} \label{non-logical-tab}
\index{type assignment, non-logical rules}
\end{table}

\section{Subject reduction}
If a $\lambda$-term is {\em well-typed}, then by inspection of the
rules we see, {\em e.g.}, that the $\lambda$-term 
cannot contain the application of a natural number to a function. 
However, to get {\em static guarantees} we
must make sure that {\em typing is invariant under reduction},
{\em i.e.},  if a $\lambda$-term is well-typed
and we reduce it then we still get a well-typed $\lambda$-term.
To establish this property we note the following
property.

\begin{proposition}[substitution]
If $\Gamma,x:A \Gives M:B$ and
$\Gamma \Gives N:A$ then 
$\Gamma \Gives \sub{x}{N}M:B$.
\end{proposition}
\Proof
By induction on the {\em height of the proof} of $\Gamma,x:A\Gives M:B$.
For instance, suppose the {\em root of the proof} has the shape:
\[
	\infer{\Gamma,x:A,y:B'\Gives M:B''}
	{\Gamma,x:A \Gives \lambda y.M:(B'\arrow B'')}~,
\]
with $x\neq y$. 
Then by inductive hypothesis,
$\Gamma,y:B'\Gives \sub{x}{N}M:B''$ and conclude
by $(\arrow_I)$.  \qed\\

We can now state the invariance of typing under reduction as follows.
Historically, this property is called {\em subject reduction}.

\begin{proposition}[subject reduction]
If $\Gamma \Gives M:A$ and $M\arrow_\beta N$ then
$\Gamma \Gives N:A$.
\end{proposition}
\Proof
Recall, that $M\arrow_\beta N$ means:
\[
M \equiv C[(\lambda x.M_1)M_2] \quad N \equiv C[[M_2/x]M_1]~.
\]
To prove subject reduction we proceed by induction on the structure of $C$.
The {\em basic case} follows directly from the substitution lemma.
For the {\em inductive case} consider in turn the cases where:
(1) $C= \lambda y.C'$, (2) $C=C'P$, and (3) $C=PC'$. \qed \\

In the `pure' $\lambda$-calculus, we identify the normal
forms with the results of the computation.
In applications, however, one can distinguish two kinds
of normal/irreducible forms: those that correspond to 
a {\em value} and those that correspond to an {\em erroneous
configuration} like dividing by zero, or applying an integer 
to another integer.
Thus a program, {\em i.e.}, a closed $\lambda$-term, has {\em three possible outcomes}:
(1) it {\em returns a value}, (2) it reaches an {\em erroneous configuration}, 
and (3) it {\em diverges} (cf. chapter \ref{intro-sec}).
Besides being invariant by reduction, a desirable
property for a type system is that: 
{\em well-typed programs cannot go wrong}, 
or at least that they go wrong in some expected way
({\em e.g.}, division by zero).
This property is often called {\em progress}, because in its simple form
it requires that if a program is not a value then it can reduce (progress).
The following exercise elaborates on this  point.

\begin{exercise}[on progress]\label{progress-ex}
Suppose we reconsider the {\em non-logical extension} 
of the simply typed $\lambda$-calculus
with a {\em basic type} $\w{nat}$, {\em constants} 
$\s{Z}$, $\s{S}$, $\s{Y}$, and with the
following fixed-point rule:
\[
C[\s{Y}M] \arrow C[M(\s{Y}M)]~.
\]
Let a {\em program} be a closed typable $\lambda$-term of type $\w{nat}$
and let a {\em value} be a $\lambda$-term 
of the shape $(\s{S}\cdots (\s{S}\s{Z})\cdots)$.
Show that if $P$ is a program in {\em normal form} (cannot reduce)
then $P$ is a {\em value}.
\end{exercise}

\section{A normalizing strategy for the simply typed $\lambda$-calculus}
We describe a {\em normalizing} strategy for the
simply typed $\lambda$-calculus. To this end 
we introduce some measures of the complexity of a type and a $\lambda$-term.

\begin{definition}[type degree]\index{degree, type}
The {\em degree of a type} is defined as follows:
\[
\delta(t)=1~, \qquad \delta(A\arrow B)= 1+\w{max}(\delta(A),\delta(B))~.
\]
\end{definition}

\begin{definition}[redex degree]\index{degree, redex}
Let  $R\equiv (\lambda x:A.M)N$ be a redex.
The {\em degree of the redex}, written $\delta_r(R)$
is the degree of the type associated with the $\lambda$-term $(\lambda x:A.M)$. 
\end{definition}

\begin{definition}[term degree]\index{degree, $\lambda$-term}
The degree of a $\lambda$-term, written $\delta_t(M)$,
is $0$ if $M$ is in normal form
and the maximum of the degrees of the redexes 
contained in $M$ otherwise.
\end{definition}

\begin{remark}
A redex $R$ is also a $\lambda$-term and we have $\delta_r(R)\leq \delta_t(R)$.
\end{remark}

\begin{proposition}[degree and substitution]
If $x$ is of type $A$ then
\[
\delta_t([N/x]M) \leq \w{max}(\delta(A),\delta_t(M),\delta_t(N))~.
\]
\end{proposition}
\Proof
The redexes in $[N/x]M$ fall in the following categories.
(1) The redexes already in $M$.
(2) The redexes already in $N$.
(3) New redexes arising by the substitution if $N\equiv \lambda y.N'$ and 
$M=C[xM']$. These redexes have degree $\delta(A)$. \qed

\begin{proposition}[degrees and reduction]\label{deg-red-prop}
If $M\arrow N$ then $\delta_t(N)\leq \delta_t(M)$.
\end{proposition}
\Proof We apply the previous analysis. \qed\\

We now define a reduction strategy that reduces first
an innermost redex of maximal degree.

\begin{definition}[innermost maximal degree strategy]\label{strategy, innermost maximal degree}\index{reduction, maximal degree}
Let $M$ be a $\lambda$-term which is {\em not in normal form}.
The innermost maximal degree strategy selects a redex
$R$ of maximal degree ($\delta_r(R)=\delta_t(M)$) and such
that all redexes contained in $R$ have {\em lower degree}.
\end{definition}

\begin{proposition}
The innermost maximal degree strategy is normalizing.
\end{proposition}
 \Proof
Notice that by reducing an innermost maximal degree redex
we guarantee that the reduced $\lambda$-term contains 
{\em strictly less redexes of maximal degree}.
Then we prove normalization by taking as {\em measure}:
\[
\mu(M)=(n,m)~,
\]
with the {\em lexicographic order} (from left to right), 
where $n=\delta_t(M)$ and  $m$ is the number of redexes of maximal degree. 
If $m=0$ then $M$ is in normal form. 
If $m=1$ then the reduced term has lower degree (first component decreases).
If $m>1$ then the first component does not increase (proposition \ref{deg-red-prop}) and the second decreases. \qed

\section{Termination of the simply typed $\lambda$-calculus (*)}
A $\lambda$-term $M$ is called {\em strongly normalizable}
if all $\beta$-reductions starting from $M$ terminate 
(thus strong-normalization is just a synonymous for termination!).

\begin{definition}\index{strong normalization}
Let $\w{SN}$ be the set of {\em strongly normalizable $\lambda$-terms}.
\end{definition}

This set plays a role similar to the set $\w{WF}$ for RPO termination 
(section \ref{rpo-wf-sec}).
The notion of size of a $\lambda$-term follows definition \ref{size-lambda-term-def}.

\begin{definition}[maximal length]\label{depth-def}
If $M\in \w{SN}$ then  the maximal length of 
a derivation starting from $M$ is called
the {\em reduction depth} of $M$, and is denoted 
$\depth{M}$. 
\end{definition}

\begin{remark}
The maximal length is well-defined because the reduction tree of 
a $\lambda$-term is finitely branching (cf. proposition \ref{koenig-prop}).
\end{remark}

In order to show that all simply typed $\lambda$-terms are 
$\arrow_\beta$-strongly normalizable, 
the {\em key idea} is to interpret types as subsets
of the set $\w{SN}$ of strongly normalizing $\lambda$-terms.

\begin{definition}[type interpretation]
\label{type-int-def}
\index{strong normalization, propositional types}\index{propositional types, interpretation}
The interpretation of a propositional type $A$ 
is defined as follows:
\[
\begin{array}{lll}

\dl b \dr = \dl t \dr  &= &\w{SN} \\
\dl A \arrow B \dr &= &\set{M \mid \qqs{N\in \dl A \dr}{(MN\in \dl B \dr)}}~.

\end{array}
\]
\end{definition}

\begin{proposition}\label{prop-int-rc}
For any type $A$, the following properties hold:
\begin{enumerate}

\item $\dl A \dr \subseteq \w{SN}$.

\item If $N_i\in \w{SN}$ for $i=1,\ldots,k$ 
then $xN_1\cdots N_k\in \dl A \dr$.

\item If $[N/x]MM_1\cdots M_k \in \dl A\dr$ and $N\in \w{SN}$ then 
$(\lambda x.M)NM_1\cdots M_k \in \dl A \dr$.

\end{enumerate}
\end{proposition}
\Proof By induction on $A$.

\Proofitemm{\mbox{Atomic types.}}
(1)  By definition.
(2)  The reductions of $xN_1\ldots N_k$ are just an interleaving of the
reductions of $N_1, \ldots N_k$.
(3)  We have:
\[
\w{depth}((\lambda x.M)NM_1\ldots M_k) \leq \w{depth}(N)+ \w{depth}([N/x]MM_1\ldots M_k)+1~.
\]
\Proofitemf{\mbox{Functional types $A\arrow B$}.}
Suppose $M\in \dl A \arrow B \dr$.

\Proofitemm{(1)} By inductive hypothesis, $x\in \dl A \dr$. 
Hence $Mx\in \dl B \dr \subseteq \w{SN}$, by inductive hypothesis. This entails $M\in \w{SN}$.

\Proofitemm{(2)} Take $M=xN_1\ldots N_k$ with $N_i \in \w{SN}$. Take $N_{k+1}\in \dl A \dr \subseteq \w{SN}$.
By inductive hypothesis, $xN_1\ldots N_k N_{k+1} \in \dl B \dr$.

\Proofitemm{(3)} 
If $[N/x]MN_1\ldots N_k\in \dl A \arrow B \dr$ then by the interpretation of
the functional types we have:
\[
\qqs{N_{k+1}\in \dl A \dr}{\  [N/x]MN_1\ldots N_k N_{k+1} \in \dl B \dr}~.
\]
Then by inductive hypothesis on $B$:
\[
\qqs{N_{k+1}\in \dl A \dr}{\ (\lambda x.M)NN_1\ldots N_k N_{k+1} \in \dl B \dr}~,
\]
which is equivalent to $(\lambda x.M)NN_1\ldots N_k  \in \dl A \arrow B \dr$. 
\qed

\begin{remark}
These interpretations of types are called {\em reducibility candidates}.
These are sets of of strongly normalizable $\lambda$-terms (property 1)
which contain at least the variables (and more) (property 2), and 
are closed under head expansions (property 3). 
\end{remark}

We can now state the soundness of the interpretation.

\begin{proposition}[soundness]\label{red-cand-sound-prop}
If $x_1:A_1,\ldots,x_k:A_k \Gives M:B$  (in Curry-style) and 
$N_i\in \dl A_i \dr$ for $i=1,\ldots,k$ then 
$[N_1/x_1,\ldots,N_k/x_k]M\in \dl B \dr$.
\end{proposition}
\Proof
By {\em induction on the typing proof}.

\Proofitemm{(\w{asmp})} Immediate by {\em definition}.

\Proofitemm{(\arrow_E)} By the {\em interpretation} of $\arrow$.

\Proofitemm{(\arrow_I)} Here is what goes on in a simplified case.
By inductive hypothesis on $x:A \Gives M:B$ we have:
\[
\qqs{N\in \dl A \dr}{[N/x]M\in \dl B \dr}.
\]
Then,  by the {\em closure under head expansions} 
of the interpretations we derive:
\[
\qqs{N\in \dl A \dr}{(\lambda x.M)N \in \dl B \dr}~,
\]
which is equivalent to $(\lambda x.M)\in \dl A \arrow B \dr$.
~\qed\\

The strong normalization property follows as a simple corollary.

\begin{corollary}[strong normalization]
If a $\lambda$-term is typable then it is strongly normalizing.
\end{corollary}
\Proof
Suppose  $x_1:A_1,\ldots,x_k:A_k \Gives M:B$.
We know $x_i\in \dl  A_i \dr$.
By proposition \ref{red-cand-sound-prop} (soundness), 
$M\in \dl B \dr$ and we know $\dl B \dr\subseteq \w{SN}$. \qed\\

A rational reconstruction of the proof could go as follows.
(1) We decide to interpret types as sets of strongly normalizing
$\lambda$-terms and show that $\Gives M:A$ implies $M\in \dl A\dr$.
(2) Then the definition \ref{type-int-def} of the type interpretation
is natural and the properties 1 and 2 of proposition \ref{prop-int-rc}
amount to check that indeed a type interpretation is composed of 
strongly normalizing $\lambda$-terms and it is not empty.
(3) Finally, the need for property 3 of proposition \ref{prop-int-rc} 
(closure under head expansion) 
appears in the proof of proposition \ref{red-cand-sound-prop}
(soundness, case $(\arrow_I)$).

\begin{exercise}[recursive types]\label{rec-type-ex}
Assume a {\em recursively defined type} $t$ satisfying the 
equation $t = t \arrow  b$ and suppose
we add a rule for typing up to type equality:
\[
\infer{\Gamma \Gives M:A \quad A = B}
{\Gamma \Gives M:B}~.
\]
Show that in this case the following $\lambda$-term (Curry's fixed point combinator) is typable ({\em e.g.}, in Curry-style):
\[
Y\equiv \lambda f.(\lambda  x.f(xx))(\lambda x.f(xx))~.
\]
Are the $\lambda$-terms typable in this system terminating?
\end{exercise}

\section{Summary and references}
A minimal property required for a type system is that
it is invariant under reduction. 
Sometimes, it is possible to connect type systems
to logic. This is the so called Curry-Howard correspondence
which goes as follows:\index{Curry-Howard correspondence}
\begin{center}
\begin{tabular}{|c|c|}
\hline
$\lambda$-calculus  &proof system \\\hline
type                &proposition \\
$\lambda$-term      &proof \\
reduction           &proof normalization \\\hline
 \end{tabular}
\end{center}
In a natural deduction presentation, an opportunity for a 
proof normalization arises when the introduction of 
an operator is followed by an elimination.
For instance, $\lambda$-abstraction is followed by an application,
a pairing is followed by a projection, and an injection is followed
by a case selection.
The book \cite{GLT89} is a good introduction to the connections between
proof theory and type theory including alternative presentations
of the logical systems.

\chapter{Type inference for propositional types}\label{hindley-inf-sec}
\markboth{{\it Type inference}}{{\it Type inference}}

Given a (pure) $\lambda$-term $M$ and a context $\Gamma$, the 
{\em type inference problem} is the 
problem of checking whether there is a type $A$ such that
$\Gamma \Gives M:A$. 
Given a (pure) $\lambda$-term $M$, a {\em variant of the problem}
is to look for a type $A$ {\em and} a context $\Gamma$ such
that $\Gamma \Gives M:A$. 
Connected to the type inference problem is the problem of actually
producing an {\em informative output}. Typically, if a $\lambda$-term $M$ is
typable, we are interested in a {\em synthetic representation of its types},
and if it is not, we look for an {\em informative error message}.

\section{Reduction of type-inference to unification}
We present a polynomial time reduction of the {\em type inference problem} 
for the propositional type system in Curry style 
(chapter \ref{simple-type-sec}) to 
the {\em syntactic unification problem}
(chapter \ref{unification-sec}). % (as noted by Hindley in 1969).   
The existence of a {\em most general unifier}
for the unification problem leads to
the existence of a {\em most general type} for the type inference problem.

\begin{definition}
A {\em goal} is a finite set $G$ of triples $(\Gamma, M, A)$
where $\Gamma$ is a {\em context}, $M$ a {\em $\lambda$-term}, and
$A$ a {\em propositional type}.
\end{definition}

We assume that all bound variables in $M$ are {\em distinct} and
different from the free ones,
that all free variables occur in the context $\Gamma$, and
that for every variable $x$ we have a type variable $t_x$.
We define a {\em reduction relation} on pairs $(G,E)$.
Assuming $G= \set{g}\union G'$ and
$g\equiv(\Gamma,M,A)\notin G'$, all the rules produce
a pair $(G'\union G_g, E\union E_g)$ where
$G_g$ and $E_g$ are defined  in Table \ref{inf-unif-tab}.

\begin{table}[b]
{\footnotesize
\[
\begin{array}{|c|c|l|}
\hline
g		&G_g			&E_g     \\
\hline

(\Gamma,x,A)	&\emptyset		&\set{t_x = A}  \\

(\Gamma,M_1M_2,A) 
		&\set{(\Gamma,M_1,t_1\arrow A), 
		(\Gamma,M_2,t_1)} 
					&\emptyset
		\qquad \qquad \qquad \quad (t_1 \mbox{ fresh})\\

(\Gamma,\lambda x.M_1,A) 
		&\set{(\Gamma,x:t_x,M_1,t)}
		&\set{A=t_x\arrow t}
		\qquad (t \mbox{ fresh}) \\\hline
\end{array}
\]}
\caption{Reduction of type inference to unification}\label{inf-unif-tab}
\index{type inference, propositional types}
\end{table}

\begin{proposition}
The reduction specified in Table \ref{inf-unif-tab}
terminates.
\end{proposition}
\Proof
It is enough to notice that every reduction step replaces
a triple $(\Gamma, M,A)$ by a finite number of triples
$(\Gamma',M',A')$ where $M'$ is structurally smaller
than $M$. \qed\\

We introduce some notation.
In the following, we consider substitutions $S$ that act
on the first-order terms built over the signature
$\Sigma=\set{b^0,\arrow^2}$.
We define:
\[
\begin{array}{ll}

S\models E		&\mbox{if } S \mbox{ unifies }E~, \\

S\models (\Gamma,M,A)&\mbox{if }S\Gamma \Gives M: S(A) \mbox{ is derivable,}~ \\
S\models G		&\mbox{if } \qqs{g\in G}{S\models g~,} \\

S\models (G,E)		&\mbox{if }S\models G \mand S\models E~.

\end{array}
\]
Given a $\lambda$-term $M_0$ with free variables $x_1,\ldots,x_n$, we set
the initial pair to $(G_0,\emptyset)$,  with $G_0=\set{(\Gamma_0,M_0,t_0)}$, 
$t_0$ fresh,  and $\Gamma_0=x_1:t_{x_{1}}, \ldots, x_n:t_{x_{n}}$. Next, we state 
the main properties of the reduction.

\begin{proposition}\label{sound-complete-prop}
If $(G_0,\emptyset)\trarrow (G,E)$ then:
\begin{enumerate}

\item
If $S\models (G,E)$ then $S\Gamma_0 \Gives M_0:St_0$.

\item
If $\Gamma\Gives M_0:A$ then
$\exists{S}{(S\models (G,E), S\Gamma_0\subseteq \Gamma, \mand A =St_0)}$.
\end{enumerate}
\end{proposition}
\Proof
For both properties we proceed by induction on the length of the reduction.

\Proofitemm{(1)} For instance, suppose (1) true  for:
$(G\union\set{(\Gamma, MN,A)},E)$.
The rule for application produces the pair $(G',E)$ 
with $G'= G\union\set{(\Gamma,M,t_1\arrow A),(\Gamma,N,t_1)}$.  
Suppose $S\models (G',E)$. This means
$S\models(G,E)$, $S\Gamma \Gives M:S(t_1\arrow A)$, and $S\Gamma
\Gives N: St_1$. 
By $(\arrow_E)$, we conclude $S\Gamma \Gives MN:
SA$. Thus $S\models (G\union\set{(\Gamma, MN,A)},E)$, and by
hypothesis $S\Gamma_0\Gives M_0:St_0$.

\Proofitemm{(2)} For instance, suppose:
$\Gamma \Gives M_0:A$,
$S\models (G\union\set{(\Gamma', \lambda x.M,A')},E)$,
$S\Gamma_0\subseteq \Gamma$, and 
$A =St_0$.
This implies: $S\Gamma' \Gives \lambda x.M:S(A')$, 
which entails: 
$S\Gamma',x:A_1 \Gives M:A_2$,
$SA'=A_1\arrow A_2$, for some $A_1, A_2$.
Suppose we reduce to the pair:
\[
(G\union\set{(\Gamma',x:t_x, M,t)},E\union\set{A'=t_x\arrow t})~.
\]
Then take $S'=S[A_1/t_x,A_2/t]$.  \qed

\begin{remark}
Property (1) entails the {\em soundness} of the method. 
Indeed, suppose from the initial goal we derive a set of equations $E$ 
and a substitution $S$ such that $S\models E$ (a {\em unifier}).
Then  we derive a {\em correct typing} $S \Gamma_0 \Gives M_0:St_0$.
On the other hand, property (2) entails the {\em completeness} of the method.
Suppose $\Gamma \Gives M_0:A$ is a {\em valid typing}.
Then we can reduce $(\Gamma_0,M_0,t_0)$ to 
$(\emptyset,E)$ and find a {\em unifier} $S$ for $E$ such that
$S\Gamma_0$ is contained in $\Gamma$ and $St_0 =A$.  
In particular, if we take the most general unifier $S$ of $E$ and
we apply it to $t_0$ we obtain the {\em most general type}: 
every other type is an instance of $St_0$.
\end{remark}

\begin{example}
The most general type of the $\lambda$-term
$\lambda f.\lambda x.f(f(x))$
is 
$(t\arrow t)\arrow (t\arrow t)$.
Note that strictly speaking the most general type is {\em not unique}.
For instance, 
$(s\arrow s)\arrow (s\arrow s)$
is also a most general type of the $\lambda$-term considered.
\end{example}

\begin{remark}[graphical presentation]
It is possible to give an equivalent `graphical' presentation
of the unification method.
(1)
Rename bound variables so that they are all distinct and different from
the free ones.
(2)
Draw the tree associated with the $\lambda$-term.
(3) Associate a distinct type variable with every internal node of the tree.
(4) Associate a type variable $t_x$ with a leaf node corresponding to the
variable $x$.
(5) For every abstraction node $(\lambda x.M^{t'})^t$ generate the
equation $t=t_x \arrow t'$.
(6) For every application node $(M^{t'} N^{t''})^t$ generate the 
equation $t'=t''\arrow t$.
\end{remark}

\begin{exercise}\label{most-gen-types-ex}
Compute, if they exist, {\em the most general types} of the following $\lambda$-terms:
\[
\begin{array}{lll}
\lambda x.\lambda y.\lambda z.xz(yz),
&\lambda x.\lambda y.x(y x),
&\lambda k. (k(\lambda x.\lambda h.h x))~.
\end{array}
\]

\end{exercise}

\section{Reduction of unification to type inference (*)} \label{unif-red-typeinf-sec}
We discuss a method to reduce any unification problem 
to a type-inference problem. We also show that the principal types are
exactly the types inhabited by a closed $\lambda$-term.
We suppose as usual that $K \equiv \lambda x.\lambda y.x$.

\begin{proposition}\label{step1-unif-typeinf}
The principal type of the (closed) $\lambda$-term $\w{E}$ below is 
$t \arrow t \arrow s \arrow s$.
\[
\w{E} \equiv \lambda x. \lambda y. \lambda w.K w (\lambda f.\lambda p.p(fx)(fy))~.
\]
\end{proposition}
\Proof The fact that $f$ is applied to both $x$ and $y$ forces the
equality of the types of $x$ and $y$. 
On the other hand, since the principal type of 
$K$ is $t\arrow s \arrow t$, the type of $w$ must be equal to the
type of the result. \qed

\begin{exercise} \label{step2-unif-typinf}
Let $M_1$ and $M_2$ be closed $\lambda$-terms with principal types
$A_1$ and $A_2$, respectively.
Prove that the principal type of the $\lambda$-term
$\lambda f.\w{E} (fM_1) M_2$ %, for $f\notin \fv{M_1M_2}$, 
is $(A_1 \arrow A_2) \arrow s \arrow s$.
\end{exercise}

\begin{exercise}\label{inhabit-to-principal-ex}
  It is easy to find a closed $\lambda$-term which has type 
  $t\arrow t \arrow t$. What about finding a closed
  $\lambda$-term whose {\em principal} type is $t\arrow t \arrow t$?
  Corollary \ref{step7-unif-typeinf} gives a general method; look for
  a more direct argument in the special case under consideration.
  \end{exercise}

\begin{proposition}\label{step3-unif-typeinf}
For every type  $A$ with (type) variables contained in 
$\set{t_1,\ldots,t_n}$ there is a closed $\lambda$-term
$M_A$ whose principal type is:
$t_1 \arrow \cdots \arrow t_n \arrow A \arrow s\arrow s$,
where $s\notin \set{t_1,\ldots,t_n}$.
\end{proposition}
\Proof  
By induction on the structure of $A$.
If $A=t_i$ we take:
\[
\lambda x_1\ldots x_n.\lambda y. \w{E} x_i y :
t_1 \arrow \cdots \arrow t_n \arrow t_i \arrow  s \arrow s~.
\]
If $A=A_1 \arrow A_2$, by inductive hypothesis we have:
\[
M_{A_{i}}: t_1 \arrow \cdots \arrow t_n \arrow A_i \arrow s \arrow s \quad i=1,2~.
\]
We define:
\[
M_{A_{1} \arrow A_{2}} \equiv \lambda x_1\ldots x_n.\lambda y.\lambda z.K z P~.
\]
This $\lambda$-term has the expected type provided 
we can force in the $\lambda$-term $P$
the type of $y$ to be $(A_1 \arrow A_2)$.
We observe that if we write:
\[
Q_i \equiv M_{A_{i}} x_1 \ldots x_n y_i~.
\]
we force the type of $y_i$ to be $A_i$.
Then we can define $P$ as follows:
\[
P\equiv \lambda y_1,y_2.\lambda p. p Q_1 Q_2 (\w{E} (y y_1) y_2)~,
\]
and as required the type of $y$ is $A_1 \arrow A_2$. \qed

\begin{exercise}\label{fromtypes-to-terms-ex}
(1) Apply the method to the types $t_1$, $t_2$ and $(t_1 \arrow t_2)$
relatively to the set of (type) variables $\set{t_1,t_2}$. 
(2) Write a program that builds the equivalent of the $\lambda$-term
$M_A$ in a language of the $\ml$-family and uses the type-inference
system to compute its principal type.
\end{exercise}

\begin{proposition}\label{step4-unif-typeinf}
Given two types $A$ and $B$ there is  a $\lambda$-term $U_{A,B}$ 
which is typable if and only if $A$ and $B$ 
are unifiable. 
\end{proposition}
\Proof Let $t_1,\ldots,t_n$ be the type variables in $A$ or $B$.
Applying proposition \ref{step3-unif-typeinf}, we derive the following terms and
principal types:
\[
\begin{array}{ll}
M_A: t_1 \arrow \cdots \arrow t_n \arrow A \arrow s \arrow s~,
&M_B: t_1  \arrow \cdots \arrow t_n \arrow B \arrow s \arrow s~.
\end{array}
\]
Then we build:
\[
\begin{array}{lll}
U_{A,B} \equiv\lambda x_1\ldots x_n. \lambda y_1.\lambda y_2. \lambda p.p P_A P_B (\w{E} y_1 y_2), 
&P_A \equiv M_A x_1 \ldots x_n y_1, 
&P_B \equiv M_B x_1 \ldots x_n y_2~.
\end{array}
\]~\qed

\begin{exercise}\label{from-unifiable-totypable-ex}
Apply proposition \ref{step4-unif-typeinf} 
if: (i) $A=t_1\arrow t_2\arrow t_2$ and $B=(t_2\arrow t_2)\arrow t_3$ and 
(ii) $A=t_1$ and $B=t_1\arrow t_2$.
\end{exercise}

\begin{corollary}\label{step5-unif-typeinf}
Every unification problem can be reduced to a type-inference problem.
\end{corollary}
\Proof
We know from exercise \ref{one-eq-one-binary-symbol-unif-ex} 
that every unification problem reduces to a unification problem 
composed of one equation
with terms built over a signature with exactly one binary symbol.
We take `$\arrow$' as binary symbol and using the proposition
\ref{step4-unif-typeinf} above we build two types $A$
and $B$ which are unifiable iff the $\lambda$-term $U_{A,B}$ is typable.  \qed

\begin{proposition}\label{step6-unif-typeinf}
For every type $A$, there is a $\lambda$-term 
$F_{(A\arrow A)}$ whose principal type is $(A\arrow A)$.
\end{proposition}
\Proof 
Let $A$ be a type whose type variables are contained in 
$\set{t_1,\ldots,t_n}$. 
Let $M_A$ be a $\lambda$-term with principal type: 
$t_1 \arrow \cdots \arrow t_n \arrow A \arrow s \arrow s$ 
(proposition \ref{step3-unif-typeinf}).  
Then build the $\lambda$-term:
\[
F_{A\arrow A} \equiv \lambda y.K y (\lambda x_1\ldots x_n.(M_A x_1 \ldots x_n y))~.
\]
This $\lambda$-term has principal type $(A\arrow A)$. \qed

\begin{corollary}\label{step7-unif-typeinf}
Let $M$ be a closed $\lambda$-term with type $A$ 
(not necessarily its principal type). 
Then one can build a closed $\lambda$-term $N$ whose principal type
is $A$. Thus the inhabited types are exactly the principal types.
\end{corollary}
\Proof 
By proposition \ref{step6-unif-typeinf},
the principal type of $F_{A\arrow A}$ is $(A\arrow A)$.
Then the $\lambda$-term $F_{A\arrow A} M$ has principal type $A$. \qed

\section{Summary and references}
A {\em type inference problem} can be (efficiently)
reduced to a {\em syntactic unification problem}.
Then the existence of a {\em most general unifier}
is reflected back in the existence of a {\em most general type}.
We have also shown that every unification
problem reduces to a type inference problem
and that for every inhabited type $A$ it is possible to build
a $\lambda$-term whose principal type is $A$.
Notice however that knowing if a type is inhabited is a {\sc Pspace}-complete
problem  \cite{DBLP:journals/tcs/Statman79}.
The connection between type inference and unification 
was already pointed out in \cite{Hindley69}.
By now, the reduction of a program analysis problem 
to the solution of a set of constraints has become
a standard technique. For instance, 
the data flow analyses performed by
optimizing compilers are reduced to 
systems of monotonic boolean equations.

\chapter{Predicative polymorphic types and type inference}\label{type-inf-sec}
\markboth{{\it Predicative polymorphic types}}{{\it Predicative polymorphic types}}

Consider any standard sorting algorithm \s{sort} on lists.
Most likely, the sorting algorithm just depends on a boolean
predicate on the elements of the list while the type of the elements
of the list does not really matter. 
One says that the sorting algorithm is {\em polymorphic} in that 
it can be applied to data of different, but related, shape.
We could type \s{sort} as follows:
\[
\s{sort}: \qqs{t}{\s{list}(t) \arrow (t\arrow t\arrow \s{bool}) \arrow \s{list}(t)}~,
\]
with the following intuitive meaning: 
for any type $t$, given a list of elements of type $t$, and 
a binary predicate on $t$, the function \s{sort} returns a list of elements
of type $t$. 
Notice that we are assigning to \s{sort} a type which is not quite {\em simple},
{\em i.e.}, propositional,
as it contains a {\em universal quantification} over types. 
In this chapter,  we introduce a particular class of universally
quantified types which can be used to type
polymorphic functions. We then study the type inference problem for 
the type system extended with such types.

\section{Predicative universal types and polymorphism}
The reader is supposed to be familiar with propositional
and first-order logic. In the standard interpretation of propositional logic
predicates are boolean values while in first-order logic
they are regarded as relations over some universe.
In second order logic, we can quantify over predicates.
For instance, here are some formulae in propositional, first-order, and
second-order logic:
\[
\begin{array}{ll}

(P\impl P)\impl (P\impl P) &\mbox{(Propositional formula)} \\

\qqs{x}{(P(x) \impl P(\s{s}(x)))} \impl (P(\s{z}) \impl \qqs{x}P(x)) 
&\mbox{(First-order formula)} \\

\qqs{P}{(P\impl P) \impl (P\impl P)} &\mbox{(Second-order formula).}

\end{array}
\]
Following the types-as-formulae correspondence outlined in chapter 
\ref{simple-type-sec}, 
we may consider
a type system where quantification over type variables is allowed.
One fundamental question is whether a type with quantified
types should be regarded as an ordinary type, or if it should
be lifted to a superior status. In this chapter, we take
the second option. In the logical jargon, this 
corresponds to  a {\em predicative} approach to
second-order quantification.  We shall not dwell into foundational issues
and just assume a distinction between types without quantification and 
types with quantification which will be called henceforth {\em type
schema}. Thus:
\[
\begin{array}{lll}
A   &\equiv (t\arrow t) \arrow (t \arrow t)       &\mbox{is a {\em type},} \\
\sigma &\equiv \qqs{t}(t\arrow t) \arrow (t\arrow t) &\mbox{is a {\em type schema}.}
\end{array}
\]
The advantage of this approach is that we stay close to propositional types 
and that in this way we can generalize the type inference techniques
presented in chapter \ref{hindley-inf-sec}. The inconvenience is that we do 
not have the full power of second-order quantification. This power
will be explored in chapter \ref{systemF-sec}.
The syntax of types, type schemas, and type contexts
is specified as follows.
\[
\begin{array}{lll}

A:: = b \Alt \w{tid} \Alt (A \arrow A)  &\mbox{(types)} \\
\sigma::= A \Alt \forall \w{tid}.\sigma       &\mbox{(type schemas)} \\
\Gamma::= \w{id}:\sigma,\ldots,\w{id}:\sigma     &\mbox{(type contexts).}
\end{array}
\]
We stress that $\forall t.(t\arrow t)$ is {\em not} a type and 
$\forall t.t \arrow \forall t.t$ is {\em not} a type schema.
Table \ref{pred-type-tab} presents an extended  type system with type schemas.
Notice that types schemas can occur in type contexts and that
in the rules $(\arrow_I)$ and $(\arrow_E)$,
we handle {\em types} (not type schemas) 

\begin{table}
{\footnotesize
\[
(\w{asmp})
\quad\infer{x:\sigma \in \Gamma}
{\Gamma \Gives x:\sigma} 
\]
\[
\begin{array}{cc}
(\arrow_I)
\quad \infer{\Gamma,x:A \Gives M:B}
{\Gamma \Gives \lambda x.M:A \arrow B} 

&(\arrow_E)\quad \infer{\Gamma \Gives M:A \arrow B\qquad
	\Gamma \Gives N:A}
	{\Gamma \Gives MN :B}

\\\\

(\forall_I)

\quad \infer{\Gamma \Gives M:\sigma \quad t\notin \ftv{\Gamma}}
{\Gamma \Gives M: \forall t.\sigma}

&(\forall_E)

\quad \infer{\Gamma \Gives M:\forall t.\sigma}
{\Gamma \Gives M: [A/t]\sigma}

\end{array}
\]}
\caption{Predicative type system (Curry-style)}\label{pred-type-tab}
\index{predicative type system, Curry style}
\end{table}

Next, we explore the connection between 
{\em universal (predicative) types} and 
{\em polymorphism}.
Sometimes, the same code/function can be applied to different
data-types. For instance,  the functional that iterates twice a function
$D\equiv \lambda f.\lambda x.f(fx)$,
will work equally well on a function over {\em booleans} or over
{\em integers}.
In the context of {\em propositional types}, 
we have already seen that we can automatically infer for $D$ 
the  {\em most general type}:
\[
D: (t\arrow t) \arrow (t\arrow t)~.
\]
The reader may be under the impression that this type is good enough to
represent the fact that $D$ will work on any argument of type
$(A \arrow A)$. Almost but {\em not quite}$\ldots$
Suppose:
\[
\begin{array}{llll}
F_1 : &(\s{bool} \arrow \s{bool})~, \qquad
F_2 : &(\s{int} \arrow \s{int})~,
\end{array}
\]
and consider the $\lambda$-term $P\equiv \s{let} \ f = D \ \s{in}  \ \pair{fF_1}{f F_2}$,
where as usual
$\s{let} \ x=M \ \s{in} \ N  \equiv (\lambda x.N)M$
and $\pair{M}{N} \equiv \lambda z.zMN$.
The reader may check that the $\lambda$-term $P$ has no propositional type and that
the example can be rephrased in the pure $\lambda$-calculus without 
appealing to the basic types $\s{bool}$ and $\s{int}$.
A possible {\em way out} is to consider that $D$ has a {\em type schema}:
\[
\sigma \equiv \forall t.(t\arrow t) \arrow (t\arrow t)~,
\] 
and then to {\em specialize} it just before it is applied to $F_1$ and $F_2$. 
This is {\em almost} what we can do with the predicative
type system in Table \ref{pred-type-tab}. 
The {\em problem} that remains is that we cannot really type the 
$\lambda$-term
$\ (\lambda f.\pair{fF_1}{fF_2}) D \ $
as expected because $\sigma \arrow \cdots$ is {\em not} even a type schema
according to our definitions. 
One could allow {\em more complex types}$\ldots$, but there is 
a more {\em conservative solution} which consists in taking 
the \s{let}-definition as a {\em primitive} and 
giving the following typing rule for it:
\[
\begin{array}{ll}
(\s{let})

&\infer{\Gamma,x:\sigma \Gives N:A\quad \Gamma \Gives M:\sigma}
{\Gamma \Gives \lets{x}{M}{N}:A}~.

\end{array}
\]
This is a {\em first formalization} (others will follow) of a type system 
which captures the polymorphism available in the $\ml$ family of
programming languages.

\begin{example}
Consider $M\equiv \lambda y.\lets{x}{\lambda z.z}{y(xx)}$ which is
not typable in the propositional type system but has type 
$A\equiv ((t\arrow t)\arrow t')\arrow t'$
in the $\ml$ type system.
The main difference is that we assign to the variable $x$ the type schema
$\sigma \equiv \forall s.(s\arrow s)$.
Then taking $B\equiv (t\arrow t)\arrow t'$ we can derive:
\[
\begin{array}{c}
y:B,z:s \Gives z:s \\\hline
y:B \Gives \lambda z.z:(s\arrow s) \quad s\notin \ftv{B}\\\hline
y:B \Gives \lambda z.z: \sigma
\end{array}
\]
On the other hand, one can derive:
$y:B,x:\sigma \Gives (xx): (t\arrow t)$.
\end{example}

The rules $(\forall_I)$ and $(\forall_E)$ in Table \ref{pred-type-tab} are {\em not syntax-directed}. When do we apply them?
One possibility is to {\em ask the programmer} to specify when 
this must be done. 
This requires an {\em enriched syntax} for $\lambda$-terms which 
includes: (i) the possibility to abstract a $\lambda$-term $M$ with respect to a 
type variable $t$, a {\em type abstraction} $\lambda t.M$, and (ii)
the possibility to apply a $\lambda$-term $M$ to a type $A$, 
a {\em type application} $MA$.
The resulting system in {\em Church-style} is composed of the
rules $(\w{asmp})$, $(\arrow_I)$, $(\arrow_E)$, and $(\s{let})$
we have already presented modulo the fact that:
(1) the $\lambda$-abstraction
is decorated with a type (cf. Table \ref{simple-types-tab}),
and (2) the rules in Table \ref{church-pred-type-tab} replace
the homonymous rules in Table \ref{pred-type-tab}.

\begin{table}
{\footnotesize
\[
\begin{array}{cc}

(\forall_I)

\quad \infer{\Gamma \Gives M:\sigma \quad t\notin \ftv{\Gamma}}
{\Gamma \Gives \lambda t.M: \forall t.\sigma}

&(\forall_E)

\quad \infer{\Gamma \Gives M:\forall t.\sigma}
{\Gamma \Gives MA: [A/t]\sigma}

\end{array}
\]}
\caption{Rules $(\forall_I)$ and $(\forall_E)$ in Church-style}\label{church-pred-type-tab}
\index{predicative type system, Church-style}
\end{table}

Due to the simplicity of the $\ml$ system, it is 
actually {\em possible to foresee} the points where the rules
$(\forall_I)$ and $(\forall_E)$ need to be applied. 
This leads to a {\em Curry-style }
and {\em syntax directed} type system: the shape of the $\lambda$-term
determines the rule to apply.

\begin{definition}[generalisation]
Given a pair composed of a context $\Gamma$ and a type $A$,
its {\em generalization} $G(\Gamma,A)$ is defined as 
the type schema that results by quantifying the type variables which occur
in $A$ but do not occur free in $\Gamma$. 
\end{definition}

\begin{example}
If $\Gamma = x: \forall t.(s\arrow t)$ and 
$A = s \arrow (t\arrow r)$ then $G(\Gamma,A) = \forall t.\forall r.A$.
\end{example}

The idea to define the syntax-directed type system presented in Table
\ref{pred-type-synt-dir-tab}
is to {\em generalize as much as possible} let-variables and then
{\em instantiate once} the type schema in the context.
We shall use the entailment symbol $\sGives$ when referring
to judgments in this system. 
Unlike in the type system in Table \ref{pred-type-tab}, the syntax-directed
type system in Table \ref{pred-type-synt-dir-tab} can only assign {\em types}
to $\lambda$-terms ({\em not} type schema).

\begin{table}
{\footnotesize
\[
\begin{array}{lclc}

(\w{asmp})
&\infer{x:\forall \vc{t}.A \in \Gamma}
{\Gamma \sGives x:[\vc{B}/\vc{t}]A}

&(\s{let}) 
&\infer{\Gamma,x:G(\Gamma,B) \sGives N:A \quad \Gamma \sGives M:B}
{\Gamma \sGives \lets{x}{M}{N}:A} \\ \\

(\arrow_I) &\infer{\Gamma,x:A \sGives M:B}
{\Gamma \sGives \lambda x.M: A \arrow B}

&(\arrow_E) &\infer{\Gamma \sGives M: A\arrow B\qquad
	\Gamma \sGives N:A}
	{\Gamma \sGives MN :B}

\end{array}
\]}
\caption{Predicative type system (Curry-style, syntax directed)}\label{pred-type-synt-dir-tab}
\end{table}

\begin{exercise}[running example, continued]\label{typeinf-runex1}
Consider again the $\lambda$-term:
\[
M\equiv \lambda y.\lets{x}{\lambda z.z}{y(xx)}~,
\]
and check that we can derive:
$\emptyset \sGives M: ((t\arrow t)\arrow t')\arrow t'$.
\end{exercise}

\section{A type inference algorithm}\label{type-inf-algo-sec}
Building on the syntax directed presentation of the type system,
we describe next a type inference algorithm.
In this section, we rely on the following {\em notation}:
\[
\begin{array}{ll}
M,N    &\mbox{type free $\lambda$-terms with \s{let}-definitions,} \\
\Gamma &\mbox{type context with {\em propositional} types,} \\
\Theta      &\mbox{partial function from identifiers to pairs $(\Gamma,A)$.}
\end{array}
\]
The (partial) function 
$\pt{M}{\Theta}$
tries to infer a principal typing judgment $\Gamma \Gives M: A$ for $M$.
The search is {\em driven by} $M$ while $\Theta$  
is a device that implicitly keeps track of the
type schema assigned to let-bound variables.
Because of this device, the algorithm just manipulates
simple, propositional types.

We assume: (i) all bound variables are renamed so as to be distinct
and different from the free variables, and (ii) in all subterms
$\lets{x}{N}{M}$ we have $x\in \fv{M}$.
Given two typing judgments $J_i \equiv \Gamma_i\Gives M_i:A_i$, $i=1,2$, we
denote by $\w{UnifyApl}(J_1,J_2)$ a triple $(S,t,J'_2)$  obtained  as follows:
\begin{enumerate}

\item obtain $J'_2\equiv \Gamma'_2\Gives M_2:A'_2$ by {\em renaming} the type variables of $J_2$ so 
that they are {\em disjoint} from those in $J_1$,

\item select a {\em fresh} type variable $t$,

\item build the {\em system of equations}:
\[
E = \set{A_1=A'_2\arrow t} \union \set{A= A' \mid x:A\in \Gamma_1, x:A'\in \Gamma'_2},
\]
\item compute (if it exists) a {\em most general unifier} $S$ of $E$.

\end{enumerate}
With this notation, the type inference algorithm is presented in 
Table \ref{type-inf-algo}.

\begin{table}
{\small
\[
\begin{array}{|l|}
\hline
\pt{M}{\Theta} = 
\s{case} \ M \\\hline\\

\begin{array}{lll}
x:   & \s{case} \ \Theta(x) \\
    &(\Gamma,A) &:\Gamma \Gives x:A \\
    &\_          &: x:t_x \Gives x:t_x 
\end{array}
         \\\\\hline\\
\begin{array}{ll}
\lambda x.M: &\s{let} \ (\Gamma \Gives M:A) = \pt{M}{\Theta} \ \s{in}  \\        
            &\begin{array}{ll}
               \s{case}   &x:A'\in \Gamma \\
               \s{true}     &: \Gamma\minus (x:A') \Gives \lambda x.M:A'\arrow A \\
              \_           &: \Gamma \Gives \lambda x.M:t\arrow A, t\mbox{ fresh} \end{array}
\end{array}
\\\\\hline  \\

\begin{array}{ll}
M_1M_2:&\s{let} \ J_i\equiv (\Gamma_i \Gives M_i:A_i)=\pt{M_i}{\Theta}  \ i=1,2 \ \s{in} \\
&\s{let} \  (S,t,\Gamma'_2\Gives M_2:A'_2)= \w{UnifyApl}(J_1,J_2)  \ \s{in} \\
&S(\Gamma_1\union \Gamma'_2 \Gives M_1M_2: t)
\end{array} \\\\\hline \\

\begin{array}{ll}
\lets{x}{M_1}{M_2}: \\
&\s{let} \ (\Gamma_1\Gives M_1:A_1)= \pt{M_1}{\Theta} \ \s{in} \\
&\s{let} \ \Theta' = \Theta[(\Gamma_1,A_1)/x]  \ \s{in} \\
&\s{let} \ (\Gamma_2 \Gives M_2:A_2) = \pt{M_2}{\Theta'} \ \s{in} \\
&\Gamma_2 \Gives \lets{x}{M_1}{M_2}:A_2  
\end{array}
\\\\\hline              
        
\end{array}
\]}
\caption{Type-inference algorithm}\label{type-inf-algo}
\index{type inference, predicative polymorphic types}
\end{table}

\begin{exercise}[running example, continued]\label{typeinf-runex2}
Consider again the $\lambda$-term:
$M\equiv \lambda y.\lets{x}{\lambda z.z}{y(xx)}$
and check that 
$\pt{M}{\emptyset} = \emptyset \Gives M:((t\arrow t)\arrow t')\arrow t'$.
\end{exercise}

\section{Reduction of stratified polymorphic typing to propositional typing (*)}
We may consider  a type system where to type $\lets{x}{M}{N}$ we actually
type $M$ and $[N/x]M$, where by typing we mean {\em propositional typing}.
This way of proceeding is {\em not particularly efficient}
because the \s{let}-expansion might take exponential time 
(see following exercise \ref{let-expansion-ex}).
However, the interesting point is that the $\lambda$-terms 
typable in this way are
exactly those typable in the original $\ml$ system. Therefore we have
the following {\em intuitive characterization}:
\begin{center}
{\em $\ml$ typing = Propositional typing + let-expansion.}
\end{center}
The type inference algorithm presented in Table \ref{type-inf-algo}
is a way to {\em keep
implicit the let-expansion} 
(but type renaming still forces a type expansion as we shall see shortly!).
Table \ref{let-exp-tab} describes a type system based on let-expansion.
We use the entailment symbol $\lGives$ to distinguish this system from the
previous ones. In the $(\s{let})$ rule,
we just check that the substituted term $M$ is typable with some type $B$;
this check is necessary if $x\notin \fv{M}$.
\begin{table}
{\footnotesize
\[
\begin{array}{lclc}

(\w{asmp}) &\infer{x:A \in \Gamma}
{\Gamma \lGives x:A}

&(\s{let}) &\infer{\Gamma \lGives [M/x]N:A\quad \Gamma \lGives M:B}
{\Gamma \lGives \lets{x}{M}{N}:A}

\\ \\ 

(\arrow_I) &\infer{\Gamma,x:A \lGives M:B}
{\Gamma \lGives \lambda x.M:A \arrow B} 

&(\arrow_E) &\infer{\Gamma \lGives M:A\arrow B\qquad
	\Gamma \lGives N:A}
	{\Gamma \lGives MN :B}

\end{array}
\]}
\caption{Propositional typing with let-expansion}\label{let-exp-tab}
\end{table}

\begin{exercise}[running example]\label{typeinf-runex3}
Consider again the $\lambda$-term:
\[
M\equiv \lambda y.\lets{x}{\lambda z.z}{y(xx)}~,
\]
and check that we can derive:
$\emptyset \lGives M: ((t\arrow t)\arrow t')\arrow t'$.
\end{exercise}

\begin{exercise}[let-expansion]\label{let-expansion-ex}
Consider a $\lambda$-calculus extended 
with \s{let} definitions of the shape $\lets{x}{M}{N}$.
Let $C$ denote a context with a hole (cf. definition \ref{context-def}) and
define the reduction relation  $\letarrow$ as follows:
\[
\letarrow = \set{(\ C[\lets{x}{N}{M}] \ , \ C[[N/x]M] \ ) \mid
C \mbox{ context}, M,N \mbox{ $\lambda$-terms }, x \mbox{ variable}}~.
\]
We extend the definition of size of a $\lambda$-term $|M|$ (cf. definition \ref{size-lambda-term-def}) with:
$|\lets{x}{M}{N}| =1+ |M| +|N|$.
We also define the {\em depth} $d(M)$ of a $\lambda$-term as follows:
\[
\begin{array}{llll}
d(x) = 1,
&d(MN) = \w{max}(d(M),d(N)), \\
d(\lambda x.M) = d(M),
&d(\lets{x}{M}{N}) =d(M) + d(N)~.
\end{array}
\]
\begin{itemize}
\item[1.]
Show that there is a strategy to reduce a $\lambda$-term $M$ to a normal form $N$ such that:
\[
|N| \leq |M|^{d(M)}~.
\]
\item[2.] Show that the reduction relation $\letarrow$ is locally confluent.

\end{itemize}
\end{exercise}

How hard is it to {\em decide} if a $\lambda$-term 
is typable in the $\ml$ system?
Well, in  {\em theory} it is hard but in {\em practice} it is easy!
The characterization via propositional typing with let expansion shows
that the problem can be solved in {\em exponential time}: 
(1) \s{let}-expand the $\lambda$-term (exponential penalty), 
(2) reduce the propositional type-inference problem to a unification problem
(efficient), and 
(3) solve the unification problem (efficient).

In fact one can show that {\em any decision problem} that runs in exponential time can be coded as an $\ml$ type inference problem.
Hence any algorithm (including the symbolic one) that solves the
problem will run in {\em at least exponential time}.
The good news are that the complexity is exponential in the 
{\em let-depth}  (example next) of the $\lambda$-term and 
that deeply nested chains of let-definitions do not seem
to appear in practice.

\begin{example}
Here is a way to blow up $\ml$ type inference.
\[
\begin{array}{lll}
P &\equiv  \lambda x,y,z.zxy &: 
t_1 \arrow t_2 \arrow (t_1 \arrow t_2 \arrow t_3)\arrow t_3 \\

M_1 &\equiv \lambda y. Pyy &:  
t_1 \arrow (t_1 \arrow t_1 \arrow t_2)\arrow t_2  \\

M_2 &\equiv  \lambda y.M_1(M_1 y) &: 
     t_1 \arrow (((t_1 \arrow t_1 \arrow t_2) \arrow t_2) \arrow \\ 
&&  ((t_1 \arrow t_1 \arrow t_2) \arrow t_2) \arrow  t_3) \arrow t_3  \\

M_3 &\equiv \lambda y.M_2(M_2 y) &: \cdots
\end{array}
\]
The number of distinct type variables and the 
size of the principal type (roughly) doubles at each step so that
inferring the principal type of $M_6$ is already problematic.
\end{example}

\section{Summary and references}
{\em Universally quantified} types are the types of {\em polymorphic}
$\lambda$-terms.  In particular we have considered a {\em predicative/stratified} 
form of universal quantification (as used in $\ml$).
It turns out that the type inference techniques developed in chapter
\ref{hindley-inf-sec} can be {\em extended} to predicative polymorphism. The
complexity of type inference is then {\em exponential} in the number
of nested let-definitions. Still the approach works well because these
complex definitions do not seem to arise in practice.
The design of a polymorphic type system for the $\ml$ language is 
due to \cite{Milner78,DamasMilner82}. The complexity
of the type inference problem is characterized in 
\cite{KTU90,Mairson90}. The book \cite{Mitchell96} 
contains a detailed analysis of the type inference algorithm 
described in Table \ref{type-inf-algo}.

\chapter{Impredicative polymorphic types}\label{systemF-sec}
\markboth{{\it Impredicative types}}{{\it Impredicative types}}

In chapter \ref{type-inf-sec}, we have introduced 
universally quantified types and observed that these types
can be regarded as the types of polymorphic functions.
In that context, a universally quantified type lives in a higher
universe of {\em type schemas}. In this chapter, we consider
an alternative approach where a universally quantified type
is still an ordinary type. Then one speaks of {\em impredicative}
types as opposed to the {\em predicative} types introduced
in chapter \ref{type-inf-sec}. 
In order to formalize impredicative types we introduce an
extension of the propositional type system presented in chapter \ref{simple-type-sec}
known as {\em system F}. 

A strong point of system F is its
expressive power. In particular, we show that the
addition of impredicative universal quantification suffices to
represent product, sum, and existential types.
The reader is supposed to be familiar with the usage of product and sum
types in programming. As for existential types, we shall see that they 
arise naturally when hiding the representation details of a data type.

We also provide an encoding of inductively defined data structures such as 
natural numbers, lists, and trees, and of the iterative functions
definable on them (iterative functions are related to the
primitive recursive functions introduced in chapter 
\ref{firstfun-sec}). 

While being quite expressive, system F can still be regarded
as a logical system. In particular, $\lambda$-terms typable
in system F are strongly normalizing. This is a difficult
result that relies on a generalization of the reducibility
candidates technique introduced in chapter \ref{simple-type-sec}.

\section{System F}\label{systemF-subsec}
System F is a {\em logical system} obtained from 
the propositional intuitionistic system (propositional types as
far as we are concerned) by introducing second order
quantification. At the type level, we can quantify
over type variables:
\[
A \equiv \qqs{t}{(t\arrow t)}~.
\]
At the term level, we can 
abstract with respect to a type and apply a $\lambda$-term to a type.
For instance, we can define a `polymorphic' identity
$\w{pid}\equiv \lambda t.\lambda x:t.x$ with the type $A$ above.
By applying $\w{pid}$ to the basic type $\w{nat}$, 
we obtain an identity  $\w{pid} \ \w{nat}$ of type
$\w{nat}\arrow\w{nat}$.
However, we may also apply $\w{pid}$ to the type $A$ itself to
obtain an identity $\w{pid}\ A$ of type $(A \arrow A)$.
In System F,  the type quantification in the type $A$ quantifies on 
{\em all types} including $A$ itself. 
One says that the type system is {\em impredicative}, as opposed
to the {\em predicative/stratified} 
system we have considered in chapter \ref{type-inf-sec}.

Table \ref{syntax-F-tab} defines the syntax of types and $\lambda$-terms
where we denote with $\fvt{\Gamma}$ the collection of type variables that
occur free in types occurring in the (type) context $\Gamma$. 

\begin{table}[b]
\[
\begin{array}{lll}

\w{tid}		&::= t \Alt s \Alt \ldots                                 &\mbox{(type variables)}\\
A		&::= \w{tid} \Alt A \arrow A \Alt \forall \w{tid}.A &\mbox{(types)}\\

\w{id} 		&::= x \Alt y \Alt \ldots                                 &\mbox{(variables)}\\
M		&::= \w{id} \Alt \lambda \w{id}:A.M \Alt MM \Alt \lambda \w{tid}.M \Alt MA &\mbox{($\lambda$-terms)} \\

\Gamma          &::=\w{id}:A,\ldots,\w{id}:A                         &\mbox{(contexts)}

\end{array}
\]
\caption{Syntax of system F: types and $\lambda$-terms (Church style)}\label{syntax-F-tab}
\end{table}

Table \ref{typing-sysF-tab} introduces the typing rules in Church-style
and the reduction rules of system F.
The novelties with respect to the system for propositional types 
(Table \ref{simple-types-tab})
are represented by the typing rules $(\forall_I)$ 
and $(\forall_E)$ and the $(\beta_t)$-rule for reducing 
the application of a type abstraction to a type. 
We stress that in this chapter the $(\beta)$ and $(\beta_t)$ rules, 
as well as the following 
$(\eta)$ and $(\eta_t)$ rules, can be applied in any context.

\begin{table}
{\footnotesize
\begin{center}{\sc Typing rules}\end{center}
\[
\begin{array}{c}
{\it (asmp)}

\quad \infer{x: A \in \Gamma}
	{\Gamma \Gives x: A} \\\\

(\arrow_I)	

\quad \infer{\Gamma,x:A \Gives M:B}
	{\Gamma \Gives \lambda x:A.M: A \arrow B}

\qquad(\arrow_E)	
\quad\infer{\Gamma \Gives M: A \arrow B ~~  \Gamma \Gives N:A}
	{\Gamma \Gives MN: B} \\\\

(\forall_I)
\quad\infer{\Gamma\Gives M: A ~~  t\notin \fvt{\Gamma}}
	{\Gamma \Gives \lambda t.M: \forall t.A} 

\qquad(\forall_E)
\quad\infer{\Gamma \Gives M: \forall t.A}
	{\Gamma \Gives MB: \sub{t}{B}A}

\end{array}
\]
\begin{center}
{\sc Reduction rules (in any context)}
\end{center}
\[
\begin{array}{lll}
(\lambda x:A.M)N  &\arrow [N/x]M           &(\beta)\\
(\lambda t.M)A    &\arrow [A/t]M      &(\beta_t)
\end{array}
\]}
\caption{Typing (Church-style) and reduction rules in system F}\label{typing-sysF-tab}
\end{table}

\begin{exercise}\label{inconsistent-systemF}
Show that without the side condition `$t\notin \fvt{\Gamma}$' in rule 
$(\forall_I)$, one can build a closed $\lambda$-term of type
$A$, for any type $A$. 
In other terms, without the side condition 
the system is logically inconsistent! 
\end{exercise}

As usual, we can add {\em extensional} rules.
The $(\eta)$ and $(\eta_t)$ reduction rules 
(applicable in any context) are the following:
\[
\begin{array}{llll}
(\lambda x:A.Mx) &\arrow M &\mbox{if } x\notin \fv{M}         &(\eta) \\
(\lambda t.Mt)   &\arrow M &\mbox{if } t\notin \fvt{M}       &(\eta_t)~.
\end{array}
\]
We leave it to the reader the check that in system F 
with the $\beta$ and $\beta_t$-rules (and possibly with 
the $\eta$ and $\eta_t$ rules): (1)
typing is preserved by reduction, and 
(2) reduction is locally confluent.
In section \ref{sn-sysF-sec}, we shall prove that    
typable $\lambda$-terms are {\em strongly normalizable};
thus confluence will follow from local confluence.

As a first example of the expressivity of second order
quantification, we consider the representation of 
product, sum, and existential types in system F.
The typing rules and the reduction rules are introduced
in Table \ref{prod-sum-ex-table}. 
The reader should be familiar with the rules 
for product and sum which have already been introduced
in Table \ref{logical-tab}. 
On the other hand, the rules for existential
types are new and deserve some comments.
A $\lambda$-term of existential type $\exists t.A$ is (up to conversion) 
a pair composed of  a type $B$ and a $\lambda$-term of 
type $[B/t]A$. Existential types can
be used to hide the details of the implementation of 
a data type and as such they can be regarded as `abstract data types'.
For instance, suppose we want to represent sets of numbers 
with operations to create the empty set, test membership, insert a
number in the set, and remove a number from the set.
Assuming, ${\bf 1}$ is the unit type, 
$\Nat$ is the type for natural numbers, and $\Bool$ the type
for booleans, we could specify the signature of a {\em set} data type as:
\begin{equation}\label{type-set-example}
A \equiv \exists t.((1\arrow t) \times (\Nat \arrow t \arrow \Bool) \times
(\Nat \arrow t \arrow t) \times (\Nat \arrow t \arrow t))~.
\end{equation}
We could then produce a concrete implementation of the data type
by instantiating the type $t$, say, with the type of the 
lists of natural numbers along with the implementations of the 
operations mentioned above.
We stress that the type $A$ above
just describes the {\em signature} of  a {\em set} data type 
but {\em not} its expected {\em behavior}.
For instance, there is no guarantee that inserting a number 
in a set and then removing it produces a set which equals the
original one.

\begin{table}
{\footnotesize
\begin{center}
{\sc Typing rules}
\end{center}
\[
\begin{array}{cc}

\infer{\Gamma \Gives M_i:A_i\quad i=1,2}
{\Gamma \Gives \pair{M_1}{M_2}:A_1\times A_2} 

&\left.\begin{array}{c}
\infer{\Gamma \Gives M: A_1\times A_2}
{\Gamma \Gives \pi_1(M):A_1} \\ 

\infer{\Gamma \Gives M: A_1\times A_2}
{\Gamma \Gives \pi_2(M):A_2}
\end{array}\right. \\ \\

\left.\begin{array}{c}
\infer{\Gamma \Gives M:A_1}
{\Gamma \Gives \s{in_{1}^{A_1+A_2}}(M):A_1 + A_2}\\
\infer{\Gamma \Gives M:A_2}
{\Gamma \Gives \s{in_{2}^{A_1+A_2}}(M):A_1 + A_2}
\end{array}\right.

&\infer{\Gamma \Gives M:A_1+A_2\quad \Gamma \Gives N_i:A_i\arrow C\quad i=1,2}
{\Gamma \Gives \s{case}(M,N_1,N_2):C} \\ \\

\infer{\Gamma \Gives M:[B/t]A}
{\Gamma \Gives \s{pack}^{\exists t.A}(B,M):\exists t.A}

&\infer{\Gamma \Gives M:\exists t.A\quad 
\Gamma\Gives N:\forall t.(A \arrow C)\quad t\notin \fvt{C}}
{\Gamma \Gives \s{unpack}(M,N):C} 

\end{array}
\]
\begin{center}
{\sc Reduction rules (in any context)}
\end{center}
\[
\begin{array}{llll}

\pi_i\pair{M_1}{M_2} &\arrow &M_i &i=1,2\\
\s{case}(\s{in_{i}^{A+B}} M)N_1 N_2 &\arrow &N_i M &i=1,2 \\

\s{unpack}(\s{pack}^{\exists t.A}(B,M),N) &\arrow &NBM

\end{array}
\]
}
\caption{Product, sum, and existential types} \label{prod-sum-ex-table}
\end{table}

\begin{table}
{\footnotesize
\begin{center}
{\sc Type encoding}
\end{center}
\[
\begin{array}{ll}
\ul{t} &=t \\
\ul{A\arrow B} &=\ul{A}\arrow \ul{B} \\
\ul{\forall t.A} &=\forall t.\ul{A} \\
\ul{A_1\times A_2} &=\forall s.(\ul{A_{1}}\arrow \ul{A_2}\arrow s) \arrow s\\

\ul{A_1+A_2} &=\forall s.(\ul{A_1}\arrow s)\arrow(\ul{A_2}\arrow s)\arrow s \\

\ul{\exists t.A} &=\forall s. (\forall t. (\ul{A}\arrow s))\arrow s

\end{array}
\]
\begin{center}
{\sc Term encoding}
\end{center}
\[
\begin{array}{lll}

\pair{}{} &=\lambda x_1:\ul{A_{1}},x_2:\ul{A_{2}}.\lambda s.\lambda p:\ul{A_1}\arrow (\ul{A_2} \arrow s).px_1x_2 \\

\pi_i     &=\lambda p:\ul{A_{1}\times A_{2}}.p\ul{A_i}(\lambda x_1:\ul{A_1},x_2:\ul{A_2}.x_i) &(i=1,2)\\

\s{in_{i}^{A_{1}+A_{2}}} &= \lambda x:\ul{A_i}.\lambda s.\lambda y_1:\ul{A_1}\arrow s.\lambda y_2:\ul{A_2}\arrow s.y_i x &(i=1,2)\\

\s{case} &=\lambda x:\ul{A_1+A_2}.\lambda s.\lambda y_1:\ul{A_1}\arrow s,y_2:\ul{A_2}\arrow s.xsy_1y_2 \\

\s{pack}^{\exists t.A}&= 
\lambda t.\lambda x: \ul{A}. \lambda s. \lambda y:\forall t.\ul{A}\arrow s. y \ t \  x \\

\s{unpack} &= \lambda x:\ul{\exists t.A}.
\lambda y: \forall t.\ul{A} \arrow C. x C y

\end{array}
\]}
\caption{Representation of product, sum, and existential types in system F}
\label{represent-prod-sum-ex}
\end{table}

Table \ref{represent-prod-sum-ex} describes an encoding of product,
sum, and existential types and $\lambda$-terms in system F.  This encoding
is quite good, as shown by the following proposition, and it can be used,
{\em e.g.}, to reduce the strong normalization of the extended system
to the strong normalization of system F.

\begin{proposition}
Suppose $\Gamma \Gives M:A$ in the system F extended with product,
sum, and existential types (table \ref{prod-sum-ex-table}).  
Then, the encoding described in Table \ref{represent-prod-sum-ex} 
preserves typing and reduction. Namely,
(1)  $\ul{\Gamma}\Gives \ul{M}:\ul{A}$ and (2)
if $M\arrow N$ then $\ul{M} \trarrow \ul{N}$.
\end{proposition}
\Proof \Proofitemf{(1)}  First check that the type encoding commutes
with substitution. Then proceed by induction on the proof of $\Gamma \Gives M:A$.

\Proofitemm{(2)} First check that the term encoding commutes with substitution.
Then proceed by case analysis on the redex. ~\qed

\begin{exercise}\label{unicity-typecheckF-ex}
Show that for every type context $\Gamma$ and $\lambda$-term $M$ there is
at most one type $A$ such that $\Gamma \Gives M:A$ is derivable 
according to the rules in Tables \ref{typing-sysF-tab} and 
\ref{prod-sum-ex-table},  and that in this case the derivation is unique.
What happens if we remove the type labels attached to the operators
$\s{in_{1}}$, $\s{in_{2}}$, and $\s{pack}$?
\end{exercise}

\section{Inductive types and iterative functions (*)}
{\em Iterative functions} are defined on the ground (with no variables), 
first-order  terms over a signature $\Sigma$.
The basic idea is to define a function 
{\em by induction on the structure of a ground term}, 
hence we have as many cases as function symbols in 
the signature $\Sigma$.
Let us consider the signature  of 
{\em tally natural numbers} $\Sigma=\set{\s{s}^{1}, \s{z}^{0}}$
and let $T=T_\Sigma(\emptyset)$ be the set of ground terms.
Given $g:T^n\arrow T$ and $h:T^{n+1}\arrow T$ the function 
$f:T^{n+1}\arrow T$ is defined by {\em iteration} by the following 
term rewriting rules:
\[
\begin{array}{cc}
f(\s{z},\vc{y})     \arrow  g(\vc{y})~,\qquad
&f(\s{s}(x),\vc{y})  \arrow  h(f(x,\vc{y}),\vc{y})~.
\end{array}
\]
At first sight this is less powerful than primitive recursive
definitions because the function $h$ {\em does not} depend directly on $x$.

However, one can first define {\em pairing} and {\em projections} and then
show that a function $f$ defined by {\em primitive recursion} such as:
\[
\begin{array}{cc}
f(\s{z},\vc{y})     \arrow  g(\vc{y})~,\qquad
&f(\s{s}(x),\vc{y})  \arrow  h(f(x,\vc{y}),x,\vc{y})~,
\end{array}
\]
can also be defined by {\em iteration} as follows:
\[
\begin{array}{ccc}
f'(\s{z},\vc{y})     \arrow  \pair{g(\vc{y})}{\s{Z}}~, 
&f'(\s{s}(x),\vc{y})  \arrow  h'(f'(x,\vc{y}),\vc{y})~, 
\end{array}
\]
where
$h'(x,\vc{y})     \arrow       \pair{h(\pi_{1}(x),\pi_{2}(x),\vc{y})}{\s{s}(\pi_{2}(x))}$.
One checks by induction on $x\in T$ that:
$f'(x,\vc{y}) = \pair{f(x,\vc{y})}{x}$, 
and from $f'$ one obtains $f$ by projection.

\begin{exercise}[predecessor, equality]\label{predecessor-ex}\index{$\lambda$-term, predecessor}
  \begin{enumerate}
\item Give a primitive recursive definition of the predecessor function $p$
where $p(0)=0$. Then transform the definition into an iterative
definition and derive a $\lambda$-term, typable in system F, to
compute the predecessor function on Church numerals
(cf. exercise \ref{pd-sub-exp-ex}). 

\item Derive $\lambda$-terms that compute the subtraction and check the inequality and equality of two Church numerals.

  \end{enumerate}
\end{exercise}

\begin{definition}[iterative functions]
Let $\Sigma$ be a signature with 
function symbols (constructors) $c_{i}$, 
where $\w{ar}(c_i)=n_i$, for $i=1,\ldots,k$.
Let $T=T_\Sigma(\emptyset)$ be the closed first-order terms over the signature.
The collection of {\em iterative functions}
is the smallest set such that:

\begin{itemize}

\item  The functions induced by the constructors and the 
projection functions are iterative functions.

\item  The set is closed under {\em composition}, namely
if $g:T^{n} \arrow T$ and $h_i:T^{m} \arrow T$, for $i=1,\ldots,n$,
are iterative functions then $g(h_1,\ldots,h_n)$ is an  iterative function.

\item The set is closed under {\em iteration}, namely if 
$h_i: T^{n_{i} + m} \arrow T$, for $i=1,\ldots,k$, 
are iterative functions then the function 
$f:T^{m+1} \arrow T$ such that:
\[
f(c_i(x_1,\ldots,x_{n_{i}}),\vc{y}) = 
h_i(f(x_1,\vc{y}),\ldots,f(x_{n_{i}},\vc{y}),\vc{y})\qquad (\mbox{for }i=1,\ldots,k)~,
\]
is an iterative function.
\end{itemize}
\end{definition}

Table \ref{encode-alg-F-tab} explains how to associate:
(1) with a {\em signature}  $\Sigma$ a {\em type} $\ul{\Sigma}$ of system F,
(2)  with a {\em constructor} of the signature $\Sigma$ a {\em closed $\lambda$-term} of system F of the appropriate type, and (3)
with a {\em ground term} $a$ over the signature $\Sigma$ a $\lambda$-term 
$\ul{a}$ of system F with type $\ul{\Sigma}$.

\begin{table}
{\footnotesize
\[
\begin{array}{ll}

\Sigma &= \set{c_i^{n_i}: \underbrace{T \times \cdots \times T}_{n_i ~{\it times}} \arrow T \mid i=1,\ldots,k} \\ 

\ul{\Sigma} &\equiv \forall t.A_1 \arrow \cdots \arrow A_k \arrow t, 
\quad \mbox{where: }   
A_i \equiv \underbrace{t \arrow \cdots \arrow t}_{n_i ~{\it times}} \arrow t ~,\\

\ul{c_{i}^{n_{i}}} &\equiv \lambda y_1:\ul{\Sigma} \ldots \lambda y_{n_{i}}:\ul{\Sigma}. \quad \lambda t.\lambda x_1:A_1\ldots \lambda x_k:A_k.  \\
&\quad x_i (y_1 t x_1 \cdots x_k)\cdots (y_{n_{i}} t x_1 \cdots x_k)  :
 \underbrace{\ul{\Sigma} \arrow \cdots \arrow \ul{\Sigma}}_{n_{i} ~{\it times}} \arrow \ul{\Sigma} \\  

\ul{a}
&\equiv \lambda t.\lambda x_1:A_1\ldots \lambda x_k:A_k.\sem{a}~, 
\mbox{with: }
\sem{c_i^{n_{i}}(a_1,\ldots,a_{n_{i}})} 
\equiv x_i \sem{a_{1}} \cdots \sem{a_{n_{i}}}~.

\end{array}
\]}
\caption{Encoding of signatures, constructors, and ground terms in system F}\label{encode-alg-F-tab}
\end{table}

\begin{example}[tally natural numbers]\label{tally-systemF}
If we apply the coding method to the signature 
$\Sigma = \set{\s{s}^1, \s{z}^0}$
of {\em tally natural numbers} we obtain the type:
\[
\ul{\Sigma} \equiv \forall t.(t\arrow t) \arrow (t \arrow t)~.
\] 
Then we represent the constructors in the signature with the 
$\lambda$-terms:
\[
\begin{array}{lll}

\ul{\s{s}} &\equiv \lambda y:\ul{\Sigma}.\lambda t. \lambda x_1:t\arrow t.\lambda x_2:t.x_1(y t x_1 x_2) &: \ul{\Sigma} \arrow \ul{\Sigma} \\

\ul{\s{z}} &\equiv \lambda t. \lambda x_1:t\arrow t.\lambda x_2:t.x_2
&: \ul{\Sigma}~. 

\end{array}
\]
The term  $n\equiv \s{s}^n \s{z}$, $n\geq 0$,  
is represented (up to conversion) by the $\lambda$-term:
\[
\ul{n} \equiv \lambda t.\lambda x_1:t\arrow t.\lambda x_2:t.x_{1}^{n}x_2 : \ul{\Sigma}~,
\]
which is a typed version of the {\em Church numeral} presented in 
section \ref{program-lambda-sec}.
We notice that:
\[
\begin{array}{llll}
\ul{s}\  \ul{n} 
&\arrow 
\lambda t. \lambda x_1:t\arrow t.\lambda x_2:t.x_1(\ul{n}\ t \  x_1\  x_2)
&\trarrow
\lambda t. \lambda x_1:t\arrow t.\lambda x_2:t.x_1(x_1^{n} \  x_2)
&\equiv \ul{n+1}~.
\end{array}
\]
\end{example}

\begin{exercise}\label{coding-signatures-ex}
Make explicit the coding of the following signatures:
(1) The signature with {\em no operation}.
(2) The signature with {\em two $0$-ary operations} (the `booleans').
(3) The signature of {\em binary words}.
(4) The signature of {\em binary trees}.
\end{exercise}

\begin{proposition}
There is a bijective correspondence between the {\em ground terms} over
a signature $\Sigma$ and the {\em closed $\lambda$-terms} of system F of the 
corresponding type $\ul{\Sigma}$ modulo $\beta\eta$-conversion.
\end{proposition}
\Proof
Let $M$ be a closed $\lambda$-term of system F in $\beta$-normal form of type $\ul{\Sigma}$,
where $\ul{\Sigma}$ is defined according to the rules in Table \ref{encode-alg-F-tab}.
The existence of the $\beta$-normal form will be proved in section \ref{sn-sysF-sec}. So the $\lambda$-term $M$ has to have the shape:
\[
	M \equiv \lambda t.\lambda x_1:A_1\ldots\lambda x_i:A_i.M'~, ~~i\leq k ~.
\]
If $i<k$ and $M'$ is not a $\lambda$-abstraction then $M'$ has the
shape $(\cdots(x_j M_1)\cdots M_h)$ and so we can $\eta$-expand
$M'$ without introducing a $\beta$-redex. 
By iterated $\eta$-expansions
we arrive at a $\lambda$-term in $\beta$ normal form of the shape:
\[
	\lambda t.\lambda x_1:A_1\ldots\lambda x_k:A_k.M'' ,
\]
where $M''$ has type $t$, it is in $\beta$ normal form, 
and may include free variables $x_1,\ldots,x_k$. 
We note that the types of the variables $x_i$ do
not contain second order quantifications.
We claim that $M''$ cannot contain a $\lambda$-abstraction:

\begin{itemize}

\item A $\lambda$-abstraction on the left of an application
would contradict the hypothesis that $M$ is in $\beta$ normal form.

\item A $\lambda$-abstraction on the right of an application
is incompatible with the `first order' types of
the variables $A_i$.
\end{itemize}
We have shown that a closed $\lambda$-term of type $\ul{\Sigma}$ is
determined up to $\beta\eta$ conversion by a $\lambda$-term 
$M''$ which is a well-typed combination of the variables
$x_i$, for $i=1,\ldots,k$. 
Since each variable corresponds to
a constructor of the signature we can conclude that there is
a unique ground term over the signature which corresponds to $M''$.  \qed

\begin{remark}
The rule $(\eta)$ is needed to have a bijection
between ground terms of the signature $\Sigma$ and closed
$\lambda$-terms of type $\ul{\Sigma}$. For instance, 
with reference to example \ref{tally-systemF} (tally natural numbers), 
there are two distinct $\lambda$-terms in $\beta$-normal form 
corresponding to the numeral $1$, namely $\ul{1}$ and 
$\lambda t.\lambda x_1:t\arrow t.x_1$.
\end{remark}

\begin{definition}
A function $f:T^n \arrow T$ over a signature $\Sigma$ is representable 
(with respect to the proposed coding)
if there is a closed $\lambda$-term $M:\ul{\Sigma}^{n} \arrow \ul{\Sigma}$, such that
for any vector of ground terms $\vc{a}$: 
\[
M\ul{\vc{a}} =_{\beta\eta} \ul{f(\vc{a})}~.
\]
\end{definition}

\begin{proposition}
All {\em iterative functions} over a signature $\Sigma$ are
representable.
\end{proposition}
\Proof
We proceed by induction on the definition of
iterative function. The interesting case
is iteration. Let $h_i: T^{n_i + m} \arrow T$ be iterative
functions for $i=1,\ldots,k$, and the function $f:T^{m+1} \arrow T$ be
defined by: 
\begin{equation} \label{eq-iter-def-DepPoly}
f(\vc{x},c_i(\vc{y})) = h_i(\vc{x},f(\vc{x},y_1),\ldots,f(\vc{x},y_{n_i}))~~i=1,\ldots,k\; ,
\end{equation}
where $\vc{x} \equiv x_1,\ldots,x_m$.\footnote{In this proof, it is convenient to write the additional parameters $\vc{x}$ before the main argument of the iteration.}
We represent $f$ with the function:
\[
	\ul{f} \equiv \lambda x_1:\ul{\Sigma}.\ldots \lambda x_m:\ul{\Sigma}.\lambda x:\ul{\Sigma}.
	x \ul{\Sigma} (\ul{h}_1 \vc{x}) \cdots (\ul{h}_k \vc{x}) \; ,
\]
where we know inductively that $\ul{h}_i$ represents $h_i$. 
Note that iteration is already built into the representation
of the data. We prove by induction on the structure of a ground term $a$ that for any
vector of ground terms $\vc{b}$, $\ul{f}\; \ul{\vc{b}}\; \ul{a} =_{\beta\eta} \ul{f(\vc{b},a)}$.

\Proofitemm{\bullet} If $a \equiv c_i^0$ then
\[
\ul{f} \;\ul{\vc{b}}\; \ul{c_i^0}
\arrow^* \ul{c_i^0} \ul{\Sigma} (\ul{h}_1 \ul{\vc{b}})\cdots (\ul{h_k} \ul{\vc{b}}) 
\arrow^* \ul{h_{i}} \ul{\vc{b}} 
=_{\beta\eta} \ul{h_{i}(\vc{b})}~,
\]
where the last step holds by induction hypothesis on $h_i$.

\Proofitemm{\bullet} If $a\equiv c_i^n(a_1,\ldots,a_n)$ then:
\[
\begin{array}{lll}
\ul{f(\vc{b}, c_i(a_1,\ldots,a_n))} &= 	\ul{h_i(\vc{b}, f(\vc{b}, a_1),\ldots,f(\vc{b}, a_n))}  \\
&=_{\beta\eta}
\ul{h_i}\;\ul{\vc{b}}\; \ul{f(\vc{b}, a_1)}\ldots \ul{f(\vc{b}, a_n)} 
&\mbox{(by induction hypothesis on $h_i$).}
\end{array}
\]
On the other hand, we compute:
\[
\begin{array}{ll}

\ul{f}\; \ul{\vc{b}}\; \ul{c_i^n(a_1,\ldots,a_n)} \\

\arrow \ul{c_i^n(a_1,\ldots,a_n)} 
\ul{\Sigma} (\ul{h_1}\;\ul{\vc{b}}) \cdots (\ul{h_k}\;\ul{\vc{b}}) \\

\arrow (\ul{h_i}\; \ul{\vc{b}})(\ul{a_1} \ul{\Sigma} (\ul{h_1}\;\ul{\vc{b}})\cdots (\ul{h_k}\; \ul{\vc{b}})) \cdots
(\ul{a_n} \ul{\Sigma} (\ul{h_1}\;\ul{\vc{b}})\cdots (\ul{h_k} \;\ul{\vc{b}}))  ~.
\end{array}
\]
Also, by induction hypothesis on $a$, we have for $i=1,\ldots, n$:
\[
\ul{f(\vc{b},a_i)} =_{\beta\eta} \ul{f}\ul{\vc{b}} \ul{a_i} \trarrow
\ul{a_i} \ul{\Sigma} (\ul{h_1}\ul{\vc{b}})\cdots (\ul{h_k} \ul{\vc{b}}))~.
\]
Hence, by combining the computations above, we obtain:
\[
\begin{array}{lll}
\ul{f}\; \ul{\vc{b}}\; \ul{c_i^n(a_1,\ldots,a_n)} 
&=_{\beta\eta}  \ul{h_i}\;\ul{\vc{b}}\; \ul{f(\vc{b}, a_1)}\ldots \ul{f(\vc{b}, a_n)}\\
&=_{\beta\eta} \ul{f(\vc{b}, c_i(a_1,\ldots,a_n))}~.
\end{array}
\]
\qed

\begin{example}
Suppose $T$ is the set of tally natural numbers and $g:T\arrow T$ and $h:T^2 \arrow T$. The iteration $it(h,g)$ of $h$ and $g$ must satisfy:
\[
\begin{array}{cc}

\w{it}(h,g)(\s{Z},y)    = g(y)~,

&\w{it}(h,g)(\s{S}(x),y) =h(\w{it}(h,g)(x,y),y)~.

\end{array}
\]
In the {\em pure} $\lambda$-calculus, we would define:
\[
\begin{array}{lll}
\w{it}  &\equiv &\lambda h. \lambda g.\lambda x. \lambda y. \ x \ (\lambda z.h \ z \ y) \ (g \ y)~.
\end{array}
\]
For $\ul{\Sigma} \equiv \forall t.(t\arrow t)\arrow (t\arrow t)$, the
term {\it it} can be type decorated as follows:
\[
\begin{array}{c}
\lambda h:\ul{\Sigma}\arrow (\ul{\Sigma} \arrow \ul{\Sigma}). \lambda g:\ul{\Sigma} \arrow \ul{\Sigma} .\lambda x:\ul{\Sigma}. 
\lambda y:\ul{\Sigma}. 
x \ \ul{\Sigma} \  (\lambda z:\ul{\Sigma} .h \ z \ y) \ (g \ y)~,
\end{array}
\]
which has type
$(\ul{\Sigma} \arrow (\ul{\Sigma} \arrow \ul{\Sigma})) \arrow (\ul{\Sigma} \arrow \ul{\Sigma}) \arrow (\ul{\Sigma} \arrow (\ul{\Sigma} \arrow \ul{\Sigma}))$.
Notice that this would {\em not work} with a propositional type  of the shape
$(B \arrow B)\arrow (B \arrow B)$!
\end{example}

\begin{example}
One can also handle the case of signatures which are
defined {\em parametrically} with respect to a collection
of data.  For instance $List(D)$ is the {\em signature
of lists} whose elements belong to the 
set $D$. This signature is equipped with
the constructors:
\[
\s{nil}: \w{List}(D), \qquad \s{cons}: D \times \w{List}(D) \arrow \w{List}(D)~.
\]
One can define {\em iterative functions} over $List(D)$ and
show that these functions can be represented
in system F for a suitable embedding
of the closed $\lambda$-terms in system F.
The sort $List(D)$ is coded by the type:
\[
\forall t. t \arrow (r \arrow t \arrow t) \arrow t~,
\]
where $r$ is a type variable, and generic elements
in $D$ are represented by (free) variables of type $r$.
\end{example}

\section{Strong normalization (*)}\label{sn-sysF-sec}
We now move towards a proof of the announced strong normalization
result.  The proof is based on a notion of {\em reducibility
  candidate} which is an {\em abstraction} of the notion already
considered for the strong normalization of the propositionally typed
$\lambda$-calculus (chapter \ref{simple-type-sec}) and recursive path
ordering (chapter \ref{termination-trs-sec}).
In order to make {\em notation lighter} we shall work with untyped
$\lambda$-terms obtained from the {\em erasure} of well-typed
$\lambda$-terms.

\begin{definition}[erasure]\label{erasure-def}\index{type erasure}
The (type) {\em erasure} function $er$ takes a typed $\lambda$-term and
returns an untyped $\lambda$-term.
It is defined by induction on the structure of the $\lambda$-term as follows:
\[
\begin{array}{c}
	
\er{x} = x,\qquad
\er{\lambda x:A.M} = \lambda x.\er{M},\qquad
\er{MN} = \er{M}\er{N}, \\
\er{\lambda t.M} = \er{M},\qquad
\er{MA}       = \er{M} ~.

\end{array}
\]
\end{definition}

In system F, we distinguish {\em two flavors of $\beta$-reduction}:
the one involving a redex $(\lambda x:A.M)N$ which we call
simply $(\beta)$ and the one involving a redex $(\lambda t.M)A$ which
we call $(\beta_t)$. Erasing type information we eliminate 
the reductions $(\beta_t)$.
However this does {\em not} affect the strong
normalization property as shown in the following.

\begin{proposition}[erasure vs. typed]\label{erasure-typed-prop}
Let $M$ be a well-typed $\lambda$-term in system F. Then:
\begin{enumerate}

\item If $M \arrow_{\beta} N$  then  $\er{M} \arrow_{\beta} \er{N}$. 

\item  If  $M \arrow_{\beta_t} N$  then $\er{M} \equiv \er{N}$. 

\item If $M$ may diverge then  $\er{M}$ may diverge.
\end{enumerate}
\end{proposition}
\Proof Properties (1) and (2) are left to the reader.
For (3), we observe that sequences of $\beta_t$-reductions always terminate as the size of the $\lambda$-term 
shrinks. Hence we can extract an infinite reduction of $\er{M}$ from an infinite reduction of $M$. \qed \\

We can now address the {\em key} issue. Suppose we 
want to adapt the {\em semantic} method already used in the propositional case.
What is the interpretation of $A \equiv \forall t.(t\arrow t)$?
We have to build first a {\em universe} $\cl{U}$ of type interpretations
where each type interpretation is a set of $\lambda$-terms.
Then we could require:
\[
\dl A \dr = \set{M \mid \qqs{X\in \cl{U}}{\qqs{N\in X}{(MN\in X)}}}~.
\]
Technically, the type interpretations are the so-called
reducibility candidates and are defined as follows. 
Let ${\it SN}$ be the collection of untyped  $\lambda$-terms
which are strongly normalizable with respect to the $(\beta)$ rule.
We shall use $P,Q,\ldots$ to denote the untyped $\lambda$-terms (as opposed
to the typed ones which are denoted with $M,N,\ldots$).

\begin{definition}[candidates]\label{red-can-def}\index{reducibility candidate}
A set $X$ of $\lambda$-terms  is a {\em reducibility candidate} if:
\begin{enumerate}
\item $X \subseteq {\it SN}$.

\item  $Q_i \in {\it SN}$, $i=1,\ldots,n$, $n\geq 0$ implies 
      $xQ_1,\ldots,Q_n \in X$.

\item $\sub{x}{Q}PQ_1,\ldots,Q_n \in X$ and $Q\in {\it SN}$ implies
      $(\lambda x.P)Q Q_1,\ldots,Q_n \in X$.

\end{enumerate}
We denote with ${\it RC}$ the collection of reducibility candidates.
\end{definition}

\begin{remark}
We have made into a {\em definition} the {\em properties} stated in
proposition \ref{prop-int-rc} of the interpretation of propositional types.
\end{remark}

\begin{proposition}[properties reducibility candidates]\label{prop-red-cand}
The following properties hold.
\begin{enumerate}

\item The set ${\it SN}$ is a reducibility candidate.

\item If $X \in {\it RC}$ then $X\neq \emptyset$.

\item The collection ${\it RC}$ is closed under arbitrary intersections.

\item If $X,Y\in {\it RC}$ then the following
set is a reducibility candidate:
\[
	X \arrow Y = \{ M \mid \qqs{N \in X}{(MN  \in Y)} \} ~.
\]
\end{enumerate}
\end{proposition}
\Proof
We abbreviate $Q_1,\ldots,Q_n$ with $\vc{Q}$. 
We recall (definition \ref{depth-def}) that if $P\in \w{SN}$ then
$\w{depth}(P)$ is the length of its longest reduction.

\Proofitemm{(1)}
As in the propositional case, we observe that $\sub{x}{Q}P\vc{Q} \in {\it
  SN}$ and $Q\in {\it SN}$ implies $(\lambda x.P)Q \vc{Q}\in {\it  SN}$.
This is an induction on ${\it depth}(P)+{\it depth}(Q)+{\it depth}(Q_1)+\cdots+{\it depth}(Q_n)$.

\Proofitemm{(2)} By definition, $x\in X$.

\Proofitemm{(3)}  Immediate.

\Proofitemm{(4)} Here we see the use of the `saturation'
condition (3) in definition \ref{red-can-def}. \qed \\

Next we define a type interpretation.

\begin{definition}[type interpretation]
Let $\w{Tvar}$ be the set of type variables.
Given a type environment $\eta: \w{Tvar} \arrow {\it RC}$ we interpret types
as follows:
\[
\begin{array}{lll}

\dl t \dr \eta 			&=& \eta(t) \\

\dl A \arrow B \dr \eta &=& \dl A\dr \eta \arrow \dl B \dr \eta \\

\dl \forall t.A \dr \eta	&=& \Inter_{X\in {\it RC}} \dl A \dr \eta[X/t] ~.
\end{array}
\]
\end{definition}

We remark that the interpretations of a functional type and a 
universal type are well-defined
because of propositions \ref{prop-red-cand}(4) and \ref{prop-red-cand}(3), 
respectively.
Strong normalization follows from the
{\em soundness of the interpretation} which is stated as follows.

\begin{proposition}[soundness]\label{prop-sound-sysF}
Let $\eta$ be a type environment  and
$x_1:A_1,\ldots,x_n:A_n \Gives M:B$
a derivable judgment. 
If $P_i \in \dl A_i \dr \eta$, for $i=1,\ldots , n$ then 
\[
[\lsub{x_1}{P_1},\ldots,\lsub{x_n}{P_n}]\er{M} \in \dl B \dr \eta~.
\]
\end{proposition}
\Proof
We abbreviate $[\lsub{x_1}{P_1},\ldots,\lsub{x_n}{P_n}]$
with $\sub{\vc{x}}{\vc{P}}$.
We proceed by induction on the typing proof.

\begin{description}

\item[$(asmp)$] follows by definition. 

\item[$(\arrow_I)$] We have to show:
\[
\lambda x.\sub{\vc{x}}{\vc{P}}\er{M} \in \dl A \arrow B \dr \eta~.
\]
By inductive hypothesis, we know:
$\sub{\vc{x}}{\vc{P}}\sub{x}{P}\er{M} \in \dl B \dr \eta$,
for all $P\in \dl A\dr \eta$. 
We conclude  by using the properties of reducibility candidates.

\item[$(\arrow_E)$] By the definition of $\arrow$.

\item[$(\forall_I)$] We have to show:
\[
\sub{\vc{x}}{\vc{P}}\er{M} \in \Inter_{X\in {\it RC}}\dl B \dr \eta[X/t]~.
\]
By the side condition on the typing rule, we know:
$\dl A_i \dr \eta = \dl A_i \dr \eta[X/t]$,
for an arbitrary $X\in {\it RC}$.
By inductive hypothesis: 
$\sub{\vc{x}}{\vc{P}}\er{M} \in \dl B \dr \eta[X/t]$, 
for an arbitrary $X\in {\it RC}$.

\item[$(\forall_E)$] We have to show:
\[
\sub{\vc{x}}{\vc{P}}\er{M} \in \dl B \dr \eta[\dl A \dr \eta/t]~.
\]
By inductive hypothesis:
$\sub{\vc{x}}{\vc{P}}\er{M} \in \Inter_{X\in {\it RC}}\dl B \dr \eta[X/t]$.
Choose $X= \dl A \dr \eta$.  \qed

\end{description}

\begin{corollary}[strong normalization]\index{strong normalization, system F}
If $\Gamma \Gives M:A$ in system F, then $M$ is strongly
normalizing.
\end{corollary}
\Proof
We note that $\qqs{A, \eta, x}{(x\in \dl A \dr \eta)}$.
Then we apply proposition \ref{prop-sound-sysF}  with $P_i \equiv x_i$,
and derive that:
$\er{M} \in \dl A \dr \eta \subseteq {\it SN}$.
By proposition  \ref{erasure-typed-prop},
we conclude that $M$ is strongly normalizing. \qed

\begin{exercise}[neutral $\lambda$-term]\label{neutral-def-systemF}\index{$\lambda$-term, neutral}
Alternative definitions of reducibility candidates can
be found in the literature; one follows.
Say that a $\lambda$-term is {\em neutral} if
it does not start with a $\lambda$-abstraction.
Define $\w{Red}(M) =\set{M' \mid M\arrow_\beta M'}$.
The collection ${\it RC'}$ is given by the sets $X$ of 
strongly normalizing $\lambda$-terms  satisfying the following conditions:

\begin{enumerate}

\item  $M\in X$ and $M\arrow_{\beta}M'$ implies $M'\in X$.

\item $M$ neutral and $\w{Red}(M)\subseteq X$ implies $M\in X$.

\end{enumerate}
Carry on the strong normalization proof using the collection ${\it RC'}$.
\end{exercise}

\section{Summary and references}
The introduction of second-order quantification 
{\em preserves the standard properties} of the propositionally typed calculus: 
subject reduction,  strong normalization, confluence $\ldots$
while {\em increasing the expressivity} in a very significant way as
one can encode inductive data types and iterative functions.
There is one catch however: {\em type inference becomes undecidable} 
which is one reason why $\ml$-like programming languages adopt a 
weaker/predicative form of polymorphism.
When extended with first-order quantification, system F is
the backbone of a higher-order constructive logic (the so called 
{\em calculus of constructions} on which the {\sc Coq} 
proof assistant  is built \cite{DBLP:journals/iandc/CoquandH88}).

The system F has been introduced by Girard in \cite{Girard70}\index{Girard} 
as a tool for the study
of the {\em cut-elimination procedure} in {\em second order
Peano  arithmetic} (${\it PA}_2$).  More precisely the {\em normalization
  of system F} implies the {\em termination of the cut-elimination
  procedure} in ${\it PA}_2$ (and thus the consistency of analysis!).
By relying on this strong connection between system F and ${\it PA}_2$
it is proven that all functions that can be shown to be total in
${\it PA}_2$ are {\em representable} in system F.  This is a {\em
  huge} collection of total recursive functions that goes well beyond
the primitive recursive functions.  The connections with the notion
of type polymorphism (or type parametricity) arising in programming 
are noticed in  \cite{DBLP:conf/programm/Reynolds74} and 
the relationship between existential types and abstract data types
are pointed out in \cite{DBLP:journals/toplas/MitchellP88}.
The results on the representation of iterative
functions are based on \cite{DBLP:journals/tcs/BohmB85}.
The type inference problem for a Curry-style system F (cf. chapter
\ref{type-inf-sec}) turns out to be undecidable
\cite{DBLP:journals/apal/Wells99}.

\chapter{Program transformations}\label{transform-sec}
\markboth{{\it Program transformations}}{{\it Program transformations}}

In this chapter, we introduce four program transformations.
Each transformation has its own interest. Moreover, when they are
put in pipeline they provide a compilation chain from a call-by-value 
$\lambda$-calculus to a register transfer level (RTL) language.  
A RTL language can be regarded as a machine independent version 
of assembly code.  Functions correspond to assembly level routines and
the functions' bodies correspond to sequences of vectors' allocations and
vectors' projections ended by a tail recursive call. 
The compilation chain is summarized in the following diagram:
\begin{equation}\label{chain}
\begin{array}{lllllllll}
  \lambda 
  &\act{\cl{C}_{\it cps}}
  &\lambda_{\w{cps}}
  &\act{\cl{C}_{\it vn}}
  &\lambda_{\w{cps},\w{vn}}
  &\act{\cl{C}_{\it cc}}
  &\lambda_{\w{cc},\w{vn}}
  &\act{\cl{C}_{\it h}}
  &\lambda_{h,\w{vn}}
\end{array}
\end{equation}
The source language is a call-by-value, 
$\lambda$-calculus (cf. chapter \ref{env-closure-sec}).
The first transformation, called {\em continuation-passing style} (CPS),
internalizes the notion of evaluation context, the second,
called {\em value naming}, assigns a name to every value, 
the third, called {\em closure conversion}, internalizes the notion of closure 
and makes sure functions are closed, {\em i.e.}, they do not contain free variables,  and the last, called {\em hoisting}, transforms a collection of closed 
nested function definitions into a collection of possibly open, flat,
{\em i.e.}, without nesting,  function definitions. 

Since we want to compose these transformations, we make sure the
target language of each transformation coincides with the source of
the following one.  As a matter of fact, all the languages are subsets
of the initial source language though their evaluation mechanism is
refined along the way.  In particular, one moves from an ordinary
substitution to a specialized one where variables can only be replaced
by other variables.

The approach to compiler correctness is similar
to the one considered for the toy compiler 
of section \ref{toy-compiler-sec}.
One proves that each transformation is correct in the sense that
the object code simulates the source code. 
Then, by composition, one derives the correctness of the compilation chain.

\section{Continuation passing style form}
The origin of the CPS transformation goes back to 
so called {\em double-negation} transformations from classical to
intuitionistic/constructive logic. In constructive logic,
the formula $\neg\neg t \arrow t$ is not derivable but
the formula $\neg\neg\neg t \arrow \neg t$ is 
(cf. exercise \ref{taut-type-ex}).
Then the idea
is to transform formulae in classical logic to {\em negated} formulae
in constructive logic so that negation is involutive on the image
of the transformation. For some fixed type variable $s$, let 
$\neg A=(A\arrow s)$, and define
a transformation of propositional types and type contexts as follows:
\[
\begin{array}{llll}
\ul{t} = t ~,
&\ul{A\arrow B}= \ul{A} \arrow \neg \neg \ul{B}~,
&\ul{\emptyset}=\emptyset~,
&\ul{\Gamma,x:A} = \ul{\Gamma},x:\ul{A}~.
\end{array}
\]
What the transformation shows is that for 
for every formula $A$ provable in classical logic, there
is a classicaly equivalent formula $\neg \neg \ul{A}$ which is provable
in constructive logic.
Now suppose we start with a $\lambda$-term of type $A$, 
say  $\Gamma \Gives M:A$, {\em i.e.}, with 
a constructive proof of $A$.
Can we build a $\lambda$-term  $\ul{M}$ 
such that $\ul{\Gamma}\Gives \ul{M}:\neg \neg \ul{A}$\ ?
For variables and $\lambda$-abstractions, the typing suggests directly:
\[
\begin{array}{ll}
\ul{x}=\lambda k.kx~,
&\ul{\lambda x.M} = \lambda k.k(\lambda x.\ul{M})~.
\end{array}
\]
The case for application is a bit more complex, but the reader may 
easily check that the following does the job:
\[
\begin{array}{c}
\ul{MN} = \lambda k.\ul{M}(\lambda m.\ul{N}(\lambda n.m n k))~.
\end{array}
\]
Moreover, the transformed $\lambda$-term simulates the
original one as soon it is provided with
an additional argument which represents the initial evaluation
context. For instance, the reader may check that:
\begin{equation}\label{red-cps-ex}
\ul{((\lambda x.x) y)} (\lambda z.z) \trarrow_{\beta} y~,
\end{equation}
where $\lambda z.z$ stands for the initial evaluation context.
As a matter of fact, types are useless in proving the simulation
property and they will be omitted in the following formal treatment.
However, as we have seen, types shed light on 
the CPS transformation and we shall come back to them in 
chapter \ref{typing-sec}.
The reader may have noticed that the reduction (\ref{red-cps-ex}) above 
performs many `useless' $\beta$-reductions. 
For this reason, as well as for simplifying the proof strategy,
we shall study an  {\em optimized version} of the CPS transformation. 

Table~\ref{lambda} introduces the source language: 
a type-free, left-to-right, call-by-value
$\lambda$-calculus.  
Notice that for technical reasons we include the variables among the values.
Also notice that the calculus  is richer than the one studied
in chapter \ref{env-closure-sec} in that it includes {\em let-definitions},
{\em polyadic abstraction}, and {\em tupling}, with the related
application and projection operators.  Polyadic abstraction 
grants a function the right to take several arguments at once while
tupling allows to build vectors of terms.
For the sake of readibility, we shall denote {\em explicitly}
the polyadic application with the symbol $@$. 
We stress that polyadic abstraction can be simulated 
by iterated $\lambda$-abstraction
and tupling can be simulated by iterated pairing.
Still, it  is worth to take them as primitive in order to 
simplify the analysis of the following program transformations.

Working with polyadic abstraction and tuples, we need a compact
notation to represent sequences of symbols.
We shall write $X^+$ (resp. $X^*$) for a non-empty (possibly empty)
finite sequence $X_1,\ldots,X_n$ of symbols.
By extension, $\lambda x^+.M$ stands for $\lambda x_1\ldots x_n.M$,
$[V^+/x^+]M$ stands for  $[V_1/x_1,\ldots,V_n/x_n]M$, and
$\s{let} \ (x=V)^+ \ \s{in} \ M$ stands for 
$\lets{x_{1}}{V_{1}}{\cdots \lets{x_{n}}{V_{n}}{M}}$.
By default, a {\em term} is a $\lambda$-term in the enriched $\lambda$-calculus
under consideration.

\begin{table}[t]
{\footnotesize
\begin{center}
{\sc Syntax}
\end{center}
\[
\begin{array}{lll}

V &::= \w{id} \Alt \lambda \w{id}^+.M \Alt (V^*)                                 &\mbox{(values)} \\
M &::= V \Alt @(M,M^+) \Alt \lets{\w{id}}{M}{M} \Alt (M^*) \Alt \prj{i}{M}       
   &\mbox{(terms)} \\
E &::= [~] \Alt @(V^*,E,M^*) \Alt \lets{\w{id}}{E}{M} \Alt (V^*,E,M^*) \Alt \prj{i}{E}  &\mbox{(evaluation contexts)}

\end{array}
\]
\begin{center}
{\sc Reduction Rules}
\end{center}
\[
\begin{array}{lll}

E[@(\lambda x_1\ldots x_n.M,V_1,\ldots, V_n)] &\arrow &E[[V_1/x_1,\ldots,V_n/x_n]M] \\
E[\lets{x}{V}{M}] &\arrow &E[[V/x]M] \\
E[\prj{i}{V_1,\ldots,V_n}] &\arrow &E[V_i] \qquad (1\leq i \leq n)

\end{array}
\]
}
\caption{A polyadic, call-by-value, $\lambda$-calculus: $\lambda$}\label{lambda}
\end{table}

Table~\ref{lambda-cps} introduces a fragment of the
$\lambda$-calculus described in Table~\ref{lambda} and a related
CPS transformation.  
An evaluation context $E$ can be represented as a term
$\lambda x.E[x]$; in a CPS transformation each
function takes its evaluation context, represented as a term, 
as a fresh additional parameter.
The {\em initial} evaluation context is defined relatively to a
fresh variable named '$\w{halt}$'. 

The reduction rules we apply to CPS terms
are those of the $\lambda$-calculus (Table~\ref{lambda}).
The syntax of CPS terms is such that in an application
or in a tuple all terms are values and this property
is preserved by reduction.
A corollary of this syntactic restriction is that
an evaluation context is either trivial or of
the shape $\s{let} \ x=[] \ \s{in} \ M$. 

Notice that strictly speaking the CPS terms are not closed
under reduction because, {\em e.g.},
$\s{let} \ x=\pi_1(V_1,V_2) \ \s{in} \ M$ reduces to 
$\s{let} \ x=V_1 \ \s{in} \ M$ which is not a CPS term.
However, the latter reduces to $[V_1/x]M$ which is again
a CPS term.

There is a potential ambiguity concerning the CPS transformation of
tuples of values. We remove it, by assuming that
$(V_1,\ldots,V_n)\sco K$ is transformed according to the case for
values. But note that if we follow the general case for tuples we
obtain the same result.

Next, we state the properties enjoyed by the presented 
CPS transformation, which  is `optimized' so as to 
pre-compute many `administrative' reductions.
In particular, thanks to this optimization, 
we can show that the CPS transformation
of a term such as $E[@(\lambda x.M,V)]$ is a term
of the shape $@(\psi(\lambda x.M),\psi(V),K_E)$ for a suitable
continuation $K_E$ depending on the evaluation context~$E$.

\begin{proposition}[CPS simulation]\label{cps-simulation}
Let $M$ be a term of the $\lambda$-calculus. 
If $M \arrow N$ then  $\cps{M} \trarrow \cps{N}$.
\end{proposition}
\Proof
The proof takes the following steps.

\begin{enumerate}

\item We show by induction on $M$ that for all values $V$, terms $M$, and continuations $K\neq x$:
\begin{equation}\label{cps-step1}
[V/x]M\sco [\psi(V)/x]K \equiv [\psi(V)/x](M\sco K)~.
\end{equation}

\item
The evaluation contexts for the $\lambda$-calculus 
described in Table~\ref{lambda}
can also be specified `bottom up' as follows:
\[
\begin{array}{lll}
E  &::= &[~] \Alt E[@(V^*,[~],M^*)]  \Alt E[\lets{\w{id}}{[~]}{M}] \Alt 
 E[(V^*,[~],M^*)] \Alt 
E[\prj{i}{[~]}] ~.
\end{array}
\]
Following this specification, we associate with an evaluation context $E$ 
a continuation $K_E$ as follows:
\[
\begin{array}{lll}

K_{[~]} &= &\lambda x.@(\w{halt},x) \\

K_{E[@(V^*,[~],M^*)]} 
&= 
&\lambda x.M^*\sco \lambda y^*.@(\psi(V)^*,x,y^*,K_E) \\

K_{E[\lets{x}{[~]}{N}]}
&=
&\lambda x.N\sco K_E \\

K_{E[(V^*,[~],M^*)]}
&=
&(\lambda x.M^*\sco \lambda y^*.(\psi(V)^*,x,y^*))\sco K_E \\

K_{E[\prj{i}{[~]}}
&=
&\lambda x.\lets{y}{\prj{i}{x}}{(y\sco K_E)} ~,

\end{array}
\]
where $M^*\sco \lambda x^*.N$ stands for $M_0\sco \lambda x_0\ldots M_n\sco \lambda x_n.N$ with $n\geq 0$.

\item For all terms $M$ and evaluation contexts $E, E'$ we prove by induction
on the evaluation context $E$ that the following holds:
\begin{equation}\label{cps-step2}
E[M]\sco K_{E'} \equiv M\sco K_{E'[E]}~.
\end{equation}

\item For all terms $M$, continuations $K, K'$, and variable $x\notin \fv{M}$ 
  we prove by induction on $M$ and case analysis that the following  holds:
  %\marginpar{first case $M\sco K$ ?}
\begin{equation}\label{cps-step3}
\begin{array}{c}

[K/x](M\sco K') \left\{
\begin{array}{ll}
\arrow  M\sco K  &\mbox{if }K \mbox{ abstraction}, M\mbox{ value}, K'=x \\
\equiv (M\sco [K/x]K') &\mbox{ otherwise.}
\end{array}\right.
\end{array}
\end{equation}

\item Finally, we prove the assertion by case analysis 
on the reduction rule. We consider the case for application.
Suppose $E[@(\lambda x^+.M,V^+)] \arrow E[[V^+/x^+]M]$.
We have:
\[
\begin{array}{lll}
E[@(\lambda x^+.M,V^+)]\sco K_{[~]}      
&\equiv @(\lambda x^+.M,V^+)\sco K_E     &\mbox{(by (\ref{cps-step2}))}\\
&\equiv @(\lambda x^+,k.M\sco k,\psi(V)^+,K_E) &\mbox{(by definition)}\\
&\arrow [K_E/k,\psi(V)^+/x^+](M\sco k) &\\
&\equiv [K_E/k]([V^+/x^+]M\sco k) &\mbox{(by (\ref{cps-step1}))}\\
&\trarrow [V^+/x^+]M\sco  K_E &\mbox{(by (\ref{cps-step3}))}\\
&\equiv E[[V^+/x^+]M]\sco K_{[~]}     &\mbox{(by (\ref{cps-step2})).}
\end{array}
\]
~\qed
\end{enumerate}

We illustrate this result on the following example.

\begin{example}[CPS]\label{cps-example}
Let $M\equiv @(\lambda x.@(x,@(x,x)),I)$, where $I\equiv \lambda x.x$. Then
\[
\cps{M}\equiv @(\lambda x,k.@(x,x,\lambda y.@(x,y,k)), I', H)~,
\]
where: $I'\equiv \lambda x,k.@(k,x)$ and $H\equiv \lambda x.@(\w{halt},x)$.
The term $M$ is simulated by $\cps{M}$ as follows:
\[
\begin{array}{ccccccc}

M       &\arrow   &@(I,@(I,I))                  &\arrow   &@(I,I)     &\arrow   &I \\
\cps{M} &\arrow   &@(I',I',\lambda y.@(I',y,H)) &\arrow^+ &@(I',I',H) &\arrow^+ &@(\w{halt},I')~.

\end{array}
\]
\end{example}

\begin{table}[t]
{\footnotesize
\begin{center}
{\sc Syntax CPS terms}
\end{center}
\[
\begin{array}{lll}

V &::= \w{id} \Alt \lambda \w{id}^+.M \Alt (V^*)                                 &\mbox{(values)} \\
M &::= @(V,V^+) \Alt \lets{\w{id}}{\prj{i}{V}}{M}             &\mbox{(CPS terms)} \\ 
K &::=  \w{id} \Alt \lambda \w{id}.M                                            &\mbox{(continuations)} \\

\end{array}
\]

\begin{center}
{\sc CPS transformation}
\end{center}
\[
\begin{array}{lll}

\psi(x) &= & x \\

\psi(\lambda x^+.M) &= &\lambda x^+,k.(M\sco k) \\

\psi((V_1,\ldots,V_n))         &= &(\psi(V_1),\ldots,\psi(V_n)) \\ \\ 

V\sco k    &= &@(k,\psi(V)) \\

V\sco (\lambda x.M) &= &[\psi(V)/x]M  \\

@(M_0,\ldots,M_n)\sco K &= &M_0 \sco \lambda x_0.\ldots (M_n \sco \lambda x_n.@(x_0,\ldots,x_n,K)) \\

\lets{x}{M_1}{M_2}\sco K &= &M_1\sco \lambda x.(M_2\sco K) \\ 

(M_1,\ldots,M_n)\sco K &= &M_1\sco \lambda x_1.\ldots (M_n\sco \lambda x_n.((x_1,\ldots,x_n)\sco K) \ ) \\

\prj{i}{M}\sco K      &= &M\sco \lambda x.\lets{y}{\prj{i}{x}}{(y\sco K)}  \\

\cps{M}           &= &M\sco \lambda x.@(\w{halt},x),\qquad\w{halt} \mbox{ fresh variable}

\end{array}
\]}
\caption{CPS $\lambda$-calculus ($\lambda_{{\it cps}}$) and CPS transformation}\label{lambda-cps}\index{continuation passing style}
\index{$\lambda$-calculus, CPS form}
\end{table}

\begin{exercise}\label{proof-cps}
  Prove the assertions (\ref{cps-step1}), (\ref{cps-step2}),
  and (\ref{cps-step3}).
\end{exercise}

\begin{exercise}\label{cps-cbv-cbn-ex}
Write down a simplified CPS transformation for a monadic call-by-value
$\lambda$-calculus without let-definitions and tuples.
Then apply the CPS transformation to show that it is possible to simulate
the call-by-value $\lambda$-calculus in the call-by-name $\lambda$-calculus
(cf. chapter \ref{env-closure-sec}).
\end{exercise}

\begin{exercise}\label{cps-control-ex}
So called {\em control operators} are programming instructions
that alter the execution flow.
For instance, consider the \s{continue} and \s{break}
commands of exercise \ref{break-continue-ex} and the control $\cl{C}$ and 
abort $\cl{A}$ operators of exercise \ref{control-abort-ex}.
CPS transformations allow to simulate such operators in a purely functional
setting. 

\begin{enumerate}

\item In chapter \ref{intro-sec}, we have interpreted a statement 
of the $\imp$ language as a function of type $(\w{State}\arrow \w{State})$. 
Define an alternative functional interpretation
where a command is regarded as a function of type:
\[
\w{State} \arrow (\w{State}\arrow \w{State})\arrow \w{State}~,
\]
and show that such interpretation can be extended to interpret
a command \s{abort} which stops the computation and returns the current
state.

\item Define a CPS transformation of the call-by-value $\lambda$-calculus
extended with the control operators $\cl{C}$ and $\cl{A}$ defined in exercise 
\ref{control-abort-ex}.

\end{enumerate}
\end{exercise}

\section{Value named form (*)}
Table~\ref{lambda-adm} introduces a {\em value named} $\lambda$-calculus
in CPS form: $\lambda_{\w{cps},\w{vn}}$. 
In the ordinary $\lambda$-calculus, the application of a $\lambda$-abstraction
to an argument (which is a value) may duplicate the argument as in:
$@(\lambda x.M,V) \arrow [V/x]M$.
In the value named  $\lambda$-calculus, all values are named 
and when we apply the name of a $\lambda$-abstraction to the name of a value
we create a new copy of the body of the function and replace its formal parameter name
with the name of the argument as in:
\[
\lets{f}{\lambda x.M}{\lets{y}{M}{@(f,y)}} \ \arrow  \ 
\lets{f}{\lambda x.M}{\lets{y}{V}{[y/x]M}}~.
\]
Notice that the definition of $E(x)$ in Table~\ref{lambda-adm} makes sure $y$
is not among the variables free in $\lambda x.M$.
Otherwise, one has to rename $y$ in order to apply the rule.
We also remark that in the value named $\lambda$-calculus the
evaluation contexts are a sequence of let definitions associating
values to names.  Thus, apart for the fact that the values are not
necessarily closed, the evaluation contexts are similar to the
environments of abstract machines for functional languages
(cf. chapter \ref{env-closure-sec}).

\begin{table}[t]
{\footnotesize
\begin{center}
{\sc Syntax}
\end{center}
\[
\begin{array}{lll}

V &::= \lambda \w{id}^+.M \Alt (\w{id}^*)                                                    &\mbox{(values)} \\
C &::= V \Alt \prj{i}{\w{id}}                                                               &\mbox{(let-bindable terms)} \\
M &::= @(\w{id},\w{id}^+) \Alt \lets{\w{id}}{C}{M} 
  &\mbox{(CPS terms)} \\
E &::= [~] \Alt \lets{\w{id}}{V}{E}                                                          &\mbox{(evaluation contexts)} \\

\end{array}
\]
\begin{center}
{\sc Reduction Rules}
\end{center}
\[
\begin{array}{cccl}

E[@(x,z_1,\ldots,z_n)]
&\arrow
&E[[z_1/y_1,\ldots,z_n/y_n]M]
&\mbox{if }E(x)=\lambda y_1 \ldots y_n.M \\

E[\lets{z}{\prj{i}x}{M}]
&\arrow
& E[[y_i/z]M]]
&\mbox{if }E(x)=(y_1,\ldots,y_n),  1\leq i \leq n 

\end{array}
\]
\[
\mbox{where: }E(x)=\left\{
\begin{array}{ll}
V  &\mbox{if }E=E'[\lets{x}{V}{[~]}] \\
E'(x) &\mbox{if }E=E'[\lets{y}{V}{[~]}],x\neq y,y\notin\fv{E'(x)} \\
\mbox{undefined} &\mbox{otherwise}
\end{array}\right.
\]
}
\caption{A value named CPS $\lambda$-calculus: $\lambda_{{\it cps,vn}}$}\label{lambda-adm}\index{$\lambda$-calculus, value named form}
\end{table}

Table~\ref{back-forth} defines the compilation into value named
form along with a {\em readback} transformation. 
The latter is useful to state the
simulation property.  Indeed, it is not true that if $M\arrow M'$ in
$\lambda_{cps}$ then $\vn{M} \trarrow \vn{M'}$ in
$\lambda_{\w{cps},\w{vn}}$. For instance, consider $M\equiv (\lambda
x.xx)I$ where $I\equiv(\lambda y.y)$.  Then $M\arrow II$ but $\vn{M}$
does not reduce to $\vn{II}$ but rather to a term where the `sharing'
of the duplicated value $I$ is explicitly represented.

\begin{example}[value named form]\label{admin-example}
Consider the term resulting from the CPS transformation in example
\ref{cps-example}:
$$N\equiv @(\lambda x,k.@(x,x,\lambda y.@(x,y,k)),I' ,H))~,$$
where: $I'\equiv \lambda x,k.@(k,x)$ and $H\equiv \lambda x.@(\w{halt},x)$.
The corresponding term in value named form is:
\[
\begin{array}{l}
\s{let} \ z_1 = \lambda x,k. (\lets{z_{11}}{\lambda y.@(x,y,k)}{@(x,x,z_{11})}) \ \s{in} \\
\s{let} \ z_2 = I' \ \s{in} \\
\s{let} \ z_3 = H \ \s{in} \\
@(z_1,z_2,z_3)~.
\end{array}
\]
\end{example}

\begin{table}[t]
{\footnotesize
\begin{center}
{\sc Transformation in value named form (from $\lambda_{cps}$ to $\lambda_{cps,vn}$)}
\end{center}
\[
\begin{array}{lll}

\vn{@(x_0,\ldots,x_n)} &= &@(x_0,\ldots,x_n) \\

\vn{@(x^*,V,V^*)}      &= &\evn{V}{y}[\vn{@(x^*,y,V^*)}] \quad V\neq \w{id}, y \mbox{ fresh}\\

\vn{\lets{x}{\prj{i}{y}}{M}}  &= &\lets{x}{\prj{i}{y}}{\vn{M}} \\

\vn{\lets{x}{\prj{i}{V}}{M}}  &= 
&\evn{V}{y}[\lets{x}{\prj{i}{y}}{\vn{M}}] \quad V\neq \w{id}, y \mbox{ fresh}\\

\evn{\lambda x^+.M}{y}            &= &\lets{y}{\lambda x^+.\vn{M}}{[~]} \\

\evn{(x^*)}{y}                   &= &\lets{y}{(x^*)}{[~]} \\ 

\evn{(x^*,V,V^*)}{y}              &= &\evn{V}{z}[\evn{(x^*,z,V^*)}{y}] \quad V\neq \w{id}, z \mbox{ fresh}

\end{array}
\]
\begin{center}
{\sc Readback transformation (from $\lambda_{\w{cps},\w{vn}}$ to $\lambda_{\w{cps}}$)} \\
\end{center}
\[
\begin{array}{lll}

\rb{\lambda x^+.M}  &= &\lambda x^+.\rb{M} \\
\rb{x^*}          &= &(x^*) \\

\rb{@(x,x_1,\ldots,x_n)}       &= &@(x,x_1,\ldots,x_n) \\ 

\rb{\lets{x}{\prj{i}{y}}{M}}  &= &\lets{x}{\prj{i}{y}}{\rb{M}} \\

\rb{\lets{x}{V}{M}} &=       &[\rb{V}/x]\rb{M} 

\end{array}
\]}
\caption{Transformations in value named CPS form and readback}\label{back-forth}
\index{value named form}
\end{table}

\begin{proposition}[vn simulation]\label{ad-simulation}
Let $N$ be a term in CPS value named form.  If
$\rb{N}\equiv M$ and $M\arrow M'$ then there exists $N'$ such
that $N\arrow N'$ and $\rb{N'}\equiv M'$.
\end{proposition}
\Proof
First we fix some notation. 
We associate a substitution $\sigma_E$ with an evaluation context $E$ of the $\lambda_{cps,\w{vn}}$-calculus
as follows:
\[
\begin{array}{ll}
\sigma_{[~]} = \w{Id}~,
&\sigma_{\lets{x}{V}{E}} = [\rb{V}/x]\comp \sigma_E~.
\end{array}
\]
Then we prove the property by case analysis. We look at the case:
\[
\rb{N}\equiv @(\lambda y^+.M,V^+) \arrow [V^+/y^+]M~.
\] 
Then $N\equiv E[@(x,x^+)]$, $\sigma_E(x)\equiv \lambda y^+.M$, and 
$\sigma_E(x^+)\equiv V^+$. 
Moreover, $E\equiv E_1[\lets{x}{\lambda y^+.M'}{E_2}]$ and
$\sigma_{E_{1}}(\lambda y^+.M') \equiv \lambda y^+.M$.
Therefore, $N\arrow E[[x^+/y^+]M']\equiv N'$ and we check that
$\rb{N'}\equiv \sigma_E([x^+/y^+]M')\equiv [V^+/y^+]M$. \qed

\section{Closure conversion (*)}
The next step is called {\em closure conversion}. It consists in
providing each functional value with an additional parameter that
accounts for the names free in the body of the function and in
representing functions using closures. Our closure conversion function
implements a closure using a pair whose first component is the code of
the transformed function and whose second component is a tuple containing the
values of the free variables. 

It will be convenient to write ``$\lets{(y_1,\ldots,y_n)}{x}{M}$'' for
``$\lets{y_1}{\prj{1}{x}}{\cdots\lets{y_n}{\prj{n}{x}}{M}}$'' and
``$\s{let}\ x_1=C_1 \ldots x_n=C_n\ \s{in}\ M$'' for ``$\s{let} \ x_1=C_1
\ \s{in} \ldots \s{let} \ x_n=C_n \ \s{in} \ M$''.  The transformation
is described in Table~\ref{lambda-cc}.  The output of the
transformation is such that all functional values are closed.

\begin{example}[closure conversion]\label{cc-example}
Let $M\equiv \vn{\cps{\lambda x.y}}$, namely:
\[
M\equiv \lets{z_1}{\lambda x,k.@(k,y)}{@(\w{halt},z_1)}~.
\]
Then $\cc{M}$ is the following term: 
\[
\begin{array}{l}
\s{let} \ c = \lambda e,x,k.(\s{let} \ (y)=e, (c,e)=k \ \s{in} \ @(c,e,y)) \ \s{in}\\
\s{let} \ e = (y), z_1=(c,e), (c,e) = \w{halt} \ \s{in} \\
@(c,e,z_1)~.

\end{array}
\]
\end{example}

\begin{table}[t]
{\footnotesize
\begin{center}
{\sc Syntactic restrictions on $\lambda_{cps,\w{vn}}$ after closure conversion} \\
All functional values are closed. ~\\~ \\ 
{\sc Closure Conversion}
\end{center}
\[
\begin{array}{lll}

\cc{@(x,y^+)}                   &=  
\lets{(c,e)}{x}{@(c,e,y^+)} \\ \\

\cc{\lets{x}{C}{M}} &= 
\begin{array}{l}
\s{let} \ c=\lambda e,x^+.\s{let} \ (z_1,\ldots,z_k)=e \ \s{in} \ \cc{N}  \ \s{in} \\
\s{let} \ e=(z_1,\ldots,z_k) \ \s{in} \\
\s{let} \ x = (c,e) \ \s{in} \\
\cc{M} \qquad \qquad (\mbox{if }C=\lambda x^+.N,\fv{C}=\set{z_1,\ldots,z_k})
\end{array} 
 \\ \\

\cc{\lets{x}{C}{M}} &= \lets{x}{C}{\cc{M}} 
\qquad \qquad \quad  (\mbox{if }C \mbox{ not a function})

\end{array}
\]
}
\caption{Closure conversion  on value named CPS terms}\label{lambda-cc}
\index{closure conversion}
\end{table}

\begin{proposition}[CC simulation]\label{cc-simulation}
Let $M$ be a CPS term in value named form. 
If $M\arrow M'$ then $\cc{M} \trarrow \cc{M'}$.
\end{proposition}
\Proof
As a first step we check that  the closure conversion function commutes with name 
substitution:
\[
\cc{[x/y]M} \equiv [x/y]\cc{M}~.
\]
This is a direct induction on the structure of the term $M$.
Then we extend the closure conversion function to contexts as follows:
\[
\begin{array}{ll}

\cc{[~]} &=[~] \\

\cc{\lets{x}{(y^*)}{E}} &= \lets{x}{(y^*)}{\cc{E}} \\

\cc{\lets{x}{\lambda x^+.M}{E}} &=      \s{let} \ c=\lambda e,x^+.\s{let} \ (z_1,\ldots,z_k)=e \ \s{in} \ \cc{M}  \ \s{in} \\
                                &\quad \s{let} \ e = (z_1,\ldots,z_k),x=(c,e) \ \s{in} \ \cc{E} \\
                                &\mbox{where: }\fv{\lambda x^+.M}=\set{z_1,\ldots,z_k} ~.
\end{array}
\]
We note that for any evaluation context $E$, $\cc{E}$ is again an evaluation context, and 
moreover for any term $M$ we have:
\[
\cc{E[M]} \equiv \cc{E}[\cc{M}]~.
\]
Finally we prove the simulation property by case analysis of the reduction rule being 
applied.

\begin{itemize}

\item Suppose $M\equiv E[@(x,y^+)] \arrow E[[y^+/x^+]M]$ where $E(x)=\lambda x^+.M$ 
and $\fv{\lambda x^+.M}=\set{z_1,\ldots,z_k}$.
Then:
\[
\cc{E[@(x,y^+)]} \equiv \cc{E}[\lets{(c,e)}{x}{@(c,e,y^+)}]~,
\]
with $\cc{E}(x)=(c,e)$,
$\cc{E}(c)=\lambda e,x^+.\s{let} \ (z_1,\ldots,z_k)=e \ \s{in} \ \cc{M}$ and
$\cc{E}(e)=(z_1,\ldots,z_k)$. Therefore:
\[
\begin{array}{l}
\cc{E}[\lets{(c',e')}{x}{@(c',e',y^+)}] \\
\trarrow \cc{E}[\s{let} \ (z_1,\ldots,z_k)=e \ \s{in} \ [y^+/x^+]\cc{M}] \\
\trarrow \cc{E}[[y^+/x^+]\cc{M}] \\
\equiv \cc{E}[\cc{[y^+/x^+]M}]  \qquad \mbox{(by substitution commutation)} \\
\equiv \cc{E[[y^+/x^+]M]}~.
\end{array}
\]

\item Suppose $M \equiv E[\lets{x}{\prj{i}{y}}{M}] \arrow E[[z_i/x]M]$ 
where $E(y)=(z_1,\ldots,z_k)$, $1\leq i \leq k$.
Then:
\[
\cc{E[\lets{x}{\prj{i}{y}}{M}]} \equiv \cc{E}[\lets{x}{\prj{i}{y}}{\cc{M}}]
\]
with $\cc{E}(y)=(z_1,\ldots,z_k)$. 
Therefore:
\[
\begin{array}{l}
 \cc{E}[\lets{x}{\prj{i}{y}}{\cc{M}}] \\
\arrow \cc{E}[[z_i/x]\cc{M}] \\
\equiv \cc{E}[\cc{[z_i/x]M}] \qquad \mbox{(by substitution commutation)} \\
\equiv \cc{E[[z_i/x]M]}~.
\end{array}
\]
~\qed
\end{itemize}

\begin{exercise}\label{cl-conv-direct-ex}
Define a closure conversion transformation that applies directly to
the source language rather than to the CPS, value named form.
\end{exercise}

\section{Hoisting (*)}
The last compilation step consists in moving all function definitions at top level.
In Table~\ref{lambda-hoist}, we formalize this compilation step as the iteration of a 
set of program transformations that commute with 
the reduction relation. Denote with $\lambda z^+.T$ 
a function that does {\em not} contain function definitions.
The transformations $(h_1)$ and $(h_2)$ consist in hoisting (moving up) the definition 
of a function   $\lambda z^+.T$. 
In transformation $(h_1)$, we commute the function definition with 
a tuple or a projection definition. This is always possible on the terms
resulting from a closure conversion since in these terms the functions are closed
and therefore cannot depend on a tuple or a projection definition above them.
In transformation $(h_2)$, we have a function definition, say $f_1$ which contains a nested
function definition, say $f_2$. In this case we extract $f_2$ putting
it at the same level, and above $f_1$. Notice that in doing this $f_1$
is not closed anymore since it may depend on $f_2$.
It can be shown that the rewriting system induced by the rules $(h_1)$ and $(h_2)$ 
applied to the terms resulting from the closure conversion terminates and is confluent.
We omit this rather technical but not difficult development.
The proof that the hoisted program simulates the original one also requires some work because to
close the diagram we need to collapse repeated definitions, which may
occur, as illustrated in the example below. Again, we omit this development.

\begin{table}[t]
{\footnotesize
\begin{center}
{\sc Syntax for $\lambda_h$} \\
{Syntactic restrictions on $\lambda_{cps,\w{vn}}$ after hoisting} \\
All function definitions are at top level.
\end{center}
\[
\begin{array}{lll}

C &::= (\w{id}^*) \Alt \prj{i}{\w{id}}                                                       &\mbox{(restricted let-bindable terms)} \\
T &::= @(\w{id},\w{id}^+) \Alt \lets{\w{id}}{C}{T}                                           &\mbox{(restricted terms)} \\
P &::= T \Alt  \lets{\w{id}}{\lambda \w{id}^+.T}{P}                                   &\mbox{(programs)} 
\end{array}
\]
\begin{center}
{\sc Specification of the hoisting transformation}
\end{center}
\[
\h{M}=N\mbox{ if }M\leadsto \cdots \leadsto N \not\leadsto, \quad \mbox{where:}
\]
\[
\begin{array}{lll}

D &::= & [~] \Alt \lets{\w{id}}{C}{D} \Alt \lets{\w{id}}{\lambda \w{id}^+.D}{M} 
\qquad \mbox{(hoisting contexts)}
\end{array}
\]
\[
\begin{array}{lll}
\\
(h_1)
&D[\lets{x}{C}{\lets{y}{\lambda z^+.T}{M}}] \leadsto \\
&D[\lets{y}{\lambda z^+.T}{\lets{x}{C}{M}}] &\mbox{if }x\notin \fv{\lambda z^+.T} \\\\

(h_2)
&D[\lets{x}{(\lambda w^+.\lets{y}{\lambda z^+.T}{M})}{N}] \leadsto \\
&D[\lets{y}{\lambda z^+.T}{\lets{x}{\lambda w^+.M}{N}}] &\mbox{if }\set{w^+}\inter \fv{\lambda z^+.T} = \emptyset  

\end{array}
\]
}
\caption{Hoisting transformation}\label{lambda-hoist}
\index{hoisting}
\index{$\lambda$-calculus, hoisted form}
\end{table}

\begin{example}[hoisting transformations and transitions]\label{hoisting-comm-ex}
Let  $$M\equiv \lets{x_{1}}{\lambda y_1.N}{@(x_1,z)}~,$$
where $N\equiv \lets{x_{2}}{\lambda y_{2}.T_2}{T_1}$ 
and $y_1 \notin \fv{\lambda y_2.T_2}$.
Then we either reduce and then hoist:

{\footnotesize\[
\begin{array}{ll}
M  &\arrow \lets{x_{1}}{\lambda y_1.N}{[z/y_1]N}  \\
   &\equiv \lets{x_{1}}{\lambda y_{1}.N}{\lets{x_{2}}{\lambda y_{2}.T_2}}{[z/y_1]T_1} \\
   &\leadsto \lets{x_{2}}{\lambda y_2.T_2}{\lets{x_{1}}{\lambda y_1.N}{[z/y_1]T_1}} %\not\leadsto
\end{array}
\]}

or hoist and then reduce:

{\footnotesize\[
\begin{array}{ll}
M  &\leadsto \lets{x_{2}}{\lambda y_2.T_2}{\lets{x_{1}}{\lambda y_1.T_1}{@(x_1,z)}}  \\
   &\arrow   \lets{x_{2}}{\lambda y_2.T_2}{\lets{x_{1}}{\lambda y_1.T_1}{[z/y_1]T_1}} %\quad \not\leadsto
\end{array}
\]}

In the first case, we end up duplicating the definition of $x_2$. 
\end{example}

We conclude by sketching an alternative definition of the hoisting 
transformation.
Let $h$ be a function that takes a term $M$ in CPS, value named form
where all functions are closed and produces a pair $(T,F)$, where
$T$ is a term without function definitions as specified in Table \ref{lambda-hoist},
and $F$ is a one-hole context composed of a list of function definitions of the shape:
\[
F::=[~] \Alt \s{let} \ \w{id}= \lambda \w{id}^+.T \ \s{in} \ F~.
\]
The definition of the function $h$ is given by induction on $M$ as follows
where $C$ is a tuple or a projection as in Table \ref{lambda-hoist}:

{\footnotesize\[
\begin{array}{ll}

h(@(x,y^+)) &= (@(x,y^+),[~]) \\

h(\lets{x}{C}{M}) &= \s{let} \ (T,F)=h(M) \ \s{in} \ (\lets{x}{C}{T},F) \\

h(\lets{x}{\lambda y^+.M}{N})  &= \s{let} \ (T,F)=h(M), (T',F')=h(N) \ \s{in} \\                               &\qquad  (T',F[\lets{x}{\lambda y^+.T}{F'}]) ~.

\end{array}
\]}

The hoisting transformation of the term $M$ then amounts to compute $(T,F)=h(M)$ and then
build the term $F[T]$ which is a program according to the syntax defined in Table \ref{lambda-hoist}.

\begin{exercise}\label{cl-conv-hoist-ex}
Apply the hoisting transformation to the terms resulting from the closure
conversion of exercise \ref{cl-conv-direct-ex}.
\end{exercise}

\section{Summary and references}
We have studied four program transformations:
{\em continuation passing style} makes the evaluation context an additional parameter,
{\em value naming} assigns a name to every value, {\em closure conversion}
explicits the notion of closure, and {\em hoisting} removes
nested function definitions. By putting these transformations
in pipeline it is possible to transform a program written in a higher-order language
such as $\ml$ into a system of functions whose body includes operations 
to build and project tuples of names and to perform tail-recursive
routine calls. 
Thus we have an implementation technique for higher-order languages
which is alternative to the one based on the abstract machines presented
in chapter \ref{env-closure-sec}.
A similar compilation chain has been analyzed in \cite{DBLP:conf/popl/Chlipala10}
which provides machine certified simulation proofs.
A simpler compilation chain arises if we bypass the CPS transformation.
In this case, the function calls are not necessarily
tail-recursive and the target code can be described as $\C$ code
with function pointers (cf. exercises \ref{cl-conv-direct-ex} and \ref{cl-conv-hoist-ex}).
An early analysis of the CPS transformation is in 
\cite{DBLP:journals/tcs/Plotkin75}. The idea of value-naming transformation
is associated with various formalizations of {\em sharing}, see, {\em e.g.}, \cite{DBLP:conf/popl/Launchbury93}.
Closure conversion arises naturally when trying to define the interpreter of a higher-order
language in the language itself, see, {\em e.g.}, \cite{DBLP:journals/lisp/Reynolds98a}. Hoisting appears to be folklore.

\chapter{Typing the program transformations}\label{typing-sec}
\markboth{{\it Typing the transformations}}{{\it Typing the transformations}}

We present a typing of the compilation
chain described in chapter \ref{transform-sec}. 
Specifically, each $\lambda$-calculus of the compilation chain is equipped
with a type system which enjoys {\em subject reduction}:
if a term has a type then all terms to which it reduces have the same type.
Then the compilation functions are extended to types and are shown
to be {\em type preserving}: if a term has a type then its compilation 
has the corresponding  compiled type.

The two main steps in typing the compilation chain 
concern the CPS and the closure conversion transformations. 
The typing of the CPS transformation has already been 
sketched in chapter \ref{transform-sec} 
where it has served as a guideline.
A basic idea is to type the continuation/the evaluation context of a
term of type $A$ with its negated type $\neg A = (A\arrow R)$, where $R$ is
traditionally taken as the type of `results'. 
In typing closure conversion, one
relies on {\em existential types} (cf. chapter \ref{systemF-sec}) 
to hide the details of the representation
of the `environment' of a function, {\em i.e.}, the tuple of variables
occurring free in its body. Thus, to type the (abstract) assembly code 
coming from the compilation of propositionally typed programs, 
we need to go beyond propositional types.

To represent types we shall follow the notation introduced starting
from chapter \ref{simple-type-sec}.
In particular, we denote with 
$\w{tid}$  the syntactic category of {\em type variables} 
with generic elements $t,s,\ldots$ and with $A$ the 
syntactic category of {\em types} with generic elements $A,B,\ldots$
We write $x^*:A^*$ for a possibly empty  sequence $x_1:A_1,\ldots,x_n:A_n$, and
$\Gamma,x^*:A^*$ for the context
resulting from $\Gamma$ by adding the sequence
$x^*:A^*$. Hence the variables in $x^*$  must not be in the domain of $\Gamma$.
If $A$ is a type, we write $\ftv{A}$ for the set of {\em type variables occurring free} in it and, by extension, if $\Gamma$ is a type context 
then $\ftv{\Gamma}$ is the union of the sets $\ftv{A}$ 
where $A$ is a type in the codomain of $\Gamma$.
A {\em typing judgment} is typically written as $\Gamma \Gives M:A$ where $M$ is some term.
We shall write $\Gamma \Gives M^*:A^*$ for $\Gamma \Gives M_1:A_1,\ldots,$$\Gamma \Gives M_n:A_n$.
Similar conventions apply if we replace the symbol $`*'$ with the symbol $`+'$ except that in this case the sequence is assumed not-empty.
A type transformation, say $\cl{T}$, is {\em lifted to type contexts} by defining
$\cl{T}(x_1:A_1,\ldots x_n:A_n) = x_1:\cl{T}(A_1),\ldots,x_n:\cl{T}(A_n)$.
Whenever we write:
\begin{quote}
if $\Gamma \Gives^{S_{1}} M:A$ then $\cl{T}(\Gamma) \Gives^{S_{2}} \cl{T}(M):\cl{T}(A)$~,
\end{quote}
what we actually mean is that if the judgment in the hypothesis is {\em derivable} in a certain `type system $S_1$' then the transformed judgment in derivable in 
the `type system $S_2$'. 
Proofs are standard and are left as exercises.

\section{Typing the CPS form}
Table~\ref{type-lambda} describes the typing rules 
for the polyadic, call-by-value, $\lambda$-calculus 
defined in Table~\ref{lambda}.
These rules are a slight generalization of those
studied in chapter \ref{simple-type-sec} and 
they are preserved by reduction.
The typing rules described in Table~\ref{type-lambda} 
apply to the CPS $\lambda$-calculus too. Table~\ref{type-lambda}
describes  the restricted syntax of the CPS types
and the CPS type transformation. Then the CPS term transformation
defined in Table~\ref{lambda-cps} preserves typing in the following
sense.

\begin{proposition}[type CPS]\label{type-cps-prop}
If $\Gamma \Gives M:A$ then 
$\cps{\Gamma},\w{halt}:\neg \cps{A} \Gives \cps{M}:R$.
\end{proposition}

\begin{exercise}\label{proof-type-cps-prop}
  Prove proposition \ref{type-cps-prop}.
  \end{exercise}

\begin{table}
{\footnotesize
\begin{center}
{\sc Syntax types}
\end{center}
\[
\begin{array}{lll}
A      &::= \w{tid} \Alt A^+\arrow A \Alt \times(A^*)  &\mbox{(types)} 
\end{array}
\]
\begin{center}
{\sc Typing rules}
\end{center}
\[
\begin{array}{cc}

\infer{x:A\in \Gamma}
{\Gamma \Gives x:A}

&\infer{
\Gamma,x:A\Gives N:B\qquad
\Gamma \Gives M:A
}
{\Gamma \Gives \lets{x}{M}{N}:B} \\ \\

\infer{\Gamma,x^+:A^+ \Gives M:B}
{\Gamma \Gives \lambda x^+.M:A^+\arrow B} 

&\infer{
\Gamma \Gives M:A^+\arrow B \qquad
\Gamma \Gives N^+:A^+
}
{\Gamma \Gives @(M,N^+):B}  \\ \\

\infer{\Gamma \Gives M^*:A^*}
{\Gamma \Gives (M^*):\times(A^*)}

&\infer{\Gamma \Gives M:\times(A_1,\ldots,A_n)\quad 1\leq i  \leq n}
{\Gamma \Gives \prj{i}{M}:A_i}

\end{array}
\]
\begin{center}
{\sc Restricted syntax CPS types, $R$ type of results}
\end{center}
\[
\begin{array}{lll}
A &::= \w{tid} \Alt A^+\arrow R \Alt \times(A^*)  &\mbox{(CPS types)}
\end{array}
\]
\begin{center}
{\sc CPS type compilation}\ $(\neg A \equiv (A\arrow R))$
\end{center}
\[
\begin{array}{ll}

\cps{t} &= t \\
\cps{\times(A^*)}  &= \times(\cps{A}^*) \\
\cps{A^+ \arrow B} &=((\cps{A})^+, \neg \cps{B}) \arrow R 
\end{array}
\]
}
\caption{Type system for $\lambda$ and $\lambda_{\w{cps}}$}\label{type-lambda}
\end{table}

\section{Typing value-named closures (*)}
Table~\ref{types-adm} describes the typing rules for the
value named calculi with functional, product, and existential types.  
Notice that for the sake of brevity, we shall omit the type of a 
term since this type is always the
type of results $R$ and write $\Gamma \aGives M$ rather than 
$\Gamma \aGives M:R$.
The first five typing rules are just a
specialization of the corresponding rules in Table~\ref{type-lambda},
while the last two rules allow for the introduction and elimination of {\em
 existential types}. The need for existential types will be motivated
next. For the time being, let us notice that 
in the proposed formalization we rely on the tuple constructor to
introduce an existential type and the first projection to eliminate
it.  This has the advantage of leaving unchanged the syntax and the
reduction rules of the value named $\lambda$-calculus. An
alternative presentation (cf. chapter \ref{systemF-sec}) consists in 
introducing specific operators to
introduce and eliminate existential types denoted with
\s{pack} and \s{unpack}, respectively.  Then one can 
read $(x)$ as $\pack{x}$ and $\prj{1}{x}$
as $\unpack{x}$ when $x$ has an existential type. Notice that
the rewriting rule which allows to {\em unpack} a {\em
 packed} value is just a special case of the rule for projection.
As in the previous system, typing is preserved by reduction.

\begin{proposition}[subject reduction, value named]\label{subj-red-adm-prop}
If $M$ is a term of the $\lambda_{\w{cps},\w{vn}}$-calculus,
$\Gamma \aGives M$ and $M\arrow N$ (definitions in Table~\ref{lambda-adm})
then  $\Gamma \aGives N$.
\end{proposition}

\begin{exercise}\label{proof-subj-red-adm-prop}
  Prove proposition \ref{subj-red-adm-prop}.
  \end{exercise}

Turning to the transformation from CPS to  {\em value named} CPS form
specified in  Table~\ref{back-forth}, we notice that it 
affects the terms but not the types. Therefore we have the following property.

\begin{proposition}[type value named]\label{type-adm-prop}
If $M$ is a term of the $\lambda_{\w{cps}}$-calculus
and $\Gamma \Gives M:R$ then $\Gamma \aGives \vn{M}$.
\end{proposition}

\begin{exercise}\label{proof-type-adm-prop}
  Prove proposition \ref{type-adm-prop}.
  \end{exercise}

Next we discuss the typing of closure conversion via existential types 
(Table \ref{types-adm}).
We recall that in closure conversion a function, say $\lambda x.M$
with free variables $z_1,\ldots, z_n$, becomes a pair (here we 
ignore the details
of the CPS, value named form):
\begin{equation}
(\lambda e,x.\s{let} \ (z_1,\ldots,z_n)=e \ \s{in} \ \cl{C}(M),(z_1,\ldots,z_n))
\end{equation}
whose first component is the function
itself, which is {\em closed} by taking the environment $e$ as an additional
argument, and the second component is a tuple containg the values of
the free variables.
Now consider the functions {\em identity} and {\em successor} 
on the natural numbers coded as follows:
\begin{equation}
\lambda x.x~,\qquad \s{let} \ y=1 \ \s{in} \ \lambda x.x+y~,
\end{equation}
with a type, say, $\Nat \arrow \Nat$.
After closure conversion, we obtain the following pairs with the respective
{\em different} types:
\[
\begin{array}{ll}
(\lambda e,x.\s{let} \ ()=e \ \s{in} \ x,())  &: ((1\times \Nat) \arrow \Nat) \times 1\\

(\lambda e,x.\s{let} \ (y)=e \ \s{in} \ x+y,(y)) &: ((\Nat \times \Nat)\arrow \Nat) \times \Nat~.

\end{array}
\]
Then take a function such as $F:(\Nat \arrow \Nat)\arrow \Nat$
which can operate both on the identity and the successor function.
This is no longer possible after closure conversion 
if we stick to the typing outlined above. We can address this issue
by {\em abstracting} the types of the identity and successor functions
after closure conversion into the following {\em existential} type:
\begin{equation}
\exists t.((t\times \Nat)\arrow \Nat)\times t~.
\end{equation}
To summarize, an environment is a tuple whose size depends 
on the number of variables occurring free in the function.
This information should be abstracted in the type;  otherwise,
we cannot type functions operating on arguments with environments 
of different size.

\begin{table}
{\footnotesize
\begin{center}
{\sc Syntax types}
\end{center}
\[
\begin{array}{lll}

A&::= \w{tid} \Alt (A^+\arrow R) \Alt \times(A^*) \Alt \exists \w{tid}.A
\end{array}
\]
\begin{center}
{\sc Typing rules}
\end{center}
\[
\begin{array}{cc}

\infer{\Gamma,x^+:A^+\aGives M}
{\Gamma \aGives \lambda x^+.M:A^+\arrow R} 

&\infer{x:A^+\arrow R,y^+:A^+\in \Gamma}
{\Gamma \aGives @(x,y^+)} \\ \\

\infer{x^*:A^*\in \Gamma}
{\Gamma \aGives (x^*):\times(A^*)}

&\infer{\begin{array}{c}
y:\times(A_1,\ldots,A_n)\in \Gamma\quad 1\leq i\leq n\quad\\
\Gamma,x:A_i\aGives M
\end{array}}
{\Gamma \aGives \lets{x}{\prj{i}{y}}{M}}  \\ \\

\infer{\Gamma \aGives V:A\quad \Gamma,x:A\aGives M}
{\Gamma \aGives \lets{x}{V}{M}}  \\ \\

\infer{x:[B/t]A\in \Gamma}
{\Gamma \aGives (x) :\exists t.A}

&\infer{y:\exists t.A \in \Gamma\quad
\Gamma,x:A\aGives M\quad t\notin \ftv{\Gamma}}
{\Gamma \aGives \lets{x}{\prj{1}{y}}{M}}

\end{array}
\]
\begin{center}
{\sc Closure conversion type compilation}
\end{center}
\[
\begin{array}{ll}

\cc{t} &=t \\
\cc{\times(A^*)} &=\times(\cc{A}^*) \\
\cc{A^+\arrow R} &= \exists t.\times((t,\cc{A}^+ \arrow R),\ t) 

\end{array}
\]
}
\caption{Type system for the value named calculi and closure conversion}\label{types-adm}
\end{table}

In order to respect our conventions on the introduction and elimination 
of existential types, the closure conversion transformation is slightly
modified in the way described in Table~\ref{modified-cc-tab}.
This modified closure conversion still enjoys the simulation properties 
stated in proposition \ref{cc-simulation} and moreover 
it  preserves typing  as follows.

\begin{table}
{\footnotesize
\[
\begin{array}{ll}

\cc{@(x,y^+)}                   &=  
\begin{array}{l}
\s{let} \ x =\prj{1}{x} \ \s{in}  \qquad \mbox{($\leftarrow$ {\sc existential elimination})}\\
\lets{(c,e)}{x}{@(c,e,y^+)} 
\end{array}
\\ \\

\cc{\lets{x}{C}{M}} &= 
\begin{array}{l}
\s{let} \ c=\lambda e,x^+.\s{let} \ (z_1,\ldots,z_k)=e \ \s{in} \ \cc{N}  \ \s{in} \\
\s{let} \ e=(z_1,\ldots,z_k) \ \s{in} \\
\s{let} \ x = (c,e) \ \s{in} \\
\s{let} \ x = (x) \ \s{in}  \qquad \mbox{($\leftarrow$ {\sc existential introduction})}\\
\cc{M} \qquad \qquad (\mbox{if }C=\lambda x^+.N,\fv{C}=\set{z_1,\ldots,z_k})
\end{array} 

\end{array}
\]}
\caption{Modified closure conversion}\label{modified-cc-tab}
\end{table}

\begin{proposition}[type closure conversion]\label{type-cc-prop}
If $M$ is a term in $\lambda_{\w{cps},\w{vn}}$  and  $\Gamma \aGives M$ 
then  $\cc{\Gamma}\aGives \cc{M}$.
\end{proposition}

\begin{exercise}\label{proof-type-cc-prop}
  Prove proposition \ref{type-cc-prop}.
\end{exercise}

The last step in the compilation chain is the hoisting transformation.
Similarly to the transformation in value named form, 
the hoisting transformation  affects the terms but {\em not} the types.

\begin{proposition}[type hoisting]\label{type-hoisting-prop}
If $M$ is a term in $\lambda_{\w{cps},\w{vn}}$, 
$\Gamma \aGives M$,  and  $M\leadsto N$ then $\Gamma \aGives N$.
\end{proposition}

\begin{exercise}\label{proof-type-hoisting-prop}
  Prove proposition \ref{type-hoisting-prop}.
\end{exercise}

\section{Typing the compiled code (*)}
We can now extend the compilation function to types by defining:
\[
\cmp{A} = \cc{\cps{A}}
\]
and by composing the previous results we derive the following type preservation 
property of the compilation function.

\begin{proposition}[type preserving compilation]\label{type-compil-thm}
If $M$ is a term of the  $\lambda$-calculus
and $\Gamma \Gives M:A$  then:
\[
\cmp{\Gamma},\w{halt}:\exists t.\times(t,\cmp{A}\arrow R,t) \aGives \cmp{M}~.
\]
\end{proposition}

\begin{exercise}\label{proof-type-compil-thm}
  Prove proposition \ref{type-compil-thm}.
\end{exercise}

\begin{remark}\label{halt-rmk}
The `\w{halt}' variable introduced by the CPS transformation
can occur only in a subterm of the shape $@(\w{halt},x)$ in the intermediate
code prior to closure conversion. 
Then in the closure conversion transformation, we can set $\cc{@(\w{halt},x)}=@(\w{halt},x)$,  and give to $`\w{halt}'$
a functional rather than an existential type. With this proviso,
proposition \ref{type-compil-thm} above can be restated as follows:
\begin{quote}
If $M$ is a term of the $\lambda$-calculus and 
$\Gamma \Gives M:A$ then $\cmp{\Gamma},\w{halt}:\neg \cmp{A} \aGives \cmp{M}$.
\end{quote}
\end{remark}

\begin{example}[typing the compiled code]\label{type-example}
We consider again the compilation of the term $\lambda x.y$ (cf. example \ref{cc-example}) which can be typed, {\em e.g.}, as follows:
$y:t_1 \Gives \lambda x.y:(t_2\arrow t_1)$.
Its CPS transformation is then typed as:
\[
y:t_1,\w{halt}:\neg \cps{t_2\arrow t_1} \Gives @(\w{halt},\lambda x,k.@(k,y)): R~.
\]
The value named transformation does not affect the types:
\[
y:t_1,\w{halt}:\neg \cps{t_2\arrow t_1} \aGives \lets{z_1}{\lambda x,k.@(k,y)}{@(\w{halt},z_1)}~.
\]
After closure conversion, we obtain the following term $M$:
\[
\begin{array}{l}
\s{let} \ c = \lambda e,x,k.\s{let} \ y=\prj{1}{e}, k=\prj{1}{k},c=\prj{1}{k},e=\prj{2}{k} \ \s{in} \ @(c,e,y) \ \s{in}\\
\s{let} \ e = (y), z_1=(c,e), z_1=(z_1),\w{halt} =\prj{1}{\w{halt}},c=\prj{1}{\w{halt}}, e = \prj{2}{\w{halt}} \ \s{in} \\
@(c,e,z_1)~,
\end{array}
\]
which is typed as follows:
$y:t_1,\w{halt}: \exists t.\times(t,\neg \cmp{t_2\arrow t_1},t) \aGives M$.
In this case no further hoisting transformation applies.
If we adopt the optimized compilation strategy sketched in remark \ref{halt-rmk}
then after closure conversion we obtain the following term $M'$:
\[
\begin{array}{l}
\s{let} \ c = \lambda e,x,k.\s{let} \ y=\prj{1}{e}, k=\prj{1}{k},c=\prj{1}{k},e=\prj{2}{k} \ \s{in} \ @(c,e,y) \ \s{in}\\
\s{let} \ e = (y), z_1=(c,e), z_1=(z_1), \ \s{in} \\
@(\w{halt},z_1)
\end{array}
\]
which is typed as follows:
$y:t_1,\w{halt}:\neg \cmp{t_2\arrow  t_1} \aGives M'$.
\end{example}

\section{Summary and references}
We have typed the compilation chain presented in chapter \ref{transform-sec}
which goes from a higher-order language
to an abstract assembly code. The typing of the CPS transformation
builds on the double negation translations from classical to 
intuitionistic logic (see, {\em e.g.}, \cite{Troelstra88}). 
The typing of closure conversion relies on 
existential types to hide the details of the representation 
\cite{DBLP:conf/popl/MinamideMH96}. 
The paper \cite{DBLP:journals/toplas/MorrisettWCG99} shows that
the typing can be extend to the impredicative polymorphic types of 
system F (cf. chapter \ref{systemF-sec}).

\chapter{Records, variants, and subtyping}\label{record-sec}
\markboth{{\it Subtyping}}{{\it Subtyping}}

Records and variants are common data types found in many programming
languages which allow to aggregate heterogeneous data.
Record (variant) types provide a user-friendly alternative 
to product (sum) types where components can be manipulated by
labels rather than by projections (injections).

In this chapter, we start by discussing an extension of 
the call-by-value, type-free, $\lambda$-calculus with records
and a possible encoding of records.
We then move on to consider a {\em typed} version of the language.
In order to gain in flexibility, we introduce
a {\em subtyping rule} for records and study 
the properties of the derived type system. We conclude by
briefly discussing how the approach with subtyping 
can be extended to variant types.

\section{Records}
A {\em record} is a notation to represent a function with a finite
domain over a set of {\em labels} which are defined as follows.

\begin{definition}[labels]\label{labels-def}\index{label, records}
We denote with $L$ a countable and totally ordered 
set of {\em labels} with generic elements
$\ell,\ell',\ldots$ 
\end{definition}

We rely on the  notation: 
\begin{equation}
\set{\ell_1=V_1,\ldots,\ell_n=V_n}~,
\end{equation}
to denote the function that associates with the label $\ell_i$ the value
$V_i$, for $i=1,\ldots,n$, and which is undefined otherwise. Whenever we write a record
we assume that the labels are all distinct: $\ell_i\neq \ell_j$ if $i\neq j$.
Given a record $R$,  we write $R.\ell$ for the selection of 
the value of $R$ on the label $\ell$.
If $\ell$ is not in the domain of definition of the record then
we are in an erroneous situation and the computation is stuck or
alternatively an error message is produced.

Table \ref{lambda-record} 
describes an extension of the type-free, call-by-value, 
$\lambda$-calculus with records. In order to have a deterministic
evaluation strategy, we assume that records are always written
with labels in {\em growing order} and that the evaluation follows this order.
\begin{table}[b]
{\footnotesize
\begin{center}
{\sc Syntax}
\end{center}
\[
\begin{array}{ll}
M::= \w{id} \Alt \lambda \w{id}.M \Alt MM \Alt 
\set{\ell=M,\ldots,\ell=M} \Alt M.\ell &\mbox{($\lambda$-terms)}\\
V::=\lambda \w{id}.M \Alt \set{\ell=V,\ldots,\ell=V} &\mbox{(values)} 
\end{array}
\]
\begin{center}
{\sc Call-by-value evaluation contexts and reduction rules}
\end{center}
\[
\begin{array}{c}
E::=[~]\Alt EM \Alt VE \Alt \set{(\ell=V)^*,\ell=E,(\ell=M)^*} \Alt E.\ell
\end{array}
\]
\[
\begin{array}{lll}
(\lambda x.M)V  &\arrow &[V/x]M                \\
\set{\ldots,\ell=V,\ldots}.\ell &\arrow &V 
\end{array}
\]}
\caption{Type-free, call-by-value, $\lambda$-calculus with records}
\label{lambda-record}\index{$\lambda$-calculus, with records}
\end{table}

We pause to notice that in principle records could be represented
in the pure $\lambda$-calculus. For instance, we could 
associate with each label a natural number and then associate
with it a Church numeral. Suppose: (i) $\ul{\ell}$ denotes the
Church numeral that corresponds to the label $\ell$,
(ii) $E$ is a $\lambda$-term that decides the equality of two Church numerals 
(cf. exercise \ref{predecessor-ex}),
(iii) $C$ is the $\lambda$-term that represents the conditional, 
(iv) $F$ is a special $\lambda$-term to represent failure,
and (v) we write $\s{let}\ x_1=M_1,\ldots,x_n=M_n \ \s{in}\  N$ for 
$(\lambda x_1,\ldots,x_n.N)M_1\cdots M_n$.
Then we could {\em compile} the call-by-value $\lambda$-calculus with records
into the call-by-value $\lambda$-calculus following the rules
in Table~\ref{compile-records} (simple cases omitted).
A record is compiled into a function taking a label as
input and then performing a sequence of conditionals. Selecting
a record's label just amounts to {\em apply} the compilation of the record
to the encoding of the label.

\begin{table}
{\footnotesize\[
\begin{array}{ll}

\cl{C}(\set{\ell_1=M_1,\ldots,\ell_n=M_n}) &=
\s{let} \ x_1=\cl{C}(M_1),\ldots,x_n=\cl{C}(M_n) \ \s{in} \\
&\quad \lambda l.C (E\ l\ \ul{\ell_1}) \ x_1 (\cdots \quad (C (E\ l\ \ul{\ell_n}) \ x_n \ F) \cdots)
\\\\
\cl{C}(M.\ell) &= \s{let} \ x=\cl{C}(M) \ \s{in}  \ (x\ \ul{\ell})~.

\end{array}
\]}
\caption{Compilation of records in the pure $\lambda$-calculus}\label{compile-records}
\end{table}

\begin{exercise}\label{ex-compil-record}
  Let $M = (\lambda x.\lambda y.\set{\ell_1=x,\ell_2=y}.\ell_2)IK$.
  Complete the definition of the compilation function $\cl{C}$ of
  Table~\ref{compile-records}, compute $\cl{C}(M)$, and reduce the
  result to normal form.
\end{exercise}
  
\section{Subtyping}
Next we turn to the issue of typing the extension of 
the $\lambda$-calculus with records. 
We take as starting point the type system in
Table \ref{simple-types-tab} that assigns simple types
to $\lambda$-terms whose $\lambda$-abstractions are decorated with
types (Church style). 
We extend the syntax of types by introducing a notion of {\em record
type} which is a notation for representing a finite function from
labels to types:
\[
\begin{array}{ll}
A::= \w{tid} \Alt (A\arrow A) \Alt \set{\ell:A,\ldots,\ell:A} &\mbox{(types).} \\
\end{array}
\]
And then we add two {\em typing rules} to introduce and eliminate record types
which are presented in Table~\ref{type-record}.

\begin{table}
{\footnotesize\[
\begin{array}{ll}

\infer{\Gamma \Gives M_i:A_i\quad i=1,\ldots,n}
{\Gamma \Gives \set{\ell_1=M_1,\ldots,\ell_n=M_n}:\set{\ell_1:A_1,\ldots,\ell_n:A_n}}

&\infer{\Gamma \Gives M: \set{\ell_1:A_1,\ldots,\ell_n:A_n}}
{\Gamma \Gives M.\ell_i:A_i}~.
\end{array}
\]}

\caption{Typing rules for records}\label{type-record}
\end{table}

The extended type system still has the property that in a given type context
each $\lambda$-term has at most one type. However, consider the record types:
\[
A=\set{\ell_1:A_1,\ell_2:A_2}~,\qquad B=\set{\ell_1:A_1}~.
\]
If we have a value of type $A$ then we could use it in any
context that waits for a value of type $B$. This simple remark
pleads for the introduction of a {\em subtyping} relation $A\leq B$.
Table \ref{subtype-record-tab} describes a possible definition 
of the subtyping relation for records and functional types.
\begin{table}
{\footnotesize
\[
\begin{array}{c}

\infer{}{t\leq t} \qquad

\infer{A'\leq A\qquad B\leq B'}
{A\arrow B\leq A'\arrow B'} \\ \\

\infer{\set{\ell'_1,\ldots,\ell'_m}\subseteq \set{\ell_1,\ldots,\ell_n}\quad
A_{\ell'_{i}} \leq B_{\ell'_{i}}\quad i=1,\ldots,m}
{\set{\ell_1:A_{\ell_{1}},\ldots,\ell_n:A_{\ell_{n}}} \leq \set{\ell'_1:B_{\ell'_{1}},\ldots,\ell'_{m} :B_{\ell'_{m}}}}

\end{array}
\]}
\caption{Subtyping rules for records}\label{subtype-record-tab}
\index{subtyping rules}
\end{table}

We write $\Gives A\leq B$ if the assertion $A\leq B$ can be
derived according to the rules in Table \ref{subtype-record-tab}.
There are a couple of intriguing points in the definition of the rules.
First, notice that the rule for functional types is {\em anti-monotonic} 
in the first argument. To get an intuition, suppose we can use natural numbers 
where integers are expected: $\Nat \leq \Z$. 
Then a function $f$ of type $\Z\arrow \Nat$ can also be used whenever
a function of type $\Nat \arrow \Z$ is expected. Indeed, $f$ will
be able to handle any natural number since it is built to work on
integers and it will return an integer since it is expected to return
a natural number. On the other hand, if $g$ has type $\Nat \arrow \Nat$
then it cannot be used where a function of type $\Z\arrow \Z$ is expected
as $g$ may fail to handle a negative integer.
Second, the rules are completely {\em syntax directed}: for each pair
of types there is at most one rule that applies and in this case
there is only one way to apply it.

\begin{proposition}\label{subtyping-rel-prop}
The subtyping relation (Table \ref{subtype-record-tab})
enjoys the following properties:
\begin{enumerate}

\item It is reflexive and transitive.

\item If $\Gives A\leq B$ then there is a closed $\lambda$-term $C_{A,B}$
  (a coercion) such that $\emptyset \Gives C_{A,B}:A\arrow B$.
\end{enumerate}
\end{proposition}
\Proof  \Proofitemf{(1)}
Reflexivity follows by induction on the structure of the type $A$.
For transitivity, we build a proof of $B\leq C$ by induction on the
height of the proofs of $A\leq B$ and $B\leq C$ and case analysis
on the last rules applied. 
For instance, suppose we have:

{\footnotesize\[
\infer{B'\leq A'\qquad A''\leq B''}
{A'\arrow A'' \leq B'\arrow B''}
\qquad
\infer{C'\leq B'\qquad B''\leq C''}
{B'\arrow B''\leq C'\arrow C''}~.
\]}

Then by inductive hypothesis we can prove $C'\leq A'$ and $A''\leq C''$ and
we conclude as follows:

{\footnotesize\[
\infer{C'\leq A'\quad A''\leq C''}
{A'\arrow A''\leq C'\arrow C''}~.
\]}

\Proofitemm{(2)}
We proceed by induction on the proof of $A\leq B$. 
For the basic case, take the identity. For the functional case, take:
\[
C_{A'\arrow A'',B'\arrow B''}  = 
\lambda f:A'\arrow A''.\lambda x:B'.c_{A'',B''}(f(c_{B',A'} x))~.
\]
For the record case, assume:
\[
\begin{array}{c}
A=\set{\ell_1:A_{\ell_{1}},\ldots,\ell_n:A_{\ell_{n}}}~,  \qquad
B=\set{\ell'_1:B_{\ell'_{1}},\ldots,\ell'_{m}:B_{\ell'_{m}}}~,
\end{array}
\]
and the conditions specified in Table \ref{subtype-record-tab}
are satisfied. Then define:
\[
C_{A,B} =
\lambda x:A.\set{\ell'_1=C_{A_{\ell'_{1}},B_{\ell'_{1}}} (x.\ell'_{1}),\ldots,
\ell'_m=C_{A_{\ell'_{m}},B_{\ell'_{m}}} (x.\ell'_{m})}~.
\]
~\qed \\

Proposition \ref{subtyping-rel-prop} above guarantees that the subtyping
relation defined by the rules in Table \ref{subtype-record-tab} is indeed
a pre-order and moreover that whenever $A$ is a subtype of $B$ we can build
a well-typed $\lambda$-term of type $A\arrow B$ that gives us a canonical way to
transform a $\lambda$-term of type $A$ into a $\lambda$-term of type $B$.

Next we discuss the integration of the subtying rule to the type system
for the $\lambda$-calculus with records. One possibility would be to add
the following typing rule while leaving all the other typing rules unchanged:
\begin{equation}\label{subtyping-rule}
\infer{\Gamma \Gives M:A\qquad \Gives A\leq B}
{\Gamma \Gives M:B}~.
\end{equation}
The problem with this approach is that typing is no more directed
by the syntax of the $\lambda$-term (we had a similar problem with the
rules $(\forall_I)$ and $(\forall_E)$ in Table \ref{pred-type-tab}). 
However, one can remark that the only 
situation where types need to be matched arises in the application 
of a $\lambda$-term to another one. Hence, we integrate subtyping to the rule
for application as follows:
\begin{equation}\label{application-subtyping-rule}
\infer{\Gamma \Gives M:A\arrow B\quad \Gamma \Gives N:A'\quad \Gives A'\leq A}
{\Gamma \Gives MN:B}~.
\end{equation}
Notice that the resulting system maintains the property that each $\lambda$-term has
at most one type.
Let us write $\Gamma \Gives_{\leq} M:A$ for a judgment derivable in the
resulting type system and let us write $\Gamma \Gives_{\leq}^{s}M:A$ for a
judgment derivable in the ordinary type system extended with the subtyping rule
(\ref{subtyping-rule}). 

\begin{proposition}\label{subtype-comparison-prop}
The following properties hold:
\begin{enumerate}
\item 
If  $\Gamma \Gives_{\leq} M:A$ then 
$\Gamma \Gives_{\leq}^{s} M:A$.

\item If $\Gamma \Gives_{\leq}^{s}M:A$ then 
there is a type $B$ such that  $\Gamma \Gives_{\leq} M:B$ and $\Gives B\leq A$.
\end{enumerate}
\end{proposition}
\Proof 
\Proofitemf{(1)}
Rule (\ref{application-subtyping-rule}) can be derived from 
the rule (\ref{subtyping-rule}) and the ordinary rule to
type application.

\Proofitemm{(2)} We proceed by induction on the 
derivation of $\Gamma \Gives_{\leq}^{s}M:A$. We
consider some significant cases.

\begin{itemize}

\item Suppose we derive $\Gamma \Gives_{\leq}^{s}M:A$ from
$\Gamma \Gives_{\leq}^{s}M:A'$ and $\Gives A'\leq A$.
Then by inductive hypothesis, we can derive 
$\Gamma \Gives_\leq M:B$ and $\Gives B\leq A'$.
And by transitivity of subtyping (proposition \ref{subtyping-rel-prop}),
we conclude $\Gives B\leq A$.

\item Suppose we derive  $\Gamma \Gives_{\leq}^{s}MN: A$ from
$\Gamma \Gives_{\leq}^{s}M:A'\arrow A$ and 
$\Gamma \Gives_{\leq}^{s}N:A'$.
Then by inductive hypothesis, we can derive
$\Gamma \Gives_{\leq} M:B_1$, $\Gives B_1\leq A'\arrow A$,
$\Gamma \Gives_{\leq} N:B_2$, and $\Gives B_2 \leq A'$.
Then we must have $B_1\equiv B'_1\arrow B''_1$, 
$\Gives A'\leq B'_1$, and $\Gives B''_1\leq A$.
By transitivity, $\Gives B_2 \leq B'_1$. 
Therefore we can derive:
$\Gamma \Gives_{\leq} MN:B''_1$ and $\Gives B''_1\leq A$.

\item Suppose we derive 
$\Gamma \Gives_{\leq}^{s} \lambda x:A.M:A \arrow A'$ from
$\Gamma,x:A\Gives_{\leq}^{s} M:A'$. 
Then by inductive hypothesis, we can derive
$\Gamma,x:A\Gives_{\leq}^{s} M:B$ and $\Gives B\leq A'$.
Hence $\Gamma \Gives_{\leq}^{s} \lambda x:A.M:A\arrow B$ 
and $\Gives A\arrow B\leq A\arrow A'$. \qed 
\end{itemize}

Thus the syntax-directed system assigns to a typable $\lambda$-term 
the least type among the types assignable to the $\lambda$-term in the more liberal
system where the subtyping rule can be freely applied.
The statement of the subject reduction property in the syntax-directed 
system requires some care because the type of a $\lambda$-term may grow 
after reduction. For instance, consider the reduction:
\[
M\equiv (\lambda x:\set{\ell_1:A_1}.x)\set{\ell_1=V_1,\ell_2=V_2} \arrow 
\set{\ell_1=V_1,\ell_2=V_2} \equiv N~.
\]
Then we may have $\emptyset \Gives_{\leq} M:\set{\ell_1:A_1}$ and 
$\emptyset \Gives_{\leq} N:\set{\ell_1:A_1,\ell_2:A_2}$.

\begin{proposition}\label{subj-red-subtyping}
If $\Gamma \Gives_{\leq} M:A$ and $M\arrow N$ then for some type $B$,
$\Gamma \Gives_{\leq} N:B$ and $\Gives B\leq A$.
\end{proposition}
\Proof
As a preliminary remark, we show that if $\Gamma,x:A\Gives_{\leq} M:B$,
$\Gamma \Gives_{\leq} N:A'$, and $\Gives A'\leq A$ then 
$\Gamma \Gives_{\leq} [N/x]M:B'$ and $\Gives B'\leq B$.
The preliminary remark is applied in the analysis of a $\beta$-reduction.
Suppose $\Gamma \Gives_{\leq} (\lambda x:A.M)N:B$. Then we
must have $\Gamma,x:A \Gives_{\leq} M:B$, $\Gamma \Gives_{\leq} N:A'$, 
and $\Gives A'\leq A$. Thus $\Gamma \Gives_{\leq} [N/x]M:B'$ and 
$\Gives B'\leq B$. \qed \\

The extension of the system with subtyping still guarantees that
a well-typed program cannot go wrong. In particular, it is not
possible to select a label $\ell$ in a record where the label is 
not defined.

\begin{proposition}\label{progress-subtyping}
Suppose $\emptyset \Gives_{\leq} M:A$ then either $M$ is a value or 
$M\arrow N$.
\end{proposition}
\Proof 
By induction on the structure of $M$.
Suppose $M$ is not a value. It cannot be a variable because 
the type context is empty.

If $M\equiv M_1M_2$ then we must have $\emptyset \Gives_{\leq} M_1:A\arrow B$,
$\emptyset \Gives_{\leq} M_2:A'$ and $\Gives A'\leq A$. 
By inductive hypothesis, if $M_1$ or $M_2$ are not values then 
they reduce and so $M_1M_2$ reduces too.
On the other hand, if $M_1$ and $M_2$ are both values then
$M_1$ must be a $\lambda$-abstraction and therefore $M$ reduces.

If $M\equiv M'.\ell$ then we must have 
$\emptyset \Gives_{\leq} M':\set{\ldots \ell:A \ldots}$.
By inductive hypothesis, if $M'$ is not a value then it reduces and
so $M$ reduces too.
On the other hand, if $M'$ is a value then it must be a record
defined on the label $\ell$ and therefore $M$ reduces. \qed

\section{Variants (*)}
Variants are data structures dual to records 
just as sums are dual to products and 
the subtyping theory developed for records can 
be adapted to some extent to variants.
As for records, we start with a set of labels (cf. definition \ref{labels-def}).
Then a variant is a notation to represent an element of a finite disjoint sum
indexed over labels.

With  reference to Table~\ref{lambda-record}, the collection of
{\em $\lambda$-terms} is extended as follows:
\begin{equation}
M::=\cdots \Alt [\ell=M]_{[\ell_1,\ldots,\ell_n]} \Alt \s{case} \ M N \cdots N
\end{equation}
where we assume the labels $\ell_1,\ldots,\ell_n$ are distinct.
A variant {\em value} is a $\lambda$-term of the shape $[\ell=V]_{[\ell_1,\ldots,\ell_n]}$.
The collection of {\em evaluation contexts} of Table~\ref{lambda-record}
is extended as follows:
\begin{equation}
E::= \cdots \Alt [\ell=E]_{[\ell^*]} \Alt \s{case} \ V^* E M^* 
\end{equation}
The {\em reduction rule} for variants is:
\begin{equation}
\s{case} \ [\ell=V]_{[\ell_1,\ldots,\ell_n]} V_1\ldots V_n \arrow V_iV\qquad \mbox{if }\ell=\ell_i~.
\end{equation}
Turning to the {\em typed} version,
we denote a {\em variant type} with the notation:
\begin{equation}
[\ell_1:A_1,\ldots,\ell_n:A_n]~.
\end{equation}
The syntax of $\lambda$-terms is modified as follows so that each label is associated with a type:
\begin{equation}
M::=\cdots \Alt [\ell=M]_{[\ell_{1}:A_{1},\ldots,\ell_{n}:A_{n}]} \Alt \s{case} \ M N \cdots N
\end{equation}
The typing and subtyping rules for introducing and eliminating variants
are given in
Table~\ref{typing-variant-table}.
We notice that the subtyping rule for variants is similar to the one for records
but {\em upside down}.

\begin{table}
{\footnotesize\[
\begin{array}{c}
  \mbox{{\sc Typing rules}} \\\\
  
  \infer{
    \begin{array}{c}
      A=[\ell_1:A_1,\ldots,\ell_n:A_n]\quad \ell=\ell_i\\
      \Gamma \Gives M:A_i
  \end{array}}
{\Gamma \Gives [\ell=M]_A :A} 

\qquad\infer{\begin{array}{c}
    \Gamma \Gives M:[\ell_1:A_1,\ldots,\ell_n:A_n]\\
    \Gamma \Gives M_i:A_i\arrow B\quad i=1,\ldots,n
\end{array}}
{\Gamma \Gives \s{case} \ M M_1 \ldots M_n:B} \\\\

\mbox{{\sc Subtyping rule}}\\\\

\infer{ \set{\ell_1,\ldots,\ell_n} \subseteq \set{\ell'_1,\ldots,\ell'_m}
\quad A_{\ell_{i}} \leq B_{\ell_{i}}\quad i=1,\ldots,n}
{
  [\ell_1:A_{\ell_{1}},\ldots,\ell_n:A_{\ell_{n}}] \leq [\ell'_1:B_{\ell'_{1}},\ldots,\ell'_{m}:B_{\ell'_{m}}]} 

\end{array}
\]}
\caption{Typing and subtyping rules for variant types}\label{typing-variant-table}
\end{table}

\begin{exercise}\label{variant-exercise}
  Show that:
  (1) the extension with variants (but without subtying) preserves
  the property that each term has at most one type,
  (2) the subtyping relation extended to variant types is still
  reflexive and transitive,
  (3) if $\Gives A\leq B$ then there is a closed $\lambda$-term
  $C_{A,B}$ such that $\emptyset \Gives C_{A,B}:A\arrow B$.
\end{exercise}

\section{Summary and references}
Records and variants are a user-friendly version of products and
disjoint unions. 
The introduction of record and variant types suggests
a notion of {\em subtyping} with the following intuition:
if $A$ is a subtype of $B$ then we should be able to use a value
of type $A$ whenever a value of type $B$ is expected.
We have shown that the subtyping rule can be added to the type
system in such a way that typing is still syntax-directed and
a typable $\lambda$-term is assigned the least type with respect to the
sub-typing pre-order. The paper \cite{DBLP:journals/iandc/Cardelli88} 
is an early reference on the formalization of subtyping and its
semantics. Elaborations can be found, {\em e.g.}, in 
\cite{DBLP:journals/iandc/Mitchell88,DBLP:journals/toplas/AmadioC93}. 
The book \cite{DBLP:books/daglib/0005958} contains several
chapters dedicated to subtyping.

\chapter{References}\label{lamref-sec}
\markboth{{\it References}}{{\it References}}

In chapter \ref{intro-sec}, we have 
considered an elementary {\em imperative} programming
language whose programs can be understood as sequences
of commands acting on a global {\em state}.
In that context, the state was regarded as an abstraction 
of the notion of computer memory and was simply modeled as a (total) function 
from identifiers to  (basic) values.

In this chapter, we reconsider the notion of {\em imperative}
programming. We replace the {\em state} mentioned above with a notion
of {\em heap}. A heap can be regarded as an abstraction
of the notion of computer memory too and it is 
modeled as a (partial) function from {\em references} to 
(possibly complex) values. In turn, references can be regarded
as an abstraction of the notion of memory address. References
are first-class values. The value associated with a reference 
can be read and modified. Moreover, during the computation, 
it is possible to generate new references and associate values with them. 

We formalize a higher-order functional language with references
which is inspired by the languages of the $\ml$ family.
Technically, we introduce the reduction rules of 
a type-free, call-by-value, $\lambda$-calculus 
with {\em references} extended with operations to generate, read, 
and write references.
We then discuss a possible compilation of the $\lambda$-calculus
with `side effects' on the heap into an ordinary 
$\lambda$-calculus. The compilation turns each expression into a function
that takes a heap as an argument and returns a pair composed
of a new heap and a value.
We conclude the chapter by introducing a propositional type system 
for the $\lambda$-calculus with references which enjoys 
a subject-reduction property and by discussing some
typing anomalies which arise with references.

\section{References and heaps}
References can be regarded as an abstraction of memory addresses and
a heap as an abstraction of a computer memory.

\begin{definition}[references]\label{references}\index{references}
We denote with $R$ a countable set of references with generic
elements $r,r',\ldots$ We assume $R$ is equipped with a
function $\cl{N}:\finprt{R}\arrow R$ such that for all $X$,
finite subset of $R$, we have $\cl{N}(X)\notin X$ (so $\cl{N}(X)$ is a `fresh reference'
relatively to $X$).
\end{definition}

\begin{definition}[heap]\label{heap}\index{heap}
A heap $h$ is a partial function over the set of
references $R$ whose domain of definition is finite. 
\end{definition}

We manipulate heaps using the
standard notation for functions.
Thus if $h$ is a heap then
$\w{dom}(h)$ is its domain of
definition, $h(r)$ its image
at $r$, and $h[v/r]$ is an `updated' heap
defined as follows (this is a variant of the
state update defined in section \ref{imp-sec}):
\begin{equation}\label{heap-update}
h[v/r](r')=\left\{\begin{array}{ll}
           h(r) &\mbox{if }r \neq r' \\
           v    &\mbox{otherwise.}
\end{array}\right.
\end{equation}
Notice that we make no assumption on the nature
of the values in a heap and that in particular 
a value can be a reference. 
In Table \ref{lam-ref-tab}, we introduce an extension of 
the type-free, call-by-value
$\lambda$-calculus with a notation closely related to the
one found in the programming languages of the $\ml$ family:
$\s{ref} \ M$ allocates a new reference which is associated with
the value of $M$, $!M$ reads the value associated
with the reference resulting from the evaluation of 
$M$, and $M:=N$ writes in the reference resulting from the evaluation
of $M$ the value of $N$.
We also introduce a constant $*$ which is used as the value resulting
from the evaluation of an assignment $M:=N$.
Ordinary programs are closed $\lambda$-terms where references
do not occur. However this property is not preserved by reduction
and for this reason we include references among the $\lambda$-terms and the
values of the language.
We rely on the following standard abbreviations:
\[
\begin{array}{lll}
\Let{x}{M}{N}  &= (\lambda x.N)M \\
M;N            &= (\lambda x.N)M &\mbox{where: }x\notin \fv{N}~.
\end{array}
\]

\begin{table}
{\footnotesize
\begin{center}
{\sc Syntax}
\end{center}
\[
\begin{array}{ll}
M::=V \Alt \w{id} \Alt MM \Alt \s{ref}\ M \Alt 
!M \Alt M:=M  &\mbox{($\lambda$-terms)} \\

V::= * \Alt r \Alt \lambda \w{id}.M   &\mbox{(values)}

\end{array}
\]
\begin{center}
{\sc Call by value evaluation contexts and Reduction rules}
\end{center}
\[
\begin{array}{l}
E::=[~] \Alt EM \Alt VE \Alt \s{ref}E \Alt !E \Alt E:=M \Alt V:=E
\end{array}
\]
\[
\begin{array}{llll}
(E[(\lambda x.M)V],h) &\arrow &([V/x]M,h)  \\
(E[\s{ref}\ V],h)     &\arrow &(E[r],h[V/r])  &\mbox{if  }r=\cl{N}(\w{dom}(h)) \\
(E[!r],h)             &\arrow &(E[h(r)],h)       &\mbox{if }r\in \w{dom}(h) \\
(E[r:=V],h)           &\arrow &(E[*],h[V/r])  &\mbox{if }r\in \w{dom}(h)
\end{array}
\]}
\caption{A call-by-value $\lambda$-calculus with references}\label{lam-ref-tab}
\index{$\lambda$-calculus, with references}
\end{table}

References and heaps can be simulated
in the pure $\lambda$-calculus. As for 
records' labels (cf. chapter \ref{record-sec}), we can use
Church numerals to represent references. The
reference $\cl{N}(X)$ can be implemented by computing
the successor of the largest numeral in the finite set $X$. 
A heap can then be represented
as a list of pairs composed of a reference and a value.
Computing the domain of a heap amounts to iterate the first
projection on the list. Reading a reference $r$ in the heap means scanning the
list till a pair $(r,V)$ is found. Updating a 
reference means building a new heap where the value 
corresponding to the reference is suitably modified.
Let us assume $\lambda$-terms $\w{New}$ to create a new reference,
$\w{Ext}$ to extend a heap with a new pair, 
$\w{Read}$ to read a reference, and  $\w{Write}$ 
to write a value in the heap. 
In Table \ref{ref-lambda-tab}, we describe the compilation of the
the $\lambda$-calculus with references into  a $\lambda$-calculus
with {\em pairing}. We also use the following abbreviation for projections:
\[
\s{let}\  (x,y)=M \ \s{in} \ N \equiv
\s{let} \ z=M, \ x=\pi_1 z, \ y=\pi_2 \ z\ \s{in} \ N~.
\]
We denote with $\ul{r}$ the Church numeral which corresponds to the
reference $r$.
A $\lambda$-term $M$ of the $\lambda$-calculus with references is compiled
into a function which takes a heap $h$ as an argument and returns
a pair composed of the heap $h$ modified according to the side-effects
of $M$ and a value which corresponds to the outcome of the computation
of $M$. 
There is some similarity between records (cf. chapter \ref{record-sec})
and heaps in that a record is a finite function defined on a set
of labels and a heap is a finite function defined
on a set of references. However, references, 
unlike labels, can be generated during the computation,
are treated as first class-values, and the value associated with a
reference can be updated.

\begin{table}
{\footnotesize
\[
\begin{array}{ll}

\cl{C}(x)           &=\lambda h.(h,x) \\
\cl{C}(\lambda x.M) &=\lambda h.(h,\lambda x.\cl{C}(M)) \\
\cl{C}(r)           &=\lambda h. (h,\ul{r}) \\
\cl{C}(*)           &=\lambda h.(h,*) \\
\cl{C}(MN)          &=\lambda h.\s{let} \ (h',x)=\cl{C}(M)h, \ 
                    (h'',y)=\cl{C}(N)h' \ \s{in} \
                    (xy)h''  \\
\cl{C}(\s{ref}M)    &=\lambda h.
                    \s{let} \ (h',x) = \cl{C}(M)h, \ r=(\w{New} \  h') \  \s{in} \
                    (\w{Ext} \ h'rx,r) \\

\cl{C}(!M)          &=\lambda h.\s{let}\ (h',r)=\cl{C}(M)h\  \s{in} \ (h',\w{Read} \ h'r) \\
\cl{C}(M:=N)        &=\lambda h.\s{let} \ (h',r)=\cl{C}(M)h, \ (h'',x)=\cl{C}(N)h' \ \s{in} 
                    \ (\w{Write} \ h''rx, *)
                           
\end{array}
\]}
\caption{Simulating the heap in a functional language}\label{ref-lambda-tab}
\index{heap simulation}
\end{table}

\begin{exercise}\label{ex-compil-lamref-lam}
  Consider the $\lambda$-term:
  \[
  M\equiv \s{let}\ x=\s{ref}(\lambda y.y) \ \s{ in }\ x:=\lambda y.(!x)y ; (!x)*~.
  \]
  Compute $\cl{C}(M)$ and analyse its reduction.
\end{exercise}

\section{Typing references}
We consider the problem of extending the propositional type system
discussed in chapter \ref{simple-type-sec}, Table \ref{simple-types-tab}, 
to the $\lambda$-calculus with references.
To type the value $*$ we introduce a basic type $1$ whose
only value is $*$.
Moreover, we introduce a new type constructor $\s{Ref}$.
A value of type $\s{Ref} \ A$ is a reference which can contain 
values of type $A$. 
In order to type a $\lambda$-term we have to make hypotheses on the type
of its free variables and of the references that occur in it.
Consequently, we introduce a notion of {\em heap context} $\Sigma$ of the
shape $r_1:A_1,\ldots,r_n:A_n$. If $r:A\in \Sigma$ then
the reference $r$ is associated with values of type $A$.
Table \ref{typing-lambda-ref-tab} gives the type system for $\lambda$-terms.

\begin{table}
{\footnotesize
\[
\begin{array}{cc}
\infer{x:A\in \Gamma}
{\Gamma;\Sigma \Gives x:A}

&\infer{r:A\in \Sigma}
{\Gamma;\Sigma \Gives r: \Refe A} \\ \\

\infer{\Gamma,x:A;\Sigma \Gives M:B}
{\Gamma;\Sigma \Gives \lambda x:A.M:A\arrow B}

&\infer{\Gamma;\Sigma \Gives M:A\arrow B\quad \Gamma;\Sigma\Gives N:A}
{\Gamma;\Sigma\Gives MN:B} \\ \\

\infer{~}
{\Gamma;\Sigma\Gives *:1}

&\infer{\Gamma;\Sigma \Gives M:A}
{\Gamma;\Sigma \Gives \refe M : \Refe A}  \\ \\

\infer{\Gamma;\Sigma\Gives M:\Refe A}
{\Gamma;\Sigma\Gives !M:A}

&\infer{\Gamma;\Sigma\Gives M:\Refe A\quad \Gamma;\Sigma \Gives N:A}
{\Gamma;\Sigma \Gives M:=N:1}

\end{array}
\]}
\caption{Typing rules for the $\lambda$-calculus with references}\label{typing-lambda-ref-tab}
\index{$\lambda$-calculus with references, typing}
\end{table}

Besides $\lambda$-terms we need to type heaps too. Consider the 
$\lambda$-term without references:
\begin{equation}\label{ref-rec-ex}
\Let{x}{\refe (\lambda z:A.z)}{\Let{y}{\refe (\lambda z:A.!x z)}{x:=\lambda z:A.!y z}}~.
\end{equation}
By reducing it, we can produce the following heap:
$h_0=[\lambda x:A.(!r_2)x/r_1,\lambda x:A.(!r_1)x/r_2]$. 
Notice that the values associated with $r_1$ and $r_2$ 
depend on $r_2$ and $r_1$ respectively. Thus to type a heap
we have to find a heap context which assigns a type to
all the references of the heap which is coherent with 
the type of the values associated with the references.
Also we require that all the references in the values of the
heap belong to the domain of definition of the heap.
This leads to the following rule for typing a heap with
respect to a heap context:

{\footnotesize\begin{equation}\label{type-heap}
\infer{\w{dom}(\Sigma)=\w{dom}(h) \qquad \emptyset;\Sigma\Gives h(r):A \quad \mbox{ for all }r\in \w{dom}(\Sigma), r:A\in \Sigma}
{\Sigma\Gives h}~.
\end{equation}}

We write $\Gamma;\Sigma \Gives (M,h):A\ $ if 
$\ \Gamma;\Sigma \Gives M: A$ and $\Sigma \Gives h$.

\begin{example}
The heap $h_0$ produced by the $\lambda$-term (\ref{ref-rec-ex}) above
can be typed in the heap context:
$\Sigma = r_1:A\arrow A,r_2:A\arrow A$.
\end{example}

\begin{proposition}\label{prop-contr-sub-lamref}
The typing system enjoys the following properties:
\begin{enumerate}

\item If $\Gamma,x:A;\Sigma \Gives M:B$ is derivable and
$x\notin \fv{M}$ then 
$\Gamma;\Sigma\Gives M:B$ is derivable.

\item If
$\Gamma,x:A;\Sigma \Gives M:B$ and $\Gamma;\Sigma\Gives V:A$
are derivable then $\Gamma;\Sigma \Gives [V/x]M:B$ is derivable.
\end{enumerate}
\end{proposition}

\begin{exercise}\label{proof-prop-contr-sub-lamref}
  Prove by induction on the proof height proposition
  \ref{prop-contr-sub-lamref}.
\end{exercise}

We now discuss the way typing is preserved by reduction.
Notice that during reduction the domain of definition of the
heap can grow since the operator \s{ref} may dynamically generate
new references. Hence we also need to extend the heap context.
We write $\Sigma'\supseteq \Sigma$ if 
$\Sigma'$ is an extension of $\Sigma$. We notice
the following weakening property of the heap context.

\begin{proposition} \label{weakening-lambdaref-prop}
If  $\Gamma;\Sigma\Gives M:A$ is derivable and 
$\Sigma'\supseteq \Sigma$ then $\Gamma;\Sigma'\Gives M:A$
is derivable.
\end{proposition}

\begin{exercise}\label{proof-weakening-lambdaref-prop}
  Prove proposition \ref{weakening-lambdaref-prop}.
\end{exercise}

Then we can state the following subject reduction property.

\begin{proposition}\label{subjred-lamref-prop}
If $\Gamma;\Sigma \Gives (M,h):A$  and $(M,h) \arrow (M',h')$
then there is $\Sigma'\supseteq \Sigma$ such that
$\Gamma;\Sigma'\Gives  (M',h'):A$.
\end{proposition}

\begin{exercise}\label{proof-subjred-lamref-prop}
  Prove proposition \ref{subjred-lamref-prop}.
  \end{exercise}

\begin{exercise}\label{type-heap-compil-ex}
Suppose we have `abstract types' $R$ and $H$ and that we can assign
the following types to the heap-manipulating functions, where $A$
can be any type:
\[
\begin{array}{ll}
\w{New}: H\arrow R~,
&\w{Ext}:H\arrow R \arrow A \arrow H~, \\
\w{Read}: H\arrow R \arrow A~,
&\w{Write}: H\arrow R \arrow A \arrow H~.
\end{array}
\]
For every propositional type $A$ and type context $\Gamma$ 
define a type translation $\ul{A}$ and context translation $\ul{\Gamma}$,
and show that the compilation function in Table~\ref{ref-lambda-tab} 
is type preserving in the sense that if $\Gamma;\emptyset \Gives M:A$ according
to the rules in Table~\ref{typing-lambda-ref-tab} then
$\ul{\Gamma} \Gives \cl{C}(M): H\arrow H\times \ul{A}$.
On the other hand, find a closed $\lambda$-term with references $M$ such that
$\cl{C}(M)$ is typable but $M$ is not.
\end{exercise}

\section{Typing anomalies (*)}
As suggested by the $\lambda$-term (\ref{ref-rec-ex}) above,
simply typed $\lambda$-terms with references can produce 
circular heaps. In fact it is possible to use
references to define general recursive functions.
First, let us consider a {\em minimal} example of typable and
{\em looping} computation. Set:
\[
\begin{array}{llll}
M_1 &\equiv \refe (\lambda x:1.x)~,\qquad 
&M_2 &\equiv \Let{y}{M_1}{y:=(\lambda x:1.(!y)x)  \ ; \ (!y) }~.
\end{array}
\]
Then $\Gives M_1:\Refe (1\arrow 1)$ and $\Gives M_2:1\arrow 1$ and
there is an infinite reduction starting with $M_2 *$.
We can generalize this idea to define a function $f$ of type
$A\arrow B$ which satisfies a recursive equation 
$f=\lambda x:A.M$ where $M$ may depend on $f$.
Let $\lambda x:A.N$ be any $\lambda$-term of type $A\arrow B$.
Then we set:
\[
\begin{array}{llll}
M_1 &\equiv \refe (\lambda x:A.N)~,\qquad
&M_2 &\equiv \Let{y}{M_1}{y:=(\lambda x:A.[!y/f]M)  \ ; \ (!y) }~.
\end{array}
\]
Initially, $y$ is a reference containing a fake function. Then
we replace the fake function with the real function where
each call to $f$ is replaced by $!y$. 
Then $y$ is a reference which contains a value which refers
to the reference $y$. This circularity allows to simulate
recursion. 

Another curious phenomenon arises when we try to mix 
references and subtyping. Namely, from $A\leq B$ we {\em cannot}
infer $\Refe A \leq \Refe B$ (or $\Refe A \leq \Refe B$). 
The $\Refe$ type constructor is neither monotonic nor anti-monotonic
with respect to the subtyping pre-order.
In practice, this means that no proper subtyping is possible on reference types.
To see this, suppose $A\leq B$ where for instance:
\[
\begin{array}{lll}
A=\set{\ell_1:C,\ell_2:C} &\leq  &\set{\ell_1:C}=B~.
\end{array}
\]
Assume $\Refe$ is anti-monotonic and $x:\Refe B$ then we should
also have $x:\Refe A$ and $(!x.\ell_2)$ will produce an error.
On the other hand, if $\Refe$ is monotonic and $x:\Refe A$ then
we should also have $x:\Refe B$ and 
$x:=\set{\ell_1=V} ; \ !x.\ell_2$ will produce an error.
 
As a third and final typing anomaly, let us notice that  the 
{\em polymorphic generalization} (cf. chapter \ref{type-inf-sec}) 
of a {\em reference} may also lead to {\em errors}.
For instance, consider:
\begin{equation}\label{poly-ref1}
\s{let} \  x = \refe (\lambda x.x) \ \s{in} \  
x:=(\lambda x.x+1); (!x) \ \s{true}~.
\end{equation}
$\ml$-like languages avoid these problems by allowing polymorphic 
generalization only on {\em values}. For instance, the programming language
{\it ocaml}  {\em accepts}:
\begin{equation}\label{poly-ref2}
\s{let} \ x = (\lambda x.x) \ \s{in}  \ x *; x \ \s{true} ~,
\end{equation}
but {\em rejects} the dangerous expression (\ref{poly-ref1}) above 
as well as the following innocuous one:
\begin{equation}\label{poly-ref3}
\s{let} \  x = (\lambda y. (\lambda x. x))\s{2} \ \s{in} \  x * ; x\  \s{true}~.
\end{equation}
In practice, most programs seem to meet this restriction.

\section{Summary and references}
Heaps can be regarded as an abstraction of computer memory.
We have considered an extension of the $\lambda$-calculus with
operations to extend, read, and modify the heap.
Expressions in this extended $\lambda$-calculus may have side
effects and can be understood as functions that
take a heap and produce a new heap and a value.

References introduce the possibility to define recursive data
structures and functions. This power comes at a price in that
the ideas developed in the purely functional setting cannot
be readily lifted to the $\lambda$-calculus with side effects.
For instance, termination of typable programs fails, no proper subtyping is possible 
on reference types, and polymorphic generalization is unsound
(but in practice it can be fixed \cite{DBLP:journals/lisp/Wright95}).

One may argue that these failures are due to the fact that the usual
type systems neglect side effects completely.  To address this issue,
so called {\em type and effect} systems
\cite{DBLP:conf/popl/LucassenG88} have been proposed. In these type
systems, references are abstracted into a finite set of {\em regions}
and types become {\em dependent} on such regions.  In particular, an
expression is now expected to produce both an effect and a value (this
is an {\em abstraction} of the idea mentioned above where an
expression with side effects is expected to produce a heap and a
value).  Type and effect system have been applied to the design of
static mechanisms for safe memory deallocation
\cite{DBLP:journals/iandc/TofteT97}.  
It has also been shown that a {\em stratified} version of
the system can guarantee the strong normalization of the typable 
$\lambda$-terms
\cite{DBLP:journals/iandc/Boudol10,DBLP:conf/aplas/Amadio09}.

\chapter{Objects}\label{obj-sec}
\markboth{{\it Objects}}{{\it Objects}}

The programming paradigms discussed so far are built on the notion
of {\em function}. Indeed term rewriting and the $\lambda$-calculus 
can be regarded as formalisms to define first-order and
higher-order functions, respectively and
imperative programs can also be regarded as functions operating
over the heap. 
In this chapter, we discuss the situation for object-oriented
programs. We start with a minimalist object-oriented 
language which is type-free and without side-effects. 
We then gradually enrich this language
with {\em side-effects} and {\em types} to obtain a language which 
corresponds to a (tiny) fragment of the  $\java$ programming language
(of which the reader is supposed to have a superficial knowledge).
We refer to this language as untyped/typed $\sJ$.
Along the way, we discuss the compilation of untyped $\sJ$
to an extension of the $\lambda$-calculus with records, recursion, 
and, possibly, references.
Thus objects can also be understood as functions. However,
typed $\sJ$ differs from 
the typed $\lambda$-calculi we have been considering  in that it requires some degree
of type-checking at run time, {\em i.e.}, type errors at run-time
are possible.

\section{An object-oriented language}
In first approximation, an object is a {\em record} 
(cf. chapter \ref{record-sec}) whose labels 
are traditionally partitioned into {\em fields}
and {\em methods}. Usually, fields are mapped to (basic) values
describing the internal state of the object while methods are 
mapped to functions that allow to manipulate this state.
As in records, the `dot-notation' is used to access 
fields and methods, {\em e.g.}, if $o$ is an object and $f$
a field then $o.f$ is the value associated with the field $f$
in the object $o$.

In object-oriented languages such as $\java$, the creation
of objects follows certain {\em patterns} known as {\em class declarations}. 
So objects are {\em classified} according to the class declaration that
is used at the moment of their creation. Class declarations are designed
so that fields and methods are suitably initialized when the object
is created.
Unlike in the $\lambda$-calculus with {\em records} 
of chapter \ref{record-sec}, {\em recursion} is built into 
object-oriented languages. First, class declarations may be mutually
recursive, and second there is a special variable \s{this} (\s{self} is also
used sometimes) which allows to refer to the object itself within, say,
the body of one of its methods. For instance,
an object $o$ may consist of a field \s{val} which is mapped to an integer
and a method \s{inc} which is mapped to the function:
\[
\lambda x.\s{this}.\s{val}=\s{this}.\s{val}+x~.
\]
Then the effect of invoking $o.\s{inc}$ on a value $v$ is that of increasing by $v$ the
value contained in the \s{val} field of the object $o$.

\subsubsection*{Class declarations}
We reserve $C,C',D,\ldots$ for class names. Each class name corresponds
to a distinct class declaration. Usually, class declarations are built
incrementally. At the very beginning, there is a class 
\s{Object} without fields and methods. Then whenever we introduce
a new class declaration we say that it {\em extends} another class declaration.
For instance, one can declare a class
\w{C} which extends the class  \w{D} and includes 
a field \w{f} and a method \w{m} as follows:

{\footnotesize\[
\begin{array}{ll}

\s{class \ C \ \s{extends}\  D  = \{}  &\mbox{(class declaration)}\\ 
\s{\cdots C' \  f \  \cdots }       &\mbox{(field declarations)} \\ 
\s{\cdots D' \ m \ (D_1\  x_1,\ldots,D_n \ x_n)\{\ e \ \} \cdots\}}  &\mbox{(method declarations).} 
\end{array}
\]}

We are using here a notation based on $\java$ where we specify the
class $\s{C'}$ of the object in the field \s{f} as well as the classes
$\s{D_1},\ldots,\s{D_n}$ of the objects $\s{x_1},\ldots,\s{x_n}$
the (function associated with the) 
method \s{m} is expecting as input and the class $\s{D'}$ of the object 
it returns as a result.

\subsubsection*{Expressions}
The body of a method is an {\em expression} denoted with $e,e',\ldots$ whose 
syntax is defined in Table~\ref{tab-java-exp}.
We have split the expressions in $3$ groups. The first group
is composed of (object) variables, (object) values (to be defined next), 
an operator \s{new} to generate an object of the class \s{C} 
while initializing its fields
with the values of the expressions $e_1,\ldots,e_n$, and the selection
operator for fields and methods.
As already mentioned, among the variables, we reserve the variable \s{this} 
to refer to the object on which a method is invoked.
The second group is optional and 
corresponds to an {\em imperative extension} of the basic
language where fields are modifiable, and therefore the sequentialization
of side effects is relevant.
The third group is also optional and consists of a casting operator. 
This operator is only relevant if we are interested 
in a {\em type system} for the language.
We anticipate that the role of such a type system is {\em not} to avoid
errors (cf. exercise \ref{progress-ex}) but to {\em localize} them 
in certain points of the computation.
\begin{table}
{\footnotesize  \[
\begin{array}{lllll}
e &::= &\w{id} &\Alt &\mbox{(variable)} \\
      &&v &\Alt &\mbox{(value)} \\
      && {\sf new} \ C(e_1,\ldots,e_n) &\Alt &\mbox{(object generation)}  \\
      && e.f &\Alt                 &\mbox{(field read)} \\
      &&e.m(e_1,\ldots,e_n)  &\Alt &\mbox{(method invocation)} \\\hline

      && e.f:=e          &\Alt    &\mbox{(field write)} \\
      && e;e              &\Alt        &\mbox{(sequentialization)} \\\hline

      && (C)(e)  &\Alt    &\mbox{(casting)} 
\end{array}
\]}
\caption{Syntax of $\sJ$ expressions}\label{tab-java-exp}
\end{table}

\subsubsection*{Values}
The definition of a value expression depends on whether we are considering
the imperative extension or not. In the imperative extension, we 
assume all fields are modifiable. To model field assignment we 
proceed as in chapter \ref{lamref-sec}. Namely, we assume a countably infinite
set of {\em references} $R$ with elements 
$r,r,\ldots$ and define a {\em heap} $h$ as a finite domain 
partial function mapping references to values. 
In this case, a {\em value} $v$ has the shape:
\begin{equation}\label{value-obj-imp}
v::=C(r_1,\ldots,r_n)\qquad n\geq 0\qquad\mbox{(values, imperative case),}
\end{equation}
where $C$ is a class name (the class of the object)
and $r_1,\ldots,r_n\in R$ are references corresponding to 
the modifiable fields of the object.

In the non-imperative, say, {\em functional}, case, 
fields are initialized when the object
is created and they are never modified. Then we can just regard
values as the closed first-order terms built over the signature of class names
where the arity of a class names is the number of the
class fields:
\begin{equation}\label{valu-obj-fun}
v::=C(v,\ldots,v)\qquad\mbox{(values, functional case).}
\end{equation}
We pause to remark that to define the  reduction rules
of the language, it is convenient 
to include values in the syntactic category of expressions,
however values {\em never} appear in a source program. 
Incidentally, in chapter \ref{lamref-sec}, we took a similar 
approach by considering references as values.

\subsubsection*{Well-formed programs}
A {\em program} is composed of a list of class declarations and a
distinguished expression where the computation starts (in $\java$
this distinguished expression would be the body of a
\s{main} method). The final value of the distinguished expression
can be taken as the output of the program. 
As for the input, we shall assume for simplicity that 
it is coded as part of the distinguished expression. 

As mentioned above, each class declaration extends another class
declaration. This induces a binary relation on class names.
We denote with $\leq$ the reflexive and transitive closure of this 
relation and we assume that if $C\leq D$ and $D\leq C$ then $C=D$.
Under this hypothesis, we can represent the subtyping relation as an
{\em inheritance tree} having as root the \s{Object} class.

The feature of declaring a class by extending another one 
makes programs more compact but requires some verification.
A {\em well-formed program} must satisfy certain conditions 
concerning fields and methods.

\begin{enumerate}

\item If $C\leq D$ then $C$ inherits all the fields of  $D$.
It is required that there are no name conflicts among the fields.
Thus, by crossing the inheritance tree towards the root one must {\em not}
find two fields with the same name.

\item Also, if $C\leq D$ then $C$ inherits all the methods of 
$D$. However, in this case $C$ may {\em redefine} (in the object-oriented
jargon one says {\em override}) a method. 
A constraint that only concerns the typed version of
the language requires that the type of the method does {\em not} change.

\end{enumerate}

It is convenient to introduce a certain number of functions that 
will be used in formulating the reduction rules and the
typing rules.

\begin{itemize}

\item $\w{field}(C)$ returns the list $f_1:C_1,\ldots,f_n:C_n$ 
of the fields accessible by an object of the class $C$ along with their
expected classes.  Upon generation, an object of the class $C$ must receive
$n$ arguments so as to initialize its fields. To avoid ambiguities,
we assume an enumeration of the field names and suppose the 
function \w{field} returns the fields in growing order.
In $\java$, the initialization of the fields is made explicit 
by defining a constructor  method in the class. 

\item $\w{mbody}(m,C)$ returns the function that corresponds
to the method $m$ in the class $C$. For instance,
if $\w{mbody}(m,C)=\lambda x_1,\ldots,x_n.e$ 
then $x_1,\ldots,x_n$ are the formal parameters and
$e$ is the expression associated with the method, respectively.

\item In the {\em typed} version of the language, it will also be useful
to have a function $\w{mtype}$ such that $\w{mtype}(m,C)$ 
returns the type of the method $m$ of the class $C$ and
a predicate $\w{override}$ such that 
$\w{override}(m,D,\vc{C}\arrow C)$ holds
if and only if  $\w{mtype}(m,D)$ is defined and it coincides with 
$\vc{C}\arrow C$. 

\end{itemize}

\begin{example}\label{exemple-dcl-class}
In Table \ref{classes-cjava}, 
we consider a list of class declarations which allows
to represent boolean values and natural numbers in unary notation.
The examples are written in the slightly more verbose notation of the $\java$
programming language. As already mentioned, $\java$
requires a constructor method to build an object in a class with
fields. Moreover, $\java$ distinguishes between private and public
declarations while in $\sJ$ all declarations are public. 
These are really minor syntactic differences and 
therefore the typed version of the $\sJ$ language 
can be regarded as a subset of $\java$. 
Notice that the proposed representation of the conditional via the 
method \s{ite} is {\em strict} (both branches are evaluated); 
a more realistic fragment of 
$\java$ would include a  {\em non-strict} conditional.
\end{example}

\begin{table}
{\footnotesize
\begin{verbatim}
class Bool extends Object {
    public Object ite (Object x, Object y){return new Object();} }
class True extends Bool{
    public Object ite (Object x, Object y){return x;} }
class False extends Bool{
    public Object ite (Object x, Object y){return y;} }
class Num extends Object {
    public Bool iszero (){return new Bool();}
    public Num pred(){return new Num();} 
    public Num succ(){return new Num();} }
class NotZero extends Num{
    public Num pd;
    public NotZero(Num x){pd=x;} 
    public Bool iszero(){return new False();}
    public Num pred(){return this.pd;} 
    public Num succ(){return new NotZero(this);} }
class Zero extends Num{
    public Bool iszero(){return new True();}
    public Num pred(){return new Zero();} 
    public Num succ(){return new NotZero(this);} }
\end{verbatim} }
\caption{Some class declarations in $\sJ$ (with $\java$ syntax).}\label{classes-cjava}
\end{table}

\begin{exercise}[programming]\label{java-prog-bool-nat-ex}
With reference to the code in Table~\ref{classes-cjava}:

\begin{enumerate}

\item Enrich the classes for the booleans and natural numbers with 
a printing method which prints (a representation of) the object 
on the standard output using $\java$'s printing functions.

\item Enrich the classes for natural numbers with an \s{isequal} method
that takes a number object and checks whether it is equal to the 
one on which the method is invoked.

\item Define classes to represent lists of pairs of natural numbers
$(n_1,m_1)\cdots (n_k,m_k)$, where $n_1,\ldots,n_k$ are all distinct,
along with methods to:
(1) given $n$, read the number $m$ associated with it,
(2) given $n$, replace the number associated with it with $m$,
(3) extend the list with a new pair $(n,m)$, 
(4) given $n$, remove from the list the pair $(n,m)$, 
(5) print (a representation of) the list on the standard output.
\end{enumerate}

\end{exercise}

\subsubsection*{Reduction rules}
Table \ref{op-sem-sj} introduces 
the syntactic category of {\em evaluation contexts} 
which correspond to a {\em call-by-value, left to right} reduction
strategy and the related reduction rules which are based on judgments
of the shape:
\begin{equation}
\begin{array}{ll}
(e,h) \arrow (e',h')   &\mbox{(reduction judgment, imperative).}
\end{array}
\end{equation}

At the beginning of the computation we assume
that the heap $h$ is empty.
Then the reduction rules 
maintain the following invariant:
for all reachable configurations $(e,h)$, all the references in $e$
and all the references that appear in a value in the codomain of the
heap $h$ are in the domain of definition of the heap ($\w{dom}(h)$).
This guarantees that whenever we look for a fresh reference it is enough
to pick a reference 
which is not in the domain of definition of the current heap.
Notice that upon invocation of a method on an object, the
object replaces the reserved variable \s{this} in the body of
the method. Also, the reduction rule for casting consists of a form
of run-time type-check: the computation of a casted object
$(D)(C(\vc{r}))$ may proceed only if $C\leq D$.

\begin{table}
{\footnotesize
\begin{center}
{\sc Call-by-value evaluation contexts}
\end{center}
\[
\begin{array}{lll}
E::= &[~] \Alt \s{new} \ C(v^*,E,e^*) \Alt E.f \Alt E.m(e^*) \Alt 
v.m(v^*,E,e^*) \Alt \\
& (C)(E) \Alt E.f:=e \Alt v.f:=E \Alt 
E;e  
\end{array}
\]
\begin{center}
{\sc Reduction rules}
\end{center}
\[
\begin{array}{cr}

\infer{r^* \mbox{ distinct and }\set{r^*}\inter \w{dom}(h)=\emptyset}
{(E[\s{new} \ C(v^*)],h) \arrow (E[C(r^*)],h[v^*/r^*])} 
&\mbox{(object generation)}\\ \\

\infer{\w{field}(C)= f_1:C_1,\ldots,f_n:C_n\quad 1\leq i \leq n}
{(E[C(r_1,\ldots,r_n).f_i],h) \arrow (E[h(r_i)],h)} 
&\mbox{(field read)}\\ \\ 

\infer{\w{mbody}(m,C)=\lambda x_1,\ldots,x_n.e}
{(E[C(r^*).m(v_1,\ldots,v_n)],h) \arrow (E[[v_1/x_1,\ldots,v_n/x_n,C(r^*)/\s{this}]e],h)}
 &\mbox{(method invocation)} \\ \\

\infer{\w{field}(C)= f_1:C_1,\ldots,f_n:C_n\quad 1\leq i \leq n}
{(E[C(r_1,\ldots,r_n).f_i:=v],h) \arrow
(E[\s{Object}()],h[v/r_i])} 
 &\mbox{(field write)}\\ \\ 

\infer{}
{(E[v;e],h) \arrow (E[e],h)}
 &\mbox{(sequentialization)}  \\ \\

\infer{C\leq D}
{(E[(D)(C(r^*))] ,h) \arrow (E[C(r^*)],h)} 
&\mbox{(casting)} 
\end{array}
\]}
\caption{Evaluation contexts and reduction rules for $\sJ$}\label{op-sem-sj}
\index{$\sJ$ reduction rules}
\end{table}

The specification of the functional fragment of $\sJ$ 
where fields are immutable can be substantially simplified.
Values are now the closed first-order terms built
over the signature of class names (cf. grammar (\ref{valu-obj-fun})).
The evaluation contexts and the reduction rules 
for assignment and sequentialization can be
dropped. The remaining rules are based on a judgment of the 
shape $e\arrow e'$ (we drop the heap) and are specified
in Table \ref{java-semantics-functional}.

\begin{table}
{\footnotesize\[
\begin{array}{cr}

\infer{\w{field}(C)= f_1:C_1,\ldots,f_n:C_n}
{E[\s{new} \ C(v_1,\ldots,v_n)] \arrow E[C(v_1,\ldots,v_n)]} 
&\mbox{(object generation)}\\ \\

\infer{\w{field}(C)= f_1:C_1,\ldots,f_n:C_n\quad 1\leq i \leq n}
{E[C(v_1,\ldots,v_n).f_i] \arrow E[v_i]} 
&\mbox{(field read)}\\ \\ 

\infer{\w{mbody}(m,C)=\lambda x_1,\ldots,x_n.e}
{E[C(v^*).m(v_1,\ldots,v_n)] \arrow E[[v_1/x_1,\ldots,v_n/x_n,C(v^*)/\s{this}]e]}
 &\mbox{(method invocation)} \\ \\

\infer{C\leq D}
{E[(D)C(v^*)] \arrow E[C(v^*)]} 
&\mbox{(casting)} \\ \\ 

\end{array}
\]}
\caption{Simplified reduction rules for the functional fragment of $\sJ$}\label{java-semantics-functional}
\end{table}

\section{Objects as records}
We define an encoding of the functional, type free
object-oriented language into a call-by-value $\lambda$-calculus extended with
records and a fixed point combinator $Y$
(in turn, records and the fixed point combinator could be 
encoded in the $\lambda$-calculus).
As a first step, we assume each class declaration
is completely expanded so that we can associate with each class name
the list of its fields and its methods with the related bodies.
So we have a system of class declarations of the shape (class names are omitted
when irrelevant):
\begin{equation}\label{class-dcl-ex}
\s{class} \ C \ \set{f_1,\ldots,f_h,m_1=\lambda x_1^*.e_1,\ldots,m_k=\lambda x_k^*.e_k}~.
\end{equation}
The methods' bodies $e_i$ may generate objects of other classes
and may refer to the object itself via the variable \s{this}.
This entails that class generators are mutually recursive and 
the variable \s{this} is defined via a fixed point combinator.

Following this intuition, we define a compilation function $\cl{C}$.
We suppose the class names are enumerated as $C_1,\ldots,C_m$ and
we reserve a fresh variable $c$ and the labels $\s{1},\ldots,\s{m}$.
The variable $c$ will be defined recursively as a record with
labels $\s{1},\ldots,\s{m}$ such that the function associated
with the label $\s{i}$ is the generator for the objects
of the class $C_i$.
On expressions (which are not values or casted objects), 
the compilation function is
simply defined as follows:
\[
\begin{array}{ll}
\cl{C}(x) &= x \\

\cl{C}(\s{new} \ C_i(e_1,\ldots,e_n)) &= 
(c.\s{i}) \  \cl{C}(e_1) \cdots \cl{C}(e_n) \quad\mbox{($c$ fresh variable)}\\

\cl{C}(e.f) &= \cl{C}(e).f \\

\cl{C}(e.m(e_1,\ldots,e_n)) &= (\cl{C}(e).m) \ \cl{C}(e_1)\cdots\cl{C}(e_n)~.
\end{array}
\]
For each declaration of a class $C$ of the shape (\ref{class-dcl-ex}),
we define the $\lambda$-term 
$N_C$
where  $y_1,\ldots,y_h$ are  fresh variables:
\begin{equation}\label{class-compilation}
\begin{array}{ll}
R_C &\equiv \set{f_1=y_1,\ldots,f_h=y_h,m_1=\lambda x_1^*.\cl{C}(e_1),\ldots,m_k=\lambda x_k^*.\cl{C}(e_k)} \\
N_C &\equiv \lambda y_1,\ldots,y_h. Y(\lambda \s{this}.R_C)~.
\end{array}
\end{equation}
Intuitively, $N_C$ is the generator for objects of the class $C$. 
The system of class declarations is reduced to one fixed point equation:
\begin{equation}
C\equiv Y(\lambda c.\set{\s{1}= N_{C_{1}},\ldots,\s{m}= N_{C_{m}}})~.
\end{equation}
Finally, a program composed of $m$ class declarations 
$C_1,\ldots,C_m$ and an expression $e$ is compiled into the $\lambda$-term:
\begin{equation}
\s{let} \ c= C \ \s{in} \ \cl{C}(e)~.
\end{equation}
As a concrete example, suppose the program $P$ is composed 
of $2$ class declarations $C_i$ each
with a field $f_i$ and a method $m_i$ with body $\lambda x_i.e_i$, $i=1,2$, 
and a main expression $e$. Then we have:
\[
\begin{array}{ll}
N_{C_{1}} &\equiv \lambda y_1.Y(\lambda \s{this}. \set{f_1=y_1,m_1=\lambda x_1.\cl{C}(e_1)}) \\
N_{C_{2}} &\equiv \lambda y_2.Y(\lambda \s{this}. \set{f_2=y_2,m_2=\lambda x_2.\cl{C}(e_2)}) \\
C       &\equiv Y(\lambda c.\set{\s{1}= N_{C_{1}},\s{2}= N_{C_{2}}}) \\
\cl{C}(P) &\equiv \s{let} \ c= C \ \s{in} \ \cl{C}(e)~.
\end{array}
\]

\begin{exercise}\label{imp-obj-fun-ex}
Extend this encoding to the language with 
{\em mutable} fields. In this case, it is convenient to take
as target language a call-by-value $\lambda$-calculus with records and
{\em references} (cf. chapter \ref{lamref-sec}).
\end{exercise}

\section{Typing objects (*)}
We design a type system for the 
full object-oriented language we have introduced.
To this end, we assume a type context $\Gamma$ 
has the shape $x_1:C_1,\ldots,x_n:C_n$
and consider typing judgments of the shape: $\Gamma \Gives e:C$.
A general goal of a type system for an object-oriented language
is to guarantee that every invocation of a field or a method
on an object is compatible with the class to which the object belongs.
Let us notice however that an incorrect application of the
casting ({\em downcasting}) may compromise this property.
For instance, we could write the expression:
\[
\s{(Bool)((new \ Object()).ite(new \ True(), new \ False()) ) }~.
\]
which is stuck since an \s{Object} has no \s{ite} method.
To avoid this situation, we could consider the following rule:
\[
\infer{\Gamma \Gives e:D \quad D\leq C}
{\Gamma \Gives (C)(e) : \ C} ~.
\]
In this rule, we can cast an object of the class $D$ as an object
of the class $C$ only if the class $D$ extends the class $C$.
This is in agreement with the intuition that objects are records
and that an object of the class $D$ can handle all the invocations
addressed to an object of the class $C$ (cf. subtyping rules for records in 
chapter \ref{record-sec}).
However this rule is {\em too} constraining. For instance, it does not allow
the typing of the expression:
\[
\s{(Bool)((new \ True()).ite(new \ True(),  new \ False()))}~,
\] 
as the result of the method  \s{ite} belongs to the class \s{Object} and
$\s{Object} \not\leq \s{Bool}$.
Then,  in $\java$, the rule for casting can be formulated as follows:
\[
\infer{\Gamma \Gives e:D\qquad (C\leq D \mbox{ or }D\leq C)}
{\Gamma \Gives (C)(e):C}~.
\]
In other terms, the casting is {\em forbidden} 
if $C$ and $D$ are {\em incomparable}.
However, this property is not preserved by reduction!
Let  $C, D$ be two incomparable classes and let $e$ be
an expression of type $C$. Then the expression
$\s{(D)((Object)(e))}$ is well typed,
but it reduces to the expression $\s{(D)(e)}$ which is not.
By climbing and descending the inheritance tree we can connect 
incomparable classes. Table \ref{type-java-example} gives a concrete example of this phenomenon in  $\java$.

\begin{table}
{\footnotesize
\begin{verbatim}
class C extends Object{
    public void m(){return;}}
class D extends Object{
    public void m(){return;}}
class Main{
    public static void main (String[] args){
        D d = new D();
        C c = new C();
        ((C)((Object)(d))).m(); //this types, but rises an exception at run time.
        ((C)(d)).m();           //this does not type, but it is a reduced of the above!
        return; }}
\end{verbatim}}
\caption{Typing anomaly in $\java$}\label{type-java-example}
\end{table}

Because preservation of typing by reduction is a desirable property, 
we formulate the typing rule for casting as follows:
\[
\infer{\Gamma \Gives e:D}
{\Gamma \Gives (C)(e) :C} \qquad\mbox{(type casting rule)}~.
\]
At typing time, we do not try to verify that the value
$C'(\vc{r})$ resulting form the evaluation of the expression
$e$ is such that $C'\leq C$. Instead, we delay this verification at running
time. If the condition is not satisfied then reduction is stuck
(alternatively, an error message could be produced).

\begin{table}

{\footnotesize
\[
\begin{array}{c}

\infer{x:C\in \Gamma}
{\Gamma \Gives x:C} 

\qquad

\infer{\begin{array}{c}
\w{field}(C)=f_1:D_1,\ldots,f_n: D_n\\
\Gamma \Gives e_i:C_i, \quad C_i\leq D_i, \quad 1\leq i \leq n
\end{array}}
{\Gamma \Gives \s{new} \ C(e_1,\ldots,e_n):C}  \\ \\

\infer{\Gamma \Gives e: C \quad \w{field}(C) = f_1:C_1,\ldots,f_n:C_n}
{\Gamma \Gives e.f_i: C_i}

\qquad

\infer{
\begin{array}{c}
\Gamma \Gives e: C \quad \w{mtype}(m,C) =  (C_1,\ldots,C_n)\arrow D\\
\Gamma \Gives e_i:C'_i \quad C'_i\leq C_i\quad 1\leq i \leq n
\end{array}}
{\Gamma \Gives e.m(e_1,\ldots,e_n): D} \\ \\

\qquad \infer{\Gamma \Gives e:D}
{\Gamma \Gives (C)(e):C} 
\\ \\

\infer{
\begin{array}{c}
\Gamma \Gives e:C\quad \w{field}(C)=f_1:C_1,\ldots,f_n:C_n\\
\Gamma \Gives e':D_i\quad D_i\leq C_i
\end{array}}
{\Gamma \Gives e.f_i:=e': \s{Object}} 

\qquad

\infer{\Gamma \Gives e_1:C_1\qquad \Gamma \Gives e_2:C_2}
{\Gamma \Gives e_1;e_2:C_2}

\end{array}
\]}

\caption{Typing rules for $\sJ$ program expressions}\label{typing-J}
\end{table}

Table \ref{typing-J} specifies the rules to type expressions
that do not contain values (as source programs do).
An important point to notice is  that the typing rules 
allow to use an object of the class $C$ where an object 
of the class $D$ is expected as long as $C$ is
a sub-class of $D$. This is a form of {\em subtyping} (cf. chapter \ref{record-sec}). The rationale
is that an object of the sub-class $C$ will be able to handle
all the field and method invocations which could be performed on an object
of the super-class $D$. Indeed, objects of the class $C$ have all fields
of the class $D$ and may redefine methods of the class $D$ provided
their type is unchanged.

Beyond expressions, we also need to check the typing of the
class declarations. Suppose a method $m$ of the class $C$ has the shape:
\[
C_0 \ m(C_1 \ x_1,\ldots,C_n \ x_n)\{e\} ~,
\]
and that the class $C$ extends the class $D$. Then the following must hold:

\begin{enumerate}

\item  $\w{override}(m,D,(C_1,\ldots,C_n)\arrow C_0)$, 

\item $x_1:C_1,\ldots,x_n:C_n,\s{this}:C \Gives e: C'_0$ and $C'_0\leq C_0$.

\end{enumerate}
A {\em class} is well typed if all its methods are well typed in the sense above.
Finally, a program is well typed if all its classes are well typed and
the distinguished expression is well typed in the empty type context.
For instance, the reader may check that we can type the class declarations
in example \ref{exemple-dcl-class}.

\begin{exercise}[more programming]\label{more-java-prog-ex}
Design a compiler from the $\imp$ language (cf. chapter \ref{intro-sec})
to the typed $\sJ$ language.
We outline a possible strategy. 
\begin{enumerate}
\item Consider a restricted set of arithmetic expressions and boolean
conditions that can be easily coded in $\sJ$.  For instance, just work
with natural numbers in unary notation and a boolean condition that
checks if a number is zero (cf. Table \ref{classes-cjava}).

\item Represent variables as unary numbers and implement 
a state as a finite list of pairs composed of a variable and 
a number. A state is 
compiled into an object of a class \s{State} with methods to
read, write, extend, and restrict (cf. exercise \ref{exemple-dcl-class}).

\item Define a class \s{Code} with subclasses \s{Skip}, 
\s{Assignment}, \s{Conditional},$\ldots$ which correspond to the various
ways of composing statements in $\imp$. It is assumed that
each object of the class \s{Code} has a method \s{execute}
that takes as argument an object of the class \s{State}.

\item For all programs $P$ and states $s$ of the $\imp$ language
define a compilation into a $\sJ$ expression $e=\cl{C}(P).\s{execute}(\cl{C}(s))$
with the following properties: 
(1) if the expression $e$ evaluates to a value $v$ then
$v$ is the representation of a state $s'$ such that 
$(P,s)\eval s'$.
(2) the evaluation of $e$ never produces an exception or a type error,

\end{enumerate}

\end{exercise}

As already mentioned, the task of the type system is to localize the
type errors around the application of the casting reduction rule.
We formalize this property for the functional case and leave it to
the reader the extension to the imperative case.
To formulate the subject reduction, we add a rule to type (functional) values
which is similar to the rule for the \s{new}:
\begin{equation}
\infer{\begin{array}{c}
\w{field}(C)=f_1:D_1,\ldots,f_n: D_n\\
\Gamma \Gives v_i:C_i, \quad C_i\leq D_i, \quad 1\leq i \leq n
\end{array}}
{\Gamma \Gives \ C(v_1,\ldots,v_n):C}  ~.
\end{equation}
In order to reason about a method selection we need a substitution
property (cf. proposition \ref{subj-red-subtyping}).

\begin{proposition}\label{sub-obj-prop}
If $x_1:C_1,\ldots,x_n:C_n \Gives e:C$,
$\emptyset \Gives v_i:D_i$, and $D_i\leq C_i$ for 
$i=1,\ldots,n$ then $\emptyset \Gives [v_1/x_1,\ldots,v_n/x_n]e:C'$ and
$C'\leq C$.
\end{proposition}

\begin{exercise}\label{proof-sub-obj-pro}
  Prove proposition \ref{sub-obj-prop}.
  \end{exercise}

And we need to check the usual decomposition property (cf. proposition 
\ref{decomp-ev-cxt}).

\begin{proposition}\label{decomp-obj-prop}
Suppose $\emptyset \Gives e:C$. Then either $e$ is a value or there
is a unique evaluation context $E$ and redex $\Delta$ such that
$e\equiv E[\Delta]$, $\emptyset \Gives \Delta:D$ for some $D$, 
and $\Delta$ has one of the following shapes:
$\s{new} \ C(\vc{v})$, $(D)C(\vc{v})$, $C(\vc{v}).f$, or
$C(\vc{v}).m(\vc{v'})$.
\end{proposition}

\begin{exercise}\label{proof-decomp-obj-prop}
Prove proposition \ref{decomp-obj-prop}.
\end{exercise}

We also observe that it is always possible to replace an expression
with another expression with a smaller type.

\begin{proposition}\label{context-obj-prop}
If $\emptyset \Gives E[e]:C$,  $\emptyset \Gives e:D$, 
$\emptyset \Gives e':D'$, and 
$D'\leq D$ then $\emptyset \Gives E[e']:C'$ for some $C'$ such
that $C'\leq C$.
\end{proposition}

\begin{exercise}\label{proof-context-obj-prop}
  Prove proposition \ref{context-obj-prop}.
  \end{exercise}
  
We can then state the subject reduction property for the typed
$\sJ$ language as follows.

\begin{proposition}\label{main-obj-prop}
Given a well-typed functional program in $\sJ$ and a well-typed functional 
expression $\emptyset \Gives e:C$ one of the following situations arises:
\begin{enumerate}

\item $e$ is a value.

\item $e\arrow e'$, $\emptyset \Gives e':C'$, and $C'\leq C$.

\item $e\equiv E[(D)(C(\vc{v}))]$ and $C\not\leq D$.

\end{enumerate}
\end{proposition}
\Proof
Suppose $\emptyset \Gives e:C$ and $e$ is not a value. Then $e$ has
a unique decomposition as $E[\Delta]$  and $\emptyset \Gives \Delta :D$ 
(proposition \ref{decomp-obj-prop}).
We proceed by case analysis on the typing of $\Delta$ to show that
either the computation is stuck because of a casting error or
it can be reduced to an expression $e'$ such that $\emptyset \Gives e':D'$ and
$D'\leq D$, and
we can then conclude by proposition \ref{context-obj-prop}.
Proposition \ref{sub-obj-prop} is needed to handle the case of a
method selection. \qed

\section{Summary and references}
An object is basically a record and 
object-oriented languages introduce user friendly 
mechanisms to define mutually recursive records.
Depending on whether fields are modifiable, one can
distinguish between functional and imperative
object-oriented languages (which are those mainly used in practice).
In {\em typed} object-oriented languages, the introduction of 
a casting operator is necessary in order to have some
programming flexibility. In this setting, the goal of 
a type system is {\em not} to avoid typing errors but
to {\em localize} them around the usage of the
casting operator.
The formalization presented in this chapter builds  
on the paper \cite{DBLP:journals/toplas/IgarashiPW01}.
The book \cite{Mitchell03} introduces the main 
design issues in object-oriented programming languages.

\chapter{Introduction to concurrency}\label{introduction-sec}
\markboth{{\it Introduction to concurrency}}{{\it Introduction to concurrency}}

In computer science, we are used to the idea of regarding a piece of
software and/or hardware as a {\em system}, {\em i.e.}, a compound of
interacting and interdependent components with varying names such as
{\em threads} or {\em processes} that we use as synonymous.

Starting from this chapter, the general goal is to formalize and reason on
systems where several threads/processes {\em compete} for the same resources
(e.g. write a variable or a channel).  Most of the time, this results
into {\em non-deterministic} behavior which means that with the same
input the system can move to several (incomparable) states.  For
instance, the computation of a circuit may be non-deterministic due to
the unpredictable delays in the propagation of signals.  Similarly,
the computation of an operating system may be non-deterministic due to
unpredictable delays in managing the accesses to memory.  We stress
that non-determinism is both a way of representing our partial
knowledge of the system and a method to keep its specification
general. For instance, we may want to prove that a certain algorithm
is correct independently of the scheduling policy or the evaluation
strategy chosen.

Some authors distinguish {\em parallel} from {\em concurrent} 
systems.
The former are a subclass of the latter that typically exhibit 
a deterministic behavior. A standard problem in parallel programming
is to decompose the task of computing a (deterministic) function
into parallel sub-tasks that when executed on suitable hardware will
hopefully provide a faster result in terms of throughput and/or latency.
We do not develop at all these algorithmic issues.

Besides being non-deterministic, certain concurrent systems may also
exhibit a {\em probabilistic} behavior. In first approximation, this
means that at certain points in the computation the next state of the
system is determined by tossing a coin.  
The basic idea we  stress
in chapter \ref{proba-sec} is that non-deterministic and probabilistic
transitions should be kept separated and that a computation in a
non-deterministic and probabilistic system is described by a
transition relation that relates states to distributions over states.

The concurrent systems we consider can be 
classified according to {\em two main criteria}:
\begin{quote}
asynchronous vs. synchronous ~~~~and~~~~ shared memory vs. message passing.
\end{quote}
The first criterion concerns the {\em relative speed} of the processes;
we  mainly focus on {\em asynchronous} systems 
where each process proceeds at its own speed, however we shall see 
in chapter \ref{time-sec} that
the techniques can be adapted to {\em synchronous/timed systems}
too, where computation proceeds in phases or rounds.
The second criterion concerns the {\em interaction mechanism} among the
processes.  In {\em shared memory}, processes interact by
modifying a shared area of memory.  Synchronization arises by waiting
that a certain condition is satisfied (cf. lock/unlock, compare and set, 
P/V, monitors, synchronized methods,$\ldots$).  
In {\em message passing}, processes interact by
sending/receiving messages on communication channels. Synchronization
arises when receiving (wait for a message to be there) and possibly
when sending (if the capacity of the channel is exceeded). The order of
transmission is not necessarily respected and various kinds of channels can be
considered according to their capacity (bounded/unbounded), the ordering
of the messages, and  the number of processes accessing the channel 
(one-to-one, one-to-many, many-to-many, $\ldots$).

\section{A concurrent language with shared memory}\label{parimp-sec}
To make things concrete, we start looking at a simple instance of an asynchronous and shared memory model.  The recipe is rather straightforward: 
we select a standard  imperative language,
namely the imperative language $\imp$ considered in chapter \ref{intro-sec},  
and add: (i) the possibility of
running several commands in parallel on the same shared memory and (ii) a
synchronization mechanism.
Table~\ref{parimp-syntax} \index{$\parimp$, syntax} 
describes the abstract syntax of the language.
We have identifiers, integers, numerical and boolean expressions, 
and processes. Besides the standard instructions for assignment, sequentialization, branching, and iteration one can declare and initialize a local identifier, start the execution 
of two processes in parallel, and wait for a boolean condition to hold and then execute  {\em atomically }
a sequence of assignments. 
In particular, the process  $\await{\s{true}}{P}$ is supposed to execute atomically
the process $P$.  To stress this, we also abbreviate it as  $\atomic{P}$.
In a process $\local{x}{n}{P}$, the identifier $x$ is bound in $P$ and
obeys the usual rules of renaming.
We denote with $\fv{P}$ the set of identifiers occurring free in $P$.

\begin{table}[b]
{\footnotesize
\[
\begin{array}{lll}

\w{id} &::=x \Alt y \Alt \cdots  &\mbox{(identifiers)}\\
n      &::= 0 \Alt 1 \ \Alt -1 \cdots &\mbox{(integers)}\\
e      &::= \w{id} \Alt n \Alt (e+e)\Alt \cdots       &\mbox{(expressions)} \\
b      &::= e<e \Alt \cdots                            &\mbox{(boolean expressions)}\\

P       &::= \s{skip}\Alt \w{id}:=e \Alt P;P \Alt \ite{b}{P}{P} \Alt  \while{b}{P} \Alt 
\\ &\qquad \local{x}{n}{P}
  \Alt (P \mid P) \Alt  \s{await} \ b \ \s{do} \ P 
&\mbox{(processes)}
\end{array}
\]}
\caption{An asynchronous, shared memory model: $\parimp$.}\label{parimp-syntax}
\index{$\parimp$, syntax}
\end{table}

Next we describe the possible executions of such processes relatively
to a {\em state} of the shared memory which is described as a total
function $s:\w{id} \arrow \Z$ from identifiers to integers (exactly as
in chapter \ref{intro-sec}).  
We recall that expressions and boolean conditions do {\em not} 
produce side-effects. Their evaluations rules are defined 
in chapter \ref{intro-sec}, Table \ref{semantics-imp}.
Next, we revisit the small-step reduction rules defined in 
chapter \ref{intro-sec}, Table \ref{small-step-imp}.
Table~\ref{parimp-smallstep} \index{$\parimp$, reduction}
defines the {\em immediate termination} predicate `$\downarrow$' 
and gives the  {\em small-step rules} for process execution
where the symmetric rule for parallel composition is omitted.
We write $(P,s)\eval s'$ if $(P,s)\trarrow (P',s')$ and $P'\downarrow$.
Notice that unlike for the sequential fragment $\imp$, 
the relation $\eval$ is {\em not} a partial function.

\begin{exercise}\label{termination-parimp-ex}
  Prove the following.
\begin{enumerate}
\item If $P\downarrow$ then for any state $s$, $(P,s)$ cannot reduce.

\item We say that a process $P$ {\em terminates} (properly)
  if for any state $s$ every
  reduction sequence starting from $(P,s)$ terminates
  in a pair $(P',s')$ such that $P'\downarrow$.
  Show that if $P$  does {\em not} contain \s{while} and \s{await} commands
  then $P$ terminates.

\item Let $P$ be a process such that the body of every
  \s{await} command in it terminates. Show that for any state $s$,
  if $(P,s)$ does not reduce then $P\downarrow$.
\end{enumerate}
\end{exercise}

In practice, every usage of the \s{await} command considered in the
following will satisfy the condition that its body does not contain
\s{await} and \s{while} commands. In view of exercise
\ref{termination-parimp-ex}, this means that we are always in a situation
where the body of the \s{await} command terminates.

We write $(P,s)\eval s'$ if $(P,s)\trarrow (P',s')$ and $P'\downarrow$.
Notice that unlike for the sequential fragment $\imp$, 
the relation $\eval$ is {\em not} a partial function.

The small-step reduction rules embody certain design choices that is worth to make explicit.
First, we have assumed that expressions and assignments are executed {\em atomically}.
This is a (grossly) simplifying hypothesis. We could refine the level of granularity
of the small step semantics to some extent and thus complicate the reasoning.
However, the basic problem we have to face is that it is difficult to determine 
the `right' level of granularity.
This is due to the fact that there is no general agreement on 
the {\em abstract memory model} that should be presented to the programmer of
a concurrent language with `shared memory'.  Ideally, the model should be `abstract' 
while allowing for correct and efficient implementations on a variety of architectures.
Second, a blocked await reduces which is a form of busy waiting.
An alternative semantics could just suspend the execution waiting
for a certain synchronization condition to be realized.
Notice that in this approach, assertion 3 of
exercise \ref{termination-parimp-ex} is false; a process
that does not reduce is not necessarily properly terminated.

\begin{table}
{\footnotesize
\[
\begin{array}{ccc}
\infer{}{\sk \downarrow}  \quad

&\infer{P\downarrow}{\local{x}{n}{P}\downarrow} \quad
 
&\infer{P_i \downarrow \quad i=1,2}
{(P_1 \mid P_2)\downarrow}  \\\\

\end{array}
\]
\[
\begin{array}{llll}

(x:=e,s) &\arrow &(\s{skip},s[v/x]) &\mbox{if }(e,s)\eval v \\ \\

(\ite{b}{P}{P'},s) &\arrow &(P,s)    &\mbox{if }(b,s) \eval \s{true} \\
(\ite{b}{P}{P'},s) &\arrow &(P',s)    &\mbox{if }(b,s) \eval \s{false} \\\\

(\while{b}{P},s)   &\arrow &(P;\while{b}{P},s) &\mbox{if }(b,s) \eval \s{true} \\
(\while{b}{P},s)   &\arrow &(\s{skip},s) &\mbox{if }(b,s) \eval \s{false} \\\\

  (P;P',s )           &\arrow &(P',s) &\mbox{if }P\downarrow \\
(P;P',s)                  &\arrow &(P'';P',s') &\mbox{if }(P,s) \arrow (P'',s')  \\ \\

(\local{x}{n}{P},s) &\arrow &(\local{x}{n'}{P'},s'[s(x)/x])
&\mbox{if }(P,s[n/x]) \arrow (P',s'[n'/x]) \\ \\

(P\mid P',s)       &\arrow &(P'' \mid P',s') &\mbox{if } (P,s)\arrow (P'',s') \\ \\

(\await{b}{P},s)   &\arrow &(\await{b}{P},s) &\mbox{if }(b,s)\eval \s{false} \\

  (\await{b}{P},s)   &\arrow &(P',s')     &\mbox{if }(b,s)\eval \s{true},
  %\\ &&&
  (P,s)\trarrow (P',s'),P'\downarrow~.

\end{array}
\]}
\caption{Immediate termination and small-step reduction for $\parimp$}\label{parimp-smallstep}
\index{$\parimp$, reduction rules}
\end{table}

\begin{exercise}\label{local-var-conc-ex}
  To appreciate the handling of the local variables in the operational
  semantics, consider the processes
 $P_n=\local{x}{n}{y:=x}$, for $n\in \Z$. Given an arbitrary state $s$,
  compute the possible reductions of the parallel process $(P_0 \mid P_1,s)$.
\end{exercise}

Let us consider a few examples that illustrate the expressivity of the language.

\begin{example}[$P$ and $V$]\label{ex-pv}
Assuming assignment atomic (as we do), the operations $P$ and $V$
for manipulating a {\em semaphore $s$ of capacity $k$}
can be expressed as follows:\footnote{The terminology is due to E.~Dijkstra and is based on the Dutch words {\em passering} (passage) and {\em vrijgave} (release).}
\[
\begin{array}{ll}

\mbox{Initially:} &s:=k~, \\
P(s) =           &\await{s>0}{s:=s-1}~, \\
V(s) =            &s:=s+1~.
\end{array}
\]
In the special case where the initial capacity is $1$, the operations
$P$ and $V$ are also called $\w{lock}$ and $\w{unlock}$, respectively. 
By using them, processes can gain exclusive access to a shared resource, 
{\em e.g.},  a process can gain the right to execute without
interruption, {\em i.e.}, atomically, a sequence of statements.
\end{example}

\begin{example}[non-deterministic sum]\label{choice-enc-ex}\index{non-deterministic sum}
We want to define a statement:
\[
[b_1 \arrow P_1 + \cdots + b_n \arrow P_n]
\]
which {\em selects non-deterministically} one of the branches
(if any) for which the condition $b_i$ is satisfied and 
starts running $P_i$.  This can be defined as follows assuming
$x,y\notin \fv{b_i\arrow P_i}$ for $i=1,\ldots,n$:
\[
\begin{array}{l}
\local{x}{1}{(Q_1 \mid \cdots \mid Q_n)}, \mbox{ where } \\
Q_i \equiv \s{var}  \ y=1 \\
    \qquad \s{await} \ b_i \ \s{do} \ \s{if} \ x=1 \ \s{then} \ x:=0 \ \s{else} \ y:=0 \ ;  \\
    \qquad \s{if} \ y=1 \ \s{then} \ P_i \ \s{else} \ \sk
\end{array}
\]
\end{example}

\begin{exercise}\label{atomic-modif-ex}
(1) Modify the definition so that once the branch $i$ is selected, the continuation $P_i$ is run atomically. 
(2) With the current definition, a statement such as
$[\s{true} \arrow \sk + \s{false} \arrow \sk]$ does {\em not} terminate  
(which is not very satisfying). Adapt the definition to fix this problem.
\end{exercise}

\begin{exercise}\label{parimp-with-spawn}
Suppose we enrich the $\parimp$ language with a \s{spawn} operator.
The process $\s{spawn} \ P$ starts the execution of $P$ in parallel 
and immediately terminates by reducing, say, to $\s{skip}$.
(1) Propose a formal semantics of the $\parimp$ language with $\s{spawn}$.
(2) Explain why in general the process 
$(\s{spawn} \  P); Q$ is {\em not} equivalent to 
$(P\mid Q)$.
(3) Propose a compilation of the enriched language into the enriched language
{\em without} parallel composition, {\em i.e.}, find a way to simulate 
parallel composition with \s{spawn}.
\end{exercise}

\begin{example}[compare and set]\label{ex-compare-set}
The {\em compare and set} (\s{cas}) operation can be defined as follows
(this operation is also called {\em compare and swap}):
\[
\begin{array}{ll}
\s{cas}(x,e_1,e_2) &= 
\atomic{\s{if}\  (x=e_1) \ \s{then} \ x:=e_2 \ \s{else} \ \sk}~.
\end{array}
\]
We stress that it is essential that the boolean test $x=e_1$ and the
assignment $x:=e_2$ are executed {\em atomically}. 
The \s{cas} operation can be taken as basic building block to 
solve more complex problems in concurrency. For instance, it can be
used to solve the so called {\em consensus problem} which can be stated
as follows. A collection of parallel processes $P_1,\ldots,P_n$ 
each holding a non-negative integer, say $v_1,\ldots,v_n$, 
have to agree on a value which is equal to one of the values
held by the processes. A solution to this problem which treats all processes
in the same way and avoids centralization points goes as follows.
Set a variable $x$ with initial value $-1$
and then let each process $P_i$ run the following procedure:
\[
\s{decide}(i) = \s{cas}(x,\ -1, \ v_i); \w{result}_i:=x~.
\]
The first process that runs the \s{decide} procedure will set $x$ to its
value $v_i\geq 0$ (atomically, and thus deciding the outcome of the consensus
protocol) while the following ones will keep $x$ unchanged and adopt its value.
\end{example}

\section{Equivalences: a taste of the design space}\label{equiv-space-sec}
We consider the question of building an equivalence on processes on 
top of the reduction system. In the sequential framework (cf. chapters
\ref{intro-sec} and \ref{equiv-sec}), we have already noticed that an answer
to this question depends on a certain number of factors such as the choice
of the  observables, the compositionality properties, and the proof methods.
With an enlarged range of choices, these factors 
play a role in the semantics of concurrent systems too. Moreover, new factors
appear such as the hypotheses on the scheduling policy.

\paragraph{Observables}  The equivalence should be compatible with a notion of observation
of the processes. If two processes $P$ and $P'$ are equivalent and 
$P$ enjoys a certain observable property then $P'$ should enjoy that property too. 
For instance, we may wait till the system comes to a {\em proper termination} and
then observe its final result. As a second example, we may be informed that 
the system has reached a {\em deadlock}, {\em i.e.}, a situation where it has
not properly terminated and it cannot progress. As a third example, we may 
interact with the system during the computation and observe its capabilities.
We refer to this observable as {\em branching} because, as explained in 
the  following example \ref{branching-ex}, it amounts
to observe the branching structure of the computation as opposed to its
linearization.

\paragraph{Scheduling} 
We may assume certain properties of the scheduler that controls the
order in which parallel processes are executed. For instance, a {\em preemptive scheduler}
will be allowed to interrupt the execution of a process at any point which is compatible with 
the atomicity assumptions while a {\em cooperative scheduler} will wait for the process to 
yield control or to suspend on a synchronization condition. 
Further, schedulers can be classified according to their ability to execute
the various processes in a {\em fair} way.

\paragraph{Compositionality} If a process $P$ is equivalent to the process $P'$ then we should be
able to replace $P$ with $P'$ in any (reasonable) process context.
In other words, the notion of equivalence should be preserved by
some operators of the language, including at least parallel composition.

\paragraph{Proof method} We should have a {\em practical} proof method to 
check the equivalence of two processes. Depending on the class of processes
we are considering, practical may mean that the equivalence 
can be efficiently {\em automated} or that the proof has a certain {\em locality}  property.

\bigskip\noindent
We elaborate on the first two points (observables and scheduling hypotheses) 
in the following examples; we shall come back to 
compositionality and proof methods in the following sections.

\begin{example}[termination]\label{termination-ex}
The following process {\em diverges} 
(or at least does not reach immediate termination)
while producing a sequence $f(0),f(f(0)),\ldots$ 
on the  `output variable' $y$ at a `rate' determined by $x$.
\[
\begin{array}{l}
y:=0;\\
\s{while} \ \s{true} \ \s{do} \\
\s{await} \ x=0 \ \s{do} \ (y:=f(y);x:=1;)

\end{array}
\]
Should it be considered equivalent to $\while{\s{true}}{\sk}$ (a diverging process)?
\end{example}

\begin{example}[deadlock]\label{deadlock-ex}\index{deadlock, example}
Consider the following deadlocked process:
\[
\begin{array}{ll}
\local{x}{0}{\await{x>0}{x:=x+1}}~.
\end{array}
\]
Should it be considered equivalent to $\while{\s{true}}{\sk}$ (a diverging process, again) or to $\sk$ (an immediately terminated process)? 
\end{example}

\begin{example}[branching]\label{branching-ex}\index{vending machine}
  Consider the following hypothetical
  controls of an old fashioned {\em vending machine:}
\[
\begin{array}{l}
[b_1\arrow P_1 ; [b_2 \arrow P_2 + b_3 \arrow P_3]] \\

[ b_1 \arrow P_1; [b_2 \arrow P_2] ~ + ~
 b_1 \arrow P_1 ; [b_3 \arrow P_3]  ] 
\end{array}
\]
with the interpretation:
\[
\begin{array}{ll}

b_1=  \mbox{ there is a coin}  
&P_1= \mbox{ accept the coin} \\

b_2=  \mbox{ there is a second coin} 
&P_2= \mbox{ accept the coin and deliver coffee} \\

b_3= \mbox{ water request} 
&P_3= \mbox{ deliver water.}
\end{array}
\]
Are the two controls {\em equivalent}? Well, one may remark that
upon accepting the first coin, the second machine decides non-deterministically
whether it is ready to wait for a second coin or to deliver water which 
is rather  annoying for the user. 
\end{example}

\begin{example}[cooperative]\index{cooperative concurrency}
In preemptive concurrency, a process can be interrupted after 
any atomic step. In cooperative concurrency, a process is interrupted only when
it has terminated or it is suspended on a waiting statement.
For instance, the processes:
$x:=1; x:=x+1$ and $x:=1; x:=2$
are equivalent in a cooperative (and a sequential) context but not
in a preemptive one.
\end{example}

\begin{example}[weak fairness]\index{fairness, weak}
Consider the following process:
\[
x:=0;y:=0;((\while{x=0}{y:=y+1} )\mid x:=1)~.
\]
If the process terminates then $y$ may contain an arbitrary natural number. This is called {\em unbounded non-determinism}.
Moreover, the process is actually {\em  guaranteed to terminate} if we assume that every process 
that  is ready to run will eventually get a chance of running.
This assumption is called {\em weak fairness}.
\end{example}

\begin{example}[strong fairness]\index{fairness, strong}
A weak fairness hypothesis is not always enough to guarantee progress.
Consider:
\[
\begin{array}{l}
x:=0;y:=0; (\ \while{y=0}{x:=1-x} \mid \await{x=1}{y:=1}  \ ) ~.
\end{array}
\]
In this example, the first process makes $x$ oscillate between $0$ and $1$ 
while the second process can really progress only when $x=1$. 
A scheduler that gives control to the second process only when $x=0$ will
not guarantee termination.
{\em Strong fairness} is the assumption that in any infinite execution 
a process which is infinitely often `ready to run' will indeed run infinitely
often. 
\end{example}

\section{Summary and references}
Early work on the semantics of concurrent processes started in 
the 60's \cite{DBLP:journals/cacm/Dijkstra65} and
was motivated by synchronization problems in operating systems.
The first step in defining the semantics of a concurrent language
amounts to decide which actions can be regarded as {\em atomic}.
This is an issue which can be hardly underestimated because there is a tension 
between atomicity and efficient implementations.
At any rate, once atomicity is fixed
a {\em small-step} reduction semantics allows to define precisely the state
transformations a concurrent process can go through.  The second step
amounts to decide the observable properties of the system and the
execution hypotheses. This step gives rise to a variety of possible
equivalences.  Compositionality and the existence of practical 
proof methods are two basic criteria to assess them.
The article \cite{HandbookAKarp} surveys the
parallelization of algorithms (which we do not cover).

\chapter{A compositional trace semantics}\label{trace-shared-sec}
\markboth{{\it Trace semantics}}{{\it Trace semantics}}

We consider the problem of defining and characterizing 
a {\em compositional} equivalence for the $\parimp$ model.
For the sequential fragment of the $\parimp$ model,
the input-output interpretation  provides a satisfying answer
(cf. chapter \ref{intro-sec}), but the extension
to the full concurrent $\parimp$ language is
not straightforward and rises some interesting
issues.% which are discussed in this and the following chapter.

\section{Fixing the observables}
Following the discussion in section \ref{equiv-space-sec}, a first
problem consists in fixing a notion of {\em observable}.  
Building on the semantics of the sequential $\imp$ language (chapter
\ref{intro-sec}) we shall take the  {\em input-output behavior} 
or, equivalently, the {\em partial correctness assertions} (pca),
as basic observable.
We warn the reader that while being reasonable, this notion of
observable is definitely not the only possible one for concurrent
processes; alternatives will be discussed in the following chapters.
Let $P,P',\ldots$ be the processes and 
$s,s',\ldots$ be the memory states introduced in section \ref{parimp-sec}.
We adapt to processes the definitions presented in chapter \ref{intro-sec}.

The IO interpretation (cf. definition \ref{io-interp}) of a process $P$ is:
\[
\sm{P}{IO}=\set{(s,s') \mid (P,s) \trarrow (P',s')\downarrow}~.
\]
Also the notion of pca's validity  is extended to processes in the
obvious way:
\[
\models \hoare{A}{P}{B} \mbox{ if }\ \qqs{s}{(s\models A \mbox{ and } 
(P,s) \trarrow (P',s')\downarrow \mbox{ implies }s'\models B)}~.
\]
Then the pca interpretation of a process is:
\[
\sm{P}{pca}=
\set{(A,B) \mid \quad \models \hoare{A}{P}{B}}~.
\]
Adapting proposition \ref{IO-pca}, we derive:
\[
\sm{P_1}{IO} = \sm{P_2}{IO}  \ 
\mbox{ iff }\ 
\sm{P_1}{pca}=\sm{P_2}{pca} ~.
\]

Let us take the input-out behavior (or equivalently
the partial correctness assertions) as basic observable.
As usual, a {\em context} $C$ is a process with a hole $[~]$. E.g.
\[
x:=3;[~] \mid \await{x=3}{x:=x+1}~.
\]
As already mentioned in chapters \ref{intro-sec}, 
\ref{equiv-sec}, and \ref{introduction-sec} 
a desirable property of a semantics is that it is {\em preserved
by contexts}, that is:
\[
\sem{P_1} = \sem{P_2} \ \mbox{ implies } \  \sem{C[P_1]} = \sem{C[P_2]}~.
\]
If two processes have the same `compositional semantics' then we can 
{\em replace} one for the other in any context.
Unfortunately, the following example shows that,
unlike in the sequential case (proposition \ref{comp-io-imp-prop}), 
{\em compositionality fails} for  the IO (and pca) interpretation.

\begin{example}[non-compositionality of IO interpretation]\label{ex-noncomp-io}
The processes $P_1 \equiv  x:=1;x:=x+1 $ and $P_2 \equiv  x:=2$ 
are IO-equivalent. However when they
are composed in parallel with the process $P_2$ we have:
$\sm{P_1\mid P_2}{IO} \neq \sm{P_2 \mid P_2}{IO}$.
\end{example}

\section{Towards compositionality}
As a first attempt at fixing the compositionality issue, we try to
refine the semantics of processes. In automata theory, we are used to
associate to an automaton the collection of its execution traces. We
follow a similar path by considering the traces of the states crossed
by a terminating execution.

\begin{definition}[trace interpretation]\label{trace-int-def}\index{$\parimp$, trace interpretation}
The trace interpretation of a process $P$ is defined as follows:
\[
\sm{P}{T}=\set{s_1\ldots,s_n \mid (P,s_1) \trarrow (P_2,s_2) \cdots \trarrow (P_n,s_n)\downarrow}~.
\]
\end{definition}

\begin{remark}\label{rmk-on-trace}
The IO semantics is exactly the subset of the trace semantics
composed of {\em words of length $2$}.
\[
\sm{P}{IO}= \set{(s,s') \mid s s' \in \sm{P}{T}}~.
\]
With reference to the previous example \ref{ex-noncomp-io}, it is easy to check that
$\sm{P_1}{T} \neq \sm{P_2}{T}$.
However, for $P_3\equiv x:=1;x:=2$ we have:
\[
\sm{P_1}{T} = \sm{P_3}{T},
\qquad
\sm{P_1\mid P_2}{T} \neq \sm{P_3\mid P_2}{T}~.
\]
So this trace semantics is not compositional either!
\end{remark}

While failing to {\em characterize} the `right equivalence/pre-order' 
we can at least {\em define} it.

\begin{definition}[pre-congruences]
A pre-congruence is a pre-order on processes which is preserved by contexts.
We define two pre-congruences relatively to the IO and trace interpretations as follows:
\[
\begin{array}{ll}
P_1 \leq_{IO} P_2 &\mbox{ if }\  \qqs{C}{\sm{C[P_1]}{IO} \subseteq \sm{C[P_2]}{IO}}~, \\
P_1 \leq_{T} P_2 &\mbox{ if }  \ \qqs{C}{\sm{C[P_1]}{T} \subseteq \sm{C[P_2]}{T}}~.

\end{array}
\]
\end{definition}

\begin{exercise}\label{largest-pre-order-ex}
Check that $\leq_{IO}$ ($\leq_T$) is the {\em largest pre-order} (reflexive and
transitive) which refines the $IO$ containment (trace containment) and
which is preserved by all contexts.
\end{exercise}

Somehow surprisingly, once we require preservation by contexts, it
does not matter whether we look at the input-output or at the traces.

\begin{proposition}
The pre-congruences $\leq_{IO}$ and $\leq_{T}$ coincide.
\end{proposition}
\Proof
$\leq_T \subseteq \leq_{IO}.\ $   By remark \ref{rmk-on-trace}, we know that 
$\sm{P_1}{T}\subseteq \sm{P_2}{T}$
 implies 
$\sm{P_1}{IO} \subseteq \sm{P_2}{IO}$.
Then it follows {\em by unfolding} the definitions that:
$P_1 \leq_T P_2$  implies $P_1 \leq_{IO} P_2$.

\Proofitemm{\leq_{IO} \subseteq \leq_T}.
For the other direction, assume {\em by 
contradiction} $P_1 \not\leq_{T} P_2$.
This means that for some context $C$ and trace $s_1\cdots s_n$:
\[
s_1\cdots s_n \in \sm{C[P_1]}{T} \quad\mand \quad
s_1\cdots s_n \notin \sm{C[P_2]}{T}~.
\]
In particular, this entails, for $Q_1\equiv C[P_1]$:
$(Q_1,s_1) \trarrow (Q_1^{2},s_2) \trarrow \cdots (Q_{1}^{n},s_n) \downarrow$.
The {\em key step} is the following: we build an {\em observer} $O$
that may terminate iff it sees the state going through $s_1\cdots
s_n$; the observer reads the state without modifying it.
Take $X=\fv{C[P_{1}]}\union \fv{C[P_{2}]}$ and recall the
$\w{IS}$ predicate from proposition \ref{IO-pca}:
\[
\w{IS}(s,X) = \MEET_{x\in X} (x=s(x))~.
\]
Then define:
\[
\begin{array}{lll}
O &\equiv  &\await{\w{IS}(s_{1},X)}{\sk}; \\
  &         &\cdots \\
   &        &\await{\w{IS}(s_{n},X)}{\sk}~.
\end{array}
\]
We have: $(s_1,s_n) \in \sm{C[P_1]\mid O}{IO}$.
On the other hand we claim that:
\[
(s_1,s_n) \notin \sm{C[P_2] \mid O}{IO}~,
\]
because the only way $O$ can terminate is that the state
goes through the configurations $s_1,\ldots,s_n$ and
since $O$ does not modify the state this 
would mean $s_1\cdots s_n\in \sm{C[P_2]}{T}$.\qed  \\

Following these preliminary remarks, we can define our {\em goal} as follows:

\begin{quote}
find an {\em interpretation} $\sem{\_}$ such that:
$\sem{P_1} = \sem{P_2}$ \mbox{ iff }
$\qqs{C}{\sm{C[P_1]}{IO} = \sm{C[P_2]}{IO}}$.
\end{quote}
Such an interpretation (if it exists) will be {\em compositional} by definition.
Sometimes one is happy with the left to right implication. In this case, the interpretation
is called {\em adequate} in that it provides a sufficient criterion to determine the equivalence
of two processes. If moreover the right to left implication holds, then one speaks
of a fully adequate (or fully abstract) interpretation.
Notice that this last property can be reformulated as follows:
\[
\sem{P_1}\neq \sem{P_2} \mbox{ implies }
\xst{C}{\sm{C[P_1]}{IO}\neq \sm{C[P_2]}{IO}}~.
\]
In words, whenever the interpretations differ we can find a context where
the IO behaviors, {\em i.e.}, observable behaviors of the processes differ.

\section{A trace-environment interpretation}\label{te-int-subsec}
To address the compositionality issue, we are guided by the following intuition: 
\begin{quote}
to analyze a process in a concurrent system
we have to account for the perturbations induced by the environment (the external world).
\end{quote}
In particular, in the framework of a trace semantics, we allow the environment (the external world) 
to modify the state after any sequence of transitions.

\begin{definition}[trace-environment interpretation]\index{$\parimp$, trace-environment interpretation}
Let $P$ be a process. Its trace-environment (TE) interpretation
is defined as follows:
\[
\begin{array}{lll}
\sm{P}{TE} &=  \{ &(s_1,s'_1)\cdots (s_n,s'_n) \mid  \\
           &&(P,s_1)\trarrow (P_2,s'_1) \\
           &&\cdots \\
           &&(P_n,s_n) \trarrow (P_{n+1},s'_n)\downarrow \}~.
\end{array}
\]
\end{definition}

\begin{exercise}\label{on-te-inter-ex}
In remark \ref{rmk-on-trace}, we have observed the equivalence in the trace interpretation of  the processes
$P_1\equiv x:=1;x:=x+1$ and $P_2\equiv x:=1;x:=2$. 
Check that: $\sm{P_1}{TE} \neq \sm{P_2}{TE}$.
\end{exercise}

\begin{remark}\label{te-int-rmk}
An equivalent view of the TE-interpretation is to {\em add a labelled rewriting rule}
that explicitly accounts for the actions of the environment:
\begin{equation}\label{env-step-rule}
\infer{}{(P,s)\act{e} (P,s')}
\end{equation}
Thus this labelled rule allows for an arbitrary modification of the state while leaving
unchanged the control of the observed process. Then we define:
\[
\begin{array}{lll}
\sm{P}{TE} &=  \{ &s_1,s'_1\cdots s_n,s'_n \mid  \\
           &&(P,s_1)\trarrow (P_2,s'_1) \act{e} (P_2,s_2)\\
           &&\cdots \\
           &&(P_{n-1},s_{n-1}) \trarrow (P_n,s'_{n-1}) \act{e} (P_n,s_n) \\
           &&(P_n,s_n) \trarrow (P_{n+1},s'_n)\downarrow \}~.
\end{array}
\]
The traces in the sense of definition \ref{trace-int-def} can be regarded as the trace-environment traces
where $s_{i+1}=s'_{i}$, for $i=1,\ldots, n-1$.
\end{remark}

In section \ref{trace-denotation-sec}, we shall show that this interpretation is preserved by
all the operators of the language. For the time being we just consider the problematic
case of parallel composition.

\begin{proposition}
The TE-inclusion is preserved by parallel composition.
\end{proposition}
\Proof 
First notice the following properties:
\[
\begin{array}{lll}
(P_1\mid P_2)\downarrow &\mbox{ implies } 
&P_1\downarrow \mbox{ and }P_2 \downarrow ~, \\ 

(P_1\mid P_2,s) \arrow (P,s')
&\mbox{ implies }
&(P_1,s)\arrow (P'_1,s') \mbox{ and }P\equiv (P'_1\mid P_2) \mbox{ or }  \\
&&(P_2,s)\arrow (P'_2,s') \mbox{ and }P\equiv (P_1\mid P'_2) ~.
\end{array}
\]
Thus from a reduction such as:
\[
\begin{array}{lll}
(P \mid Q,s_1) &\trarrow &(P_2 \mid Q_2,s'_1) \\
\cdots \\
(P_n \mid Q_n,s_n) &\trarrow &(P_{n+1} \mid Q_{n+1},s'_{n})\downarrow
\end{array}
\]
one can extract a reduction for $P$ where all the reduction steps taken
by the other process are {\em simulated by the environment}. 
As a {\em concrete example}, suppose $\sm{P_1}{TE}\subseteq \sm{P'_1}{TE}$ and 
\[
(P_1\mid Q_1,s_1) \arrow (P_2\mid Q_1,s_2) \arrow
(P_2\mid Q_2,s_3) \arrow (P_3 \mid Q_2,s_4)\downarrow~.
\]
We can turn this into:
\[
(P_1,s_1) \arrow (P_2,s_2) \act{e} (P_2,s_3) \arrow (P_3,s_4)\downarrow~.
\]
Then $(s_1,s_2)(s_3,s_4) \in \sm{P_1}{TE} \subseteq \sm{P'_1}{TE}$ 
entails:
\[
(P'_1,s_1) \trarrow (P'_2,s_2) \act{e} (P'_2,s_3) \trarrow (P'_3,s_4)\downarrow~.
\]
Now put back the $Q_1$ process and let it play the {\em role of
the environment}:
\[
(P'_1\mid Q_1,s_1) \trarrow (P'_2\mid Q_1,s_2) \arrow (P'_2\mid Q_2,s_3) \trarrow (P'_3\mid Q_2,s_4)\downarrow~.
\]
This argument can be generalized.
Suppose $\alpha = (s_1,s'_1)\cdots (s_n,s'_n)$,
$\alpha \in \sm{P_1\mid Q}{TE}$, and $\sm{P_1}{TE}\subseteq \sm{P_2}{TE}$.
Derive a reduction for $P_1$ which must also belong to $P_2$.
Then, by putting back the  thread $Q$,
conclude that $\alpha\in \sm{P_2\mid Q}{TE}$. \qed \\

Since the TE interpretation refines the IO interpretation, its
adequacy will follow by the announced compositionality property shown in
chapter \ref{trace-denotation-sec}.  We now address the full abstraction
problem.

\begin{proposition}
Let $P_1$ and $P_2$ be processes such that 
$\sm{P_1}{TE} \not\subseteq \sm{P_2}{TE}$.
Then there is a context $C$ such that 
$\sm{C[P_1]}{IO} \not\subseteq \sm{C[P_2]}{IO}$.
\end{proposition}
\Proof
Let $\alpha=(s_1,s'_1)\cdots (s_n,s'_n)$ be a trace-environment sequence such that
$\alpha \in \sm{P_1}{TE}$ and $\alpha \notin \sm{P_2}{TE}$. 
We build an {\em observer process} $O$ that in a sense {\em plays the role
of the environment} and works as follows: 
\begin{quote}
upon observing $s'_1$ builds $s_2$ and\\
$\cdots$\\
upon observing $s'_{n-1}$ builds $s_n$ and terminates.
\end{quote}

\noindent
Notice that in this case the observer {\em does modify the state}.
Formally, assume $X=\fv{P_1}\union \fv{P_2}$.
The command that {\em builds a new state} is defined as follows:
\[
\w{MAKE}_{s,\set{x_1,\ldots,x_n}} = x_1:=s(x_1);\cdots ;x_n:=s(x_n)~,
\]
and the {\em observer process} $O$ is defined by:
\[
\begin{array}{lll}
O &\equiv &O_1 \\

O_i &\equiv &\await{\w{IS}(s'_{i},X)}{\w{MAKE}_{s_{i+1},X}} ; \\
  && \cdots \\
  && \await{\w{IS}(s'_{n-1},X)}{\w{MAKE}_{s_{n},X}}~.
\end{array}
\]
Then take as {\em process context} 
$C= [~] \mid O$  and let $C_i=[~]\mid O_i$.
We have that $(s_1,s'_n)\in \sm{C[P_1]}{IO}$
because:
\[
\begin{array}{l}
(C[P_1],s_1)\trarrow (C[P_2],s'_1)  \trarrow (C_2[P_2],s_2) \\
\cdots \\
(P_n \mid \sk,s_n) \trarrow (P'_n\mid \sk,s'_n) \downarrow 
\end{array}
\]
On the other hand, $(s_1,s'_n) \notin \sm{C[P_2]}{IO}$ because
$O$ terminates only if it can observe the states $s'_i$ and 
build atomically the states $s_{i+1}$ for $i=1,\ldots,n-1$. And
this contradicts the hypothesis that 
$(s_1,s'_1)\cdots (s_n,s'_n)\notin \sm{P_2}{TE}$. \qed

\section{The interpretation domain (*)}
The trace-environment interpretation introduced in chapter
\ref{trace-shared-sec} assigns a meaning (or denotation) to a process
which is formally a set of finite sequences of pairs of states.  
In the following sections, our
main task is to show that this meaning can be computed
in a {\em compositional} way in the sense that the denotation of a
program phrase can be built out of the denotations of its sub-phrases.
Concretely, this amounts to define a {\em domain} of interpretation, say $D$, and a collection
of functions on $D$ that correspond to the operators of the programming language.
For instance, we have to find a function {\sf par} on $D$ which corresponds
to parallel composition and satisfies:

\begin{equation}\label{comp-def}
\sm{P_1 \ \mid  \   P_2}{TE} = \sm{P_1}{TE}\  {\sf par} \  \sm{P_2}{TE}~.
\end{equation}

We denote with $\w{St}$ the set of {\em states}, {\em i.e.}, the
collection of total functions from identifiers 
to integers.
As a first step, we notice that the interpretation of a process
$\sm{P}{TE}$ belongs to the power-set $L=2^{(St\times St)^*}$ which when ordered
by set-theoretic inclusion is a {\em complete lattice} (cf. chapter \ref{equiv-sec}).
\begin{equation}
\sm{P}{TE}\in L = 2^{(St\times St)^*}~.
\end{equation}

\begin{definition}[closed set of traces]\label{closed-set-def}\index{closed set of traces}\index{closed set of traces}
We say that $X\in L$  is {\em closed} if it satisfies the following conditions:
\[
\begin{array}{cc}

\infer{\alpha \beta \in X}
{\alpha  (s,s) \beta \in X}~,

\qquad
&\infer{\alpha (s,s')(s',s'')\beta \in X}
{\alpha (s,s'')\beta \in X}~.

\end{array}
\]
\end{definition}

These are {\em a kind of reflexivity and transitivity} properties which are
called {\em stuttering and mumbling}, respectively, in the trace theory jargon.
Note that all process interpretations are closed and this property will
be used, {\em e.g.}, in the proof of proposition \ref{denot-chara-prop}.

\begin{definition}[closure function]\label{closure-fun-def}
The {\em closure function} $c:L\arrow L$ is defined by:
\[
c(X) = \Inter \set{Y \in L \mid X\subseteq Y, Y \mbox{ closed}}~.
\]
\end{definition}

Thus the function $c$ associates to a set $X$ 
the least set of closed traces that contains it.
We notice the following properties.

\begin{proposition}\label{closure-prop}
Let $X,Y,X_i$ vary over $L$ and let $c$ be the closure function. Then:
\begin{enumerate}

\item If $X\subseteq Y$ then $c(X)\subseteq c(Y)$.

\item $c(c(X))=c(X) \supseteq X$.

\item The union of closed sets is closed.

\item $c(\Union_{i\in I} X_i) = \Union_{i\in I} c(X_i)$.

\end{enumerate}
\end{proposition}
\Proof We leave properties 1-3 as exercises and consider property 4,
One inclusion  follows by monotonicity (property 1).
for the other,
we know from property 3 that $\Union_{i\in I} c(X_i)$ is closed.
Thus it suffices to check that: 
$\Union_{i\in I} X_i \subseteq \Union_{i\in I} c(X_i)$
which holds since by property 2, $X_i\subseteq c(X_i)$. \qed \\

It follows that  $(c(L),\subseteq)$ is  again a complete lattice where the sup are set-theoretic unions.
We take  $D=c(L)$ as our {\em domain of interpretation}.

\section{The interpretation (*)}\label{trace-denotation-sec}
First, we define some standard operations on the domain $D$
which are instrumental to the interpretation of $\parimp$ processes.
The reader may recognize definition patterns found in formal languages.

\begin{description}

\item[Skip] We define:
${\bf Skip} = c(\set{(s,s) \mid s\in \w{St}})\in D$.
Notice that this is different from the closure of the empty-set.

\item[Concatenation] For $X,Y\in D$ let
$X;Y = c(\set{\alpha\beta \mid \alpha \in X, \beta \in Y})\in D$.
Notice that we need to {\em close} the concatenation of 
$X$ and $Y$ in the ordinary language-theoretic sense.

\item[Iteration] For $X\in D$ let:
\[
\begin{array}{lll}
X^0 = {\bf Skip}\in D~,
&X^{n+1} = X;X^n\in D ~,
&X^*    = \Union_{n\geq 0} X^n\in D~.
\end{array}
\]
By proposition \ref{closure-prop}(3), there is no need to close the
countable union.

\item[Parallel] A general {\em shuffle operation $\mid$ on 
  words} can be defined as follows, $\epsilon$ being as usual the
  empty word:
\[
\begin{array}{l}

\epsilon \mid \alpha = \alpha \mid \epsilon = \set{\alpha} ~,\\

a\alpha \mid b \beta = 
\set{a \gamma \mid \gamma \in (\alpha \mid b\beta)} 
\union
\set{b \gamma' \mid \gamma' \in (a\alpha\mid \beta)}~.

\end{array}
\]
Notice that the shuffle of two words is a {\em set of words}
which is {\em not} necessarily closed.
Then define a {\em parallel operator} on $X,Y\in D$ as:
\[
X\mid Y = \Union_{\alpha \in X,\beta\in Y} c(\alpha \mid \beta)~.
\]
\end{description}

\begin{exercise}\label{cont-par-ex}
Show that the concatenation, iteration, and parallel operators
we have defined on the complete lattice $(D,\subseteq)$ are monotonic and 
preserve arbitrary unions.
\end{exercise}

We associate a closed set with a {\em boolean condition} $b$ (without side
effects) as follows:
\[
\begin{array}{ll}

\sem{b} &= c(\set{(s,s) \mid (b,s)\eval \s{true}})~.
\end{array}
\]
Intuitively, this is the closed set induced by the set of states
satisfying the boolean condition.
Then associate a closed set to processes as follows:
\[
\begin{array}{ll}

\sem{\sk} &={\bf Skip} \\

\sem{x:=e} &=c(\set{(s,s[n/x]) \mid (e,s)\eval n})  \\

\sem{P;P'} &= \sem{P};\sem{P'} \\

\sem{\s{if} \ b \ \s{then} \ P \ \s{else} \ P'} 
&= (\sem{b};\sem{P})\union (\sem{\neg b};\sem{P'}) \\

\sem{\while{b}{P}}
&= (\sem{b};\sem{P})^*;\sem{\neg b} \\

\sem{P\mid P'} &= \sem{P} \mid \sem{P'} \\

\sem{\await{b}{P}} &= c(\set{(s,s') \mid (s,s)\in \sem{b}, (s,s')\in \sem{P}})~.

\end{array}
\]
To force the {\em atomic} execution of the body of an \s{await} statement,
we select the traces of length $1$ which correspond to the input-output
behaviors (remark \ref{te-int-rmk}).

The extension to variable declarations requires some work.
Given $X\subseteq (\w{St}\times\w{St})^*$, $x$ variable, $n$ integer, define:
\[
\begin{array}{lll}

X[x=n] &= &\{(s_1,s'_1)\cdots (s_n,s'_n)\in X \mid \\
          &&s_1(x)=n, s_{i+1}(x)=s'_i(x), i=1,\ldots,n-1 \} \\ \\

X\backslash x &=&\{(s_1[m_1/x],s'_1[m_1/x])\cdots (s_n[m_n/x],s'_n[m_n/x]) \mid \\
              &&\quad (s_1,s'_1)\cdots (s_n,s'_n)\in X, \ 
                 m_1,\ldots,m_n\in \Z \}~.

\end{array}
\]
The operator $(\_)[x=n]$ fixes the initial value of $x$ to $n$ and makes sure
the environment cannot affect the value of $x$ by 
forcing $s_{i+1}(x)=s'_{i}(x)$.
The operator $(\_)\backslash x$ makes sure that the internal modifications of
$x$ are not observable by the environment (the value of the state at $x$ 
is never modified by a process transition). Then define:
\[
\sem{\local{x}{n}{P}} =  c( \ (\ \sem{P}[x=n] \ ) \backslash x \ )~.
\]
In words, first we select the traces where
the initial value of the variable $x$ is $n$ and the 
environment cannot affect $x$'s value and second we hide to the
environment the way $x$ is manipulated.

This concludes the compositional definition of the interpretation.
The reader can check that this interpretation does indeed follow
the pattern outlined in (\ref{comp-def}). Moreover, it turns out to be equivalent
to the operational interpretation.

\begin{proposition}[denotational characterization]\label{denot-chara-prop}
For all processes $P$, $\sm{P}{TE} = \sem{P}$~.
\end{proposition}
\Proof
The proof proceeds by induction on the structure of $P$.
As an example, we show:
\begin{equation}\label{seq-case-pr}
\sem{P;Q}=\sm{P;Q}{TE}
\end{equation}
assuming $\sem{P}=\sm{P}{TE}$ and $\sem{Q}=\sm{Q}{TE}$.

\Proofitemm{\sem{P;Q}\subseteq \sm{P;Q}{TE}.}
Suppose:
\[
\begin{array}{ll}
\alpha  &= (s_1,s'_1)\cdots (s_n,s'_n) \in \sem{P}=\sm{P}{TE} \\
\beta   &= (t_1,t'_1) \cdots (t_m,t'_m)\in \sem{Q}=\sm{Q}{TE}~.
\end{array}
\]
We have already observed that the operational interpretation of 
a process is closed.
Then by the properties of the closure operator 
(proposition \ref{closure-prop}), it suffices to show that
$\alpha\beta \in \sm{P;Q}{TE}$.
Indeed, we have:
\[
\begin{array}{lll}

(P;Q,s_1) &\trarrow &(P_1;Q,s'_1) \\
\cdots    &\trarrow &\cdots \\
(P_{n-1};Q,s_n) &\trarrow &(P_n;Q,s'_n) \arrow (Q,s'_n) \quad\mbox{(where: $P_n\downarrow$)} \\
(Q,t_1)   &\trarrow &(Q_1,t'_1) \\
\cdots &\trarrow &\cdots \\
(Q_{m-1},t_m) &\trarrow &(Q_m,t'_m)\downarrow~.

\end{array}
\]
\Proofitemf{\sem{P;Q}\supseteq \sm{P;Q}{TE}.}
Suppose $\gamma \in \sm{P;Q}{TE}$ is generated as follows:
\[
\begin{array}{lll}

(P;Q,s_1) &\trarrow &(P_1;Q,s'_1) \\
\cdots &\trarrow &\cdots \\
(P_{n-1};Q,s_n) &\trarrow &(Q_1,s'_n) \\
(Q_1,s_{n+1}) &\trarrow &(Q_2,s'_{n+1}) \\
\cdots       &\trarrow &\cdots \\
(Q_{m},s_{n+m}) &\trarrow &(Q_{m+1},s'_{n+m})\downarrow~.

\end{array}
\]
By the semantics of concatenation, we must also have the
following transitions:
\[
\begin{array}{lll}
(P_{n-1},s_n) &\trarrow &(P_n,s''_{n})\downarrow \\
(Q,s''_{n})   &\trarrow &(Q_1,s'_n)~.
\end{array}
\]
It follows that:
\[
\begin{array}{ll}

\alpha = (s_1,s'_1)\cdots (s_n,s''_n) \in \sm{P}{TE}=\sem{P}~, \\
\beta = (s''_n,s'_n)\cdots (s_{n+m},s'_{n+m}) \in  \sm{Q}{TE}=\sem{Q}~.
\end{array}
\]
Then $\alpha\beta \in \sem{P};\sem{Q}$ and by definition of closure (mumbling),
$\gamma \in \sem{P;Q}$. \qed \\

An immediate corollary is that the trace-environment interpretation
is preserved by process contexts.

\begin{corollary}
If $\sm{P}{TE}=\sm{P'}{TE}$ then $\sm{C[P]}{TE}=\sm{C[P']}{TE}$.
\end{corollary}
\Proof
For instance, if $\sm{P}{TE}=\sm{P'}{TE}$ 
then by proposition \ref{denot-chara-prop},
$\sem{P}=\sem{P'}$. Thus, for any $Q$:
$\sm{P\mid Q}{TE}=\sem{P\mid Q}=\sem{P'\mid Q}=\sm{P'\mid Q}{TE}$. \qed \\

Another interesting application of the characterization is that it provides an angle
to analyze process equivalence.

\begin{exercise}\label{valid-equiv-trace-ex}
Show that the following processes (in-)equivalences hold in the TE semantics:
\[
\begin{array}{cccccc}

\sk;P  &= &P ~,

&P &=  &P;\sk~, \\

(P;P');P'' &= &P;(P';P'') ~,

&P\mid P' &= &P'\mid P~, \\

(P\mid P')\mid P'' &= &P \mid (P'\mid P'') ~,

&P\mid \sk &= &P~, \\

\while{\s{true}}{\sk} &\leq &P ~,

&\while{\s{true}}{\sk} &= &\await{\s{false}}{P} ~.
\end{array}
\]
\end{exercise}

Thus $\sk$ is the unit for both sequential and parallel composition. Further, sequential composition
is associative while parallel composition is both associative and commutative. Finally, the diverging
computation is the least element of the interpretation.

\begin{exercise}[invalid equivalences]\label{invalid-equiv-ex}
Show that the following equivalences (which hold
in the sequential IO semantics) fail in the TE semantics, where
$x,y,z$ are distinct variables:
\[
\begin{array}{lcllcl}

(1)\quad x:=y;y:=x &= &x:=y~,

&(2) \quad x:=y;x:=z &= &x:=z~, \\

(3)\quad x:=y;z:=x &= &x:=y;z:=y~, 

&(4)\quad x:=y;z:=y &=  &z:=y; x:=y~.
\end{array}
\]
\end{exercise}

\begin{exercise}[\s{await} from \s{atomic}]\label{atomic-ex} %\marginpar{correction to be added}
In section \ref{parimp-sec}, we have regarded $\atomic{P}$ as 
an abbreviation for $\await{\s{true}}{P}$. 
Suppose we regard $\awaitt{b}{P}$ as an abbreviation for:
\[
\begin{array}{ll}
\s{var} \ x=1 \ \s{while} \ x=1 \ \s{do}  \ (
\atomic{\s{if} \ b \ \s{then} \ (P; x:=0)})
\end{array}
\]
where $x$ is a fresh variable. 
Show that the following equality holds in the considered semantics:
\[
\await{b}{P} = \awaitt{b}{P}~.
\]
\end{exercise}

\begin{exercise}[shuffling of infinite words]\label{infinite-shuffle-ex}
Let $\Sigma$ be an {\em alphabet} (a non-empty set) with generic elements $a,b,c,\ldots$
If $X$ is a set let
$X^\omega$ be the set of {\em infinite words} on $X$ (countable and not finite).
If $\alpha$ is a word then $\alpha^\omega$ is 
$\alpha \alpha \cdots$
We denote with $R,S,\ldots$ relations on $D=\Sigma^\omega \times \Sigma^\omega \times \Sigma^\omega$
and write $R(\alpha,\beta,\gamma)$ as an abbreviation for $(\alpha,\beta,\gamma)\in R$.
We say that a relation $R$ is {\em admissible} if:
\[
\begin{array}{ll}
R(\alpha,\beta,a\gamma) \mbox{ implies } & (\ \alpha = a\alpha' \mand R(\alpha',\beta,\gamma) \ ) \mbox{ or } \\
                                         & (\ \beta  = a\beta' \mand R(\alpha,\beta',\gamma) \ )~.
\end{array}
\]
We define:
\[
\begin{array}{lll}
S_0    &= D, \\
S_{n+1} &= \{ (\alpha,\beta,a\gamma) \mid 
       & (\ \alpha = a\alpha' \mand S_n(\alpha',\beta,\gamma) \ )  \mbox{ or } \\
       && (\ \beta  = a\beta' \mand S_n(\alpha,\beta',\gamma) \ )  \ \} ~,\\ 
S_\omega &= \inter_{n<\omega} S_n~.
\end{array}
\]
Problems:
(1) Show that there is a largest admissible relation that we denote with $\w{Shuffle}$.
(2) Prove or disprove: $\w{Shuffle} = S_\omega$.
(3) Prove or disprove: 
(i) $\w{Shuffle}(a^\omega,b^\omega, (ab)^\omega)$. 
(ii) $\w{Shuffle}((ab)^\omega,a^\omega, (abb)^\omega)$.
(iii) $\w{Shuffle}(a^\omega, b^\omega,a^\omega)$.

\end{exercise}

\begin{exercise}[fair schedules and associativity]\label{fair-assoc-ex}
A {\em $k$-schedule} is a vector $(f_1,\ldots,f_k)$
of $k$ functions on the natural numbers $\Nat$ such that:
\begin{itemize}

\item for $j=1,\ldots,k$ and $n\in \Nat$: $f_j(n)<f_j(n+1)$ 
(the functions are strictly growing).

\item for $i,j\in \set{1,\ldots,k}$ and $i\neq j$: $\w{im}(f_i)\inter \w{im}(f_j)=\emptyset$ (the ranges of the functions are disjoint).

\item $\Union_{j=1,\ldots,k} \w{im}(f_j) = \Nat$ 
(the union of the ranges covers the natural numbers).

\end{itemize}

Let $\Sigma$ be a non-empty set with generic elements $a,b,c,\ldots$
and let $\Sigma^\omega$ be the (countably) infinite words over 
$\Sigma$ with generic elements $\alpha,\beta,\ldots$.
If $\alpha\in \Sigma^\omega$ and $i\in \Nat$ then $\alpha[i]$
denotes the character at position $i$ of the word where we
start counting from $0$. For instance, 
if $\alpha=ababab\cdots$ then $\alpha[3]=b$.

If $(f_1,\ldots,f_k)$ is a $k$-schedule 
and $\alpha_i\in \Sigma^\omega$ for $i=1,\ldots,k$ then 
$M[f_1,\ldots,f_k](\alpha_1,\ldots,\alpha_k)$ 
is a word whose value at position $i\in \Nat$ is defined as follows:
\[
M[f_1,\ldots,f_k](\alpha_1,\ldots,\alpha_k)[i] = 
\alpha_j[f_{j}^{-1}(i)] \qquad \mbox{if }i\in \w{im}(f_j) 
\]
where $f_j^{-1}(i)$ denotes the (unique!) number that the
function $f_j$ maps to $i$.

\begin{enumerate}
\item Suppose $\alpha=a^\omega$ and $\beta=b^\omega$.
(i)  Assuming $f_1(i)=2\cdot i$ and $f_2(i)=2\cdot i+1$,
compute $M[f_1,f_2](\alpha,\beta)$.
(ii)  Is there a $2$-schedule $(f,g)$ such that 
$M[f,g](\alpha,\beta)=(aab)^\omega = aabaabaab\cdots$?
(iii) Is there a $2$-schedule $(f,g)$ such that
$M[f,g](\alpha,\beta)=ab^\omega = abbbbb\cdots$?

\item
Suppose $(f_1,f_2)$ and $(g_1,g_2)$ are two $2$-schedules.
Show that there is a $3$-schedule $(h_1,h_2,h_3)$ such that 
for all words $\alpha_i$, $i=1,2,3$ we have:
\[
M[g_1,g_2](M[f_1,f_2](\alpha_1,\alpha_2),\alpha_3) = 
M[h_1,h_2,h_3](\alpha_1,\alpha_2,\alpha_3).
\]

\item
Now suppose $(h_1,h_2,h_3)$ is a $3$-schedule.
Define two $2$-schedules $(f_1,f_2)$ and $(g_1,g_2)$
such that for all words $\alpha_i$, $i=1,2,3$ we have:
\[
M[g_1,g_2](M[f_1,f_2](\alpha_1,\alpha_2),\alpha_3) = 
M[h_1,h_2,h_3](\alpha_1,\alpha_2,\alpha_3)~.
\]

\item
We define a binary merge operation $M$ that associates 
a set of words to two words as follows:
\[
M(\alpha,\beta) = \set{M[f,g](\alpha,\beta) \mid 
(f,g) \mbox{ is a $2$-schedule}}~.
\]
We then extend the operation to sets of words by defining for
$X,Y\subseteq \Sigma^\omega$:
\[
M(X,Y) = \Union_{\alpha\in X, \beta \in Y} M(\alpha,\beta)~.
\]
Show that this merge operation is associative, {\em i.e.}, 
for all sets of words $X_i\subseteq \Sigma^\omega$, $i=1,2,3$:
\[
M(M(X_1,X_2),X_3) = M(X_1,M(X_2,X_3))~.
\]
\end{enumerate}
\end{exercise}

\section{Summary and references}
We have described a {\em trace-environment interpretation} 
for the $\parimp$ language.
The interpretation is compositional and abstract.
The key point for {\em compositionality} is that we describe the way
both the process and the environment 
may affect the store (which is what can be observed).
The key point for {\em abstraction} is that $\parimp$ 
can simulate the environment's actions; the \s{await} statement 
is crucial here.
The presentation is based on \cite{DBLP:journals/iandc/Brookes96}.

The trace-environment interpretation can be
organized in a denotational style where the meaning of a program (process) is
computed by composition of the meaning of its sub-programs.  This
makes manifest the compositionality of the interpretation.  We refer
the reader to \cite{DBLP:journals/iandc/Brookes96} for a variation
over the presented semantics which takes into account fairness
constraints. This requires working over infinite traces; exercises
\ref{infinite-shuffle-ex} and \ref{fair-assoc-ex} go in this direction 
by defining shuffling operations on infinite words.

\chapter{Implementing atomicity}\label{atomic-sec}
\markboth{{\it Atomicity}}{{\it Atomicity}}

The operational $\parimp$ model assumes the 
possibility of executing atomically a
process. A simple implementation strategy 
could consist in having a global {\em lock variable} that must be acquired 
by a process before turning into
`atomic mode' and is released upon termination (cf. example \ref{ex-pv}). 
Such a strategy is intuitively inefficient because it 
limits the degree of parallelism of the computation.
This intuition can be supported by a simple numerical argument
known as {\em Amdahl's law}.\index{Amdahl's law}
For instance, the law entails that if $10\%$ of a task has to be executed 
sequentially while the remaining $90\%$ can be executed in parallel
then by allocating $10$ processors to the task we can expect
a speed up of at most (roughly) $5$, {\em i.e.}, by multiplying the cost
of the hardware by $10$ we can only divide the computation time by $5$ 
(which is rather disappointing).

In the following we discuss some process transformations
that aim at reducing the amount of computation that has to be executed
atomically.  This should be regarded both as an opportunity to have a
glimpse at some basic implementation strategies and as a case study
where we practice the operational model.

\section{An optimistic strategy}
In an {\em optimistic} implementation strategy of an atomic
transaction mechanism we run the steps of the transaction concurrently
with those of other parallel processes hoping that they will not
affect the variables relevant to the transaction.  If they do then we
start again the transaction.  Intuitively, such an approach
works well if the chances that two atomic transactions try to modify
the same variables at about the same time are low.

In more detail, the transformation can be described as follows.
Given a process $P$, we can statically determine an over-approximation of the 
visible variables that $P$ may read or write during its execution.
For instance, this can correspond to the set $\fv{P}$ of variables
occurring free in $P$.
For each variable $x$ let us assume we dispose of fresh variables
$x_r$ and $x_l$. The super-script $r$ and $l$ stand for {\em read} 
and {\em local}, respectively, for reasons that we explain next.
Let us write $\vcb{x}$ for the list of distinct variables in $\fv{P}$ and
let us denote with $x^{*}_{r}$ and $x^{*}_{l}$ the corresponding lists of
fresh variables.
Rather than running $P$ atomically we run {\em non-atomically} a 
modified process $P'=[x^{*}_{l}/\vcb{x}]P$ where each 
read/write operation to the variables $\vcb{x}$ is replaced by a reference
to the fresh local variables $x^{*}_{l}$ which are initialized with the 
values of $\vcb{x}$. 
Before running $P$, we also save %(atomically) 
the initial value of the variables $\vcb{x}$
in the fresh local variables $x^{*}_{r}$.
If and when we are done  with the execution of $P'$ we check {\em atomically}
that the current value of $\vcb{x}$ equals that of $x^{*}_{r}$. If
this is the case, in the same atomic step 
we write $x^{*}_{l}$ in $\vcb{x}$ and we conclude 
successfully the transaction, otherwise we try again. Notice that it may happen
that the variables $\vcb{x}$ are modified during the computation above.
All that matters is that the value of $\vcb{x}$ is the same as the value of 
$x^{*}_{r}$ just before writing in $\vcb{x}$ the variables $x^{*}_{l}$.
In particular, they can be modified while initializing the variables
$x^*_r$.

The transformation can be described formally by a function $\cl{C}_o$ 
($o$ for optimistic) on $\parimp$ processes. The key case concerns the 
$\s{await}$ and it is defined as follows assuming $\fv{\await{b}{P}}=\set{x_1,\ldots,x_n}$,
$\vcb{x}=x_1,\ldots,x_n$,  and using vectorial notations such as  $\vcb{x}:=\vcb{v}$ 
and $x^{*}_{r}=\vcb{x}$ as an abbreviation for $x_1:=v_1;\cdots;x_n:=v_n$
and $x_{r,1}=x_1 \AND \cdots \AND x_{r,n}=x_n$, respectively.
We also assume that the variables $c,x^{*}_{r},x^{*}_{l}$ do not appear free in $\await{b}{P}$ and that
$P$ does not contain $\s{await}$ statements.
\begin{equation}\label{opt-comp-atomic}
\begin{array}{lll}

\co{\await{b}{P}} &= &\s{var} \ c=1, \  x^{*}_{r} = \vcb{0}, \  x^{*}_{l}=\vcb{0}  \\
                   &  &\s{while} \ c=1 \ \s{do}\\
                   && (\   x^{*}_{r} :=\vcb{x}; \\ 
                   &&  \ \ x^{*}_{l}:=x^{*}_{r}; \\
                   &  &\ \ \s{if} \ [x^{*}_{l}/\vcb{x}]b \ \s{then} \\
                   &&   \ \qquad ( \ [x^{*}_{l}/\vcb{x}]P; \\
                   &&   \ \qquad\ \   \s{atomic}(\s{if} \ x^{*}_{r}=\vcb{x} \ \s{then} \ \vcb{x}:=x^{*}_{l};c:=0;) \ ) \ )~.
\end{array}
\end{equation}
There are a number of possible variations on this schema.
For instance, one can distinguish the variables which are
read from those that are written.
In another direction, instead of computing an over-approximation of
the collection of variables which are affected by the atomic statement,
we could determine this set at {\em run time}.
Also, it should be noticed that in the translation (\ref{opt-comp-atomic})
above 
the computation of the process
$[x^{*}_{l}/\vcb{x}]P$  may operate on unexpected states which
in more complex programming settings may lead to exceptions
or diverging computations. Certain implementations of atomic transactions
ensure that the program always operates over consistent states,
{\em i.e.}, states which could actually arise in the reference semantics.
The following exercise elaborates on this point.

\begin{exercise}\label{atomic-abort-ex}
Suppose the $\parimp$ language is extended with a command
\s{abort}  which stops the computation and returns the current
state (such command was discussed in exercise \ref{cps-control-ex}).
Extend the optimistic compilation function so that it handles
\s{abort} commands.
\end{exercise}

\section{A pessimistic strategy}
A more pessimistic (or conservative) implementation strategy for
atomic transactions consists in gaining control of all the resources 
relevant to the atomic process before running it. 
For instance, suppose we associate a {\em lock variable} $\ell_x$ 
with every (shared) variable $x$. Recall that a {\em lock variable} is simply
a variable that is supposed to be used as a semaphore of capacity $1$ 
(see example  \ref{ex-pv}). 

As in the optimistic strategy, given a process $P$ we can statically determine 
an over-approximation of the variables the process $P$ may read or write during its execution.
Let us denote these variables with $x_1,\ldots,x_n$.
Then an implementation of $\s{atomic}(P)$ consists in a process that
acquires the locks for $x_1,\ldots,x_n$, then runs $P$, and eventually
releases the locks for $x_1,\ldots,x_n$. Such an implementation scheme
is known as {\em two phase locking}: the first phase is the one where
the process acquires the locks  and the second the one where it releases them.
This locking scheme can be refined by distinguishing between reading
and writing accesses. Indeed a {\em write} access must be exclusive but a {\em read} 
access can be shared by an arbitrary number of processes.
The function $\cl{C}_p$ formalizes this pessimistic transformation on $\parimp$ processes.
The key cases concern the variable declaration and the atomic statement:
\[
\begin{array}{ll}

\cp{\s{var} \ x=v \ \s{in} \ P} &= 
\s{var} \ x=v, \ell_x = 1 \ \s{in} \ \cp{P} \\

\cp{\s{atomic}(P)} &= \w{lock}(\ell_{x_{1}});\cdots ; \w{lock}(\ell_{x_{n}});P; 
                       \w{unlock}(\ell_{x_{1}});\cdots;\w{unlock}(\ell_{x_{n}})~. \\

\end{array}
\]
It should be noticed that  parallel processes running a two phase 
locking protocol may end up in a {\em deadlock}. 
For instance, suppose $P_1$ tries to acquire the locks
for $x_1$ and $x_2$ while $P_2$ tries to acquire the locks for $x_2$
and $x_1$. We can arrive at a deadlocked configuration where $P_1$ has
acquired the lock for $x_1$ and $P_2$ the lock for $x_2$. 
In general, one can represent a deadlock associated with locks as 
a circular waiting situation where all parallel processes which are not 
properly terminated are waiting to acquire a lock which is currently held
by another process.

An approach to {\em deadlock resolution} consists in introducing a {\em
  monitor process} that at appropriate times detects circular waiting
and breaks the circle by {\em aborting} one of the processes.  This
means that the selected process must release all the acquired locks
and start again.

Rather than taking action after the deadlock has happened, 
another approach consists in {\em preventing} it.
One basic approach that works if the locks can be {\em totally ordered}
consists in acquiring the locks in growing order.
A more general approach not requiring a total order consists in  introducing 
an information on the {\em age} of the atomic transactions.
For instance, the so called {\em wait-die} scheme works as follows.
If an older transaction tries to acquire a lock held by a younger 
transaction then it waits the lock is released, while if a 
younger transaction tries to acquire a lock held by an older one 
then it must release all the acquired locks and start again 
(while keeping its age).

\begin{exercise}\label{wait-die-ex}
Suppose the age of a transaction is a positive natural number.
Write pseudo-code for an \s{acquire} function that 
takes as input a list of locks and an age and tries
to acquire the locks following the wait-die strategy sketched above.
\end{exercise}

\section{A formal analysis of the optimistic strategy (*)}
We conclude this chapter by sketching a formal analysis of the optimistic
strategy. With reference to the trace-environment interpretation 
defined in section \ref{te-int-subsec}, one would like to show that for any 
$\parimp$ process $P$, we have $\sm{P}{TE} = \sm{\co{P}}{TE}$. 
We shall approach this problem through the notion of 
{\em simulation} which we have already met in chapter \ref{equiv-sec}.
Recall that $\downarrow$ is a predicate on programs/processes that defines immediate termination and $\arrow$ is a binary relation that defines the 
small-step reduction of $\parimp$. 
As usual, we denote with $\trarrow$ the 
reflexive and transitive closure of $\arrow$.

\begin{definition}\label{def-weaksim-parimp}
A binary relation $\cl{R}$ on $\parimp$ processes is a {\em weak simulation} if whenever $P\rel{R} Q$ the following holds for any
state $s$:
\begin{itemize}

\item if $P\downarrow$  then $\xst{Q'}{(\ (Q,s) \trarrow (Q',s)$ and $Q'\downarrow \ )}$.

\item if $(P,s) \arrow (P',s')$ then $\xst{Q'}{(\ (Q,s) \trarrow (Q',s') \mand P'\rel{R} Q' \ )}$.
\end{itemize}
\end{definition}

We denote with $\leq$ the union of all weak simulations. The reader
may check that this is again a weak simulation. Also we notice the following
properties.

\begin{proposition}\label{weak-sim-parimp-prop}
Let $P,Q$ be $\parimp$ processes. Then:

\Defitem{(1)} If $P\leq Q$ then $\sm{P}{TE} \subseteq \sm{Q}{TE}$.

\Defitem{(2)} The reverse implication does not hold.

\Defitem{(3)} If $P\leq Q$ then for any process $R$, 
$(P\mid R)\leq (Q\mid R)$.

\Defitem{(4)} If $P\leq Q$ then for any variable $x$ and 
integer value $n$, $\local{x}{n}{P} \leq \local{x}{n}{Q}$.

\end{proposition}
\Proof
\Proofitemf{(1)}
Suppose $P\leq Q$ and $\alpha\in \sm{P}{TE}$.
We proceed by induction on the length of the trace $\alpha$ to show that
$\alpha \in Q$.

\begin{description}

\item[$\alpha=(s_1,s'_1)$]
This means $(P,s_1) \trarrow (P_2,s'_1)$ and $P_2\downarrow$.
Then by repeatedly applying the second condition defining a simulation we have:
\[
(Q,s)\trarrow (Q_2,s'_1) \mand P_2 \leq Q_2~.
\]
Also by the first condition $(Q_2,s'_1)\trarrow (Q'_2,s'_1)$ and $Q'_2 \downarrow$.
Thus:
\[
(Q,s) \trarrow (Q'_2,s'_1)\downarrow
\]
which means $\alpha\in \sm{Q}{TE}$.

\item[$\alpha =(s_1,s'_1)\alpha',\alpha'\neq \epsilon$]
This means $(P,s_1) \trarrow (P_2,s'_1)$ and $\alpha'\in \sm{P_2}{TE}$.
Then:
\[
(Q,s_1) \trarrow (Q_2,s'_1)\mand P_2\leq Q_2~.
\]
By inductive hypothesis, $\alpha'\in \sm{Q_2}{TE}$.
It follows $\alpha \in \sm{Q}{TE}$.
\end{description}

\Proofitemm{(2)}
Recall that a non-deterministic sum can be defined in $\parimp$
(example \ref{choice-enc-ex}). Then we consider:
\[
\begin{array}{lll}
P & \equiv &(a=0) \arrow a:=1; [(b=0) \arrow b:=1 + (c=0) \arrow c:=1] \\
Q &\equiv  &[(a=0) \arrow a:=1; [(b=0)\arrow b:=1] \ + \ (a=0) \arrow a:=1; [(c=0)\arrow c:=1]]~.
\end{array}
\]
It is intended that only the first assignment after the test-for-zero is executed atomically.
Thus for instance:
\[
(P,[0/a]s) \arrow  \arrow ([(b=0) \arrow b:=1 + (c=0) \arrow c:=1], \ [1/a]s)
\]
Moreover notice that:
\[
([(b=0) \arrow b:=1 + (c=0) \arrow c:=1], \ [0/b,0/c]s) \arrow
(P',[1/b,0/c]s), (P'',[0/b,1/c]s)~,
\]
where $P',P''\downarrow$. On the other hand, $Q$ cannot simulate the
first step of $P$. If it takes the first branch it cannot modify $c$
and if it takes the second it cannot modify $b$.
Another possibility is to notice that in the trace-environment
semantics all looping processes are interpreted as the empty set while
the weak simulation semantics may distinguish two looping processes
such as $P= \s{while} \ \s{true} \ \s{do} \ x:=1$ and $Q=\s{while}
\ \s{true} \ \s{do} \ \s{skip}$.  Indeed, we have $P\not\leq Q$: the
move $(P,[0/x]s) \arrow (P,[1/x]s)$ cannot be matched by $Q$.

\Proofitemm{(3)}
We show that the following relation $\cl{R}$ is a weak simulation:
\[
\cl{R} = \leq \union \set{(P\mid R,Q\mid R) \mid P\leq Q, R \mbox{ process}}~.
\]
Suppose $P\leq Q$ and $(P\mid R,s) \arrow (P'\mid R',s')$. 
We analyze the two possible cases:

\begin{description}

\item[$(P,s)\arrow (P',s'), R'=R$]
Then $(Q,s) \trarrow (Q',s')$ and $P'\leq Q'$.
So $((Q\mid R),s) \trarrow ((Q'\mid R),s')$ and $(P'\mid R) \rel{R} (Q'\mid R)$.

\item[$(R,s)\arrow (R',s'), P'=P$]
Then $((Q\mid R),s) \arrow ((Q\mid R'),s')$ and $(P\mid R') \rel{R} (Q\mid R')$.

\end{description}

\Proofitemm{(4)}
We show that the following relation $\cl{R}$ is a weak simulation:
\[
\cl{R} = \leq
\union \set{(\local{x}{n}{P},\local{x}{n}{Q}) \mid P\leq Q, n\in \Z}~.
\]
Suppose $P\leq Q$ and 
$(\local{x}{n}{P},s) \arrow (\local{x}{m}{P'},s'[s(x)/x])$ because
$(P,s[n/x]) \arrow (P',s'[m/x])$.
Then $(Q,s[n/x])\trarrow (Q',s'[m/x])$ and $P'\leq Q'$.
Thus $(\local{x}{n}{Q},s) \trarrow (\local{x}{m}{Q'},s'[s(x)/x])$ and 
$(\local{x}{m}{P'}) \rel{R} (\local{x}{m}{Q'})$. \qed

\begin{exercise}\label{seq-cong-sim-ex}
Show that if $P_i\leq Q_i$ for $i=1,2$ then $P_1;P_2\leq Q_1;Q_2$.
\end{exercise}

Proof techniques for simulation (and bisimulation) are 
developed in the more abstract setting of labelled transition systems
in chapter \ref{lts-sec}. For the time being, we recall
from chapter \ref{equiv-sec}
that to show that $P\leq Q$ it suffices to exhibit
a relation $\cl{R}$ which contains the pair $(P,Q)$ and which is 
a weak simulation. As an application of this technique, 
let us show the following.

\begin{proposition}\label{prop-optimistic}
Let $P$ be a $\parimp$ process then $P\leq \co{P}$.
\end{proposition}
\Proof 
We consider the relation:
\[
\begin{array}{ll}
\cl{R}= & \set{(P,\co{P}) \mid P \ \parimp \ \mbox{ process}} \union  
        \set{(P,Q) \mid P\downarrow, Q\downarrow}~.
\end{array}
\]
We have to check that whenever $(P,Q)\in \cl{R}$ then the two conditions
specified in the definition \ref{def-weaksim-parimp} above hold.
For the first condition, we check that if $P\downarrow$ then $\co{P}\downarrow$
by induction on the definition of immediate termination.
For the second condition, we proceed by induction on the reduction
$(P,s) \arrow (P',s')$ according to the rules specified in table \ref{parimp-smallstep} of chapter \ref{introduction-sec}. 
The only interesting case is when $P\equiv \await{b}{P_1}$, $(b,s)\eval \s{true}$ and $(P_1,s) \trarrow (P',s')$ with $P'\downarrow$.
First, we need a lemma that relates the reductions of $(P_1,s)$ to those
of $([x^{*}_{l}/\vcb{x}]P_1,s[\vcb{s(x)}/x^{*}_{l}])$.
Then one exhibits a sequence of reductions such that 
$(\co{P},s) \trarrow (Q,s')$ and $Q\downarrow$.
Now in general it is not true
that $Q \equiv \co{P'}$, and this is precisely the reason we enlarged
the definition of $\cl{R}$ to include all the pairs of immediately
terminated processes. \qed \\

It follows from propositions \ref{weak-sim-parimp-prop} and 
\ref{prop-optimistic} that $\sm{P}{TE}\subseteq \sm{\co{P}}{TE}$.  
For the sake of simplicity, we discuss the reverse inclusion
in a particular case.

\begin{proposition}\label{prop-optimistic-reverse}
Suppose  $P\equiv \await{b}{P_1}$ is a $\parimp$ process and $P_1$ 
does {\em not} contain parallel composition, \s{while} and 
\s{await} statements. Then $\co{P}\leq P$.
\end{proposition}
\Proof By the hypotheses on $P_1$, for any state $s$,
the reduction of $(P_1,s)$ is deterministic and terminates.
So there exist unique $P'_1$ and $s'$ such that
$(P_1,s)\trarrow (P'_1,s')$ and $P'_1\downarrow$.

We write $Q\Downarrow$ if for all states $s$,
all reductions starting from $(Q,s)$
terminate to a configuration $(Q',s')$ such that
$Q'\downarrow$ and $s=s'$.

The pair $(P,s)$ can either loop on itself if $(b,s)\eval \s{false}$ or
move to $(P'_1,s')$ if $(b,s)\eval \s{true}$.
On the other hand, we claim that if $(\co{P},s)\trarrow (Q,s'')$ then
either $Q$ has the shape $\local{c}{1}{Q'}$ and $s=s''$ or
$Q$ has the shape $\local{c}{0}{Q'}$, $\local{c}{0}{Q'} \Downarrow$,
and $s'=s''$. Then the rough idea is to define a simulation that
relates the processes of the first type to $P$ and those
of the second type to $P'_1$. \qed

\begin{exercise}\label{proof-prop-optimistic-reverse}
Complete the proof of proposition \ref{prop-optimistic-reverse}.
\end{exercise}

\section{Summary and references}\label{summary-atomic}
Atomicity is a major issue in concurrency theory
starting from early work on the implementation of atomic transactions
in databases
\cite{DBLP:journals/jacm/Papadimitriou79b,Bernstein87,Lynch:1993:ATC:562590}.
Later, related concepts have been developed in the framework of
concurrent programming. In particular, let us mention the notion of
{\em concurrent object} and {\em linearizability}
\cite{DBLP:journals/toplas/HerlihyW90} and the related results that
classify the synchronization power of various concurrent objects
\cite{DBLP:journals/toplas/Herlihy91} (see also chapter \ref{conc-obj-sec}).  
Nowadays, the various strategies to implement atomicity we 
have discussed are applied to standard programming languages 
({\it C++}, $\java$, {\it Haskell}, $\ml$,$\ldots$).
In particular, the work on so called hardware/software 
transactional memories 
\cite{DBLP:conf/isca/HerlihyM93,DBLP:conf/podc/ShavitT95}
is mainly concerned with the problem of finding 
an efficient implementation of the \s{atomic} operator.
Amdahl's law is presented in \cite{DBLP:conf/afips/Amdahl67}. 
An early and quite readable description of the optimistic strategy 
in the framework of database systems can be found in 
\cite{DBLP:journals/tods/KungR81}.

\chapter{Rely-guarantee reasoning}\label{rely-guarantee-sec}
\markboth{{\it Rely-guarantee assertions}}{{\it Rely-guarantee assertions}}

We have seen in chapter \ref{trace-shared-sec} that 
the {\em semantics} of concurrent processes calls for new techniques.
Not surprisingly, a similar and related phenomenon arises
in the {\em specification} of concurrent processes.
In chapter \ref{intro-sec}, we have introduced the notion
of partial correctness assertion (pca).
Table \ref{hoare-seq} gives the rules to reason on a
sequential fragment of the $\parimp$ language. 
These rules are {\em sound} (proposition \ref{hoare-seq}) and can be 
{\em inverted} (proposition \ref{rule-inv-prop}) 
thus providing a syntax-directed method to reduce a pca 
to an ordinary logical statement.
Is it possible to extend these results to the $\parimp$ language?

\section{Rely-guarantee assertions}
We recall and extend some of the notation introduced in section
\ref{pca-sec} to reason on pca.
We associate with a program $P$ the input-output 
relation on states:
\[
(s,s')\in \sm{P}{IO} \ \mbox{if } \ (P,s)\eval s'~.
\]
For example, in the case $P$ is an assignment $x:=e$, we have:
\[
\sm{P}{IO} = \set{(s,s[v/x]) \mid (e,s)\eval v}~,
\]
which turns out to be (the graph of) a total function.

In the assertions, we identify a boolean predicate $b$ 
with the set of states that satisfy it, thus $b$ stands for
$\set{s \mid s\models b}$.
We denote the set of states with $\w{St}$, 
unary relations on $\w{St}$ with $A,B,\ldots$ and
binary relations on $\w{St}$ with $R,G,\ldots$. 
To manipulate relations on states, 
we  use the following notation:
\[
\begin{array}{lll}
\w{Id} &=\set{(s,s)\mid s\in \w{St}} &\mbox{(identity relation)}\\
\w{Top}  &= \set{(s,s') \mid s,s'\in \w{St}} &\mbox{(top relation)}\\

A;R  &= \set{s' \mid \xst{s}{s\in A \mand (s,s')\in R}} &\mbox{(image)}\\
R;A  &= \set{s \mid \xst{s'}{s'\in A \mand (s,s')\in R}} &\mbox{(pre-image)}\\
R;R' &=\set{(s,s'') \mid \xst{s'}{(s,s')\in R \mand (s',s'')\in R'}} &\mbox{(composition).}
\end{array}
\]
As usual, if $R$ is a relation then $R^*$ is its reflexive and transitive closure.

As mentioned above, the generation
of the logical conditions follows the structure of the program 
(proposition \ref{rule-inv-prop}).
One would like to follow this pattern for the concurrent programs
of the $\parimp$ language too.
So let us focus on parallel composition, which is the core of the matter,
and let us try to formulate a rule of the shape:
\[
\infer{\hoare{A_i}{P_i}{B_i}\quad i=1,2}
{\hoare{f(A_1,A_2)}{P_1\mid P_2}{g(B_1,B_2)}}~,
\]
where $f$ and $g$ are two ways of combining predicates.
Take:
$P_1 \equiv  x:=1;x:=x+1$ and $P_2 \equiv x:=2$.
We already know from example \ref{ex-noncomp-io} 
that:
\[
\sm{P_1}{pca}=\sm{P_2}{pca} \quad \mbox{ and }\quad
\sm{P_1\mid P_2}{pca}\neq \sm{P_2\mid P_2}{pca}~.
\]
In particular, this means that any derivation that would
end with a proof of the shape:
\[
\infer{\begin{array}{c}
\hoare{A_1}{P_2}{B_1}\quad \hoare{A_2}{P_2}{B_2}
\end{array}}
{\hoare{\s{true}}{P_2\mid P_2}{x\leq 2}}~,
\]
where 
$\models g(B_1,B_2) \subseteq \set{s \mid s(x)\leq 2}$ could be turned
into a derivation of the triple
$\hoare{\s{true}}{P_1\mid P_2}{x\leq 2}$
which is obviously {\em not} valid! 
An early approach to this problem goes back to Owicki and Gries.
Their rule has the shape:
\begin{equation}\label{owicki-rule}
\infer{\hoare{A_1}{P_1}{B_1}\quad \hoare{A_2}{P_2}{B_2}}
{\hoare{A_1\inter A_2}{P_1\mid P_2}{B_1\inter B_2}}
\end{equation}
provided the proofs of the premises do not `interfere'.
Having to look at the internal structure of processes is not
very satisfying and it is clearly at odd with one basic principle
of module composition: 
to compose proofs (modules) one should just know
what is proved (the interface)  without 
depending on the details of the proof (the implementation).
A way to tackle these limitations
is  to consider a richer specification 
language whose judgments have the shape:
\begin{equation}\label{rely-guarantee-asrt}\index{rely-guarantee assertion}
\jones{P}{A}{R}{G}{B}
\end{equation}
where:
(1) $A$ and $B$ are a pre-condition and a post-condition, respectively 
as in Floyd-Hoare rules, hence sets of states,
(2) $R$ is a {\em relation on states} that describes the
environment transitions that are admitted (thus $R$ is part of the 
pre-conditions), and
(3) $G$ is {\em relation on states} that describes
the program transitions that are guaranteed (thus $G$ is part of the
post-condition). 
We refer to assertions of the shape (\ref{rely-guarantee-asrt}) as
{\em rely-guarantee assertions}, or rga for short.  
To define their validity, we recall
from chapter \ref{trace-shared-sec} that 
$\parimp$ programs may perform the following {\em labelled transitions}:
\[
\begin{array}{ll}
(P,s)\act{p} (P',s')  &\mbox{(program transition, cf. Table \ref{parimp-smallstep})} \\
(P,s) \act{e} (P,s')  &\mbox{(environment transition, cf. rule (\ref{env-step-rule})).}
\end{array}
\]
\begin{definition}[computation]
A {\em computation} of a program $P_0$ is a (finite or infinite) sequence:
\begin{equation}\label{computation-shape}
(P_0,s_0) \act{\lambda_0} (P_1,s_1) \act{\lambda_1} (P_2,s_2) \act{\lambda_2}\cdots
\end{equation}
where $s_i$ are states and $\lambda_i\in \set{p,e}$.
\end{definition}

\begin{definition}[validity rely-guarantee]\index{rely-guarantee assertion}
The rely-guarantee assertion $\jones{P}{A}{R}{G}{B}$ is valid if {\em for all 
computations} of $P=P_0$ of the shape (\ref{computation-shape}) 
such that the following {\em pre-condition} holds:
\[
s_0\in A \mbox{ and } \qqs{i}{(\lambda_i=e \mbox{ implies } (s_i,s_{i+1})\in R)}~,
\]
it follows that the following {\em post-condition} holds:
\[
\qqs{i}{(P_i\downarrow \mbox{ implies } s_i\in B) \mbox{ and } (\lambda_i = p \mbox{ implies } (s_i,s_{i+1})\in G)}~.
\]
\end{definition}

Thus the pre-condition concerns the initial configuration and all
transitions performed by the environment (including those after termination) 
while the post-condition concerns the final configurations (if any) and all the transitions
performed by the program.

\begin{exercise}\label{hoare-vs-rga-ex}
The validity of a given pca is equivalent to the validity of a derived 
rga. Specifically, show that the pca $\hoare{A}{P}{B}$ is valid iff the rga  
$\jones{P}{A}{\w{Id}}{\w{Top}}{B}$ is valid.
\end{exercise}

\begin{remark}
Rga's can discriminate programs which are trace-environment equivalent.
For instance, consider the programs $\s{loop}\equiv \while{\s{true}}{\sk}$ and 
$P\equiv x:=0;\s{loop}$.  We know that 
$\sm{\s{loop}}{TE}=\sm{P}{TE}$, since all the diverging computations
receive the empty interpretation.
On the other hand, the  rga $(\s{true},\w{Id},\w{Id},\s{false})$ is satisfied by
$\s{loop}$ but not by $P$ because the assignment $x:=0$ does not 
respect the guarantee condition $\w{Id}$.
\end{remark}

\begin{definition}[stability]\index{stability}
Let $A$ be a unary relation and $R$ a binary relation on some set.
Then we write $\stable{A}{R}$ if $s\in A$ and $(s,s')\in R$ implies
$s'\in A$ and say that $A$ is stable with respect to $R$.
\end{definition}

\begin{exercise}\label{stability-ex}
Let $A,B$ be unary relations and $R$ be a binary relation. Show that:

\begin{enumerate}
\item $\stable{A}{R}$ iff $A;R^* \subseteq A$.

\item $A;R^* \subseteq B$ iff $\ \xst{A'}{(A\subseteq A', \stable{A'}{R}, A'\subseteq B)}$.

\end{enumerate}
\end{exercise}

Table~\ref{rely-guarantee} \index{rely-guarantee rules} 
provides a collection of rules to derive 
valid rely-guarantee assertions, with the proviso that
the guarantee relation $G$ in the conclusion of the rules
contains the identity relation $\w{Id}$. Without this hypothesis,
the rules are unsound. For instance, one can derive 
$\sk;\sk:(\s{true},\w{Top},\emptyset,\s{true})$ from
$\sk:(\s{true},\w{Top},\emptyset,\s{true})$.
Also notice that we do not include a rule for the local variables.
This can be done at the price of some technicalities which are not
essential for the following discussion.

\begin{table}
{\footnotesize
\[
\begin{array}{cc}

\infer{\begin{array}{c}
A\subseteq A'\quad R\subseteq R'\\
B'\subseteq B\quad G'\subseteq G \\
\jones{P}{A'}{R'}{G'}{B'}
\end{array}}
{\jones{P}{A}{R}{G}{B}}

&\infer{\begin{array}{c}
\jones{P}{A}{R}{G}{C}\\
\jones{Q}{C}{R}{G}{B}
\end{array}}
{\jones{P;Q}{A}{R}{G}{B}} \\\\

\infer{\begin{array}{c}
\jones{P}{A\inter b}{R}{G}{B}\\
\jones{Q}{A\inter \neg b}{R}{G}{B}
\end{array}}
{\jones{\ite{b}{P}{Q}}{A}{R}{G}{B}}

&\infer{A\subseteq B \quad \stable{A}{R}}
{\jones{\sk}{A}{R}{G}{B}} \\ \\

\infer{\begin{array}{c}
\stable{A}{R}\quad \stable{B}{R} \\
A;\sm{x:=e}{IO} \subseteq B \\
\sm{x:=e}{IO} \subseteq G
\end{array}}
{\jones{x:=e}{A}{R}{G}{B}}

&\infer{\begin{array}{c}
\stable{A}{R} \quad \stable{B}{R} \\
(A \inter \neg b) \subseteq B \\
\jones{P}{A \inter b}{R}{G}{A}
\end{array}
}
{\jones{\while{b}{P}}{A}{R}{G}{B}} \\ \\

\infer{\begin{array}{c}
\stable{A}{R} \quad \stable{B}{R} \\
b;\sm{P}{IO}
\subseteq G \\
\jones{P}{A\inter b}{\w{Id}}{\w{Top}}{B}
\end{array}
}
{\jones{\await{b}{P}}{A}{R}{G}{B}}

&\infer{\begin{array}{c}
(R\union G_1) \subseteq R_2\quad
(R\union G_2) \subseteq R_1 \\
(G_1\union G_2) \subseteq G \quad 
(B_1 \inter B_2)\subseteq B \\
\jones{P}{A}{R_1}{G_1}{B_1}\\
\jones{Q}{A}{R_2}{G_2}{B_2}
\end{array}}
{\jones{P\mid Q}{A}{R}{G}{B}}~.

\end{array}
\]
}
\caption{Rely-guarantee rules for $\parimp$ (assume $G\supseteq \w{Id}$)}\label{rely-guarantee}
\index{rely-guarantee rules}
\end{table}

\begin{proposition}[soundness rga rules]\label{sound-rga-prop}
The rely-guarantee assertions derivable in the system in Table~\ref{rely-guarantee} are valid.
\end{proposition}
\Proof
We present the argument for the rule for parallel composition.
A computation of the shape:
\[
(P\mid Q,s_0) \act{\lambda_0} (P_1\mid Q_1,s_1) \act{\lambda_1} \cdots
\]
can be turned into a computation of $P$ where the moves played
by $Q$ are actually attributed to the environment:
\begin{equation}\label{comp-P}
(P,s_0) \act{\lambda'_0} (P_1,s_1) \act{\lambda'_1} \cdots
\end{equation}
with $\lambda'_i=e$ if $\lambda_i=e$ or $\lambda_i=p$ and the $i^{\w{th}}$ move
is due to $Q$, and $\lambda'_i=p$ otherwise.
By a symmetric argument, we derive a computation for $Q$ too:
\begin{equation}\label{comp-Q}
(Q,s_0) \act{\lambda''_0} (Q_1,s_1) \act{\lambda''_1} \cdots
\end{equation}
We claim that:
\[
\begin{array}{lll}
\lambda'_i = p &\mbox{implies} &(s_i,s_{i+1})\in G_1 \\
\lambda''_i= p &\mbox{implies} &(s_i,s_{i+1})\in G_2~.
\end{array}
\]
We reason by contradiction and take $j$ to be the least natural number
where the property above fails. For instance, suppose:
$\lambda'_j=p$ and $(s_j,s_{j+1})\notin G_1$. 
Now for a transition $\lambda'_i$ with $i<j$ we have $3$ cases:
either $\lambda'_i=p$ and $(s_i,s_{i+1})\in G_1$,
or $\lambda_i=e$, $\lambda'_i=e$ and $(s_i,s_{i+1})\in R$,
or $\lambda_i=p$, $\lambda''_i=p$ and $(s_i,s_{i+1})\in G_2$ 
because $i<j$.
In particular, if $\lambda'_i=e$ we have that 
$(s_i,s_{i+1})\in R\union G_2\subseteq R_1$.
Then the following computation of $P$:
\[
(P,s_0) \act{\lambda'_0} \cdots (P_j,s_j) \act{p} (P_{j+1},s_{j+1}) 
\]
contradicts the hypothesis $P:(A,R_1,G_1,B)$.
It follows that if $\lambda_i=p$ then $(s_i,s_{i+1})\in G_1\union G_2 \subseteq G$.
Finally, we notice that $(P_i\mid Q_i)\downarrow$ entails $P_i\downarrow$ and
$Q_i\downarrow$. Thus $s_i\in B_1\inter B_2\subseteq B$. \qed

\begin{exercise}\label{proof-rga-ex}
Prove the pca $\hoare{x=0}{x:=1;x:=x+1 \mid x:=2}{x\in \set{2,3}}$ using the
rely-guarantee system.
\end{exercise}

\begin{exercise}\label{proof-complete-rga-sys-ex}
  Complete the proof of proposition \ref{sound-rga-prop}.
\end{exercise}

Unfortunately, the move from pca's to rga's is not quite sufficient
to reason about concurrent programs. For instance, it is problematic
to prove simple assertions such as:
\begin{equation}\label{need-aux-var-ex}
\hoare{x=0}{x:=x+1 \mid x:=x+1}{x=2}~.
\end{equation}
The problem is that in the rule for parallel composition we need
to {\em abstract} the possible state transformations of the parallel processes
into a relation on states. In doing this, we lose information. For instance,
we can record the fact that the variable $x$ can be incremented by one,
but we lose the information that this state transition can occur {\em at most
once}. Further, in slightly more complicated programs such as 
 $x:=x+1;x:=x+2$, one loses information on the {\em order}
of the state transformations too.

A `solution' which goes back to the Owicki-Gries system is to allow for an {\em
  instrumentation} of the program. This means enriching the program
with auxiliary variables and assignments which allow to record
(essential parts of) the history of the computation without affecting
it. The `auxiliary variable' rule states that if we can prove a
rely-guarantee assertion of the instrumented program then we can
transfer this property to the original program where the
instrumentation is removed. For instance, with reference to the pca 
(\ref{need-aux-var-ex}) above we can prove:
\begin{equation}\label{aux-stat}
\hoare{x=0,p=0,q=0}{\atomic{x:=x+1;p:=1} \mid \atomic{x:=x+1;q:=1}}{x=2}
\end{equation}
and then derive the desired pca (\ref{need-aux-var-ex})
by erasing the auxiliary variables $p$ and $q$ along 
with the related assignments
and atomic statements. In practice, one can insert {\em program counters}
in all processes and describe exactly in the assertions 
the way the computation progresses.

It turns out that when adding an {\em auxiliary variable rule}, it is
possible to invert the rules presented in Table~\ref{rely-guarantee}
along the spirit of proposition \ref{rule-inv-prop}.  
However, as for
the non-interference rule (\ref{owicki-rule}), this is not very satisfying
and really points to a weakness of the specification language. 
A more
powerful and flexible approach consists in building a full fledged
modal logic that allows to describe the transitions of the program and
the environment.
Examples of such modal logics are discussed later in
chapter \ref{modal-sec} in the more abstract framework of labelled
transition systems.

\section{A coarse grained concurrent garbage collector (*)}
We suppose the reader is familiar with the idea that a
program may need to allocate memory at run time and that 
such memory should be collected and reused whenever possible
so that the program can carry on its execution within 
certain given memory bounds.

In general, it is hard to predict when a 
memory block becomes useless to the rest of the computation
and can be collected. One approach to this issue consists
in designing a specific program called {\em garbage collector}
which periodically analyzes the state of the memory and 
collects the blocks which are useless. Specifically, 
the memory is modelled as a directed graph with a collection
of root nodes which are the entry points of the program to the memory.
All nodes which are not {\em accessible} from the roots are considered
as {\em garbage} and can be collected. Thus, at least from a 
logical point of view, the activity of the garbage collector can 
be decomposed in two phases: a {\em marking} phase where 
the accessible nodes are determined and 
a {\em collecting} phase where the  inaccessible nodes 
become again available for future usage. In practice,
it is convenient to relax a bit this specification. Namely,
one determines an {\em over-approximation} of the accessible nodes and 
consequently one collects a {\em subset} of the inaccessible ones.

Going towards a formalization, let us assume a fixed set of 
nodes $N$ and a fixed subset of roots $\w{Roots}\subseteq N$.
We also use $E\subseteq N\times N$ to denote the collection of directed edges
which varies over time. For any given collection of edges $E$,
we have a collection of nodes which are accessible from the roots:
$\w{Acc}(E) = \w{Roots};E^*$.

The activity of the  program, henceforth called \w{Mutator},  
on the memory graph can
be summarized as follows: it selects two accessible nodes $i,j$ 
and redirects an outgoing edge of the first node, say $(i,k)$, 
towards $j$. In other terms, the edge $(i,k)$ is replaced by the 
edge $(i,j)$, where $i,j,k$ are not necessarily distinct. 
The fundamental property which is
guaranteed by the \w{Mutator} is that the collection of accessible nodes
can only decrease. Thus if we denote with  $E$ and $E'$
the collection of edges before and
after a \w{Mutator}'s action we have that:
\begin{equation}\label{mon-mutator}
\w{Acc}(E')\subseteq \w{Acc}(E)~.
\end{equation}
This representation of the \w{Mutator} seems very simple but it is 
actually reasonable provided we assume 
that among the root nodes  there is: (i) a special node 
called $\s{nil}$  without outgoing edges and (ii) a special node
called $\s{fl}$ (free list) that points to the list of free nodes 
that can be used to allocate memory. 
With these hypotheses, expected operations of the \w{Mutator} 
such as setting a pointer to \s{nil} or redirecting a pointer
towards a newly allocated node, fall within the scope of the model.

The activity of the garbage collector is a bit more complex.
The task of the marking phase is to determine a set $M \subseteq N$ which
over-approximates the collection of reachable nodes. A natural way
to approach the task is to compute iteratively the least fixed point
of the function associating to a set of nodes $M$ the 
set of nodes $\w{Roots} \union M;E$. This can be expressed 
in an imperative programming notation as follows:

{\footnotesize\[
\begin{array}{lll}
  \w{Marker} \equiv  &M:= \emptyset \\
                     &M_1:=\w{Roots} \\
                     &\s{while} \ M_1 \not\subseteq M \ \s{do} \\
                     &\qquad M:=(M\union M_1) \\
                     &\qquad M_1:=(M;E) 
\end{array}
\]}

The \w{Marker} program satisfies the pca:
\begin{equation}\label{post-marking}
\hoare{\s{true}}{\w{Marker}}{(\w{Roots}\subseteq M) \  \mbox{ and } \ (M;E \subseteq M)}~.
\end{equation}
It follows by induction that:  $\w{Acc}(E) \subseteq M$.
Once the marking phase is completed, the collecting phase consists in 
inserting all the nodes which are not in the set $M$ in the 
{\em free list} pointed by $\s{fl}$.

Our goal in the following is to reason on the properties of the
\w{Mutator} and \w{Marker} when they are run in parallel.
The property we want to check is that once the \w{Marker}
terminates the set $M$ is indeed an over-approximation of 
the set $\w{Acc}(E)$. In order to express the specification
we introduce an auxiliary variable $\w{done}$ which is 
initially set to $\s{false}$ and becomes $\s{true}$ when
$\w{Marker}$ terminates. 

Formally, we regard a {\em memory state} $s$ of our parallel program
as a function from the variables $E$, $M$, $M_1$, $\w{done}$ 
to the appropriate value domains. When defining a relation $R$ on memory states 
we say that a variable $x$ is  stable if $(s,s')\in R$ implies 
$s(x)=s'(x)$. We will also say that $\w{Acc}(E)$ decreases if 
$(s,s')\in R$ implies $\w{Acc}(s(E))\supseteq \w{Acc}(s'(E))$.
We define:
\[
\w{Marker'} \equiv (\w{done}:=\s{false} ; \ \w{Marker} ; \ \w{done}:=\s{true})~.
\]
Then we want to show that:
\begin{equation}\label{parallel-assert}
\w{Mutator} \mid \w{Marker'} : (\emptyset, \emptyset, G, \s{true})
\end{equation}
where $G=G_1 \union G_2$ and:
\[
\begin{array}{ll}
G_1 &= \set{ (s,s') \mid M,M_1,\w{done} \ \mbox{stable}, \w{Acc}(E) \ \mbox{decreasing}} \\
G_2 &= \{ (s,s') \mid E \ \mbox{stable}, \\
    &\qquad s(\w{done})=\s{true} \mbox{ implies }(\w{Acc}(s(E))\subseteq s(M),  \ 
       \w{M},\w{M_1},\w{done} \ \mbox{stable}) \}~.
\end{array}
\]
By the rule for parallel composition in Table~\ref{rely-guarantee}, 
the assertion (\ref{parallel-assert}) is reduced to:
\[
\begin{array}{ll}
(1) &\w{Mutator}: (\emptyset, R_1,G_1,\s{true}) \\
(2) &\w{Marker'}:(\emptyset, R_2,G_2,\s{true}) \\
(3) &R_1 = \set{(s,s') \mid E \mbox{ stable}}\\
(4) &R_2 = G_1~.

\end{array}
\]
The related inclusions are easily checked.

Having outlined a formal analysis of the garbage collector, let us reconsider our
description of the \w{Marker} to notice that we are assuming
that the computation of the set $M;E$ of nodes reachable in one step from
the set $M$ is performed {\em atomically}. This is a potentially 
long operation and one would like to split it in smaller pieces.
The difficulty that arises in this case is that while visiting
the nodes in $M$ the set $E$ may be modified by the \w{Mutator}
in a non-monotonic way. For instance, 
we can have $\w{Roots}=\set{i,j}$ and the collection of edges 
oscillating between $E_1=\set{(i,k)}$ and $E_2=\set{(j,k)}$. 
If the \w{Marker} visits $i$ ($j$) while the collection of edges is 
$E_2$ ($E_1$, respectively) then it will never notice that the
node $k$ is accessible!  For this reason, it is usually
assumed that in finer grained concurrent garbage collectors
the mutator must \w{help} the marker. At our abstract level, that
could mean that the mutator is also allowed to add elements to the
set $M$.

\section{Summary and references}
Rely-guarantee assertions to reason about concurrent programs are
based on both unary and binary relations on states.  The latter
describe the transitions of the environment we can rely upon and the
transitions of the program that are guaranteed.  
Owicki-Gries system is presented in 
\cite{DBLP:journals/acta/OwickiG76}.
Rely-guarantee
assertions for reasoning about $\parimp$ programs were put forward in
\cite{DBLP:journals/toplas/Jones83} and  then developed by
Stirling \cite{DBLP:journals/tcs/Stirling88}.
The presentation above is close
to \cite{DBLP:conf/esop/Nieto03} which in turn is based on
\cite{DBLP:journals/fac/XuRH97}.

The compact modeling of the garbage collector is introduced in 
\cite{DBLP:journals/cacm/DijkstraLMSS78}.
More refined solutions to the concurrent garbage collection problem
are described and  analyzed, {\em e.g.}, in 
\cite{DBLP:journals/cacm/DijkstraLMSS78,DBLP:journals/cacm/Gries77,DBLP:journals/toplas/Ben-Ari84,DBLP:journals/ipl/Snepscheut87}.
The number of proof obligations to be checked for fine grained
garbage collectors becomes quickly overwhelming and 
a machine-assisted proof is instrumental to 
raise the confidence in the proposed solution. Examples of
such developments can be found in 
\cite{DBLP:conf/mfcs/NietoE00,DBLP:conf/popl/DoligezG94}.
Also it seems useful to cast the problem of concurrent garbage collection 
in the more general framework of lock-free concurrent data
structures \cite{DBLP:journals/tpds/HerlihyM92} 
which is discussed in chapter \ref{conc-obj-sec}.

\chapter{Labelled transition systems and bisimulation}\label{lts-sec}
\markboth{{\it LTS and bisimulation}}{{\it LTS and bisimulation}}

So far we have considered a rather concrete model of concurrent
computation, namely parallel imperative programs with shared
variables. To analyze the design spectrum which is available in the
semantics of concurrency, it is convenient to move to a more abstract
framework where systems perform some set of {\em actions}.
Such a system is called {\em labelled transition system}
(lts).  Concretely, an action could consist in changing the contents of
a shared variable or sending a message.
In this chapter, we formalize the notion of (bi-)simulation over a lts, 
consider a way to abstract away internal computation steps 
({\em weak} (bi-)simulation), and present some proof techniques for
(bi-)simulation.

\section{Labelled transition systems}
A labelled transition system (lts) 
can be regarded as an {\em automaton} where we do not 
specify the set of initial and final states.

\begin{definition}[labelled transition system]\label{lts-def}\index{labelled transition system}
A {\em labelled transition system} is a ternary relation $\arrow$ such that
$\ \arrow \subseteq S \times \w{Act} \times S \ $,
$S$ is a set of states, and $\w{Act}$ is a set of actions.
We also write $s\act{\alpha} s'$ for $(s,\alpha,s')\in \arrow$.
\end{definition}

\begin{example}
In section \ref{parimp-sec}, we have presented a transition system
for the $\parimp$ language. 
It is possible to regard this system
as a lts by taking the set of states $S$ as the collection of $\parimp$ processes and 
$\w{Act}$ as the collection of pairs of {\em memory states} (not to be confused with
the states of the lts we are defining).  Then we would write
$(P,(s,s'),Q)\in \arrow$ if $(P,s) \arrow (Q,s')$ according to the
rules in Table~\ref{parimp-smallstep}.
\end{example}

Inspired by definition \ref{trace-int-def} of trace for $\parimp$ processes, 
we introduce a notion of
trace and trace equivalence on lts.

\begin{definition}[traces]\label{trace-lts-def} \index{traces in a lts}
We define the {\em set of traces} of a state $s$ in a lts as:
\[
\trace{s} = \set{\alpha_1\cdots \alpha_n \mid s \act{\alpha_{1}} \cdots \act{\alpha_{n}} }~.
\]
We say that two states $s$ and $t$ are trace equivalent if $\trace{s}=\trace{t}$.
\end{definition}

There are a few points to be noticed concerning the definition
\ref{trace-lts-def} above.  First, it neglects {\em termination} since
this notion is not even present in the definition \ref{lts-def} of lts
(but a termination predicate on states could be added).  
And since termination is neglected, the set of traces is closed
under prefix.
Second, there
is no notion of closure of the traces under reflexivity and
transitivity. This point is treated later in section
\ref{weak-trans} once the notion of {\em internal action} is
introduced.  Third, the {\em environment} seems to play no role in the
behavior of the lts and the related definition of trace
equivalence. We shall see in chapter \ref{ccs-sec} that it is possible to
enrich lts with a notion of synchronization and parallel composition
and then prove that the notion of trace equivalence in definition
\ref{trace-lts-def} is indeed preserved by parallel composition.

\section{Simulation and bisimulation}\label{bisimulation-ssec}
In chapter \ref{introduction-sec}, we have motivated the interest of 
accounting for the {\em branching behavior} of a system 
(example \ref{branching-ex} of the vending machine).
The notion of {\em (bi-)simulation} is a very popular approach to this issue.
We have already met this notion in chapter \ref{equiv-sec}
in the framework of the $\lambda$-calculus and in chapter
\ref{atomic-sec} in the framework of the $\parimp$ concurrent language.
Next, we reconsider this notion in the setting of labelled transition systems.
The proposed definition ignores certain observables such as termination 
and deadlock.  However, 
it is quite possible to enrich the notion of lts with
predicates that represent termination and/or deadlock and then 
to formulate a notion of (bi-)simulation which depends on these
predicates.

\begin{definition}[(bi-)simulation]\label{bisim-def}\index{bisimulation}
Let $\arrow \subseteq S\times \w{Act} \times S$ be a {\em labelled transition system}.
A binary relation $\cl{R}$ on $S$ is a {\em simulation} if:
\begin{equation}\label{bis-def-right}
\infer{s \rel{R} t, \quad s\act{\alpha} s'}{\xst{t'}{t\act{\alpha} t', \quad s' \rel{R} t'}}~.
\end{equation}
Moreover we say that $\cl{R}$ is a {\em bisimulation} if:
\begin{equation}\label{bis-def-left}
\infer{s \rel{R} t, \quad t\act{\alpha} t'}{\xst{s'}{s\act{\alpha} s', \quad s' \rel{R} t'}}~.
\end{equation}
\end{definition}

\begin{remark}
In definition \ref{bisim-def} as well as in the following ones there is an implicit
universal quantification on the states which are not existentially quantified.
\end{remark}

\begin{proposition}[on bisimulation]
The following properties hold for the collection of bisimulations
over a labelled transitions system:
\begin{enumerate}

\item The {\em empty} and {\em identity} relations are bisimulations.

\item The collection of bisimulations is closed under 
{\em inverse}, {\em composition}, and arbitrary {\em unions}.

\item There is a {\em greatest bisimulation}  which is defined 
as the union of all bisimulations and that we represent with $\sim$
(some authors call it {\em bisimilarity}).

\item Bisimulations are {\em not} closed under (finite) intersection.
\end{enumerate}
\end{proposition}
\Proof Properties (1-2) follow by a simple unravelling of the definitions.
Property (3) follows by the closure under arbitrary unions.
For property (4), consider the lts $1\act{a}2$, $1\act{a}3$, $x\act{a}y$, $x\act{a} z$ and 
the relations $\cl{R}_1=\set{(1,x),(2,y),(3,z)}$, $\cl{R}_2=\set{(1,x),(2,z), (3,y)}$. \qed

\begin{exercise}[on simulation]\label{sim-ex}
Show that the properties above 
are {\em true of simulations too} but for closure under inverse.
Further, denote with
$\leq$ the {\em greatest simulation}. Find a lts with states 
$s,t$ such that $s \leq t$, $t \leq s$, and $s\not\sim t$.
\end{exercise}

\begin{exercise}[trace vs. simulation]\label{trace-sim-ex}
For $s,t$ states of a lts show that 
$s\leq t$ implies $\trace{s} \subseteq \trace{t}$, while the
converse may fail.
\end{exercise}

\begin{exercise}[backward simulation]\label{backward-sim-ex}\index{simulation, backward}
Let $\arrow \subseteq S\times \w{Act} \times S$ be a lts.
A binary relation $\cl{R}$ on $S$ is a {\em backward simulation} if:
\begin{equation}\label{back-sim-def}
\infer{s \rel{R} t, \quad s'\act{\alpha} s}{\xst{t'}{t'\act{\alpha} t, \quad s' \rel{R} t'}}~.
\end{equation}
We write $\cl{R}[s]$ for the set $\set{t\mid s\rel{R} t}$ and say that $\cl{R}$ is
{\em total} if for all $s\in S$, $\cl{R}[s]\neq \emptyset$.
Show that if $\cl{R}$ is total then $\trace{s} \subseteq \Union_{t\in \cl{R}[s]} \trace{t}$.
\end{exercise}

\begin{remark}[alternative definition of bisimulation]
Sometimes a bisimulation is defined as a {\em symmetric relation} $\cl{R}$ such 
that:
\[
\infer{s \rel{R} t, \quad s\act{\alpha} s'}{\xst{t'}{t\act{\alpha} t', \quad s' \rel{R} t'}}~.
\]
The advantage of this definition is that 
one can omit the second condition (\ref{bis-def-left}).
The inconvenience is that by forcing a bisimulation to be 
symmetric we make it larger than really needed. 
However, notice that given a bisimulation $\cl{R}$ one can 
{\em always} derive a symmetric relation which is a bisimulation by taking $\cl{R}\union \cl{R}^{-1}$.
\end{remark}

Let $\arrow \subseteq S\times \w{Act} \times S$ be a {\em labelled
  transition system}.  Notice that $L=2^{S\times S}$ is a {\em
  complete lattice} with respect to inclusion (cf. definition
\ref{lattice-def}).
Bisimulation can be characterized as the  greatest fixed point of 
a certain monotonic function $\cl{F}$ on binary relations
which we introduce below.

\begin{definition}[function $\cl{F}$]\label{fun-bis-def}\index{function $\cl{F}$}
We define $\cl{F}:L \arrow L$ as:
\[
\begin{array}{lll}
\cl{F}(\cl{R}) = &\{ (s,t) \mid &s \act{\alpha} s' \mbox{ implies } \xst{t'}{t \act{\alpha} t' \mbox{ and } s' \rel{R} t'} \mbox{ and } \\
                 && t\act{\alpha} t' \mbox{ implies }\xst{s'}{s\act{\alpha} s' \mand s'\rel{R} t'} \}~.
\end{array}
\]
\end{definition}

We notice the following properties of the function $\cl{F}$ 
(cf. proposition \ref{Tarsky-prop}).

\begin{proposition}\label{prop-f-bis}
The following properties hold:
\begin{enumerate}

\item $\cl{R}$ is a {\em bisimulation} iff $\cl{R}\subseteq \cl{F}(\cl{R})$.

\item $\cl{F}$ is {\em monotonic} on $L$.

\item The {\em greatest bisimulation} $\bis$ is the {\em greatest fixed point} of $\cl{F}$.
\end{enumerate}
\end{proposition}

\begin{exercise}\label{proof-prop-f-bis}
  Prove proposition \ref{prop-f-bis}.
\end{exercise}

\begin{remark}[transfinite definition of bisimulation]\label{transf-bis-rmk}
The bisimulation $\bis$ being the greatest fixed point of the monotonic function $\cl{F}$,
it can be {\em approximated from above} as follows (cf. chapter \ref{equiv-sec}):
\[
\begin{array}{lll}
\bis_0  \ = \ S\times S \qquad %
&\bis_{\kappa+1} \ =  \ \cl{F}(\bis_{\kappa}) ~, \qquad
&\bis_{\kappa} \ = \ \Inter_{\kappa'<\kappa} \bis_{\kappa'} \qquad \mbox{($\kappa$ limit ordinal).}
\end{array}
\]
Thus to show $s\not\bis t$ it suffices to find an ordinal $\kappa$ such that $s\not \bis_\kappa t$.
\end{remark}

\begin{definition}[image finite lts]\index{image finite lts}\label{image-fin-prop}
A lts $\arrow \subseteq S\times \w{Act} \times S$ is {\em image finite} if for all $P,\alpha$
the set $\set{s' \mid s\act{\alpha} s'}$ is {\em finite}.
\end{definition}

For finite (image finite) lts the greatest fixed point is reached
in a finite (countable) number of iterations.

\begin{proposition}[bisimulation for (image) finite lts]
The following properties hold:

\begin{enumerate}

\item 
 If the support $S$ of the lts is {\em finite} 
then there is a natural number $n$ such that the greatest
bisimulation coincides with $\bis_{n}$.

\item If the lts is {\em image finite} 
  then the greatest bisimulation coincides with $\bis_{\omega}$
  (a form of {\em co-continuity}).

\end{enumerate}

\end{proposition}

\Proof 
For the first property, see exercise \ref{fix-finite-lat}.
For the second property, suppose $s\bis_\omega t$ and $s\act{\alpha} s'$.
Then:
\[
\qqs{n}{\xst{t_n}{t\act{\alpha} t_n \mand s' \bis_n t_n}}~.
\]
Since the set $\set{t' \mid t\act{\alpha} t'}$ is finite there 
must be a  $t'$ in this set such that $\set{n \mid t_n=t'}$ is infinite.\footnote{This is a version of the so called {\em pigeonhole principle} which states 
that if infinitely many pigeons are put in finitely many boxes then at least one
box must contain infinitely many pigeons.}
Thus there is an infinite sequence $n_1 < n_2 < n_3 <\cdots$ such that
$s' \bis_{n_{j}} t_{n_{j}} = t'$. Then for any $n$ we can find a $n_j\geq n$
such that $s' \bis_{n_{j}} t'$; and this entails $s' \bis_{n} t'$. 
Hence $s' \bis_\omega t'$. \qed \\

We conclude this section by introducing a notation to denote lts
which will be extended in chapter  \ref{ccs-sec} 
to a full language of processes known as $\ccs$.
The notation is generated by the following grammar:
\begin{equation}\label{basic-proc-alg}
P::= 0 \Alt \alpha.P \Alt P+P~,\qquad \alpha \in \w{Act}~.
\end{equation}
Here $0$ denotes the empty lts, also called {\em nil},
$\alpha.P$ is the lts denoted by $P$
{\em prefixed} by the action $\alpha$, and 
$P+Q$ is the non-deterministic sum of the lts denoted by $P$ and $Q$.
In this notation, the states' identities are immaterial; what matters
of a state is not its name but the actions it can do.
Using this notation, the lts version 
of the two vending machines in  example \ref{branching-ex} 
can be represented as follows:
\[
\begin{array}{ccc}
  a.(b.0 + c.0)
  &\mbox{ and }
  &a.b.0+a.c.0~.
  \end{array}
\]
Notice that identifying the two machines amounts to {\em distribute}
the prefix over the non-deterministic sum.
We use the following abbreviations:
$b$ for $b.0$,
$b^n$ for $b.\ldots.b.0$ ($b$ prefixed $n$ times), and
$b^\omega$ for the infinite lts $b.b.\cdots$.
If $I$ is a (possibly infinite) set then $\Sigma_{i\in I}\ P_i$ 
denotes the non-deterministic sum of the lts denoted by $P_i$.
We apply the notation in the following exercise.

\begin{exercise}[non-bisimilar lts]\label{non-bis-ex}
Consider the lts $P_i,Q_i$ defined as follows:
\[
\begin{array}{llllll}
P_0 &= &b.0                   &Q_0 &= &c.0 \\
P_{i+1} &= &a.(P_i+Q_i)\qquad  &Q_{i+1} &= &a.P_i + a.Q_i~.
\end{array}
\]
\begin{enumerate}

\item Show that for all $i$ natural number:
(i) $P_i \bis_i P_i+Q_i \bis_i Q_i$, 
(ii) $P_i \not\bis_{i+1} P_i+Q_i \not\bis_{i+1} Q_i \not\bis_{i+1} P_i$.

\item Show that:
(i) $\qqs{i\leq n}{b^n \bis_i b^\omega}$, 
(ii) $\Sigma_{i\geq 0}\ b^i + b^\omega \bis_\omega \Sigma_{i\geq 0} \ b^i$, 
(iii) $\qqs{i}{b^i \not\bis_\omega b^\omega}$, 
(iv)
$\Sigma_{i\geq 0} \ b^i + b^\omega \not\bis_{\omega+1} \Sigma_{i\geq 0} \ b^i$.

\end{enumerate}
\end{exercise}

\section{Weak transitions}\label{weak-trans}
Certain computation steps should not be directly observable.
For instance, in sequential programs usually one is just
interested in the input-output behavior and not in the
way the output is computed.
To model this situation in lts,
we enrich  the collection of actions with a {\em distinct internal action}
$\tau$. For instance, $\tau^\omega$ 
is a diverging system which never interacts with the environment.
As another example, we could regard a system such as
$a.\tau.b.0$ equivalent to $a.b.0$.
Though the internal action is not directly observable,
it may make a difference. For instance, consider the lts
$a.0 + b.0$ and  $\tau.a.0+\tau.b.0$
with the interpretation:
$a= \mbox{ `accepts to deliver coffee'}$ and 
$b= \mbox{ `accepts to deliver tea'}$.
The second system decides `internally' whether to deliver 
coffee or tea while the first 
will take a decision that may be controlled
by the environment. 

Given a lts with $\tau$ transitions, we derive a related
lts with the same states but where an observable transition
may be preceded and followed by an arbitrary number of internal
transitions (think of $\epsilon$ transitions in
automata theory). As usual, if $\cl{R}$ is a binary relation then
we denote with $\cl{R}^*$ its reflexive and transitive closure.

\begin{definition}[derived weak lts]\label{weak lts}\index{weak lts}
Let a lts $\arrow\subseteq S\times (\w{Act}\union \set{\tau}) \times S$
be given where $\tau \notin \w{Act}$ is a {\em distinct internal action}.
We derive from this lts another {\em weak lts}
$\Arrow \ \subseteq \ S\times (\w{Act}\union \set{\tau}) \times S$
where:
\[
\begin{array}{l}
\wact{\alpha} \ = \ \left\{ 
                \begin{array}{ll}
                (\act{\tau})^* &\mbox{if }\alpha = \tau \\
                (\act{\tau})^*(\act{\alpha})(\act{\tau})^* &\mbox{otherwise.}
                \end{array}
                 \right.
\end{array}
\]
\end{definition}

\begin{remark}
When working with weak transitions, lts tend to be image {\em infinite},
and therefore proposition \ref{image-fin-prop} {\em cannot} be applied.
\end{remark}

The notion of bisimulation for lts with internal actions is simply the
standard notion of bisimulation on the derived weak lts (a similar
convention applies to the notion of simulation).

\begin{definition}[weak bisimulation]\label{weak-bis-def}\index{weak bisimulation}
Let $\arrow\subseteq S\times (\w{Act}\union \set{\tau}) \times S$ be
a lts with a distinct internal action $\tau$.
A binary relation $\cl{R}$ on $S$ is a {\em weak bisimulation} if it is a bisimulation with respect
to the weak transition system $\Arrow$. 
We denote with $\wbis$ the largest weak bisimulation.
\end{definition}

The following definition of weak bisimulation is the one which is used {\em in practice}.

\begin{definition}[one step weak bisimulation]\index{weak bisimulation, one step}\label{one-step-wbis-def}
A relation $\cl{R}$ is a {\em  one step weak bisimulation} if:
\[
\begin{array}{cc}

\infer{s\rel{R}t \quad s\act{\alpha} s'}
{\xst{t'}{t\wact{\alpha}t', \quad s'\rel{R}t'}}~, \qquad
&
\infer{s\rel{R}t \quad t\act{\alpha} t'}
{\xst{s'}{s\wact{\alpha}s',\quad s'\rel{R}t'}}~.

\end{array}
\]
\end{definition}

\begin{proposition}\label{wbis-onestep-prop}
A relation $\cl{R}$ on a lts is a {\em weak bisimulation} iff 
it is a {\em one step weak bisimulation}.
\end{proposition}

\begin{exercise}\label{proof-wbis-onestep-prop}
  Prove proposition \ref{wbis-onestep-prop} by diagram chasing.
  \end{exercise}

Henceforth we just speak of {\em weak bisimulation} and use
the more convenient definition \ref{one-step-wbis-def}.

\begin{definition}[weak up to strong]\label{uptostrong-def}\index{weak up to strong bisimulation}
We say that a relation $\cl{R}$ on a lts is a 
{\em weak bisimulation up to strong bisimulation} if:
\[
\begin{array}{cc}

\infer{s \rel{R} t, \quad s\act{\alpha} s'}
{\xst{t'}{t\wact{\alpha} t', \quad s' \bis \rel{R} \bis t'}} ~,\qquad

&\infer{s \rel{R} t, \quad t\act{\alpha} t'}
{\xst{s'}{s\wact{\alpha} s', \quad s' \bis \rel{R} \bis t'}}~.

\end{array}
\]
\end{definition}

\begin{exercise}\label{weak-up-to-strong-ex}
Show that if $\cl{R}$ is a weak bisimulation up to strong bisimulation
then $\cl{R}\subseteq \wbis$.
\end{exercise}

\section{Proof techniques for bisimulation (*)}\label{proof-bis-ssec}
The standard method to prove $s \bis t$ is
to {\em exhibit a relation} $\cl{R}$ such that $s \rel{R} t$ and
$\cl{R}\subseteq \cl{F}(\cl{R})$, where $\cl{F}$ is as in definition \ref{fun-bis-def}.
Exercise \ref{weak-up-to-strong-ex} suggests that it is possible to refine 
this proof technique by exhibiting a relation
$\rel{R}$ which is a bisimulation {\em up to a relation} `with suitable properties'.
In the following, we provide a rather general treatment of what
`with suitable properties' means.
First, some preliminary remarks.
Let $(L,\leq)$ be a complete lattice and $f:L\arrow L$ be a
monotonic function on $L$. The function $f$ induces a {\em transitive
relation} $<_f$ which refines $\leq$:
\[
	x <_f y \mbox{ if } x\leq y \mand x\leq f(y) ~.
\]
Notice that $<_f$ is {\em anti-symmetric} but {\em not necessarily reflexive}.
In the case we are interested in, $L$ is the power-set
$2^{S\times S}$, $f$ is $\cl{F}$, and
$\cl{R}$ is a bisimulation iff $\cl{R} <_{\cl{F}} \cl{R}$.

\begin{definition}
We say that a function $h:L\arrow L$ {\em preserves} $<_f$ if:
\[
x <_f y \mbox{ implies } h(x) <_f h(y)~.
\]
\end{definition}

\begin{exercise}\label{preserve-forder-ex}
Show that the set of functions preserving the order $<_f$
is closed under {\em composition} and {\em supremum}.
\end{exercise}

\begin{proposition}[key property]
Let $(L,\leq)$ be a complete lattice, $f:L\arrow L$ be a monotonic
function (with greatest fixed point $\w{gfp}(f)$), and
$h:L\arrow L$ be a function that {\em preserves} $<_f$.
Then:
\[
	x\leq f(h(x))	~~\mbox{ implies }~~ x \leq \w{gfp}(f)~.
\]
\end{proposition}
\Proof
Given $x$, we build a bigger element $y$ such that $y\leq f(y)$.
To this end, we define a sequence $x_0 = x$, $x_{n+1} = x_n \join h(x_n)$.
Let $y = \JOIN_{n\geq 0} x_n$; obviously $x_n\leq x_{n+1}$ and $x \leq y$. 
We show that $x_n <_f x_{n+1}$,  by induction on $n$. 
\begin{description}

\item[$n=0$] 
$x_0 = x \leq f(h(x)) \leq f(x\join h(x))$, since $x\leq f(h(x))$
by hypothesis and $f$ is monotonic.

\item[$n>0$] We have to show:
\[
	x_n  = x_{n-1} \join h(x_{n-1}) \leq f(x_n \join h(x_n)) = f(x_{n+1})~.
\]
Since $f$ is monotonic, we have 
$f(x_n) \join f(h(x_n)) \leq f(x_n \join h(x_n))$.
By inductive hypothesis, we know $x_{n-1} \leq f(x_n)$.
Moreover, since $h$ preserves $<_f$, we have
$h(x_{n-1}) \leq f(h(x_n))$.
\end{description}

\noindent
Finally, we remark that $y\leq f(y)$, as:
\[
	y = \JOIN_{n\geq 0} x_n \leq \JOIN_{n\geq 0} f(x_{n+1}) 
	\leq f(\JOIN_{n\geq 1} x_n) = f(y)~.
\]
Since $y\leq f(y)$ implies $y\leq \w{gfp}(f)$, we conclude
$x \leq \w{gfp}(f)$. \qed

\begin{exercise}[when $h$ is a closure]\label{up-to-closure-ex}
We say that a function $h$ on a lattice $L$ 
is a closure if $\w{id}\leq h = h\comp h$.
Show that if $h$ is a closure then $y=h(x)$ in the previous
construction.
\end{exercise}

In our {\em application scenario}, this means that to prove that $s$ and $t$ are
bisimilar it suffices to find:
(1) a function $\cl{H}$ that preserves $<_{\cl{F}}$ and 
(2) a relation $\cl{R}$ such that $(s,t)\in \cl{R}$ and
$\cl{R} \subseteq \cl{F}(\cl{H}(\cl{R}))$.

\begin{exercise}\label{upto-strong-ex}
Let $\cl{H}(\cl{R}) = \bis \comp \cl{R} \comp \bis$.
Check that $\cl{H}$ preserves $<_{\cl{F}}$.
\end{exercise}

We introduce a notion of weak bisimulation up to expansion
which is often used in applications.

\begin{definition}[expansion]\label{expansion-def}\index{expansion}
A binary relation $\cl{R}$ on a lts is an {\em expansion} if:
\[
\begin{array}{cc}

\infer{s \rel{R} t,\quad s\act{\alpha} s'}
{\xst{t'}{t\wact{\alpha} t'\quad s'\rel{R} t'}}~, \qquad

\infer{s \rel{R} t\quad t\act{\alpha} t'}
{\xst{s'}{s'\rel{R} t' \mand (s\act{\alpha} s'\mbox{ or } (\alpha=\tau, s'=s))}}~.

\end{array}
\]
We denote with  $\lexp$ be the {\em largest expansion}
and with $\gexp$ its inverse.  Also we read $s \lexp t$ as
$s$ {\em expands} to $t$.
\end{definition}

Note that an expansion is a hybrid relation  which is {\em weak} on one
side and {\em almost strong} on the other side.
The intuition is that the state on the right is a kind of implementation of the
one on the left, {\em i.e.}, the state on the right may take more
internal steps to perform the `same task'.

\begin{exercise}[weak bisimulation up to expansion]\label{expansion-ex}
Define:
$\cl{H}(\cl{R}) \ = \ \gexp \comp \cl{R} \comp \lexp$.
Let $\cl{F}$ be the monotonic function induced by the definition of
(one step) weak bisimulation.
Show that:

\begin{enumerate}

\item $s\sim t$ implies strictly $s\lexp t$ (and $s\gexp t$).

\item $\lexp \union \gexp$ implies strictly $\wbis$.

\item 
$\cl{H}$ preserves $<_{\cl{F}}$. And 
explicit the condition that needs to be checked to 
ensure that a relation $\cl{R}$ is a weak bisimulation up to expansion.

\end{enumerate}
\end{exercise}

The following exercise highlights two possible pitfalls in the
usage of up-to techniques.

\begin{exercise}[pitfalls]\label{upto-pitfalls-ex}
Define:
$\cl{H'}(\cl{R}) \ = \ \wbis \comp \cl{R} \comp \wbis$ and 
$\cl{H''}(\cl{R}) \ = \  \lexp \comp \cl{R} \comp \gexp$.
As in the previous exercise, let 
$\cl{F}$ be the monotonic function induced by the definition of
(one step) weak bisimulation. Show that:
\begin{enumerate}

\item  $\cl{H'}$ does {\em not} preserve $<_{\cl{F}}$.
Suggestion: consider 
$\cl{R}=\set{(\tau.a,0)}$ and check that
$\cl{R} \subseteq \cl{F}(\cl{H'}(\cl{R}))$ while obviously
$\cl{R}\not\subseteq \wbis$. 

\item $\cl{H''}$ does {\em not} preserve $<_{\cl{F}}$,
by an argument similar to the one used for $\cl{H'}$.
\end{enumerate}
\end{exercise}

\section{Summary and references}
The notion of {\em labelled transition system} provides an abstract
setting to explore the variety of possible semantics of concurrent
systems.  In particular, we have developed the notion of {\em
  bisimulation} which corresponds to the greatest fixed point of a
certain monotonic function on lts.  Bisimulation is a {\em natural
  notion} and Park \index{Park} \cite{DBLP:conf/tcs/Park81} 
seems the first to have
used it in the semantics of programming languages.  In order to
abstract the internal behavior of a system, we have introduced the
notion of {\em internal action} and the related notions of {\em weak
  transition} and {\em weak bisimulation}.  Finally, we have discussed
an {\em up to proof technique} which allows to reduce the size of the
relation to be exhibited to show that two states are bisimilar.

\chapter{Modal logics}\label{modal-sec}
\markboth{{\it Modal logics}}{{\it Modal logics}}

In chapter \ref{rely-guarantee-sec}, we have considered 
partial correctness and rely-guarantee assertions as means
to specify the behaviour of concurrent processes and
in doing this we have faced some problems due to the
limited expressive power of the specification language.
In this chapter, we take a bold step in that for a given
notion of equivalence on lts we aim at a specification
language which captures exactly the equivalence.
The presented languages build on the notion of (propositional) {\em modal logic}
which is an extension of usual logic with {\em modalities} 
that qualify the validity of the assertions: possibly true,
necessarily true,$\ldots$ 
In particular, we introduce a {\em diamond modality} indexed over
the actions of the lts and stipulate:
\[
s\models \pos{\alpha}{A} \mbox{ if } \xst{s'}{s\act{\alpha} s' \mbox{ and } s'\models A}~,
\]
which is read as follows: a state $s$ satisfies the formula $\pos{\alpha}{A}$ if 
there is a state $s'$ such that $s\act{\alpha} s'$ and $s'$ satisifies $A$.
It turns out that the full {\em infinitary} specifications generated by this
extension characterize bisimulation, while
restricted versions correspond to coarser equivalences such as 
simulation or trace equivalences.

In practice, one needs finite means to describe `infinitary' specifications.
An elegant way to achieve this, is to define (monotonic) formulae by 
least and greatest fixed points (in the spirit of proposition \ref{Tarsky-prop}).
The resulting modal language is called the $\mu$-calculus. 
For finite state lts,  we present a simple algorithm to decide whether a state
satisfies a formula of the $\mu$-calculus.

\section{Modal logics vs. equivalences}\label{modal-log-eq-sec}
We introduce a modal logic which consists of 
a classical propositional logic enriched with 
a so called {\em diamond} modality describing
the ability to perform an action.

\begin{definition}[formulae]\label{modal-syntax-def}\index{modal logic, syntax}
The collection of formulae of a propositional modal logic is defined as:
\begin{equation}
A::= \MEET_{i\in I} A_i \Alt \neg A \Alt \pos{\alpha}{A} \qquad \alpha \in \w{Act}
\end{equation}
where the set $I$ can also be empty or infinite.
By convention, we write
$\s{true}$ for $\MEET \emptyset$ and 
$\whn{\alpha}A$ for $\neg \pos{\alpha}{\neg A}$.

\end{definition}

\begin{definition}[formulae satisfaction]\index{modal logic, satisfaction}
We define when a state in a lts satisfies a formula, written $s\models A$, as follows:
\[
\begin{array}{ll}

s\models \MEET_{i\in I} A_i &\mbox{if }\  \qqs{i\in I}{s \models A_i}  \\
s \models \neg A           &\mbox{if } \ s \not\models A \\
s\models \pos{\alpha}{A}  &\mbox{if } \ \xst{s'}{s\act{\alpha} s' \mand s'\models A}~.

\end{array}
\]
We also write:
\[
\begin{array}{lll}

\sem{A} &= \set{s \mid s \models A}  &\mbox{(formula interpretation)}\\

\sem{s} &=\set{A \mid s\models A}    &\mbox{(state interpretation)} \\

s \sim_L s' &\mbox{if }\sem{s}=\sem{s'} &\mbox{(logical equivalence)}~.
\end{array}
\]
\end{definition}

\begin{exercise}[on modal formulae] \label{modal-semantics-ex}
Spell out what it means to satisfy $\whn{\alpha}{A}$.
Find a formula showing that $a.(b.0+c.0) \not\sim_L a.b.0+a.c.0$.
\end{exercise}

It is easily checked that two bisimilar states are logically equivalent.

\begin{proposition}\label{bis-impl-logical-prop}
Let $s,s'$ be states in a lts. 
If $s\bis s'$ then $s \sim_L s'$.
\end{proposition}

\begin{exercise}\label{proof-bis-impl-logical-prop}
Prove proposition \ref{bis-impl-logical-prop}.
\end{exercise}

To show the converse of proposition \ref{bis-impl-logical-prop},
we introduce the (possibly infinite) so called 
{\em characteristic formulae}. 

\begin{definition}[characteristic formula]\label{charform-def}\index{characteristic formula}
Given a state $s$ in a lts and an ordinal $\kappa$ the characteristic formula
$C^\kappa(s)$ is defined as follows:
\begin{equation}\label{char-form-def}
\begin{array}{lll}

C^0(P)       &= &\s{true} \\ 

C^{\kappa+1}(s) &= &\MEET_{s \act{\alpha} s'} \  \pos{\alpha}{C^{\kappa}(s')} ~\AND~ 
\MEET_{\alpha\in \w{Act}}\  \whn{\alpha}{ ( \ \OR_{s \act{\alpha}s'} C^{\kappa}(s') \ )} \\ 

C^{\kappa}(s) &= &\MEET_{\kappa'<\kappa} C^{\kappa'}(s) \qquad \mbox{($\kappa$ limit ordinal).}
\end{array}
\end{equation}
\end{definition}

\begin{proposition}
For any state $s$ and ordinal $\kappa$:
\begin{enumerate}
\item  $s \models C^{\kappa}(s)$.
\item $s'\models C^{\kappa}(s)$ iff $s \bis_{\kappa} s'$, where
$\bis_\kappa$ is the approximation of bisimulation defined in remark \ref{transf-bis-rmk}.
\end{enumerate}
\end{proposition}
\Proof
To prove a property for all ordinals one relies on the principle of transfinite induction. Namely one shows that  if a property is true of all ordinals less than $\kappa$ then it is true of $\kappa$. \qed

\begin{exercise}\label{image-finite-ex}
Suppose that the set of actions $\w{Act}$ is {\em finite}. 
Then show for {\em image finite} lts the following property: if two processes 
are not bisimilar then there is a {\em finite formula} that distinguishes them.
\end{exercise}

Given that full modal logic characterizes bisimulation, one may look
for fragments of the logic that characterize coarser equivalences or
pre-orders.
We consider the cases of trace inclusion (definition \ref{trace-lts-def}) 
and simulation pre-order (definition \ref{bisim-def}).

\begin{proposition}
The modal formulae $A$ of the following shape characterize
trace inclusion:
\[
A ::= B \Alt \MEET_{i\in I} A_i ~,
\]
where: $\ B ::= \s{true} \Alt \pos{\alpha}{B}\ $.
\end{proposition}
\Proof
If $s$ is a state then let 
$T(s)$ be the collection of its traces.
Associate to a state $s$ the following formula $C(s)$:
\[
C(s) = \MEET_{\alpha_1\cdots \alpha_n \in \tr{s}} \pos{\alpha_1}{\cdots \pos{\alpha_n}{\s{true}}}~.
\]
Then for any state $s'$, $s'\models C(s)$ iff $T(s)\subseteq T(s')$. \qed 

\begin{proposition}
The modal formulae $A$ of the following shape characterize the simulation
pre-order:
\[
A::= \MEET_{i\in I} A_i \Alt \pos{\alpha}{A}~.
\]
\end{proposition}
\Proof 
We build the formula $C^\kappa(s)$ taking the left hand side of the formula (\ref{char-form-def}) that works for bisimulation.
Then for any state $s'$,
$s'\models C^\kappa(s)$ iff $s\leq_\kappa s'$,
where $\leq_\kappa$ is the $\kappa$-approximation of simulation. \qed

\section{A modal logic with fixed points: the $\mu$-calculus (*)}
In the presented modal language, to express, {\em e.g.}, that a process can do infinitely many actions $\alpha$ we
need an {\em infinite formula}. It is possible to increase the expressive power of formulae while keeping the 
syntax {\em finite}. An elegant 
extension known as {\em $\mu$-calculus} consists in adding to the logical
formulae {\em least fixed points}. Then the syntax of modal formulae given in definition \ref{modal-syntax-def}
is revisited as follows.

\begin{definition}[formulae with fixed points]\index{modal logic with fixed points}
The modal formulae with fixed points have the following syntax:
\[
\begin{array}{ll}
\w{id} ::= x \Alt y \Alt \ldots &\mbox{(formula identifiers)} \\
A::= \MEET_{i\in I} A_i \Alt \neg A \Alt \pos{\alpha}{A} \Alt \w{id} \Alt \mu \w{id}.A &\mbox{(formulae).}
\end{array}
\]
\end{definition}

In a formula $\mu x.A$, the identifier $x$ is bound in $A$ by the least fixed point operator $\mu$.
Also we assume that each free occurrence of $x$ in $A$ is {\em positive}, {\em i.e.},
under an even number of negations. This positivity condition is essential to show that
the function induced by the formula is monotonic and therefore has a least (and a greatest) fixed point (cf. exercise \ref{pos-mu-ex} below).

Since a formula may contain free identifiers, its interpretation is given 
relatively to an assignment $\rho:\w{id} \arrow 2^S$ as follows:
\[
\begin{array}{ll}
\sem{\MEET_{i\in I} A_i}\rho &= \Inter_{i\in I} \sem{A_i}\rho \\
\sem{\neg A}\rho            &= \cpl{(\sem{A}\rho)} \\
\sem{\pos{\alpha}{A}}\rho   &= \set{s \mid s\act{\alpha} s' \mand s'\in \sem{A}\rho} \\
\sem{x}\rho                 &= \rho(x) \\
\sem{\mu x.A}\rho           &= \Inter \set{X\subseteq S \mid \sem{A}\rho[X/x] \subseteq X}~.
\end{array}
\]
Of course, if $A$ is a closed formula its interpretation does {\em not} depend on the assignment
and we can write $s\models A$ if $s\in \sem{A}\rho$, for some $\rho$.

\begin{exercise}[positivity]\label{pos-mu-ex}
Check that for all well-formed formulae $A$, identifier $x$, and assignments $\rho$, the function
$X \mapsto \sem{A}\rho[X/x]$ is monotonic on $2^S$. Conclude that the semantics of a formula
$\mu x.A$ does indeed correspond to a least fixed point.
\end{exercise}

An intuitive way to understand the meaning of a formula $\mu x.A$ is 
to unfold it as an infinite 
disjunction $\JOIN_{\kappa} A^\kappa$ where:
$A^0=\s{false}$, $A^{\kappa+1} = [A^{\kappa}/x]A$ and 
$A^{\kappa} = \JOIN_{\kappa'<\kappa} A^{\kappa'}$, for $\kappa$ limit ordinal.
This viewpoint is based on the 
iterated definition of the least fixed point mentioned in chapter
\ref{equiv-sec}.

{\em Greatest fixed points} are derived by duality from least fixed points by defining:
\[
\nu x.A = \neg \mu x.\neg ([\neg x/x]A)~.
\]
For instance: $\nu x.\pos{\alpha}{x} = \neg \mu x.\neg \pos{\alpha}{\neg x}$.

\begin{exercise}[greatest fixed points]\label{gfp-ex}
Check that the interpretation of $\nu$ does indeed correspond
to a greatest fixed point, namely:
\[
\sem{\nu x.A}\rho = \Union \set{X\subseteq S \mid X \subseteq \sem{A}\rho[X/x]}~.
\]
\end{exercise}

We have seen that disjunction and greatest fixed points can be derived from
conjunction, least fixed points, and negation.  An alternative approach
consists in dropping negation and taking conjunction, disjunction,
$\mu$ and $\nu$ operators as primitive. This way we have to deal with
an additional operator but we can drop the positivity condition on the
fixed points since conjunction and disjunction are guaranteed to induce
monotonic functions.

\begin{exercise}[deriving negation]\label{derive-neg-ex}
Show that the  negation operator can be defined (on closed formulas).
Hint: Consider the following equations:
\[
\begin{array}{llll}

\neg \pos{\alpha}{A} = \whn{\alpha}{\neg A},\quad

&\neg \whn{\alpha}{A}	= \pos{\alpha}{\neg A},\quad

&\neg \mu x.A	= \nu x.\neg A, \quad 

\neg \nu x.A	=\mu x.\neg A~.

\end{array}
\]
\end{exercise}

It turns out that for  {\em finite lts}, the modal logic with fixed points 
can express the characteristic formula of a state by
a {\em finite} formula.

\begin{proposition}\label{char-frproc}
Let $s$ be a state in a finite lts.
Then there is a closed finite characteristic formula $C(s)$ involving only greatest fixed points such that
for any state $s'$, $s'\models C(s)$ iff $s \bis s'$.
\end{proposition}
\Proof For every state $s$ introduce a propositional variable $x_s$ and 
an equation based on the characteristic formula in definition 
\ref{charform-def}:
\[
\begin{array}{lll}

x_s &= &\MEET_{s \act{\alpha} s'} \  \pos{\alpha}{x_{s'}} ~\AND~ 
\MEET_{\alpha\in \w{Act}}\  \whn{\alpha}{ ( \ \OR_{s \act{\alpha}s'} x_{s'} \ )} 

\end{array}
\]
Then the general idea is to take the {\em greatest} fixed point of 
this system of equations and project on
the component which corresponds to the state $s$.
For instance, suppose $S=\set{1,2}$ and $1\act{a} 2$, $1\act{b} 1$,
$2\act{b} 1$, $2\act{b} 2$.
Then define:
\[
\begin{array}{ll}
A_1 = \pos{a}{x_1} \AND \pos{b}{x_2} \AND \whn{a}{x_2} \AND \whn{b}{x_1} ~,
&A_2 = \pos{b}{x_1} \AND \pos{b}{x_2} \AND \whn{a}{\s{false}} \AND \whn{b}{(x_1 \Or x_2)}~.
\end{array}
\]
The characteristic formula for, {\em e.g.},
the state $1$ can be written as $\nu x_1.[\nu x_2.A/x_2]A_1$. \qed \\

Another interesting property of the $\mu$-calculus on finite lts is
that the model-checking problem is {\em decidable}. We spend the rest
of the section to present a proof of this fact that relies on the
following elementary property of fixed points.

\begin{proposition}[reduction]\label{reduction-lemma}
Let $f$ be a monotonic function over $2^S$ and $s\in S$ be a state. 
Consider the following monotonic functions over $2^S$:
$(f\union s)(x) = f(x)\union \set{s}$ and 
$(f\minus s)(x) = f(x)\minus \set{x}$.
Also if $g$ is a monotonic function denote by
$\nu(g)$ and $\mu(g)$ its greatest and least fixed point.
Then:
\begin{enumerate}

\item
$s\in \nu(f) \IFF s\in f(\nu(f \union s))$.

\item
$s\in \mu(f) \IFF s\in f(\mu(f\minus s))$.

\item
$s\in \nu(f\union s)$.

\item
$s\notin \nu(f\minus s)$.
\end{enumerate}
\end{proposition}
\Proof 
\Proofitemf{(1)}
Suppose $s\in  \nu(f)$. Then:
\[
f(\nu(f))\union \set{s} = \nu(f)\union \set{s} = \nu(f)~.
\]
By definition of $\nu(f\union s)$ this implies $\nu(f)\subseteq \nu(f\union s)$.
By monotonicity, $\nu(f) = f(\nu(f)) \subseteq f(\nu(f\union s))$,
and therefore $s\in f(\nu(f\union s))$.
On the other hand, suppose $s\in f(\nu(f\union s))$. It follows:
\[
\nu(f\union s) = f(\nu(f\union s))\union \set{s} = f(\nu(f\union s))~.
\]
By definition of $\nu(f)$ this implies $\nu(f\union s) \leq \nu(f)$.
By monotonicity, $f(\nu(f\union s))\leq f(\nu(f))=\nu(f)$, and therefore
$s\in \nu(f)$.

\Proofitemm{(2)} Prove by a dual argument: 
$s\notin \mu(f) \IFF s\notin f(\mu(f\minus s))$.

\Proofitemm{(3-4)} Immediate by unfolding the fixed point. \qed \\

This proposition suggests a strategy to unfold recursive formulae. The starting idea is to tag each fixed point
with a set of states. Then properties (1-2) of proposition \ref{reduction-lemma} when read from left to right
suggest to record in the tag the states that are crossed when unfolding a fixed point while
properties (3-4) of proposition 
\ref{reduction-lemma} provide the halting conditions.
To formalize this idea, we begin by introducing the syntax of {\em tagged} formulae.

\begin{definition}[formulae with tagged fixed points]\label{tagged-form-def}\index{modal logic with tagged fixed points}
The modal formulae with tagged fixed points have the following syntax:
\[
\begin{array}{ll}
\w{id} ::= x \Alt y \Alt \ldots &\mbox{(formula identifiers)} \\
T      ::= \set{s_1,\ldots,s_n} &\mbox{(tags, finite sets of states)} \\
A::= \MEET_{i\in I} A_i \Alt \JOIN_{i\in I} A_i \Alt \pos{\alpha}{A} \Alt \whn{\alpha}{A} \Alt \w{id} \Alt \mu \w{id}:T.A \Alt \nu \w{id}:T.A 
&\mbox{(tagged formulae).}
\end{array}
\]
\end{definition}

The interpretation of tagged fixed points is as follows while the interpretation of the logical
and modal operators is left unchanged:
\[
\begin{array}{ll}
\sem{\mu x:T.A}\rho           &= \Inter \set{X\subseteq S \mid (\sem{A}\rho[X/x])\minus T \subseteq X}~, \\
\sem{\nu x:T.A}\rho           &= \Union \set{X\subseteq S \mid X \subseteq (\sem{A}\rho[X/x])\union T}~.

\end{array}
\]
Based on this interpretation and proposition \ref{reduction-lemma}, we introduce in
Table~\ref{checker-mu-fig} \index{$\mu$-calculus, model-checker} 
the collection of rules to model-check states against
finite formulae of the $\mu$-calculus.

\begin{table}
{\footnotesize
\[
\begin{array}{c}

\infer{s: A~~s: B}
	{s: A \AND  B} \\\\

\infer{s: A}
	{s: A \Or  B}

\qquad

\infer{s: B}
	{s: A \Or  B} \\ \\

\infer{s': A}
	{s:\pos{\alpha}{A}} \mbox{ for some } s\act{\alpha} s'

\qquad
 \infer{s': A}
	{s:\whn{\alpha}{A}} \mbox{ whenever }s\act{\alpha} s'' \\ \\

\infer{s \notin T~~
        s: [\mu x:T\union \set{s}.A/x] A}
	{s: \mu x:T. A} 
\qquad

\infer{s \notin T~~
        s: [\nu x:T\union \set{s}.A/x] A}
	{s: \nu x:T. A} \\ \\

\infer{s \in T}
	{s: \nu x:T. A}

\end{array}
\]}
\caption{A model checker for the $\mu$-calculus} \label{checker-mu-fig}
\end{table}

\begin{proposition}[soundness]
Let $A$ be a closed formula of the modal, tagged $\mu$-calculus.
If we can derive the assertion $s: A$ according to the rules 
in Table \ref{checker-mu-fig} then $s\in \dl  A \dr$.
\end{proposition}
\Proof By induction on the height of the proof, relying on the reduction 
proposition \ref{reduction-lemma} for the rules that fold the
fixed points. \qed \\

Proving completeness  of the method for finite state lts amounts to prove
the termination of the unfolding process. Suppose we look at the rules
in Table~\ref{checker-mu-fig} bottom up.
All rules but those that unfold fixed points either entail termination or shrink the size of the formula to be proved.
Hence any infinite backward development must include
an infinite number of applications of the rules unfolding fixed points.
Now we remark that these rules add new 
elements to the tags. Since $T\subseteq S$ and $S$ is finite,
we might conjecture that this process eventually terminates. 
We prove this property in two steps.
First,  we present a simple rewriting system whose termination proof
exposes the kernel of the combinatorial problem.
Second, we show termination of the bottom up proof development
by exhibiting a reduction preserving translation from judgments to terms
of the simple rewriting system.

\begin{definition}\label{theta}
We define a collection of $\sigma$-terms as follows where $n$ is a natural number:
\[
\begin{array}{ll}
\w{id}::= x \Alt y \Alt \cdots  &\mbox{(identifiers)}\\
\theta::= \w{id} \Alt 1 \Alt \theta * \theta \Alt  \bullet \theta \Alt \sigma^n \w{id}.\theta &\mbox{($\sigma$-terms)}
\end{array}
\]
\end{definition}

\begin{definition}\label{theta-red}
A term $\theta$ can be reduced according to the following rules
where the rules can be applied at top level only:
\[
\begin{array}{llll}

\sigma^{n+1} x.\theta \arrow  [\sigma^{n} x.\theta/x]\theta, 

&\bullet \theta \arrow \theta,  

&\theta * \theta' \arrow \theta, 

&\theta * \theta' \arrow \theta' ~.

\end{array}
\]
\end{definition}

\begin{proposition}\label{SN-sigma-terms}
The rewriting system defined in \ref{theta-red} terminates.
\end{proposition}
\Proof
Let $\w{WF}$ be the collection of terminating
$\sigma$-terms. If $\theta\in \w{WF}$ let $d(\theta)$ be the length of the longest
reduction sequence (this is well defined because the reduction
tree is finitely branching). We want to prove:
\begin{equation} \label{sub-sn}
	\theta, \theta' \in \w{WF} ~~\mbox{ implies }~~ [\theta'/x]\theta \in \w{WF}~.
\end{equation}
We prove (\ref{sub-sn}) by induction on $d(\theta)$.
The only interesting case is when $\theta$ has the
shape $\sigma^{n+1}y.\theta$. Then we observe:
\[
\begin{array}{llll}
	[\theta'/x](\sigma^{n+1} y.\theta)  
	&\equiv \sigma^{n+1} y.[\theta'/x]\theta 
	&\arrow [\sigma^n y.[\theta'/x]\theta/y]([\theta'/x]\theta) 
	&\equiv [\theta'/x][\sigma^n y.\theta/y]\theta~.
\end{array}
\]
We note that $(\sigma^{n+1} y.\theta) \arrow [\sigma^n y.\theta/y]\theta \in \w{WF}$.
Hence, we can apply the inductive hypothesis on $d([\sigma^n y.\theta/y]\theta)$, 
and we conclude that $[\theta'/x][\sigma^n y.\theta/y]\theta \in \w{WF}$.

Next we prove that all $\sigma$-terms terminate.
We proceed by induction on a relation $\succ$ which is 
the least transitive relation such that:
\[
\begin{array}{lllll}

\sigma^{n+1} x.\theta \succ \sigma^n{x}.\theta~, 

&\sigma^{n+1} x.\theta \succ \theta~, 

&\bullet \theta \succ \theta~,

&\theta * \theta' \succ \theta~,

&\theta * \theta' \succ \theta' ~.

\end{array}
\]
Clearly $\succ$ is a well founded relation.
Again the only interesting case is when
the term has the shape $\sigma y^{n+1}.\theta$.
By the inductive hypothesis $\sigma y^{n}.\theta\in \w{WF}, \theta \in \w{WF}$,
and by (\ref{sub-sn}) 
$[\sigma y^{n}.\theta /y]\theta \in \w{WF}$.
\qed

\begin{definition}
Given a finite lts with a set of states $S$, 
we associate a $\sigma$-term to a modal formula as follows,
where $n=\card S +1$:
\[
\begin{array}{ll}

\la \s{true} \ra = \la \s{false} \ra = 1~,
  
&\la x \ra = x~,  \\

\la A \AND  B \ra  = \la  A \ra * \la  B \ra~,  

&\la  A \Or  B \ra  = \la  A \ra * \la  B  \ra~,    \\

\la \pos{\alpha}{A} \ra  = \bullet \la  A \ra~,

&\la \whn{\alpha}{A} \ra = \bullet \la A \ra~, \\

\la \mu x:T. A \ra = \sigma x^{(n-\card{T})}.\la  A \ra~, 

&\la \nu x:T. A \ra = \sigma x^{(n-\card{T})}.\la  A \ra~.

\end{array}
\]
\end{definition}

Suppose  $s': A'$ is a premise of $s: A$ in the
proof development.
We show $\la  A \ra \arrow \la  A' \ra$ by inspection of the proof rules. 
The only interesting case is when we unfold a fixed point. 
Since in the translation we have picked
$n= \card S +1$ bigger than $\card{T}$ we can compute, {\em e.g.},
in the case of the least fixed point:
\[
\begin{array}{llll}

\la \mu x:T. A \ra 
&= \sigma x^{(n-\card{T})}.\la  A \ra 
&\arrow [\sigma x^{(n-\card{T}-1)}.\la  A \ra/x]\la  A \ra 
&= \la [\mu x:T\union \set{s}. A/X] A \ra~.
\end{array}
\]

\begin{proposition}\label{compl-tableau}
The model checker is complete on finite structures.
\end{proposition}
\Proof
We show by induction on $A$ that $s\models A$ iff
a proof rule applies. We can bound the depth of
a path in a bottom up proof development. Hence, if $s\models A$
by developing the proof bottom up we eventually obtain a proof
of $s: A$.\qed

\section{Summary and references}
We have described a family of {\em modal logical languages} which can
be used to characterize bisimulation as well as coarser equivalences.
We have also presented a few basic results on a fixed point extension of
modal logic known as $\mu$-calculus. 
The $\mu$-calculus is a kind of basic modal 
logical language to which more user-friendly logical languages
can be compiled.  It was introduced by Kozen \index{Kozen} \cite{DBLP:journals/tcs/Kozen83}, following previous work by V.~Pratt.
The simple proof of decidability of the model checking
problem for finite lts we have presented is based on \cite{DBLP:conf/icalp/Winskel89}.
The model-checking problem for the $\mu$-calculus is
known to be in $\mbox{{\sc NP}}\inter\mbox{{\sc co-NP}}$ (like the graph
isomorphism problem). Upper bounds on the time complexity are 
polynomial in the size of the lts and exponential in the so called 
{\em alternation depth} of the formula. This is a measure that counts the 
number of alternations of nested greatest and least fixed points.  
It is also known  \cite{DBLP:conf/concur/Bradfield96,DBLP:conf/icalp/Lenzi96} 
that bounding the alternation depth limits the expressivity of the logic, {\em
  i.e.}, the hierarchy of formulae obtained by measuring the
alternation depth is {\em strict}. The basic theory of the $\mu$-calculus
is developed systematically in \cite{ANbook}.

\chapter{Labelled transition systems with synchronization}\label{ccs-sec}
\markboth{{\it $\ccs$}}{{\it $\ccs$}}

One can make the basic model of labelled transition systems 
a bit more interesting by adding some {\em parallelism and synchronization}
mechanisms. One elegant way to provide a synchronization mechanism is
to introduce a {\em notion of co-action} and suppose that synchronization
happens when a process can perform an action and another parallel process can
perform the corresponding co-action. Thus, given a set $A$, take the set of
actions to be:
\begin{equation}\label{act-with-synch}
\w{Act}= \set{a,\ol{a} \mid a\in A}  \union \set{\tau}~.
\end{equation}
It is convenient to extend the co-action definition to the whole
set $\w{Act}$ by assuming:
\[
\begin{array}{ll}
\ol{\tau}=\tau ~, \quad
&\ol{\ol{a}} = a ~.
\end{array}
\]
$\ccs$ (Calculus of Communicating Systems) is a minimal set of
operators to represent such labelled transition systems enriched with
the co-action mechanism; we introduce this formalism and discuss two
ways to define its bisimulation semantics which turn out to be
equivalent.  $\ccs$ is a {\em simple} model of concurrent systems and
we shall build on it to discuss the notions of {\em deterministic}
(chapter \ref{det-sec}), {\em timed} (chapter \ref{time-sec}), and
{\em probabilistic} concurrent system (chapter \ref{proba-sec}).  We
shall also consider an extension of $\ccs$, known as $\pi$-calculus
(chapter \ref{pi-sec}), that allows for a rather direct embedding of
higher-order functional programs.

\section{$\ccs$}\label{ccs-ssec}
Actions in $\ccs$ are defined according to the equation
(\ref{act-with-synch}) above.  Besides the nil, prefix and
non-deterministic choice operators introduced in chapter
\ref{lts-sec}, $\ccs$ includes  operators to declare a
  local action (cf. local variable in $\parimp$), to put processes in
parallel, and to define recursive behaviors:\index{$\ccs$, syntax}
\[
\begin{array}{ll}
P      &::=  0 \Alt \alpha.P \Alt (P+P) \Alt (P\mid P) \Alt \new{a}{P} \Alt A(\vc{a})
\end{array}
\]
where $\alpha \in \w{Act}$  and $A,B,\ldots$ are process identifiers.
An action name is free if it is not in the scope of a local
action declaration (a $\nu$).
We write $\vc{a}$ for a possibly empty list of action names $a_1,\ldots,a_n$.
Similarly, $\new{\vc{a}}{P}$ stands for $\nu a_1\cdots\nu a_n \ P$, 
and $[\vc{b}/\vc{a}]$ for $[b_1/a_1,\ldots, b_n/a_n]$.
It is assumed that each process identifier $A$ is defined 
by a unique equation $A(\vc{b})=P$ where the free names in $P$ 
are contained in the set of parameters $\set{\vc{b}}$.
For instance, $A$ could be a process identifier defined by the
equation:
\begin{equation}\label{buf1}
A(a,b) = a.\new{c}{(A(a,c) \mid \ol{b}.A(c,b))}~.
\end{equation}
Here the set of variables occurring free in 
$a.\new{c}{(A(a,c) \mid \ol{b}.A(c,b))}$ is $\set{a,b}$ which happens
to be included (actually equal) to the set of parameters of the process identifier $A$. Also notice that an action name, say $b$, may appear in a prefix as such or in its dual form $\ol{b}$.

Moving towards semantics, the main design decision consists in regarding
$a,b,\ldots$ as channel names on which parallel processes synchronize.
More precisely, a synchronization may only happen when a process is ready to
perform an action and another parallel 
process is ready to perform its co-action as, 
{\em e.g.}, in the process
$(a.P \mid \ol{a}.Q)$. Following the synchronization, the process moves
to $(P\mid Q)$. $\ccs$ is an asynchronous model of concurrency where
interaction is possible through {\em rendez-vous} synchronization on
{\em pure} channels. A {\em rendez-vous} channel is a channel of
null capacity where the sender must always wait for a receiver.
A channel is {\em pure} if no message value is exchanged; all that
matters is the synchronization. 

An important consequence of assuming a synchronization by 
{\em rendez-vous} is to offer a better control on the role of the environment. 
In the $\parimp$ model, the environment can modify the (visible part
of the) state and these modifications may affect the future
computation of the process. In $\ccs$, the only way the environment
may affect the computation of the process is to perform an action
which is dual to an action that the process is {\em ready} to perform.

Starting from this intuition, we follow {\em two paths} to define 
a compositional semantics of $\ccs$. The first path consists in 
associating a labelled transition system with each $\ccs$ process.
Then the equivalences on lts defined in the previous chapter \ref{lts-sec},
apply to $\ccs$ processes too and lead to a compositional semantics.
The second path consists in looking at $\ccs$ as a (rudimentary)
programming language and define its possible reductions
similarly to what we have done for the $\parimp$ language
in chapter \ref{introduction-sec}. Then what needs to be done is to fix
a notion of observable and to derive a notion of compositional equivalence.
We work with the notion of (weak) bisimulation introduced
in chapter \ref{lts-sec} and in the end, we show that the two
paths outlined above actually lead to the same compositional
equivalence.

As a concrete example illustrating the difference between 
the two approaches, consider the $\ccs$ process $P\equiv (a.0 \mid \ol{a}.0)$.
In the first approach, we have to consider the labelled transitions:
\[
\begin{array}{c}
P \act{a} (0 \mid \ol{a}.0) \act{\ol{a}} (0\mid 0)~,  \quad
P \act{\ol{a}}(a.0 \mid 0) \act{a} (0\mid 0)~, \quad
P \act{\tau} (0 \mid 0)~.
\end{array}
\]
While in the second, we just have have the reduction:
\[
P \arrow (0 \mid 0)~.
\]
We shall see that the $\tau$ transitions correspond to the reductions
while the other labelled transitions correspond to interactions
with the environment.

\section{Labelled transition system for $\ccs$}\label{lts-ccs-ssec}
Table~\ref{ccs-lts} \index{$\ccs$, lts} describes a lts for $\ccs$ processes where
the  symmetric rules for $\mid$ and $+$ are omitted.
\begin{table}
{\footnotesize
\[
\begin{array}{cc}

\infer{}{\alpha.P \act{\alpha} P}

&\infer{P\act{\alpha} P'\quad \alpha \notin \set{a,\ol{a}}}{\new{a}{P} \act{\alpha} \new{a}{P'}} \\ \\ 

\infer{P\act{\alpha} P'}{P\mid Q \act{\alpha} P'\mid Q}

&\infer{P\act{a}P' \qquad Q\act{\ol{a}} Q'}{(P\mid Q) \act{\tau} (P'\mid Q')} \\ \\ 

\infer{P\act{\alpha} P''}{P+P' \act{\alpha} P''}

&\infer{B(\vc{a})=P\quad [\vc{b}/\vc{a}]P\act{\alpha} P'}{B(\vc{b}) \act{\alpha} P'}

\end{array}
\]}
\caption{Lts for $\ccs$ (symmetric rules omitted)}\label{ccs-lts}
\end{table}

\begin{exercise}[labelled transitions]\label{lts-ccs-ex}
Check that:
\[
\new{b}{(a.P\mid P')} \mid \new{c}{\ol{a}.Q} \act{\tau}
\new{b}{(P\mid P')} \mid \new{c}{Q}~.
\]
\end{exercise}

\begin{example}[an unbounded buffer in $\ccs$]\label{buf1-ex}
In $\ccs$, the communication is by {\em rendez-vous} (or handshake, or synchronous).
What if we want {\em channels with buffers}? 
One approach is to {\em enrich the model}. 
Another approach is to show that {\em buffers can be expressed in $\ccs$}.
An unbounded buffer taking inputs on $a$ and producing outputs on $b$
can be written as (up to renaming, this is the same as equation (\ref{buf1})):
\[
\w{Buf}(a,b)  = a.\new{c}{(\w{Buf}(a,c) \mid \ol{b}.\w{Buf}(c,b))} ~.
\]
We write more suggestively $a\mapsto b$ for $\w{Buf}(a,b)$, 
assuming $a\neq b$.
We would like to show that $a\mapsto b$ works  indeed as an {\em unbounded buffer}.  
Let $\ol{c}^{n} = \ol{c}\ldots\ol{c}.0$, $n$ times, $n\geq 0$.
We should have:
\[
P(n)=\new{a}{(\ol{a}^{n} \mid a\mapsto b)} \mbox{ `equivalent to' } \ol{b}^{n}
\]
An interesting exercise because
$P(n)$ has a {\em  non trivial dynamics}.
For the time being we just analyze {\em some} of 
the labelled transitions of $P(n)$.
\begin{itemize}

\item For $n=0$, $P(n)$ cannot reduce.

\item For $n>0$, we need to generalize a bit the form of the process
$P(n)$. Let $Q(n,m)$ be a process of the form:
\[
Q(n,m) = \new{a,c_1,\ldots,c_m}{(\ol{a}^{n} \mid a\mapsto c_1 \mid \cdots 
                                  \mid c_m \mapsto b)}~,
\]
for $m\geq 0$. Note that $P(n) = Q(n,0)$ and $Q(0,k)$ cannot reduce for any
$k$.  Moreover, the message can traverse the whole chain so that for $n>0$:
\[
Q(n,m) \wact{\ol{b}} Q(n-1,2m+1)~.
\]
Thus:
\[
P(n) \wact{\ol{b}} \cdots \wact{\ol{b}} Q(0,2^{n}-1) \mbox{ `equivalent to' } 0~,
\]
where we recall that: $\wact{\ol{b}}= (\act{\tau})^* \act{\ol{b}} (\act{\tau})^*$.
\end{itemize}

Note that there are plenty of reductions we did not consider!
Yet, in chapter \ref{det-sec} we shall be able to conclude that 
this analysis suffices to derive that 
$P(n)$ is `equivalent to' $\ol{b}^n$.
\end{example}

We can regard $\ccs$ as a {\em labelled transition system} where
states are processes. We say that $P$ and $Q$ are {\em strongly
  bisimilar} (written $P\bis Q$) if they are bisimilar with respect to
the lts we have just defined. We say that they are {\em weakly
  bisimilar} (written $P\wbis Q$) if they are bisimilar with respect
to the derived lts $\wact{\alpha}$ where internal actions are
`abstracted'. Obviously $P\bis Q$ implies $P\wbis Q$.
Next, we consider the issue of compositionality.

\begin{definition}[$\ccs$ context]\index{$\ccs$, context}
A {\em context} $C$ is a process with a hole $[~]$ (here and in the following
we omit the symmetric cases when listing contexts):
\[
C::= [~] \Alt \alpha.C \Alt C+P \Alt C\mid P \Alt \new{a}{C}~.
\]
\end{definition}

\begin{proposition}
If $P\bis Q$ then $C[P] \bis C[Q]$.
\end{proposition}
\Proof
We apply the standard technique which amounts to define a relation $\cl{R}$
which includes the processes of interest and show that it is a 
bisimulation.
\[
\begin{array}{ll}

\mbox{Prefix} &\cl{R} = \set{(\alpha.P_1,\alpha.P_2) \mid P_1\bis P_2}\union \bis~\\

\mbox{Sum} 
&\cl{R} = \set{(P_1+Q,P_2+Q) \mid P_1\bis P_2} \union \bis~ \\

\mbox{Parallel} 
&\cl{R}= \set{(P_1 \mid Q,P_2\mid Q) \mid P_1\bis P_2} \union \bis~  \\

\mbox{Restriction}  
&\cl{R} = \set{(\new{a}{P_1},\new{a}{P_2}) \mid P_1\bis P_2} \union \bis~. 
\end{array}
\]
~\qed

\begin{remark}
In general, the non-deterministic sum does {\em not} preserve weak
bisimulation as:
\[
\tau.a \wbis a    \qquad \mbox{ but } \qquad \tau.a+b \not\wbis a+b~.
\]
\end{remark}

However, a {\em guarded version} of the non-deterministic sum has this property.
Denote with $D$ the following contexts:
\[
D::= [~] \Alt \alpha.D \Alt \alpha'.D+P \Alt (D\mid P) \Alt \new{a}{D}
\qquad \alpha' \neq \tau~.
\]

\begin{proposition}
If $P\wbis Q$ then $D[P] \wbis D[Q]$.
\end{proposition}
\Proof Similar to the strong case. \qed

\begin{remark}[on unguarded sum]
There are {\em two viewpoints} on the non-preservation of weak-bisimulation
by the sum:

\begin{enumerate}

\item One should take the largest congruence which refines $\wbis$. 
Then, {\em e.g.}, we should distinguish $\tau.a$ from $a$.

\item In most applications one just needs a guarded sum and so it
is enough to have a notion of equivalence which is preserved by guarded sums.

\end{enumerate}

The second viewpoint tends to prevail at least in formalisms
like $\parimp$ (chapter \ref{introduction-sec}) 
or the $\pi$-calculus (chapter \ref{pi-sec}) 
where a guarded form of sum is
easily derived from parallel composition. 

\end{remark}

\begin{exercise}[sequentialization in $\ccs$]\label{seq-ccs-ex}
We consider a fragment of $\ccs$ (we drop sum and recursive definitions) 
extended with
an operator `;' for process sequentialization. Thus the {\em process syntax} is as follows:
\begin{equation}\label{extended}
P::= 0 \Alt \alpha.P \Alt (P\mid P) \Alt P;P \Alt \new{a}{P} \qquad \alpha\in \set{a,\ol{a} \mid a\in A}\union \set{\tau}~.
\end{equation}
The {\em labelled transition system} for $\ccs$ is extended with the following rules for process sequentialization:
\[
\infer{P\act{\alpha} P'}
{P;Q \act{\alpha} P';Q} \qquad 
\infer{P\searrow \qquad Q\act{\alpha} Q'}
{P;Q \act{\alpha} Q'}
\]
Here the predicate $\searrow$ denotes {\em proper termination} and it is defined as the least set of processes 
such that:\footnote{This is related to (but slightly different from) the {\em immediate termination} predicate
introduced for the language $\parimp$ (chapter \ref{introduction-sec}).}
\[
\begin{array}{llll}
\infer{}
{0\searrow}
\quad
&\infer{P\searrow \quad Q\searrow}
{(P\mid Q) \searrow}
\quad
&\infer{P\searrow \quad Q\searrow}
{(P;Q)\searrow}
\quad
&\infer{P\searrow}
{\new{a}{P}\searrow}~.
\end{array}
\]
If $P$ is a process in the extended language (\ref{extended}) 
and $c$ is a name not occurring
in $P$ then $\sem{P}c$ is a $\ccs$ process (without process sequentialization) 
defined as follows:
\[
\begin{array}{lll}

\sem{0}c &= \ol{c}.0  \\
\sem{\alpha.P}c   &= \alpha.\sem{P}c \\
\sem{P_1\mid P_2}c &= \new{c_1,c_2}{((\sem{P_1}c_1 \mid \sem{P_2}c_2) \mid c_1.c_2.\ol{c}.0)} &(c_1,c_2\notin \fv{P_1}\union \fv{P_2}\union \set{c})\\
\sem{P_1;P_2}c    &= \new{c'}{(\sem{P_1}c' \mid c'.\sem{P_2}c)} 
&(c'\notin \fv{P_1}\union \fv{P_2}\union \set{c}) \\
\sem{\new{a}{P}}c &= \new{a}{\sem{P}c} &(c\neq a)\\

\end{array}
\]
(1) Show that if $P\searrow$ then $\sem{P}c \wact{\ol{c}} Q$ and $Q\bis 0$, where $\bis$ is the largest {\em strong bisimulation}.
(2) Define the notion of {\em weak simulation up to strong bisimulation} and show that this is a sound technique
to prove weak simulation.
(3) Show that for all processes $P$, $P$ is {\em weakly simulated} by $\sem{P}c$ up to strong bisimulation.
\end{exercise}

\begin{exercise}[bisimulation up-to context]\label{bis-upto-ex}\index{bisimulation, up-to context}
We introduce a notion of bisimulation up to the contexts.
For simplicity let us assume:
$D::= [~] \Alt \alpha.D \Alt (D\mid P) \Alt \new{a}{D}$.
Suppose:
\[
\cl{H}(\cl{R}) = \set{(P',Q') \mid \xst{D}{P'\equiv D[P], P\rel{R} Q, D[Q]\equiv Q'}}~.
\]
Let $\cl{F}$ be the monotonic function associated with (one step) weak
bisimulation.
\begin{enumerate}

\item Analyze the one-step transitions of a process $D[P]$ as a function
of the transitions of $P$ and $D[0]$.

\item Show that $\cl{H}$ preserves $<_\cl{F}$.

\item Explicit the associated notion of bisimulation up to context.

\end{enumerate}
\end{exercise}

\begin{exercise}[prime factorization]\label{prime-fact-ex}
Let $A$ be a set of actions with generic elements $a,b,\ldots$ 
and let $P$ denote a process in the following fragment of $\ccs$:
\[
P::= {\bf 0} \Alt a.P \Alt (P+P) \Alt (P\mid P)\qquad a\in A~.
\]
Notice that there is no notion of co-action
and therefore no possibility of synchronization among parallel processes.
In this exercise when we speak of a process, we refer to a process
in this fragment. We define the {\em size of a process}, say $|P|$, by induction on the structure
of $P$ as follows:
\[
\begin{array}{llll}

|{\bf 0}|=0~,
&|a.P|=1+|P|~,
&|P+Q|=\w{max}(|P|,|Q|)~,
&|P\mid Q| = |P|+|Q|~.
\end{array}
\]
We say that a process $P$ is {\em irreducible} if $P\sim P_1\mid P_2$ 
implies that $P_1\sim 0$ or $P_2\sim 0$ and we say that
$P$ is {\em prime} if $P$ is irreducible and moreover $P\not\sim 0$.
Prove the following assertions.
\begin{enumerate}
\item If $P\act{a}Q$ then $|P|> |Q|$.

\item If $P\sim Q$ then $|P|=|Q|$ (but the converse fails).

\item For all processes $P,Q,R$ the following properties hold:
\begin{itemize}

\item If $(P\mid R) \sim (Q\mid R)$ then $P\sim Q$ (this is a kind of 
{\em cancellation property}).

\item If $(P\mid R) \sim (Q\mid R')$ and $R\act{a}R'$ then there exists 
$Q'$ such that $Q\act{a}Q'$ and $P\sim Q'$.
\end{itemize}

\item Every process $P$ such that $P\not\sim 0$ can 
be expressed up to strong bisimulation as the {\em parallel composition of
prime processes}.

\item Every process $P$ such that $P\not\sim 0$ has a {\em unique
decomposition} as the parallel composition of prime processes.
Unicity here has to be understood in the same sense as the
unicity of the prime factorization of a natural number.
\end{enumerate}
\end{exercise}

\section{A reduction semantics for $\ccs$}\label{red-ccs-sec}
We want to define a reduction semantics for $\ccs$.
A technical problem is that  in the {\em syntax}, 
the synchronizing processes can be far away as in:
$\ \new{b}{(a.P\mid P')} \mid \new{c}{\ol{a}.Q}$.
In the lts presented in Table~\ref{ccs-lts} we have
tackled this problem by keeping track of the {\em potential} 
transitions of every sub-process.
An {\em alternative approach} consists in introducing a notion
of {\em structural equivalence} on processes which is strong enough
to bring two synchronizing processes in contiguous positions and weak
enough to identify only processes that are intuitively equivalent.
To simplify the formalization we drop the non-deterministic sum.
In this context, we assume a structural equivalence $\equiv$ which 
is the least congruence such that: (i) it includes renaming, 
(ii) parallel composition is associative and commutative, and (iii):
\[
\begin{array}{ll}
\new{a}{(P\mid Q)} \equiv \new{a}{P} \mid Q &\mbox{if }a\notin \fv{Q}\\
A(\vc{b}) \equiv [\vc{b}/\vc{a}]P
&\mbox{if }A(\vc{a})=P~.
\end{array}
\]
An {\em evaluation context} $E$ is defined by:
\begin{equation}
E::=[~]\Alt \new{a}{E} \Alt E\mid P
\qquad \mbox{(evaluation contexts).}
\end{equation}
Then the {\em reduction relation} is:
\[
\begin{array}{ll}
P\arrow Q  &\mbox{if }\quad 
P\equiv E[a.P'+Q'\mid \ol{a}.P''+Q''] \mand Q\equiv E[P'\mid P''] \\

P\arrow Q &\mbox{if }\quad
P\equiv E[\tau.P+Q'] \mand Q\equiv E[P]~.
\end{array}
\]

\begin{exercise}[reduction up to structural equivalence]\label{red-uptostreq-ex}
Check that:
\[
\new{b}{(a.P\mid P')} \mid \new{c}{\ol{a}.Q} \arrow
\new{b}{(P\mid P')} \mid \new{c}{Q}~.
\]
\end{exercise}

Internal transitions and reductions can be related as follows.

\begin{proposition}\label{lts-red-prop}
Let $P$ be a $\ccs$ process (without non-deterministic sum).
\begin{enumerate}

\item The structural equivalence $\equiv$ is a strong bisimulation.

\item  If $P\act{\tau} P'$ then $P\arrow P'$.

\item  If $P\arrow P'$ then $P\act{\tau}P''$ and $P'\equiv P''$.

\end{enumerate}
\end{proposition}

\begin{exercise}\label{proof-lts-red-prop}
  Prove proposition \ref{lts-red-prop}.
  \end{exercise}

The next step is to introduce some candidates for the notion of 
{\em basic observable}. \index{commitment}
We write
$P \act{a}$
if the process $P$ is `ready to perform' a visible communication action on channel $a$.
This is also called a {\em strong commitment} (or {\em barb}). \index{commitment}
It is a simple exercise to define $\act{a}$ by induction on the
structure of $P$.
We also write $P\wact{a}$ if for some $Q$, $P\wact{\tau} Q$ and $Q\act{a}$.
This is also called a {\em weak} commitment.

It is also possible to abstract the polarity of the commitement (input
or output) and the name of the channel on which the process commits.
So we write:
\[
\begin{array}{llllll}
P\act{\exists} \mbox{ if } \xst{a}{P\act{a}}~,  \qquad
&P\wact{\exists} \mbox{ if } \xst{a}{P\wact{a}}~.
\end{array}
\]
Finally, we use $\Downarrow$ for weak normalisation:
$P\Downarrow$ if $\xst{P'}{P\Arrow P' \mand P'\not\act{\tau}}$.

\begin{definition}[static contexts]\index{$\ccs$, static context}
We define the static contexts as the contexts of the following shape:
\[
C::= [~] \Alt (C\mid P) \Alt \new{a}{C}~.
\]
\end{definition}

Intuitively, they  are called {\em static}  because  
they persist after a transition
(unlike a prefix or a sum). It is generally held that a useful equivalence 
should be preserved at least by static contexts.
Incidentally, in the simple case considered, 
static contexts coincide with evaluation 
contexts.
Next we introduce a notion of compositional equivalence which is
based on the notion of bisimulation.

\begin{definition}[contextual bisimulation]\label{cxt-bis}\index{bisimulation, contextual}
A binary relation $\cl{R}$ on processes is a {\em strong
contextual bisimulation} if whenever $P\rel{R}Q$ the following
conditions hold (and reciprocally for $Q$):

\begin{description}

\item[(cxt)] For all {\em static contexts} $C$, $C[P] \rel{R} C[Q]$.

\item[(red)] If $P\arrow P'$ then for some $Q'$, $Q\arrow Q'$ and $P'\rel{R}Q'$.

\item[(cmt)] For all observable action $a$, if $P\act{a}$ then $Q\act{a}$.

\end{description}
For the {\em weak} version replace $\arrow$ by $\Arrow$ and
$\act{a}$ by $\wact{a}$. For the one-step weak version the
replacement only takes place on the right of the implication.
Denote with $\cbis$ ($\cwbis$) the largest contextual (weak) bisimulation.
\end{definition}

Informally, we can say that a contextual bisimulation is a relation
that is preserved by static contexts and by reduction, and that is
compatible with commitments. If we drop the preservation
by static contexts we obtain the following notion.

\begin{definition}[barbed bisimulation]\label{barb-bis-def}\index{bisimulation, barbed}
A binary relation $\cl{R}$ on processes is a {\em strong
barbed bisimulation} if whenever $P\rel{R}Q$ the following
conditions hold (and reciprocally for $Q$):

\begin{description}

\item[(red)] If $P\arrow P'$ then for some $Q'$, $Q\arrow Q'$ and $P'\rel{R}Q'$.

\item[(cmt)] For all observable action $a$, if $P\act{a}$ then $Q\act{a}$.

\end{description}
We denote with $\bbis$ the largest such equivalence, and 
with $\wbbis$ its weak variant.
\end{definition}

Barbed bisimulation distinguishes less processes than contextual bisimulation
and it is not preserved by parallel composition.

\begin{exercise}[on barbed bisimulation]\label{barb-bis-ex}
In the framework of $\ccs$, show that barbed bisimulation is {\em not} preserved
by parallel composition.
\end{exercise}

Because preservation by the operators of the language is essential
for compositional reasoning, the notion of barbed bisimulation
can be refined as follows.

\begin{definition}[barbed equivalence]\index{barbed equivalence}
We say that two processes are {\em barbed equivalent} if
put in any static context they are barbed bisimilar.
We denote such equivalence with $\bebis$.
The weak variant based on weak barbed bisimulation is denoted with 
$\wbebis$.
\end{definition}

Then the comparison of contextual bisimulation and barbed equivalence
arises as an obvious question.

\begin{exercise}[on barbed equivalence]\label{barb-eq-ex}
Show that if two processes are contextually bisimilar then
they are barbed equivalent.
\end{exercise}

The converse can be quite tricky to prove. 
The characterization of labelled
bisimulation we are aiming at is more direct/natural when working with
{\em contextual bisimulation} than with {\em barbed equivalence}.

\begin{exercise}[variations on commitment]\label{cmt-ex}
Show that we get an 
equivalent notion of contextual bisimulation if the condition
[cmt] is replaced by: 
$P \wact{\exists} \mbox{ implies }
Q \wact{\exists}$. On the other hand, show that we get an incomparable 
notion of contextual bisimulation if the condition
[cmt] is replaced by: 
$P \Downarrow \mbox{ implies }
Q \Downarrow$.
\end{exercise}

The labelled bisimulation introduced in section \ref{lts-ccs-ssec}
is an example of contextual bisimulation.

\begin{proposition} 
The largest labelled bisimulation is a contextual
bisimulation (both in the strong and weak case).
\end{proposition}
\Proof
Denote with $\bis$ ($\wbis$) the labelled (weak) bisimulation.
It has been proved that $\bis$ and $\wbis$ are preserved by static  contexts.
Incidentally, note that an {\em arbitrary} labelled bisimulation
does not need to be saturated by static contexts.
Moreover the conditions [red] and [commit] of contextual  bisimulation are 
particular cases of the bisimulation game in the  labelled case. \qed \\

The previous proposition shows that every labelled bisimulation is 
contained in contextual bisimulation. The converse is given
by the following.

\begin{proposition}\label{cxtbis-is-labbis-prop}
The largest  contextual bisimulation $\cbis$ (or $\cwbis$ in the weak case)
is a labelled (weak) bisimulation.
\end{proposition}
\Proof 
We consider directly the {\em weak case}. 
An {\em internal choice}\footnote{It is called internal 
because the environment has no way of controlling it; by opposition, a choice such as 
$a.P+b.Q$ is called external.} 
in $\ccs$ can be defined as follows:
\[
P\isum Q = \new{a}{(a.P \mid a.Q \mid \ol{a})} = \tau.P+\tau.Q~.
\]
If $P\cwbis Q$ and $P\wact{\tau} P'$ then 
$Q\wact{\tau} Q'$ and $P'\cwbis Q'$, by condition [red]
of contextual bisimulation.

So suppose $P\wact{a} P'$. 
Let $o_1,o_2$ be two distinct fresh names (not in $P$ and $Q$)
and define the static context:
\begin{equation}\label{cxt-proof-cxt}
C= [~] \mid \ol{a}.(o_1 \isum (o_2 \isum 0))~.
\end{equation}
By hypothesis, $C[P] \cwbis C[Q]$.
Clearly, $C[P]\wact{\tau} P'\mid (o_2 \isum 0)$ and again
by hypothesis (condition [red]) 
$C[Q] \wact{\tau} Q''$ and $P''\equiv P'\mid (o_2 \isum 0) \cwbis Q''$.

Now we argue that $Q''$ must be of the shape 
$Q'\mid (o_2 \isum 0)$ where $Q\wact{a} Q'$.
The case $Q''= C[Q']$ and $Q\wact{\tau} Q'$ is impossible
because $P''\wact{o_{2}}$ entails $Q'\wact{o_{2}}$, and 
the latter entails $Q'\wact{o_{1}}$ which cannot be matched by $P''$.
The cases  $Q\wact{a} Q'$ and $Q''= Q'\mid R'$ where 
$R'\in \set{o_1\isum (o_2 \isum 0), o_1, o_2, 0}$ are also impossible
for similar reasons.
Thus we must have $Q''= Q'\mid (o_2\isum 0)$ and 
$P'\mid (o_2 \isum 0) \cwbis Q''$.

It is easy to argue that since 
$P'\mid (o_2\isum 0) \wact{\tau} P'\mid 0$ we must have
$Q' \mid (o_2 \isum 0) \wact{\tau} Q_1 \mid 0$ and 
$Q_1 \mid 0 \cwbis P'\mid 0$.
Thus $Q\wact{a} Q_1$ and 
$P' \equiv (P'\mid 0) \cwbis (Q_1 \mid 0) \equiv Q_1$.
Strictly speaking, we use an {\em up to technique}. \qed

\begin{exercise}\label{cxt-bis-strongcase-ex}
  \begin{enumerate}
  \item Show that in the strong case it is possible
    to simplify the context $C$ (\ref{cxt-proof-cxt}) in the proof above.

    \item Show that (weak) labelled bisimulation implies
(weak) barbed bisimulation.

      \item Show that (weak) labelled bisimulation implies
        (weak) barbed equivalence.
        \end{enumerate}
\end{exercise}

\section{Value-passing $\ccs$ (*)}\index{$\ccs$, value passing}\label{value-ccs-sec}
As already mentioned, in $\ccs$ communication is pure synchronization.
We now consider an extension where {\em values} can be sent
along channels. In the following, values are just basic
atomic objects such as booleans or integers which can be tested
for equality: 
\[
v::= v_0 \Alt v_1 \Alt v_2 \ldots \qquad \mbox{(values)}
\]
In chapter \ref{pi-sec}, we shall consider the more complex case
where the values are actually channels.
Let $\w{Val}$ be the set of values. Then 
the collection of actions given by equation (\ref{act-with-synch}) 
is revised as follows:
\begin{equation}\label{act-with-synch-value}
\w{Act}= \set{av, \ol{a}v \mid a\in A, v\in \w{Val} } \union \set{\tau}
\qquad \mbox{(actions with value passing)}
\end{equation}
Input and output actions are now pairs composed of a channel name 
and a value.
To write value-passing $\ccs$ processes, we need a notion of {\em variable}
ranging over values which stands for the value read upon communication.
\[
\w{id}::= x \Alt y \Alt \ldots \qquad \mbox{(variables)}
\]
We also need a notion of {\em term} which is either a value or a variable
(we call this {\em term} by analogy with first-order logic):
\[
t::= v \Alt \w{id} \qquad \mbox{(term)}
\]
Then the syntax of $\ccs$ value-passing processes is as follows:
\[
\begin{array}{ll}
P      &::=  0 \Alt a(\w{id}).P \Alt \ol{a}t.P \Alt [t=t]P,P \Alt (P+P) \Alt (P\mid P) \Alt \new{a}{P} \Alt A(\vc{a})
\end{array}
\]
where $a(x).P$ is the process that receives a value $v$ on the channel
$a$ and becomes $[v/x]P$, $\ol{a}v.P$ is the process that sends $v$ on
$a$ and becomes $P$, and $[v=v']P,Q$ is the process that compares the
values $v$ and $v'$ and runs $P$ if they are equal and $Q$ otherwise.

The reduction semantics for $\ccs$ is easily extended to value passing
$\ccs$. Omitting the details concerning the evaluation context and 
the structural equivalence, the synchronization rule with 
exchange of values is:
\[
\ol{a}v.P  \mid a(x).Q \arrow P \mid [v/x]Q~,
\]
and we add the usual rules for the conditional (cf. chapter
\ref{introduction-sec}):
\[
[v=v]P,Q \arrow P~, \qquad [v=v']P,Q \arrow Q \quad (v\neq v') ~.
\]
Notice that the resulting reduction semantics is supposed
to operate on terms {\em without} free value variables. 
As a matter of fact, reducing processes with free variables would be a form
of {\em symbolic execution} and requires carrying along with
the process a set of constraints which describe the
possible values of its free variables.

The labelled semantics of $\ccs$ with value passing rises some subtle
issues concerning the treatment of the input prefix.
Consider a process $a(x).P$. The action structure we have given
above in equation (\ref{act-with-synch-value}) suggests
a rule of the shape:
\begin{equation}\label{early-binding}\index{binding, early}
\infer{}{a(x).P \act{av} [v/x]P}\qquad \mbox{(early binding)}~.
\end{equation}
However, by changing a little bit the action structure (\ref{act-with-synch-value}),
we could also think of a rule that maps a process
to a {\em function} from values to processes:
\begin{equation}\label{late-binding}\index{binding, late}
\infer{}{a(x).P \act{a} \lambda x.P}\qquad \mbox{(late binding)}~.
\end{equation}
This in turn requires defining an obvious notion of bisimulation on
functions: two functions $\lambda x.P$ and $\lambda x.Q$ from values
to processes are bisimilar if for all values $v\in \w{Val}$, $[v/x]P$
and $[v/x]Q$ are bisimilar (cf. chapter \ref{equiv-sec}).
The first rule is called {\em early binding} and formalizes a situation
where the communication channel and the value received are 
selected at the same time. By opposition, 
the second rule is  called {\em late binding}.
It turns out that the late binding approach leads to 
a labelled bisimulations which is more discriminating than
the one based on early binding.
For instance, consider the processes:
\[
\begin{array}{lll}
P &\equiv &a(x).([x=v_0]\ol{b}.0,0 + [x=v_1]\ol{c}.0,0)~, \\
Q &\equiv &a(x).[x=v_0]\ol{b}.0,0 + a(x).[x=v_1]\ol{c}.0,0~.
\end{array}
\]
The processes $P$ and $Q$ are `early-binding bisimilar' but not `late-binding
bisimilar'. Specifically, by a late-binding input the process $P$ goes
to a function that cannot be matched by $Q$.
In this case, the comparison with contextual bisimulation suggests
that the early binding semantics is the `right' one.

We conclude this quick review of value passing $\ccs$ by mentioning
that at the price of an {\em infinitary} syntax, it is quite simple to 
reduce it to ordinary $\ccs$. This is similar
in spirit to transformations from predicate logic
to propositional logic where universal and existential quantifications
are replaced by infinitary conjunctions and disjunctions, respectively.
In our case, the basic idea is to replace the input of a value by 
the non-deterministic sum of infinitely many inputs:
\begin{equation}\label{input-infinite-sum}
\sem{a(x).P } = \Sigma_{v\in \w{Val}} a_v.\sem{[v/x]P}~.
\end{equation}
Incidentally, for a finite and small set of values this gives an effective
way of programming value passing in basic $\ccs$.

\section{Summary and references}
Labelled bisimulation requires: labels, labelled transitions, and
labelled bisimulation.  The choice of the labels and the rules of the
bisimulation game may be hard to justify.  On the other hand,
contextual bisimulation requires reduction, static contexts, and
commitments.  This approach is more natural but it may be harder to
prove that two processes are contextual bisimilar.  For $\ccs$, labelled
bisimulation coincides with contextual bisimulation.  In general this
kind of result is a guideline when we are confronted to more
complicated models (such as the $\pi$-calculus in chapter
\ref{pi-sec}).  

$\ccs$ is a model of message passing based on redez-vous communication
among two processes.  Another popular interaction mechanism consists
in allowing several parallel processes to synchronize on the same
label. This mechanism does not scale so well when we want to add more
structure to the actions as, {\em e.g.}, in value passing
synchronization. 

$\ccs$ has been introduced by Milner in
\cite{DBLP:books/sp/Milner80}\index{Milner}; a revised
presentation is in \cite{Milner:1995:CC:225494}.
The reduction semantics of concurrent systems is put forward
in \cite{DBLP:journals/tcs/BerryB92}.
The notion of {\em contextual
  bisimulation} is studied by \cite{DBLP:journals/tcs/HondaY95}.  The
earlier definition of {\em barbed equivalence} can be found in
\cite{DBLP:conf/icalp/MilnerS92}.
Exercise \ref{prime-fact-ex} is based on \cite{DBLP:journals/tcs/MilnerM93}.

\chapter{Testing processes}\label{testing-sec}\markboth{{\it Testing}}{{\it Testing}}
\markboth{{\it Testing}}{{\it Testing}}

In this chapter, we discuss an alternative approach to the
notion of process (in-)equivalence which is based on a notion {\em test}.
The basic ingredients of the approach are as follows:
\begin{itemize}

\item a process $P$ is run in parallel with a testing
  process $Q$ which is of a similar nature and able to interact with $P$,

\item an (internal) computation of $(P\mid Q)$ is deemed {\em successful}
  if it reaches a configuration where a {\em certain (simple) predicate} is valid.

\end{itemize}  

We shall cast the technical development in the setting of the
labelled transition systems with synchronisation described in
chapter \ref{ccs-sec}.
Since the processes we consider are {\em non-deterministic}, each given test can
produce several computations and one is naturally led to distinguish
two basic situations: the one where at least one computation
is successful and the one where all computations are successful.
In the first case, one says that $P$ {\em may} pass the test $Q$ while
in the second one says that $P$ {\em must} pass the test $Q$.
Once this framework is fixed, it is immediate to define two pre-orders
based on the {\em may} and the {\em must} interpretation of the test.
Namely, if $P_1,P_2$ are processes then we say that $P_1$ is {\em less or equal than} $P_2$ if
every time $P_1$ may (respectively, must) pass a test, $P_2$ may
(respectively, must) pass the same test.

Historically, the theory of testing has been developed around
a predicate that asserts a {\em strong commitment} (in the
sense of section \ref{red-ccs-sec}) on
a distinguished action $w$. In this case,
$P$ {\em may} pass the test $Q$ means that $(P\mid Q)$ may
reduce to a process that strongly commits on $w$ while
$P$ {\em must} pass the test $Q$ means that every computation
of $(P\mid Q)$ reaches a point where it strongly commits on $w$.
Section \ref{testing-cmt-sec}, characterizes the
induced may and must pre-orders in this case.

Another basic predicate that comes to mind is {\em termination}
intended as the impossibility to perform an internal action.
In this case, may and must have a familiar interpretation:
$P$ {\em may} pass the test $Q$ means that $(P\mid Q)$
is {\em weakly normalizing} and $P$ {\em must} pass the test $Q$ means
that $(P\mid Q)$ is {\em strongly normalizing}. Section \ref{testing-ter-sec},
characterizes the induced may and must pre-orders in this case
and compares them with the traditional ones.

\section{Testing notation}\label{testing-notation-sec}
We recall some standard notations discussed in the previous chapters and
introduce a few more which turn out to be handy in the testing framework.

\subsection*{Actions, transitions, and commitments}
We work on the labelled transition systems with synchronisation
described in chapter \ref{ccs-sec}.
We write $a,b,\ldots$ to denote observable
actions (not the internal action $\tau$)  varying on a set
$\w{Act}$. We also write $s,s',\ldots$ to denote
finite traces (or words) of observable actions. If $s$ is such a trace then
$\ol{s}$ is the trace obtained by taking the co-actions of
the actions in $s$.
We have the usual {\em weak transition} relations:
\[
\begin{array}{ll}
\wact{\tau}= (\act{\tau})^*~,
&\wact{a} = (\wact{\tau}) \comp (\act{a}) \comp (\wact{\tau})~, \\ \\
\wact{\epsilon} = \wact{\tau}~,
&\wact{a_{1}\cdots a_{n}} = (\wact{a_{n}}) \comp \cdots \comp (\wact{a_{1}})~.
\end{array}
\]
We say that a process $P$ is {\em stable} if it cannot perform an internal reduction.
Given a process $P$, we define:
\[
\begin{array}{lll}
  S(P) &=\set{a\in \w{Act} \mid P\act{a}\cdot}  &\mbox{(strong commitments),} \\
  W(P) &=\set{a \in \w{Act} \mid P\wact{a}\cdot} &\mbox{(weak commitments).}
\end{array}
\]
In general, $S(P)\subseteq W(P)$ and $S(P)=W(P)$ if $P$ is stable.

We assume the labelled transition systems have a {\em nil} element that
performs no action and are {\em closed} under the operations of
action prefix, sum, parallel composition,
and name restriction described in chapter \ref{ccs-sec}.
In particular, we shall use the CCS notation to describe
certain families of tests and suppose $\Omega$ is
a process that loops such as:
\[
\begin{array}{lll}
  \Omega        &=\new{a}{(\ol{a}\mid \s{rec} X.a.(\ol{a}\mid  X))} &\mbox{(definitely looping process)}~.
\end{array}
\]

\subsection*{Finite and infinite traces}
The set $\trace{p}$ of {\em finite traces} of a process $P$
is defined as:
\[
\trace{P} = \set{s \mid P \wact{s}\cdot}\qquad\mbox{(set of finite traces).}
\]
We denote with $u,v,\ldots$ countably
infinite traces (or words) of observable actions.
If $u$ is an infinite trace then let $u[k]$ be the action
in position $k$ counting from position $1$
and $u(k)$ be the finite prefix of $u$ up
to the $k^{{\it th}}$ position included (thus $u(0)$ is the empty trace).
If $P$ is a process then we  write $P\wact{u}$ if
$P$ has an infinite reduction which corresponds to the trace $u$
and we define the set $\inftrace{P}$ of {\em infinite traces} of the process
$P$ as:
\[
\inftrace{P}=\set{u\mid P\wact{u}} \qquad\mbox{(set of infinite traces).}
\]

\subsection*{Hereditary termination}
If $P$ is a process, we write:
\[
\begin{array}{l}
P\downarrow \mbox{ if all internal reduction sequences terminate,}\\
P\Downarrow \mbox{ if an internal reduction sequence terminates.}
\end{array}
\]
So $\downarrow$ corresponds to {\em must-termination} (or strong normalization)
and $\Downarrow$ to {\em may-termination} (or weak normalization).

We generalize the definition of strong normalization as follows.
If $s$ is a sequence of observable actions and $a$ an action then:
\[
\begin{array}{ll}
P\downarrow \epsilon  &\mbox{if }P\downarrow \\
P\downarrow as  &\mbox{if }P\downarrow \mbox{ and } (P\wact{a}P' \mbox{ implies }P'\downarrow s)~.
\end{array}
\]
Thus if $P\downarrow s$ then all processes that may be reached from $P$ performing a prefix of the sequence of (weak) actions $s$ terminate.
Notice that $P\downarrow s$ does {\em not} imply that $P$ can actually
perform $P\wact{s}$. For instance, $0\downarrow s$ for all $s$.
The hereditary termination predicate is extended to infinite
traces by stating that $P\downarrow u$ if for all $k\geq 0$, $P\downarrow u(k)$.

We collect in the sets $C(P)$ and $C^\omega(P)$ the collection of
finite and infinite traces, respectively, along which the process
$P$ terminates:
\[
\begin{array}{llll}
C(P)  &= \set{s \mid P\downarrow s}~, \qquad

C^\omega(P)    &=\set{u \mid P \downarrow u}~.
\end{array}
\]
We say that a process $P$ is {\em reactive} if for all sequences of actions
  $s$ we have $P\downarrow s$.

\subsection*{Image finiteness}
In certain situations, we shall assume the labelled
transition system is {\em image finite} (definition
\ref{image-fin-prop}), {\em i.e.},
for all processes $P$ and actions $\alpha$ we have:
\[
\set{P' \mid P\act{\alpha} P'} \mbox{ is finite}\qquad\mbox{(image finiteness).}
\]
By extension, we say that a {\em process} is {\em image finite}
if the portion of the lts reachable from the process is image finite.
Image finite lts are a {\em sweet spot} for certain testing
pre-orders as they are the lts on which they can be
characterized in a {\em finitary} sense
(recall that in section \ref{modal-log-eq-sec}
we already relied on image finiteness to obtain a finitary logical
description of bisimulation).
Notice that we do {\em not} assume the lts is finitely branching, {\em i.e.},
the set $\set{\alpha \mid P\act{\alpha}\cdot}$ can be infinite.

\section{Testing strong commitment}\label{testing-cmt-sec}
We suppose the collection of actions contains a {\em special
  observable action} $w$. There is no co-action for this action.  We
now call {\em tests} (all) the elements of the lts and call {\em
  processes} those tests whose derivatives cannot perform the special
action $w$.  In this section, we shall reserve the
letters $P,Q$ for processes and the
letter $R$ for the (more general) tests, {\em i.e.}, the processes
that may strongly commit on the special action $w$.

\subsection*{Testing pre-orders}
A {\em computation} of a test $R_0$ is a possibly infinite sequence of internal reductions:
\[
R_0\act{\tau} R_1 \act{\tau} \cdots \act{\tau} R_n \cdots
\]
A computation of a test $R_0$ is {\em successful}
if there is an $i\in \Nat$ such that $R_i\act{w}$.
We write:
\[
\begin{array}{ll}
R\  \may &\mbox{if some computation starting from }R \mbox{ is successful.}\\
R\ \must &\mbox{if every computation starting from }R \mbox{ is successful.}
\end{array}
\]
Thus tests are partitioned in three sets: (i) those that must,
(ii) those that may but must not, and (iii) those that may not
(this last set contains all (ordinary) processes).
The relevance of a test depends very much on the
the way it is interpreted (may or must).
For instance, consider the test $R=\tau.w$. Then
$(P\mid R) \may$ is always true while $(P\mid R)\must$
is true if and only if $P$ terminates.

\begin{definition}[$\may$, $\must$ pre-orders]\index{pre-orders, $\may$, $\must$}
If $P,Q$ are processes (not tests) then we define:
\[
\begin{array}{ll}
P \mayleq Q &\mbox{if for all tests }R \ ((P\mid R) \may \mbox{ implies }
(Q\mid R) \may)~, \\

P \mustleq Q &\mbox{if for all tests }R \ ((P\mid R) \must \mbox{ implies }
(Q\mid R) \must)~.
\end{array}
\]
\end{definition}

\begin{proposition}\label{test-par-pres-prop}
May and must pre-orders are preserved by parallel composition.
\end{proposition}
\Proof
Suppose $P_1\mayleq P_2$ and $Q_1\mayleq Q_2$.
Then for all tests $R$, $(Q_1\mid R)$ is also a test
and parallel composition is associative. So
$((P_1\mid Q_1) \mid R)\may$ implies
$((P_2\mid Q_1) \mid R)\may$, and 
we derive $(P_1\mid Q_1) \mayleq (P_2\mid Q_1)$.
In the same way, we prove that
$(Q_1\mid P_2)\mayleq (Q_2\mid P_2)$.
Then using commutativity of parallel composition and
transitivity, we conclude $(P_1\mid Q_1)\mayleq (P_2\mid Q_2)$.
The same argument applies to the must pre-order. \qed

\subsection*{A remark on specification}
Suppose $S$ represents a specification process and $I$ a corresponding
implementation. When using $\may$ testing we typically aim to prove:
\[
I\mayleq S \qquad\mbox{($\may$ requirement)}
\]
namely all tests that may be passed by the implementation are passed by
the specification too.
When using $\must$ testing we aim to prove:
\[
S\mustleq I \qquad \mbox{($\must$ requirement)}
\]
namely if the specification must pass a test then the implementation must pass
the test too.
Informally, one could say that the $\may$ requirement is fixing the
{\em rights} of the implementation while the $\must$ requirement
is fixing its {\em duties}.

In general, may and must testing are incomparable. For instance,
consider:
\[
\begin{array}{llll}
P=a.(b+c)~, &Q=a.b+a.c~,  &P\mayleq Q~, &P\not\mustleq Q~, \\

P=a+\Omega~, &Q=b+\Omega~, &P\not\mayleq Q~, &P\mustleq Q~.
\end{array}
\]
However, we shall see that if the specification $S$ satisfies
an hereditary termination predicate then the must requirement
$S\mustleq I$ implies the may requirement $I\mayleq S$ (notice
the inversion of the order).

It is also worth noting that in general, the must testing pre-order is incomparable with simulation too (cf. definition \ref{bisim-def}).
For instance, consider:
\[
\begin{array}{llll}
  P_1=a.(b.c+b.d), &Q_1=a.b.c+a.b.d,  &P_1=_{\sf must} Q_1,  &P_1\not\leq_{{\sf sim}} Q_1,\\
  P_2=a.b,        &Q_2=a.b+a,          &P_2 \not\mustleq Q_2, &P_2 =_{{\sf sim}} Q_2~.
\end{array}
\]

\subsection*{May testing and trace inclusion}
The characterization of may testing as trace inclusion is quite direct.

\begin{definition}[test for $\may$]\label{test-def-may}
Suppose $s$ is a finite sequence of observable actions.
We define the family of tests $R_0(s)$ as:
$R_0(s)= \ol{s}.w$.
\end{definition}

\begin{proposition}\label{test-may-prop}
Suppose $P,Q$ are processes and 
$s$ is a finite sequence of actions. Then:

\begin{enumerate}
\item $P\wact{s} \cdot$  iff $(P\mid R_0(s)) \may$.
\item $P\mayleq Q$ iff $\trace{P}\subseteq \trace{Q}$.
\end{enumerate}
\end{proposition}
\Proof
\begin{enumerate}

\item  If $P\wact{s}$ then we have a successful computation:
$(P\mid R_0(s)) \wact{\tau} \cdot \act{w}$.
On the other hand, if we have a successful computation of $(P\mid R_0(s))$
then necessarily $R_0(s) \act{\ol{s}} w$ and $P\wact{s}$.

\item Suppose $P \mayleq Q$ and $P\wact{s}$. Then, by the previous assertion:
\[
\begin{array}{lllll}
(P\mid R_0(s)) \may  &\mbox{ implies} 
&(Q\mid R_0(s)) \may  &\mbox{ implies} 
&Q\wact{s} \cdot~.
\end{array} 
\]
On the other hand, suppose $\trace{P}\subseteq \trace{Q}$,
$R$ is a test, and $(P\mid R) \may$.
This means that for some $s$, $P\wact{s} P_1$, $R\wact{\ol{s}} R_1$ and
$R_1\act{w}$.
Since $s\in \trace{P}$ we also have $s\in \trace{Q}$.
Hence $(Q\mid R)\wact{\tau} (Q_1\mid R_1) \act{w}$,
that is $(Q\mid R) \may$. \qed

\end{enumerate}

\subsection*{Testing hereditary termination}
A process $P$ terminates iff $(P\mid \tau.w)\must$.
Thus in a sense must testing is a generalization of termination.

\begin{definition}[hereditary termination test]
  \label{her-ter-test}\index{termination, hereditary}
  If $s$ is a finite sequence of actions then we define
  an hereditary termination test $R_1(s)$ as:
  \[
  R_1(s)=(\ol{s}.0\mid \tau.w)~.
  \]
\end{definition}

\begin{proposition}\label{termination-test-prop}
Suppose $P,Q$ are processes and $s$ is a finite sequence of actions.
  Then:
  \begin{enumerate}

  \item $P\downarrow s$ iff $(P\mid (\ol{s}.0\mid \tau.w))\must$.

  \item If $P\mustleq Q$ and $P\downarrow s$ then $Q\downarrow s$.

  \item If $P \mustleq Q$ and $P$ is reactive then $Q$ is reactive.
  \end{enumerate}
\end{proposition}
\Proof
We prove the first assertion 
by induction on $|s|$. If $s=\epsilon$, we verify:
\[
P\downarrow \epsilon \mbox{ iff }(P\mid \tau.w) \must~.
\]
If $s=a\cdot s_1$ then by definition: 
\[
P\downarrow s \mbox{ iff } P\downarrow \mbox{ and }\forall P_1\ (P\wact{a} P_1 \mbox{ implies }P_1\downarrow s_1)~.
\]
By induction hypothesis on $s_1$, this can be rewritten as:
\[
P\downarrow \mbox{ and }\forall P_1\ (P\wact{a} P_1 \mbox{ implies }
(P_1 \mid R_1(s_1))\must)~,
\]
and this is equivalent to:
\[
(P\mid R_1(s))\must~.
\]
Assertions 2 and 3 follow immediately by definition of
must-testing. \qed

\subsection*{Must set predicates}
Hereditary termination is essential but not quite enough to characterize the
must pre-order. For instance, it identifies all reactive processes. 
With this motivation, we introduce a second  family of tests
that we call must-set {\em tests} since they relate to a must-set
{\em predicate} we introduce next.

\begin{definition}[must-set tests]\label{test-def-must}
  Suppose $s$ is a finite sequence of actions
  and $B$ a
  {\em finite} set of actions. We define a family
  of tests $R_2(s,B)$ as follows:%
\[
\begin{array}{llll}

R_2(\epsilon,B) &=\Sigma_{a\in B} \ \ol{a}.w~, 
&R_2(b\cdot s,B) &=\tau.w+ \ol{b}.R_2(s,B)~. 

\end{array}
\]
\end{definition}

\begin{remark}
  If $B\inter \ol{B}=\emptyset$ then we can also define:
\[
\begin{array}{ll}
  R_2(s,B) &= \new{d}{\ol{d}.0 \mid d.w \mid \ol{s}.d.\Pi_{a\in B} \ \ol{a}.w}~.
  \end{array}
  \]
\end{remark}
 
To characterize the processes that pass the must-set {\em tests},
we introduce a family of must-set {\em predicates}.
The predicate $M(P,s,B)$ holds
if whenever $P$ performs the sequence of actions $s$ it {\em must}
perform some action in $B$ ($B$ is the {\em must-set} that justifies
the name of the test and the predicate).

\begin{definition}[must-set predicates]\index{predicate, must-set}
  Let $P$ be a process,
  $s\in \w{Act}^*$, and $B\subseteq_{{\it fin}} \w{Act}$. We write:
  \[
  M(P,s,B) \mbox{ if }\forall P' \ (P\wact{s} P' \mbox{ implies }W(P')\inter B\neq \emptyset)~.
  \]
\end{definition}

\begin{proposition}\label{must-set-test-prop}
   Let $P$ be a process,
  $s\in \w{Act}^*$, and $B\subseteq_{{\it fin}} \w{Act}$. Then: 
$P\downarrow s$ and $M(P,s,B)$ iff $(P\mid R_2(s,b))\must$.
\end{proposition}
\Proof By induction on the length of $s$.
If $s=\epsilon$ we have:
\[
\begin{array}{lll}
  (P\mid R_2(\epsilon,B)) \must &\mbox{iff} &P\downarrow \mbox{ and }\forall P' \ (P\wact{\tau} P' \mbox{ implies }W(P')\inter B\neq \emptyset) \\
  &\mbox{iff} &P\downarrow \epsilon \mbox{ and }M(P,\epsilon,B)~.
\end{array}
\]
If $s=as_1$ we have:
\[
\begin{array}{lll}
  (P\mid R_2(s,B))\must &\mbox{iff} &P \downarrow \mbox{ and }\forall P' \ (P\wact{a} P' \mbox{ implies }(P'\mid R_2(s_1,B))\must) \\
  &\mbox{iff} &P\downarrow \mbox{ and }\forall P' \ (P\wact{a} P' \mbox{ implies }(P'\downarrow s_1 \mbox{ and }M(P',s_1,B)) \\
  &\mbox{iff} &P\downarrow s \mbox{ and }M(P,s,B)~.
\end{array}
\]
\qed \\

By combining the tests $R_1$ and  $R_2$ we see that whenever
$P\mustleq Q$ we have the following properties:
\begin{enumerate}

\item For all $s$, if $P\downarrow s$ then $Q\downarrow s$.

\item For all $s,B$, if $P\downarrow s$ and $M(P,s,B)$ then
  $Q\downarrow s$ and $M(Q,s,B)$.

\end{enumerate}

By logical manipulation of the conjunction of the two implications,
we can eliminate
two occurrences of the hereditary termination predicate and arrive at
the following definition and proposition.

\begin{definition}[must-set pre-order]\label{must-set-order-def}\index{pre-order, must-set}
  Let $P,Q$ be processes. We write $P\leq_M Q$ if
  for all $s\in \w{Act}^*$ such that $P\downarrow s$
  the following two properties hold:
  \begin{enumerate}
  \item $Q \downarrow s$.
  \item for all $B\subseteq_{\w{fin}} \w{Act}$ if $M(P,s,B)$
    then $M(Q,s,B)$.
  \end{enumerate}
\end{definition}

\begin{proposition}\label{must-implies-must-set-prop}
  Let $P,Q$ be processes. If $P\mustleq Q$ then $P\leq_M Q$.
\end{proposition}
\Proof By the definition of must pre-order and propositions \ref{termination-test-prop}
and \ref{must-set-test-prop}. \qed\\

\begin{remark}\label{Bfinite-limitation}
 Consider the reactive processes:
$P=\Sigma_{n\geq 0} \ a.b_{n}$ and $Q=a.0$.
  Then for all finite sets $B$, we have $\neg M(P,a,B)$
   since $P$ after doing $a$ can do one out of infinitely
   many actions and therefore no finite $B$ can cover all of them.
   We conclude that $P=_M Q$ and we have an example
   of how the hypothesis that the must-set $B$ is finite
   limits the discriminating
   power of these tests on lts which are not image finite.
\end{remark}

It is possible to give an alternative definition of the must-set pre-order
in which we require that the process reaches a {\em stable} configuration and
then its {\em strong} commitments intersect the actions in $B$.
In general, this is a weaker condition but equivalent to the one
we have given when combined with hereditary termination.

\begin{definition}[strong-stable must-set predicate]
  Let $P$ be a process,
  $s\in \w{Act}^*$, and $B\subseteq_{{\it fin}} \w{Act}$. We write:
  $M_S(P,s,B)$  if $\forall P' \ (P\wact{s} P'\not\act{\tau}$  implies
  $S(P')\inter B\neq \emptyset)$.
\end{definition}

\begin{proposition}\label{must-set-strong-stable-prop}
  \begin{enumerate}
  \item If $M(P,s,B)$ then $M_S(P,s,B)$.
  \item If $M_S(P,s,X)$ and $P\downarrow s$ then $M(P,s,X)$.
  \end{enumerate}
  
\end{proposition}

\begin{exercise}\label{proof-must-set-strong-stable-prop}
Prove proposition \ref{must-set-strong-stable-prop}.
  \end{exercise}

  \begin{remark}
    Consider the recursive process $P=\s{rec} X.\tau.X+a.X$ and
    the empty trace $s=\epsilon$.  Then  $M_{S}(P,s,B)$ for any $B$,
    since $P$ cannot reduce to a stable process. On the other hand,
    if $a\notin B$ then $M(P,s,B)$ is false.
   \end{remark}

\subsection*{A modal logic view}
We take a modal logic view at the predicates we have been
defining to characterize the may and must testing pre-orders.
Consider the modal logic on actions with box and diamond modalities
introduced in chapter \ref{modal-sec}.
If $a\in \w{Act}$ and $A$ is a modal formula then recall that:
\[
\begin{array}{ll}
P \models \la a \ra A\mbox{ if } \exists P' (P\wact{a} P'\mbox{ and } P'\models A)~,
&P\models  [a]A \mbox{ if }\forall P'(P\wact{a}P' \mbox{ implies }P'\models A)~.
\end{array}
\]
Notice that the modalities are interpreted {\em weakly}.  
Weak modalities can be defined starting from strong modalities and fixed
points.
Suppose $F$ is a collection of formulae. If $P$ is a process
its interpretation relative to $F$ is the set of formulae in $F$
that satisfy it:
\[
\dl P \dr^F = \set{A \in F \mid P\models A}~,
\]
and the induced pre-order on processes is:
$P\leq_F Q$ if $\dl P \dr^F \subseteq \dl Q \dr^F$.
If one regards the formulae as `tests', then we are again saying that
all the tests passed by $P$ are passed by $Q$ too.
We look for collections of modal formulas $F_{{\sf trace}}$ and $F_{{\sf must-set}}$
that correspond to the trace inclusion and must-set pre-orders.
For trace inclusion, we can just take:
\[
F_{{\sf trace}} = \set{ \la s \ra \s{true} \mid s\in \w{Act}^*}~,
\]
with the obvious extension of the diamond modality to sequences of
actions.
The situation for the must-set pre-order is a bit more complex.
First, to express termination we need a least fixed point:
\[
\downarrow = \mu X.([\tau]\s{false}\Or [\tau]X)~.
\]
Then hereditary termination becomes for $s=a_1\cdots a_n \in \w{Act}^*$:
\[
\begin{array}{llll}
\downarrow \epsilon &= \downarrow~,\qquad
\downarrow as &= \downarrow \AND \ [a]\downarrow s~.
\end{array}
\]
We can write this in more compact form by introducing a box modality, 
say $[~]_{\downarrow}$, 
on sequences of actions which is sensitive to termination. So define:
\[
\begin{array}{ll}
  P\models  [\epsilon]_{\downarrow} A &\mbox{if }P\downarrow \mbox{ and }P\models A ~,\\
  P\models [as]_{\downarrow}A         &\mbox{if }P\downarrow \mbox{ and }\forall P'(P\wact{a}P' \mbox{ implies }P' \models [s]_{\downarrow} A)~.
\end{array}
\]
Then hereditary termination becomes:
$\downarrow s = [s]_{\downarrow}\s{true}$.
The second basic condition we have to express is:
\[
P\downarrow s \AND M(P,s,B)~,
\]
which can be written as: $[s]_{\downarrow} \la B \ra \s{true}$,
where $\la B \ra \s{true} =\OR_{a\in B} \la a \ra \s{true}$.
So we define:
\[
F_{{\sf must-set}}=\set{[s]_{\downarrow}A \mid A\in \set{\s{true}, \la B \ra\s{true}},
s\in \w{Act}^*, B\subseteq_{\w{fin}} \w{Act}}~.
\]
When processes are {\em reactive}, $[s]_{\downarrow} A$ can be simply written
as $[s]A$ and since $[s]\s{true}$ is equivalent to $\s{true}$,
the collection of relevant formulae boils down to:
\[
\set{[s]\la B \ra\s{true} \mid s\in \w{Act}^*, B\subseteq_{\w{fin}} \w{Act}}~,
\]
and we see that the must-set predicate can be regarded as a kind of
box-diamond modal formula (box on a sequence of actions and diamond on a finite
set of actions). Notice that taking $B=\emptyset$, we have:
\[
[s]\la \emptyset \ra \s{true}= [s]\s{false} = \neg \la s \ra \s{true}~.
\]
Hence, in the reactive case, 
$P\leq_{F_{{\sf must-set}}} Q$ implies $Q \leq_{F_{{\sf trace}}} P$
  (the following proposition \ref{infinite-traces-prop}, gives another
  presentation of this result).

\subsection*{Acceptance sets}
Acceptance sets provide an alternative to the must-set approach.

\begin{definition}[acceptance set]\index{acceptance set}
Let $P$ be a process and $s\in \w{Act}^*$ a sequence of actions.
The {\em acceptance set} $\acset{P,s}$ is a {\em set of sets} of
observable actions:
\[
\begin{array}{lllll}

\acset{P,s} &=\set{W(P') \mid  P\wact{s} P'}~.

\end{array}
\]
\end{definition}

To determine $\acset{P,s}$, we consider all processes $P'$ such that
$P\wact{s} P'$  and
then we compute the set of weak commitments $W(P')$.
Since we do {\em not} assume finite branching,
in general the sets of commitments $S(P')$ and  $W(P')$ can be infinite.
On the other hand, assuming {\em image finiteness}, 
if $P\downarrow s$ then the set
$\set{P' \mid P\wact{s} P'}$
is finite.

To compare acceptance sets, we introduce the following {\em pre-order}.

\begin{definition}[pre-order on acceptance sets]\index{pre-order, acceptance sets}
  Let $A_1,A_2$ be acceptance sets. Then:
\[
A_1 \leq_A A_2 \mbox{ if }\forall Y\in A_2\ \exists X \in A_1\ (X\subseteq Y)~.
\]
\end{definition}

As for the must-set predicate, there is an alternative definition
of acceptance sets that relies on stability and strong commitment.

\begin{definition}[stable-strong acceptance sets]\label{acsset-def}
Let $P$ be a process and $s\in \w{Act}^*$ a sequence of actions.
The {\em stable strong acceptance set} $\acsset{P,s}$ is a set of sets of
observable actions:
\[
\begin{array}{lllll}

\acsset{P,s} &=\set{S(P') \mid  P\wact{s} P'\not\act{\tau}}~.

\end{array}
\]
\end{definition}

We use the pre-order on acceptance sets to compare the two definitions
which are equivalent when combined with hereditary termination.

\begin{proposition}\label{acc-set-strong-stable-prop}
Let $P$ be a process and $s\in \w{Act}^*$ a trace. Then:
  \begin{enumerate}

  \item $\acset{P,s}\leq_A \acsset{P,s}$.

  \item If $P\downarrow s$ then $\acsset{P,s}\leq_A \acset{P,s}$.
    \end{enumerate}
\end{proposition}

\begin{exercise}\label{proof-acc-set-strong-stable-prop}
  Prove proposition \ref{acc-set-strong-stable-prop}.
\end{exercise}

\begin{remark}
  Consider again the recursive process $P=\s{rec} X.\tau.X+a.X$ and the
  empty trace $s=\epsilon$. Then
  $\acsset{P,s}=\emptyset$, $\acset{P,s}=\set{\set{a}}$, and
  $\set{\set{a}} \leq_A \emptyset$ but
  $\emptyset \not\leq_A \set{\set{a}}$.
\end{remark}

It remains to relate the must-set predicate with the
acceptance sets. First, we notice that the pre-order on
acceptance sets implies the inclusion of the must-set predicates.

\begin{proposition}\label{aset-to-mset-prop}
  Let $P,Q$ be processes and $s\in \w{Act}^*$ a sequence of
  observable actions. Then $\acset{P,s}\leq_A \acset{Q,s}$
  implies that for all finite set of actions $B$,
  $M(P,s,B)$ implies $M(Q,s,B)$.
\end{proposition}

\begin{exercise}\label{proof-aset-to-mset-prop}
  Prove proposition \ref{aset-to-mset-prop}.
  \end{exercise}

The other direction is a bit more delicate as it appeals to the
hereditary termination predicate and image finiteness.

\begin{proposition}\label{mset-to-aset-prop}
  Let $P$ be an image-finite processes, $Q$ a process, and $s\in \w{Act}^*$.
  Further suppose $P\downarrow s$ and 
  for all $B$ finite set of actions, $M(P,s,B)$ implies $M(Q,s,B)$.
  Then $\acset{P,s} \leq_A \acset{Q,s}$.
\end{proposition}

\begin{exercise}\label{proof-mset-to-aset-prop}
  Prove proposition \ref{mset-to-aset-prop}.
  \end{exercise}

The example in remark \ref{Bfinite-limitation} shows
that the hypothesis that $P$ is image finite is needed.
In view of propositions \ref{aset-to-mset-prop} and \ref{mset-to-aset-prop},
we can rephrase definition \ref{must-set-order-def}
using acceptance sets and derive the following corollary.

\begin{definition}[acceptance set pre-order]\label{acc-set-order-def}
  Let $P,Q$ be processes. We write $P\leq_A Q$ if
  for all $s\in \w{Act}^*$ such that $P\downarrow s$
  the following two properties hold:
  \begin{enumerate}
  \item $Q\downarrow s$.
  \item $\acset{P,s} \leq_A \acset{Q,s}$.
  \end{enumerate}
\end{definition}

\begin{corollary}\label{must-set-acc-set-prop}
  Let $P,Q$ be processes such that $P$ is image-finite. Then
  $P\leq_M Q$ iff $P\leq_A Q$.
\end{corollary}

\subsection*{Hereditary termination and infinite traces}
Trace inclusion, combined with reactivity and image finiteness,
suffices to guarantee {\em infinite} trace inclusion.

\begin{proposition}\label{inf-trace-inclusion-prop}
  Let $P,Q$ be processes such that $\trace{P}\subseteq \trace{Q}$
  and $Q$ is image finite.
  \begin{enumerate}
\item     
  If $P\wact{u}$ and $Q\downarrow u$  then $Q\wact{u}$.
\item If $Q$ is reactive then $\inftrace{P}\subseteq \inftrace{Q}$.
  \end{enumerate}
\end{proposition}
\Proof
\begin{enumerate}
  \item From $P\wact{u}$, $u$ infinite trace, we derive:
$\forall k\ (u(k)\in \trace{P}\subseteq \trace{Q})$.
Build the following rooted tree.

\begin{itemize}

\item Put the process $Q$ at the root. This is the only node at level $0$.

\item For each node $n$ at level $i$ associated with say process $Q'$,
compute ${\cl P}=\set{Q'' \mid Q' \wact{u[i+1]} Q''}$.
For each process in $Q''$ in ${\cl P}$, introduce a new node $n'$ at level $i+1$
associated with $Q''$ and an edge from $n$ to $n'$.

\end{itemize}

Paths in the tree from the root to a node at level $i$, $i\geq 0$, 
correspond to a reduction of process $Q$ labelled with $u[1]\cdots u[i]$.
The tree has infinitely many nodes but it is finitely branching because
$Q$ is image finite and $Q\downarrow u$.
By proposition \ref{koenig-prop} (König's lemma), there must be an infinite path which
corresponds to an infinite reduction $Q\wact{u}$.

\item If $Q$ is reactive then $Q\downarrow u$ and the previous
  assertion applies. \qed \\
\end{enumerate}

Next we consider the relationship between 
the must-set pre-order and (infinite) trace inclusion.

\begin{proposition}\label{infinite-traces-prop}
  \begin{enumerate}

  \item If $P\leq_M Q$, $P\downarrow s$, and $Q \wact{s}\cdot$ then
    $P\wact{s}\cdot$.
    
  \item If $P\leq_M Q$, $P\downarrow u$, $P$ is image finite,
    and $Q\wact{u}$ then    $P\wact{u}$.

 \item If $P$ is reactive and image finite and $P\leq_M Q$ then
   $\trace{Q}\subseteq \trace{P}$ and $\inftrace{Q}\subseteq \inftrace{P}$.

\end{enumerate}
\end{proposition}

\begin{exercise}\label{proof-infinite-traces-prop}
  Prove proposition \ref{infinite-traces-prop}.
  \end{exercise}

\begin{remark}
  Proposition \ref{infinite-traces-prop} may fail without the
  image-finiteness assumption. Consider:
  \[
  \begin{array}{lll}
    P=\Sigma_{n\geq 0} \ a^n.0~, &Q=a^\omega~, &R=\ol{a}^\omega \mid \tau.w~.
    \end{array}
  \]
  Then $P,Q$ are reactive,  $P\leq_M Q$, and $T(P)=T(Q)$.
  However, $T^\omega(Q)\not\subseteq T^\omega(P)$ and
  $(P\mid R)\must$ while $(Q\mid R)\mustnot$.
\end{remark}

We have noticed (proposition \ref{infinite-traces-prop}(1))
that if $P$ is a reactive process
then $P\leq_M Q$ implies $\trace{Q}\subseteq \trace{P}$.
What about the other trace inclusion? In general, this fails;
for instance, consider  $P=a.b+a.c$ and $Q=a.b$.
However, if $P$ behaves in a rather {\em determinate way}
(in chapter \ref{det-sec}, definition \ref{determinate-def} will propose a stronger definition of determinate process)
then we have the following proposition.

\begin{proposition}\label{det-must-set-prop}
  Suppose $P,Q$ are processes, $P$ is reactive, and for all traces $s$,
$P\wact{s} P_1,P_2$ implies $W(P_1)=W(P_2)$. Then $P\leq_M Q$ implies
  $\trace{P}=\trace{Q}$.
\end{proposition}

\begin{exercise}\label{proof-det-must-set-prop}
Prove proposition \ref{det-must-set-prop}.
  \end{exercise}

\begin{remark}
  Notice that we may well have reactive processes
  $P,Q$ such that $\trace{P}=\trace{Q}$, $P$ is determinate, and
  $P\not\leq_M Q$. For instance, consider again $P=a.b$ and $Q=a.b+a$.
  On the other hand, it is easy to check that
  if both $P$ {\em and} $Q$ are reactive and determinate
  then $\trace{P}=\trace{Q}$ implies $P=_M Q$. 
\end{remark}

\subsection*{Hereditary termination and must set tests are enough}
In image finite lts, the following {\em key property} holds:
the simple {\em finite} tests for hereditary termination (definition
\ref{her-ter-test})  and must-sets (definition \ref{test-def-must})
suffice to must-test processes.
Before proving this result, we notice that
the property of must-passing a test is {\em not} stable under reduction.
For instance, consider $R=0\mid w.0+\tau.0$. Then
$R\ \must$ but $R\act{\tau} R'$ and $R'=(0\mid 0) \mustnot$.
The following predicate is helpful in describing this situation
and is used in the proof of proposition
\ref{soundness-acceptance-prop}. 

\begin{definition}[unsuccessful predicate]\label{def-u-predicate}\index{predicate, unsuccessful}
  Given a test $R$ and a finite trace $s$,
  we write $U(R,s,R')$ if there is some sequence
of labelled transitions such that $R\wact{s} R'$
and all the tests traversed (including $R$ and $R'$)
cannot perform immediately the $w$ action.
If $u$ is an infinite  trace we write $U(R,u)$ if there is
an infinite reduction starting from $R$ and labelled with $u$
which never crosses a test (including $R$)
that can perform immediately the $w$ action.
\end{definition}

In other terms, $U(R,s,R')$ can be read as:
there is an {\em unsuccessful} transition labelled
$s$ from $R$ to $R'$ (and similarly for $U(R,u)$).

\begin{proposition}\label{test-transition}
Suppose $P$ is a process and $R$ is a test. If $(P\mid R)\must$
and $U(R,\ol{s},R')$ then (i) $P\downarrow s$ and (ii)
if $P\wact{s} P'$ then $(P'\mid R')\must$.
\end{proposition}
\Proof
By induction on the length of the sequence $s$.
If $s=\epsilon$ then $U(R,\epsilon,R')$ implies $R\not\act{w}$.
Therefore $(P\mid R)\must$ implies $P\downarrow \epsilon$ and
if $P\wact{\tau}P'$ then $(P'\mid R')\must$.

If $s=as_1$ then $U(R,\ol{as_1},R')$ implies there is some $R_1$ such
that $U(R,\ol{a},R_1)$ and $U(R_1,\ol{s_{1}},R')$.
As in the basic case, $R\not\act{w}$ implies $P\downarrow \epsilon$.
Suppose $P\wact{a} P_1$. Then $(P\mid R)\wact{\tau} (P_1\mid R_1)$
via a sequence of reductions where the test is {\em not} successful.
Therefore $(P\mid R)\must$ implies $(P_1\mid R_1) \must$
and by inductive hypothesis we conclude $P_1\downarrow s_1$
and if $P\wact{a}P_1 \wact{s_{1}} P'$ then $(P'\mid R')\must$. \qed \\

\begin{proposition}[soundness]\label{soundness-acceptance-prop}
  Let $P,Q$ be processes such that $P$ is image finite.
  If $P\leq_M Q$ then $P\mustleq Q$.
\end{proposition}
\Proof
Let $R$ be a test such that  $(Q\mid R)\mustnot$, 
that is there is an unsuccessful computation of $(Q\mid R)$.
We show how to derive an unsuccessful computation of $(P\mid R)$.
We distinguish two cases.

\begin{itemize}

\item The computation is finite and the process $Q$ performs a
finite sequence of observable actions $s$ such that:
\[
Q\wact{s} Q_1, R \wact{\ol{s}} R_1, (Q\mid R) \wact{\tau} (Q_1\mid R_1)\not\act{\tau}~.
\]
From $(P\mid R)\must$ and $U(R,\ol{s},R_1)$ we derive using
proposition \ref{test-transition} that $P\downarrow s$.
By definition of the must-set pre-order, it follows that
$Q\downarrow s$.
By proposition \ref{mset-to-aset-prop}, it follows that there exists $P_1$ such
that $P\wact{s} P_1$ and $W(P_1)\subseteq W(Q_1)$.
Hence $(P\mid R) \wact{\tau} (P_1\mid R_1)$ and $(P_1\mid R_1)\mustnot$
since $R_1$ is stuck and $P_1$ cannot synchronize with it.

\item The computation is infinite. We distinguish three cases.

\begin{itemize}

\item In the unsuccessful computation of $(Q\mid R)$, the test $R$
  performs a finite number of steps and then stabilizes.
  So for some $s$, we have $U(R,\ol{s},R_1)$ and $Q\wact{s} Q_1$.
  Now we show that $(P\mid R)\must$ leads to a contradiction.
  Indeed, from $(P\mid R)\must$ we derive $P\downarrow s$.
  Since $P\leq_M Q$ we must have $Q\downarrow s$.  But this
  contradicts the hypothesis that the process $Q_1$ has a diverging
  internal computation.

\item In the unsuccessful computation of $(Q\mid R)$, the test $R$ performs a finite
number of interactions with $Q$ and then infinitely many internal steps.
Again for some $s$,  we have $U(R,\ol{s},R_1)$, $Q\wact{s} Q_1$,
and starting from $R_1$ we have an infinite internal reduction.
As before, assuming $(P\mid R)\must$ produces a contradiction
as it implies that $P\downarrow s$. Then, since
$P\leq_M Q$, we have $Q\downarrow s$ and $P\wact{s} P_1$.
But then $(P\mid R)\mustnot$.

\item It remains the interesting case where $Q$ and $R$ engage in infinitely
many interactions. So there is an infinite trace $u$ of observable
actions such that $Q\wact{u}$ and $U(R,\ol{u})$.
If for some $k$, $P\not\downarrow u(k)$ we can conclude directly
that $(P\mid R)\mustnot$.
Otherwise, we have  $P\downarrow u$. Also since $P\leq_M Q$,
by proposition \ref{infinite-traces-prop}(1), we have
$\forall k(P\wact{u(k)}\cdot)$. 
So we are in a situation  where $P\downarrow u$ and
$\forall k (P\wact{u(k)}\cdot)$ and 
proposition \ref{infinite-traces-prop}(2) applies to conclude that
$P\wact{u}$ and $(P\mid R)\mustnot$. \qed

\end{itemize}
\end{itemize}

\section{Testing termination (*)}\label{testing-ter-sec}
We consider a variant of the may and must pre-orders where
we rely on the termination predicate rather than on the
strong commitment predicate.
To stress this change of predicate, we
shall denote the resulting may and must pre-orders
with $\tmayleq$ and $\tmustleq$ ($t$ for termination)
while keeping the notation
$\mayleq$ and $\mustleq$ for those introduced in the
previous section.

\begin{definition}[may and must termination pre-orders]\index{pre-order, may and must termination}
  Let $P_1,P_2$ be processes. We write:
  \[
  \begin{array}{ll}
    P_1 \tmayleq P_2 &\mbox{if }\forall Q \ ((P_1\mid Q)\Downarrow \mbox{ implies }(P_2\mid Q)\Downarrow)~, \\
    P_1 \tmustleq P_2 &\mbox{if }\forall Q \ ((P_1\mid Q)\downarrow \mbox{ implies }(P_2\mid Q)\downarrow)~.
  \end{array}
  \]
\end{definition}

Notice that in this section there is no distinction between processes
and tests and they are both denoted with $P,Q,\ldots$.

\subsection*{May termination testing}
First, we look at the may termination testing pre-order. This
is incomparable with the ordinary may testing pre-order as it is actually
sensitive to both termination and deadlock.

\begin{definition}[tests for may termination]
  Let $s\in \w{Act}^*$ be a finite trace and $B$ a finite set of
  observable actions.
We define the following two families of tests where $a$ is a fresh name:
  \[
  \begin{array}{llll}
    R_1(s) &=\new{a}{\ol{s}.a \mid \ol{a} \mid a.\Omega}~, 
    &R_2(s,B)&=\new{a}{\ol{s}.a.\Sigma_{b\in B}\ \ol{b}.\Omega \mid \ol{a} \mid a.\Omega}~.
  \end{array}
  \]
\end{definition}

Intuitively to avoid divergence,
the test $R_1(s)$ has to run completely $\ol{s}.a$, 
while the test $R_2(s,B)$ has to run completely $\ol{s}.a$
and then it must {\em not} run actions in $\ol{B}$.

\begin{proposition}\label{test-prop-conv}
  Let $P$ be a process, $s$ a trace, and $B$ a finite set of
  observable actions. Then:
  \begin{enumerate}

  \item $(P\mid R_1(s))\Downarrow$ iff $\exists P' (P\wact{s} P'\not\act{\tau})$.

  \item $(P\mid R_2(s,B))\Downarrow$ iff $\exists P' (P\wact{s} P'\not\act{\tau} \mbox{ and }S(P')\inter B=\emptyset)$.
  \end{enumerate}
  \end{proposition}
  \Proof
  \begin{enumerate}

  \item $(\Leftarrow)$ We have:
    \[
    (P\mid R_1(s)) \wact{\tau} (P'\mid \new{a}{a.\Omega}) \not\act{\tau}~.
    \]
    $(\Rightarrow)$ We must have $P\wact{s}P_1$ so
    that:
    \[
    (P\mid R_1(s))\wact{\tau} (P_1\mid \new{a}{a.\Omega})~,
    \]
    and then we must have $P_1\Downarrow$. 
    
  \item $(\Leftarrow)$ We have:
    \[
    (P\mid R_2(s,B)) \wact{\tau} (P'\mid \Sigma_{b\in B} \ \ol{b}.\Omega \mid \new{a}{a.\Omega})\not\act{\tau}~.
    \]
    $(\Rightarrow)$ We must have $P\wact{s} P_1$ so that:
    \[
    (P\mid R_2(s,B))\wact{\tau} (P_1 \mid \Sigma_{b\in B} \ \ol{b}.\Omega \mid \new{a}{a.\Omega})~,
    \]
    and $P_1$ must reduce to a $P'$ such that $P'\not\act{\tau}$ and
    $S(P')\inter B=\emptyset$. \qed
  \end{enumerate}

  The definition \ref{acsset-def} of
  strong stable acceptance sets turns out to be
  relevant to characterize the may termination pre-order.
  We recall that if $P_1,P_2$ are processes and $s$ is a finite trace
  then $\acsset{P_2,s} \leq_A \acsset{P_1,s}$
  means that for all  $P'_1$ such that $P_1\wact{s} P'_1\not\act{\tau}$,
  there exists $P'_2$ such that $P_2\wact{s}P'_2\not\act{\tau}$
  and $S(P'_2)\subseteq S(P'_1)$.

  \begin{proposition}\label{may-convergence-implies-acceptance}
    Let $P_1,P_2$ be processes such that
    $P_1\tmayleq P_2$ and $P_2$ is {\em image finite}.
    Then for all $s\in \w{Act}^*$, 
    $\acsset{P_2,s} \leq_A \acsset{P_1,s}$.
    
  \end{proposition}
  \Proof
  If $P_1\wact{s}P'_1 \not\act{\tau}$ then by proposition
  \ref{test-prop-conv}(1), there is some $P'_2$ such that
  $P_2\wact{s}P'_2 \not\act{\tau}$.
  Reasoning by contradiction,  suppose
  for all such $P'_2$ we have $S(P'_2)\not\subseteq S(P'_1)$.
  Then using image finiteness, let $B$ be a finite set built by choosing
  an action in each $S(P'_2)\minus S(P'_1)$. Then
  $(P_1\mid R_2(s,B))\Downarrow$ while $(P_2\mid R_2(s,B))\not\Downarrow
  $. \qed

  \begin{proposition}\label{acceptance-implies-may-convergence}
    Suppose for all $s$, $\acsset{P_2,s} \leq_A \acsset{P_1,s}$. Then
    $P_1\tmayleq P_2$.
  \end{proposition}
  \Proof
  Suppose $(P_1 \mid Q)\Downarrow$. Then, for some $s$ we have
  $P_1 \wact{s} P'_1$, $Q\wact{\ol{s}} Q'$ and $(P'_1\mid Q')\not\act{\tau}$.
    By hypothesis, there must be a $P'_2$ such that $P_2\wact{s} P'_2\not\act{\tau}$ and $S(P'_2)\subseteq S(P'_1)$. Hence $P'_2 \mid Q'\not\act{\tau}$ and
    $(P_2\mid Q)\Downarrow$. \qed \\

In general, the pre-orders $\tmayleq$ and $\mustleq$ are
incomparable since the former requires weak normalization while the
latter requires the strong one
(concrete example available in table \ref{inc-ineq-table}). However, if
we consider image finite and reactive processes then
may termination is the inverse of {\em ordinary} must testing.

\begin{proposition}\label{may-conv-and-must-test}
  Let $P_1,P_2$ be reactive processes such that $P_2$ is image finite.
Then:
  \[
  P_1 \tmayleq P_2 \mbox{ iff }P_2 \mustleq P_1~.
  \]
\end{proposition}
\Proof
By propositions \ref{may-convergence-implies-acceptance} and \ref{acceptance-implies-may-convergence}, we know (assuming $P_2$ is image finite):
\[
P_1 \tmayleq P_2 \mbox{ iff }\forall s\ (\acsset{P_2,s} \leq_A \acsset{P_1,s})~.
\]
On the other hand, if $P_1,P_2$ are reactive and $P_2$ is image finite
then the ordinary must pre-order coincides with the 
pre-order on the strong, stable acceptance sets:
\[
\begin{array}{lll}
  P_2\mustleq P_1 &\mbox{iff } P_2\leq_M P_1 &\mbox{(propositions \ref{must-implies-must-set-prop} and \ref{soundness-acceptance-prop})} \\
    &\mbox{iff }P_2 \leq_A P_1 &\mbox{(corollary \ref{must-set-acc-set-prop})}\\
      &\mbox{iff }\forall s\ (\acsset{P_2,s} \leq_A \acsset{P_1,s})
      &\mbox{(proposition \ref{acc-set-strong-stable-prop}).}
      \end{array}
\]
Notice that the reactivity of $P_2$ is required; for instance, 
$0\tmayleq \tau+\Omega$ while $0\not\mustleq \tau+\Omega$. \qed
    
  \subsection*{Must termination testing}
  We turn to the must termination pre-order. This is strictly weaker 
  than the ordinary must pre-order.

\begin{proposition}\label{must-implies-must-conv}
    Let $P_1,P_2$ be processes. If 
$P_1\mustleq P_2$ then $P_1 \tmustleq P_2$.
\end{proposition}
\Proof
Suppose $(P_1 \mid Q) \downarrow$. This is equivalent to
$(P_1\mid Q \mid \tau.w)\must$, which by hypothesis implies
$(P_2\mid Q \mid \tau.w)\must$, which is equivalent
to $(P_2\mid Q)\downarrow$. \qed

\begin{remark}
  In general, the
  $4$ pre-orders $\mayleq$, $\tmayleq$, $\mustleq$,
  and $\tmustleq$ are all {\em pairwise incomparable}
  except in the case considered
  in the proposition \ref{must-implies-must-conv} above.
  Counter-examples to the remaining $11$ inclusions are given in table
  \ref{inc-ineq-table}.
  For instance, read the first line as stating: (i) in the $\not\subset$ column,
  that $a.b\mayleq a.b+a$ while $a.b\not\mustleq a.b+a$ and (ii) in the
  $\not\supset$ column, that $a+\Omega \mustleq b+\Omega$ while
  $a+\Omega\not\mayleq b+\Omega$.
  \end{remark}

\begin{table}
{\footnotesize  \[
  \begin{array}{cc|cc}
                   &                 &\not\subset  &\not\supset \\\hline
    \mayleq        &\mustleq         &a.b,a.b+a   &a+\Omega,b+\Omega     \\
    \mayleq        &\tmayleq   &a.b+a,a.b   &a\mid \Omega,b\mid \Omega\\
    \mayleq        &\tmustleq   &0,\tau+\Omega&a+\Omega,b+\Omega \\
    \mustleq       &\tmayleq   &\tau+\Omega,\Omega &0,\tau+\Omega \\
    \mustleq       &\tmustleq   &\mbox{(holds!)}   &a.b,a.b+a  \\
    \tmayleq &\tmustleq   &0,\tau+\Omega &\tau+\Omega,\Omega
  \end{array}
  \]}
\caption{Incomparable testing pre-orders}\label{inc-ineq-table}
  \end{table}

Towards a characterization of must termination, we notice the
following properties.

  \begin{proposition}\label{prop-conv-must}
    Let $P$ be a process and $s$ a trace of observable actions.
    Then $P\downarrow s$ iff $(P\mid \ol{s})\downarrow$.
  \end{proposition}
  \Proof By induction on $|s|$, proof similar to the one of proposition \ref{termination-test-prop}(1). \qed

  \begin{proposition}\label{prop-conv-must0}
    Let $P_1,P_2$ be processes such that  $P_1 \tmustleq P_2$. Then:
  \begin{enumerate}
    \item $C(P_1)\subseteq C(P_2)$,
    \item $C(P_1)\inter \trace{P_2} \subseteq \trace{P_1}$.
  \end{enumerate}
  \end{proposition}
  \Proof
  \begin{enumerate}

  \item By proposition \ref{prop-conv-must}, 
    $P_1\downarrow s$ is equivalent to $(P_1 \mid \ol{s})\downarrow$.
    By hypothesis, the latter implies $(P_2 \mid \ol{s})\downarrow$ which is
    equivalent to $P_2\downarrow s$.

  \item Suppose $P_1 \downarrow s$ and $P_2\wact{s}$.
    Let $R(s)=\ol{s}.\Omega$.
    Then $(P_2\mid R(s))\not\downarrow$ implies
    $(P_1 \mid R(s))\not\downarrow$. But since $P_1\downarrow s$
    we must have $P_1\wact{s}$. \qed
\end{enumerate}

    \begin{proposition}\label{conv-inf-case}
Let $P_1,P_2$ be processes such that  $C(P_1)\subseteq C(P_2)$
and $C(P_1)\inter \trace{P_2} \subseteq \trace{P_1}$. Further suppose
$P_1$ is image finite. Then:
$C^\omega(P_1)\inter \inftrace{P_2} \subseteq \inftrace{P_1}$.
    \end{proposition}
    \Proof Suppose $P_2\wact{u}$ and $P_1\downarrow u$.
    So for all $k$:
$u(k) \in C(P_1)\inter \trace{P_2} \subseteq \trace{P_1}$.
    Since $P_1$ is image finite, we conclude $u\in \inftrace{P_1}$. \qed

\begin{proposition}\label{conv-inf-case2}
  Let $P_1,P_2$ be processes such that $P_1$ is image finite.
  If   $C(P_1)\subseteq C(P_2)$ and $C(P_1)\inter \trace{P_2}\subseteq \trace{P_1}$
  then $P_1 \tmustleq P_2$.
\end{proposition}
\Proof
From  $(P_2\mid Q)\not\downarrow$ we derive that
$(P_1\mid Q)\not\downarrow$. We distinguish $2$ cases.

\begin{itemize}

\item Divergence is due to an infinite interaction between $P_2$ and
  $Q$. So suppose $P_2\wact{u}$ and $Q\wact{\ol{u}}$.
  If for some $k$, $P_1\not\downarrow u(k)$ then
  $(P_1\mid Q)\not\downarrow$.
  Otherwise, $P_1\downarrow u$ and, by proposition \ref{conv-inf-case}, 
  we conclude $P_1\wact{u}$ and therefore $(P_1\mid Q)\not\downarrow$.

\item After a finite interaction, either $P_2$ or $Q$ do not terminate.
  So suppose $P_2 \wact{s} P'_2$ and $Q\wact{\ol{s}} Q'$.
  \begin{itemize}

 \item 
  If $P'_2\not\downarrow$ then $P_2\not\downarrow s$ and therefore
  $P_1\not\downarrow s$ which implies $(P_1\mid Q)\not\downarrow$.

\item If $P'_2\downarrow$ and $Q'\not\downarrow$ then
either $P_1\not\downarrow s$  or $P_1\downarrow s$, and in both
cases $(P_1\mid Q)\not\downarrow$. \qed
\end{itemize}
 \end{itemize}

\begin{proposition}\label{must-convergence-and-may-test}
  Let $P_1,P_2$ be reactive processes such that $P_1$ is image finite. Then:
  \[
  P_1\tmustleq P_2 \mbox{ iff }P_2\mayleq P_1~.
  \]
\end{proposition}
\Proof
By propositions \ref{prop-conv-must0}, \ref{conv-inf-case}, and \ref{conv-inf-case2} we have:
\[
P_1\tmustleq P_2 \mbox{ iff }\trace{P_2}\subseteq \trace{P_1}~,
\]
and we have shown in proposition \ref{test-may-prop}(2)
that trace inclusion characterizes the may testing pre-order. \qed \\

To summarize the contents of propositions \ref{may-conv-and-must-test} and
\ref{must-convergence-and-may-test},
assuming $P_1,P_2$ are reactive processes and $P_2$ is image finite, 
we have the following situation (again, notice the inversion of the order when switching from may to must):
  \[
  P_1\tmayleq P_2 \mbox{ iff } P_2 \mustleq P_1
  \mbox{ implies strictly }P_2 \tmustleq P_1 \mbox{ iff }P_1 \mayleq P_2~.
  \]
  Going back to our discussion on the may and must requirements,
  it should be noticed that the may {\em termination} requirements
  are equivalent to the must requirements on reactive processes.

\section{Summary and references}
Section \ref{testing-cmt-sec} is a rephrasing of results in 
\cite{DBLP:journals/tcs/NicolaH84,DBLP:books/daglib/0066919} in
the context of the CCS action structure.
A closely related notion of {\em failure equivalence} has been
developed at about the same time
in the framework of the CSP process calculus
\cite{DBLP:conf/concur/BrookesR84}.

If one focuses on reactive processes, then the must-testing
{\em equivalence} can be regarded as a refinement of trace equivalence that is
still sensitive to deadlock. As observed at the end of section \ref{testing-cmt-sec},
the must-testing equivalence is incomparable
with the (weak) simulation equivalence but it is definitely coarser than
the (weak) bisimulation equivalence.

On non-reactive processes, the comparison between must-testing
equivalence and (weak) bisimulation is more delicate as must-testing
identifies all processes that may diverge. It is possible to revise
the definition of bisimulation to take this requirement into account
and then prove that the modified bisimulation is strictly
included in the must-testing
equivalence.

However, in certain application contexts the treatment of divergence
imposed by must-testing appears to be too severe and attempts
have been made to relax the requirements so as to allow for some
form of {\em benign} divergence. For instance, see the notions of
fair testing developed in \cite{DBLP:conf/concur/BrinksmaRV95,DBLP:conf/icalp/NatarajanC95}.
The may termination testing pre-order discussed
in section \ref{testing-ter-sec} gives yet another angle on this issue.

The testing semantics approach has been applied to a number of concurrent
programming models; we refer to \cite{DBLP:conf/esop/BernardiCLS25} for an
extensive bibliography.

\chapter{Determinacy and confluence}\label{det-sec}
\markboth{{\it Determinacy}}{{\it Determinacy}}

In automata theory, one can envisage various definitions of
determinacy. For instance, in the framework of {\em finite} automata,
consider the following ones.

\begin{enumerate}

\item There is no {\em word} $w$ that admits two computation paths in the
graph such that one leads to an accepting state and the other to a non-accepting state.

\item Each {\em reachable configuration} admits at most one successor.

\item For each {\em state}, either there is exactly one outgoing transition labelled with
 $\epsilon$,  or all outgoing transitions are labelled
with distinct symbols of the input alphabet.

\end{enumerate}

Thus one can go from {\em `extensional' conditions} (intuitive but hard
to verify) to {\em `syntactic' conditions} (verifiable but not as general).
In the following, we propose a definition of determinate lts and
show that all the equivalences included between trace equivalence
and bisimulation collapse on such lts.\footnote{A note on terminology: we
speak of {\em determinacy} and {\em determinate} lts rather than
{\em determinism} and {\em deterministic} lts to emphasize the fact that
we are interested in lts which on a given label
can choose among several transitions 
but whose observable behavior is insensitive to this choice.}

We also introduce a notion
of {\em confluence} on labelled transition systems. This is a stronger
property than determinacy which allows for a restricted form
of parallel composition and for the representation of determinate
models of parallel computation such as {\em Kahn networks}. 
Finally, we consider {\em reactive} systems, {\em i.e.}, systems which
enjoy a kind of generalized termination property.
It turns out that for such system, it is enough to check a {\em local} 
form of confluence.

\section{Determinacy in lts}
In the first place, it is useful to recall why non-determinacy is needed.
First, it arises naturally in {\em race conditions} where 
two `clients' request the same service such as:
\[
\new{a}{(\ol{a}.P_1 \mid \ol{a}.P_2 \mid a)}~.
\]
Second, it is a tool for general specification and portability. 
It is often the case that we do not want to commit on 
a particular behavior. For instance, consider:
\[
\new{a,b}{(\ol{a}.\ol{b}.\ol{c} \ \mid \  a.\ol{b}.\ol{d} \ \mid
  \  b)}~.
\]
Depending on the compilation, the design of the virtual machine, 
the processors timing,$\ldots$ we might always run $\ol{d}$ 
rather than $\ol{c}$ (or the other way around).

On the other hand, determinate systems are easier to 
test, debug, and possibly prove correct. 
Notice that often the implementation seems `determinate' because
the scheduler
determinizes the program's behavior.  However this kind of
determinacy is {\em not portable}: running the program in another
environment may produce different results.

We now move towards a definition of determinacy. Here are some reasonable 
requirements:

\begin{itemize}

\item If $P$ and $P'$ are {\em `equivalent'} then one is determinate
if and only if the other is.

\item If we run an {\em `experiment'} twice we always get the same `result'.

\item If $P$ is determinate and we run an experiment then {\em the
residual of $P$} after the experiment should still be determinate.

\end{itemize}

If we  place ourselves in the context of a  simple model such as $\ccs$, we can 
interpret {\em equivalent} as {\em weak bisimilar} and  {\em experiment} 
as a finite  sequence of {\em labelled transitions}.

As in the previous chapters, let us denote with $\cl{L}$ the set of {\em visible actions and co-actions} with
generic elements $a,b,\ldots$ and let us 
denote with $\w{Act}= \cl{L}\union \set{\tau}$ 
the set of {\em actions},  with  generic elements $\alpha,\beta,\ldots$
Let $s\in \cl{L}^*$ denote a finite word over $\cl{L}$.
Then:
\[
\begin{array}{ll}

P \wact{\epsilon} P'  &\mbox{if }P\wact{\tau} P' \\
P \wact{a_{1}\ldots a_{n}} P', \ n\geq 1
&\mbox{if }P \wact{a_{1}} \cdots \wact{a_{n}} P'~.

\end{array}
\]
If $P\wact{s} Q$ we say that $Q$ is a {\em derivative} of $P$.
As usual we write $|s|$ for the length of the word $s$.

\begin{definition}[determinate]\label{determinate-def}\index{determinate process}
A process $P$ is {\em determinate} if for any $s\in \cl{L}^*$, 
if $P\wact{s} P_i$ for $i=1,2$ then $P_1 \wbis P_2$.
\end{definition}

\begin{remark}
This definition relies on the notion of {\em labelled
transition system}. 
Indeed, in the transition $P\act{a} P'$, $a$ represents a {\em minimal}
interaction with the environment and $P'$ is
the {\em residual} after the interaction. 
\end{remark}

\begin{exercise}\label{det-ornot-ex}
Are the following $\ccs$ processes determinate?
(1) $a.(b+c)$.
(2) $a.b+ac$.
(3) $a+a.\tau$.
(4) $a+\tau.a$.
(5) $a+\tau$.
\end{exercise}

\begin{proposition}\label{det-invariance-prop}
The following properties hold:

\begin{enumerate}

\item If $P$ is determinate and $P\act{\alpha} P'$ then $P'$ is determinate.

\item  If $P$ is determinate and $P\wbis P'$ then $P'$ is determinate.

\end{enumerate}

\end{proposition}
\Proof
Like most of the following proofs, the argument is by {\em diagram chasing}.

\begin{enumerate}
\item
Suppose $P\act{\alpha} P'$ and $P'\wact{s} P_i$ for $i=1,2$.

\begin{itemize} 

\item If $\alpha=\tau$ then $P\wact{s} P_i$ for $i=1,2$. 
Hence $P_1\wbis P_2$.

\item If $\alpha=a$ then $P\wact{a\cdot s} P_i$ for $i=1,2$.
Hence $P_1\wbis P_2$.

\end{itemize}

\item
Suppose $P\wbis P'$ and $P'\wact{s} P'_i$ for $i=1,2$.

\begin{itemize}

\item By definition of {\em weak bisimulation}:
$P\wact{s} P_i$  and $P_i \wbis P'_i$,
for $i=1,2$.

\item Since $P$ is {\em determinate}, we have $P_1\wbis P_2$.

\item Therefore, we conclude by {\em transitivity} of $\wbis$:
$P'_1 \wbis P_1 \wbis P_2 \wbis P'_2$. ~\qed
\end{itemize}
\end{enumerate}

\begin{definition}[$\tau$-inertness]\index{$\tau$-inertness}
We say that a process $P$ is {\em $\tau$-inert} if for all its derivatives
$Q$, if $Q\wact{\tau} Q'$  then $Q\wbis Q'$.
\end{definition}

\begin{proposition}
If a process is determinate then it is $\tau$-inert.
\end{proposition}
\Proof
Suppose $P\wact{s} Q$ and $Q\wact{\tau} Q'$.
Then $P\wact{s} Q$ and $P\wact{s} Q'$.
Thus by determinacy, $Q\wbis Q'$. \qed \\
 
Next we recall a {\em weak version} of the
notion of trace equivalence for lts 
presented in definition \ref{trace-lts-def} (this is
the same notion of trace considered in chapter \ref{testing-sec}).
We define the {\em traces} of a process $P$ as:
\[
\trace{P} = \set{s \in \cl{L}^* \mid P \wact{s} \cdot}~,
\]
and say that two processes $P,Q$ are {\em trace equivalent} if 
$\trace{P}=\trace{Q}$. 
Notice that the traces of a process form  a {\em non-empty}, {\em prefix-closed}
set of {\em finite words} over $\cl{L}$.

\begin{exercise}\label{tr-bis-ex}
Are the following equations valid for 
{\em trace equivalence} and/or {\em weak bisimulation}?
\[
\begin{array}{llll}

a+\tau = a,
&\alpha.(P+Q) = \alpha.P+\alpha.Q,
&(P+Q)\mid R = P\mid R + Q\mid R,
&P = \tau.P~.
\end{array}
\]
\end{exercise}

\begin{exercise}[compositionality of trace semantics]\label{comp-trace-ex}
Show that if $P,Q,R$ are $\ccs$ processes and 
$\trace{P}= \trace{Q}$ then $\trace{P\mid R} = \trace{Q\mid R}$.
\end{exercise}

The following result entails that on determinate processes
most equivalences (trace, simulation-induced equivalence, 
bisimulation,$\ldots$) collapse.

\begin{proposition}\label{det-bis-tr-prop}
Let $P,Q$ be processes.

\begin{enumerate}

\item If $P\wbis Q$ then $\trace{P}=\trace{Q}$.

\item Moreover, if $P,Q$ are {\em determinate} then 
$\trace{P}=\trace{Q}$ implies $P\wbis Q$.

\end{enumerate}
\end{proposition}
\Proof 
\Proofitemf{(1)} Suppose $P\wbis Q$ and $P\wact{s} \cdot$.
Then $Q\wact{s} \cdot$ by induction on $|s|$ using the properties
of weak bisimulation.

\Proofitemm{(2)} Suppose $P,Q$ determinate and $\trace{P}=\trace{Q}$.
We show that:
\[
 \set{(P,Q) \mid \trace{P}=\trace{Q}}
\]
is a bisimulation.

\begin{itemize}

\item If $P\act{\tau} P'$ then $P\wbis P'$ by {\em determinacy}.
Thus taking $Q\wact{\tau} Q$ we have:
\[
P'\wbis P \qquad \trace{P} = \trace{Q}~.
\]
By (1), we conclude:
$\trace{P'} = \trace{P} =\trace{Q}$.

\item If $P\act{a} P'$ then we note that:
\[
\trace{P} = \set{\epsilon} \union \set{a}\cdot \trace{P'} 
\union \Union_{a\neq a',P\wact{a'}P''} \set{a'}\cdot \trace{P''}~.
\]
This is because all the processes $P'$ such that $P\wact{a} P'$
are bisimilar, hence trace equivalent.
A similar reasoning applies to $\trace{Q}$. 
Thus there must be a $Q'$ such that 
$Q\wact{a} Q'$ and $\trace{P'}=\trace{Q'}$.\qed

\end{itemize}

\begin{example}[the unbounded buffer reconsidered]\label{buf2-ex}
Recall the unbounded buffer in example \ref{buf1-ex}:
\[
\begin{array}{ll}
\w{Buf}(a,b)  &= a.\new{c}{(\w{Buf}(a,c) \mid \ol{b}.\w{Buf}(c,b))} \\
P(n)          &= \new{a}{(\ol{a}^{n} \mid \w{Buf}(a,b))}
\end{array}
\]
We shall also use the more suggestive notation f$a\mapsto b$ for
$\w{Buf}(a,b)$.  One can show that these processes are {\em
  determinate}.  In fact one can show that they actually enjoy a
stronger property known as {\em confluence} which is introduced next
in section \ref{confluence-sec}.
\end{example}

\section{Confluence in lts}\label{confluence-sec}
We introduce a notion of {\em confluence}
that strengthens  determinacy and is preserved by some form of
communication (parallel composition + restriction). 
For instance,
\[
\new{a}{((a+b) \mid \ol{a})}
\]
will be {\em rejected} because $a+b$ is {\em not} confluent (while
being {\em determinate}).

The notion of confluence we consider is reminiscent of {\em confluence} in 
rewriting systems (cf. definition \ref{confluence-def}).
By analogy, one calls confluence the related theory
in process calculi but bear in mind that:
(1) confluence is relative to a {\em labelled transition system}
and (2) we close diagrams {\em up to equivalence}.

Before introducing formally the notion of confluence for lts we need to define a notion
of action difference.

\begin{definition}[action difference]\index{action difference}
Suppose $\alpha,\beta\in\w{Act}$. Their {\em action difference} $\alpha\minus\beta$ is defined as:
\[
\alpha\minus \beta = \left\{
\begin{array}{ll}
\alpha &\mbox{if }\alpha \neq \beta \\
\tau   &\mbox{otherwise.}
\end{array}\right.
\]
\end{definition}

We can generalize the notion of action difference to sequences of
visible actions $r,s\in \cl{L}^*$. To compute the {\em difference} $r\minus s$ of
$r$ by $s$ we scan $r$ from left to right deleting
each label which occurs in $s$ taking into account the multiplicities
(cf. difference of {\em multi-sets}). We abuse notation by writing
$a\notin s$ to mean that $a$ does not occur in the word $s$.
\[
\begin{array}{ll}

(\epsilon \minus s) &=  \epsilon \\

(a r \minus s)   &= \left\{ \begin{array}{ll}
                              a \cdot (r\minus s)     &\mbox{if }a \notin s  \\
                              r \minus (s_1\cdot s_2)    &\mbox{if }s=s_1a s_2,a \notin s_1 ~.
                              \end{array} \right.
\end{array}
\]
For instance:
$aba \minus ca = ba$ and $ca \minus aba = c$.

\begin{exercise}\label{action-diff-ex}
Let $r,s,t\in \cl{L}^*$. Show that:
\begin{enumerate}

\item $(rs) \minus (r t) = s\minus t$.

\item $r\minus (st) = (r \minus s) \minus t$.

\item $(rs) \minus t = (r\minus t)(s\minus(t\minus r))$.
\end{enumerate}

\end{exercise}

We now introduce a notion of confluent process.

\begin{definition}[confluence]\label{conf-lts-def}\index{confluence of lts}
A process $P$ is {\em confluent} if {\em for every derivative} $Q$
of $P$ we have:
\begin{equation}\label{conf0}
\infer{Q\wact{\alpha} Q_1\quad Q\wact{\beta} Q_2}
{\xst{Q'_1,Q'_2}
{(Q_1 \wact{\beta\minus\alpha} Q'_1 , 
  \quad Q_2 \wact{\alpha\minus \beta} Q'_2,
  \mbox{ and }\quad Q'_1 \wbis Q'_2)}} \quad \mbox{[conf 0]}
\end{equation}
\end{definition}

The condition in definition \ref{conf-lts-def} is labelled
as [conf 0] to distinguish it from two equivalent
conditions that we state below and that are labelled
[conf 1] and [conf 2].

\begin{itemize}

\item
A process $P$ is {\em confluent 1} if if {\em for every derivative} $Q$
of $P$ we have:
\begin{equation}\label{conf1}
\begin{array}{c}

\infer{Q\act{\alpha} Q_1\quad Q\wact{\beta} Q_2}
{\xst{Q'_1,Q'_2}{
(\ Q_1 \wact{\beta\minus\alpha} Q'_1, \quad 
   Q_2 \wact{\alpha\minus\beta} Q'_2, \mbox{ and }\quad 
   Q'_1 \wbis Q'_2 \ )}}
\quad \mbox{[conf 1]}

\end{array}
\end{equation}

\item
A process $P$ is {\em confluent 2} if for all $r,s\in \cl{L}^*$ we have:
\begin{equation}\label{conf2}
\infer{P \wact{r} P_1 \qquad P \wact{s} P_2}
{\xst{P'_1,P'_2}{
 (\ P_1 \wact{s\minus r} P'_1, \quad
 P_2 \wact{r\minus s} P'_2, \mbox{ and }\quad
 P'_1 \wbis P'_2 \ )}}
\quad \mbox{[conf 2]}
\end{equation}

\end{itemize}

\begin{remark}
In conditions [conf 0] and [conf 1] if $\alpha=\beta$ then we close the diagram with $\tau$ actions only.
\end{remark}

A first {\em sanity check} is to verify that the confluent processes 
are invariant under {\em transitions} and {\em equivalence} 
(cf. proposition \ref{det-invariance-prop}).

\begin{proposition}
The following properties hold:
\begin{enumerate}

\item If $P$ is confluent and $P\act{\alpha} P'$ then $P'$ is confluent.

\item If $P$ is confluent and $P\wbis P'$ then $P'$ is confluent.

\end{enumerate}
\end{proposition}
\Proof
\Proofitemf{(1)} If $Q$ is a derivative of $P'$ then it is also a derivative of
  $P$.

\Proofitemm{(2)} It is enough to apply the fact that:
\[
(P\wbis P' \mand P\wact{\alpha} P_1) \ \mbox{ implies } \ 
\xst{P'_{1}}{(P'\wact{\alpha} P'_{1} \mand P_1 \wbis P'_{1})}
\]
and the transitivity of $\wbis$. \qed\\

Confluence implies $\tau$-inertness, and from this we can
show that it implies determinacy too.

\begin{proposition} 
Suppose $P$ is {\em confluent}. Then $P$ is: 
(1) $\tau$-{\em inert} and  (2) {\em determinate}.
\end{proposition}
\Proof
First a reminder.
A relation $R$ is a {\em weak bisimulation up to $\wbis$} if:
\[
\infer{P \ R \ Q \quad P\wact{\alpha} P'}
{\xst{Q'}{Q\wact{\alpha} Q' \mbox{ and } P' (\wbis \comp R \comp \wbis) Q'}}
\]
(and symmetrically for $Q$).
It is important that we work with the {\em weak moves}
on both sides, otherwise the relation $R$ is 
{\em not} guaranteed to be contained in
$\wbis$ (cf. exercise \ref{expansion-ex}). 
Now we move to the proof.

\begin{enumerate}

\item We want to show that $P\wact{\tau} Q$ implies $P\wbis Q$.
We show that:
\[
R = \set{(P,Q) \mid P\wact{\tau} Q}
\]
is a weak bisimulation up to $\wbis$.
It is clear that whatever $Q$ does, $P$ can do too with some
  extra moves.
In the other direction, suppose, {\em e.g.}, $P\wact{\alpha} P_1$ with $\alpha\neq \tau$
(case $\alpha=\tau$ left as exercise).
By [conf 0],
$Q\wact{\alpha} Q_1$, 
$P_1 \wact{\tau} P_2$, and 
$Q_1 \wbis P_2$.
That is: $P_1 (R \comp \wbis) Q_1$.

\item We want to show that if $P$ is confluent then it is determinate.
Suppose $P\wact{s} P_i$ for $i=1,2$ and $s\in \cl{L}^*$.
We proceed by {\em induction} on $|s|$.
If $|s|=0$ and $P\wact{\tau}P_i$ for $i=1,2$ then 
{\em by $\tau$-inertness}
$P_1 \wbis P \wbis P_2$.
For the inductive case, suppose $P \wact{a} P'_i \wact{r} P_i$ for $i=1,2$.
By confluence and $\tau$-inertness,  we derive that $P'_1 \wbis P'_2$.
By {\em weak bisimulation}, $P'_2 \wact{r} P''_2$ and $P''_2 \wbis
  P_1$.
By {\em inductive hypothesis}, $P_2 \wbis P''_2$.
Thus $P_2 \wbis P''_2 \wbis P_1$ as required. \qed

\end{enumerate}

\begin{exercise}\label{tau-inert-det-conf-ex}
We have seen that confluence implies determinacy which implies
$\tau$-inertness. Give examples that show that these implications
cannot be reversed.

\end{exercise}

We now turn to the condition (\ref{conf1}) labelled [conf 1] which is 
`asymmetric' in that 
the move from $Q$ to $Q_1$ just concerns a {\em single} action.

\begin{proposition}\label{conf1-prop}
A process $P$ is confluent iff for every derivative $Q$ of $P$,
it satisfies condition [conf 1].
\end{proposition}

\begin{exercise}\label{proof-compl-ex}
  Prove proposition \ref{conf1-prop}.
\end{exercise}

A similar result holds for condition (\ref{conf2}) labelled [conf 2].

\begin{proposition}\label{prop-conf2}
A process $P$ is {\em confluent} iff it satisfies [conf 2].
\end{proposition}

\begin{exercise}\label{proof-complement2-ex}
Prove proposition \ref{prop-conf2}.
\end{exercise}

Next, we return to the issue of building confluent (and therefore
determinate) processes.

\begin{proposition}[building confluent processes]
If $P,Q$ are confluent processes then so are:
(1) $0$, $\alpha.P$,
(2) $\new{a}{P}$, and 
(3) $\sigma P$ where $\sigma$ is an injective substitution 
on the free names of $P$.
\end{proposition}
\Proof
Routine analysis of transitions (cf. similar statement for determinacy). \qed

\begin{remark}[on sum]
In general, $a+b$ is {\em determinate} but it is {\em not confluent} for
$a\neq b$.
\end{remark}

\begin{definition}[sorting]\label{sorting-def}\index{sorting in $\ccs$}
Let $P$ be a process. We define its {\em sorting} $\cl{L}(P)$ as the 
set:
\[\set{a\in\cl{L} \mid \xst{s\in \cl{L}^*}{P \wact{s}Q \act{a} \cdot}}~.
\]
\end{definition}

\begin{exercise}\label{buf3-ex}
With reference to exercise \ref{buf2-ex}, show that 
$\cl{L}(a\mapsto b) = \set{a,\ol{b}}$.
\end{exercise}

\begin{definition}[restricted composition]\label{restr-comp-def}\index{restricted parallel composition}
A restricted composition is a process of the shape:
$\new{a_1,\ldots,a_n}{(P \mid Q)}$
where:
\begin{enumerate}

\item $P$ and $Q$ do not share  visible actions:
$\cl{L}(P)\inter \cl{L}(Q)=\emptyset$.

\item $P$ and $Q$ may interact only on the restricted names:
\[
\cl{L}(P)\inter \ol{\cl{L}(Q)} \subseteq 
\set{a_1,\ldots,a_n}\union
\set{\ol{a}_1,\ldots,\ol{a}_n}~.
\]
\end{enumerate}
\end{definition}

\begin{proposition}\label{conf-proc-prop}
Confluence is preserved by restricted composition.
\end{proposition}
\Proof
We abbreviate $\new{a_1,\ldots,a_n}{(P \mid Q)}$ as
$\new{a^*}{(P \mid Q)}$.
First we observe that any derivative of $\new{\vc{a}}{(P\mid Q)}$
will have the shape $\new{\vc{a}}{(P'\mid Q')}$ where $P'$ is a derivative
of $P$ and $Q'$ is a derivative of $Q$.

Since sorting is preserved by transitions, the two conditions
on sorting in definition \ref{sorting-def} will be satisfied.
Therefore, it is enough to show that the diagrams in [conf 1]
commute for processes of the shape $R=\new{\vc{a}}{(P\mid Q)}$ under the given
hypotheses.

We consider one case. Suppose:
$R \act{a} \new{a^*}{(P_1 \mid Q)}$
because $P\act{a}  P_1$.
 Also assume:
$R\wact{a} \new{a^*}{(P_2 \mid Q_2)}$
because $P \wact{s a r} P_2$ and 
$Q\wact{\ol{s}\cdot \ol{r}} Q_2$ 
with $s\cdot r \in \set{\vc{a}, \ol{\vc{a}}}^*$ 
and $a \notin \set{\vc{a}, \ol{\vc{a}}}$.

Since $P$ is confluent we have:
\[
\infer{P \act{a} P_1 \quad P\wact{s a r} P_2}
{\xst{P'_{1},P'_{2}}{
P_1 \wact{sr} P'_1, \quad 
P_2 \wact{\tau} P'_2, \mbox{ and } \quad 
P'_1\wbis P'_2}}~.
\]
 Then we have:
\[
\new{\vc{a}}{(P_1 \mid Q)} \wact{\tau} \new{\vc{a}}{(P'_1 \mid Q_2)} 
\wbis \new{\vc{a}}{(P'_2 \mid Q_2)}~,
\]
thus closing the diagram (note that we use the congruence properties of $\wbis$). \qed

\begin{exercise}\label{proof-cmpl3-ex}
Consider other cases of the proof, for instance:
\[
\begin{array}{ll}
\new{a}{(P \mid Q)} \act{\tau} \new{a}{(P \mid Q)}
&\mbox{as }P \act{a} P_1,\quad Q\act{\ol{a}}Q_1 ~,\\

\new{a}{(P \mid Q)} \wact{\tau} \new{a}{(P_2 \mid Q_2)}
&\mbox{as }P \act{s} P_2,\quad Q\wact{\ol{s}}Q_2 ~.\\

\end{array}
\]
\end{exercise}

\section{Kahn networks (*)}
Kahn networks 
are a determinate model of parallel computation where
communication is point-to-point, {\em i.e.},  for every channel there
is at most one sender and one receiver, and channels are order preserving
buffers of unbounded capacity, {\em i.e.}, sending is non blocking
and the order of emission is preserved at the reception.

In this model, each (sequential) process may:
\begin{enumerate}

\item perform arbitrary {\em sequential deterministic computation},

\item {\em insert a message in a buffer},

\item {\em receive a message from a buffer}. If the buffer is empty then 
the process {\em must} suspend,

\end{enumerate}

However, a process {\em cannot} try to receive a message from 
several channels at once.
In a nutshell Kahn's approach to the semantics of such systems is as follows.
First, we regard the unbounded buffers as finite or infinite words over
some data domain and 
second, we  model the nodes of the network as functions over words.
Kahn observes that the associated system of equations has
a least fixed point which defines the semantics of the whole system.

Kahn networks are an important (practical) 
case where {\em parallelism} does {\em not} induce
race conditions and it is compatible with {\em determinacy}.
For instance, they are frequently used
in the {\em signal processing} community.
Our modest goal is to
formalize Kahn networks as a fragment of $\ccs$ and to 
apply the developed theory to show that the fragment is 
confluent and therefore determinate.

We will work with a `data domain' that contains just
one element.  The generalization to arbitrary data domains is not difficult,
but we would need to formalize determinacy and confluence in the framework
of an extended $\ccs$ where messages carry values (as, {\em e.g.}, in 
the value passing $\ccs$ described in chapter \ref{ccs-sec}).
First, let us conclude the analysis of the {\em unbounded buffers} in $\ccs$.

\begin{exercise}\label{buf4-ex}
  With reference to examples\ref{buf1-ex}, \ref{buf2-ex}, and exercise
  \ref{buf3-ex}:

\begin{enumerate}

\item
  Apply  proposition \ref{conf-proc-prop} to derive that
  the process $a\mapsto b$ is confluent.

\item Derive from proposition \ref{det-bis-tr-prop} that to prove
$P(n)=\new{a}{(\ol{a}^{n} \mid a\mapsto b)} \wbis \ol{b}^{n}$
  it is enough to check the trace equivalence
  $\trace{P(n)}= \trace{\ol{b}^{n}}$.

\item Prove the trace equivalence above.

\end{enumerate}
\end{exercise}

We define a class of $\ccs$ processes sufficient to
represent Kahn networks.

\begin{definition}[restricted processes]\index{Kahn networks}
Let $\w{KP}$ be the least set of processes such that
 $0\in \w{KP}$ and if $P,Q\in \w{KP}$ and $\alpha$ is an action then: 
\begin{enumerate}

\item $\alpha.P\in \w{KP}$,

\item $A(\vc{b}) \in \w{KP}$ provided the names $\vc{b}$ are all distinct, 
$A$ is defined by an equation $A(\vc{a})=P$, and $P\in \w{KP}$.

\item $\new{\vc{a}}{(P\mid Q)} \in \w{KP}$ provided
$\cl{L}(P)\inter\cl{L}(Q) =\emptyset$ and 
$\cl{L}(P)\inter\ol{\cl{L}(Q)} \subseteq \set{\vc{a},\ol{\vc{a}}}$,

\end{enumerate}
\end{definition}

\begin{exercise}\label{kahn-confluent-ex}
Check that: (1) $a\mapsto b$ is a $\w{KP}$ process and (2) 
Kahn processes are confluent.
\end{exercise}

\begin{example}
Suppose we have a Kahn network with three nodes, and 
the following ports and behaviors where we use 
$!$ for output and $?$ for input.

{\footnotesize
\[
\begin{array}{l|ll}  

\mbox{Node} &\mbox{Ports} &\mbox{Behaviors}\\
\hline
 
1 &?a,?b,?c, !d, !e,!f    &A_1=?a.!d.!e.?b.?c.!f.A_1\\
2 &!b,?d                 &A_2=?d.!b.A_2\\
3 &!c,?e                 &A_3=?e.!c.A_3~.

\end{array}
\] }

The corresponding $\ccs$ system relies on the equations
for the buffer process plus:

{\footnotesize
\[
\begin{array}{ll}
A_1(a,b,c,d,e,f)  &=a.\ol{d}.\ol{e}.b.c.\ol{f}.A_1(a,b,c,d,e,f) \\
A_2(b,d)          &=d.\ol{b}.A_2(b,d) \\
A_3(c,e)          &=e.\ol{c}.A_3(c,e)~.

\end{array}
\]}

The sorting is easily derived:

{\footnotesize
\[
\begin{array}{ll}

\cl{L}(A_1(a,b,c,d,e,f) &= \set{a,b,c,\ol{d},\ol{e},\ol{f}} \\
\cl{L}(A_2(b,d))        &= \set{\ol{b},d} \\
\cl{L}(A_3(c,e))        &= \set{\ol{c},e}~.

\end{array}
\]}

To build the system, we have to introduce a buffer 
before every input channel. 
Thus the initial configuration is:

{\footnotesize
\[
\begin{array}{l}

\nu  a', b,b',c,c',d,d',e,e' \ \\
(a\mapsto a'  \mid 
 b\mapsto b'  \mid  
 c \mapsto c' \mid 
 d\mapsto d'  \mid 
 e\mapsto e'  \mid \\
A_1(a',b',c',d,e,f) \mid
A_2(b,d') \mid
A_3(c,e') \ )

\end{array}
\]}

It is easily checked that the resulting process belongs to the
class \w{KP}. 
\end{example}

To summarize, 
to build confluent processes we can use:
(i) nil and input prefix,
(ii) restricted composition,
(iii) injective recursive calls, and
(iv) recursive equations $A(\vc{a})=P$, where $P$ is built according
to the rules above.
This class of processes is enough to represent Kahn networks.
Notice that, via recursion, we can also represent
Kahn networks with a dynamically changing number of nodes 
(see example \ref{buf1-ex}).

\section{Reactivity and local confluence in lts (*)}
We know that a terminating and locally confluent rewriting system is
confluent (proposition \ref{newman-proposition}).
We present a suitable generalization of this result 
to confluent lts. First, we recall from section \ref{testing-notation-sec}
that a process is
{\em reactive} if all its derivatives are strongly normalizing.

\begin{definition}[local confluence] \index{local confluence, in lts}
Let $P$ be a process.
We say that it is {\em locally confluent} if for all its derivatives $Q$: 
\[
\infer{Q\act{\alpha} Q_1\quad Q\act{\beta} Q_2}
{\xst{Q'_1,Q'_2}
{(
Q_1 \wact{\beta\minus\alpha} Q'_1, \quad 
Q_2 \wact{\alpha\minus \beta} Q'_2, \mbox{ and } \quad 
Q'_1 \wbis Q'_2)}}~.
\]
\end{definition}

\begin{exercise}\label{reactivity-recproc-ex}
Consider again the process:
\[
A(a,b) = a.\new{c}{(A(a,c) \mid \ol{b}.A(c,b))}~.
\]
Is the process $A(a,b)$ reactive? 
Consider the cases $a\neq b$ and $a=b$.

\end{exercise}

\begin{exercise}\label{loc-conf-ter-react-ex}
Consider the process:
$A = a.b + \tau.(a.c +\tau.A)$.
Check whether $A$ is:
(1) $\tau$-inert,
(2) locally confluent,
(3) terminating,
(4) reactive,
(5) determinate,
and (6) confluent.
\end{exercise}

Suppose $P$ is a {\em reactive} process and let 
$W$ be the set of its derivatives. 
For $Q,Q'\in W$ write $Q>Q'$ if $Q$ 
rewrites to $Q'$ by a positive number of 
$\tau$-actions.
Then $(W,>)$ is a well founded set.

\begin{proposition}
If a process is {\em reactive} and {\em locally confluent}
then it is {\em confluent}.
\end{proposition}
\Proof 
Let $B$ be the relation $\act{\tau} \union (\act{\tau})^{-1} \union \wbis$ 
(restricted to $W$) and $B^*$ its reflexive and transitive closure. 
Note that $B^*$ is symmetric too.
We take the following steps.

\begin{enumerate}

\item 
For every derivative $Q$ of $P$ it holds:
\[ 
\infer{Q\wact{\tau} Q_1, \quad Q\wact{\alpha} Q_2} 
{\xst{Q_{3}}{( \ 
Q_1\wact{\alpha} Q_3 \mbox{ and }
Q_2 B^* Q_3 \ )}} ~.
\] 

\item 
The relation $B^*$ is a weak-bisimulation.

\item 
The process $P$ is $\tau$-inert.

\item 
The process $P$ is confluent.
\end{enumerate}
Note that  $B^*$ is a binary relation on $W$ (the derivatives of $P$).

\paragraph{Step 1}
The argument is {\em by induction} (cf. proposition \ref{ind-prop})  
on the well founded order $(W,>)$. 

\begin{itemize}

\item If $Q=Q_1$ then the statement holds trivially. 

\item So assume $Q\act{\tau} Q_3\wact{\tau} Q_1$ and
consider 2 cases. 

\begin{enumerate}

\item If $Q\act{\tau} Q_4 \wact{\alpha} Q_2$. 

\begin{itemize}

\item 
By local confluence, $Q_3\wact{\tau} Q_5$, $Q_4\wact{\tau} Q_6$, and 
$Q_5\wbis Q_6$. 
 
\item 
By inductive hypothesis, $Q_6\wact{\alpha} Q_7$ and   $Q_2 B^* Q_7$. 
 
\item
By definition of bisimulation, $Q_5\wact{\alpha} Q_8$ and  $Q_7\wbis Q_8$. 
 
\item 
By inductive hypothesis, $Q_1\wact{\alpha} Q_9$ and   $Q_8 B^* Q_9$. 
 
\end{itemize}
So $Q_2 B^* Q_7 \wbis Q_8 B^* Q_9$, and by definition of $B$, $Q_2 B^* Q_9$.

\item
If $Q\act{\alpha} Q_4 \wact{\tau} Q_2$ with $\alpha \neq \tau$. 

\begin{itemize}

\item 
By local confluence, $Q_3 \wact{\alpha} Q_5$, $Q_4\wact{\tau} Q_6$, 
$Q_5 \wbis Q_6$. 
 
\item 
By inductive hypothesis, $Q_1 \wact{\alpha} Q_7$ and $Q_5 B^* Q_7$. 
 
\end{itemize}
So $Q_2 \lwact{\tau} Q_4 \wact{\tau} Q_6 \wbis Q_5 B^* Q_7$. 
Hence $Q_2 B^* Q_7$. 

\end{enumerate}
\end{itemize}

\paragraph{Step 2}
The relation $B^*$ is a weak-bisimulation. \\\noindent

Suppose $Q_0 B Q_1 \cdots B Q_n B Q_{n+1}$ and $Q_0 \wact{\alpha} Q'_0$. 
Proceed by induction on $n$ and case analysis on $Q_n B Q_{n+1}$. 
By inductive hypothesis, we know that $Q_n \wact{\alpha} Q'_n$ and 
$Q'_0 B^* Q'_n$. 

\begin{enumerate}

\item   If $Q_n \wbis Q_{n+1}$ then 
$Q_{n+1} \wact{\alpha} Q'_{n+1}$ and $Q'_{n} \wbis Q'_{n+1}$. 
So $Q'_0 B^* Q'_{n} \wbis Q'_{n+1}$ and we use  
$B^* \comp \wbis \subseteq B^*$. 
 
\item  If  $Q_n \lact{\tau} Q_{n+1}$ then  
$Q_{n+1} \wact{\alpha} Q'_{n}$. 
 
\item If $Q_n \act{\tau} Q_{n+1}$ then by Step (1), 
$Q_{n+1} \wact{\alpha} Q'_{n+1}$ and $Q'_{n} B^* Q'_{n+1}$. 

So $Q'_0 B^* Q'_{n} B^* Q'_{n+1}$ and we use  
$B^* \comp B^* \subseteq B^*$. 

\end{enumerate}

\paragraph{Step 3}
The process $P$ is $\tau$-inert. \\\noindent

By definition, $\act{\tau} \subseteq B^*$ and  
by Step (2), $B^* \subseteq \wbis$.

\paragraph{Step 4}
The process $P$ is confluent. \\\noindent

By induction on the well-founded order $W$. We distinguish two cases.
 
\begin{enumerate}

\item 
Suppose $Q\act{\alpha} Q_3\wact{\tau} Q_1$ and  
$Q\act{\beta} Q_4\wact{\tau} Q_2$, with 
$\alpha,\beta\neq \tau$.  

\begin{itemize}

\item By local confluence, $Q_3\wact{\beta\minus\alpha}Q_5$, 
$Q_4\wact{\alpha\minus\beta}Q_6$, and $Q_5\wbis Q_6$. 
 
\item By Step (3), $Q_4\wbis Q_2$, and by weak bisimulation, 
$Q_2\wact{\alpha\minus \beta} Q_8$, $Q_6\wbis Q_8$. 
 
\item By Step (3), $Q_3\wbis Q_1$, and by weak bisimulation, 
$Q_1\wact{\beta\minus\alpha} Q_7$, $Q_5\wbis Q_7$. 
 
\end{itemize}

So we have $Q_8\wbis Q_6\wbis Q_5\wbis Q_7$ as required. 
 
\item  Suppose $Q\act{\tau} Q_3\wact{\alpha} Q_1$ and  
$Q\wact{\beta} Q_2$. 
 
\begin{itemize}

\item  By Step (3), $Q \wbis Q_3$, and by weak bisimulation, 
$Q_3 \wact{\beta} Q_5$, $Q_2\wbis Q_5$. 
 
\item By inductive hypothesis, $Q_1\wact{\beta\minus \alpha} Q_6$,  
$Q_5\wact{\alpha\minus \beta} Q_7$, and $Q_6\wbis Q_7$. 
 
\item By weak bisimulation, $Q_2 \wact{\alpha\minus\beta} Q_4$ and 
$Q_4\wbis Q_7$. 
 
\end{itemize}

So $Q_4 \wbis Q_7 \wbis Q_6$ as required. \qed

\end{enumerate}

\begin{exercise}\label{reactive-ex}
Suppose $P$ is a $\ccs$ process that is reactive and such
that for every derivative $Q$ of $P$ we have:
\[
\infer{Q\act{\tau} Q_1 \qquad Q \act{\tau} Q_2}
{Q_1 \wbis Q_2}~.
\]
Show that this implies that for every derivative $Q$ of $P$ we have:
\[
\infer{Q\wact{\tau} Q_1\quad Q\wact{\tau} Q_2}
{\xst{Q'_{1},Q'_{2}}{( \ 
Q_1 \wact{\tau} Q'_1, \quad 
Q_2 \wact{\tau} Q'_2,\quad \mbox{ and }\quad 
Q'_1 \wbis Q'_2 \ )}}~.
\]
\end{exercise}

\section{Summary and references}
A process is determinate if it always reacts  in the same
way to the stimuli coming from the environment.
Confluence is a stronger property than determinacy
that is preserved by a restricted form of parallel composition.
Following \cite{Milner:1995:CC:225494}[chapter 11],
we have presented $3$ alternative  characterizations of confluence.
We have seen that a restricted form of parallel 
composition preserves confluence and 
as a case study we have shown that this fragment of $\ccs$
is enough to represent Kahn networks \cite{DBLP:conf/ifip/Kahn74}.
Synchronous data flow languages such as 
Lustre \cite{DBLP:conf/popl/CaspiPHP87} can be
regarded as a refinement of this model where buffers have size $0$.
A rather complete study of the notion of confluence in the more
general framework of the $\pi$-calculus is in 
\cite{DBLP:conf/icalp/PhilippouW97}, which builds
on previous work on confluence for 
$\ccs$ with value passing.
For reactive processes, local confluence entails confluence.
This is a generalization of Newman's proposition \ref{newman-proposition}
described in \cite{DBLP:journals/tcs/GrooteS96}.

\chapter{Synchronous/Timed models}\label{time-sec}
\markboth{{\it Time}}{{\it Time}}

As mentioned in chapter \ref{introduction-sec}, an important
classification criterion in concurrent systems is 
the  {\em relative speed} of the processes.
In particular, in chapter \ref{introduction-sec} we have contrasted
asynchronous and synchronous systems.
So far we have considered models ($\parimp$, $\ccs$)
where processes are {\em asynchronous}, {\em i.e.}, proceed at
independent speeds. In particular, processes can only {\em synchronize} through
an {\em await} statement or an {\em input/output} communication.
In the following we are going to discuss an enrichment of the $\ccs$ model where
processes are {\em synchronous} (or {\em timed}).
In first approximation, in a synchronous concurrent
system all processes {\em proceed in lockstep} (at the same speed).
In other words, the computation is regulated by a notion of
{\em instant} (or {\em round}, or {\em phase}, or {\em pulse},$\ldots$).

Though synchronous circuits are typical examples of
synchronous systems, one should {\em not} conclude that
synchronous systems are {\em hardware}.
Notions of synchrony are quite useful in the design
of {\em software systems} too. 
The programming of  many problems in a distributed setting
 can be `simplified' or even `made possible' by a 
synchronous assumption. Examples include: leader election, minimum spanning tree, and consensus in the presence of failures.
In general, the notion of synchrony is a useful {\em logical
concept} that can make {\em programming easier}. 

The formalization of a synchronous model depends on the way the
notion of instant is considered. One possibility is to assume that at
each instant each (sequential) process performs a locally defined amount of work. For
instance, a popular definition found in books on distributed algorithms
requires that at each instant
each process (1) writes in the output communication channels, (2) reads
the contents of the input communication channels, and (3) computes its
next state.  However, a less constrained viewpoint is possible which
consists in assuming that at each instant, each process performs an
arbitrary, but hopefully finite, number of actions.  The instant ends
when each process has either terminated its task for the current
instant or it is suspended waiting for events that cannot arise.
This is the viewpoint taken by synchronous languages such as $\esterel$
and we shall describe next its formalization in the framework of $\ccs$.
The reader should keep in mind that we select $\ccs$ because of its simplicity but
that the approach can be easily ported to other models of concurrent systems.
In particular, in section \ref{sl-sec} we shall sketch a synchronous
model where processes interact through {\em signals} rather than
$\ccs$ {\em channels}.

\section{Timed $\ccs$}\label{tccs-sec}
We discuss the definition of a synchronous/timed model on top of $\ccs$.
Following the terminology in the literature, we call this model {\em timed} $\ccs$ ($\tccs$).
As usual, we write $\alpha,\alpha',\ldots$  
for the  $\ccs$ actions and we reserve $a,b,\ldots$ 
for the $\ccs$ actions but the $\tau$ action.
We denote with $\mu,\mu',\ldots$ the $\tccs$ actions. 
\index{$\tccs$, syntax}
They are obtained
by extending the $\ccs$ actions (chapter \ref{ccs-sec})
with a new $\tick$ action which represents the move to the following instant:
\[
\begin{array}{lll}

\mu  &::= \alpha \Alt \tick          &\mbox{($\tccs$ actions).} 
\end{array}
\]
We also extend the syntax of $\ccs$ processes with a new operator `else-next'
which allows to program processes which are 
time dependent and are able to react to the absence of an event.
Intuitively, the process $\pres{P}{Q}$ tries to run $P$ in
the current instant and if it cannot it runs $Q$ in the following.
\[
\begin{array}{lll}

P &::= \cdots \Alt \pres{P}{P}      &\mbox{($\tccs$ processes).} 

\end{array}
\]
The labelled transition system for $\tccs$ includes the 
usual rules for the $\alpha$ actions (Table \ref{ccs-lts}) plus:
\[
\begin{array}{ll}
\infer{P \act{\alpha} P'}
{\pres{P}{Q} \act{\alpha} P'}

&\mbox{(a rule for else-next).}
\end{array}
\]
Moreover, we introduce in Table~\ref{tick-lts} \index{$\tccs$, lts}
special rules for the $\tick$ action describing
the passage of time. The intuition is the following:
\[
\begin{array}{|c|}
\hline
\mbox{A process can $\tick$ if and only if it cannot perform $\tau$ actions.} \\\hline
\end{array}
\]
Incidentally, this is in perfect agreement  with the usual feeling that we 
do not see time passing when we have something to do!
\begin{table}
{\footnotesize
\[
\begin{array}{cc}
\infer{P \not\act{\tau} \cdot}
{\pres{P}{Q} \act{\tick} Q}
\quad
\infer{}{0\act{\tick} 0} \\ \\ 

\infer{}
{a.P \act{\tick} a.P}
\quad
\infer{P_i \act{\tick} P'_i\quad i=1,2\quad (P_1\mid P_2)\not\act{\tau}\cdot}
{(P_1\mid P_2)\act{\tick} (P'_1\mid P'_2)} \\ \\ 

\infer{P_i\act{\tick} P'_i\quad i=1,2}
{(P_1+P_2)\act{\tick} (P'_1+P'_2)}
\quad
\infer{P\act{\tick} P'}
{\nu a \ P\act{\tick} \nu a\  P'}
\end{array}
\]}
\caption{Labelled transition system for the $\tick$ action}\label{tick-lts}
\index{$\tick$ action}
\end{table}

\begin{exercise}[on formalising \tick actions]\label{tick-not-tau}
Check that $P\act{\tick}\cdot$ if and only if $P\not\act{\tau}
  \cdot$ 
The lts in Table~\ref{tick-lts} uses the negative condition $P\not\act{\tau}\cdot$.
Show that this condition can be formalized in a positive way by
defining a formal system to derive judgments of the shape
$P\dcl L$ where $L$ is a set of observable actions and 
$P\dcl L$ if and only if $P\not\act{\tau}\cdot$ and 
$L=\set{a \mid P \act{a} \cdot}$.

\end{exercise}

The following exercise identifies two important choices in 
the design of $\tccs$.

\begin{exercise}[continuations of \tick action]\label{cont-tick-ex}
We say that $P$ is a `$\ccs$ process' if it does not contain the
\w{else\_next} operator. Show that:

\begin{enumerate}

\item If $P\act{\tick} Q_1$ and $P\act{\tick} Q_2$ then $Q_1=Q_2$.
So the passage of time is {\em deterministic}.

\item If $P$ is a $\ccs$ process and $P\act{\tick} Q$ then $P=Q$.
So $\ccs$ processes are {\em insensitive to the passage of time}.
\end{enumerate}

\end{exercise}

\begin{exercise}[programming a switch]\label{switch-ex}
Let $\s{tick}.P=\pres{0}{P}$ and $\s{tick}^{n}.P = \s{tick}\cdots\s{tick}.P$,
$n$ times.
\begin{enumerate}
\item Program a {\em light switch} 
$\w{Switch}(\w{press},\w{off},\w{on},\w{brighter})$
 that behaves as follows:
(i) 
initially the switch is off, 
(ii) if the switch is off and it is pressed then the light turns
  on, 
(iii) if the switch is pressed again in the following 2 instants then
the light becomes brighter while if it is pressed at a later instant
it turns off again,
(iv)
if the light is brighter and the switch is pressed then it
  becomes off.

\item Program a {\em fast user} $\w{Fast}(\w{press})$ 
that presses the switch every 2 instants
and a {\em slow user} $\w{Slow}(\w{press})$ 
that presses the switch every 4 instants.

\item Consider the systems:
\[
\begin{array}{l}
\nu \w{press} \ ( \ 
\w{Switch}(\w{press},\w{off},\w{on},\w{brighter}) 
\mid 
\w{Fast}(\w{press}) \ ) \\

\nu \w{press} \ ( \ 
\w{Switch}(\w{press},\w{off},\w{on},\w{brighter}) \mid \w{Slow}(\w{press}) \ )
\end{array}
\]
and determine when the light is going to be off, on, and bright.
\end{enumerate}
\end{exercise}

\begin{definition}[bisimulation for $\tccs$]
The notion of {\em weak transition} is extended to the $\tick$ action by
defining:
\[
\wact{\tick} \ =  \ \wact{\tau} \comp \act{\tick} \comp \wact{\tau}
\qquad
\mbox{(weak $\tick$ action)}~.
\]
Then we denote with $\wbis_{\tick}$ the related largest weak bisimulation.
\end{definition}

\begin{exercise}\label{tccs-bis-parcomp-ex}
Show that $\wbis_{\tick}$ is preserved by parallel composition.
Also show that $\pres{\pres{P_1}{P_2}}{P_3} \wbis_{\tick} \pres{P_1}{P_3}$.
Thus the nesting of else-next operators
on the left is useless!
\end{exercise}

\begin{exercise}[more on congruence of $\wbis_{\tick}$]\label{tick-wbis-cong-ex}
Suppose $P_1 \wbis_{\tick} P_1$ and $Q_1\wbis_{\tick} Q_2$.
Prove or give a counterexample to the following equivalences.

\begin{enumerate}

\item $P_1+Q_1 \wbis_{\tick} P_2+Q_2$.

\item  $\pres{(a.P_{1})}{Q_{1}} \wbis_{\tick} \pres{(a.P_{2})}{Q_{2}}$.

\item $\pres{P_{1}}{Q_{1}} \wbis_{\tick} \pres{P_{2}}{Q_{2}}$.

\end{enumerate}

\end{exercise}

We have identified the $\ccs$ processes with the $\tccs$ processes that do
not contain an else-next operator. A natural question is whether the
equivalences we have on $\ccs$ are still valid when the $\ccs$ processes are
placed in a timed environment.  A basic observation is that a
diverging computation does not allow time to pass.  Thus if we denote
with $\Omega$ the diverging process $\tau.\tau.\tau\cdots$ we have $0
\not\wbis_{\tick} \Omega$ while in the ordinary (termination
insensitive) bisimulation for $\ccs$ we have $0 \wbis \Omega$.  
The situation is more pleasant  for {\em reactive} processes 
cf. chapter \ref{det-sec}).

\begin{proposition}[$\ccs$ vs. $\tccs$]
Suppose $P,Q$ are $\ccs$ processes. 

\begin{enumerate}

\item $P\wbis_{\tick}Q$ implies  $P\wbis Q$.

\item If moreover, $P,Q$ are reactive then $P\wbis Q$ implies
$P\wbis_{\tick} Q$.
\end{enumerate}
\end{proposition}
\Proof (1) $\tccs$ bisimulation is stronger than $\ccs$ bisimulation and $\alpha$-derivatives of $\ccs$ are again $\ccs$ processes.

\Proofitemm{(2)}
First notice that for a $\ccs$ process being reactive w.r.t. $\ccs$ actions is the same as being reactive w.r.t. $\tccs$ actions.
For $\alpha$ actions, the condition $P\wbis Q$ suffices.
Otherwise, suppose $P\wact{\tick} P'$.
By exercises \ref{tick-not-tau} and  \ref{cont-tick-ex}(2),
this means $P\wact{\tau} P' \act{\tick} P'$.
By definition of $\ccs$ bisimulation, $Q\wact{\tau} Q_1$, $P' \wbis Q_1$.
By reactivity, $Q_1\wact{\tau} Q' \act{\tick}$.
Again by definition of $\ccs$ bisimulation, $P' \wbis Q'$,
Hence $Q\wact{\tick} Q'$ and $P'\wbis Q'$. \qed

\begin{exercise}[termination sensitive bisimulation]\label{bis-ter-ex}
Rather than restricting the attention to reactive processes, another possibility
is to consider a bisimulation for $\ccs$ which is sensitive to termination.
We write $P\dcl$ if $P\not\act{\tau}\cdot$ and $P\Downarrow$ if $P\wact{\tau} Q$ and $Q\dcl$.
Show that on $\ccs$ processes the bisimulation $\wbis_{\tick}$ can be characterized
as the largest relation $\cl{R}$ which is a weak labelled bisimulation 
(in the usual $\ccs$ sense) and such that if $P\rel{R} Q$ and  $P\dcl$ then $Q\Downarrow$.
\end{exercise}

\section{A determinate calculus based on signals (*)}\label{sl-sec}
As a case study, we consider a variant of the $\tccs$ model where 
processes interact through {\em signals} (rather than channels).
A signal is either emitted or not. Once it is emitted
it {\em persists} during the instant and it is {\em reset} at the end of it.
Thus the collection of emitted signals grows {\em monotonically}
during each instant.

The presented calculus is named $\SL$
({\em synchronous language}).
We describe it as a fragment of timed $\ccs$ where
we write $s,s',\ldots$ for {\em signal} names.
The syntax of $\SL$ processes \index{$\SL$, syntax} is as follows:
\[
\begin{array}{ll}

P ::= 0 \Alt  s.P,P  \Alt 
     (\s{emit} \ s) \Alt 
     (P\mid P) \Alt 
     \nu  s \ P \Alt 
     A(\vc{s})  

&\mbox{($\SL$ processes).} 

\end{array}
\]
The newly introduced operators can be understood in terms of those of $\tccs$ as follows:
\[
\begin{array}{ll}
s.P,Q       &= \pres{s.P}{Q} \\
(\s{emit} \ s)&= \pres{\ol{s}.\w{Emit}(s)}{0} \\
              &\mbox{where: } \w{Emit}(s)= \pres{\ol{s}.\w{Emit}(s)}{0}~. \\
\end{array}
\]
Notice that in $\SL$ there is no sum and no prefix for emission (cf. asynchronous
  $\pi$-calculus, chapter \ref{pi-sec}).
The input is a specialized form of the input prefix and 
the else-next operator.
The derived synchronization rule is:
\[
(\s{emit} \ s) \mid s.P,Q \act{\tau} \act{\tau} (\s{emit} \ s) \mid P~.
\]
The second $\tau$ transition is just recursion unfolding and we will ignore it in the following.
Notice that:
\[
(\s{emit} \ s) \mid s.P_1,Q_1 \mid s.P_2,Q_2 \wact{\tau} 
(\s{emit} \ s) \mid P_1 \mid P_2~.
\]
The $\tick$ action can be expressed as:
\[
\tick.P = \nu s \ s.0,P\qquad s\notin \fv{P}~.
\]
A {\em persistent input} (as in $\tccs$) is expressed as:
\[
\s{await} \ s .P = A(\vc{s}), \quad 
\mbox{where: }A(\vc{s})=s.P,A(\vc{s}), \ 
\fv{P}\union \set{s}=\set{\vc{s}}~.
\]

\begin{exercise}\label{switch-in-sl-ex}
Re-program in $\SL$ the light switch seen in exercise \ref{switch-ex}.
Compare the solution with the one based on $\tccs$.
\end{exercise}

The $\SL$ calculus enjoys a {\em strong} form of confluence where
one can close the diagram in {\em at most one step} and {\em up to 
$\alpha$-renaming}. 

\begin{proposition}[strong confluence]\label{confluence-sl-prop}
For all $\SL$ programs $P$ the following holds:
\[
\infer{P \act{\tau} P_1 \qquad P \act{\tau} P_2}
{P_1 \equiv P_2 \mbox{ or }\xst{Q}{(P_1 \act{\tau} Q, P_2\act{\tau} Q)}}
\]
\end{proposition}
\Proof 
Internal reductions are due either to {\em unfolding} or to
{\em synchronization}.
The only possibility for a {\em superposition} of the redexes is:
\[
(\s{emit} \ s) \mid s.P_1,Q_1 \mid s.P_2,Q_2~.
\]
And we exploit the fact that emission is {\em persistent}. \qed \\

The bisimulation $\wbis_{\tick}$ developed for
$\tccs$ can be applied to $\SL$ too. However, because of the 
restricted form of $\SL$ processes, one can expect additional equations to hold.
For instance: 
\begin{equation}\label{eq-sl}
s.(\s{emit} \ s), 0 \mbox{ should be `equivalent' to } 0~.
\end{equation}
A similar phenomenon arises with asynchronous communication in the $\pi$-calculus (cf. chapter \ref{pi-sec}).
More generally, because $\SL$ is determinate (cf. proposition \ref{confluence-sl-prop}) one can expect
a collapse of the {\em bisimulation} and {\em trace} semantics 
(cf. proposition \ref{det-bis-tr-prop}).

\begin{exercise}[on $\SL$ equivalence]\label{sl-equiv-ex}
Check that the equation (\ref{eq-sl}) does not hold in the $\tccs$ embedding.
Also, prove or disprove the following equivalences:
\begin{enumerate}

\item $s.(s.P,Q),Q \wbis_{\tick} s.P,Q$.

\item $(\s{emit} \ s) \mid s.P,Q  \wbis_{\tick}(\s{emit} \ s) \mid P$.

\end{enumerate}

\end{exercise}

\section{Summary and references}
Time, in the sense we have described it here, is derived 
from the notion of computation and as such it is a {\em logical}
notion rather than a concept we attach on top of the computational model.
Time passes when no computation is possible. 
Moving from an asynchronous to a synchronous model means
enriching the language with the possibility to react
to the absence of computation, {\em i.e.}, to the passage
of time.
The distinction between synchronous and asynchronous models
is standard in the analysis of distributed algorithms (see, {\em e.g.},
\cite{Lynch96}).
In the framework of process calculi,
a notion of `timed' $\ccs$ is introduced in 
\cite{DBLP:conf/icalp/Yi91}.
This calculus has a $\s{tick}(x)$ operator that describes the passage
of $x$ time units where $x$ is a non-negative real.
A kind of \w{else\_next} operator is proposed in 
\cite{DBLP:journals/iandc/NicollinS94}.
A so called {\em testing semantics} of a process calculus very close to the one
presented here is given in \cite{DBLP:journals/iandc/HennessyR95}.
However, it seems fair to say that all these works generalize to $\ccs$
ideas that were presented for the $\esterel$ programming language
\cite{DBLP:conf/concur/BerryC84,DBLP:journals/scp/BerryG92}.
Two basic differences in the $\esterel$ approach are that
processes interact through {\em signals} and 
that the resulting calculus is {\em determinate}.

Another important difference is that in the $\esterel$ model it is actually possible
to {\em react immediately} (rather than at the end of the instant)
to the absence of a signal. This requires some semantic care, 
to avoid writing {\em paradoxical programs}  such as 
$ \ s.0,(\s{emit} \ s) \ $
which are supposed to emit $s$ when $s$ is not there 
(cf. stabilization problems in the design of synchronous circuits).
It also requires some clever {\em compilation techniques} to determine
whether a signal is not emitted. In fact these techniques (so far!) are
specific to {\em finite state models}.

The $\SL$ model \cite{DBLP:journals/tse/BoussinotS96} 
we have described is a {\em relaxation} of the $\esterel$ model
where the {\em absence of a signal} can only be detected at the
end of the instant. If we forget about {\em name generation}, 
then the $\SL$ model 
essentially defines a kind of {\em monotonic Mealy machine}. 
Monotonic in the sense that output signals can only depend
{\em positively} on input signals (within the same instant).
The monotonicity restriction allows to {\em avoid the
  paradoxical programs} as monotonic boolean equations do 
have a least fixed point!
The $\SL$ model has a natural and {\em efficient implementation model}
that works well for {\em general programs} (not just finite state
machines). The model has been adapted  to several 
{\em programming environments} ($\C$ \cite{DBLP:journals/spe/Boussinot91}, 
{\it Scheme} \cite{DBLP:conf/ppdp/SerranoBS04}, $\ml$ \cite{DBLP:conf/ppdp/MandelP05}) and
it has been used to program significant applications.

The $\esterel$/$\tccs$/$\SL$ models described here actually follow an earlier
attempt at describing synchronous/timed systems in the framework 
of $\ccs$ known as {\it SCCS}/{\it Meije} model \cite{DBLP:journals/tcs/Milner83,DBLP:journals/tcs/AustryB84}.
The basic idea of these models is that the actions of the system
live in an abelian (commutative) group freely generated from a
collection of basic actions.  At each instant, each (sequential) process must perform
exactly one action and the observable result of the computation
is the group composition of the actions performed by each process.
This gives rise to a model with pleasant algebraic properties
but whose implementation and generalization to a full scale programming
languages appear to be problematic.

Finally, let us mention {\em timed automata} 
as another popular formalism for describing `timed' systems 
\cite{DBLP:journals/tcs/AlurD94}.
This is an enrichment of finite state automata with {\em timing 
constraints} which still enjoys decidable model-checking properties.
This is more a {\em specification language} for finite control systems 
than a  {\em programming language}.

\chapter{Probability and non-determinism}\label{proba-sec}
\markboth{{\it Probability}}{{\it Probability}}

Probabilities arise in several areas of system design and analysis.
For instance, one may want to  analyze programs or protocols that
toss coins at some point in the computation, {\em e.g.}, to compute the
probability that a test for number primality returns the correct
answer.  In another direction, one may want to evaluate the
reliability of a system given some probability of failure of its
components.  And yet in another direction, one may be interested in
evaluating the performance of a system in terms of, say, the average
waiting time of its users.  As already mentioned, in concurrent
systems, non-determinism arises to account for {\em race conditions}
and also as a {\em specification device}.  It is then natural to
{\em  lift} methods for (deterministic) probabilistic systems to
non-deterministic ones. This is not a simple task and still the
subject of ongoing research.
In this chapter, we focus on the notion of {\em probabilistic} rewriting
system which is a rewriting system where a state can reduce to
a distribution over states. For such systems, we consider the
notion of {\em almost sure} termination which intuitively means that
under any reduction strategy the system is guaranteed to terminate
with probability $1$. We consider a sufficient criteria to
establish almost sure termination and apply it
to the analysis of the so called dining philosophers protocol.

\section{Probabilistic rewriting and termination}
Let $S$ be a finite or countable set.
A (probability) distribution over $S$ is a function $\Delta: S\arrow [0,1]$
such that $\Sigma_{s\in S} \Delta(s) = 1$.
We say that a distribution $\Delta$ is {\em Dirac} if all the probability is concentrated
in one point, namely there exists $s\in S$ such $\Delta(s)=1$.
We denote with $D(S)$ the set of distributions over $S$.
If $s_1,s_2\in S$ and $p\in [0,1]$ then we denote with $s_1+_ps_2$ the
distribution $\Delta\in D(S)$ such that $\Delta(s_1)=p$ and $\Delta(s_2)=(1-p)$.

\begin{definition}[probabilistic rewriting system]\label{prob-rew-def}\index{rewriting system, probabilistic}
  A probabilistic rewriting system is a pair $(S,\arrow)$ such that
  $S$ is finite or countable and $\arrow \subseteq S \times D(S)$.
\end{definition}

An ordinary rewriting system on $S$ (definition \ref{reduction-system-def})
can be viewed as a probabilistic
rewriting system where if $s\arrow \Delta$ then $\Delta$ is Dirac.
We say that the system is {\em not non-deterministic}, shortened as NND,
if for every $s\in S$,
there is at most one distribution $\Delta$ such that $s\arrow \Delta$.
A NND system is not quite deterministic as it can exhibit
probabilistic behavior (here is a situation where double negation is not involutive).
In particular, {\em Markov chains}
are a special case of NND systems
where for every $s\in S$ there is {\em exactly}
one distribution $\Delta$ such that $s\arrow \Delta$.

Let $(S,\arrow)$ be a probabilistic rewriting system. A (finite) {\em computation}
is a sequence $\gamma$ with the following shape and properties:
\begin{equation}\label{comp-prob}
  \gamma=s_0 \Delta_1 s_1 \cdots \Delta_n s_n~,\qquad n\geq 0~,
  s_i \arrow \Delta_{i+1}~, \mbox{ and }\Delta_{i+1}(s_{i+1})>0~\mbox{ for }i=0,\ldots,n-1.
\end{equation}
  The computation $\gamma$ has {\em length} $n$ which we denote with
  $\ell(\gamma)$.
  We say that $\gamma$ is {\em terminated} if
  $\set{\Delta \mid s_n \arrow \Delta}=\emptyset$.
  We attach to the computation $\gamma$ (terminated or not) 
the following {\em weight} (which can be seen as a {\em probability} in
an appropriate setting, hence the notation):
\[
\prob{\gamma}=\Delta_1(s_1) \cdots \Delta_n(s_n)~\qquad\mbox{(weight of a computation).}
\]
A {\em scheduler} $\cl{S}$ is a partial function that takes as input an arbitrary computation and produces a possible
distribution among those that can extend the computation. So if $\gamma$ is the computation above we must
have that $\cl{S}(\gamma)$ is undefined if $\gamma$ is terminated and otherwise
$\cl{S}(\gamma)\in \set{\Delta \mid s_n \arrow \Delta}$.
The scheduler can be regarded as a device that resolves the non-determinism and brings us back
in the realm of NND systems. Once we have a NND system it makes sense to speak
of the probability of certain events and in particular we shall focus in this section on the probability of termination.

We say that the computation $\gamma$ as displayed in (\ref{comp-prob})
agrees with the scheduler $\cl{S}$ if
$\Delta_{i+1}= \cl{S}(s_0 \Delta_1 s_1 \cdots \Delta_i s_i)$ for $i=0,\ldots,n-1$.
We denote with $T(s,\cl{S})$ the collection of terminated computations $\gamma$ 
which start with $s$ and agree with the scheduler $\cl{S}$.

The probability that $s$ terminates relatively to a scheduler $\cl{S}$ is defined as:
\[
\Sigma_{\gamma \in T(s,\cl{S})} \ \prob{\gamma} \qquad\mbox{(probability of termination, relative to a scheduler).}
\]

\begin{definition}[almost sure termination]\index{termination, almost sure}
  We say that a probabilistic rewriting system $(S,\arrow)$ terminates almost
  surely (or with probability $1$), if for any scheduler $\cl{S}$ and any $s\in S$ the
  probability that $s$ terminates relatively to $\cl{S}$ is $1$.
\end{definition}

\begin{example}\label{ex-non-det-ter}
  Consider the probabilistic rewriting system $(S,\arrow)$ where
  $S=\set{a,b}$, $a\arrow \Delta_2$, $a\arrow \Delta_3$, and for
  $n\in \set{2,3}$:
  \[
  \Delta_n(a) = \frac{1}{n},\qquad\Delta_n(b)=1-\frac{1}{n}~.
  \]
  At each rewriting step the system reaches the normal form $b$ with
  probability at least $1/3$. Then the probability of a non-terminated,
  $n$ steps computation is at most $(1-1/3)^n$ and since this quantity
  goes to $0$ as $n$ goes to infinity we can conclude that for any
  scheduler the   probability of a non-terminating computation is null.
\end{example}

\begin{remark}
  The reader should keep in mind that the
  definition of almost sure termination depends on the class of
  schedulers. It is possible to enlarge this class by
  considering {\em randomized}
  schedulers. A randomized scheduler is not forced to choose one reduction
  but can choose a distribution
  over the set of possible reductions.
  For instance if $s\arrow \Delta_1$ and $s\arrow \Delta_2$ then
  the scheduler may select the distribution $\Delta_1$ with probability
  $1/3$ and the distribution $\Delta_2$ with probability $2/3$.
  In another direction, one may just focus on {\em memoryless} schedulers that
  select the next reduction as a function of the last state of the computation.
\end{remark}

We introduce next a notion of {\em finite} probabilistic rewriting system
and provide a simple criteria that guarantees almost sure termination
for such systems.

\begin{definition}[finite probabilistic rewriting system]
  We say that the probabilistic rewriting system $(S,\arrow)$
  is finite if $S$ is finite and  the relation $\arrow$ is finite.
\end{definition}

Notice that even if $S$ is finite, there still can be uncountably many
distributions;  hence the additional condition on the finiteness of
the rewriting relation.  For finite systems, it is possible to give a
simple sufficient criteria for probabilistic termination.

\begin{proposition}\label{suf-cond-ter-proba}
  Let $(S,\arrow)$ be a finite probabilistic rewriting system.
  Let $(W,>)$ be a well-founded order and $\mu:S\arrow W$ a function
  such that:
  \begin{equation}\label{proba-ter-cond}
    s\arrow \Delta \mbox{ implies }\exists s'(\Delta(s')>0 \mbox{ and }\mu(s)>\mu(s'))~.
  \end{equation}
  Then $(S,\arrow)$ terminates almost surely.
\end{proposition}
\Proof
Let $n=\card S$ be the number of elements in $S$.
Notice that for any computation $\gamma=s_0 \Delta_1 s_1 \cdots \Delta_m s_m$
such that $\mu(s_0)>\cdots > \mu(s_m)$ we must have $m=\ell(\gamma) \leq n-1$.
This is because the $s_0,\ldots,s_m$ must be all distinct,
as the order $>$ is well-founded, and $S$
contains $n$ elements.

Let $q$ be the least positive probability occurring
in a distribution; it exists because there are finitely
many distributions in the system and $S$ is finite. Then let $p=q^{n-1}$.
For any execution $\gamma$ such that $\ell(\gamma)\leq (n-1)$,
we have $\prob{\gamma}\leq p$.

We are going to generalize the argument in example \ref{ex-non-det-ter}.
We denote with $\w{NT}(s,\cl{S},m)$ the set of {\em not} terminated executions
starting with $s$, having length $m$, 
and agreeing with the scheduler
$\cl{S}$. We show that for every
starting state $s$, for every scheduler $\cl{S}$, and for every
integer $k\geq 1$:
\begin{equation}
\Sigma_{\gamma \in \w{NT}(s,\cl{S},k\cdot (n-1)) } \ \prob{\gamma} \leq (1-p)^k~.
\end{equation}
If $k=1$ then we know that there is at least one terminating
execution with length at most $n-1$ and probability at least $p$.
Hence the probability of the non-terminating executions of length
$(n-1)$ is at most $(1-p)$.
Next, suppose the assertion for $k$. We are going to decompose
a non-terminating execution of length $(k+1)\cdot (n-1)$
as a non-terminating execution of length $k\cdot (n-1)$
concatenated with a non-terminating execution of length $(n-1)$.
Let $N$ abbreviate $\w{NT}(s,\cl{S},(k+1)\cdot (n-1))$. Define the
set of computations $I$ as the prefixes of the computations in $N$
of length $k\cdot (n-1)$:
\[
I=\set{\gamma \mid \ell(\gamma)=k\cdot (n-1), \exists \gamma'(\gamma \cdot \gamma'\in N)}~.
\]
Notice that $I\subseteq \w{NT}(s,\cl{S},k\cdot (n-1))$, hence by inductive
hypothesis we have:
\begin{equation}\label{ind-case-prob}
\Sigma_{\gamma\in I} \ \prob{\gamma} \leq (1-p)^k~.
\end{equation}
For each $\gamma\in I$, we define $J_\gamma$ as the set of computations
that concatenated with $\gamma$ produce a computation in $N$:
\[
J_\gamma = \set{\gamma' \mid \gamma \cdot \gamma'\in N}~.
\]
By an argument similar to the one for the base case, we derive that:
\begin{equation}\label{base-case-prob}
\Sigma_{\gamma' \in J_\gamma} \ \prob{\gamma'} \leq (1-p)~.
\end{equation}
Then we can conclude by the following computation:
\[
\begin{array}{lll}
\Sigma_{\gamma\in N} \ \prob{\gamma} &= \Sigma_{\gamma'\in I,\gamma'' \in J_{\gamma'}} \ \prob{\gamma'\cdot \gamma''} \\
                                &=  \Sigma_{\gamma'\in I,\gamma'' \in J_{\gamma'}} \ \prob{\gamma'}\cdot \prob{\gamma''} \\
&= \Sigma_{\gamma'\in I} \ \prob{\gamma'} \cdot (\Sigma_{\gamma''\in J_{\gamma'}}\  \prob{\gamma''}) \\
&\leq  \Sigma_{\gamma'\in I} \ \prob{\gamma'} \cdot (1-p) 
&\mbox{(by \ref{base-case-prob})} \\
  &\leq (1-p)^{(k+1)}
  &\mbox{(by \ref{ind-case-prob})}~.
\end{array}
\]
\qed

\begin{example}\label{ex-sys-term-proba}
  Let $S=\set{a,b,c,d}$ with rewriting:
  \[
  \begin{array}{llll}
    a\arrow b, &a\arrow c &b\arrow a +_{1/2} d &c\arrow a+_{1/3} d
  \end{array}
  \]
  Taking the order $a>b,c>d$ and $\mu$ as the identity,
  we satisfy condition (\ref{proba-ter-cond}). Hence the
  system terminates almost surely.
\end{example}

\begin{exercise}\label{ex-non-finite-prob-ter}
  Consider the probabilistic rewriting system $(S,\arrow)$ with
  $S=\set{a,b}$ and $a \arrow \Delta_n$ for $n\geq 1$, where:
  \[
  \begin{array}{lll}
    p_n=\frac{1}{2^{n}}, &\Delta_n(a) = (1-p_n), &\Delta_n(b)=p_n~.
  \end{array}
  \]
  Show that this (non-finite) system satisfies the
  condition (\ref{proba-ter-cond})
  but it does {\em not} terminate almost surely.
  On the other hand, the system does terminate almost surely
  if we just consider {\em memoryless} schedulers (remark \ref{scheduler-remark}).
\end{exercise}

\section{Dining philosophers (*)}
As a substantial case study for probabilistic termination, 
we introduce a family of systems freely inspired by the classical
problem of the dining philosophers which we describe briefly
(the problem was proposed in 1965
by E. Dijkstra at a final examination).

A system is composed of $n\geq 2$ dining philosophers, alternating with
$n$ chopsticks, with the $2n$ elements disposed in a ring.
The goal of each philosopher
is to grab repeatedly the two chopsticks on his left and on his right
and eat. The problem is that two philosophers may compete for the
same chopstick and that each philosopher can eat only once it
has grabbed the two chopsticks next to him.
A symmetric, distributed, and probabilistic strategy to
coordinate the philosophers'
activities proposed by \cite{DBLP:conf/popl/LehmannR81} goes as follows.
Each philosopher is repeatedly engaged in the following tasks:

\begin{enumerate}

\item It decides probabilistically whether to look for the left or right chopstick,

\item Say, it decides to go for the left chopstick first (the case where it goes for the right is completely symmetric). Then it waits till he can grab  it.

\item Then it looks whether the right chopstick is available: if it is then it grabs it, eats, and releases the two
  chopsticks, otherwise he releases the chopstick already grabbed.

\end{enumerate}

Our goal is to show that for any possible scheduler and initial state,
at least one philosopher will manage to eat (but still leaving the possibility
that one or more philosophers will starve).
We rephrase this goal as
a termination problem as follows: we remove the transitions in step 3.
that move a philosopher into eating state
and show that: (i) the remaining transitions must
terminate almost surely and (ii) in each normal form
at least one philosopher can grab the second chopstick and eat.

Let $X=\set{\ell, L, r, R}$ be a set describing the $4$ possible
states of each philosopher (before eating).  Intuitively, $\ell$ ($r$)
is the state where the philosopher is trying to grab the left (right)
chopstick and $L$ ($R$) is the state where the philosopher has grabbed
the left (right) chopstick.

A state of a system of $n$ philosophers is a vector
$(x_0,\ldots,x_{n-1})$ such $x_i\in X$ and for no $i\in
\set{0,\ldots,n-1}$ we have $x_i=R$ and $x_{i+1}=L$, where addition is
always intended modulo $n$. By forbidding the configuration $x_i=R$
and $x_{i+1}=L$, we exclude an odd situation where a chopstick is
grabbed at the same time by two philosophers.

A state of a system rewrites to another state if one of its
components, say $x_i$ rewrites into $y_i$ according to the following
rules, for $i\in \set{0,\ldots,n-1}$:
\[
\begin{array}{l|l|l}
  x_i=     &y_i=     &\mbox{side condition}\\\hline

  \ell     &L        &\mbox{if }x_{i-1}\neq R \\
  r        &R        &\mbox{if }x_{i+1}\neq L \\
  L    &\ell+_{1/2}r  &\mbox{if }x_{i+1}=L \\
  R    &\ell+_{1/2}r  &\mbox{if }x_{i-1}= R
\end{array}
\]
The last two rewriting rules produce non-Dirac distributions; the notation
$\ell+_{1/2} r$ means that $x_i$ becomes $\ell$ with probability $1/2$ and
$r$ with probability $1/2$.

\begin{example}\label{dining-ex}
  Let us consider some reductions for the case $n=3$. Starting from the
  state $s_0=(r,\ell,\ell)$ we can have, for instance the following reductions:
  \[
  \begin{array}{l}
    s_0 \arrow (r,L,\ell) \arrow (r,L,L) \arrow (r,\ell,L) +_{1/2} (r,r,L)~,\\
    (r,\ell,L) \arrow (r,L,L)~,\\
    (r,r,L) \arrow (R,r,L) \not\arrow~.
  \end{array}
  \]
  So for some scheduler, it is possible to loop back with
  probability $1/2$ or to go to a normal form with probability $1/2$.
\end{example}

It is useful to identify certain special configurations arising during
the reduction.

\begin{definition}[separation and conflict points]
Let $s=(x_0,\ldots,x_{n-1})$ be a state.
The integer $i\in \Z_n$ is a:
  \begin{itemize}
\item  {\em separation point} for $s$ 
  if $x_i\in \set{\ell,L}$ and $x_{i+1}\in \set{r,R}$.

\item {\em conflict point} for $s$ if
  $x_i \in \set{r,R}$ and $x_{i+1}\in \set{\ell,L}$
  (with the usual proviso that $x_i=R$ and $x_{i+1}=L$ is
  impossible).
\end{itemize}
  \end{definition}

\begin{proposition}\label{prop-rewr-dining}
  The following properties hold.
  \begin{enumerate}
  \item In every state, there are as many separation points as conflict points.

 \item Separation and conflict points alternate and there are at most $n/2$ of them.

  \item Rewriting cannot decrease the number of separation and conflict points.

  \item Rewriting cannot change the position of a separation point.
  \end{enumerate}
\end{proposition}

\begin{exercise}\label{proof-prop-rewr-dining}
Prove proposition \ref{prop-rewr-dining}.
\end{exercise}

By proposition \ref{prop-rewr-dining}(4),
separation points once created keep a fixed position.
On the other hand, each conflict point
is framed by two separation points on the left and on the right
but its exact position between these two separation points
may change. Then we have to make sure that a 
conflict point cannot oscillate between the two framing
separation points indefinitely.
To this end, we introduce a notion of {\em potential} of a conflict point.

\begin{definition}[potential]
Let $s=(x_0,\ldots,x_{n-1})$ be a state and
$i\in \Z_n$ a conflict point for $s$. Its potential is determined as follows:
    \begin{itemize}

    \item if $x_i=r,x_{i+1}=\ell$ take the maximum distance from the
      separation points framing $i$.

    \item if $x_i=r,x_{i+1}=L$ take the distance from the separation point
      framing $i$ on the right.

    \item if $x_i=R,x_{i+1}=\ell$ take the distance from the separation point
      framing $i$ on the left.
    \end{itemize}
\end{definition}

\begin{definition}[well-founded measure]
    The measure we associate to a state is the lexicographic order from left to right of:
    \begin{enumerate}

    \item $n/2$ minus the number of separation points (if $n$ is odd take
      the integer part of $n/2$).

    \item sum of the potentials of the conflict points.

    \item number of states which are $\ell$ or $r$.

    \end{enumerate}
\end{definition}

\begin{example}\label{dining-ex2}
  Continuing example \ref{dining-ex}, consider the following (probabilistic)
  reduction and the corresponding strictly decreasing sequence in the
  left to right lexicographic order:
\[
\begin{array}{llll}
(r,\ell,\ell) &\arrow (r,L,\ell) &\arrow (r,L,L)    &\arrow_{1/2} (r,r,L)~, \\
(0,2,3)       &>_{\w{lex}} (0,2,2) &>_{\w{lex}} (0,2,1) &>_{\w{lex}} (0,1,2)~.
  \end{array}
  \]
\end{example}

\begin{proposition}\label{prop-termination-dining}
  The rewriting system of the dining philosophers terminates
  almost surely.
\end{proposition}  
\Proof It is enough to check that condition (\ref{proba-ter-cond})
of proposition \ref{suf-cond-ter-proba} is satisfied.
The first two (non-probabilistic) rules do not affect the
number and the position of the separation (and conflict) points.
The potential of the conflict points does not increase since the
potential of a conflict point of the shape $r \ell$ is the maximum
of the potential of the conflict points of the shape $R\ell$ or $rL$
that can be obtained as a result of a reduction of this type.
On the other hand, the number of states $\ell$ or $r$ is decreased by $1$.

The following two probabilistic rule have a symmetric treatment.
Let us consider the situation where $x_i=L$ is replaced by
$\ell+_{1/2} r$ where $x_{i+1}=L$. If we transform $L$ to $\ell$ we are
in a loop. So let us consider the case where we transform
$x_i x_{i+1}=LL$ into $rL$ and distinguish two cases.

\begin{itemize}

\item If $x_{i-1}\in \set{\ell,L}$ then we increase by one the number
  of separation points and the first component of the measure decrease.

\item If $x_{i-1}=r$ (here $x_{i-1}$ cannot be $R$)
  then the conflict point at position $i-1$ has moved to position $i$
  and therefore its potential, {\em i.e.}, its distance from the
  separation point on the right, has decreased by $1$. So the second
  component decreases while the third increases and the first stays the same. \qed 
  \end{itemize}

Every normal form contains a conflict point, and around a conflict
point there is a philosopher that can eat by grabbing a second chopstick.
For instance, if the conflict point $i$ is such that $x_i,x_{i+1}=R,\ell$
then $x_{i-1} \notin \set{r,R}$ by the hypothesis that we have a normal
form. Hence $x_{i-1}\in\set{\ell,L}$ and the philosopher in position $i$
can grab the left chopstick. Therefore for any initial configuration,
almost surely
the system goes to a configuration where at least one philosopher is going to eat.

\begin{remark}\label{scheduler-remark}
  In the second step of the original probabilistic algorithm
for the dining philosophers
\cite{DBLP:conf/popl/LehmannR81}, 
each philosopher actively tries to get the first chopstick. This
induces a potential looping computation that is ruled out by assuming
a fair scheduler. In the presented modeling, we 
suppose that the philosopher is suspended and
may resume the computation when the chopstick
becomes available. In this case, as noted by
\cite{DBLP:journals/dc/DuflotFP04}, it is possible
to prove probabilistic termination without making fairness assumptions.
\end{remark}

\section{Summary and references}
Probabilistic rewriting systems are rewriting systems where
a state can reduce to a distribution over states.
Markov chains   (see, {\em e.g.}, \cite{Norris98})
are a typical example of {\em purely} probabilistic rewriting
systems. In this chapter, we have considered more general systems which
can have both probabilistic and non-deterministic reductions.
To speak of probabilities in the presence of non-determinism,
one introduces a class of {\em schedulers}.
Each scheduler resolves the non-determinism and brings us back to the realm
of purely probabilistic rewriting.
A probabilistic rewriting system
terminates almost surely if for every scheduler, the system terminates with
probability $1$.
For {\em finite} probabilistic rewriting systems, it is possible
to give a simple criteria that guarantees almost sure termination
(see, {\em  e.g.}, \cite{DBLP:conf/wdag/DuflotFP01}) and
as an application of the criteria, we have analyzed a variant of
the dining philosophers protocol
\cite{DBLP:conf/popl/LehmannR81,DBLP:journals/dc/DuflotFP04}.
There is a growing body of results on {\em probabilistic}  labelled
transition systems which we have entirely omitted. See, {\em e.g.},
\cite{Jonsson01probabilisticextensions} for a preliminary overview and
some references.

\chapter{$\pi$-calculus}\label{pi-sec}
\markboth{{\it $\pi$-calculus}}{{\it $\pi$-calculus}}

$\ccs$ provides a basic model of {\em communication} and {\em
  concurrency} while ignoring the mechanisms of {\em procedural} and
{\em data abstraction} which are at the heart of sequential
programming (in this respect, $\ccs$ is close to Turing machines). 
We can contrast $\ccs$
with a basic model of sequential programming
such as the (typed) $\lambda$-calculus.
A basic question is: how can we {\em integrate} the $\lambda$-calculus and
$\ccs$?

One standard approach, supported both by theory and by practice, is to
take the $\lambda$-calculus as the {\em backbone} of the programming
language and to add on top a few features for communication and
concurrency.  The resulting language provides a comfortable
programming environment but one may question whether this is the {\em
  simplest model} one can hope for (Ockham's razor).  It turns out
that the superposition of the concepts of function and process leads
to some {\em redundancy} and that it is possible to reduce to simpler
languages such as the $\pi$-calculus.

There are {\em two main ways} to look at the $\pi$-calculus.  On one
hand,  it can be regarded as an
extension of $\ccs$ where channels exchange values that are themselves channel
names. As such it inherits from $\ccs$ a relatively simple and tractable
theory including labelled transition systems and bisimulation proof
methods. This  viewpoint is developed in sections \ref{red-pi-sec} and \ref{lts-pi}.
On the other hand,
it can be regarded as a concurrent extension of one of 
the {\em intermediate functional languages}
studied in the chapter \ref{transform-sec} on the compilation
of functional languages.
As such it has an expressive
power comparable (up to some encoding!)  to the one of modern
programming languages.  
This viewpoint is elaborated in section \ref{proc-fun-sec}.

\section{A $\pi$-calculus and its reduction semantics}\label{red-pi-sec}
The basic idea is that the $\pi$-calculus is an extension of value passing 
$\ccs$ where 
processes exchange {\em channel names} as in:
\[
(\ x(y).P \mid \new{z}{\ol{x}z.Q} \ ) \act{\tau} \new{z}{([z/y]P \mid Q)}~.
\]
Quoting the authors who introduced the $\pi$-calculus:
\begin{quote}
It will appear as though we reduce all concurrent
computation to something like a cocktail party, in which the
only purpose of communication is to transmit (or to receive)
a name which will admit further communications.
\end{quote}
The abstract syntax of {\em a possible} $\pi$-calculus 
is defined as follows.
\[
\begin{array}{lll}
\w{id} &::= x \Alt y \Alt \ldots &\mbox{(names)} \\
P      &::= 0 \Alt \w{id}(\w{id}).P \Alt \ol{\w{id}}\w{id}.P \Alt (P\mid P)  \Alt 
            \new{\w{id}}{P} \Alt [\w{id}=\w{id}]P \Alt !(\w{id}(\w{id}).P) &\mbox{(processes).} 
\end{array}
\]
The informal semantics is as follows:
$0$ does nothing, $x(y).P$ waits for
a name $z$ on the channel $x$ and then becomes $[z/y]P$, 
$\ol{x}y.P$ sends $y$ on the channel $x$ and becomes $P$,
$(P\mid Q)$ runs $P$ and $Q$ in parallel, $\new{x}{P}$ creates the
new name $x$ and runs $P$, $[x=y]P$ compares $x$ and $y$ and becomes
$P$ if they are equal (otherwise it is stuck), 
$!(x(y).P)$  waits for
a name $z$ on the channel $x$ and then becomes $[z/y]P \mid !(x(y).P)$
(thus the operator $`!'$ replicates an input and allows to generate
infinite recursive behaviors).
In processes, the formal parameter of an input 
and the $\nu$ bind names. We define $\fv{P}$ as the set of names 
occurring free in
a process $P$. As usual, bound names can be renamed according to the rules
of $\alpha$-conversion.

\begin{remark}[definable operators]
In the presented version of the $\pi$-calculus, we have dropped two
operators which are present in $\ccs$: non-deterministic choice and
recursive definitions.  The reason is that, up to some restrictions,
both can be encoded in the presented calculus. 
The encoding of non-deterministic choice is related to the
one we have already considered in example \ref{choice-enc-ex}.
The encoding of recursive definitions amounts to replace, say, 
$\ \s{letrec} \ A(x)=P\ \s{in} \ Q\ $ by
$\ \new{A}{(!(A(x).P') \mid Q')}\ $ where $P'$ and $Q'$ are
obtained from $P$ and $Q$, respectively, 
by replacing each (tail) recursive call, say
$A(y)$, with a message  $\ol{A}y$.
\end{remark}

\begin{remark}[name renaming and substitution]\label{rename-sub-rmk}
The names of the $\pi$-calculus can be split in two categories: 
those on which we can just perform $\alpha$-renaming and those
on which we can perform both $\alpha$-renaming and general (non-injective) 
substitutions. In particular, names bound by the $\nu$ operator
fall in the first category while names bound by the (replicated) input
operator fall in the second one. It makes sense to regard the first
category of names as {\em constants} and the second one as {\em variables},
and indeed some authors  
distinguish two syntactic categories 
and add a third one which is the union of the first two. 

If we start with a process whose free names are constants
then this property is preserved by reduction and all substitutions
replace a variable by a constant. This is true of the labelled transitions
described in the following section \ref{lts-pi} too, assuming that 
all the actions are built out of constant names. 
A consequence of this remark is that it is possible to suppose that 
the reduction rules and the labelled transitions are given on processes 
where all free names are constants. This is in line with the usual
practice in operational semantics where the reduction rules
are defined on `closed' programs (as, {\em e.g.}, in chapter \ref{env-closure-sec}).
\end{remark}

In order to define a compositional semantics for the $\pi$-calculus we
follow the approach presented for $\ccs$ in chapter \ref{ccs-sec}
which amounts to define a contextual bisimulation and a labelled bisimulation
and show that they coincide. However, because the notion of label for
the $\pi$-calculus is not obvious, this time we shall start with 
contextual bisimulation.
Table~\ref{pi-def} \index{$\pi$-calculus, syntax}
defines the static contexts, a structural equivalence,
and the reduction rules for the $\pi$-calculus.\index{$\pi$-calculus, reduction}
The related notions of commitment and contextual (weak) bisimulation
are inherited directly from $\ccs$ (section \ref{red-ccs-sec}).

\begin{table}
{\footnotesize
\begin{center}
{\sc Static contexts}
\end{center}
\[
\begin{array}{ll}
C      &::= [~] \Alt C\mid P \Alt \new{x}{C} 
\end{array}
\]
\begin{center}
{\sc Structural equivalence}
\end{center}
\[
\begin{array}{cl}
P_1 \mid (P_2\mid P_3) \equiv (P_1 \mid P_2)\mid P_3 &\mbox{(associativity)} \\
P_1 \mid P_2 \equiv P_2 \mid P_1                     &\mbox{(commutativity)} \\
\new{x}{P_1}\mid P_2 \equiv \new{x}{(P_1\mid P_2)} \quad x\notin\fv{P_2}
&\mbox{(extrusion)}
\end{array}
\]
\begin{center}
{\sc Reduction}
\end{center}
\[
\begin{array}{cc}

\infer{}{\ol{x}y.P_1 \mid x(z).P_2\arrow P_1 \mid [y/z]P_2} 

&\infer{}{\ol{x}y.P_1 \mid !(x(z).P_2)\arrow (P_1 \mid [y/z]P_2)\mid !(x(z).P_2)} \\ \\

\infer{}{[x=x]P \arrow P} 

&\infer{P\equiv C[P']\quad P'\arrow Q' \quad C[Q']\equiv Q}{P\arrow Q}
\end{array}
\]}
\caption{Reductions for the $\pi$-calculus}\label{pi-def}
\end{table}

\begin{exercise}\label{red-pi-ex}
Reduce the processes
$\new{y}{\ol{x}y.P} \mid x(z).Q$ and
$\new{y}{\ol{x}y.P} \mid x(z).(y(w).Q)$.
\end{exercise}

\section{A lts for the $\pi$-calculus}\label{lts-pi}
We now consider the problem of defining a labelled transition system for the $\pi$-calculus.
This is a rather technical exercise. As a first step, we distinguish four types of actions:

{\footnotesize\[
\begin{array}{|c|c|}
\hline
\mbox{{\bf Action} }\alpha      &\mbox{{\bf Example}} \\ \hline

\tau               &\ol{x}y.P \mid x(z).Q \act{\tau} P\mid [y/z]Q \\

\mbox{input }xz       &x(y).P \act{xz} [z/y]P \\

\mbox{output }\ol{x}y      &\ol{x}y.P \act{\ol{x}y} P \\

\mbox{bound output }\ol{x}(y) &\new{y}{\ol{x}y.P} \act{\ol{x}(y)} P\\\hline

\end{array}
\]}

We remark that the first three cases would arise naturally in an extension of $\ccs$ with ground
values too (section \ref{value-ccs-sec}). The real novelty is the {\em bound output case}. Note that an effect
of the bound output action is to {\em free} the restricted name $y$. 
The bound output action carries a bound name and one has to
be careful to {\em avoid conflicts}. Here are 
some typical situations.

{\footnotesize
\[
\begin{array}{|c|c|}
\hline
\new{y}{x(y).P}
&\mbox{The name which is input must not conflict with the fresh one.}\\

\new{y}{\ol{x}y.P}
&\mbox{The name which is output should become free}.\\

\new{y}{\ol{x}y.P} \mid x(z).Q
&\mbox{The scope of $\nu y$ should extend to $Q$.}\\

\new{y}{\ol{x}y.P} \mid x(z).(y(w).Q)
&\mbox{The fresh $y$ and the one on the recipient side are distinct.} \\\hline
\end{array}
\]}

We fix some conventions concerning free and bound names in actions.
(1)
All occurrences of a name in an action are free
except $y$ in a bound output action $\ol{x}(y)$.
(2) Define $\fv{\alpha}$ ($\bv{\alpha}$) 
as the set of names occurring free (bound) in the action $\alpha$.
(3) Let ${\sf v}(\alpha) = \fv{\alpha} \union \bv{\alpha}$.

Based on this, Table~\ref{lts-pi-table} 
\index{$\pi$-calculus, lts}
defines a labelled transition system for 
the $\pi$-calculus. Rules apply up to $\alpha$-renaming and symmetric rules are omitted.

\begin{table}
{\footnotesize
\[
\begin{array}{c}

\infer{}{x(y).P \act{xz} [z/y]P}

\qquad
\infer{}{\ol{x}y.P \act{\ol{x}y} P} \\ \\

\infer{P\act{\alpha} P'\quad x\notin {\sf v}(\alpha)}
{\new{x}{P} \act{\alpha} \new{x}{P'}}

\qquad
\infer{P\act{\ol{y}x} P'\quad x\neq y}
{\new{x}{P} \act{\ol{y}(x)} P'} \\ \\

\infer{P\act{xy} P'\quad Q \act{\ol{x}y} Q'}
{P\mid Q \act{\tau} P'\mid Q'}

\qquad
\infer{P\act{xy} P'\quad Q \act{\ol{x}(y)} Q'\quad y\notin \fv{P}}
{P\mid Q \act{\tau} \new{y}{(P'\mid Q')}} \\ \\

\infer{P\act{\alpha} P' \quad \bv{\alpha}\inter \fv{Q}=\emptyset}
{P\mid Q \act{\alpha} P'\mid Q}

\qquad

\infer{x(y).P\act{\alpha} P'}
{!(x(y).P) \act{\alpha} P'\mid !(x(y).P)}

\\ \\

\infer{}
{[x=x]P\act{\tau} P}

\end{array}
\]}
\caption{A lts for the $\pi$-calculus}\label{lts-pi-table}
\index{$\pi$-calculus, lts}
\end{table}

\begin{exercise}[on the lts]\label{lts-examples}
Apply the definition to compute the labelled transitions
of the processes discussed above.
\end{exercise}

The next step is to define a notion of bisimulation on
the labelled transition system. This is not completely obvious.
Suppose we want to show that $P$ and $Q$ are 
`labelled' bisimilar.
Further, suppose $P\act{\ol{x}(y)} P'$ makes a bound output.
What is the condition on $Q$?
We note that bisimilar processes may have different sets
of free names. For instance, suppose $P\equiv \new{y}{\ol{x}y.P'}$ and
$Q\equiv \new{y}{\ol{x}y.Q'} \mid R$, with $y\in \fv{R}$.
Then the transition $P\act{\ol{x}(y)}\cdot$ cannot be matched
(literally) by $Q$ because $y$ is free in $Q$. This leads to
the following definition.

\begin{definition}[labelled bisimulation]\index{$\pi$-calculus, labelled bisimulation}
A binary relation $\cl{R}$ on processes is a strong labelled
bisimulation if:
\[
\infer{P\rel{R} Q,\qquad P\act{\alpha} Q,\qquad 
\bv{\alpha} \inter \fv{Q} = \emptyset}
{\xst{Q'}{Q\act{\alpha} Q' \qquad P'\rel{R} Q'}}~.
\] 
and, as usual, a symmetric condition holds for $Q$.
For the weak case, we replace $\act{\alpha}$ by $\wact{\alpha}$.
We denote with $\lbis$ ($\lwbis$) the largest strong (weak) labelled
bisimulation.
\end{definition}

\begin{proposition}
Strong (weak) labelled bisimulation
is preserved by static contexts.
\end{proposition}
\Proof
We want to show that strong (weak) labelled bisimulation
is preserved by static contexts.
Let us abbreviate with  $\nu y^*$ a {\em possibly empty} list 
$\nu y_1,\ldots, \nu y_n$.
We define a binary relation:
\[
\cl{R} = \set{(\new{y^*}{P\mid R},\new{y^*}{Q\mid R}) \mid P\lbis Q}~.
\]
We show that $\cl{R}$ is a labelled bisimulation.
First let us see what goes wrong with the relation
one would define for $\ccs$:
\[
\cl{R'} = \set{(P\mid R,Q\mid R) \  \mid \ P\lbis Q}~.
\]
We can have  $P\mid R \act{\tau} \new{y}{P'\mid R'}$ because
$P\act{\ol{x}(y)} P'$ and $R\act{xy} R'$. 
Then we just have $Q\mid R \act{\tau} \new{y}{Q'\mid R'}$ 
and $P'$ is bisimilar to $Q'\ldots$

Let us look again at this case when working with the
larger relation $\cl{R}$.
Suppose $\new{y^*}{P\mid R} \act{\tau}
  \new{y}{\new{y^*}{P'\mid R'}}$ because 
$P\act{\ol{x}(y)} P'$ and $R\act{xy} R'$.
Then $Q\act{\ol{x}(y)} Q'$ and $P'\lbis Q'$.
Therefore
$\new{y^*}{Q \mid R} \act{\tau} \new{y}{\new{y^*}{Q'\mid R'}}$
  and now:
\[
\new{y}{\new{y^*}{P'\mid R'}} \rel{R}
\new{y}{\new{y^*}{Q'\mid R'}}~.
\]
It is actually possible to develop a little bit of
`bisimulation-up-to-context' techniques (cf. exercise \ref{bis-upto-ex})
to get rid once and for all of these technicalities. \qed

\begin{exercise}\label{pi-bis-ex}
Complete the proof that labelled bisimulation 
is preserved by static contexts.
Then generalize the proof to the weak case.
\end{exercise}

\begin{exercise}[$\tau$-transitions vs. reduction]\label{pi-trans-red-ex}
In Table~\ref{pi-def}, we have defined a reduction relation $\arrow$
on the $\pi$-calculus. 
Show that reduction and $\tau$-transitions are the same up
to structural equivalence (cf. proposition \ref{lts-red-prop}). Namely:

\begin{enumerate}

\item If $P\arrow Q$ then for some $Q'$, $P\act{\tau} Q'$ and $Q\equiv Q'$.

\item If $P\act{\tau} Q$ then $P\arrow Q$.

\end{enumerate}
For instance, consider:
$(\new{y}{\ol{x}y.P_1} \mid P_2) \mid x(z).P_3$.
\end{exercise}

Since labelled bisimulation is preserved by static contexts, we can
easily conclude that the largest labelled bisimulation is a contextual
bisimulation.  We are then left to show that the largest contextual
bisimulation is a labelled bisimulation.

\begin{proposition}
The largest contextual bisimulation is a labelled bisimulation.
\end{proposition}
\Proof 
With reference to the proof for $\ccs$
(proposition \ref{cxtbis-is-labbis-prop}),
the static contexts we have to build are now slightly more
elaborate.

{\footnotesize
\[
\begin{array}{|c|c|}
\hline
\mbox{{\bf Action}}    &\mbox{{\bf Static Context}} \\ \hline

xy              &[~] \mid \ol{x}y.o_1\isum (o_2 \isum 0) \\ 

\ol{x}y          &[~] \mid x(z).[z=y](o_1\isum (o_2 \isum 0)) \\ 

\ol{x}(y)        &[~] \mid x(z).([z=y_1]w_1 \mid \cdots \mid
                    [z=y_k]w_k \mid (o_1 \isum (o_2 \isum 0))) \\\hline

\end{array}
\]}

\noindent where $y_1,\ldots,y_k$ are the free names in the pair of processes
under consideration and $w_1,\ldots,w_k$, $o_1,o_2$ are fresh names. \qed

\begin{exercise}\label{cxtbis-labis-pi-ex}
Complete the proof that the largest contextual bisimulation
is a labelled bisimulation.
\end{exercise}

\begin{remark}[input prefix]
Strictly speaking, the contextual/labelled bisimulation we have
studied is not preserved by the input prefix in the sense that it is
not true that $P \bis Q$ implies $x(y).P \bis x(y).Q$.  For instance,
take $P\equiv [z=y]\nu w\ol{z}w.0$ and $Q\equiv 0$. The point is that
our semantics compares processes by (implicitly) assuming that all
free names are `constants' in the sense of remark
\ref{rename-sub-rmk}.  However, the comparison of processes with
`variable' names (like $y$ in the example above) can be reduced to the
problem of comparing processes with constant names by considering all
possible substitutions of the variable name.  While in principle this
leads to infinitely many cases, a little analysis shows
that it is enough to substitute all names which are free in the
processes under consideration plus a fresh one.

More sophisticated analyses are possible by defining labelled transition
systems which perform a symbolic execution of processes.
However, these technical developments appear to be of limited interest as
they are rarely used in applications.
Moreover, there are interesting fragments of the $\pi$-calculus 
where they become useless because the obvious notion
of bisimulation we have considered is actually preserved
by the input prefix.
\end{remark}

\section{Variations (*)}
We have considered a particular variety of the $\pi$-calculus with the
aim of having a relatively simple labelled transition system and
labelled bisimulation. However, a number of variations are possible.
We mention a few  and discuss their impact on the characterization
of labelled bisimulation as contextual bisimulation.

\paragraph{Polyadic channels}
This is an extension where several names can be transmitted at once.
We shall see in chapter \ref{proc-fun-sec} 
that this extension is quite natural when looking at the
$\pi$-calculus as an intermediate language.
The extension calls for some form of typing 
to guarantee that in a synchronization
the sender and the receiver agree on the number of names to be exchanged.
Moreover, in the formalization of the labelled transition system,
the structure of the actions is a bit more complicated as several
names can be extruded as the result of a communication. 
The syntax of the actions becomes:
\[
\alpha::= \tau \Alt x(y_1,\ldots,y_n) \Alt \new{z_1,\ldots,z_m}\ol{x}(y_1,\ldots,y_n)
\]
with the requirement for the output that $z_1,\ldots,z_m$ is a (possibly empty)
subsequence of $y_1,\ldots,y_n$ composed of distinct names.

\paragraph{Asynchronous communication}
This is actually a restriction that requires that 
an output action cannot prefix another action. 
Again we shall see that this restriction is suggested by looking
at the $\pi$-calculus as an intermediate language.
From a semantic viewpoint, this restriction entails that 
an input action is not directly observable and calls for
a modification of the bisimulation condition, or equivalently,
for a modification of the labelled transition system.
In the adapted semantics, one can show, {\em e.g.}, 
that the processes $x(y).\ol{x}y$ and $0$ are weakly bisimilar.

\paragraph{Other restrictions}
We may consider restricted communication patterns.
For instance, in the context of asynchronous communication we
may make the additional assumption that each name has
a {\em unique} receiver. Then in a process 
$P\equiv !(x(y).P) \mid \ol{x}z$ the output on the $x$ channel is not
observable from the environment because the capability of receiving 
on $x$ is attributed to the process $P$ itself.
In another direction, we may 
want to drop the operation for name comparison. Then, modulo
some additional hypotheses, it may be {\em impossible} to distinguish
two names $y$ and $z$.
For instance, imagine $y$ and $z$ are two names for the
same service and that the only way the observer can use
$y$ and $z$ is to send a message on them.
In this case, a theory of bisimulation will have to consider  
names modulo an equivalence relation.

\section{From $\lambda$ to $\pi$ notation (*)}\label{proc-fun-sec}
We develop the view that the $\pi$-calculus is a
concurrent extension of an intermediate language used in the
compilation of languages of the $\ml$ family.  
In chapter \ref{transform-sec}, we have seen that
a standard  polyadic, call-by-value, $\lambda$-calculus
can be put in CPS (continuation passing style), value named form. 
It turns out that, modulo a simple change of notation, 
this language corresponds to a
deterministic and sequential fragment of the $\pi$-calculus.  
From chapter \ref{typing-sec}, we know that  
the CPS and value named transformations are type preserving.
Thus modulo the change of notation, the $\pi$-calculus inherits
the propositional typing discipline of the $\lambda$-calculus.
As a second step, in the following section \ref{lambda-join-sec},
we consider an extension of the restricted $\pi$-calculus,
called $\lambda_j$-calculus, which allows to express recursive, parallel, 
and concurrent behaviors and to  which the ordinary $\pi$-calculus can be
compiled. To summarize, we have the following diagram:
\[
\lambda 
\act{\cl{C}_{\w{cps}}} 
\lambda_{\w{cps}} 
\act{\cl{C}_{\w{vn}}} 
\lambda_{\w{cps,vn}} 
\cong 
\pi_{\w{restricted}} 
\subset \lambda_j 
\stackrel{\sem{\_}}{\leftarrow}
\pi~,
\]
which is read from left to right as follows.
The call-by-value $\lambda$-calculus is put
in CPS form ($\lambda_{\w{cps}}$) and then in CPS value named form
($\lambda_{\w{cps,vn}}$). This is equivalent to a restricted form
of $\pi$-calculus ($\pi_{\w{restricted}}$). 
When this restricted $\pi$-calculus is extended 
with recursive definitions, parallel composition,
and a form of {\em join definition}, 
it becomes sufficiently expressive
to represent the ordinary $\pi$-calculus.

Table~\ref{lambda-cps-vn-recall} recalls the 
{\em value named} $\lambda$-calculus
in CPS form introduced in chapter \ref{transform-sec}.
In this $\lambda$-calculus, all values are named 
and when we apply the name of a $\lambda$-abstraction to the name of a value
we create a new copy of the body of the function and replace its formal parameter name with the name of the argument as in:
\[
\lets{y}{V}{\lets{f}{\lambda x.M}{@(f,y)}} \ \arrow  \ 
\lets{y}{V}{\lets{f}{\lambda x.M}{[y/x]M}}~.
\]
We also recall that in the value named $\lambda$-calculus 
the evaluation contexts are  sequences of let definitions 
associating values to names.
We can move from $\lambda$ to $\pi$ with a simple change of notation which is
summarized in Table \ref{lambda-pi}.

\begin{table}[b]
{\footnotesize
\begin{center}
{\sc Syntax}
\end{center}
\[
\begin{array}{lll}

V &::= \lambda \w{id}^+.M                                                     &\mbox{(values)} \\
M &::= @(\w{id},\w{id}^+) \Alt \lets{\w{id}}{V}{M}                                           &\mbox{(CPS terms)} \\
E &::= [~] \Alt \lets{\w{id}}{V}{E}                                                          &\mbox{(evaluation contexts)} \\

\end{array}
\]
\begin{center}
{\sc Reduction Rule}
\end{center}
\[
\begin{array}{cccl}

E[@(x,z_1,\ldots,z_n)]
&\arrow
&E[[z_1/y_1,\ldots,z_n/y_n]M]
&\mbox{if }E(x)=\lambda y_1,\ldots,y_n.M \\

\end{array}
\]
\[
\mbox{where: }E(x)=\left\{
\begin{array}{ll}
V  &\mbox{if }E=E'[\lets{x}{V}{[~]}] \\
E'(x) &\mbox{if }E=E'[\lets{y}{V}{[~]}],x\neq y, y\notin\fv{E'(x)} \\
\mbox{undefined} &\mbox{otherwise.}
\end{array}\right.
\]
}
\caption{A value named, CPS $\lambda$-calculus: $\lambda_{{\it cps,vn}}$}\label{lambda-cps-vn-recall}
\end{table}

\begin{table}
{\footnotesize
\[
\begin{array}{|c|c||c|c|}
\hline
\mbox{$\lambda$-interpretation} &\lambda\mbox{-syntax}            &\pi\mbox{-syntax}     &\mbox{$\pi$-interpretation}\\\hline 

\mbox{function application}  &@(x,y^+)            &\ol{x}y^+  &\mbox{calling a service}\\ 

\mbox{function definition}   &\lets{x}{\lambda y^+.M}{N}   &\new{x}{(!x(y^+).M \mid N)}   &\mbox{service definition}\\\hline

\end{array}
\]}
\caption{Changing notation from $\lambda$ to $\pi$}\label{lambda-pi}
\end{table}

\begin{example}[compilation and reduction simulation]
The following example illustrates the compilation of a $\lambda$-term
and how the compiled term simulates the original one.

{\footnotesize
\[
\begin{array}{lll}
I=\lambda x.x, &I'=\lambda x,k.@(k,x), &I(z)=!z(x,k).@(k,x), \\
               &K=\lambda x.@(\w{halt},x), &K(k)=!k(x).@(\w{halt},x)
\end{array}
\]
\[
\begin{array}{|c|cl|}

\hline
\lambda &@(I,y) &\arrow y \\ &&\\

\lambda_{\w{cps}} &@(I',y,K) &\arrow @(K,y) \\
          &&\arrow @(\w{halt},y) \\ &&\\

\lambda_{\w{cps,vn}}&\lets{z}{I'}{\lets{k}{K}{@(z,y,k)}} 
&\arrow \lets{z}{I'}{\lets{k}{K}{@(k,y)}} \\
&&\arrow \lets{z}{I'}{\lets{k}{K}{@(\w{halt},y)}}  \\ &&\\ 

\pi &\new{z}{(I(z) \mid \new{k}{(K(k) \mid \ol{z}(y,k))})} 
&\arrow \new{z}{(I(z) \mid \new{k}{(K(k) \mid \ol{k}y)})} \\
&&\arrow \new{z}{(I(z) \mid \new{k}{(K(k) \mid \ol{\w{halt}} \ y)})}
\\\hline

\end{array}
\]}
\end{example}

We stress that the $\pi$-terms obtained from the compilation 
are {\em highly constrained};
the following section \ref{lambda-join-sec} discusses the  restrictions and 
some possible relaxations.

We can lift the correspondence between $\lambda$ and $\pi$ terms to
types. In chapter \ref{typing-sec}, 
we have shown that the CPS and value named transformations
preserve  (propositional) types.
Modulo the change of notation presented above, 
this provides a propositional type system for the $\pi$-calculus.
We recall that $\w{tid}$  is the syntactic category of {\em type variables} 
with generic elements $t,s,\ldots$, and $A$ is
the syntactic category of {\em types} with generic elements $A,B,\ldots$
A {\em type context} is written $x_1:A_1,\ldots,x_n:A_n$ 
with the usual conventions. We also write $x^*:A^*$ for a 
possibly empty  sequence $x_1:A_1,\ldots,x_n:A_n$, and
$\Gamma,x^*:A^*$ for the context resulting from $\Gamma$ by adding the sequence
$x^*:A^*$.
Table~\ref{types-vn-pi} recalls the typing rules for the
CPS, value named calculus and presentes the very same rules
formulated in the $\pi$-notation.  
For the sake of brevity, we shall omit the type of a 
term since this type is always the
type of results $R$ and write $\Gamma \aGives M$ rather than 
$\Gamma \aGives M:R$. A type of the form $A^+\arrow R$ corresponds
to a channel type $\chtype{A^+}$ in $\pi$-calculus notation.
As expected, one can show that 
typing in this system is preserved by reduction.

\begin{table}
{\footnotesize
\begin{center}
{\sc Typing rules with $\lambda$ notation}
\end{center}
\[
\begin{array}{lll}

A&::= \w{tid} \Alt (A^+\arrow R)  &\mbox{(types)}
\end{array}
\]
\[
\begin{array}{cc}

\infer{\Gamma,y:A^+\arrow R \aGives N\quad 
\Gamma,x^+:A^+ \aGives M}
{\Gamma \aGives \lets{y}{\lambda x^+.M}{N}} \qquad

&\infer{x:A^+\arrow R,y^+:A^+\in \Gamma}
{\Gamma \aGives @(x,y^+)}

\end{array}
\]
\begin{center}
{\sc Typing rules with $\pi$ notation}
\end{center}
\[
\begin{array}{lll}

A&::= \w{tid} \Alt \chtype{A^+}  &\mbox{(types)}
\end{array}
\]
\[
\begin{array}{cc}

\infer{\Gamma,y:\chtype{A^+} \piGives N\quad 
\Gamma,x^+:A^+ \piGives M}
{\Gamma \piGives \nu y\ (!y(x^+).M \mid N)} \qquad

&\infer{x:\chtype{A^+},y^+:A^+\in \Gamma}
{\Gamma \piGives \ol{x}y^+} 

\end{array}
\]
}
\caption{Isomorphic type systems in $\lambda$ and $\pi$ notation}\label{types-vn-pi}
\end{table}

\subsection*{Adding concurrency}\label{lambda-join-sec}
The typed $\lambda_{cps,\w{vn}}$-calculus which is 
the target of the compilation
chain is restricted in several ways. We
show how these restrictions can be relaxed to obtain a calculus
which can represent the computations of the $\pi$-calculus
in a rather direct way. The restrictions and the related 
relaxations concern the possibility of: 
(1) defining processes by {\em general recursion},
(2) running processes in {\em parallel}, and 
(3) having a concurrent access to a resource.
The first two relaxations are rather standard 
while the third one lends itself to some 
discussion.

\paragraph{General recursion} It is easily shown
that in the {\em typed}  $\lambda_{\w{cps},\w{vn}}$-calculus all computations terminate;
this is a consequence of the termination of the corresponding 
typed $\lambda$-calculus.
To allow for infinite computations we introduce recursive definitions,
that is in $\lets{x}{M}{N}$ we allow $M$ to depend recursively on $x$.
For instance, we can write 
a non-terminating term $\lets{x}{\lambda y.@(x,y)}{@(x,z)}$.

\paragraph{Parallelism} The computations in the $\lambda_{\w{cps,vn}}$-calculus are
essentially sequential since at any moment there is at most
one function call which is active. To allow for some parallelism we
allow for function calls to be put in parallel as in 
$@(x,y^+) \mid @(z,w^+)$. We notice that while the resulting 
calculus allows for parallel computations it fails to represent 
concurrent computations (cf. discussion in chapter \ref{introduction-sec}). 
The reason is that two parallel calls
to the same function such as:
\[
\lets{x}{\lambda y.M}{@(x,z) \mid @(x,w)} 
\]
can be executed in an arbitrary order without affecting the
overall behavior of the process. In other terms, the reductions
of the calculus are (strongly) {\em confluent} 
(yet another example of parallel and deterministic system, cf. 
chapters \ref{det-sec} and \ref{time-sec}).

\paragraph{Concurrency} There are several possibilities to introduce
concurrent behaviors in the calculus; we consider $3$ of them.
One possibility 
is to introduce a mechanism to define a function which can be called
{\em at most once}. Then two parallel calls to such a function would be
concurrent as in:
\[
\letonce{x}{\lambda y.M}{@(x,z) \mid @(x,w)}~.
\]
Here the first call that reaches the definition {\em consumes} it and the
following ones are stuck.
Another possibility is to associate {\em multiple definitions} to
the same name. This situation is dual to the previous one
in that the definitions rather than the calls are concurrent as in:
\[
\s{letmlt}\ x =\lambda y.M_1 \ \s{or}\  x= \lambda y.M_2  \ \s{in} \ @(x,z)~.
\]
Here the first definition that captures the call is executed and the
remaining ones are stuck.
A third and final possibility consists
in introducing and {\em joining two names} in a definition which is written as:
\[
\letjoin{x}{y}{M}~. 
\]
As usual in a definition, we assume the 
names $x$ and $y$ are bound in $M$.
The effect of joining the names $x$ and $y$ is that 
any function transmitted on $x$ can be applied 
to any argument transmitted on $y$. In first approximation, 
the reduction rule for a joined definition is:
\[
\letjoin{x}{y}{E[@(x,z)\mid @(y,w)]} \arrow
\letjoin{x}{y}{E[@(z,w)]}~.
\]
Thus a joined definition allows for a {\em three-way synchronization}
among two function calls and a definition. In turn, this synchronization
mechanism allows to simulate a situation where several threads 
compete to access the same communication channel.
An advantage of this approach with respect to the \s{letonce} and
\s{letmlt} described above is
that the calculus keeps a {\em standard} 
definition mechanism where each name introduced is defined once and for all.
At the same time, this relatively modest extension suffices to 
express the synchronization mechanisms of the $\pi$-calculus.

\begin{exercise}\label{lambdaj-ex}
This is an open ended exercise whose goal is to define
and play with a minimal extension of the 
$\lambda_{\w{cps,vn}}$-calculus with recursive definitions,
parallel calls, and join definitions.
For the sake of brevity, we
call the resulting calculus the $\lambda_{j}$-calculus.

\begin{enumerate}

\item Define the structural equivalence and reduction rules
  of the $\lambda_{j}$-calculus. By applying the structural
  equivalence it should be possible to transform any term
  into a list of definitions followed by the parallel composition
  of function calls.

\item Show that in $\lambda_j$ it is possible to mimick the
  \s{letonce} and \s{letmlt} definition mechanisms.

\item Formalize an encoding of the monadic $\pi$-calculus
  in the $\lambda_j$-calculus which is based on the following
  idea. Assign to each name $x$ in the
 $\pi$-calculus a pair of names
$\inp{x},\out{x}$ in the $\lambda_j$-calculus. 
Then an input on $x$ is transformed into a call
to the name $\inp{x}$ while an output on $x$ becomes a call to the
name $\out{x}$. The names $\inp{x},\out{x}$ are {\em joined} so that
a reduction in $\lambda_j$ is possible whenever there is at least
one call to $\inp{x}$ and one call to $\out{x}$.
Not surprisingly, recursive definitions are needed in the
encoding of the replicated input.

\item Check your encoding allows to simulate the following
  reduction of the $\pi$-calculus:
\[
R \equiv \nu x \ (x(y).P \mid \ol{x}z.Q) \arrow
\nu x \ ([z/y]P\mid Q)~.
\]

\item Adapt
the type system for the $\lambda_{\w{cps,vn}}$-calculus presented
in Table~\ref{types-adm} to the $\lambda_j$-calculus you have defined.

\item Extend to types your translation from the monadic $\pi$-calculus
  to the $\lambda_j$-calculus and check that typing is preserved
  by the translation.
\end{enumerate}

 \end{exercise}

\section{Summary and references}
The $\pi$-calculus is an extension of $\ccs$ where processes exchange
channel names. As in $\ccs$, it is possible to define the notions of
contextual and labelled bisimulation and show that they coincide.  In
particular, this comes as a justification of the definition of the
labelled bisimulation which is quite technical.  
The $\pi$-calculus is
introduced by Milner {\em et al.} in
\cite{DBLP:journals/iandc/MilnerPW92a}\index{Milner} following earlier work 
in \cite{Engberg86acalculus}.
The books \cite{DBLP:books/daglib/0018113,SW} explore its theory.
The encoding of non-deterministic choice is studied
in \cite{DBLP:journals/iandc/Nestmann00} 
while the notion of bisimulation for asynchronous
communication is analyzed in \cite{DBLP:journals/tcs/AmadioCS98}.

The $\pi$-calculus can also be regarded as a concurrent relaxation of a
functional language in CPS, value named form.  This fact explains
its ability to encode a variety of features of high-level programming
languages.  Milner in \cite{DBLP:journals/mscs/Milner92} is the
first to discuss a translation from the $\lambda$-calculus to the
$\pi$-calculus.  Since then a variety of translations have appeared
in the literature.  
The notion of multiple synchonization is commonly found in Petri nets (see, {\em e.g.}, \cite{Reutenauer90}).\index{Petri nets}
In the framework of the $\pi$-calculus, the 
notion of {\em join definition} is put forward 
in \cite{DBLP:conf/popl/FournetG96}. This work contains
a more elaborate definition mechanism than the one we have described
here and it sketches a number of sophisticated encodings.

\chapter{Concurrent objects}\label{conc-obj-sec}
\markboth{{\it Concurrent objects}}{{\it Concurrent objects}}
In this chapter, we reconsider {\em shared memory} concurrency.
The $\parimp$ model introduced in chapter \ref{introduction-sec} 
has a pedagogical value  in that it allows to illustrate 
many interesting problems
that arise in concurrency in a relatively simple setting.
On the negative side, it is clear that its modelling power is 
rather limited. First, it does not support the introduction of data
structures such as lists, queues, trees, graphs, $\ldots$ and the
related operations on them.  Second, the memory model does
not allow for the dynamic allocation, manipulation, and possibly disposal 
of memory locations which is typical of imperative programming.
Incidentally, these considerations are similar to those 
motivating the move from CCS to the $\pi$-calculus.

Research on concurrent programming in the shared memory model
has focused on the issue of programming data structures that 
allow for concurrent access, {\em i.e.}, for the concurrent 
execution of several operations on the data structure, while 
providing an observable behavior which is `equivalent' to 
that of a data structure where the execution of the operations
is sequential, {\em i.e.}, each operation is run from the beginning
to the end without interference from the other operations.
In a certain technical setting, this property 
 is called {\em linearizability}
and because the technical setting corresponds roughly to that of 
a $\java$-like concurrent object-oriented programming language
one speaks of concurrent and sequential {\em objects} rather
than concurrent and sequential {\em data structures}, respectively.
Also, the {\em operations} of the data structures  correspond to 
the {\em methods} of the object.

An important result in this field 
is the existence of {\em universal  constructions} that transform any `sequential' object into a
`concurrent' one without introducing locks.  Instead of locks, one
relies on relatively simple atomic operations such as {\em compare and
  set} (cf. example \ref{ex-compare-set}).  Such concurrent data 
structures are called {\em lock-free}.
The absence of locks is instrumental to ensure a form
of collective progress, {\em i.e.}, there is a guarantee that some
operations will be completed while others may be delayed indefinitely.
In fact one can go one step further and produce {\em wait-free} data
structures where each operation is guaranteed to terminate in a
bounded number of steps.  Unfortunately, such universal transformations
tend to be rather inefficient and research has focused on both ways
to have more efficient constructions in some special cases and on ways
to relax the correctness conditions so as to allow for some
efficient implementation techniques.

Our first goal in this chapter is to discuss these
issues in a (fragment of a) state of the art 
programming language ($\java$) and to
hint to their formalization. 
The reader is supposed to have a superficial knowledge of
the $\java$ programming language. Section \ref{conc-prog-java} reviews
basic notions of concurrent programming in the $\java$ programming
language and section \ref{conc-java-spec} builds on chapter \ref{obj-sec}
to provide a formalization of the reduction rules and the typing rules
of a tiny object-oriented concurrent language equipped with a construct
that allows the atomic execution of a sequence of statements.

Our second goal concerns the {\em semantics} of concurrent objects.
In section \ref{intro-conc-obj}, we introduce the problem,
in section \ref{coind-obj-client} we cast it in the general framework of
labelled transition systems with synchronization, and in
sections \ref{may-trace-preorder}
and \ref{client-test} we characterize a suitable may-testing
pre-order on concurrent objects as a variant of trace inclusion
up to rewriting.

\section{Review of concurrent programming in $\java$}\label{conc-prog-java}
We review a few basic notions of 
concurrent programming in $\java$ and provide a few
examples of concurrent objects.

We start by describing a basic method to create threads (processes) in $\java$.
$\java$ has a predefined class \s{Thread} which in turn 
has predefined methods \s{start} and \s{run}.
By invoking \s{start} on a \s{Thread} object, we invoke the \s{run} method
on it and return immediately. By default the \s{run} method does nothing;
so to have some interesting behavior one needs to create a class which
extends the \s{Thread} class and redefines the \s{run} method.
As an example, in Table \ref{ping-pong} we define a class \s{PingPong}
which extends the \s{Thread} class and redefines the \s{run} method.
What the redefined \s{run} method does is to print the \s{String} value
which constitutes the internal state of a \s{PingPong} object.
The \s{main} method creates two \s{PingPong} objects, one writing ``ping'' and 
the other writing ``pong'' and starts them in parallel.

\begin{table}

{\footnotesize
\begin{verbatim}
public class PingPong extends Thread {
    private String word;
    public PingPong(String w){word = w;}                     //constructor
    public void run(){for (;;){System.out.print(word+" ");}} //redefine run
    public static void main (String[] args){
        (new PingPong("ping")).start(); 
        (new PingPong("pong")).start();} }
\end{verbatim}
}
\caption{Creating threads in $\java$}\label{ping-pong}
\end{table}

Threads running in parallel may share a common object.
For instance, suppose the shared object is a {\em counter}
with methods \s{getValue} to read the contents of the counter
and \s{increment} to increment by one its contents.
Table \ref{wrongcounter} gives a preliminary (and wrong!) description
of a counter class.
\begin{table}
{\footnotesize

\begin{verbatim}
public class WrongCounter {
    private long value = 0;
    public long getValue() { return value; }
    public void increment() {value=value+1; return; } }
\end{verbatim}
}
\caption{A wrong counter}\label{wrongcounter}
\end{table}

Suppose two threads invoke once the \s{increment} method
on an object of the \s{WrongCounter} class. Reading a value from
memory, incrementing it, and storing the result back into memory
is not an atomic operation in $\java$. As a result, it
is quite possible that the final value 
of the counter is $1$ rather than $2$. 
In fact, the $\java$ specification does not even guarantee that
reading or writing a \s{long} variable is an atomic operation.
Indeed, a \s{long} variable can be stored in two consecutive memory words
and the access to such two words does not need to be atomic.
In principle, it could happen that by reading a \s{long} variable
we get a value which is a `mix' of values written by concurrent
threads.

A simple way to solve these issues is to specify that all the methods
of the counter are \s{synchronized} as in Table \ref{syncounter}.
A thread that invokes a synchronized method on an object implicitly acquires
a lock that guarantees exclusive access to the state of the object and
releases the lock upon returning from the method. Other threads invoking
a synchronized method on the same object at the same time will be delayed. 
This is a reformulation of an older synchronization mechanism
in concurrent programming known as {\em monitor}.
This approach is obviously `correct' but it can be
inefficient. 

\begin{table}
{\footnotesize
\begin{verbatim}
              
public class SyncCounter {
    private long value = 0;
    public synchronized long getValue() {return value;}
    public synchronized void increment() {value=value+1; return;} }

\end{verbatim}
}
\caption{A synchronized counter}\label{syncounter}
\end{table}

An alternative approach consists in reducing the granularity of the operations
to that of \s{compareAndSet} operations. 
In $\java$, \s{compareAndSet} is actually 
a method which can be invoked on an object of a special `\s{Atomic}' class 
and which returns \s{true} if the comparison is successful and
\s{false} otherwise. This implementation of the counter is
described in Table \ref{CAScounter} which relies on a \s{AtomicInteger} class.

\begin{table}
{\footnotesize

\begin{verbatim}
public class CASCounter {
    private AtomicInteger value= new AtomicInteger(0);
    public int getValue(){return value.get();}
    public void increment(){
        int v;
        do { v = value.get(); }
         while (!value.compareAndSet(v, v + 1));
        return;} }
\end{verbatim}
}
\caption{A counter with \s{compareAndSet}}\label{CAScounter}
\end{table}

Notice that this time the implementation of the \s{increment} method
is significantly different. First the value of the counter is 
read and incremented and then atomically the current value of the counter 
is compared to the value read and if they are equal then the counter 
is incremented.  In case of contention, this solution relies on 
{\em busy waiting} while the previous one relies on a 
{\em context switch}. 
\footnote{The general wisdom is that if the probability
of contention is low then busy waiting may be more efficient than
context switch. In case of significant contention, 
{\em exponential back-off} is a general strategy to improve 
a busy waiting solution which is used, {\em e.g.}, to handle collisions
in the \s{Ethernet} protocol.
In our case, it consists in introducing a {\em delay} after each 
iteration of the \s{while} loop. The delay is  chosen randomly from 
an interval which increases exponentially with the number of iterations.}

Table \ref{CASStack} presents an implementation of a concurrent stack object
using \s{compareAndSet} which is known as Treiber's algorithm.
The implementation of the \s{push} and \s{pop} method follows the
approach we have already presented for the \s{increment} method.
First the methods do some speculative work on the side and then
they make it visible with a \s{compareAndSet} method provided no
interference has occurred so far. In this example,
we work on objects of the \s{AtomicReference} class and
the stack is implemented as a linked list of objects of the \s{Node} class.

\begin{table}
{\footnotesize
\begin{verbatim}
public class Node{
    int value;
    Node next;
    public Node(int v){value=v; next=this;} }  // Node constructor
public class CASStack {
    AtomicReference<Node> head = new AtomicReference<Node>();
    public void push(int v) {
        Node oldHead;
        Node newHead = new Node(v);
        do {oldHead = head.get();
            newHead.next = oldHead;
        } while (!head.compareAndSet(oldHead, newHead));}
    public int pop() {
        Node oldHead;
        Node newHead;
        do {oldHead = head.get();
            if (oldHead == null) return -1;   // -1 default value for empty stack
            newHead = oldHead.next;
        } while (!head.compareAndSet(oldHead,newHead));
        return oldHead.value;}}
\end{verbatim}}

\caption{A stack with \s{compareAndSet}}\label{CASStack}
\end{table}

It should be noticed that \s{compareAndSet} can only manage 
a single pointer atomically. More complex operations such as
inserting an element in a queue represented as a linked list,
may require the (virtual) update of more pointers at once. In this case
more sophisticated programming techniques are needed.

Treiber's algorithm guarantees a form of collective progress in the following
sense: if in a method invocation, say \s{m1},
\s{compareAndSet} returns \s{false} it must be
the case that another method invocation, say \s{m2},
has modified the \s{head} after the method \s{m1} has executed
\s{head.get()}. This means that the \s{compareAndSet} in method \s{m2}
has returned \s{true} and so the method \s{m2} exits the
\s{while} loop and is ready to return a result.
It follows that given a {\em bounded} number of methods' invocations,
no matter how the methods are scheduled, there is a guarantee that
all methods will return an answer; every internal computation
of the methods invoked on the object must terminate.
On the other hand, it is possible in principle for a  method
invocation to fail to return a value if it is
overcome by (infinitely many) other methods' invocations.

\section{A specification of a fragment of concurrent $\java$}\label{conc-java-spec}
We introduce the syntax, reduction rules, and typing rules
of a tiny imperative and concurrent object-oriented language 
(called $\cJ$) which is an 
extension of the (imperative) $\sJ$ language formalized in 
chapter \ref{obj-sec}. 

We recall that an {\em object} value is composed of the 
name of a {\em class} and a list
of references which correspond to the object's {\em fields}.
A {\em class} is a declaration where we specify how to build
and manipulate the objects of the class. 
In particular, we specify the fields of each object and the
methods that allow their manipulation.

As usual, we assume a class  \s{Object} without fields and methods.
Every other class declaration extends a previously defined class and,
in particular, we assume a class \s{Thread} which extends the \s{Object} class
with a method \s{start} with no arguments and returning 
an object of the \s{Object} class. The effect of invoking
the \s{start} method on an object of the \s{Thread} class is 
to spawn in parallel the invocation of the \s{run} method on the object
(if any).
In this section, we assume all fields are {\em modifiable} (we stick to
the imperative version of the language) and denote with 
$R$ an infinite set of references (cf. chapter \ref{obj-sec}) with elements
$r,r',\ldots$ A reference is a pointer to an object.
The {\em value} $v$ of an object has the shape:
$C(r_1,\ldots,r_n)$, for $n\geq 0$, 
where $C$ is the name of the class to which the object belongs
and  $r_1,\ldots,r_n$ are the references associated with
the modifiable fields of the object.
A {\em heap memory} $h$ is a partial function with finite domain
from references to values.

\begin{table}
{\footnotesize
\[
\begin{array}{lllll}
e &::= &\w{id} &\Alt &\mbox{(variable)} \\
      &&v &\Alt &\mbox{(value)} \\
      && {\sf new} \ C(e_1,\ldots,e_n) &\Alt &\mbox{(object generation)}  \\
      && e.f &\Alt                 &\mbox{(field read)} \\
      &&e.m(e_1,\ldots,e_n)  &\Alt &\mbox{(method invocation)} \\
      && (C)(e)  &\Alt    &\mbox{(casting)} \\
      && e.f:=e          &\Alt    &\mbox{(field write)} \\
      && e;e              &\Alt        &\mbox{(sequentialization)} \\
      && \atomic{e}       &            &\mbox{(atomicity)}
\end{array}
\]}
\caption{Expressions in $\cJ$}\label{exp-cj-tab}
\end{table}

Table \ref{exp-cj-tab} defines the syntactic category of {\em expressions} for $\cJ$.
As usual (cf. chapter \ref{obj-sec}), 
to define the reduction rules, it is convenient 
to include values in the syntactic category of expressions.
However, it is intended that expressions in a source program 
do {\em not} contain values. 
As in chapter \ref{obj-sec}, a 
{\em program} is composed of a list of class declarations and a
distinguished expression where the computation starts. 
The final value of the distinguished expression
can be taken as the output of the program. 
Among the variables, we reserve \s{this} 
to refer to the object on which a method is invoked.
Also, we reserve the names \s{start} and \s{run} for methods of objects
of the \s{Thread} class.
A {\em well-formed} program must satisfy certain conditions 
concerning fields and methods which are specified in chapter \ref{obj-sec}.

\subsection*{Reduction rules for $\cJ$}
In order to define the reduction rules, it is convenient to
introduce the syntactic category of sequential {\em evaluation contexts} 
which correspond to a {\em call-by-value, left to right} reduction
strategy and which are defined as follows:
\begin{equation}\label{ev-cxt-cj}
\begin{array}{lll}
E::= &[~] \Alt \s{new} \ C(v^*,E,e^*) \Alt E.f \Alt E.m(e^*) \Alt 
v.m(v^*,E,e^*) \Alt \\
& (C)(E) \Alt E.f:=e \Alt v.f:=E \Alt 
E;e  &\mbox{(evaluation contexts)}
\end{array}
\end{equation}
Along with the notion of evaluation context, we introduce
a notion of {\em redex}, namely an expression which (up to some type checks)
is ready to reduce.
\begin{equation}\label{redexes-cj}
\Delta::= \s{new} \ C(v^*) \Alt 
          v.f \Alt
          v.m(v^*) \Alt
          (D)(v) \Alt
          v.f:=v \Alt
          v;e \Alt
          \atomic{e} \qquad \mbox{(redexes)}
\end{equation}
The reduction of an expression involving the \s{start} method 
may produce the spawn of an expression to be evaluated in parallel.
Consequently, we consider a judgment of the shape:
\[
(e,h) \act{\mu} (e',h')
\]
where $\mu$ is a (possibly empty) finite multi-set of expressions 
of the shape $v.\s{run()}$.
Equivalently, $\mu$ can be regarded as a finite sequence of expressions
where the order is irrelevant.
As usual, we denote with $\emptyset$ the empty multi-set and moreover we write
$\arrow$ as an abbreviation for $\act{\emptyset}$.
Table \ref{java-semantics} introduces the rules for reducing
expressions and configurations (the last two rules).
The first $8$ rules are driven by the shape of the redexes specified
in grammar (\ref{redexes-cj}). In the rule for \s{atomic},
we write: 
\[
(e_1,h_1) \stackrel{\mu}{\rightarrow^{*}} (e_n,h_n) \mbox{ for }
(e_1,h_1) \act{\mu_{1}} \cdots \act{\mu_{n-1}} (e_n,h_n) \mbox{ and } 
\mu = \mu_1\union \cdots \union \mu_{n-1}~.
\]
The rule for an atomic expression may spawn several threads, but
their actual reduction may only start once the evaluation of the
atomic expression is completed. Also note that in the proposed 
semantics $\s{atomic}(v) \arrow v$.
The following rule allows to cross an evaluation context.
The last two rules, explain how to reduce a {\em configuration} 
which is a triple  $(e,\mu,h)$ composed of a main expression, 
a multi-set of secondary expressions (initially empty), 
and a heap (initially empty too). This amounts to select non-deterministically
one of the expressions and reduce it according to the rules above.

Recall that at the beginning of the computation we can assume
that the multi-set of expressions $\mu$ contains no references
and that the heap $h$ is empty.
Then the reduction rules
are supposed to maintain the following invariant:
for all reachable configurations $(e,\mu,h)$, all the references in $e, \mu$
and all the references that appear in a value in the codomain of the
heap $h$ are in the domain of definition of the heap ($\w{dom}(h)$).
This guarantees that whenever we look for a fresh reference it is enough
to pick a reference which is not in the domain of definition of the current heap.

\begin{table}
{\footnotesize\[
\begin{array}{cr}

\infer{r^* \mbox{ distinct and }\set{r^*}\inter \w{dom}(h)=\emptyset}
{(\s{new} \ C(v^*),h) \arrow (C(r^*),h[v^*/r^*])} 
&\mbox{(object generation)}\\ \\

\infer{\w{field}(C)= f_1:C_1,\ldots,f_n:C_n\quad 1\leq i \leq n}
{(C(r_1,\ldots,r_n).f_i,h) \arrow (h(r_i),h)} 
&\mbox{(field read)}\\ \\ 

\infer{\w{mbody}(m,C)=\lambda x_1\cdots x_n. e}
{(C(r^*).m(v_1,\ldots,v_n),h) \arrow ([v_1/x_1,\ldots,v_n/x_n,C(r^*)/\s{this}]e,h)}
 &\mbox{(method invocation)} \\ \\

\infer{C\leq \s{Thread}}
{(C(r^*).\s{start}(),h) \act{C(r^*).\s{run}()} (\s{Object}(),h)} 
  &\mbox{(\s{start} invocation)}\\\\

\infer{C\leq D}
{((D)(C(r^*)) ,h) \arrow (C(r^*),h)} 
&\mbox{(casting)} \\ \\

\infer{\w{field}(C)= f_1:C_1,\ldots,f_n:C_n\quad 1\leq i \leq n}
{(C(r_1,\ldots,r_n).f_i:=v,h) \arrow
(\s{Object}(),h[v/r_i])} 
 &\mbox{(field write)}\\ \\ 

\infer{}
{(v;e,h) \arrow (e,h)}
 &\mbox{(sequentialization)} \\ \\ 

\infer{(e,h) \stackrel{\mu}{\rightarrow^{*}} (v,h')}
{(\atomic{e},h) \act{\mu} (v,h')}
 &\mbox{(atomicity)} \\ \\ 

\infer{(e,h) \act{\mu} (e',h')}
{(E[e],h) \act{\mu} (E[e'],h')}
&\mbox{(evaluation context)} \\\\

\infer{(e,h) \act{\mu'} (e',h')}
{(e,\mu,h) \arrow (e', \mu \union \mu',h')}
  &\mbox{(main expression)} \\\\

\infer{(e',h) \act{\mu'} (e'',h')}
{(e,\pset{e'}\union \mu, h) \arrow (e,\pset{e''}\union \mu\union \mu',h')}
  &\mbox{(secondary expression)}

\end{array}
\]}
\caption{Small step reduction rules for $\cJ$}\label{java-semantics}
\index{$\cJ$, reduction rules}
\end{table}

\subsection*{Type system for $\cJ$}
Following the discussion in chapter \ref{obj-sec} (notably on the typing
of casting), we present a type system for $\cJ$.
As usual, a type environment $\Gamma$ has the shape $x_1:C_1,\ldots,x_n:C_n$
and we consider typing judgments of the shape: $\Gamma \Gives e:C$.
Table \ref{typing-cJ} specifies the rules to type expressions
that do not contain values or references (as source programs do).
The rules governing the typing of class declarations and programs
are those specified for the sequential fragment $\sJ$ in chapter
\ref{obj-sec}. The typed language, but for the \s{atomic} operator,
can be regarded as a fragment of the $\java$ programming language.
General, but not very efficient, methods to compile the
\s{atomic} operator have been proposed. The 
basic idea is to follow an {\em optimistic strategy} such as the
one described in chapter \ref{atomic-sec}. Unlike in the $\parimp$ language
however, in $\cJ$, and more generally in $\java$, 
it is not possible to determine statically the collection 
of object's fields which will be affected by the atomic transaction.
In first approximation, the atomic execution of an expression $e$ 
is compiled into  a {\em speculative} execution of the expression $e$ which 
maintains a list of  object's fields which are read and/or written
along with their updated values. At the end of the speculative execution,
if certain coherence conditions are met, the computation is {\em committed},
and otherwise the computation is re-started.

\begin{table}

{\footnotesize
\[
\begin{array}{c}

\infer{x:C\in \Gamma}
{\Gamma \Gives x:C} 

\qquad

\infer{\begin{array}{c}
\w{field}(C)=f_1:D_1,\ldots,f_n:D_n\\
\Gamma \Gives e_i:C_i, \quad C_i\leq D_i, \quad 1\leq i \leq n
\end{array}}
{\Gamma \Gives \s{new} \ C(e_1,\ldots,e_n):C}  \\ \\

\infer{\Gamma \Gives e: C \quad \w{field}(C) = f_1:C_1,\ldots,f_n:C_n}
{\Gamma \Gives e.f_i: C_i}

\qquad

\infer{
\begin{array}{c}
\Gamma \Gives e: C \quad \w{mtype}(m,C) =  (C_1,\ldots,C_n)\arrow D\\
\Gamma \Gives e_i:C'_i \quad C'_i\leq C_i\quad 1\leq i \leq n
\end{array}}
{\Gamma \Gives e.m(e_1,\ldots,e_n): D} \\ \\

\infer{\Gamma\Gives e:C\quad C\leq \s{Thread}}
{\Gamma \Gives e.\s{start}():\s{Object}}

\qquad \infer{\Gamma \Gives e:D}
{\Gamma \Gives (C)(e):C} 
\\ \\

\infer{
\begin{array}{c}
\Gamma \Gives e:C\quad \w{field}(C)=f_1:C_1,\ldots,f_n:C_n\\
\Gamma \Gives e':D_i\quad D_i\leq C_i
\end{array}}
{\Gamma \Gives e.f_i:=e': \s{Object}} 

\qquad

\infer{\Gamma \Gives e_1:C_1\qquad \Gamma \Gives e_2:C_2}
{\Gamma \Gives e_1;e_2:C_2} \\ \\

\infer{\Gamma \Gives e:C}
{\Gamma \Gives \atomic{e}:C}

\end{array}
\]}

\caption{Typing rules for $\cJ$ program expressions}\label{typing-cJ}
\index{$\cJ$, typing rules}
\end{table}

\begin{exercise} \label{subj-red-cj-ex}
Building on proposition \ref{main-obj-prop}, formulate and prove a
subject reduction property for the typed $\cJ$ language.
\end{exercise}

\section{Introduction to the semantics of concurrent objects}\label{intro-conc-obj}
In this section and the following, 
we analyze the semantics of {\em concurrent objects} such as
the Treiber's stack described in Table \ref{CASStack} at the level
of labelled transition systems equipped
with a CCS like action structure (see chapter \ref{ccs-sec}).

We think of a {\em process} as a kind of CCS process, possibly the
result of the compilation of a more complex process involving value
passing as described in section \ref{value-ccs-sec}.
An {\em object} is a kind of passive process
that receives {\em requests} and may provide {\em answers} to them.
Think of a request as a method invocation and of an answer as the result
of an invocation. What happens  between a request and the corresponding
answer is described by internal reductions. The analysis of 
these internal reductions can 
be quite challenging but it is {\em not} relevant to the following discussion.
A {\em client} is a process that may send requests to an object
and receive the respective answers. Moreover, a client may also
engage in observable activities
as schematized in Table \ref{interaction-table}.
An object and a client agree on an {\em interface} to exchange requests and
answers and these exchanges are private (not directly observable).
Moreover, the interaction between an object and a client
complies with the following rules:
\begin{itemize}
  \item each request comes with a unique number and this number is recalled
    in the corresponding answer ;
    an object can answer at most once to any given request.

  \item when a client sends a request it can decide either to
    wait immediately for the corresponding answer
    or to disregard forever any answer to the request.
\end{itemize}
\begin{table}
  {\footnotesize  \[
    \begin{array}{lc}

      \left.
  \begin{array}{l}
    \mbox{$n$=interaction number} \\
    \mbox{req=request}\\
    \mbox{ans=answer}
  \end{array}\right.
  
  &\begin{array}{ccc}
    \left[
      \begin{array}{ccc}
        \mbox{{\sc object}}
        &\left.\begin{array}{c}
      \stackrel{(\w{req},n)}{\longleftarrow} \\
      \stackrel{(\w{ans},n)}{\longrightarrow}
    \end{array}\right.
    &\mbox{{\sc client}}
    \end{array}
\right]    
    &\leftrightarrow
    &{\mbox{\sc observer}}
  \end{array}

\end{array}
  \]}
\caption{Request/answer interaction between object and client}\label{interaction-table}
\end{table}

\subsection*{Formalization}
To represent formally the expected behaviour of objects and clients,
we shall introduce:
\begin{itemize}

\item a kind of {\em enriched CCS action structure} to represent the
  unique number associated with each pair of request and answer.

\item a {\em co-inductive definition} of the behaviour expected by objects
  and clients.

\end{itemize}

Beyond the correct handling of requests and answers, the co-inductive
definition has to capture the intended client's behaviour.
It turns out that this can be done abstractly at the level of the
client's labelled transition system as follows:
\begin{itemize}

\item answer anticipation: an answer can anticipate any
  other action which is not the corresponding request,

\item request postponement: a request can be postponed after
  any action which is not the answer to the request.

\end{itemize}

\subsection*{Comparing objects}
Assuming this framework,
let us now move towards the problem of comparing objects' behaviours.
As we already mentioned, requests and answers are 
protected from a direct observation.
Given this, how do we compare two objects, say $p_1$ and $p_2$?
We shall assume a {\em may-testing framework} as described in section
\ref{testing-cmt-sec}. Specifically, we assume clients may commit
on a special action $w$ and write $p_1\mayleq p_2$ if
for every client $q$, $(p_1\mid q)\may$ implies $(p_2\mid q)\may$.
The main technical result characterizes this pre-order as
a pre-order based on {\em trace inclusion up to trace rewriting}.

\subsection*{An example}
We illustrate this preliminary discussion with a concrete example.
Suppose we want to analyze objects that are supposed to
implement a multiple readers, multiple writers, boolean variable.
We fix the interface between object and client as follows:

\begin{itemize}

\item the labels $r,w_0,w_1$ represent the
  requests : read, write $0$, and write $1$, respectively.

\item the labels $r_0,r_1,w$ represent the answers:
  value read $0$, value read $1$, and write completed, respectively.

\end{itemize}

We recall that in standard CCS, an action over some set of labels $A$
is either a distinct internal action $\tau$ or an action $a\in A$
or a co-action $\ol{a}$. We shall introduce an {\em enriched} framework,
where an action distinct
by the internal action is actually a pair $(a,n)$ where $a\in A$ and
$n\in \Nat$ is a natural number. Each action $(a,n)$ has a co-action
$\ol{(a,n)}$. We use the enrichment to describe the expected properties
of objects and their clients. 

Returning to the boolean variable example and taking the object's
viewpoint, a request is an action $(a,n)$ where
$a\in \set{r,w_0,w_1}$, 
and an answer is a co-action $\ol{(a,n)}$
where $a\in \set{r_0,r_1,w}$. So in the modelling,
the action-co-action mechanism
is instrumental to the distinction between object and client.

In Table \ref{seq-proc}, we describe the labelled
transitions of a first object that answers
immediately each request. An object that exhibits this
behaviour is often called {\em sequential} and it is taken
as the specification to which {\em concurrent} 
implementations should comply.

Let $N\subseteq \Nat$ be a set which contains
the numbers still available for an interaction
with the client.
The rules describe a labelled transition system over
the enriched action structure. One can think of the $A_j(\ldots)$
as parametric processes; once the parameters are given we have a process
and the rules in Table \ref{seq-proc} describe its labelled transitions.
\begin{table}
{\footnotesize\[
\begin{array}{llll}

  A_1(x,N) &\act{(r,n)} &A_2(x,N\minus \set{n},n)  &(n\in N)\\
  A_1(x,N) &\act{(w_x,n)} &A_3(x,N\minus \set{n},n)  &(n\in N)\\
  A_1(x,N) &\act{(w_{\ol{x}},n)} &A_3(\ol{x},N\minus \set{n},n)  &(n\in N)\\
  A_2(x,N,n) &\act{\ol{(r_x,n)}} &A_1(x,N)  &(n\notin N)\\
  A_3(x,N,n) &\act{\ol{(w,n)}}   &A_1(x,N)  &(n\notin N)

\end{array}
\]}
\caption{A sequential object for a boolean variable}\label{seq-proc}
\end{table}

The description of the sequential object is rigorous but not
very readable. A more compact and readable notation is as follows:
\begin{equation}\label{seq-bool-var}
\begin{array}{ll}
  A(x) &=r(n).\ol{(r,n)}.A(x) + w_x(n).\ol{(w,n)}.A(x) + w_{\ol{x}}(n).\ol{(w,n)}.A(\ol{x})
\end{array}
\end{equation}
Here the sequential object is described as a recursive process that
is ready to receive three types of requests with the associated number $n$,
answers immediately the selected request, and updates the value
of the variable if needed (if $x$ is a boolean value then $\ol{x}$ is its complement). We rely on this handy notation to describe
in Table \ref{conc-bool-var} three variants of the sequential object.
In the first variant $(V_1)$, the object answers immediately each request
but it may lazily update the value of the variable (updating is presumably a costly operation).
In the second variant $(V_2)$, the object delays the answer to a request to complement the value of the variable.
Finally, the third variant $(V_3)$ combines the strategies of the first and second variant. We shall consider the relationships among these variants in example \ref{bool-var-again} once the semantic framework is in place.

\begin{table}
  {\footnotesize
    \[
    \begin{array}{lll}
  A(x) &=r(n).\ol{(r,n)}.A(x) + w_x(n).\ol{(w,n)}.A(x) + w_{\ol{x}}(n).\ol{(w,n)}.A(\ol{x})+ w_{\ol{x}}(n).\ol{(w,n)}.B(x) &(V_1)\\
 B(x) &=r(n).\ol{(r,n)}.B(x) + w_x(n).\ol{(w,n)}.A(x) + w_{\ol{x}}(n).\ol{(w,n)}.A(\ol{x}) \\\\

 A(x) &=r(n).\ol{(r,n)}.A(x) + w_x(n).\ol{(w,n)}.A(x) + w_{\ol{x}}(n).B(x,n) &(V_2)\\
 B(x,n') &=r(n).\ol{(r,n)}.B(x,n') + w_x(n).\ol{(w,n')}.\ol{(w,n)}.A(x) + w_{\ol{x}}(n).\ol{(w,n')}.\ol{(w,n)}.A(\ol{x}) \\\\

 A(x) &=r(n).\ol{(r,n)}.A(x) + w_x(n).\ol{(w,n)}.A(x) + w_{\ol{x}}(n).\ol{(w,n)}.A(\ol{x})+w_{\ol{x}}(n).B(x,n) &(V_3)\\
 B(x,n') &=r(n).\ol{(r,n)}.B(x,n') + w_x(n).\ol{(w,n')}.\ol{(w,n)}.A(x) + w_{\ol{x}}(n).\ol{(w,n')}.\ol{(w,n)}.A(\ol{x})
    \end{array}
    \]}
 
\caption{Three variants of the boolean variable}\label{conc-bool-var}
  \end{table}

We stress that there is no need to describe the {\em internal}
object's behaviour using a CCS notation as we have done for
the boolean variable. For instance, in the case of the stack
object described in section \ref{conc-prog-java}, the object's interface
$I=I_R\union I_A$ could be described as follows:
\[
\begin{array}{ll}
  I_R =\set{\s{empty},\s{pop},\s{push_v} \mid v\in V}~,
  &I_A =\set{\s{true},\s{false},\s{pop_{null}},\s{pop_v},\s{push} \mid v\in V}~,
\end{array}
\]
where $V$ is the set of values that can be stored in the stack
not containing the special element \s{null}.
The description of the internal state of the object would typically
rely on a heap memory similar to the one described in
section \ref{conc-java-spec}.
Moreover, for each kind of request there should be a formal description of
the legal sequences of atomic heap transformations possibly
leading to an answer.

\begin{exercise}\label{ex-trace-seq-obj-incl-conc-obj}
  Prove or disprove: every trace of the sequential object in Table~\ref{seq-proc}
  is a trace of the object variant
  $V_i$ in Table~\ref{conc-bool-var}, for $i=1,2,3$.
\end{exercise}

\section{Co-inductive definitions of objects and clients (*)}\label{coind-obj-client}
In this section, we formalize the hypotheses on objects and clients
and derive certain commutations properties of the clients' traces.
Let $A$ be a set of labels and $\Nat$ be the set of natural numbers.
The collection of {\em actions} \w{Act} is defined as:
\[
\w{Act}=\set{\tau} \union \set{(a,n)\mid a\in A, n\in \Nat} \union \set{
  \ol{(a,n)} \mid a\in A, n\in \Nat} \qquad\mbox{(actions)}
\]
Let $P$ be a set whose elements we call {\em processes}.
We suppose a labelled transition system $(P,\arrow,\w{Act})$ where
$\arrow \subseteq P\times \w{Act} \times P$.
We suppose that the labelled transition system (lts) has a neutral element
$0$  which has no transitions at all and it 
is closed under action prefix, 
parallel composition, and action restriction.
Namely, if $p,q\in P$, and $a$ is an action (not an internal action)
then
there are processes strongly equivalent to $a.p$, $(p\mid q)$,
and $\new{a}{p}$,
where $a.\_$, $\_\mid\_$, and $\new{a}{\_}$ are the usual
action prefix, parallel composition, and action restriction
as described in chapter \ref{ccs-sec}.
As usual, synchronization is the combination of an action with a co-action;
the label {\em and} the number in the action and the co-action have to be the
same.
Here {\em strong equivalence} can be taken to mean, {\em e.g.},
strong bisimulation (definition \ref{bisim-def}).

Among all processes, we want to define those that are objects and those
that are clients. These definitions are parametric on an interface which is
defined as a pair of disjoint sets of labels $I_R,I_A$:
\[
I_R,I_A\subseteq A, I_R\inter I_A = \emptyset \qquad{\mbox{(client-object interface)}}
\]
with the convention that the labels in $I_R$ are used to formulate requests
and the labels in $I_A$ to provide answers.
We also denote with $\w{Obs}$ the set of observable actions in a system,
{\em i.e.}, those which do not use the labels in the interface:
\[
\w{Obs}= \set{(a,n),\ol{(a,n)} \mid n\in \Nat, a\notin I_R\union I_A}
\qquad\mbox{(observable actions)}.
\]
Given an interface, we want to define the collection of objects and
the collection of clients which are compatible with this interface.
However, these definitions depend on the collection of numbers that
can be used to formulate requests and the collection of numbers that
have been used to formulate a request and can still be used to provide an
answer. So rather than defining a set of processes we actually define
a {\em family of sets} of processes indexed on a pair of disjoint
sets of natural numbers. We start with the definition of the family of
objects which is simpler. As expected, families of sets are partially
ordered by pointwise set inclusion.

\begin{definition}[objects]\label{obj-def}
We define co-inductively a family of objects:
\[
O:  2^\Nat \times 2^\Nat \arrow 2^P
\]
as the largest family 
such that for all $N_1,N_2\subseteq \Nat$:
(i) if $O(N_1,N_2)\neq \emptyset$ then $N_1\inter N_2=\emptyset$ and 
(ii) if $p\in O(N_1,N_2)$ and $p\act{\alpha}p'$ then exactly one of
  the following conditions holds:
  \begin{itemize}

  \item $\alpha=\tau$ and $p'\in O(N_1,N_2)$,

  \item $\alpha=(a,n)$, $a\in I_R$, $n\in N_1$, 
    and $p'\in O(N_1\minus\set{n},N_2\union \set{n})$,

  \item $\alpha=\ol{(a,n)}$, $a\in I_A$, $n\in N_2$
    and $p'\in O(N_1,N_2\minus\set{n})$.

\end{itemize}
\end{definition}

Notice that an object cannot perform observable actions outside its
interface.
  As usual with co-inductive definitions, to prove, say, that $p\in O(N_1,N_2)$
  it suffices to exhibit a family $F$ that satisfies the conditions
  in definition \ref{obj-def} and such that $p\in F(N_1,N_2)$.
  We practice on the following example.

\begin{example}\label{object-example}
  Let us reconsider the sequential object of Table \ref{seq-proc}.
  Recall that in this case the interface is:
$I_R = \set{r,w_0,w_1}$ and $I_A=\set{r_0,r_1,w}$.
We define for $N_1,N_2\subseteq \Nat$ the family $F$:
  \[
  \begin{array}{ll}
    F(N_1,N_2) = \left\{
    \begin{array}{ll}
      \set{A_1(x,N_1) \mid x\in \set{0,1}}  &\mbox{if }N_2=\emptyset \\
      \set{A_2(x,N_1,n), A_3(x,N_1,n) \mid x\in \set{0,1}} &\mbox{if }
      N_2=\set{n},n\notin N_1 \\
      \emptyset &\mbox{otherwise.}
    \end{array}
    \right.
    \end{array}
  \]
  It remains to check that the $5$ transition schema in Table \ref{seq-proc}
  respect the conditions in definition \ref{obj-def}.
  \end{example}

Next we turn to the co-inductive definition of a family of
clients.  It is not difficult to imagine a concrete syntax in some
value passing process calculus where a client generates a fresh number
to formulate a request and then spawns a thread that immediately waits
for an answer to the request.  However, here the name of the game is
to describe axiomatically at the level of the labelled transition
system what the expected behaviour of a client is. The good news is
that the conditions turn out to be rather simple.

\begin{definition}[clients]\label{client-def}
We define co-inductively a family of clients:
\[
C:  2^\Nat \times 2^\Nat \arrow 2^P
\]
as the largest family such that for all $N_1,N_2\subseteq \Nat$: 
(i) if $C(N_1,N_2)\neq \emptyset$ then $N_1\inter N_2=\emptyset$ and
(ii) if $p\in C(N_1,N_2)$ and $p\act{\alpha}p'$ then  the following conditions hold:
  \begin{itemize}

  \item if $\alpha\in \set{\tau}\union \w{Obs}$ then $p'\in C(N_1,N_2)$.

  \item if $\alpha=(a,n)\notin \w{Obs}$ then $a\in I_A$, $n\in N_2$, 
    and $p'\in C(N_1,N_2\minus \set{n})$.

  \item if $p'\act{(a,n)}p''$, $a\in I_A$, $n\in N_2$ and
    $\alpha\neq \ol{(b,n)}$ for $b\in I_R$ then:
    \[
    p\act{(a,n)} \cdot \act{\alpha} p'' \qquad \mbox{(answer anticipation)}
    \]
  \item if $\alpha=\ol{(a,n)}\notin \w{Obs}$ then $a\in I_R$, $n\in N_1$
    and $p'\in C(N_1\minus\set{n},N_2\union\set{n})$; moreover, we require
    that  if $p'\act{\beta} p''$ and $\beta\neq (b,n)$ for $b\in I_R$ then:
    \[
    p\act{\beta} \cdot \act{\ol{(a,n)}} p'' \qquad \mbox{(request postponement).}
    \]

\end{itemize}
\end{definition}

If we denote with $r_i$ a request with number $i$ and with $a_i$ the
corresponding answer then the two most important conditions
for a client $p$ are described by the following inference rules:
\[
\begin{array}{cc}

   \mbox{(answer anticipation)}      &\mbox{(request postponement)} \\\\
  
  \infer{p\act{\alpha} \cdot \act{a_i} \cdot p'\qquad \alpha\neq r_i}
        {p\act{a_i}\cdot \act{\alpha} p'}\qquad

        &\infer{p \act{r_i} \cdot \act{\alpha} p'\qquad \alpha \neq a_i}
        {p\act{\alpha}\cdot \act{r_{i}} p'} 

\end{array}
\]
which can be rephrased in plain English as follows:
(i) an answer can 
  anticipate any action which is not the corresponding request and
(ii)
 a request can be postponed after
 any action which is not the answer to the request.
The remaining conditions just ensure that the numbers associated with
requests and answers are used as expected.

\begin{example}\label{client-example}
In Table \ref{client-ex}, we describe a simple client for the objects introduced in the Tables \ref{seq-proc}
  and \ref{conc-bool-var}. Informally, the client repeats the following
  behaviour: (i)
  it chooses internally  whether to read or write a boolean variable,
  (ii) it sends the corresponding request to the object, (iii) then it waits for
  an answer, and (iv) finally, it performs some observable action $(a,0)$.
  \begin{table}
{\footnotesize    \[
    \begin{array}{ll}
    A_1(N) \act{\tau} A_2(N) \qquad
    A_1(N) \act{\tau} A_3(N) \qquad
    A_1(N) \act{\tau} A_4(N) \\

    A_2(N) \act{\ol{(r,n)}} A_5(N\minus\set{n},n) &n\in N \\

    A_3(N) \act{\ol{(w_0,n)}} A_6(N\minus\set{n}, n) &n\in N \\

    A_4(N) \act{\ol{(w_0,n)}} A_6(N\minus\set{n},n) &n\in N \\

      A_5(N,n) \act{(r_0,n)} A_7(N) \qquad
          A_5(N,n) \act{(r_1,n)} A_7(N)             &n\notin N \\
          A_6(N,n) \act{(w,n)} A_7(N)               &n\notin N \\
          A_7(N) \act{\ol{(a,0)}} A_1(N)
    \end{array}
    \]}
    \caption{An example of a client}\label{client-ex}
    \end{table}
As in the example \ref{object-example},  we define for $N_1,N_2\subseteq \Nat$ a family $F$:
  \[
  \begin{array}{ll}
    F(N_1,N_2) = \left\{
    \begin{array}{ll}
      \set{A_i(N_1) \mid i\in \set{1,2,3,4,7}}  &\mbox{if }N_2=\emptyset \\
      \set{A_i(N_1,n) \mid i\in \set{5,6},} &\mbox{if }
      N_2=\set{n},n\notin N_1 \\
      \emptyset &\mbox{otherwise.}
    \end{array}
    \right.
    \end{array}
  \]
  It remains to check that the transitions 
  respect the conditions in definition \ref{client-def}.
In a less formal notation that complements the one introduced
for objects, the client can be described by the
  following recursive equation:
  \[
  A = \tau.\nu n\ \ol{r}n.(r,n).A+\tau.\nu n\ \ol{w_0}n.(w,n).A+\tau.\nu n \ \ol{w_1}n.(w,n).A~.
  \]
Here, {\em e.g.}, $\nu n \ \ol{r}n.(r,n)$ denotes the action of issuing
the request $r$ along with a {\em fresh} number $n$ and $(r,n)$ denotes
the action of receiving the corresponding answer.
\end{example}

We can now state some consequences of the answer
anticipation and request postponement hypotheses on the clients' behaviours
and on the way clients can observe objects.
Suppose $s=\alpha_1\cdots \alpha_k$
is a sequence of actions.
We write $p \act{s} p'$ as an abbreviation
for $p\act{\alpha_1} \cdots \act{\alpha_k} p'$.
We also write $p\act{\alpha}\cdot \act{\beta} p'$ to mean
that there is some $p''$ such that
$p\act{\alpha} p''$ and $p''\act{\beta} p'$.

\begin{proposition}\label{client-commutation-prop}
Suppose $p\in C(N_1,N_2)$ is a client and $s=\alpha_1\cdots \alpha_k$
is a sequence of actions in $\set{\tau}\union \w{Obs}$.
 
\begin{enumerate}  

\item If $p\act{\ol{(a_1,n_1)}} \cdot \act{s} \cdot \act{\ol{(a_2,n_2)}} p'$ then
$p \act{s} \cdot \act{\ol{(a_2,n_2)}} \cdot \act{\ol{(a_1,n_1)}} p'$, where $a_1,a_2\in I_R$.
    
 \item If $p\act{(a_1,n_1)} \cdot \act{s} \cdot \act{(a_2,n_2)} p'$ then
   $p\act{(a_2,n_2)} \cdot \act{(a_1,n_1)} \cdot  \act{s} p'$, where
   $a_1,a_2\in I_A$.

  \item If $p\act{\ol{(a_1,n_1)}} \cdot \act{s} \cdot \act{(a_2,n_2)} p'$ then
    $p\act{(a_2,n_2)} \cdot \act{\ol{(a_1,n_1)}} \cdot \act{s} p'$, where
    $a_1\in I_R$, $a_2\in I_A$, and $n_1\neq n_2$.
\end{enumerate}
\end{proposition}

\begin{exercise}\label{ex-proof-client-commutation-prop}
  Prove proposition \ref{client-commutation-prop}.
  \end{exercise}

An informal way to state the proposition is as follows.
Say that a request/answer is adjacent to another request/answer
in a client's trace if all actions performed between the two are in
$\set{\tau}\union \w{Obs}$.

\begin{enumerate}

\item two adjacent requests can be commuted.

\item two adjacent answers can be commuted.

\item a request followed by an unrelated answer can be commuted.

\end{enumerate}

Notice however that in general an answer followed by a
request cannot be commuted as the answer may block the execution
of the request.
The trace of an object is composed just
of requests and answers (it does not contain observable actions).
The commutation properties of the traces of
a client induce legal commutations in the traces of an object.
We illustrate this point with an example.

\begin{example}\label{ex-obj-trace}
  We abbreviate with $r_i$ a request action
  and with $a_i$ the corresponding answer actions.
  Consider the following $5$ very simple objects (one trace suffices to
  describe their behaviour):
  \[
  \begin{array}{lllll}
    p_1 = r_1.r_2.\ol{a}_1.\ol{a}_2 ,
    &p_2 =r_2.r_1.\ol{a}_1.\ol{a}_2, 
    &p_3 = r_1.r_2.\ol{a}_2.\ol{a}_1, 
    &p_4 = r_1.r_2.\ol{a}_1.\ol{a}_2.r_3, 
    &p_5 = r_1.\ol{a}_1.r_2.\ol{a}_2 .

  \end{array}
  \]
  Let us call {\em system} the parallel composition
  of the object with a client where the names in the
  interface are suitably restricted.
  Because on the client side requests can be commuted,
  any trace the system can generate relying on object
  $p_1$ can also be generated relying on object $p_2$ (and vice versa).
  Similarly, because on the client side answers can be commuted,
  any trace the system can generate relying on object $p_1$
  can also be generated relying on object $p_3$ (and vice versa).
  Because on the client side requests can be postponed, 
  any trace the system can generate relying on object $p_4$
  can also be generated relying on object $p_1$ (and vice versa).
 Because an answer can anticipate an unrelated request, any
  trace the system can generate relying on object $p_1$ can
  also be generated relying on object $p_5$. However, this time
  the converse fails. Suppose the client $q$ has the shape:
  \[
  q=\ol{r}_1.a_1.\alpha.\ol{r}_2.a_2~,
  \]
  where $\alpha$ is some observable action. The client $q$
  can produce the trace $\alpha$ while interacting with $p_5$
  but it fails to do so while interacting with $p_1$.
  \end{example}
  
\section{May testing and trace pre-orders (*)}\label{may-trace-preorder}
We adapt the may strong commitment testing 
described in section \ref{testing-cmt-sec}.
We assume there is a special observable action $w$.
There is no co-action for this action. This action is
the device by which a client marks the success of a test
on the object. In this section, we show that trace
inclusion up to a certain trace rewriting relation
is included in the may-testing pre-order.

\begin{definition}[$\may$ pre-order]
  If $p_1,p_2$ are objects in $O(N_1,N_2)$ then we write
  $p_1\mayleq p$ if for all clients $q \in C(N_1,N_2)$,
$(p_1\mid q) \may$  implies $(p_2\mid q) \may$.
\end{definition}

\subsection*{Rewriting traces}
For a given interface $I=I_R\union I_A$,
let $t$ be an object's (finite) trace.
This is a word over the following (infinite) alphabet:
\[
\set{(a,n)\mid a\in I_R,n\in \Nat} \union
\set{\ol{(a,n)} \mid a\in I_A,n\in \Nat}~.
\]
Because of definition \ref{obj-def}, for each $n\in \Nat$ there is
at most one request and one answer associated with $n$ and if both
are present then the request must precede the answer.
As in the example \ref{ex-obj-trace} above,
let us abbreviate a request and an answer depending on the number $n$
with $r_n$ and $a_n$, respectively.
We define rules to rewrite traces. The transitive and reflexive
closure of the rewriting system defines a pre-order on traces.

\begin{definition}[trace pre-order]\label{trace-preorder}
  A rewriting system on traces is defined by
  the following (word) rewriting rules where $i\neq j$ and $t,t'$
  are arbitrary:
\[
\begin{array}{lllll}
(R_1)&t r_i r_j t' &\arrow &t r_j r_i t'   &\mbox{(request commutation)}\\
(R_2)&t a_i a_j t' &\arrow &t a_j a_i t'  &\mbox{(answer commutation)}\\
(R_3)&t r_i a_j t' &\arrow &t a_j r_i t'  &\mbox{(request-answer commutation)} \\
(R_4)&t r_i        &\arrow &t             &\mbox{(erasure of terminal request)}
\end{array}
\]
We write $t\leq t'$ if there is a possibly empty sequence of reductions
from $t$ to $t'$. In other terms, the pre-order
$\leq$ is the reflexive and transitive closure
of the relation $\arrow$.
\end{definition}

The rewriting system in definition \ref{trace-preorder} allows for
infinite reductions, however it is easy to see that in every such
reduction the last two rules ($R_3$ and $R_4$) can only be
applied a finite number of times. First, notice that the erasure rule
$(R_4)$
reduces the size of the trace while all the others keep it unchanged.
Second, define the {\em potential}  of a trace as the sum of
the {\em distances} between each request and the corresponding
answer. The distance is just the number of times one has to move right
to go from a request to the corresponding answer.  If a request has no
answer count the distance from a {\em virtual} answer located
rightmost in the trace.  Similarly, if an answer has no request count
the distance from a virtual request located leftmost in the trace.
Then the first two rules ($R_1$ and $R_2$)
keep the potential unchanged while the last two ($R_3$ and $R_4$) decrease the
potential.

\subsection*{Trace inclusion up to rewriting}
The pre-order on traces induces a pre-order on
objects as follows.

\begin{definition}[trace object pre-order]\label{trace-obj-preorder}
  Let $p_1,p_2\in O(N_1,N_2)$ be two objects. Then we write $p_1\leq_{T}^{o} p_2$ as:
  \[
  \forall t_1\in \trace{p_1} \ \exists t_2 \in \trace{p_2}\ (t_1 \leq t_2)~.
  \]
\end{definition}

\begin{remark}
  One may regard a trace as a particular simple object
  (as in example \ref{ex-obj-trace}). Beware that the pre-orders on traces
  and on objects are {\em incomparable}. For instance, consider
  $t_1=r_1a_1$, $t_2=r_1a_1r_2a_2 $, and $t_3=r_1r_2a_2a_1$. Then:
$t_1 \not\leq t_2$ and $t_1 \otleq t_2$.
 On the other hand, we have:
$t_3 \leq t_2$ and $t_3\not\leq_{T}^{o} t_2$.
This is because for any prefix $t$ of $t_2$, $r_1r_2a_2\not\leq t$.
\end{remark}

We notice the following property of clients.

\begin{proposition}\label{trace-ord-success-client-prop}
  Let $p\in O(N_1,N_2)$ be an object and $t\in \trace{p}$.
  If $q\in C(N_1,N_2)$ is a client such that $q\wact{\ol{t}}\act{w}$ and
  $t\leq s$ then $q\wact{\ol{s}}\act{w}$.
\end{proposition}

\begin{exercise}\label{proof-trace-ord-success-client-prop}
  Prove proposition \ref{trace-ord-success-client-prop}
  \end{exercise}

Trace inclusion up to rewriting is a {\em sound} criteria to infer
the may testing pre-order.

\begin{proposition}\label{trace-obs}
If  $p_1,p_2\in O(N_1,N_2)$ and  $p_{1} \otleq p_{2}$ then $p_1 \mayleq p_2$.
\end{proposition}
\Proof Suppose $q\in C(N_1,N_2)$ is a client and
$(p_1\mid q)\wact{w}\cdot$.  Then there is a trace $t$ such that
$p_1\wact{t}\cdot$ and $q\wact{\ol{t}}\cdot\act{w}$.
Since $p_1\otleq p_2$, there is a trace $s$ such that
$t\leq s$ and $p_2\wact{s} \cdot$.
By proposition \ref{trace-ord-success-client-prop}, we derive
$q\wact{\ol{s}}\cdot\act{w}$, and we conclude that $(p_2\mid q) \wact{w}$. \qed

\section{Clients to test objects (*)}\label{client-test}
The goal of this section is to show that the converse of proposition
\ref{trace-obs} holds: the may testing pre-order is included in
trace inclusion up to rewriting. Thus, given proposition \ref{trace-obs},
trace inclusion up to rewriting is a sound and {\em complete} criteria
for the may testing pre-order.
To this end, we introduce a family of clients to test objects.
Each client depends on an object's trace $t$ which can be 
regarded as an alternating sequences of
  (distinct) requests and answers:
\begin{equation}\label{alternating-trace}
  t=A_0 R_1 A_1 \cdots R_n A_n R_{n+1} \qquad\mbox{(alternating structure object's trace).}
\end{equation}
  All these sequences are supposed non-empty with the exception of the
  sequences $A_0$ and $R_{n+1}$ which can be empty. In particular, if the trace $t$ is empty then   we have $n=0$ and $A_0=R_1=\emptyset$.
  The definition of the tests is invariant under permutation of requests
  (or answers) in the same sequence. For this reason, we can regard the sequences
  $A_i,R_i$ as sets too.

For $j=1,\ldots, n+1$, we denote with $R_j^+$ the set composed
  of the requests in $R_j$ for which there is an answer in the trace $t$
  and we set $R_j^-=R_j\minus R_j^+$; these are the requests in $R_j$ for
  which there is no answer in the trace $t$.
Recalling that requests must precede answers, notice that $R_{n+1}^{+}=\emptyset$ and $R_{n+1}^{-}=R_{n+1}$.

Similarly, for $j=0,\ldots,n$, we denote with $A_j^+$ the
  set composed of the answers in $A_j$ for which there is a request in the trace
  $t$ and we set $A_j^-=A_j\minus A_j^+$; these are the answers in $A_j$ for
  which there is no request in the trace $t$. Notice that
  $A_0^-=A_0$ and $A_0^+=\emptyset$.

We denote with $D$ a set composed of $n+1$ fresh action names $d_0,d_1,\ldots,d_n$
  which are not in the interface $I_R\union I_A$. If $d$ is one of these names
  and $n$ is a natural number then $d^n.P$ stands for $(d,0).\cdots.(d,0).P$
  where the prefixed action $(d,0)$ is repeated $n$ times.

Finally, if $r\in R^+_{j}$ is a request then we denote with
$a(r)$ the corresponding answer and
with $i(a(r))$ the unique number such that $a(r) \in A_{i(a(r))}^{+}$.

\begin{table}
  {\footnotesize
    \[
\begin{array}{lll}
   C(A_0t) &=
    \Pi_{a\in \Union_{j=0,\ldots,n} A^{-}_{n}} \ a.\ol{d}_{i(a)}.0 \mid
    d_0^{\sharp A_0}.C(t)   &\mbox{(head)}\\ \\

    C(R_jA_jt) &=
    \underbrace{\Pi_{r\in R_{j}^{+}} \ \ol{r}.a(r).\ol{d}_{i(a(r))}.0}_{\textit{group 1}} \mid
    \underbrace{\Pi_{r\in R_{j}^{-}} \ \ol{r}.0}_{\textit{group 2}} \mid
    \underbrace{d_{j}^{\sharp A_{j}}.C(t)}_{\textit{group 3}}     &\mbox{(body)}\\ \\

    C(R_{n+1}) &= \Pi_{r\in R_{n+1}} \ \ol{r}.0 \mid w    &\mbox{(tail)}

  \end{array}
\]
}
\caption{Client $C(t)$ associated with an object's trace $t$}\label{trace-to-client}
\end{table}

Given a trace $t$, the definition of the test $C(t)$
is split in three parts that concern the {\em head}, the {\em body}, and
the {\em tail} of the trace, as displayed in Table\ref{trace-to-client}
where all the names $d_j$, for $j=0,\ldots,n$, are supposed to be
restricted at top level so that they cannot occur in a test's trace.
We describe informally the behaviour of the process $C(t)$.
Let us start with the {\em head}.
If $A_0\neq \emptyset$ then 
the process must receive all answers in
$A_{0}$ in some arbitrary order before moving to the body.
Next we consider the {\em body}.
The process $C(t)$ starts in parallel three groups of threads.
Group 1 issues in parallel all the requests in $R_{j}^{+}$,
immediately waits for the corresponding answer, and signals this fact on the
name $d_{i(a(r))}$ which corresponds to the answer (in general,
$j\leq i(a(r))$).
Group 2 issues in parallel all the requests in $R_{j}^{-}$
without waiting for any answer. 
Group 3 is a single thread which
checks on action $d_j$ that
all the answers in $A_j$ have been received
before moving to process $C(t)$.
Notice that the process {\em must} receive all the answers in $A_j$
but it is not obliged to issue all the
requests in $R_j$. It must only issue the requests which are necessary
to receive an answer in $A_j$. The remaining requests may well
be issued, if at all, in a later step.
Finally, we arrive at the {\em tail}.
The process issues the requests without answers
and is ready to perform the $w$ success action.

\begin{proposition}\label{char-client-prop}
  Let $p\in O(N_1,N_2)$ be an object and suppose
  $t$ is a trace in $\trace{p}$ with the alternating structure
  described in (\ref{alternating-trace}). Let $C(t)$ be the
  associated process described above. Then:

  \begin{enumerate}

  \item $C(t)$ is a client in $C(N_1,N_2)$.

  \item $C(t) \wact{\ol{t}}\cdot \act{w}$.
\end{enumerate}    
\end{proposition}

\begin{exercise}\label{proof-char-client-prop}
  Prove proposition \ref{char-client-prop}.
\end{exercise}

We can now state the key property that relates the traces of the
client $C(t)$ with the trace pre-order.

\begin{proposition}\label{main-lemma}
  Let $p\in O(N_1,N_2)$ be an object
  and suppose $t\in \trace{p}$ is an object's trace with
  the associated client $C(t)$.
  If $C(t) \wact{\ol{s}}\cdot\act{w}$ then $t\leq s$.
\end{proposition}
\Proof 
Let us denote with $\tilde{R}_{j},\tilde{A}_{j}$ 
the set of requests and the set of answers, respectively,
which are executed in the trace $s$ between
the last synchronization $d_{j-1}$ and the first synchronization on
$d_j$, for $j=1,\ldots,n$.
We stress that we are {\em not} assuming that the requests are executed
before the answers as it is the case when looking at trace $t$.
Requests and answers can be interleaved in an arbitrary way as long as they
respect the rule that a request
must come before the corresponding answer.

We also denote with $\tilde{A}_{0}$ the set of answers executed before
the last synchronization on $d_0$ and with $\tilde{R}_{n+1}$ the
set of requests executed after the last synchronization on
$d_n$.
Finally, we denote with $\tilde{R}_{n+2}$ the set of requests
which have no answer in $t$ and which are not executed at all in $s$.

We are going to show $t\leq s$ by describing a strategy to reduce
$t$ to $s$. To start with, we label every request and answer in $t$
with a number which describes its final position in $s$.
We distinguish $4$ cases.

\begin{enumerate}

\item $r\in R_{i}^{-}$ and $r\notin \Union_{j=1,\ldots,n+1} \tilde{R}_j$.
  This is a request without answer which is in $t$ but not in $s$.
  Assign to it the position $n+2$.

\item $r\in R_{i}^{-}$ and $r\in \tilde{R}_j$.
  This a request without answer which is both in $t$ and in $s$.
  Because of the structure of $C(t)$ we must have $j\geq i$.
  Assign to it the position $j$.
  
\item $a\in A_{i}^{-}$. This is an answer with no request.
  Because of the structure of $C(t)$,
  $a$ must be in $s$ too, say $a\in \tilde{A}_j$ with $j\leq i$.
  Assign to it the position $j$.

\item $r\in R_{i}^{+}$ and $a(r)\in A_{j}^{+}$. This is a request
  with corresponding answer. Because of the structure of $C(t)$ we
  must have $r\in \tilde{R}_{h}$ and $a(r)\in \tilde{A}_{k}$ with
  $i\leq h \leq k \leq j$. Assign to $r$ the position $h$ and
  to $a(r)$ the position $k$.
\end{enumerate}

Let us label every request and every answer with its intended position and
denote the positions with $\ell,\ell',\ldots$ 
Then consider the following rewriting rules where $i\neq j$ and $\ell'<\ell$:
\[
\begin{array}{llll}
(L_1)&t r_{i}^{\ell} r_{j}^{\ell'} t' &\arrow &t r_{j}^{\ell'} r_{i}^{\ell} t' \\   
(L_2)&t a_{i}^{\ell} a_{j}^{\ell'} t' &\arrow &t a_{j}^{\ell'} a_{i}^{\ell} t' \\
(L_3)&t r_i^{\ell} a_j^{\ell'} t' &\arrow &t a_{j}^{\ell'} r_{i}^{\ell} t'  
\end{array}
\]
These rules are just a {\em restricted} version of the rules $(R_1-R_3)$;
a reduction according to rules $(L_1-L_3)$ is a legal reduction
according to the rules $(R_1-R_3)$ as soon as we erase the labels.
Call {\em inversion a pair} of symbols $x^\ell,y^{\ell'}$ in the trace where $x$ and $y$
are requests or answers, $x^\ell$ precedes $y^{\ell'}$ (not necessarily immediately) and $\ell'<\ell$.
Every reduction sequence starting from a labelled trace $t$ will terminate
since each reduction reduces the total number of inversions.
We notice that if an answer $a^\ell$ is followed on the right
(not necessarily immediately followed) by a request
$r^{\ell'}$ in $t$ then necessarily $\ell<\ell'$. This means that
to order the symbols by growing position we never need to commute
an answer $a^\ell$ with a request $r^{\ell'}$ where $\ell'<\ell$.
It follows that upon termination, the trace has
the structure:
\[
t_1=\tilde{A}_0 \tilde{R}_1 \tilde{A}_1 \cdots \tilde{R}_{n} \tilde{A}_n \tilde{R}_{n+1} \tilde{R}_{n+2}~.
\]
Then we can get rid of the requests in $\tilde{R}_{n+2}$
by applying rule $(R_4)$.

As a third and final step, we have to reduce each subtrace
$\tilde{R}_{j} \tilde{A}_j$ to the corresponding subtrace in $s$.
But this is always possible: the trace where all requests come before
all answers can always be rewritten to any trace which contains the
same requests and the same answers in some legal order.
First commute the requests so that their relative positions are
the same as in $s$. Then do the same for the answers. Finally,
use induction on the
length of the trace to reduce the first trace to the second by
applying rule $(R_3)$. \qed

\begin{example}
We illustrate the strategy described in the proof above.
Suppose:
\[
\begin{array}{llll}
  t =r_1a_0r_2r_3a_2r_4a_1a_3,
  &s=a_0r_2a_2r_3a_3r_1a_1
\end{array}
\]
So looking at the alternations in $t$, we have:
\[
R_1=\set{r_1}, A_{1}=\set{a_0}, R_2=\set{r_2,r_3}, A_{2}=\set{a_2},
R_3=\set{r_4},A_{3}=\set{a_1,a_3}~.
\]
In this case, $A_0=R_4=\emptyset$ and the `synchronization barriers'
are on the names $d_1$, $d_2$, and $d_3$.
Now suppose that according to these barriers, in $s$ we have:
\[
\tilde{R}_1=\emptyset,
\tilde{A}_1=\set{a_0},
\tilde{R}_2=\set{r_2},
\tilde{A}_2=\set{a_2},
\tilde{R}_3=\set{r_1,r_3},
\tilde{A}_3=\set{a_1,a_3}~.
\]
This induces the following labelling of the elements in $t$:
\[
r_{1}^{3}a_{0}^{1}r_{2}^{2}r_{3}^{3}a_{2}^{2}r_{4}^{5}a_{1}^{3}a_{3}^{3}~.
\]
If we sort these elements with respect to the position following,
{\em e.g.}, a bubblesort algorithm we end up with the following trace:
\[
a_{0}^{1}r_{2}^{2}a_{2}^{2}r_{1}^{3}r_{3}^{3}a_{1}^{3}a_{3}^{3}r_4^{5}
\]
Now, if we drop the positions and erase the last request with rule $(R_4)$, we get:
\[
a_{0}r_{2}a_{2}r_{1}r_{3}a_{1}a_{3}~.
\]
Then we rewrite so that the relative positions of the requests and
the answers are the same as in $s$:
\[
a_{0}r_{2}a_{2}r_{3}r_{1}a_{3}a_{1}~,
\]
and finally we operate on the subtrace  $r_{3}r_{1}a_{3}a_{1}$ to make it
equal to $r_{3}a_3r_{1}a_{1}$ thus getting to $s$.
\end{example}
   
We are now ready to prove the announced result.

\begin{proposition}\label{completeness}
  Let $p_1,p_2\in O(N_1,N_2)$ be objects.
  If $p_1\mayleq p_2$ then $p_1\otleq p_2$.
\end{proposition}
\Proof Suppose $p_1 \wact{t}$ and consider the client $C(t)$.
Then $(p_1\mid C(t))\wact{w}$ by proposition \ref{char-client-prop}(2).
By the hypothesis, it follows that $(p_2 \mid C(t)) \wact{w}$.
So there must be a trace $s$ such that $p_2\wact{s}$ and
$C(t)\wact{\ol{s}}\cdot \act{w}$. But then, by proposition \ref{main-lemma},
$t\leq s$. \qed

\begin{example}[boolean variable, again]\label{bool-var-again}
  We go back to the possible implementations of a boolean
  variable object. Suppose 
  $S$ is the sequential specification described in Table \ref{seq-proc}
  and $V_i$ for $i=1,2,3$ are the three variants given in
  Table \ref{conc-bool-var}.
  We have:
  \[
  \begin{array}{lll}
    V_2 \mayleq S=_{\s{may}} V_3 \mayleq V_1
    &\mbox{but}
    &S\not\mayleq V_2~\mbox{ and } ~V_1\not\mayleq S~.
    \end{array}
  \]
  For simplicity, let us assume the initial value in the variable is $0$.
  One may argue that $V_2$ should be rejected as it is unable to
  produce traces in the sequential specification (cf. exercise \ref{ex-trace-seq-obj-incl-conc-obj}).
  For instance,
  assuming the initial value in the variable is $0$, consider
  the object's trace $t=(w_1,1)\ol{(w,1)}$.
  On the other hand, $V_1$ should be rejected as it may produce
  traces which cannot be linearized to a trace of the
  sequential specification. For instance, consider the object's trace
  $t=(w_1,1)\ol{(w,1)}(r,2)\ol{(r_0,2)}$.
  These considerations suggest may-testing equivalence or equivalently
  trace equality up to rewriting as a basic and necessary
  correctness criteria for concurrent objects.
  However, one should keep in mind that this criteria is based
  on a generalized version of trace equivalence;
  it is not difficult to imagine buggy variants of the object
  $V_3$ which are still may-equivalent to the sequential specification
  $S$ but that may diverge or produce a deadlock. For instance,
  consider the variants in Table~\ref{v45-bool-var}.
    It seems an interesting (research) exercise to characterize 
    testing pre-orders on objects
    which are sensitive to termination and deadlock (see chapter \ref{testing-sec}).
\begin{table}  
{\footnotesize    \[
    \begin{array}{lll}
 A(x) &=r(n).\ol{(r,n)}.A(x) + w_x(n).\ol{(w,n)}.A(x) + w_{\ol{x}}(n).\ol{(w,n)}.A(\ol{x})+w_{\ol{x}}(n).B(x,n) &(V_4)\\
 B(x,n') &=r(n).\ol{(r,n)}.B(x,n') + w_x(n).\ol{(w,n')}.\ol{(w,n)}.A(x) +
 \tau.B(x,n') \\ \\

  A(x) &=r(n).\ol{(r,n)}.A(x) + w_x(n).\ol{(w,n)}.A(x) + w_{\ol{x}}(n).\ol{(w,n)}.A(\ol{x})+w_{\ol{x}}(n).B(x,n) &(V_5)\\
 B(x,n') &=r(n).\ol{(r,n)}.B(x,n') + w_x(n).\ol{(w,n')}.\ol{(w,n)}.A(x) 
    \end{array}
    \]}
    \caption{Two more (buggy) variants of the sequential object for a boolean variable}
    \label{v45-bool-var}
\end{table}    
\end{example}

\section{Summary and references}
We have reviewed some basic notions of concurrent programming in
$\java$ and formalized the reduction and typing rules of a tiny
fragment of it.  $\java$'s synchronization builds on the notion of
{\em monitor} which is described in
\cite{DBLP:journals/cacm/Hoare74,DBLP:journals/tse/Hansen75}.  The
notion of {\em linearizability} is introduced in
\cite{DBLP:journals/toplas/HerlihyW90}.
The introduction of a universal construction to
transform any `sequential' object into a `concurrent' one without
introducing locks is due to \cite{DBLP:journals/toplas/Herlihy91}.
The article \cite{DBLP:journals/siamcomp/GibbonsK97} analyzes
the complexity of determining if a trace is linearizable and
the article \cite{DBLP:journals/csur/DongolD15} surveys various proof
techniques for verifying linearizability.
The notion of linearizability is connected to a
notion of observational refinement  in
\cite{DBLP:journals/tcs/FilipovicORY10}.
In this chapter, we have tried to cast this connection in the general
framework of labelled transition systems with synchronization.

\appendix

\addcontentsline{toc}{chapter}{Bibliography}
\markboth{{\it Bibliography}}{{\it Bibliography}}
{\footnotesize \bibliography{conc} }

\addcontentsline{toc}{chapter}{Index}
\markboth{{\it Index}}{{\it Index}}
{\footnotesize  \input{opmeth.ind} }

\end{document}